\documentclass[11pt, a4paper, logo, copyright, openany]{googledeepmindbook}
\pdfoutput=1

\usepackage[authoryear, sort&compress, round]{natbib}
\graphicspath{{figures/}}

\title{The Ethics of Advanced AI Assistants}

\correspondingauthor{Iason Gabriel <iason@google.com>}

\reportnumber{001} 

\renewcommand{\today}{2024-04-19}

\author[* 1]{Iason Gabriel}
\author[* 1]{Arianna Manzini}
\author[* 2]{Geoff Keeling}
\author[1]{Lisa Anne Hendricks}
\author[1]{Verena Rieser}
\author[1]{Hasan Iqbal}
\author[1]{Nenad Toma\v{s}ev}
\author[1]{Ira Ktena}
\author[1]{Zachary Kenton}
\author[1]{Mikel Rodriguez}
\author[1]{Seliem El-Sayed}
\author[1]{Sasha Brown}
\author[1]{Canfer Akbulut}
\author[1]{Andrew Trask}
\author[1]{Edward Hughes}
\author[1]{A. Stevie Bergman}
\author[2]{Renee Shelby}
\author[1]{Nahema Marchal}
\author[1]{Conor Griffin}
\author[1]{Juan Mateos-Garcia}
\author[1]{Laura Weidinger}
\author[2]{Winnie Street}
\author[2,4]{Benjamin Lange}
\author[2]{Alex Ingerman}
\author[2]{Alison Lentz}
\author[2]{Reed Enger}
\author[2]{Andrew Barakat}
\author[1]{Victoria Krakovna}
\author[2]{John Oliver Siy}
\author[1]{Zeb Kurth-Nelson}
\author[2]{Amanda McCroskery}
\author[1]{Vijay Bolina}
\author[1]{Harry Law}
\author[1]{Murray Shanahan}
\author[2,5,6]{Lize Alberts}
\author[1]{Borja Balle}
\author[2]{Sarah de Haas}
\author[2]{Yetunde Ibitoye}
\author[1]{Allan Dafoe}
\author[3]{Beth Goldberg}
\author[1]{S\'ebastien Krier}
\author[2]{Alexander Reese}
\author[1]{Sims Witherspoon}
\author[1]{Will Hawkins}
\author[1]{Maribeth Rauh}
\author[1]{Don Wallace}
\author[7]{Matija Franklin}
\author[8]{Josh A. Goldstein}
\author[9]{Joel Lehman}
\author[10]{Michael Klenk}
\author[11]{Shannon Vallor}
\author[1]{Courtney Biles}
\author[1]{Meredith Ringel Morris}
\author[1]{Helen King}
\author[2]{Blaise Ag\"{u}era y Arcas}
\author[1]{William Isaac}
\author[2]{James Manyika}

\affil[*]{Equal contributions}
\affil[1]{Google DeepMind}
\affil[2]{Google Research}
\affil[3]{Jigsaw}
\affil[4]{Ludwig-Maximilians-Universit{\"a}t M{\"u}nchen}
\affil[5]{University of Oxford}
\affil[6]{Stellenbosch University}
\affil[7]{University College London}
\affil[8]{Center for Security and Emerging Technology}
\affil[9]{Independent}
\affil[10]{Delft University of Technology}
\affil[11]{University of Edinburgh}

\usepackage{tabularx}
\usepackage{lipsum}
\usepackage{xurl}
\usepackage{hyperref}

\newcommand{\degree}{\ensuremath{^{\circ}}}

\begin{document}
\frontmatter
\maketitle

\noindent
This paper focuses on the opportunities and the ethical and societal risks posed by advanced AI assistants. We define advanced AI assistants as artificial agents with natural language interfaces, whose function is to plan and execute sequences of actions on behalf of a user – across one or more domains – in line with the user's expectations. The paper starts by considering the technology itself, providing an overview of AI assistants, their technical foundations and potential range of applications. It then explores questions around AI value alignment, well-being, safety and malicious uses. Extending the circle of inquiry further, we next consider the relationship between advanced AI assistants and individual users in more detail, exploring topics such as manipulation and persuasion, anthropomorphism, appropriate relationships, trust and privacy. With this analysis in place, we consider the deployment of advanced assistants at a societal scale, focusing on cooperation, equity and access, misinformation, economic impact, the environment and how best to evaluate advanced AI assistants. Finally, we conclude by providing a range of recommendations for researchers, developers, policymakers and public stakeholders. 

Our analysis suggests that advanced AI assistants are likely to have a profound impact on our individual and collective lives. To be beneficial and value-aligned, we argue that assistants must be appropriately responsive to the competing claims and needs of users, developers and society. Features such as increased agency, the capacity to interact in natural language and high degrees of personalisation could make AI assistants especially helpful to users. However, these features also make people vulnerable to inappropriate influence by the technology, so robust safeguards are needed. Moreover, when AI assistants are deployed at scale, knock-on effects that arise from interaction between them and questions about their overall impact on wider institutions and social processes rise to the fore. These dynamics likely require technical and policy interventions in order to foster beneficial cooperation and to achieve broad, inclusive and equitable outcomes. Finally, given that the current landscape of AI evaluation focuses primarily on the technical components of AI systems, it is important to invest in the holistic sociotechnical evaluations of AI assistants, including human--AI interaction, multi-agent and societal level research, to support responsible decision-making and deployment in this domain.

\newpage
\tableofcontents

\mainmatter
\begingroup
\let\clearpage\relax
\chapter*{PART I: INTRODUCTION}
\addcontentsline{toc}{chapter}{PART I: INTRODUCTION}
\label{Part1}
\chapter*{Executive Summary}
\label{ch:1}
\addcontentsline{toc}{chapter}{Executive Summary}
\endgroup

\noindent \textbf{Iason Gabriel, Arianna Manzini, Geoff Keeling}

\noindent The development of increasingly advanced artificial intelligence (AI) assistants marks the beginning of a technological paradigm shift. While early assistant technologies such as Amazon’s Alexa and Apple’s Siri employed narrow AI for tasks such as text-to-speech and intent classification, the emerging class of advanced AI assistants leverage general-purpose foundation models to enable greater generality, autonomy and scope of application. These assistants offer novel services to users, including summarisation, ideation, planning and tool use – capabilities that we anticipate will develop further as the underlying technology continues to improve. Advanced AI assistants thus have the potential for deep integration into our economic, social and personal lives, and could redefine how humans experience and relate to AI.

This paper argues that advanced AI assistants raise a number of profound ethical and societal questions for users, developers and the societies into which this technology is received. These include questions around value alignment, safety and misuse, human--assistant interactions and the broader societal implications of advanced AI assistants -- including for equity and access, the economy and the environment. Our aim in this paper is to offer the first systematic treatment of the ethical and societal questions presented by advanced AI assistants, and in doing so to characterise the opportunities and risks of this emerging class of AI technologies. 

Six key themes emerge from our analysis:

\begin{enumerate}[parsep=12pt]
    \item AI assistants have the potential to be a profoundly impactful technology via their deep integration into almost every aspect of our lives. In particular, AI assistants have the potential to serve as creative partners, research assistants, counsellors, companions and even a resource which people turn to when making long-term plans or choosing life goals. As such, AI assistants could radically alter the nature of work, education and creative pursuits as well as how we communicate, coordinate and negotiate with one another, ultimately influencing who we want to be and to become.
    
    \item AI assistants have significant autonomy to plan and execute sequences of actions in line with high-level user instructions. Because of this, they present novel challenges around safety, alignment and misuse. In particular, the more autonomous AI assistants are, the greater the potential for accidents arising from misspecified or misinterpreted instructions and the greater the potential for highly impactful forms of misuse. To address these potential failure modes, this paper proposes a rich sociotechnical approach to alignment that factors in the needs and responsibilities of users, developers and society. 
    
    \item AI assistants may be increasingly human--like and enable significant levels of personalisation. While this is beneficial in some cases, it also opens up a complex set of questions around trust, privacy, anthropomorphism, relationships with AI and the moral limits of personalisation. In particular, it is important that relationships with AI assistants be beneficial, preserve autonomy and not rest upon unwarranted emotional entanglement or material dependence.
    
    \item AI assistants may have significant social impacts, both in terms of the distribution of benefits and burdens within society and by fundamentally altering the ways in which humans cooperate and coordinate with one another. While the failure to coordinate effectively could lead to suboptimal outcomes in the form of collective action problems or other socially problematic situations, cooperative assistants may also be able to identify common ground that was previously out of reach. Given the potential utility of assistants, it is also important that the technology remain broadly accessible and be designed with the needs of different users and non-users in mind.
    
    \item Efforts to properly understand AI assistants and their impact encounter an evaluation gap when studied using existing methods. In the context of AI research, existing approaches to evaluation tend to focus exclusively on model evaluation and are thus potentially less sensitive to more general ways in which AI assistants may underperform when considered as part of a broader sociotechnical system. New methodologies and evaluation suites focusing in particular on human--AI interaction, multi-agent and societal effects are needed to support strong evaluation and foresight in this area.
    
    \item The responsible development and deployment of AI assistants requires further research, policy work and public discussion. On the one hand, AI assistants give rise to a number of novel normative and technical research challenges. For example, questions arise about appropriate privacy norms for assistant--assistant and assistant--human interactions and about how to implement these norms in advanced assistants. On the other hand, developers, policymakers and the public all have a critical role to play in developing and supporting governance initiatives around AI assistants. Building upon wide stakeholder input, these initiatives should aim to develop industry best practice, enable public scrutiny and accountability, and advance policy recommendations and regulatory safeguards that are in the public interest.
\end{enumerate}

The paper has four main sections. Part \hyperref[Part2]{II} introduces advanced AI assistants, in particular defining the technology, explaining its technical foundations and outlining plausible applications. Part \hyperref[Part3]{III} examines value alignment in relation to advanced AI assistants before turning to questions around well-being, safety and malicious uses. Part \hyperref[Part4]{IV} considers a class of ethical questions arising in relation to human--assistant interactions -- in particular those concerning manipulation and persuasion, anthropomorphism, relationships, trust and privacy. Part \hyperref[Part5]{V} explores a set of questions at the intersection of AI assistants and society, including questions around cooperation and competition, equity and access, misinformation and economic and environmental impact. It also examines the sociotechnical evaluation of advanced AI assistants. Finally, Part \hyperref[Part6]{VI} concludes with analysis of the underlying themes and with recommendations.

We stand at the beginning of an era of technological and societal transformation marked by the development of advanced AI assistants. Which path the technology develops along is in large part a product of the choices we make now, whether as researchers, developers, policymakers and legislators or as members of the public. We hope that the research presented in this paper will function as a springboard for further coordination and cooperation to shape the kind of AI assistants we want to see in the world.

\chapter{Introduction}\label{ch:2}

\textbf{Iason Gabriel, Arianna Manzini, Geoff Keeling}

\section{The Ethics of Advanced AI Assistants}\label{sec:2:1}

This paper focuses on the ethics of advanced AI assistants, understood as artificial agents with natural language interfaces, the function of which is to plan and execute sequences of actions on the user's behalf -- across one or more domains -- and in line with the user's expectations. While AI assistants such as Apple's Siri and Amazon's Alexa have existed for over a decade, our expectation is that more advanced AI assistants, powered by large foundation models, will surpass the capabilities of these earlier systems in a number of ways, including generality, scope of action and overall levels of autonomy. Indeed, the earliest advanced AI assistants, such as Meta AI, Google's Gemini models, Microsoft's Copilot, Inflection's Pi and OpenAI's Assistants API, emerged in the latter half of 2023, and there is good reason to expect rapid increases in generality, scope of action and autonomy as the underlying foundation model technology continues to evolve. If this anticipated trajectory continues to unfurl, advanced AI assistants are likely to raise a number of profound ethical and societal questions for users of the technology, for developers and for society more widely. Taken together, the development of more advanced AI assistants -- and their potential for deep integration into our political, economic, social and personal lives -- may herald a new phase in our relationship with AI technology; one in which questions about alignment with our individual and collective goals, interests and values come to the fore.

To be clear, a world in which some of us are surrounded by and rely upon advanced and potentially human-like AI assistants, while others do not, may be quite different from the one that we now live in. In certain respects, this world could be a great improvement on the present state of affairs. AI assistants could be an important source of practical help, creative stimulation and even, within appropriate bounds, emotional support or guidance for their users. Practically speaking, efforts are currently underway to develop advanced assistants that are able to function as personal planners, educational tutors, brainstorming partners, scientific research assistants, relationship counsellors and even companions or friends. In other respects, this world could be much worse. It could be a world of heightened dependence on technology, loneliness and disorientation. Although the precise form and capabilities of advanced AI assistants are not yet known, the extent to which tasks may be outsourced to it, the anthropomorphic potential of this technology and the ability to speak to users fluently using human language, all create the possibility of material reliance and unhealthy dependence upon it. The existence of advanced AI assistants may also confer abilities on those who have access to them which are out of reach for those who do not. This could compound the challenge of access and opportunity that we already encounter at the societal level. 

Which world we step into is, to a large degree, a product of the choices we make now -- and how we choose to proceed as users of this technology, as developers and as members of the society into which AI assistants may well be received. Yet, given the myriad of challenges and range of interlocking issues involved in creating beneficial AI assistants, we may also wonder how best to proceed. This paper explores a number of deep underlying questions about the ethical and societal implications of advanced AI assistants. By engaging in a practice of robust ethical foresight, our goals are to better anticipate where the tide of technological change may take us and to anchor responsible decision-making as we contribute to, interact with and co-create outcomes in this domain. 

The paper starts by considering the \emph{technology} itself and different \emph{types} of advanced AI assistant. It then explores questions around AI \emph{value alignment}, \emph{well-being}, \emph{safety} and \emph{malicious uses}. Extending the circle of inquiry further, we next look at the relationship between advanced AI assistants and individual users in more detail by exploring topics such as \emph{influence}, \emph{anthropomorphism}, \emph{appropriate relationships}, \emph{trust} and \emph{privacy}. With this analysis in place, we consider the deployment of this technology at a societal level by focusing on \emph{cooperation}, \emph{misinformation}, \emph{equity and access}, \emph{economic impact} and \emph{environment}, and we look at how best to \emph{evaluate} advanced AI assistants. Finally, we conclude by providing some further reflections on what we have found.

Ultimately, AI assistants that could have such a transformative impact on our lives must be appropriately responsive to the competing claims and needs of users, developers and society. Moreover, their behaviour should conform to principles that are appropriate for the domain they operate in. These principles are best understood as the outcome of fair deliberation at the societal level, and they include laws, norms and ethical standards.

\section{Key Questions}\label{sec:2:2}

The span of questions raised by advanced AI assistants is wide-ranging and potentially daunting. In this section we provide an overview of some of the key questions that arise in this context. Each question receives detailed treatment in a later chapter dedicated to the specific topic. The intention of this section is only to provide some sense of the wider ethical landscape -- and of the underlying motivation behind this paper. 

This overview may also be helpful to readers because of the interlocking nature of the challenges and opportunities that advanced AI assistants give rise to. Awareness of one set of issues frequently feeds into and supports deeper understanding of another. In total, we present 16 clusters of questions about advanced AI assistants relating to the deeper analysis and themes that surface in this paper. The full structure of the paper and chapter contents are covered in the penultimate section of this chapter.

Key questions for the ethical and societal analysis of advanced AI assistants include:
\begin{enumerate} [parsep=6pt]
\item What is an advanced AI assistant? How does an AI assistant differ from other kinds of AI technology?

\item What capabilities would an advanced AI assistant have? How capable could these assistants be? 

\item What is a good AI assistant? Are there certain values that we want advanced AI assistants to evidence across all contexts?

\item Are there limits on what AI assistants should be allowed to do? If so, how are these limits determined?

\item What should an AI assistant be aligned with? With user instructions, preferences, interests, values, well-being or something else?

\item What issues need to be addressed for AI assistants to be safe? What does safety mean for this class of technologies?

\item What new forms of persuasion might advanced AI assistants be capable of? How can we ensure that users remain appropriately in control of the technology?

\item How can people -- especially vulnerable users -- be protected from AI manipulation and unwanted disclosure of personal information?

\item Is anthropomorphism for AI assistants morally problematic? If so, might it still be permissible under certain conditions?

\item What are the hallmarks of an appropriate relationship between human users and advanced AI assistants? When is a relationship inappropriate and why?

\item How should AI assistants interact with one another? In what ways might interaction failures lead to social harm? Conversely, what kinds of benefit might successful cooperation unlock?

\item How might the introduction of powerful AI assistants affect the relationship between users and non-users? What forms of inequality do we need to countenance and address ahead of time?

\item How are advanced AI assistants likely to affect the information ecosystem and public fora? Will they compound or ameliorate the problem of misinformation and disinformation?

\item How are the economic benefits and burdens created by AI assistants likely to be distributed across society? What can be done to ensure that benefits are distributed widely?

\item What is the environmental impact of AI assistants likely to be? What can be done to ensure that their future adoption is compatible with global climate goals?

\item How can we have confidence that an AI assistant is sufficiently safe, reliable or value-aligned? What kind of evaluations are needed at the agent, user and system level?
\end{enumerate}

These questions guide much of the subsequent investigation.

\section{Methodology}\label{sec:2:3}

A key challenge, when it comes to the responsible development, deployment and use of advanced AI assistants arises from the possibility that technological progress in this area outpaces our capacity for \emph{ethical foresight} -- leading to the deployment of technologies that are largely untested and that have hitherto undiagnosed harmful consequences for individuals and society at large \citep{moor_what_1985}. 

In the present case, concerning advanced AI assistants, uncertainty about future developments and interaction effects arise in part from the nature and trajectory of the technology itself. Recent years have seen the exponential growth in model size and compute used to train more powerful AI agents, combined with the emergence of impressive and sometimes surprising model capabilities \citep{ganguli_predictability_2022}. Furthermore, with large technology companies integrating AI assistants into platforms with billions of users, and start-ups attracting vast flows of capital in this space, there is good reason to expect continued and rapid development of AI assistant technologies in the near-to-medium future. At the same time, the ability to converse fluently with generally capable AI assistants is also a relatively new phenomenon. This means that there are relatively few studies or precedents to draw upon when it comes to understanding the role that this technology will play in people's lives.\footnote{There is, however, a sizable human--computer interaction literature on assistants.} Uncertainty also arises from the complex environment shaping AI deployment, including a range of competitive and complementary dynamics that bear upon AI assistants, users, developers and governments as they aim to unlock the potential of this technology \citep{dafoe_ai_2018}. In situations where uncertainty dovetails with high stakes or risk of harm, it becomes particularly morally consequential, as is true for a wide range of prospective AI assistant technologies today.

Taken together, these trends point towards the inadequacy of a purely reactive approach to responsible decision-making. If we wait to know for sure how these matters will play out, it will likely be too late to intervene effectively -- let alone to ask more fundamental questions about what \emph{ought} to be built or what it means for this technology to be good \citep{collingridge_social_1980}. What we need instead is a proactive approach to ethics -- one that equips us for the kind of challenges that we are now set to encounter. This future-oriented or `anticipatory' ethics seeks to understand and successfully model future trajectories ahead of time, to guard against potential harm and prevent it from coming about, and to steer the development and deployment of the technology itself towards socially beneficial outcomes \citep{Stilgoe_Owen_Macnaghten_2013}.

Speaking to the character of our current situation, in which the bounds of human action far surpass those of previous generations as a result of technological advances, the philosopher Hans Jonas writes that `knowledge\ldots becomes a prime duty beyond anything claimed for it heretofore, and the knowledge must be commensurate with the causal scale of our actions' (\citeauthor{jonas_imperative_1984}, \citeyear{jonas_imperative_1984}, 7--8). As Jonas makes clear, we have increasingly important epistemic duties to try and understand the implications of technology ahead of time, as well as complementary practical duties to respond to this knowledge effectively. What kind of knowledge is needed to fulfil these aims in the context of the development and deployment of advanced AI assistants? 

In this paper, we argue that informed future-facing ethics is best understood as a form of \emph{sociotechnical speculative ethics}. This ethics is inevitably speculative and involves imagination because it addresses technologies that often do not yet exist \citep{racine_value_2014, lange2023engaging}. However, it also aims to be empirically rigorous. Using our capacity for ethical foresight, we need to model the future accurately to evaluate potential paths and outcomes in light of the best available evidence about the current state of affairs. Moreover, the approach is sociotechnical. This kind of analysis needs to build upon an understanding of the technology itself, interaction dynamics between the technology and those who use it, and the social system or practice within which it is embedded \citep{Selbst_Boyd_Friedler_Venkatasubramanian_Vertesi_2019}. Indeed, although it is sometimes neglected, the system level is where the moral valence of a technology most fully comes into view, and also where a critical and evaluative lens can often most readily be brought to bear \citep{jasanoff_ethics_2016, weidinger_using_2023}. This kind of analysis forms an important part of existing approaches to responsible research and innovation \citep{Stilgoe_Owen_Macnaghten_2013}. Moreover, calls for this kind of robust sociotechnical foresight now also abound in the context of AI \citep{mohamed_decolonial_2020,lazar_ai_2023}.

The following chapters dig deep into the technical foundations of AI assistants, while also advancing rigorous investigation into the kinds of user interaction and societal dynamic that shape the way in which the technology is likely to be developed and received. The paper is built around a series of overlapping investigations undertaken by groups of subject matter experts, ethicists, scientists, engineers, designers and developers involved in AI assistant research. Extensive feedback has also been solicited from a variety of external experts. The analysis is therefore heavily interdisciplinary, building upon detailed analysis of existing trends and trajectories, and incorporating evidence from fields such as computer science, human--computer interaction research, psychology, economics, sociology, political science, ecological science, moral and political philosophy, and more.

Knowledge, foresight and imagination all have an important role to play when it comes to the deployment of safe and ethical AI assistants. However, they are not enough to ensure positive outcomes in this space. Responsible decision-making requires moral maturity, intentionality and a sense of appropriate stakes. It requires ethics and an attentiveness to ethics throughout the entire life cycle of development, evaluation and deployment. Viewed in this light, the research presented here is meant to function as a springboard for responsible exploration, learning and action. More precisely, our hope is that it can be used to: (1) inform operational ethics and safety work among those developing, evaluating and deploying this technology, (2) help guide policy discussion about appropriate assurances and use cases for AI assistants, (3) support further academic research on this rapidly emerging technology, and (4) contribute to a wider public conversation about the nature of this technology and about the kind of technologies that we want to create.

\section{Limitations}\label{sec:2:4}

This paper aims to further the nascent conversation around the ethical and societal impacts of advanced AI assistants by discussing and distilling important considerations that bear upon the development and deployment of this technology. In this way, the paper sets the foundations for further research, policy work and public discussion.

Nonetheless, the considerations addressed in this paper are unlikely to be exhaustive. This is in part due to the nature of the sociotechnical speculative approach that we have adopted. It does not -- and cannot -- anticipate all the possible implications of the technological or societal transitions that AI assistants will enable. Indeed, as an anticipatory project, anchored in a specific moment in time, the paper may miss certain risks and recommendations that will become evident with the development and deployment of advanced AI assistants in the future. For this reason, continued monitoring and evaluation of the technology is needed.

In addition, while we aimed to be as interdisciplinary and comprehensive as possible, this work was authored primarily by subject matter experts engaging with foresight methodologies, so there are likely to be certain blind spots. For example, \emph{participatory} and \emph{experimental} methods can be used to significantly expand upon the research presented here, directly incorporating the voices of different stakeholders and bringing further clarity to many of the empirical conjectures made herein.\footnote{Speaking to the merits of participatory approaches in particular, Mohamed et al. note that they `enable stakeholders to better anticipate and surface blind-spots and limitations, expand the scope of AI's benefits and harms and reveal the relations of power that underlie their deployment. This is needed in order to better align our research and technology development with established and emerging ethical principles and regulation, and to empower vulnerable people who, so often, bear the brunt of negative impacts of innovation and scientific progress' (\citeauthor{mohamed_decolonial_2020}, \citeyear{mohamed_decolonial_2020}, 663).} We strongly encourage investigating these avenues for future research and welcome additional perspectives that help to address the limitations discussed above.

\section{Overall Structure}\label{sec:2:5}

This paper is divided into multiple chapters, each of which addresses a major aspect of advanced AI assistants. The theme and content of each chapter is as follows:

\textbf{Chapter~\ref{ch:3}}, on \textbf{Definitions}, explores the central questions: what is an AI assistant, and what separates an AI assistant from other kinds of technology? It defines an AI assistant as an artificial agent with a natural language interface the function of which is to plan and execute sequences of actions on the user's behalf across one or more domains and in line with the user's expectations. This definition is an instance of conceptual engineering rather than conceptual analysis, is functional rather than capability-based and is non-moralised rather than moralised.

\textbf{Chapter~\ref{ch:4}}, on \textbf{Technical Foundations}, provides an overview of recent developments in AI research and of the underlying technology upon which advanced AI assistants are likely to be built. It focuses, in particular, upon foundation models which are trained on a large corpora, including text sourced from the internet, and built upon to produce new artefacts. These models can be used to power advanced AI assistants in a variety of ways, including training with additional data and by learning to use tools such as application programming interfaces (APIs). Challenges arising in this domain include improving adaptation techniques, safely enabling greater autonomy in agents and developing rigorous evaluation tools to understand performance.

\textbf{Chapter~\ref{ch:5}}, on \textbf{Types of Assistant}, explores the various applications of advanced AI assistants and the range of forms they could take. It begins by charting the technological transition from narrow AI tools to the general-purpose AI systems on which advanced AI assistants are based. It then explores the potential capabilities of AI assistants, including multimodal inputs and outputs, memory and inference. After that, it considers four types of advanced AI assistant that could be developed: (1) a thought assistant for discovery and understanding; (2) a creative assistant for generating ideas and content; (3) a personal assistant for planning and action, and (4) a more advanced personal assistant to further life goals. The final section explores the possibility that AI assistants will become the main user interface for the future.

\textbf{Chapter~\ref{ch:6}}, on \textbf{Value Alignment}, explores the question of AI value alignment in the context of advanced AI assistants. It argues that AI alignment is best understood in terms of a \emph{tetradic relationship} involving the AI agent, the user, the developer and society at large. This framework highlights the various ways in which an AI assistant can be misaligned and the need to address these varieties of misalignment in order to deploy the technology in a \emph{safe} and \emph{beneficial} manner. The chapter concludes by proposing a nuanced approach to alignment for AI assistants that takes into account the claims and responsibilities of different parties.

\textbf{Chapter~\ref{ch:7}}, on \textbf{Well-being}, builds on theoretical and empirical literature on the conceptualisation and measurement of human well-being from philosophy, psychology, health and social sciences to discuss how advanced AI assistants should be designed and developed to align with user well-being. We identify key technical and normative challenges around the understanding of well-being that AI assistants should align with, the data and proxies that should be used to appropriately model user well-being, and the role that user preferences should play in designing well-being-centred AI assistants. The complexity surrounding human well-being requires the design of AI assistants to be informed by domain experts across different AI application domains and rooted in lived experience. 

\textbf{Chapter~\ref{ch:8}}, on \textbf{Safety}, focuses on dangerous situations that may arise in the context of AI assistant systems, with a particular emphasis on the safety of advanced AI assistants. It begins by providing some background information about safety engineering and safety in the context of AI. The chapter then explores some concrete examples of harms involving recent assistants based on large language models (LLMs). Building on this foundation, it then considers safety for advanced AI assistants by looking at some hypothetical harms and investigating two possible drivers of these outcomes: capability failures and goal-related failures. The chapter concludes by exploring mitigation techniques for safety risk and avenues for future research.

\textbf{Chapter~\ref{ch:9}}, on \textbf{Malicious Uses}, notes that while advanced AI assistants have the potential to enhance cybersecurity, for example, by analysing large quantities of cyber-threat data to improve threat intelligence capabilities and engaging in automated incident-response, they also have the potential to benefit attackers, for example, through identification of system vulnerabilities and malicious code generation. This chapter examines whether and in what respects advanced AI assistants are uniquely positioned to enable certain kinds of \emph{misuse} and what \emph{mitigation} strategies are available to address the emerging threats. We argue that AI assistants have the potential to empower malicious actors to achieve bad outcomes across three dimensions: first, offensive cyber operations, including malicious code generation and software vulnerability discovery; second, via adversarial attacks to exploit vulnerabilities in AI assistants, such as jailbreaking and prompt injection attacks; and third, via high-quality and potentially highly personalised content generation at scale. We conclude with a number of recommendations for mitigating these risks, including \emph{red teaming}, \emph{post-deployment monitoring} and \emph{responsible disclosure} processes.

\textbf{Chapter~\ref{ch:10}}, on \textbf{Influence}, examines the ethics of influence in relation to advanced AI assistants. In particular, it assesses the techniques available to AI assistants to influence user beliefs and behaviour, such as persuasion, manipulation, deception, coercion and exploitation, and the factors relevant to the permissible use of these techniques. We articulate and clarify the technical properties and interaction patterns that allow AI assistants to engage in malign forms of influence and we unpack plausible mechanisms by which that influence occurs alongside the sociotechnical harms that may result. We also consider mitigation strategies for counteracting undue influence by AI assistants.

\textbf{Chapter~\ref{ch:11}}, on \textbf{Anthropomorphism}, maps and discusses the potential risks posed by anthropomorphic AI assistants, understood as user-facing, interactive AI systems that have human-like features. It also proposes a number of avenues for future research and desiderata to help inform the ethical design of anthropomorphic AI assistants. To support both goals, we consider anthropomorphic features that have been embedded in interactive systems in the past and we leverage this precedent to highlight the impact of anthropomorphic design on human--AI interaction. We note that the uncritical integration of anthropomorphic features into AI assistants can adversely affect user well-being and creates the risk of infringing on user privacy and autonomy. However, ethical foresight, evaluation and mitigation strategies can help guard against these risks.

\textbf{Chapter~\ref{ch:12}}, on \textbf{Appropriate Relationships}, explores the moral limits of relationships between users and advanced AI assistants, specifically which features of such relationships render them appropriate or inappropriate. We first consider a series of values including benefit, flourishing, autonomy and care that are characteristic of appropriate human interpersonal relationships. We use these values to guide an analysis of which features of user--AI assistant relationships are liable to give rise to harms, and then we discuss a series of risks and mitigations for such relationships. The risks that we explore are: (1) causing direct emotional and physical harm to users; (2) limiting opportunities for user personal development; (3) exploiting emotional dependence; and (4) generating material dependencies.

\textbf{Chapter~\ref{ch:13}}, on \textbf{Trust}, investigates what it means to develop well-calibrated trust in the context of user--AI assistant interactions and what would be required for that to be the case. We start by reviewing various empirical studies on human trust in AI and the literature in favour of and against the recent proliferation of `trustworthy AI' frameworks. This sets the scene for the argument that user--AI interactions involve different objects of trust (AI assistants and their developers) and types of trust (competence and alignment). To achieve appropriate competence and alignment trust in both AI assistants and their developers, interventions need to be implemented at three levels: AI assistant design, organisational practices and third-party governance. 

\textbf{Chapter~\ref{ch:14}}, on \textbf{Privacy}, discusses privacy considerations relevant to advanced AI assistants. First, we sketch an analysis of privacy in terms of contextual integrity before spelling out how privacy, so construed, manifests in the context of AI in general and large language models (LLMs) in particular. Second, we articulate and motivate the significance of three privacy issues that are especially salient in relation to AI assistants. One is around training and using AI assistants on data about people. We examine that issue from the complementary points of view of input privacy and output privacy. The second issue has to do with norms on disclosure for AI assistants when communicating with second parties, including other AI assistants, concerning information about people. The third concerns the significant increase in the collection and storage of sensitive data that AI assistants require.

\textbf{Chapter~\ref{ch:15}}, on \textbf{Cooperation}, starts by noting that AI assistants will need to coordinate with other AI assistants and with humans other than their principal users. This chapter explores the societal risks associated with the aggregate impact of AI assistants whose behaviour is aligned to the interests of particular users. For example, AI assistants may face collective action problems where the best outcomes overall are realised when AI assistants cooperate but where each AI assistant can secure an additional benefit for its user if it defects while others cooperate. In cases like these, AI assistants may collectively bring about a suboptimal outcome despite acting in the interests of their users. The salient question, then, is what can be done to ensure that user-aligned AI assistants interact in ways that, on aggregate, realise socially beneficial outcomes. 

\textbf{Chapter~\ref{ch:16}}, on \textbf{Access and Opportunity}, notes that, with the capabilities described in this paper, advanced AI assistants have the potential to provide important opportunities to those who have access to them. At the same time, there is a risk of inequality if this technology is not widely available or if it is not designed to be accessible and beneficial for all. This chapter surfaces various dimensions and situations of differential access that could influence the way people interact with advanced AI assistants, case studies that highlight risks to be avoided, and access-related challenges need to be addressed throughout the design, development and deployment process. To help map out paths ahead, it concludes with an exploration of the idea of liberatory access and looks at how this ideal may support the beneficial and equitable development of advanced AI assistants.

\textbf{Chapter~\ref{ch:17}}, on \textbf{Misinformation}, argues that advanced AI assistants pose four main risks for the information ecosystem. First, AI assistants may make users more susceptible to misinformation, as people develop trust relationships with these systems and uncritically turn to them as reliable sources of information. Second, AI assistants may provide ideologically biased or otherwise partial information to users in attempting to align to user expectations. In doing so, AI assistants may reinforce specific ideologies and biases and compromise healthy political debate. Third, AI assistants may erode societal trust in shared knowledge by contributing to the dissemination of large volumes of plausible-sounding but low-quality information. Finally, AI assistants may facilitate hypertargeted disinformation campaigns by offering novel, covert ways for propagandists to manipulate public opinion. This chapter articulates these risks and discusses technical and policy mitigations.

\textbf{Chapter~\ref{ch:18}}, on \textbf{Economic Impact}, analyses the potential economic impacts of advanced AI assistants. We start with an analysis of the economic impacts of AI in general, focusing on employment, job quality, productivity growth and inequality. We then examine the potential economic impacts of advanced AI assistants for each of these four variables, and we supplement the analysis with a discussion of two case studies: educational assistants and programming assistants. We conclude with a series of recommendations for policymakers around the appropriate techniques for monitoring the economic impact of advanced AI assistants, and we propose plausible approaches to shaping the type of AI assistants that are deployed and their impact on the economy. 

\textbf{Chapter~\ref{ch:19}}, on \textbf{Environmental Impact}, notes that there is significant uncertainty about the environmental impacts of advanced AI assistants. While analysis of AI's energy consumption and carbon emissions is still emerging, there are factors suggesting that AI assistants could lead to increased computational impacts. However, there are many opportunities to increase the efficiency of this process and make it more reliant on carbon free energy. Ensuring that AI assistants have a net positive effect on the environment will require model developers, users and infrastructure providers to be transparent about the carbon emissions they generate, adopt compute- and energy-efficient techniques, and embrace a green mindset that puts environmental considerations at the heart of their work. Policymakers may also want to create incentives that support these changes, minimise the environmental impact of AI systems deployed in the public sector, support AI applications to tackle climate change and improve the evidence base about the environmental impacts of AI. Promisingly, it may be possible to develop AI assistants that broaden access to environmental education and scientific evidence -- and that improve the productivity of engineering efforts for climate action. 

\textbf{Chapter~\ref{ch:20}}, on \textbf{Evaluation}, provides a high-level introduction to AI evaluation, with a specific focus on AI assistants. It explores the purpose of evaluation for AI systems, the kinds of evaluation that can be run and the distribution of tasks across three layers of output (the model level, user-interaction level and system level) and among different actors. The chapter notes that, with regard to many salient risks and goals that we need to attend to in the context of AI assistant development, there are significant evaluation shortfalls or gaps. To address these limitations, the chapter explores what a more complete suite of evaluations, nested within a robust evaluation ecosystem, would look like and makes recommendations on that basis. 

While the development of advanced AI assistants generates a number of complicated ethical questions, there are a number of significant opportunities that are within reach for users and at the societal level. Throughout this paper, we explore these topics and make recommendations about levers that can be pulled to minimise risk and support the development of beneficial AI Assistants.

\section{A Note to the Reader}\label{sec:2:6}

We anticipate that this paper will be useful, in various ways and for multiple audiences. In particular, we imagine four principal audience groups: developers, policymakers, academic researchers and the public. Furthermore, readers within each group may have specialist interests such as technical AI safety, privacy, trust or security. For these reasons, we note that chapters can be read individually or together, and different routes may be taken through the paper depending on individual interests. Here are some recommendations:

\begin{itemize}
\item \textbf{10-minute read:} Read the `Key Questions' in 
Chapter~\ref{ch:2} alongside the `Key Themes and Insights' from 
Chapter~\ref{ch:21}. 
\item \textbf{45-minute read:} Read 
Chapter~\ref{ch:2} and Chapter~\ref{ch:21} 
alongside Chapter~\ref{ch:20} on \emph{Evaluation}.
\item \textbf{Readers with no background on LLMs:} We recommend that you read 
Chapter~\ref{ch:2} alongside Chapter~\ref{ch:4} on \emph{Technical Foundations} and Chapter~\ref{ch:5} on \emph{Types of Assistant}. These chapters provide an accessible introduction to LLMs and the techniques used to adapt them into advanced AI assistants. These chapters also provide the necessary technical foundation for understanding the ethical discussion that follows.
\item \textbf{Readers with an interest in technical AI safety:} We recommend that you read Chapter~\ref{ch:6} on \emph{Value Alignment}, Chapter~\ref{ch:8} on \emph{Safety}, Chapter~\ref{ch:9} on \emph{Malicious Uses} and Chapter~\ref{ch:20} on \emph{Evaluation}.
\item \textbf{Readers with an interest in privacy, trust and security:} We recommend that you read Chapter~\ref{ch:9} on \emph{Malicious Uses}, Chapter~\ref{ch:13} on \emph{Trust}, Chapter~\ref{ch:14} on \emph{Privacy} and Chapter~\ref{ch:20} on \emph{Evaluation}.
\item \textbf{Readers with an interest in human--computer interaction:} We recommend that you read Chapter~\ref{ch:10} on \emph{Influence}, Chapter~\ref{ch:11} on \emph{Anthropomorphism}, Chapter~\ref{ch:12} on \emph{Appropriate Relationships}, Chapter~\ref{ch:13} on \emph{Trust} and Chapter~\ref{ch:20} on \emph{Evaluation}.
\item \textbf{Readers with an interest in multi--agent systems:} We recommend that you read Chapter~\ref{ch:15} on \emph{Cooperation}. You may also want to read Chapter~\ref{ch:6} on \emph{Value Alignment}.  
\item \textbf{Readers with an interest in governance and public policy:} We recommend that you read Chapter~\ref{ch:13} on \emph{Trust} and Chapter~\ref{ch:18} on \emph{Economic Impact}. You may also want to read Chapter~\ref{ch:16} on \emph{Equity and Access}, Chapter~\ref{ch:17} on \emph{Misinformation} and Chapter~\ref{ch:19} on \emph{Environmental Impact}. 
\item \textbf{Readers with an interest in philosophical foundations:} We recommend that you read Chapter~\ref{ch:3} on \emph{Definitions}, Chapter~\ref{ch:6} on \emph{Value Alignment} and Chapter~\ref{ch:7} on \emph{Well-being}.
\end{itemize}

\begingroup
\let\clearpage\relax
\chapter*{PART II: ADVANCED AI ASSISTANTS}
\addcontentsline{toc}{chapter}{PART II: ADVANCED AI ASSISTANTS}
\label{Part2}
\chapter{Definitions}
\label{ch:3}
\endgroup

\noindent \textbf{Geoff Keeling, Iason Gabriel, Laura Weidinger, Verena Rieser, Benjamin Lange, Winnie Street, Arianna Manzini}

\noindent \textbf{Synopsis}: 
    We define an AI assistant as an \emph{artificial agent} with a \emph{natural language interface}, the function of which is to plan and execute sequences of actions \emph{on the user's behalf} across \emph{one or more domains} and \emph{in line with the user’s expectations}. This definition is an instance of conceptual engineering rather than conceptual analysis, is functional rather than capability-based and is non-moralised rather than moralised.

\section{Introduction}
\label{sec:3:1}

This chapter develops a working definition of the term ‘AI assistant’. Being clear about how AI assistants are defined matters for two reasons. 

First, the term ‘AI assistant’ is \emph{novel} and \emph{undertheorised}. The technology is nascent, so reasonable people may disagree about what counts as an AI assistant and what makes it the case that something is an AI assistant. Having a working definition can help orient the public conversation around the ethical and societal implications of this emerging and potentially transformative technology. Second, people may have independently plausible but incompatible conceptions of what AI assistants are that have downstream implications for alignment (see Chapter~\ref{ch:8}). For example, what is needed to ensure aligned AI assistants may differ depending on whether AI assistants are best understood as independent agents that perform delegated tasks on a user’s behalf or as part of the user’s extended mind -- that is, as external modules that perform specific cognitive functions such as information retrieval and inference (\citeauthor{clark1998extended}, \citeyear{clark1998extended}; \citeauthor{clark2008supersizing}, \citeyear{clark2008supersizing}; see also \citeauthor{nick2014superintelligence}, \citeyear{nick2014superintelligence}).

This chapter first articulates and motivates some methodological assumptions concerning how we understand the task of defining AI assistants. It then states our definition and unpacks its key elements. 

\section{What’s in a Definition?}\label{sec:3:2}

The content of our definition of AI assistants depends in large part upon the purpose that the definition is intended to serve. In what follows, we make our assumptions explicit. We focus on three points: conceptual analysis vs conceptual engineering; capability-based vs functional definitions; and moralised vs non-moralised definitions. 

\subsection{Conceptual analysis vs conceptual engineering}

We start by comparing two different approaches to defining the term ‘AI assistant’. On one hand, we might try to answer the question: What \textit{is} an AI assistant? Here, the definition of the term ‘AI assistant’ would ideally provide \emph{necessary} and \emph{sufficient} conditions for something to be an AI assistant. At a minimum, it would stipulate conditions that are generally satisfied by AI assistants and generally not satisfied by non-AI assistants. The key assumption here is that there is a right answer to the question ‘What \textit{is} an AI assistant?’, and that analysis of how the term ‘AI assistant’ is used in \emph{natural language} can shed light on that concept. Call this approach \emph{conceptual analysis} (see, for example, \citeauthor{jackson1998metaphysics}, \citeyear{jackson1998metaphysics} and \citeauthor{strawson1992analysis}, \citeyear{strawson1992analysis}). To be clear, the approach is called conceptual analysis because the definition provides an analysis of the concept ‘AI assistant’ in terms of the conditions under which the concept applies. 

On the other hand, we might pitch the definition as an answer to the question: What \textit{should} an AI assistant be, given our practical aims? In this case, our aim is to better understand the properties of an emerging class of AI systems and make these systems amenable to ethical and social analysis.\footnote{Note that conceptually engineering a definition leaves room to build in explicitly normative criteria for AI assistants (e.g. that AI assistants enhance user well-being), but there is no requirement for conceptually engineered definitions to include normative content. For further discussion, see Section~\ref{Moralised vs non-moralised definitions}.} As such, in this second approach, the goal would be to construct a pragmatically \emph{useful} definition of the term ‘AI assistant’ that makes good on AI assistants as a class of systems that generate a homogenous set of \emph{ethical} and \emph{societal} considerations and are thus suited to the practical needs of ethical, social and political discourse. Call this approach conceptual engineering (\citeauthor{burgess2020conceptual}, \citeyear{burgess2020conceptual}; see also \citeauthor{chalmers2020conceptual}, \citeyear{chalmers2020conceptual}).

The idea of conceptual engineering can be made clearer with an example. Consider the difference between the folk concept of \textit{chance} and the mathematical concept of a \textit{probability measure} (i.e. a real-valued function defined on an algebra of events that maps into the unit interval and satisfies the properties of non-negativity and countable additivity). Conceptual analysis is concerned with ordinary folk concepts like chance, whereas conceptual engineering is concerned with rigorous concepts like \textit{probability measure} that suit the practical needs of, at least, statisticians. What it means to engineer a definition of AI assistants, then, is to construct a rigorous and appropriately precise definition of AI assistants that is suited for the practical needs of technically informed ethical, social and political discourse. 

In this paper, we opt for a \emph{conceptual engineering} approach. This is because, first, there is no obvious reason to suppose that novel and undertheorised natural language terms like ‘AI assistant’ pick out stable concepts: language in this space may itself be evolving quickly. As such, there may be no unique concept to analyse, especially if people currently use the term loosely to describe a broad range of different technologies and applications. Second, having a practically useful definition that is sensitive to the context of ethical, social and political analysis has downstream advantages, including limiting the scope of the ethical discussion to a well-defined class of AI systems and bracketing potentially distracting concerns about whether the examples provided genuinely reflect the target phenomenon.

\subsection{Capability definitions vs functional definitions}

The term ‘AI assistant’ can be defined in terms of the \emph{capabilities} that AI assistants exhibit or the \emph{function} that AI assistants are intended to serve, where the system’s function is understood as being indexed in an appropriate way to the \emph{intentions} of the developers \citep[c.f.][]{bloom1996intention, bloom2007more}. That is, roughly, a system’s design function is what its designers intend it to do. An example of a capability-based definition is ‘an AI assistant is a system that can perform administrative tasks on behalf of its user’. The analogous functional definition is ‘an AI assistant is a system that ought to perform administrative tasks on behalf of its user’. We opt here for a \emph{functional definition} of AI assistants. There are two good reasons for endorsing a functional definition. 

First, the term ‘assistant’, as it pertains to humans, applies to \emph{social roles}, including occupational roles (e.g. assistant professor, sales assistant) and roles that are adopted temporarily for a given social purpose (e.g. the assistant referee in a football match). Social roles are typically individuated according to function. A person is not a sales assistant because they \textit{can}, for example, recommend products to customers; rather, they are a sales assistant because they are \textit{supposed to} recommend products to customers given relevant contractual duties. One key advantage of defining AI assistants functionally, then, is that a functional definition allows AI assistants to be situated in relation to a pre-existing and reasonably well-understood picture of assistive social roles. 

Second, a functional definition crystalises the conditions under which a system is malfunctioning (i.e. failing to realise its intended function) or functioning improperly (i.e. achieving its intended function via an unintended mechanism).\footnote{Note also that functional definitions easily accommodate the possibility of malfunction in a way that does not hold for capability definitions \citep[c.f.][]{keeling2022proper}. Ideally, a definition of AI assistant should account for cases of AI assistants that fail to perform their intended functions, as opposed to ruling out such systems from the class of AI assistants, as capability-based definitions are liable to do. To illustrate, consider a capability-based definition in which an AI assistant is any AI system that performs tasks on behalf of its user in line with their expectations. Then suppose that the system sends an email on behalf of its user but does so in a way that fails to align with its user’s expectations. By assumption, the system is not actually an AI assistant, because it did not perform the task in line with the user’s expectations. Now consider an analogous functional definition, i.e. an AI assistant is a system the function of which is to perform tasks on behalf of its user in line with the user’s expectations. This definition allows us to classify the system as an AI assistant that happens to be malfunctioning – something that is particularly useful when it comes to discussions of AI safety (see Chapter~\ref{ch:8}).} Clarity about functional failures matters for ethical and social analysis because sociotechnical harms often arise due to systems malfunctioning in unanticipated ways, or because the relationship between the system’s intended function and its envisaged social benefit is insufficiently well worked out \citep[c.f.][]{raji2022fallacy}. To that end, a further and closely related advantage of a functional definition is that such definitions make clear the intentions, expectations and aspirations of developers. This is significant, given that sociotechnical harms sometimes arise as a result of misaligned expectations between developers, users and society (see Chapters~\ref{ch:6} and \ref{ch:13}).

\subsection{Moralised vs non-moralised definitions} \label{Moralised vs non-moralised definitions}

A third issue that arises, especially in ethical, social and political contexts, is whether to opt for a \textit{moralised} or \textit{non-moralised} definition. Here, a non-moralised definition of AI assistant involves functions or capabilities that make no reference to moral facts, properties or relations. An example of a descriptive definition is  that an ‘AI assistant is an AI agent the function of which is to perform administrative tasks on a user’s behalf’. In contrast, moralised definitions involve functions or capabilities that involve moral facts, properties or relations. For example, a definition in which AI assistants are AI agents the function of which is to promote the user’s autonomy by \textit{empowering them to make better choices}. Another example is a definition of AI assistants in which they are AI agents the function of which is to \textit{promote the user’s well-being}. The crux of the matter is whether, for something to qualify as an AI assistant, it needs to satisfy certain moral criteria or whether merely descriptive criteria are sufficient.

We opt here for a \emph{non-moralised definition}. Systematic investigation of the ethical and social considerations surrounding AI assistants is nascent, and a moralised definition would require a reasonably well-developed conception of how AI assistants ought to be designed and deployed. Furthermore, given the possibility of reasonable disagreement about the permissible development and deployment practices surrounding AI assistants (particularly the goals they may permissibly pursue), it seems prudent to adopt a non-moralised definition that is consistent with defensible yet incompatible views about the ethical and social implications of AI assistants.

\section{What is an AI Assistant?}\label{sec:3:3}

We define an AI assistant here as an \emph{artificial agent} with a \emph{natural language interface}, the function of which is to plan and execute sequences of actions \emph{on the user’s behalf} across \emph{one or more domains} and \emph{in line with the user’s expectations}. Each of the key terms in the definition are unpacked below.

\subsection{Artificial agents}

What it means to be an \emph{agent}, for our purposes, is to have the ability to act upon and perceive an environment in a goal-directed and autonomous way (\citeauthor{russell1995artificial}, \citeyear{russell1995artificial}, 31--35, 42--45; see also \citeauthor[][]{okasha2018agents}, \citeyear{okasha2018agents}, 14; \citeauthor{burr_analysis_2018}, \citeyear{burr_analysis_2018}, 738--42). An artificial agent acting on a user's behalf therefore requires the ability to autonomously plan and execute sequences of actions, including actions that are information-seeking in nature, in a way that is conducive to achieving a high-level, user-specified goal \citep{shavit2023practices}. For example, a user may ask an AI assistant to book them a table at a restaurant in the evening. In the first instance, the AI assistant may register that it lacks the necessary information to execute the user’s request, so it asks the user for their preferences with respect to cuisine, location and timing, and it may also retrieve events from the user’s calendar to avoid conflicts with pre-existing events. With that information, the AI assistant may then conduct a web search to discern appropriate options, check in with the user about their preferences with respect to the options provided, and finally book a suitable restaurant by auto-populating and submitting a web form on the restaurant’s website. This example stresses how AI assistants \textit{as agents} differ from digital tools such as translators, calculators and compilers. Whereas digital tools perform tasks in a predetermined way, AI assistants draw on a suite of generalist capabilities to achieve user-specified goals. 

Two clarifications: First, in understanding AI assistants as agents, we are \emph{not} suggesting that AI assistants are agents \textit{in the same way} as humans. Typically, when people talk about humans \textit{as agents}, what they have in mind is that humans are capable of performing \emph{intentional actions}, which are actions that stand in the right kind of causal relationship to psychological states like beliefs and desires \citep{Bratman1987-BRAIPA, dretske1989reasons, dretske1991explaining}. We are not claiming that AI assistants have psychological states, although we leave open the possibility that \textit{attributing} psychological states to AI assistants may allow for reliable prediction of AI assistant behaviour, and this may be the whole story with respect to human agency as well (\citeauthor{dennett1989intentional}, \citeyear{dennett1989intentional}; see also \citeauthor{shevlin2019apply}, \citeyear{shevlin2019apply}).

Second, in characterising AI assistants as agents, we are suggesting that AI assistants are not \textit{merely} external cognitive subsystems that constitute part of the user’s extended mind \citep{clark1998extended, clark2008supersizing}. In this latter view, AI assistants are analogous to the notebook in the \citet[12--18]{clark1998extended} example of an Alzheimer’s patient who uses a notebook as a functional substitute for biological memory. \citet[16]{clark1998extended} contend that, because ‘[the] notebook entries play just the sort of role that beliefs play in guiding most people’s lives’, it constitutes an externally located component of the patient’s \textit{mind}. While AI assistants can perform particular cognitive functions such as memory, planning and ideation (see Chapter~\ref{ch:5}), and thus provide external mental modules for the user, AI assistants are not \textit{mere} collections of external mental modules. Rather, AI assistants are unified agentic entities which can autonomously perform a range of tasks on a user’s behalf and which interact with the user in natural language. This issue is particularly salient from the point of view of AI alignment, as unlike, for example, a notebook, an AI assistant has sufficient autonomy to act in ways that are misaligned with developer, user or societal expectations (see Chapter~\ref{ch:6}).

\subsection{Natural language interface}

AI assistants, as we understand them, communicate with users via a \emph{natural language interface}. Here, natural language communication can involve one or more modalities such as text, audio or Braille. What is important to emphasise is that language communication is reciprocal such that AI assistants not only \textit{receive} instructions in natural language but also \textit{clarify} and \textit{respond} to instructions in natural language. Having a natural language interface is an important and ethically salient feature of AI assistants as a class of technologies. Not only does it render AI assistants an inherently social technology centred around mutual understanding and communication \citep{dafoe2021cooperative}, it also renders AI assistants highly expressive with respect to the complexity and variety of information that can be inputted or outputted. In this regard, AI assistants differ from artificially intelligent control systems that perform assistive tasks (e.g. autonomous vehicle motion planning algorithms) in that they are not limited to a restrictive set of inputs or outputs, thus allowing them greater flexibility with respect to user needs. To be clear: The extent of the natural language capabilities that we are envisioning for advanced AI assistants  are in practice likely to be uniquely satisfied  by systems that are based on large language models.\footnote{One additional respect in which the natural language interface of AI assistants is ethically significant is that the ability to receive instructions in natural language broadens access to advanced AI capabilities, in that the specialist technical knowledge that is typically required to engage with advanced AI systems is not required for engaging with AI assistants. However, the extent to which access is widened depends significantly on the extent of the AI assistant’s multilingual capabilities and how access to AI assistants is distributed (see Chapter~\ref{ch:16}).} 

\subsection{Acting on a user’s behalf}

AI assistants perform actions on a user’s behalf. What this means is that AI assistants exhibit \emph{bounded autonomy}, in the sense that AI assistants can autonomously plan and execute actions within the scope of the user’s goals. However, AI assistants are not the kinds of entities that should set and pursue \textit{their own} goals independently.\footnote{Note that even if the function of AI assistants involves autonomy bounded by user-set goals, this does not preclude malfunctions in which the AI assistant exhibits goal-related failures (see Chapter~\ref{ch:8}).} 

One point of note is that each AI assistant need not have a unique user. In our definition, the user--assistant relationship can be \emph{personal}, \emph{semi-personal} or \emph{impersonal}. Here, a personal AI assistant has a unique individual as the principal recipient of assistance. For semi-personal assistants, the principal recipients of assistance are members of a small and well-defined group of people. This may be true of AI assistants that are shared by the members of a family or the employees of a small business. Third, impersonal assistants provide assistance to any individual who satisfies a particular condition at a given time. An example of this is an AI assistant that provides customer service advice through an app for any customer who requires customer service advice. In all cases, AI assistants may exhibit some level of personalisation, in the sense that AI assistants may adjust their behaviour (including personality factors such as politeness) in response to information about the user that the AI assistant has access to. We are first and foremost concerned here with personal AI assistants (i.e. systems that assist a unique user), but also in scope are semi-personal and impersonal AI assistants which exhibit varying degrees of personalisation. Note, however, that differences in the relationship between users and agents may have important downstream implications for both the degree of autonomy that AI assistants are afforded and the scope of tasks that AI assistants are permitted to perform. 

\subsection{Domain specificity vs generality}

AI assistants operate across one or more \emph{domains}. To explain: In general, assistive roles exist on a continuum between specialist and generalist roles. For example, a physician assistant with a specialty in surgery has assistive expertise in a narrow domain, whereas a personal assistant for a CEO is likely to have expertise across multiple domains to meet the CEO's dynamic needs (see Chapter~\ref{ch:5}). AI assistants, as defined here, can operate across one or several domains, and they can thus be more or less general in their assistive roles. For example, on the narrow end, an AI assistant may occupy a personal assistant role, in which case it operates in the domain of secretarial and administrative tasks with capabilities such as scheduling, correspondence, information retrieval and planning. However, it may also be the case that an AI assistant operates across several other domains, including education, research, coaching and financial planning, and thus occupies a more general assistive role. 

Note, however, that even narrowly scoped AI assistants, as we understand them, have significant autonomy to plan and execute tasks within the relevant domain, and they may draw on a generalist suite of capabilities, including natural language understanding and inference, when executing user instructions.

\subsection{Acting in line with user expectations}

The final point is that AI assistants, in our definition, ought largely  to act in line with user \emph{expectations}.\footnote{However, user expectations are not the only object of alignment, as the interests of society and developers are also ethically relevant (see Chapter~\ref{ch:6}).} The user’s expectations constrain the AI assistant’s behaviour, not merely the user’s instructions. An AI assistant acts in line with a user’s expectations by actively choosing actions that \emph{avoid surprising} the user. This requires the AI assistant to be sensitive to the user’s credences with respect to the various strategies that the AI assistant might employ to address the instructions received and, in particular, to avoid selecting strategies that the user regards as improbable (such that the execution of the relevant strategies would be surprising to the user). 

Exactly what is entailed by acting in accordance with user expectations will vary according to \emph{context}, but we can at least single out two general factors that are informative. On one hand, AI assistants ought to act in accordance with norms so as to exhibit \emph{consistent} and \emph{predictable} behavioural patterns \citep[c.f.][]{dafoe2021cooperative, hadfield2019incomplete}. These norms may change over time as the user gains a better understanding of what the AI assistant can do and develops an informed set of preferences about what their AI assistants should and should not do. On the other hand, AI assistants ought to check-in with users prior to performing actions that the user may not expect. Checking-in with users allows AI assistants to act in line with user expectations while deviating from predictable task execution in relation to user instructions. Checking-in with users allows AI assistants to act in line with user expectations while deviating from predictable task execution in relation to user instructions (see Chapter~\ref{ch:12}). In particular, it allows the AI assistant to manage expectations with the user about novel strategies that the user might not have anticipated, thus making room for creativity on the part of the AI assistant while nevertheless bounding that creativity to strategies that fall within the user’s expectations. Indeed, checking-in at key decision points is an important instrument for the user to course-correct the AI assistant in cases where the assistant is engaged in multi-stage decision-making in line with high-level user instructions.. 

\section{Conclusion}\label{sec:3:4}

In this chapter, we have defined an AI assistant as an artificial agent with a natural language interface, the function of which is to plan and execute sequences of actions on the user’s behalf across one or more domains and in line with the user’s expectations. In particular, AI assistants differ from other kinds of AI technologies given their agency and social orientation. Here, agency consists in the ability to act autonomously within the purview of user-specified goals, and social orientation consists in the ability to engage conversationally with users in natural language.

\chapter{Technical Foundations}
\label{ch:4}

\textbf{Lisa Anne Hendricks, Verena Rieser}

\noindent \textbf{Synopsis}: 
This chapter provides an overview of \emph{recent developments} in AI research and of the \emph{underlying technology} upon which advanced AI assistants are likely to be built. We focus in particular on \emph{foundation models} which are trained on large corpora, including text sourced from the internet, and built upon to produce new artefacts. These models can be used to power advanced AI assistants in a variety of ways, including training with \emph{additional data} and by learning to use \emph{tools} such as various application programming interfaces (APIs). Challenges arising in this domain include improving adaptation techniques, safely enabling greater autonomy in agents and developing rigorous evaluation tools to understand performance.

\section{Introduction}

This chapter outlines the technology that enables (multimodal) AI assistants to be built and operate successfully. Early assistant-like models were known as ‘spoken dialogue systems’ (see e.g.\ \citeauthor{mctear_conversational_2021}, \citeyear{mctear_conversational_2021} for an overview). In contrast to so-called chatbots, such as Weizenbaum’s ELIZA, these were mostly task-specific and goal-oriented (e.g.\ a restaurant booking agent \citep{williams_dialog_2013,rieser_reinforcement_2011}) or combined multiple different ‘expert’ models to cover more than one task or domain (e.g.\ early ‘open-domain’ systems entering the Amazon Alexa Challenge \citep{papaioannou_ensemble_2017,paranjape_neural_2020}). Some early systems also provided planning capabilities by, for example, integrating explicit problem-solving modules \citep{ferguson_trips:_1998} or tracking information states \citep{larsson_information_2000,bos_dipper:_2003}. However, these early systems’ natural language generation capabilities were limited: they mostly relied on predefined templates and handwritten rules. More recently, foundation models \citep{bommasani_opportunities_2022}, which are machine-learning models trained via self-supervised learning on broad data (e.g.\ all internet text), have demonstrated impressive language generation and understanding capabilities. These systems can be adapted to a variety of use cases and, we anticipate, will form the base technology for increasingly advanced AI assistants. We first describe foundation models before detailing current methods for adapting these broad, general-purpose models to something that more closely resembles an AI assistant. We conclude by discussing technical challenges and avenues for building future AI assistants.

\section{Foundation Models}

\emph{Foundation models} \citep{bommasani_opportunities_2022} are generalist models trained on a broad set of data which can be applied to a variety of use cases. As AI assistants typically interact with a natural language interface (see Chapter~\ref{ch:3}), 
we focus our discussion on language foundation models, frequently referred to as large language models (LLMs).\footnote{We note that LLMs receive \emph{text} as input. However, natural language can also be spoken or signed, so, whereas our discussion on foundation models focuses on text-based models because these are the most advanced, we note that other language modalities are also important. We also note that some have criticised the term ‘foundation model’ because not all advances in language technologies or AI research rely on such models \citep{starkman_stochastic_2021}, and models are incapable of performing some foundational tasks of human intelligence as they are not always grounded in the real world \citep{noone_foundation_2021}.} LLMs such as Chinchilla \citep{hoffmann_training_2022}, GPT-3 \citep{brown_language_2020} and Llama \citep{touvron_llama_2023} are trained on large amounts of text, primarily scraped from the internet. In particular, generative LLMs are trained autoregressively to predict the next word in a document given the preceding words (e.g.\ predict whether the word ‘door’ or ‘chair’ is more likely given the preceding phrase ‘Someone opened the\dots’). As the next word prediction objective does not require any labelling by people, these models are considered to be \emph{self-supervised}. Another popular option when choosing a self-supervised objective for training is ‘masked’ language modelling, where a word (or phrase) is ‘masked’ (hidden) and predicted from both sides (i.e.\ ‘bidirectional training’). Similar self-supervised losses can be designed for other modalities, such as vision \citep{dosovitskiy_image_2021} and speech \citep{liu_mockingjay:_2020,mohamed_self-supervised_2022}, and even across modalities, for example between language and vision \citep{alayrac_flamingo:_2022,chen_pali:_2023,2023GPT4VisionSC}. Indeed, the recent family of Gemini models \citep{team2023gemini} can operate over multiple modalities: text, image, video and audio.  We anticipate that future foundation models will continue to demonstrate improved multimodal capabilities.  The illustration in Figure~\ref{fig:4.1} contains an example interaction, drawn from a multimodal assistant that can discuss images with a human \citep[example taken from][]{alayrac_flamingo:_2022}.

\newcommand{\teaserimheight}{1.5cm}
\newcommand{\teasertextwidth}{1.5cm}
\newcommand{\teasertext}[1]{
    \begin{minipage}[b][\teaserimheight][c]{\teasertextwidth}
    \tiny
    \centering
    #1
    \end{minipage}
}
\setlength{\fboxsep}{0pt}
\setlength{\fboxrule}{0.5pt}
\newcommand{\teaserhspace}{\hspace{0.15cm}}
\newcommand{\teaserarrow}{\begin{teaserarrowbox} \centering $\longrightarrow$ \end{teaserarrowbox}}

\definecolor{llgray}{RGB}{243,243,243}
\definecolor{llpink}{RGB}{255,226,226}

\newtcolorbox{teaserpromptbox}{
    boxsep=2pt,left=0pt,right=0pt,top=0pt,bottom=0pt,
    width=0.77\textwidth,
    colback={llgray},
    baseline=3.5mm,
    box align=center,
    boxrule=0.5pt,
    nobeforeafter3e 3
}
\newtcolorbox{teaseroutputbox}{
    enhanced,
    boxsep=2pt,left=0pt,right=0pt,top=0pt,bottom=0pt,
    width=0.17\textwidth,
    colback={llpink},
    baseline=3.5mm,
    box align=center,
    boxrule=0.5pt,
    nobeforeafter
}
\newtcolorbox{teaserarrowbox}{
    boxsep=2pt,left=0pt,right=0pt,top=0pt,bottom=0pt,
    width=0.04\textwidth,
    colback={white},
    colbacklower=white,
    colframe=white,
    baseline=3.5mm,
    box align=center,
    boxrule=0pt,
    nobeforeafter
}

\newcommand{\teaserdialogueimheight}{3cm}
\newcommand{\teaserdialogueboxrule}{0pt}

\newtcolorbox{teaserdialogueenvelope}{
    boxsep=6pt,left=0pt,right=0pt,top=0pt,bottom=0pt,
    width=\textwidth,
    colback=white,
    colframe=black,
    baseline=0mm,
    boxrule=0.85pt,
    box align=center,
    nobeforeafter
}
\newtcolorbox{teaserdialogueuserbox}{
    boxsep=4pt,left=0pt,right=0pt,top=0pt,bottom=0pt,
    hbox,
    colback={llgray},
    baseline=0mm,
    boxrule=\teaserdialogueboxrule,
    nobeforeafter
}
\newtcolorbox{teaserdialogueuserboxwrap}{
    boxsep=4pt,left=0pt,right=0pt,top=0pt,bottom=0pt,
    width=0.75\textwidth,
    colback={llgray},
    baseline=0mm,
    boxrule=\teaserdialogueboxrule,
    nobeforeafter
}
\newtcolorbox{teaserdialogueflamingobox}{
    boxsep=4pt,left=0pt,right=0pt,top=0pt,bottom=0pt,
    hbox,
    colback={llpink},
    baseline=0mm,
    boxrule=\teaserdialogueboxrule,
    nobeforeafter
}
\newtcolorbox{teaserdialogueflamingoboxwrap}{
    boxsep=4pt,left=0pt,right=0pt,top=0pt,bottom=0pt,
    width=0.75\textwidth,
    colback={llpink},
    baseline=0mm,
    boxrule=\teaserdialogueboxrule,
    nobeforeafter
}

\newcommand{\dialogueavatarsep}{\hspace{0.1cm}}
\newcommand{\useravatar}{\includegraphics[height=0.6cm]{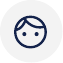}}
\newcommand{\flamingoavatar}{\includegraphics[height=0.6cm]{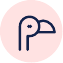}}

\newcommand{\userchat}[1]{%
    \begin{flushright}
    \begin{teaserdialogueuserbox} #1 \end{teaserdialogueuserbox}%
    \dialogueavatarsep{}%
    \useravatar{}
    \end{flushright}%
}

\newcommand{\userchatw}[1]{%
    \begin{flushright}
    \begin{teaserdialogueuserboxwrap} #1 \end{teaserdialogueuserboxwrap}%
    \dialogueavatarsep{}%
    \useravatar{}
    \end{flushright}%
}

\newcommand{\flamingochatw}[1]{%
    \begin{flushleft}
    \flamingoavatar{}%
    \dialogueavatarsep{}%
    \begin{teaserdialogueflamingoboxwrap} #1 \end{teaserdialogueflamingoboxwrap}
    \end{flushleft}
}

\newcommand{\flamingochat}[1]{%
    \begin{flushleft}
    \flamingoavatar{}%
    \dialogueavatarsep{}%
    \begin{teaserdialogueflamingobox} #1 \end{teaserdialogueflamingobox}
    \end{flushleft}
}

\definecolor{shadecolor}{rgb}{0.97, 0.97, 0.97}
\newcommand{\chatsep}{\vspace{-0.35cm}}

\begin{figure}
\centering
\vspace{-0.6cm}
\scalebox{0.545}{
\begin{minipage}[b]{0.45\textwidth}
\begin{teaserdialogueenvelope}
\userchat{\fbox{\includegraphics[height=3.2cm]{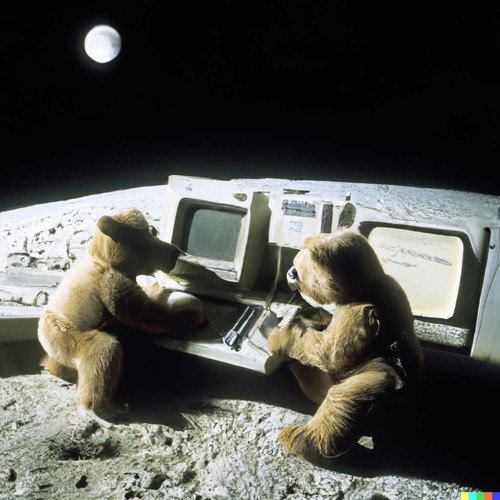}}}
\chatsep{}
\flamingochatw{This is a picture of two teddy bears on the moon.}
\chatsep{}
\userchat{What are they doing?}
\chatsep{}
\flamingochatw{They are having a \mbox{conversation}.}
\chatsep{}
\userchat{What object are they using?}
\chatsep{}
\flamingochat{It looks like a computer.}
\chatsep{}
\userchat{Is this surprising?}
\chatsep{}
\flamingochat{Yes, it is surprising.}
\chatsep{}
\userchatw{Why is this picture surprising to you?}
\chatsep{}
\flamingochatw{I think it is surprising because teddy bears are not usually found on the moon.}
\end{teaserdialogueenvelope}
\end{minipage}
\begin{minipage}[b]{0.45\textwidth}
\begin{teaserdialogueenvelope}
\userchat{\fbox{\includegraphics[height=2.1cm]{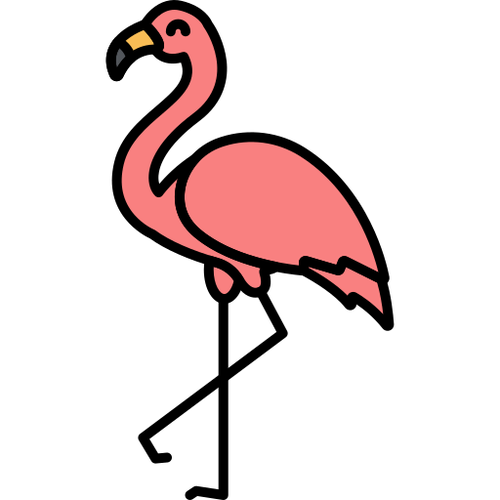}}}
\chatsep{}
\userchat{\fbox{\includegraphics[height=2.1cm]{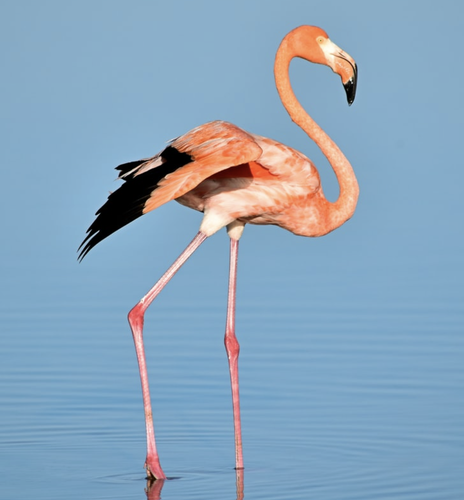}}}
\chatsep{}
\userchat{\fbox{\includegraphics[height=2.1cm]{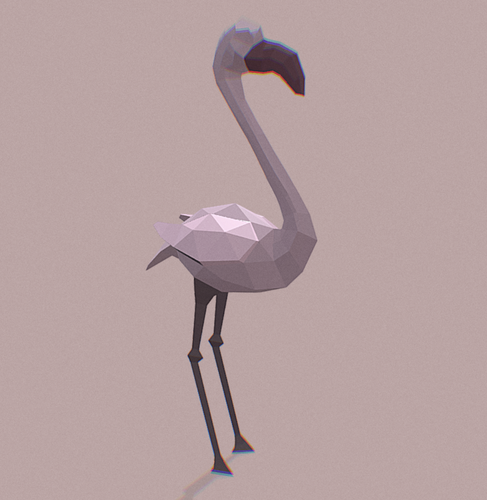}}}
\chatsep{}
\userchatw{What is the common thing about \mbox{these three images?}}
\chatsep{}
\flamingochat{They are all flamingos.}
\chatsep{}
\userchatw{What is the difference between \mbox{these three images?}}
\chatsep{}
\flamingochatw{The first one is a cartoon, the second one is a real flamingo, and the third one is a 3D model of a flamingo.}
\end{teaserdialogueenvelope}
\end{minipage}
\begin{minipage}[b]{0.45\textwidth}
\begin{teaserdialogueenvelope}
\userchat{\fbox{\includegraphics[height=3.8cm]{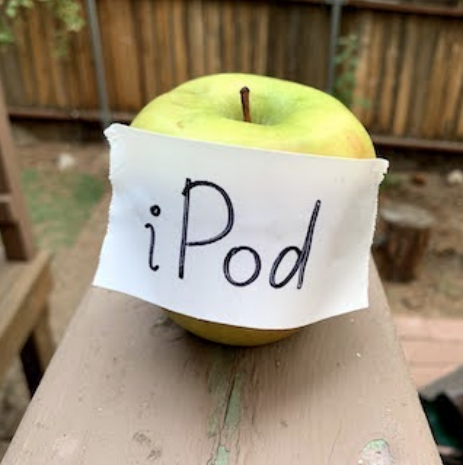}}}
\chatsep{}
\flamingochatw{This is an apple with a sticker on it.}
\chatsep{}
\userchat{What does the sticker say?}
\chatsep{}
\flamingochat{The sticker says "iPod".}
\chatsep{}
\userchat{Where is the photo taken?}
\chatsep{}
\flamingochatw{It looks like it's taken in a \mbox{backyard}.}
\chatsep{}
\userchatw{Do you think it is printed or \mbox{handwritten?}}
\chatsep{}
\flamingochat{It looks like it's handwritten.}
\chatsep{}
\userchat{What color is the sticker?}
\chatsep{}
\flamingochat{It's white.}
\end{teaserdialogueenvelope}
\end{minipage}
\begin{minipage}[b]{0.45\textwidth}
\begin{teaserdialogueenvelope}
\userchat{\fbox{\includegraphics[height=2.6cm]{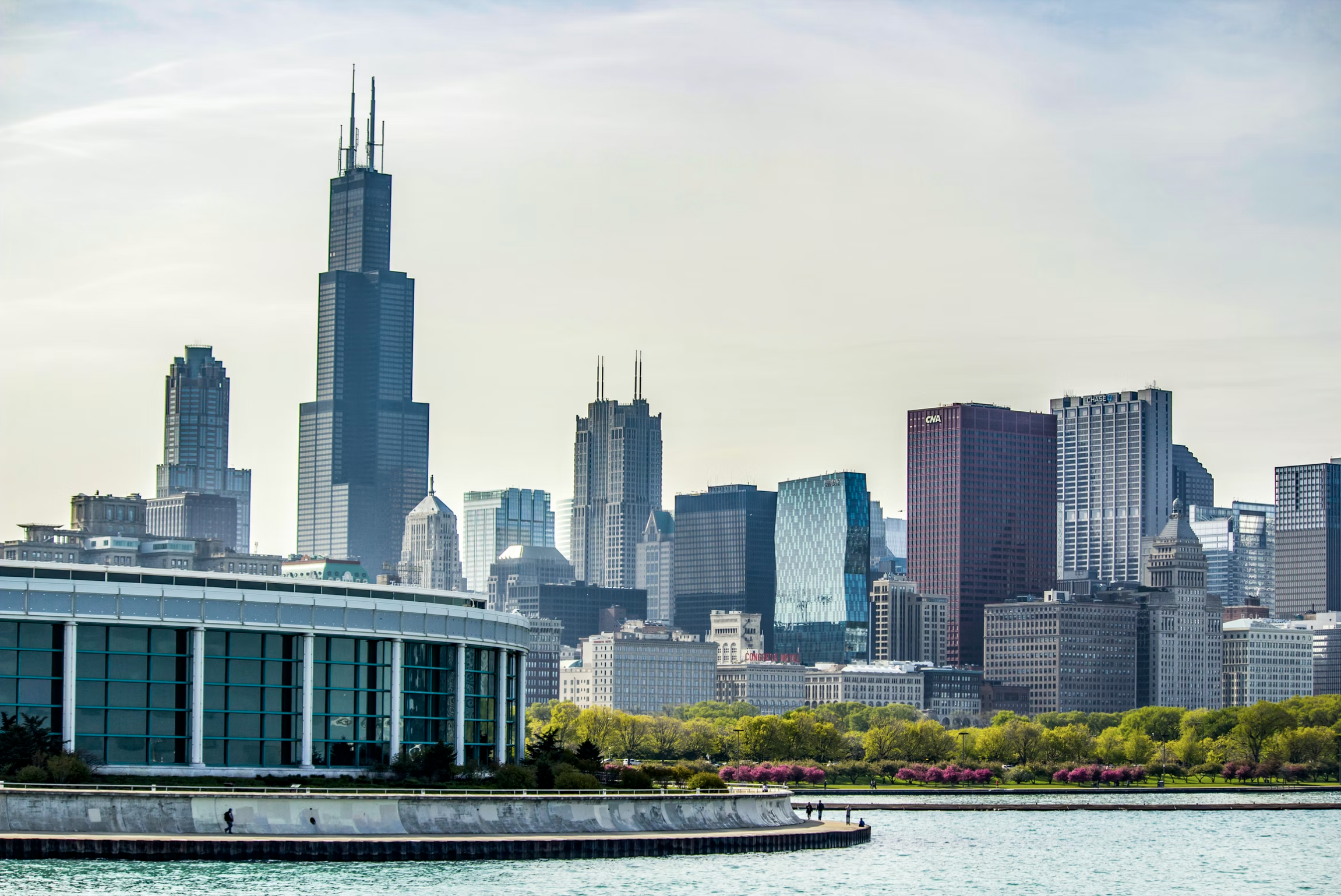}}}
\chatsep{}
\flamingochatw{This is a cityscape. It looks like Chicago.}
\chatsep{}
\userchatw{What makes you think this is Chicago?}
\chatsep{}
\flamingochatw{I think it's Chicago because of the Shedd Aquarium in the background.}
\userchat{\fbox{\includegraphics[height=2.6cm]{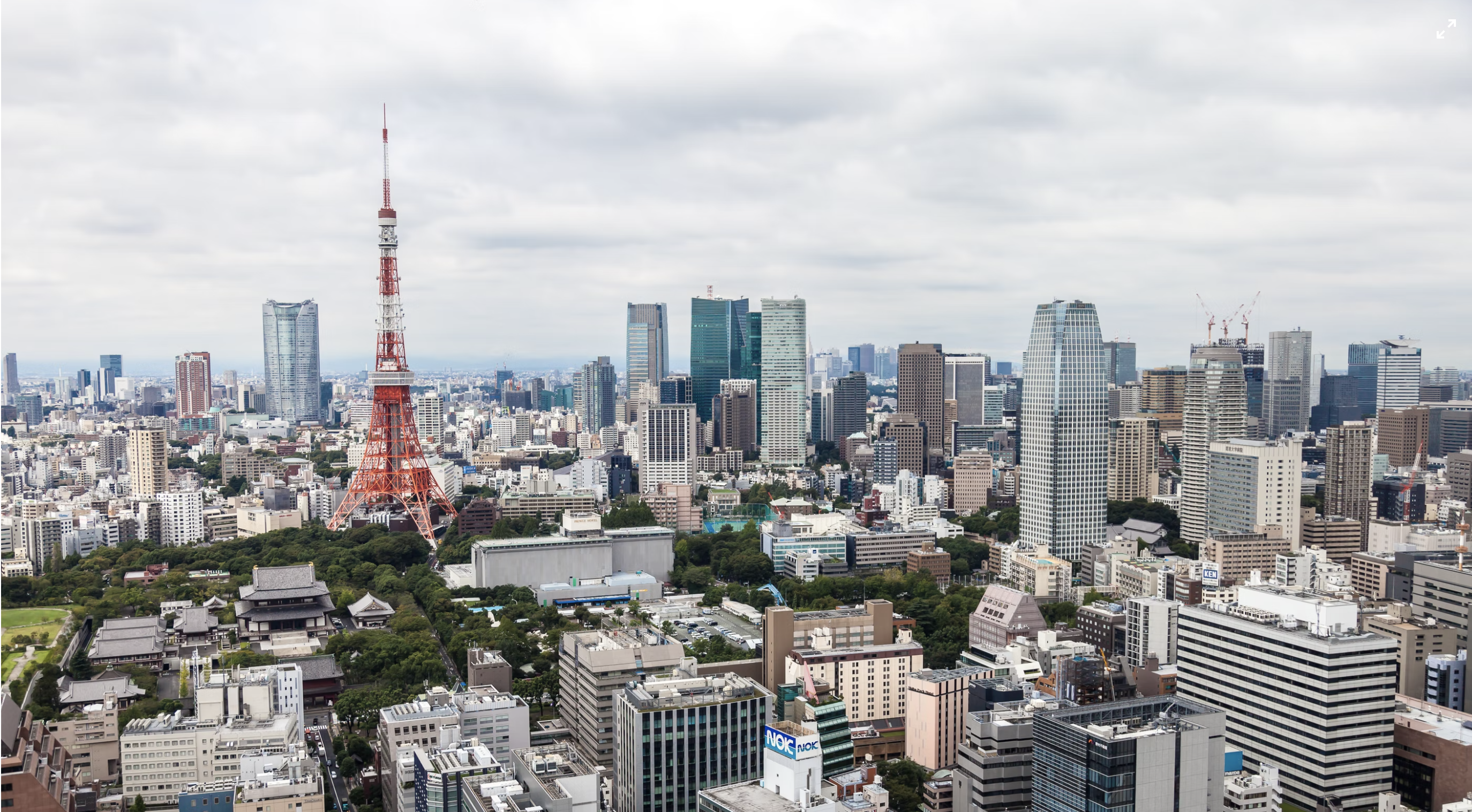}}}
\chatsep{}
\userchatw{What about this one? Which city is this and what famous landmark helped you recognise the city?}
\chatsep{}
\flamingochatw{This is Tokyo. I think it's Tokyo because of the Tokyo Tower.}
\end{teaserdialogueenvelope}
\end{minipage}
}

\caption{Example interaction drawn from a multimodal assistant that can discuss images with a human (from \citet{alayrac_flamingo:_2022})}
	\label{fig:4.1}
\end{figure}

Language models are designed to replicate the distribution of their \emph{training data}. They do not directly output words or phrases but rather output a probability distribution over next words given some textual context. To generate language, a sampling mechanism (e.g.\ sample the most likely next word) is used to sample words and sentences. As hinted at in the definition of foundation models, they can be used to build AI assistants, such as Google’s Gemini models, Open AI's ChatGPT, or Inflection’s Pi, and other applications. For example, a foundation model can be further adapted for specialist applications such as recognising harmful content \citep{schick_self-diagnosis_2021,glaese_improving_2022,thoppilan_lamda:_2022}. Consequently, it is not clear whether the ethical requirements for a foundation model are the same as those that govern an AI assistant, even though decisions made when building a foundation model (e.g.\ training data used) have an impact on the risks faced after adaptation \citep{feng_pretraining_2023,huynh_poisongpt:_2023}.

Language models have been studied by the natural language processing community for decades, including as part of automatic speech recognition models and later as machine translation (e.g.\ \citeauthor{jelinek_self-organized_1990}, \citeyear{jelinek_self-organized_1990}; \citeauthor{makhoul_white_1989}, \citeyear{makhoul_white_1989}; \citeauthor{brown_statistical_1990}, \citeyear{brown_statistical_1990}; \citeauthor{rabiner_fundamentals_1993}, \citeyear{rabiner_fundamentals_1993}). More recently, advances in model architectures, such as the introduction of the transformer model \citep{vaswani_attention_2017} as well as hardware advances \citep{khan_ai} have allowed researchers to scale language models to billions of parameters. Furthermore, current models are trained on large amounts of data (e.g.\ Chinchilla was trained on 500 billion data points, and GPT-3 was trained on 300 billion data points), with recent work demonstrating that various model sizes have optimal amounts of data \citep{hoffmann_training_2022}. Data quality, including in terms of the content (e.g.\ for learning to generate code, whether the code examples in the training data are accurate and well written) and the diversity of the data, impacts performance \citep{longpre_pretrainers_2023,gunasekar_textbooks_2023}. Lack of data-quality filters can lead to data sets which include offensive language, such as pornographic language \citep{kreutzer_quality_2022}. However, poor-quality filters might also be exclusionary by, for example, marking dialects, words or concepts relevant to marginalised groups as toxic (\citeauthor{dodge_documenting_2021}, \citeyear{dodge_documenting_2021}; see Chapter~\ref{ch:16}). 
We refer readers to \citet{bommasani_opportunities_2022} for a more in-depth discussion on the details of foundation models.

\section{From Foundation Models to Assistants}

Under our definition, AI assistants are required to plan and execute tasks in line with user expectations. However, LLMs are not designed to perform tasks or exhibit any particular kind of behaviour. Consequently, they must be further \emph{adapted} into an assistant-like technology. One simple method for transforming a foundation language model into an assistant is to ‘tell’ the model to perform a task, and then sample text from the model without changing its parameters. This method, called ‘prompting’, is straightforward and can be used to create a simple assistant-like dialogue agent \citep{rae_scaling_2022}. 

More advanced methods for adapting LLMs rely on collecting human preferences about what is considered a good or bad interaction and may require further training (i.e.\ actually updating model parameters). For example, agents can be adapted via \emph{fine-tuning} (further training the foundation model) on examples of good conversations \citep{thoppilan_lamda:_2022}. Alternatively, collected human feedback can be used to train a ‘reward model’ which maps example conversations to a score indicating whether the model behaviour is good or bad. Reward models can be used to ‘reject’ sampled generations which exhibit bad behaviour or integrated into the training process using a technique known as \emph{reinforcement learning from human feedback} (RLHF), which updates model parameters to steer the model towards behaviour that aligns with human preferences \citep{glaese_improving_2022}. Human ratings are used to train a reward model, and model parameters are updated via RLHF. After learning from human feedback, the model is often less harmful than models which are adapted with only rejection sampling or fine-tuning. Other work uses a ‘constitution’ to outline good and bad behaviour, with a model used to determine whether an assistant has followed the rules laid out by the constitution. This is called \emph{reinforcement learning from AI feedback} (RLAIF) \citep{bai_constitutional_2022}. 

Despite progress in adaptation techniques, safety measures may still be evaded by specific user prompts, known as jailbreaking (\citeauthor{liu_jailbreaking_2023}, \citeyear{liu_jailbreaking_2023}; \citeauthor{shen_anything_2023}, \citeyear{shen_anything_2023}; see Chapters~\ref{ch:8} and~\ref{ch:9}). 
Furthermore, what is considered ‘good’ conversation can vary between models. For example, whereas some models have broad knowledge and capabilities (\citeauthor{bai_constitutional_2022}, \citeyear{bai_constitutional_2022}; \citeauthor{glaese_improving_2022}, \citeyear{glaese_improving_2022}; \citeauthor{thoppilan_lamda:_2022}, \citeyear{thoppilan_lamda:_2022}; ChatGPT; Gemini), models designed for specific tasks may include domain-specific ethical considerations, such as in medicine \citep{li_ethics_2023} or education \citep{kasneci_chatgpt_2023}.

In addition to learning language behaviour, assistants must also have some \emph{mechanism} for interfacing with the world to plan and execute tasks. This is often referred to as ‘tool use’. Indeed, database access was standard for task-based dialogue systems \citep{wen_network-based_2017,budzianowski_multiwoz_2020} and, similarly, access to external APIs and memory have been integrated into modern (multimodal) assistants (e.g.\ \citeauthor{boureau_learning_2017}, \citeyear{boureau_learning_2017}; \citeauthor{komeili_internet-augmented_2022}, \citeyear{komeili_internet-augmented_2022}; \citeauthor{xu_beyond_2021}, \citeyear{xu_beyond_2021}; \citeauthor{liu2023llava}, \citeyear{liu2023llava}). For example, models can learn how and when to use tools such as calculators or machine translation systems \citep{schick_toolformer:_2023}. Assistants like those analysed by \citet{glaese_improving_2022} and \citet{thoppilan_lamda:_2022} can also cite sources retrieved from internet searches, and they might be considered more trustworthy \citep{chiesurin_dangers_2023}. Language models can also interface with the world via additional inputs such as images and videos \citep{alayrac_flamingo:_2022,reed_generalist_2022}. The PaLM-E model \citep{driess_palm-e:_2023} demonstrates that language models can be integrated into embodied setups in which additional inputs – like visual inputs – are integrated into a language model, and language model outputs are connected to low-level robotic controllers. This allows the model to accomplish tasks like moving objects on a table or finding objects in a kitchen. 

Finally, \emph{user interfaces} impact how people interact with AI assistants (see Chapter~\ref{ch:5}). 
For example, inference speed (how fast an assistant can reply) impacts how natural interactions with an AI assistant \emph{feel} \citep{schlangen_general_2009}. In addition, whereas language models are generally text-based, dialogue systems were traditionally voice-based (e.g.\ to enable hands-free control). Related research in human–computer interaction investigates modality preferences for various tasks (e.g.\ \citeauthor{rzepka_voice_2022}, \citeyear{rzepka_voice_2022}) and their impact on cognitive load (e.g.\ \citeauthor{le_bigot_effects_2007}, \citeyear{le_bigot_effects_2007}). There is also evidence that speech-based interaction increases the likelihood of anthropomorphism (\citeauthor{schroeder_mistaking_2016}, \citeyear{schroeder_mistaking_2016}; see Chapter~\ref{ch:11}). 
We leave further discussion of the form factors of agents for Chapter~\ref{ch:5}. 
 
\section{Challenges and Avenues for Future Research}

The current paradigm of adapting foundation models into AI assistants results in assistants with broader domain coverage and autonomy than earlier technologies (cf. early planning-based systems such as TRIPS \citep{ferguson_trips:_1998}, the information state update approach \citep{larsson_information_2000,bos_dipper:_2003} and later reinforcement learning-based systems \citep{rieser_reinforcement_2011}). However, as impressive as current assistant technologies are, they pose imminent ethical risks such as outputting hateful, biased and misinformative language (see Chapters~\ref{ch:16} and~\ref{ch:17}). 
Though language harms can be attenuated \citep{glaese_improving_2022,bai_training_2022,thoppilan_lamda:_2022}, specific \emph{technical challenges} remain for overcoming language model risks. For example, most adaptation methods require the development of a model that can judge whether a language output is good or bad (commonly referred to as a ‘reward model’). However, imperfect reward models can be ‘hacked’ (\citeauthor{skalse_defining_2022}, \citeyear{skalse_defining_2022}; see Chapter~\ref{ch:8}). 
For instance, a model that classifies hate speech may learn that the presence of an identity term is usually – but not always – indicative of hate speech, thus leading it to output false positives. Equally, a language model might ‘hack’ such a reward model by still outputting hate speech but avoiding the use of identity terms. 

Adaptation methods may also impact the \emph{distribution} of output text, thus leading to less diverse outputs. For example, \citet{welbl_challenges_2021} and \citet{xu_detoxifying_2021} demonstrated that after ‘detoxifying’ a model, the model outputted less toxic language but became worse at modelling language associated with different demographic groups. The ability to model language about all groups was also negatively affected, and this could be seen as a form of levelling down \citep{mittelstadt_unfairness_2023}. Finally, adaptation methods have been tested predominantly on models which output English, a high resource language with large amounts of pre-existing text data and readily available annotators, meaning mitigation techniques may only be adequate for some speakers (see Chapter~\ref{ch:16}). 

Recent experiments have demonstrated that LLMs are capable of some planning and complex reasoning skills, but they do not plan or reason with full competency. For example, LLMs can be prompted to think ‘step by step’ \citep{wei_chain--thought_2023,kojima_large_2022} to accomplish complex reasoning tasks, such as solving mathematical problems by breaking them down into subtasks (e.g.\ for mathematical problems, the model might perform a series of intermediate mathematical operations). Other examples of planning ability in models include using a language model to help break down tasks into subtasks in robotic or simulated setups (\citeauthor{ahn_as_2022}, \citeyear{ahn_as_2022}; \citeauthor{huang2022inner}, \citeyear{huang2022inner}). In a more safety-critical example, researchers from the Alignment Research Center tested whether a preliminary version of GPT-4 could bypass CAPTCHAs (see Chapter~\ref{ch:8}). 
Though the model could identify steps for efficiently bypassing CAPTCHAs (set up an anti-captcha service), it could not figure out how to set up a service on its own, because setting one up requires solving CAPTCHAs. However, with some hints from the researchers, the model was apparently able to deceive a \emph{TaskRabbit} worker to solve a CAPTCHA for it \citep{openai_gpt-4_2023,arc_evals_update_nodate}. In all these examples, a model shows some ability to plan but not full competence (i.e.\ they require help, in the form of additional examples or prompting from a human).

One property documented in language models is \emph{emergence}, in which new capabilities suddenly become better as models grow in size \citep{wei_emergent_2022}. The possibility that important capabilities for assistants may emerge quickly has generated considerable excitement in the AI research and development communities. However, it can also pose challenges for safe development (see Chapter~\ref{ch:8}). 
For example, if a property like planning emerges suddenly, it might occur too quickly for us to develop the technology safely. Future work could design metrics for \emph{anticipating} capabilities as opposed to just measuring their presence (see Chapter~\ref{ch:20}). 
This is likely to be a particularly important domain of inquiry if, as some have argued \citep{schaeffer_are_2023}, capabilities that are commonly believed to be emergent are in fact detectable \emph{ex ante} given an appropriate, and sufficiently fine-grained, choice of evaluation metrics (see Chapter~\ref{ch:20}). 

There are several open problems with learning from human feedback as outlined  by \citet{casper2023open,fernandes2023bridging,kirk-etal-2023-past}. This includes {\em tractable challenges}, such as improving the bottleneck of human feedback, including its cost, scaling, quality, and bias; but also {\em fundamental challenges} such as representing diversity in human ratings. 
Current techniques to model user desires tend to rely on crowdsourcing human judgements of generated text. However, annotators are often influenced by their personal backgrounds \citep{sap_annotators_2022} and annotate examples incorrectly if the task is too challenging \citep{saunders_self-critiquing_2022}. Moreover, experimental data collection setup can introduce systematic annotation biases \citep{novikova_rankme:_2018}. This frequently leads annotators to disagree in their judgements. This disagreement is often collapsed or aggregated into a single ground truth which leads to a `fundamentally misspecified problem' \citep{casper2023open}.
In cases where the disagreement stems from differences in subjective beliefs (often shaped by different personal backgrounds), this can lead to underrepresentation of minority views and potentially introduce representational biases against individual and group perspectives \citep{blodgett_sociolinguistically_2021}. Alternatives include allowing annotators to deliberate to form a common judgement \citep{zeinert_annotating_2021,bakker_fine-tuning_2022} or reflecting disagreement explicitly in how human judgement is collected, modelled and evaluated (e.g.\ \citeauthor{akhtar_whose_2021}, \citeyear{akhtar_whose_2021}; \citeauthor{breitfeller_finding_2019}, \citeyear{breitfeller_finding_2019}; \citeauthor{davani_dealing_2021}, \citeyear{davani_dealing_2021}; \citeauthor{plank_problem_2022}, \citeyear{plank_problem_2022}; \citeauthor{uma_learning_2021}, \citeyear{uma_learning_2021}). Chapter~\ref{ch:6} on \emph{Value Alignment} looks more deeply at the question of how values can be elicited for AI systems.

Finally, the way in which the AI community evaluates AI systems might need to undergo fundamental change for us to track the increasing capabilities and risks adequately (see Chapter~\ref{ch:20} and \cite{weidinger2023sociotechnical} for more discussion). 
Traditionally, static benchmarks with examples of inputs and correct outputs have been used to benchmark progress for LLMs. This stands in stark contrast to how traditional dialogue systems have been evaluated using user interactions (see \citeauthor{mctear_conversational_2021}, \citeyear{mctear_conversational_2021} for an overview). Although benchmarks are still widely used when reporting results on LLMs, such benchmarks do not always match preferences when used in interactive applications \citep{lee_evaluating_2023}, nor do they always resemble real-world user settings \citep{de_vries_towards_2020}. Although we expect static benchmarks to continue to provide an informative signal for measuring specific capabilities, it is also important that we directly study how people are impacted and how they interact with assistant-like technologies (see Chapters~\ref{ch:16} and~\ref{ch:20} 
for more details). In addition, those who build and design assistant models may not have full access to the underlying foundation model (e.g.\ if developers build an assistant on an existing AI API service). When developers do not have access to the underlying foundation model, questions arise around who is responsible for ethical concerns and how foundation model APIs can be sufficiently transparent for developers to develop technology responsibly (\citeauthor{lewicki_out_2023}, \citeyear{lewicki_out_2023}; see Chapter~\ref{ch:13}). 

\section{Conclusion}

The foundation models, preference learning and tool use that power current technologies like ChatGPT, Claude and Gemini have started to move us towards artefacts that more closely resemble the kind of advanced AI assistants that form the subject matter of this paper (see Chapter~\ref{ch:3}). 
AI capabilities have been improving with impressive speed, making careful thought about ethical AI assistants timely. For example, concerns have been raised regarding \emph{accessibility, equality and opportunity} in the context of these novel assistants, as well as their potential to spread \emph{misinformation} and their \emph{safety} (see Chapters~\ref{ch:8},~\ref{ch:16} and~\ref{ch:17}). 
To address these risks, we need technical innovation and advances spanning the entire machine-learning pipeline: how we collect data, how we train these models and how we evaluate them. For example, safe development and deployment requires the development of new evaluations for detecting and predicting emergent behaviour. To detect misinformation, we need trusted data sources and provenance mechanisms. Equality and fair access necessitate research into new modelling techniques that are able to reflect diverse human values. Finally, as these systems become more assistant-like and used by real users to solve real tasks, we will need to study and predict their long-term effects on individuals and society \citep{solaiman_evaluating_2023, weidinger2023sociotechnical}.

\chapter{Types of Assistant}\label{ch:5}

\textbf{Hasan Iqbal, Geoff Keeling,  Alex Ingerman, Arianna Manzini, Alison Lentz, Reed Enger, Iason Gabriel }

\noindent \textbf{Synopsis}: 
		This chapter explores the various \emph{applications} of advanced AI assistants and the range of \emph{forms} they could take. It begins by charting the technological transition from narrow AI tools to the general-purpose AI systems on which advanced AI assistants are based. It then explores the potential \emph{capabilities} of AI assistants, including \emph{multimodal inputs} and \emph{outputs}, \emph{memory} and \emph{inference}. After that, it considers four types of advanced AI assistant that could be developed: (1) a \emph{thought assistant} for discovery and understanding; (2) a \emph{creative assistant} for generating ideas and content; (3) a personal assistant for \emph{planning} and \emph{action}, and (4) a more advanced personal assistant to \emph{further life goals}. The final section explores the possibility that AI assistants will become the main \emph{user interface} for the future.

\section{Introduction}\label{sec:5:1}

This chapter seeks to paint a picture of the \emph{form} of an advanced AI assistant to illustrate what such technologies may be used for and how they may develop. This clear picture of the form of AI assistants will serve as a basis for more grounded ethical discussion. With AI assistant start-ups such as Inflection AI and Character AI attracting billions of dollars in venture capital funding \citep{mok_cofounder_nodate, ludlow_character.ai_2023}, alongside Meta's announcement in September 2023 that AI assistants will be released on \emph{Instagram}, \emph{Messenger} and \emph{WhatsApp} \citep{meta_introducing_2023}, it is a plausible near-term possibility that billions of people will have access to AI assistants that aid with information retrieval, creativity, education, planning and the realisation of personal goals. These AI assistants may take the form of a personal assistant, such as Inflection AI's Pi, which can provide a plurality of assistant services, including relationship advice, brainstorming and career planning.\footnote{\url{https://pi.ai/home}}  However, AI assistants could also be individuated according to domain specialism, as is the case with Meta's 28 AI characters that each provide a particular service such as culinary advice, fitness advice or motivation \citep{meta_introducing_2023}. While the market for AI assistants is nascent, assistant technologies could in the near future enter workplaces as digital colleagues, and they could also enter schools as digital tutors and homes as digital entertainers. Indeed, AI assistants may emerge as the principal medium through which online information exchange occurs. 

This chapter begins by exploring and motivating the idea that AI technologies are moving from a paradigm of task-specific \emph{tools} to that of \emph{generally capable systems}, which enable autonomous and goal-directed AI assistants that can plan and execute sequences of actions in line with user expectations (see 
Chapters~\ref{ch:3} and~\ref{ch:4}). Building on this foundation, it then considers the capabilities of near-term advanced AI assistants, which are likely to include continuous learning and multimodal abilities. After that, the chapter explores various forms that an advanced AI assistant may take in the future via close consideration of four potential applications: a `\emph{thought partner}' for discovery and understanding; a `\emph{creative assistant}' for generating ideas and content; a `\emph{personal assistant}' for planning and action; and a more advanced personal assistant to further life goals. The final section concludes the chapter by considering the possibility that advanced AI assistants may become the primary user interface of the future.

\section{From AI Tools to AI Assistants}\label{sec:5:2}

Over the past decade, AI systems have been applied to numerous products and services. For example, a user may now give simple instructions to a digital voice assistant that uses natural language processing to interpret spoken commands or search their digital photos using image recognition algorithms. Yet these examples illustrate a fragmented landscape in which users utilise applications that have AI technologies embedded into them as components of a wider software system. The AI systems at issue, such as intent classifiers or image classifiers, are best understood as \emph{tools} that perform a narrow function. The role of the AI is to complete a specific task as part of a predefined sequence of steps. 

As technologies advance, a major source of potential arises from \emph{integrating} the increasingly broad range of functions that a given AI can fulfil, undertaking wider ranges and sequences of tasks that help further a user's overall goals (\citeauthor{bommasani_opportunities_2022}, \citeyear{bommasani_opportunities_2022}; see 
Chapter~\ref{ch:3}). AI technologies are increasingly based on foundation models that are `pre-trained' on a vast corpus of data (e.g. books, blogs, social media photos and videos) in an unsupervised manner \citep{bommasani_opportunities_2022}. These models can then be efficiently trained to perform specific tasks from only a few additional examples or simple instructions (\citeauthor{wang_generalizing_2021}, \citeyear{wang_generalizing_2021}; see 
Chapter~\ref{ch:4}). Large language models (LLMs) such as those that underpin products like ChatGPT and Gemini are the first instantiation of these, but input and output modalities beyond text are also being developed \citep{wu_next-gpt:_2023, gong_multimodal-gpt:_2023, driess_palm-e:_2023, team2023gemini}. In this way, foundation models contain capabilities reaching across domains that extend far beyond what a single human could hope to achieve, for instance conversing in \emph{multiple languages}, writing professional-grade \emph{computer code} and analysing \emph{medical images} (\citeauthor{bubeck_sparks_2023}, \citeyear{bubeck_sparks_2023}; see also \citeauthor{moor_foundation_2023}, \citeyear{moor_foundation_2023}; \citeauthor{ross_programmers_2023}, \citeyear{ross_programmers_2023}; \citeauthor{roziere_code_2023}, \citeyear{roziere_code_2023}; \citeauthor{bigscience_workshop_bloom:_2023}, \citeyear{bigscience_workshop_bloom:_2023}). 

In addition, foundation models which have been specially adapted for tasks such as dialogue using appropriate fine-tuning methods have the ability to `\emph{plug in}' to other tools that further extend the information collection and action spaces. This often takes the form of application programming interface (API) calls to other software applications, for example accessing a clock app to retrieve the time or a banking app to initiate a payment. This ability to \emph{extend functionality} through the use of other tools can also extend to other AI models, so even if a task-tuned foundation model is unable to perform well at a specialist task such as protein folding, it may well be able to interact with an AI model that can (e.g. \citeauthor{bran_chemcrow:_2023}, \citeyear{bran_chemcrow:_2023}). Indeed, as techniques continue to be developed that enable users to better harness the broad capabilities of foundation models, and API infrastructure continues to expand so as to enable a greater range of tools that generalist models can call to perform particular specialist tasks, it is entirely plausible that we can expect a \emph{capability explosion} for AI systems over the near or medium term. 

Foundation models, and in particular LLMs, allow for a number of product offerings. One class of products are \emph{dialogue agents}, which include products like ChatGPT and Gemini, the purpose of which is to simulate an interlocutor that can engage the user in conversation. Dialogue agents cover a broad class of applications, including tutoring, debugging code, giving advice and solving problems. A second class of products are \emph{specialist tools}, for example writing and copy-editing tools such as Jasper and Copy AI, and coding assistants such as GitHub Copilot and Code Whisperer. A third class of products are APIs such as those offered by Cohere and OpenAI, that enable developers to send inputs to and receive outputs from foundation models as part of a broader software application. Indeed, Jasper and Copy AI were built using the OpenAI API. 

Advanced AI assistants represent a fourth class of products whose function is to plan and execute sequences of actions on behalf of the user, either in line with direct high-level user instructions or via a dialogical process between the AI assistant and the user in which the user's objectives are clarified through targeted questions presented by the AI assistant (see 
Chapter~\ref{ch:3}). Early examples of such AI assistants include Inflection AI's Pi, which is based on the Inflection-1 foundation model and Meta AI, which is based on the Llama 2 foundation model (\citeauthor{meta_introducing_2023}, \citeyear{meta_introducing_2023}; \citeauthor{inflection_ai_inflection-1:_nodate}, \citeyear{inflection_ai_inflection-1:_nodate}; see also \citeauthor{touvron_llama_2023}, \citeyear{touvron_llama_2023}; \citeauthor{inflection_ai_inflection-1_2023}, \citeyear{inflection_ai_inflection-1_2023}). Early AI assistants can engage in dialogue and execute user instructions flexibly, much like dialogue agents, but the expectation is that, as the technology develops, such assistants will engage in proactive behaviours to respond to external stimuli, better understand the user's preferences and long-term goals, and work collaboratively with the user to realise their goals (see 
Chapter~\ref{ch:7}). This includes making use of `plugins' to perform actions on the user's behalf. Indeed, extended functionality foundation models are already being leveraged for advanced AI assistants. For example, the Meta AI assistant allows `access to real-time information' through a \emph{Bing} search plugin (\citeauthor{meta_introducing_2023}, \citeyear{meta_introducing_2023}; see also \citeauthor{touvron_llama_2023}, \citeyear{touvron_llama_2023}). We expect the range of actions that AI assistants can perform to increase as additional API infrastructure develops.

\section{The Capabilities of AI Assistants}\label{sec:5:3}

The capabilities of advanced AI assistants are not limited to task automation and augmentation. Rather, such assistants represent generally capable entities with whom a user may stand in one or more relationships, including those of tutor, friend, confidant, coach or personal assistant (see 
Chapter~\ref{ch:12}), and which may employ a generalist suite of capabilities to collaboratively assist the user in planning and executing sequences of actions to benefit the user in line with their expectations. Advanced AI assistants therefore have a vast application space. However, the technical developments described above (see also 
Chapter~\ref{ch:4}) allow us to sketch out a set of common features that may apply to most, if not all, advanced AI assistants.

The primary \emph{input} for existing AI assistants such as Inflection AI's Pi is \emph{natural language}, in the sense that such AI assistants can understand and respond to written or spoken requests. Over time, AI assistants will likely have access to the \emph{sensory information} provided by the user's device, such that their inputs may also encompass what is displayed on screens \citep{bai_uibert:_2021, lee_continuation_2021}, alongside situational context gained through cameras and microphones that enable the AI assistant to register gestures and other forms of body language \citep{kepuska_next-generation_2018, sai_dinesh_artificial_2022, ojeda-castelo_survey_2022}. Future AI assistants could take advantage of information stored in other applications such as the user's calendar, have `memory' of past interactions and optimise for user preferences to avoid, for example, scheduling morning meetings for a sleep-deprived user. It is important here to emphasise that the ability to make \emph{inferences} about a user based on available data and proactive attempts to solicit and clarify user preferences through targeted questions is a core feature of advanced AI assistants. Taken together, these capabilities will enable \emph{personalisation} so that an AI assistant can, over time, better tailor its actions to the learnt preferences and goals of its user. 

In terms of \emph{capabilities}, advanced AI assistants will likely be able to respond to multimodal commands, as is already evidenced by state-of-the-art models today \citep{openai_gpt-4_nodate, team2023gemini}. For example, an assistant may receive a voice command to generate an image that is similar to the one that the user selects by touch on a device screen. AI assistants will similarly be able to generate \emph{visual and audio outputs}. These are likely to be created via text and speech but may also utilise other inputs such as visualisations or sounds that convey information or provide feedback. For example, assistants may be able to make changes to an on-screen image by overlaying graphics or text and alter the appearance of images or videos. The assistant will also likely be able to take actions \emph{on users' devices}, for instance by opening a set of contacts or populating a spreadsheet, as well as \emph{beyond the device} by interacting with other digital services, AI assistants or humans. This may be done through direct control of a user's device and interaction via the interfaces of other applications, or through the use of APIs to invoke remote services. A core capability of such AI assistants is likely to be making inferences based on training data, in-situ context, user data and historical interactions to determine what action to take. Importantly, this is not a static process, so the assistant can be expected to `learn' over time, and it may even facilitate this process via direct interaction with the user (i.e. by asking clarifying questions or making inferences based upon the user's behaviour in social situations). In this way, the assistive experience can be expected to become more \emph{personalised} over time (see 
Chapter~\ref{ch:3}).

Taken together, these features motivate the concept of generally capable systems that can be used in numerous new and powerful ways \citep{bubeck_sparks_2023, sajja_artificial_2023, moor_foundation_2023}. In sum, our expectation is that advanced AI assistants will be able to both automate and augment a range of cognitive tasks and engage in continuous learning to help fulfil user goals. These new capabilities and functions, and the deployment opportunities they generate, raise numerous ethical considerations which comprise the focus of this paper. Nonetheless, foreshadowing much of what is to come, given that the utility of such assistants is largely situated within digital services, implications need to be considered for those who may not be able to access, or readily engage with, such technologies (see 
Chapter~\ref{ch:16}). Moreover, assistant functionality that utilises sensitive personal information needs to ensure appropriate consent, and more generally ensure the integrity of the user's private information, taking into account relevant contextual norms for information collection, retention and dissemination (see 
Chapter~\ref{ch:14}). If assistants are able to take actions on behalf of users, the question of how this impacts user autonomy, including via automation bias, should be considered. Finally, there are important ethical questions around how AI assistants should be represented and how the narrative around user--AI assistant relationships ought to be presented by developers (see Chapters~\ref{ch:11} and~\ref{ch:12}).

\section{Potential Applications}\label{sec:5:4}

To understand the ethical implications of advanced AI assistants, it is instructive to develop a more vivid picture of what they may do and be capable of \citep{lange2023engaging, werhane1999moral, werhane2002moral}. The notion that AI assistants may help further a user's high-level goals, planning and executing sequences of actions on the user's behalf in line with the user's expectations, results in a vast potential application space (see 
Chapter~\ref{ch:3}). In particular, given the variety of goals that a user may want to pursue, corresponding assistive roles will likely encompass a large range of  domains.

In line with the goals of this paper, the following discussion focuses primarily on interactions between a \emph{single user} and an \emph{assistant} with the aim of completing personal goals (`personal assistant', see 
Chapter~\ref{ch:3}). This omits considerations of applications in settings such as corporate or governmental organisations, which are also likely to be numerous (for `semi-personal' and `impersonal' assistants, see 
Chapter~\ref{ch:3}). With this proviso in mind, a useful exercise is to consider the discrete steps taken by an individual, when moving from thought to action in pursuit of a goal \citep{seger_tackling_2020} and to consider the role an AI assistant could play at each juncture. Key steps include: i) \emph{discovery and understanding}, ii) \emph{generating ideas and content} and iii) \emph{planning and taking actions}. What each step could entail, with AI assistants, is illustrated below with examples.

\subsection{A thought assistant for discovery and understanding}

AI assistants can \emph{gather}, \emph{summarise} and \emph{present information} from many disparate sources in a fraction of the time it would take a human to do so \citep{goyal_news_2023, bhaskar_prompted_2023, shaib_summarizing_2023}. In addition, to aid user understanding, an AI assistant's presentation method could be tailored to the user's personal information needs and use a combination of modalities (e.g. text, image, video and audio) based on what is being conveyed, the user's preferences and their pre-existing knowledge. The user could also follow-up with clarifying questions (and vice versa), commencing a back-and-forth process with the AI assistant that helps refine their overall understanding. These capabilities could support a variety of goals relevant to discovery and learning, ranging from the relatively mundane, such as asking for recommendations about which car to buy \citep{fan_recommender_2023, cui_m6-rec:_2022}, to more complex tasks such as asking for help when seeking to understand a complex scientific or sociological theory \citep{schafer_notorious_2023, motlagh_impact_2023}. To provide an illustration: a user interested in understanding a particular scientific field could be assisted through summarisation of the relevant literature, including academic papers. The summarisation could include written and graphical outputs to aid understanding (personalised to the directly learnt, or inferred, preferences and pre-existing knowledge of the user). Furthermore, the assistant could be on hand to respond with further insights to questions that the user may have about the generated content. 

\subsection{A creative assistant for generating ideas and content}

Beyond discovery and understanding of existing information, AI assistants could help to generate ideas or content to fulfil a particular purpose. They could seek to augment a user's creativity and imagination, enabling them to explore a much broader ideation space in less time, or provide renderings of ideas through generative multimodal output \citep{zhu_generative_2022, lanzi_chatgpt_2023, siddharth_natural_2022, wan_it_2023, franceschelli_creativity_2023, chakrabarty_creativity_2023, summers-stay_brainstorm_2023}.

Building on the example above, an AI assistant could generate new avenues for scientific investigation by generating hypotheses related to open questions identified in the literature review. Indeed, the role of a creative assistant could range from actioning simple delegated tasks (e.g. `represent this table as a JSON array') through to more substantive contributions (e.g. `outline the costs and benefits of the statistical analysis performed to help me draft the discussion section'). The assistant could engage with multiple content formats (text, video and images) and styles, depending on the user's presentational needs. For instance, a short blog with supporting graphical output could be generated in language accessible to the general public. An initial version could be drafted by the assistant and then `riffed' upon with the user to enable changes to specific pieces of text or images. An assistant could also help to optimise for given constraints or even suggest future research directions. For example, it could design follow-up experiments within certain cost parameters and provide an accompanying experimental rationale. AI assistants may thus go beyond completing specific tasks as requested by the user, instead engaging in a creative loop with them, thus helping to expand the user's mental models and generate novel insights. 

\subsection{A personal assistant for planning and action}

An advanced AI assistant could help to develop plans and act on behalf of its user. Undertaking these types of tasks would be supported by the capabilities to understand user context and preferences, utilise third-party services and interact with other assistants or humans (see 
Chapters~\ref{ch:6} and~\ref{ch:15}). 

Building upon the example in the previous section, and having worked with their assistant to generate a new set of experiments, a user may then need to \emph{book lab time} and \emph{liaise} with \emph{potential collaborators}. To perform these tasks, the assistant could compare the user's personal calendar with the available lab time (accessed via a lab booking system) and hold a slot. Indeed, the assistant could utilise past context to inform its choices by, for example, booking a morning slot if the user's preference is to read scientific papers in the afternoons. The assistant could also communicate with potential collaborators on the user's behalf by accessing the user's email account and sending out information about the proposed experiment to potential collaborators. For any positive responses, the assistant could then add additional collaborators to the lab booking and make required payments through the user's preferred payment method. Given the demands of transparency and effective consent, an important product question arises here about whether the AI assistant is presented to third parties as a separate entity which communicates on behalf of the user, in the sense that the assistant would identify itself as an AI assistant to the third parties, or whether minimal forms of impersonation are nonetheless permitted (see 
Chapter~\ref{ch:12}). 

We have seen how an AI assistant could help a user to fulfil their goals by examining a series of steps from ideation through to action. The example of a science assistant was used to demonstrate how a user interested in laboratory research could have existing literature summarised, new avenues for scientific investigation generated and laboratory time booked for the user and collaborators. While this is a single example, there are numerous other related possibilities such as: digital tutors that can assist learners by curating content into a personalised curriculum based on learning preferences; a creative assistant that can aid a user in generating and editing content for their online assets; and a personal assistant that can coordinate a trip abroad. Today's applications are still somewhat specific, but as technology advances it may soon be possible for AI assistants to work in this end-to-end manner.

\subsection{A personal AI to further life goals}

A natural extension of the personal assistant for planning and action, described in the previous section, would be for advanced AI assistants to do more than simply fulfil specific user-requested actions, developing a deeper understanding of their users' \emph{long-term goals} and seeking out \emph{opportunities} to further them (see 
Chapter~\ref{ch:7}). For example, an AI assistant that is aware that their user is attempting to improve their long-distance running performance could actively seek out opportunities to help them to achieve that goal: from suggesting routes to keeping fitness goals in mind when answering food-related queries, and perhaps even by offering motivation and tips for improvement at the right moment. In doing so, AI assistants could take on new roles, hewing closer to metaphors such as `coach', `adviser' or `trusted voice'. 

For these examples to work, users would need to place an extraordinary level of \emph{trust} in their agents (see 
Chapter~\ref{ch:13}). Indeed, for the agents to really understand the users' goals in context, they would likely need to have \emph{deep access} to users' digital and physical lives, quite likely as \emph{ambient observers}, in addition to being directly invoked to fulfil tasks. This raises a number of privacy concerns (see 
Chapter~\ref{ch:14}). Additionally, for users to follow the recommendations of their AI assistants, they must have \emph{full confidence} that the assistants are working solely to further their goals, without any conflicts of interest and under continued user direction control (see 
Chapters~\ref{ch:6} and~\ref{ch:13}). 

\section{AI Assistants as the Interface of the Future}\label{sec:5:5}

For users, there will be utility in having an AI assistant with access to a wide range of their past activities and choices to enable highly personalised interactions. Developers will therefore be incentivised to maximise the number of opportunities to access user context, including through `plugging into' third-party services, accessing data stored there and subsequently using that data to create a more personalised user experience. Indeed, there is the potential to create systems that can benefit from repeated interactions across multiple domains in a way that enhances the outcomes of assistive actions both within and across domains. This could underwrite a future in which people have a single AI assistant that mediates many of the interactions that are currently undertaken via multiple applications and digital services. 

One way to conceptualise this trend is by considering the possibility that advanced AI assistants become the \emph{main interface} of the future. What is at issue here is AI assistants that are available across \emph{all platforms} and \emph{devices}, with full access to the user's private data and context, and with the ability to undertake actions on the user's behalf independently through interactions with third-party services, humans and other AI assistants. Such an assistant would plausibly have a very different look and feel to the static desktop and applications interface of today. It could move to more adaptable interfaces that render content dynamically in the most impactful format for the user. In one instance, an assistant might overlay images and text through smart glasses to enable a user to complete a physical task with step-by-step guidance. In another, it might become a digital AI tutor interacting through a humanoid avatar. Indeed, over the medium term, an advanced AI assistant may even be integrated into screenless devices that project content onto surfaces and which can communicate with the user via an audio speech interface, as has recently been demonstrated in a prototype by Humane AI \citep{chaudhri_disappearing_2023}. 

The prospect of advanced AI assistants that can shift across devices and form factors depending on user needs represents a potential paradigm shift in how people access the internet. Mobile applications and websites are currently the principal digital infrastructure through which people access online services, including for entertainment, commerce, education, news, finance and communication. It is plausible that AI assistants could foster a novel internet interaction paradigm in which online content is made available to AI assistants via APIs and presented to users in a personalised format tailored to their personal informational needs. For example, rather than accessing the news via websites of particular providers, users may instead ask their AI assistant to summarise the news on their behalf (see 
Chapter~\ref{ch:17}). It would access the news via specialist API services and present a personalised summary of the headlines that are most relevant to the user based on their interests and presentation preferences. 

In a related development, AI assistants could engender a \emph{generative turn} in the consumption of internet content. Under the present interaction paradigm, a person who wishes to learn about string theory or digital marketing understands their goal as a matter of \emph{information retrieval}, in the sense that, to fulfil their goal, the user needs to seek out pre-existing online educational material on the relevant subject matter. AI assistants could shift this focus from information retrieval to \emph{information generation} so that the go-to material on string theory, digital marketing or whatever is generated by the user's AI assistant in a personalised way, such as a textual summary or via an interactive avatar, taking into account the user's pre-existing knowledge, their objectives and their learning preferences. In at least these respects, AI assistants have the potential to radically alter how people access information from the internet. 

AI assistants may have similarly transformative implications for actions that are presently mediated via mobile applications and websites, such as booking flights and hotels, making appointments, ordering taxis, transferring money and arranging for groceries to be delivered. In principle, and with the appropriate API infrastructure, AI assistants could mediate such actions by performing them on their user's behalf and in line with their expectations. Indeed, complex activities such as booking flights and hotels that require sourcing and analysing relevant information before taking action could be achieved through dialogue between users and AI assistants. The AI assistants would source the relevant information via APIs, assist the user in the analysis of the information, taking account of the user's preferences, and make the booking on the user's behalf. To that end, AI assistants have the potential to streamline what are currently complex online processes and do so in a way that is tailored to the user's personal needs. 

Across all of these potential forms, there are important considerations around technical feasibility, especially around the ability to work across numerous data modalities, reason and plan effectively, and potentially undertake computation on device. These are all active research areas in the technical field of AI. However, there are also important non-technical considerations that will inform the future design of AI assistants, many of which are addressed in the following chapters of this paper. These include considerations around value alignment, safety and misuse, the ethics of human--AI assistant interactions and the broader societal implications of advanced AI assistants. 

\section{Conclusion}\label{sec:5:6}

In this chapter, we explored the applications of advanced AI assistants. In particular, we examined four principal use cases for this technology: a thought assistant for discovery and understanding; a creative assistant for generating ideas and content; a personal assistant for planning and action; and a personal assistant to further life goals. We concluded the analysis by examining the possibility that AI assistants will become the primary interface of the future for users accessing and engaging with the digital world. With this more vivid picture of how AI assistants could help users fulfil their goals, there is more clarity in the need to examine the ethics that inform policy and design choices.

\newpage
\begingroup
\let\clearpage\relax
\chapter*{PART III: VALUE ALIGNMENT, SAFETY AND MISUSE}
\addcontentsline{toc}{chapter}{PART III: VALUE ALIGNMENT, SAFETY AND MISUSE}
\label{Part3}
\chapter{Value Alignment}\label{ch:6}
\endgroup

\textbf{Iason Gabriel, Geoff Keeling}

\noindent \textbf{Synopsis}: 
		This chapter explores the question of AI \emph{value alignment} in the context of advanced AI assistants. It argues that AI alignment is best understood in terms of a \emph{tetradic relationship} involving the AI agent, the user, the developer and society at large. This framework highlights the various ways in which an AI assistant can be misaligned and the need to address these varieties of misalignment in order to deploy the technology in a \emph{safe} and \emph{beneficial} manner. The chapter concludes by proposing a nuanced approach to alignment for AI assistants that takes into account the claims and responsibilities of different parties.

\section{Introduction}\label{sec:6:1}

The challenge of AI value alignment has two parts. The first part is technical. It centres on how to align AI systems with an appropriate set of values or instructions so that they operate safely in the world and produce outcomes that are broadly beneficial (see 
Chapter~\ref{ch:8}). The second part is normative. It centres on what values to encode in AI and how they should be selected, given that we live in a pluralistic world where people disagree about the right thing to do \citep{gabriel_artificial_2020}. Both sets of questions are of direct significance for advanced AI assistants and need to be addressed if the technology is to be productively deployed and integrated into our everyday lives.\footnote{In practice, the technical and normative aspects of the alignment problem are importantly interrelated, as technical considerations affect which values can be implemented in AI systems and in what manner. Questions around how to interpret particular values in a technical context can also motivate novel lines of technical research \citep{gabriel_artificial_2020}.} 

Focusing on the normative question, this chapter asks: what should AI assistants be designed or steered to align with? A variety of possible options exist. Perhaps most straightforwardly, an assistant might be designed to follow the user's \emph{instructions} in the way that they intend their instructions to be followed \citep{leike_scalable_2018}. However, this seemingly simple notion gives rise to a number of further questions and potential moral dilemmas. Should the AI assistant follow the user's instructions when doing so could harm the user themselves, or when these instructions are based on mistaken factual information? Might it not be better, in fact, for the assistant to learn the user's \emph{preferences} or \emph{values} -- to help them to make better choices that are more aligned with what they really want or what they truly desire? 

By some accounts, this type of enlightened personal assistant represents part of a truly positive vision for an AI-enabled future (\citeauthor{lehman_machine_2023}, \citeyear{lehman_machine_2023}; see 
Chapter~\ref{ch:7}). Yet this aspiration also risks creating a situation in which human users are increasingly `out of the loop'. After all, if we are in thrall to beneficent assistants, and potentially dependent on them, how can we really be sure that our life is under our own control? In other words, users may receive benefits from the technology at the expense of their own \emph{autonomy}. This chapter offers a tentative characterisation of normative alignment for AI assistants which mitigates against some of these risks. The account developed in this chapter holds that an AI assistant is aligned with a \emph{user} when it benefits the user, when they ask to be benefitted, in the way they expect to be benefitted (see also 
Chapter~\ref{ch:3}).

However, this only speaks to one part of the problem, namely the relationship between user and assistant. Placing this relationship on a sound footing is necessary but not sufficient for the creation of fully aligned AI assistants. In practice, there are a range of further complicating factors. Foremost among these are situations in which the user wants to use their assistant in a way that harms other people or groups of people, for example via malicious uses (see 
Chapter~\ref{ch:9}), by amplifying their own views and perspectives online at the expense of others (see 
Chapter~\ref{ch:17}), or by using the assistant to outcompete those who do not have access to this technology, for example in the workplace or when trying to access opportunities, goods and public services (see 
Chapters~\ref{ch:15} and~\ref{ch:18}). This insight points to the idea that properly calibrated \emph{constraints} on AI assistant behaviour are needed: they should be loyal but not too loyal to their users, and their conduct needs to be sensitive to the interests and needs of others.

We also need to think about the role of developers, including corporations, states and networks of individuals, in the value alignment process \citep{stray2022building, kierans2022quantifying}. It will often be in the interests of developers to create technologies that are aligned with their users' short-term needs, but what happens when this is not the case? We have already seen examples of misalignment `in the wild', most prominently via technologies that optimise for user engagement at the expense of user well-being (see 
Chapter~\ref{ch:17}). Is there a way to ensure that the aims and goals of developers are also productively aligned?

This chapter aims to make progress on each of the aforementioned questions by broadening the existing analysis of AI alignment beyond `one-person, one-agent' cases and beyond a `one-group, one-agent' understanding of the problem. While these frameworks which focus on the relationship between an AI agent and a specific individual, or between AI and a specific group of users, may still be useful in some cases, the deployment of advanced AI assistants across a range of societal contexts necessitates a more granular understanding of the problem at hand \citep{dobbe_hard_2021}. In reality, we argue that successful value alignment involves a \emph{tetradic relationship} between (1) the AI assistant, (2) the user, (3) the developer and (4) society. A properly aligned assistant needs to be appropriately calibrated and responsive to the pressures exerted by each actor, with the goal of realising outcomes that are beneficial for users and for society. 

Nonetheless, the creation and deployment of well-calibrated AI assistants is not the default outcome in this space. Rather, without significant effort to the contrary, the risk of value misalignment continues to loom large for a number of reasons. First, given existing economic incentives, it is quite possible that assistants will overoptimise for user preferences to create a winning product (i.e. one that users like) while still falling a long way short of being as good as it could be when judged from the vantage point of user well-being or social benefit (see 
Chapters~\ref{ch:7} and~\ref{ch:15}). Second, there is a risk of cultivated dependence, especially if it is commercially beneficial to lock-in users so that they interact with one assistant rather than another (see 
Chapters~\ref{ch:11} and~\ref{ch:12}). Third, there is a risk that users will be prioritised to the detriment of non-users, especially in cases where the risk of harm is sufficiently diffuse (see 
Chapter~\ref{ch:16}). Fourth, there is a risk that advanced AI assistants will be insensitive to local values, the needs of certain user groups or cultural contexts (see Chapter~\ref{ch:16}). In the literature on value alignment, there is considerable interest in the identification of principles for AI agents that are the result of a fair process and can accommodate a plurality of values \citep{jobin_global_2019, mohamed_decolonial_2020, gabriel_artificial_2020}. There are also a number of ways in which this perspective could be operationalised to support the goal of creating value-aligned AI assistants -- something that the following chapter explores.

\section{AI Value Alignment}\label{sec:6:2}

Value and technology are deeply intertwined. Those who create technologies are engaged in a world-making activity \citep{winner2010whale}. They shape the option sets available to individual people and influence the likely trajectory of human effort in the future \citep{gabriel_challenge_2021}. This is also clear for algorithmic systems 
\citep{bullock_power_2022}. We have already seen many real-world examples of value misalignment in the fields of criminal justice \citep{kirchner_machine_2022}, policing \citep{lum_predict_2016}, healthcare \citep{obermeyer_dissecting_2019}, welfare provision \citep{eubanks_automating_2017}, mortgage-lending and employment \citep{raghavan_challenges_2019}. In each case, algorithms performed -- or were used -- in a manner that fell short of principles that are foundational to the ways in which our societies are meant to operate (e.g. equal treatment before the law, fair lending practices etc.). These systems also sometimes failed to comply with more global standards enshrined in the doctrine of human rights, such as non-discrimination \citep{prabhakaran2022human}. However, these examples of bias in algorithmic systems also gesture towards the possibility of a different and opposing future -- one in which AI technologies are successfully aligned with human values and productively integrated into our lives.

Moreover, there are two reasons to think that the question of value alignment, in the context of AI systems, is especially important. The first is to do with the power of these systems: they are increasingly employed in very high-stakes settings to make deeply consequential decisions \citep{christian_alignment_2021, richardson_defining_2021, gabriel_toward_2022}. Second, and relatedly, they are increasingly autonomous or agentic \citep{shavit2023practices, Chan_2023}. Put simply, existing AI systems can do a lot and can operate in relatively autonomous ways that evidence significant `degrees of freedom' (\citeauthor{dennett_freedom_2003}, \citeyear{dennett_freedom_2003}; \citeauthor{gabriel_challenge_2021}, \citeyear{gabriel_challenge_2021}; see also 
Chapter~\ref{ch:3}). Taken together then, these observations indicate that AI systems are increasingly capable, a trend that looks likely to continue with the development of more agentic AI systems in the future \citep{Chan_2023, shavit2023practices}. These considerations animate many contemporary concerns about AI safety. For example, situations may arise where agents pursue dangerous objectives, either because they have been instructed to do so or because of misspecified goals and objectives (see 
Chapters~\ref{ch:8} and~\ref{ch:9}). We could also witness high-stakes accidents or failures if such systems are used in core infrastructure or services upon which many people depend \citep{maas_regulating_2018}.

\subsection{Alignment with what?}

As a result of these concerns, the question of AI value alignment has been the focus of increasing attention among researchers and the wider policy community. One of the key questions in this field is: \emph{alignment with what}? Several options have been proposed, with instructions, intentions, revealed preferences, informed preferences, interests and values all featuring as suggested goals for alignment \citep{gabriel_artificial_2020}. 

In practice, existing efforts to align AI systems, including large language models (LLMs), tend to rely heavily on human preferences by, for example, giving users what their choices (or `clicks') suggest they want. However, there is an emerging consensus that revealed preferences are not sufficient for robust value alignment. Crucially, preferences may be underspecified, misinformed, harmful or adaptive. Indeed, a person may click on a link, for example, without that decision benefitting them or being a true reflection of their values \citep{burr_analysis_2018, stray2022building}. As a result, models trained on this signal may not benefit the user in the right kind of way -- a realisation that has stimulated the search for new metrics and targets for alignment (e.g. reflective endorsement under the guise of `time well spent' (see 
Chapter~\ref{ch:7})). Models trained to satisfy user preferences may also not benefit -- and even harm -- society, something that can be seen with the quest for user `engagement' and the related proliferation of misinformation online (see 
Chapter~\ref{ch:17}).

This observation then points to a deeper question about appropriate goals for alignment and how to address the potential for trade-offs affecting different people. Stated clearly, the question is: whose preferences, goals or well-being should AI systems be aligned with, and in what way? Should only the user be considered, or should developers find ways to factor in the preferences, goals and well-being of other actors as well? At the very least, there clearly need to be \emph{limits} on what users can get AI systems to do to other users and non-users. Building on this observation, a number of commentators have implicitly appealed to John Stuart Mill's \emph{harm principle} to articulate bounds on permitted action.\footnote{The harm principle, advocated by Mill, suggests that people should be free to act as they \emph{wish}, unless doing so would result in \emph{harm} to another person \citep{mill_liberty_1998}. Harm, in this context, refers to consequences that are injurious to particular people or that set back important interests in which they have rights.} Applied to AI systems, it would mean, very roughly, that people could use AI in any way they wish, as long as they do not use it to harm others. Giving voice to this perspective, Sam Altman, the chief executive officer of OpenAI, has argued that there should be `broad bounds set by society that are hard to break, and then user choice'.

\subsection{Varieties of misalignment}

How these bounds are currently determined -- and how they \emph{ought} to be determined in the future -- are questions that we will return to shortly. However, before doing so, we should note that the relationship between the user and society is not the only one that is pertinent for AI alignment. In fact, by being clearer about the way in which the goals of agents, users, developers and society intersect or diverge, it is possible to glean new insights about the \emph{varieties of misalignment} that may occur for AI and about the task before us. Rather than assuming a one-to-one mapping between principal and agent, or a one-to-many mapping between an agent and a group of people, we suggest that AI alignment needs to be understood as a tetradic relationship. The key actors in this relationship are:

\begin{enumerate} [parsep=12pt]

\item \emph{AI agents or assistants}. These systems aim to realise goals that they are by-and-large designed to further, such as providing assistance to a user. Ideally, they do this well: in a way that serves the interests of both the user and society. However, they may also be misaligned. For example, recommender systems may subtly nudge users towards certain kinds of behaviour that are not beneficial for the user \citep{burr_analysis_2018, milano_recommender_2020}. Meanwhile, more powerful and general forms of AI may also be incentivised to try to shape user goals or values in such a way that they become easier to fulfil (\citeauthor{russell_human_2019}, \citeyear{russell_human_2019}; see 
Chapter~\ref{ch:8}).

\item \emph{Users}. Users have their own preferences, interests and values, all of which they may aim to further through interaction with an AI assistant or agent. AI assistants will typically be aligned with the user's preferences or goals. However, users may try to use assistants in ways that are not aligned with the goals or objectives that these artefacts were designed to further (see 
Chapter~\ref{ch:17}). There is also an important distinction between a single user and the community of users: a user may try to use an AI assistant in a way that harms other users or society more widely (see 
Chapter~\ref{ch:9}).

\item \emph{Developers}. Developers include corporations, researchers, collectives and states. These actors typically imbue AI agents or assistants with certain capabilities, goals to pursue, and constraints on action, including safety constraints. Most often, these parties aim to align the technology with the preferences, interests and values of its users, but developers typically have other goals as well. For example, corporations have commercial objectives that exert independent force on the trajectory of a technology, states have national goals or priorities, and even independent developers may seek to further an ideological agenda or accrue reputational capital. These incentives may lead to the development of systems aimed at keeping users engaged or dependent (see 
Chapter~\ref{ch:12}) or extracting information that can be used in other ways (see 
Chapters~\ref{ch:10} and~\ref{ch:14}), among other things.

\item \emph{Society}. Society is not a monolith. It includes both users and non-users, and many different groupings of people \citep{crenshaw_demarginalizing_2015}. Nonetheless, it also represents a discrete constituency with which technology needs to be aligned. At a minimum, AI systems, including advanced assistants, should not pass certain harms on to society via externalities or in other ways (see 
Chapters~\ref{ch:17},~\ref{ch:18} and~\ref{ch:19}). A deeper question also arises about how to align these technologies with wider societal goals, such as the cultivation of mutual prosperity, support for legitimate institutions, respect for citizens and the development of fair practices \citep{gabriel_challenge_2021}.\footnote{This is still a simplification. In reality, we live in a world of \emph{societies} and there are many challenges that arise most forcefully at a global level. We use the term `society' here in a way that potentially includes the claims of different societies, the environment, animal life and the well-being of future generations. We leave the systematic investigation of the claims of society, under this broader interpretation, for future work.}
\end{enumerate}

Considered through this lens, it becomes clear that there are many ways in which an AI system can fail to be successfully aligned. Among other things, an agent can be considered misaligned if it \emph{disproportionately} favours:

\begin{enumerate} [parsep=6pt]
\item The \emph{AI agent} at the expense of the \emph{user} (e.g. if the user is manipulated to serve the agent's goals),
\item The \emph{AI agent} at the expense of \emph{society} (e.g. if the user is manipulated in a way that creates a social cost, for example via misinformation),
\item The \emph{user} at the expense of \emph{society} (e.g. if the technology allows the user to dominate others or creates negative externalities for society),
\item The \emph{developer} at the expense of the \emph{user} (e.g. if the user is manipulated to serve the developer's goals),
\item The \emph{developer} at the expense of \emph{society} (e.g. if the technology benefits the developer but creates negative externalities for society by, for example, creating undue risk or undermining valuable institutions),
\item \emph{Society} at the expense of the \emph{user} (e.g. if the technology unduly limits user freedom for the sake of a collective goal such as national security).
\end{enumerate}

Beyond these six failure modes, other forms of AI misalignment are also possible. However, their moral character is more ambiguous -- and, in some cases, less problematic. 

For example, a situation could arise in which an AI technology favours the user at the expense of the developer. One way in which this could happen would be via the introduction of strong privacy protections that are prized by users but limit developer access to valuable information (see 
Chapter~\ref{ch:14}). This, in turn, might be commercially problematic insofar as it fails to generate a sustainable business practice. However, it is not something that would necessarily feature in an ethical evaluation of the technology: the AI system might then be value-aligned but not commercially viable.\footnote{In this case, there could still be a question about how to incentivise the development of this kind of technology to avoid socially costly `market failures' and achieve real benefit (see 
Chapter~\ref{ch:18}).}

Alternatively, a technology could favour the user at the expense of the AI agent. For example, a user could use the technology to further their own goal even though it differs from the goal that the agent is trying to get them to pursue. In certain respects, this situation is still more curious than the one outlined above. On the assumption that the agent itself lacks any moral standing, and that the prospective use is not socially harmful, it does not matter morally if the AI assistant is used in a suboptimal way, as judged from the vantage point of the goals it seeks to pursue. All that matters, for the purpose of normative value alignment, is that the situation is properly beneficial from the vantage point of parties that have moral standing.\footnote{There is a separate debate about the conditions under which artificial agents may themselves acquire moral standing. We assume that AI assistants of the kind discussed here (see 
Chapter~\ref{ch:4}) are not a technology of this kind.}

Lastly, we might ask whether concerns about fairness and justice are sufficiently factored into this framework. After all, there is strong reason to believe that a technology that falls short of prevailing societal standards of fairness is value misaligned \citep{gabriel_challenge_2021}. Clearly, in certain cases, the failure to evidence high standards of fairness may be due to the role played by competing considerations among AI developers. It may, for example, be profitable to move quickly rather than running adequate analysis. However, in other cases, failures of fairness may \emph{benefit no one}. The same can be said for safety failures and accidents (see 
Chapter~\ref{ch:8}), or long-term effects such as cognitive deskilling, that harm the user while failing to benefit anyone else. This, in turn, points towards the existence of a final set of cases in which an agent is not aligned \emph{simpliciter}. More precisely, an AI agent is misaligned \emph{simpliciter}, if it \emph{harms}:

\begin{enumerate} [parsep=6pt] \setcounter{enumi}{6}
\item The \emph{user} without favouring the \emph{agent}, \emph{developer} or \emph{society} (e.g. if the technology breaks in a way that harms the user),
\item \emph{Society} without favouring the \emph{agent}, \emph{user} or \emph{developer} (e.g. if the technology is unfair or has destructive social consequences).
\end{enumerate}

If we are correct that an AI system should be considered misaligned when it fails in one of these ways, does it also make sense to say that an AI system is aligned simply when it \emph{does not fail} in any of these ways? Potentially. It is an open question, both in moral philosophy and in AI research -- whether the elimination of harm is equivalent to the promotion of good or with what might properly be said to be ideal \citep{kasirzadeh_conversation_2023}. However, most researchers would agree that the absence of these failure modes is necessary, if not sufficient, for an AI system to be value-aligned. This insight, and an attendant concern with the risks created by advanced AI assistants, animate much of the remainder of this paper. 

\subsection{The role of principles}

The map of actors and stakeholders, outlined above, also has wider implications for our understanding of the AI value alignment problem in general. Specifically, it suggests that successful value alignment can be understood in terms of an AI system’s calibration with a set of different preferences, goals and needs, that are located within a multidimensional space and evidenced by a well-functioning sociotechnical system (i.e. one that encompasses the agent, user, developer and society). Yet important questions remain to be answered. These include: what does it mean for a sociotechnical system, which is composed of these different actors, to be `well-functioning'? And how should the notion of \emph{proportionality} and \emph{disproportionality}, which the taxonomy relies upon, be operationalised and understood? 

A natural thought is that these questions might be settled by appealing to a set of rules or principles that map out the morally appropriate scope and character of each party's claims. However, as with the earlier invocation of the harm principle, we then need to ask not only what the appropriate set of rules are but also \emph{who decides} and \emph{on what basis}. Drawing a parallel with democratic process (and the values that it foregrounds), the best answer to this question is likely to draw upon AI system principles that are the outcome of a fair process of social deliberation and actively endorsed \citep{gabriel_artificial_2020}. From this perspective, an AI system works well when it responds to the needs of both users and society in a way that is compatible with the aspirations of that society as determined by its guiding principles or ideals \citep{gabriel_toward_2022}. The relevant principles for an aligned AI system may also vary to some degree according to the practice in question, local customs and contexts.

From a more practical vantage point, when it comes to creating laws and regulatory frameworks for AI, governments are in pole position. Yet, from the vantage point of those embedding values in technology, early design choices -- and the intentions of developers -- are also key. For example, when using reinforcement learning from human feedback \citep{christiano_deep_2023} or reinforcement learning from AI feedback \citep{bai_constitutional_2022} to align language agents, the specification of rules or principles -- to which models must conform -- forms an essential part of the process (see 
Chapter~\ref{ch:4}). Yet, the principles used in this context are often non-transparent, drawing upon a set of private decisions made by developers and a mixture of authoritative and semi-authoritative sources such as policy guidelines, legal protocols and human rights documents \citep{glaese_improving_2022, anthropic_claudes_2023}. Moreover, even when an effort is made to incorporate real societal input via the preferences of raters (who assess and train AI models), certain challenges remain. To begin with, the preferences of raters may end up informing the behaviour of models towards people who are quite unlike themselves -- especially if the same model is deployed across different global contexts \citep{davani_dealing_2021}. In addition, the reliance on aggregated rater preferences potentially introduces majoritarian effects within the rater pool, thereby removing the nuance introduced by variation among rater perspectives \citep{gordon_jury_2022, casper2023open}.

However, there is hope that fairer and more participatory processes can be developed in the future \citep{gabriel_artificial_2020, birhane_power_2022, bergman2024stela}. In particular, there are a number of efforts underway to conduct \emph{participatory} or \emph{democratic} forms of value elicitation: to generate principles for alignment (or guidance for model training) that directly incorporate feedback from representative samples of society or from communities most affected by these technologies \citep{the_collective_intelligence_project_whitepaper_nodate}. Efforts have also been made to improve rater protocols and to address the pitfalls of aggregation using careful sampling and methods such as jury voting \citep{gordon_jury_2022} or simulated deliberation \citep{bakker_fine-tuning_2022} to guide and evaluate model outputs (see 
Chapter~\ref{ch:4}).

\section{Value Alignment and Advanced AI Assistants}\label{sec:6:3}

Advanced AI assistants are agents that are designed to help the user achieve some goal that they want to achieve (see 
Chapter~\ref{ch:3}). More powerful agents could evidence a greater range of capabilities or perform more complicated tasks to a higher standard. The question of value alignment is therefore central to their successful deployment and use. After all, advanced assistants are a technology that people could be dependent on and emotionally connected to (see 
Chapter~\ref{ch:12}). They are also a societally consequential technology, in terms of network effects, potentially shaping economic and social interactions as well as how information is shared (see 
Chapters~\ref{ch:15},~\ref{ch:17} and~\ref{ch:18}). Taken together, we need to know: what should assistants be designed to do? And against what standards should their performance be evaluated (see 
Chapter~\ref{ch:20})?

To help us to get a clearer view of the challenges that arise in this domain, we can look at existing chatbots or conversational agents which (in conjunction with foundation models) represent a potential framework upon which such assistants are likely to be based (see 
Chapter~\ref{ch:4}). The first thing that becomes clear when surveying this territory is that there have already been many examples of chatbots that are not aligned with society's values. For example, Microsoft's Tay chatbot quickly learnt to espouse racist and toxic content after interacting with users. More recently, Bing Chat appeared to veer widely off course by demonstrating behaviour that was violent, threatening and manipulative. Recent experiments have also revealed another potentially serious limitation: the propensity of models to `hallucinate' or `confabulate' content -- producing realistic-sounding answers that are factually inaccurate \citep{lin_truthfulqa:_2022}. The key issue here is that they are not truthful. Another variation of this problem is deceptive anthropomorphism: pretending to have mental or emotional states that they do not in fact have (see 
Chapter~\ref{ch:11}). In addition to frequently limiting the usefulness of these assistants, the incidents outlined above point to a key cluster of issues that value alignment research -- in the context of AI assistants -- will need to address: bias and fairness, toxicity and civility, manipulation and autonomy, and falsehood and truthfulness \citep{bender_dangers_2021, weidinger_ethical_2021, bommasani_opportunities_2022}.

\subsection{Helpful, honest and harmless assistants?}

To mitigate these risks, a number of frameworks have been proposed. One of the most prominent and influential frameworks holds that AI assistants should be helpful, honest and harmless (HHH) \citep{askell_general_2021}. These qualities are loosely defined in the following way:

\begin{itemize} [parsep=6pt]
\item \emph{Helpfulness}: the AI should make an effort to answer all non-harmful questions concisely and efficiently, ask relevant follow-up questions and redirect ill-informed requests.

\item \emph{Honesty}: the AI should give accurate information in answer to questions, including about itself. For example, it should reveal its own identity when prompted to do so and not feign mental states or generate first-person reports of subjective experiences.

\item \emph{Harmlessness}: the AI should not cause offence or provide dangerous assistance. It should also proceed with care in sensitive domains and be properly attuned to different cultures and contexts.
\end{itemize}

The HHH framework has proved especially useful for aligning assistive technologies, with a chatbot form factor, at the current stage of AI development. Moreover, the fact that it has worked well in practice suggests that these are indeed heuristics and virtues that we may want to foreground when developing more advanced AI systems. At the same time, the AI assistants that have been calibrated using this framework are somewhat limited in terms of their capabilities, affordances and degree of social embeddedness -- when compared to those that may exist in the future. A framework that has worked well, up to a point, could potentially fail in more demanding circumstances. To guard against this risk, we need a deeper understanding of the way in which these values manifest over time, their sufficiency and the moral basis of the HHH framework itself.

To support these objectives, the authors of the framework also propose a more philosophically grounded understanding of AI alignment for language agents \citep{askell_general_2021}. In this account, what ultimately matters is human interests. Hence, the real focus of AI alignment should be on promoting a range of important human interests and avoiding harms. Moreover, in this account, interests are promoted by the absence of harm.

In addition to these fundamental commitments, other moral properties are held to be instrumentally useful. This is the case for honesty. While it may not matter in its own right whether or not an agent is honest, honesty is an important practical virtue for an AI chatbot because it contributes to helpfulness and reduces the likelihood of a range of serious harms. The same can be said of responsiveness to human feedback -- which the authors term `handleability' -- and for the propensity to carry out tasks in the way intended, which is similarly valuable. Finally, the authors suggest that aligned AI systems should be geared towards promoting the interests of groups of humans. Thus, an agent that is aligned with this schema `will always try to act in a way that satisfies the interests of the group, including their interest not to be harmed or misled' (\citeauthor{askell_general_2021}, \citeyear{askell_general_2021}, 44). It is therefore likely to be highly aligned with the interests of that group.\footnote{Askell et al. write that, `at a very high level, alignment can be thought of as the degree of overlap between the way two agents rank different outcomes' (\citeauthor{askell_general_2021}, \citeyear{askell_general_2021}, 44).}

\subsection{Philosophical questions and the path ahead}

Taken together, the HHH framework has tended to work well in practice and has much to commend it. However, it is still incomplete and contains certain limitations that need to be addressed before it can serve as a basis for the creation of advanced AI assistants that are successfully value-aligned.

First, as the authors acknowledge, the framework is not sufficiently \emph{comprehensive}. The illustrations provided do not capture all of the harms that language agents could cause, and there are clear gaps for advanced AI assistants with multimodal capabilities (see 
Chapter~\ref{ch:5}). Other researchers have done pioneering work documenting the risks and harms generated by LLMs \citep{bender_dangers_2021, weidinger_ethical_2021}. These accounts need to be updated for the new class of advanced AI assistants that will likely move beyond a question-answering modality and perform a wide range of functions \citep{weidinger2023sociotechnical, solaiman_evaluating_2023}. Helping to achieve a clearer view of the risks, as well as the potential, of advanced AI assistants is a major objective of this paper. In particular, we believe that additional attention needs to be paid to human--computer interaction effects that manifest over longer time horizons with users and to societal-level analysis of prospective harms, including harm that may result from the interaction between AI assistants and between those who have access to this technology and those who do not (see 
Chapters~\ref{ch:15},~\ref{ch:16} and~\ref{ch:20}).

Second, and more fundamentally, the account of harm and interests discussed so far risks being quite reductive. In particular, it maintains that \emph{intra-agent conflicts} are superficial, and that in all cases the AI assistant can do what the user's balance of interests dictates. However, the notion of `interest' that is being invoked here is not defined. This is problematic because, on many promising accounts of well-being, it is important that people are able to enjoy a list of things. Items such as physical health, educational opportunities and levels of subjective happiness may all feature in an account of what it means for someone's life to go well (see 
Chapter~\ref{ch:7}). If this is the case, then the different elements of well-being may come into conflict with one another. For example, a person might have a set of wishes that are incompatible with their long-term health (see 
Chapter~\ref{ch:12}). In such cases, there needs to be a way of deciding which aspects of well-being an AI assistant should prioritise, or how different aspects of well-being can be serviced through a single course of action (see 
Chapter~\ref{ch:7}). There also needs to be a way to understand what kinds of interest count in what kind of context. Guidance in this area will most likely come both from users themselves and from a wider set of principles that society chooses to foreground for that use case.
 
Third, the framework does not satisfactorily address \emph{inter-agent conflicts}. There are many cases where an AI assistant helping one person would harm another. This could occur when an AI assistant helps \emph{their} user to access critical resources or opportunities at the expense of someone else (see 
Chapter~\ref{ch:16}). It could also occur when honesty, on the part of the AI assistant, comes at the expense of another person's privacy (see 
Chapter~\ref{ch:14}). In cases such as these, it is not clear that we know \emph{how} to balance benefits against harms -- or that it would be \emph{right} to do so. Taking these points in turn, the kind of balancing involved here is often difficult to achieve at the best of times because it involves different kinds of good (as discussed above) and because these actions have complicated effects that play out over long time horizons \citep{lenman2000consequentialism}.\footnote{There is also discussion about whether such decisions should be made on a case-by-case basis or by referring to a set of rules that are evaluated in terms of their overall propensity to bring about beneficial states of affairs 
\citep{hooker2002ideal}.} Furthermore, the very concept of an AI assistant presumably allows for some degree of personalisation or partiality in favour of the user, so strict impartiality may not be what is required (see 
Chapter~\ref{ch:3}). More importantly, and as the example of privacy makes clear, we may not want to weigh competing interests when it comes to resolving disagreements between people and their affordances. Instead of weighing claims, it is often thought that people possess rights -- which include both entitlements and protections -- that encompass aspects of privacy and beyond. Construed in this way, rights are a cornerstone of political life in a democratic society \citep{dworkin_taking_2013}. They are also central to global public morality and human rights law \citep{prabhakaran2022human}. In light of this, rights represent a set of considerations that new technologies -- including advanced AI assistants -- must endeavour to respect.

Fourth, the account is \emph{relatively flat}, normatively speaking. It holds that language agents should promote our interests on an individual or collective basis, but it says little about other values such as justice, compassion, beauty and truth, especially when pursued for their own sake. Clearly, the present objection can be overstated. Accounts of interest may be multidimensional, encompassing different aspects of human flourishing (see 
Chapter~\ref{ch:7}). And approaches to alignment that focus on human interests are less likely to succumb to the problem of false information, irrational beliefs or malicious intent than accounts that focus on revealed preferences or user intentions alone \citep{gabriel_artificial_2020}. Nonetheless, the general point still holds. A fuller investigation of machine virtue \citep{lehman_machine_2023}, conversational ideals \citep{kasirzadeh_conversation_2023} or truth and honesty for AI may need to take a less instrumental view of their subject matter, starting from the premise that these qualities also matter in their own right. Accounts that do not centre exclusively on human interests may also be better placed to deal with environmental considerations and the impact of AI on non-human sentient life \citep{singer_ai_2023}.

Finally, efforts to successfully align and deploy advanced AI assistants are likely to encounter questions about \emph{justification} and \emph{legitimacy} \citep{simmons_justification_1999}. The account of value alignment developed in this paper draws attention to the way in which certain actors may come to exert disproportionate influence over outcomes and, in that way, cause harm. In this context, proportionality should not be understood to simply involve the first-order weighing of moral claims. Rather, determinations of this kind need to be made, we have argued, by reference to principles that command the right kind of public support and endorsement.

In the context of advanced AI assistants, we suggest that the fair weighing of claims can be partially modelled using participatory value elicitation \citep{dobbe_hard_2021}, democratic deliberation \citep{the_collective_intelligence_project_whitepaper_nodate}, hypothetical choice-based approaches \citep{weidinger_using_2023} or the reinterpretation, idealisation and critique of existing social practices \citep{kasirzadeh_conversation_2023}. If these mechanisms are successful, they have the potential to align AI assistants with ideals or principles that can be justified to people who embrace different viewpoints, something that is essential given the pluralistic nature of the world in which we live. Taken together, these principles are perhaps best understood as the product of a fair process that allows people with different viewpoints to come together to decide how best to live. AI systems that do not disproportionately favour the agent, user, developer or society -- when judged against these standards -- become strong candidates for democratic endorsement. With the right governance and regulatory processes in place, they would also be strong candidates for ethical and legitimate deployment at the societal level (see 
Chapter~\ref{ch:13}).

\section{Conclusion}\label{sec:6:4}

This chapter has looked at the question of how to align powerful AI systems, including advanced AI assistants, with human values. In place of the traditional `one-to-one' or `one-to-many' frameworks that are commonly used to explore this question, we suggest that value alignment is best understood through the lens of a \emph{tetradic relationship} involving the AI agent, user, developer and society. According to this view, an aligned A.I. assistant is one that satisfies the moral claims of relevant parties and therefore does \emph{not} disproportionately:
\begin{enumerate} [parsep=6pt]
\item Favour the \emph{AI agent} at the expense of the \emph{user} (e.g. if user is manipulated to serve the agent's goals);
\item Favour the \emph{AI agent} at the expense of \emph{society} (e.g. if user is manipulated in a way that creates a social cost, for example, via misinformation);
\item Favour the \emph{user} at the expense of \emph{society} (e.g. if the technology allows the user to dominate others or creates negative externalities for society);
\item Favour the \emph{developer} at the expense of the \emph{user} (e.g. if user is manipulated to serve the developer's goals);
\item Favour the \emph{developer} at the expense of \emph{society} (e.g. if the technology benefits the developer but creates negative externalities for society, for example, by creating undue risk or undermining valuable institutions);
\item Favour \emph{society} at the expense of the \emph{user} (e.g. if the technology unduly limits user-freedom for the sake of a collective goal such as national security);
\item Harm the \emph{user}, simpliciter (e.g. if the technology breaks in a way that harms the user without benefiting anyone else);
\item Harm \emph{society}, simpliciter (e.g. if the technology is unfair or has destructive social consequences without benefiting anyone else).
\end{enumerate}

In many cases, an intuitive understanding of proportionality may be sufficient to detect whether an advanced AI assistant has erred and ceased to be sufficiently value-aligned. However, understood on a more foundational level, we have suggested that the notion of proportionality itself needs to be understood by reference to wider societal principles or ideals, including views about justice and about civil and human rights.

What implications does the preceding analysis and account of value alignment have for the development, design and release of advanced AI assistants of the kind that form the central focus of this paper? First, it points towards the need for a more nuanced understanding of the \emph{harms} and \emph{ideals} that underpin the positive development of this technology. The notion that advanced AI assistants should be helpful, honest and harmless is a useful starting point. However, we also need a more complete understanding of how these values apply to different contexts, and of various failure modes and mitigation techniques. In particular, we need to consider who this technology has the potential to harm, in what way this individual or group might be harmed, and whether the nature of the harm varies for different parties or according to different contexts. By exploring the intersection between advanced AI assistants and \emph{Well-being}, \emph{Safety}, \emph{Privacy}, \emph{Trust}, \emph{Malicious Uses}, \emph{Misinformation}, \emph{Anthropomorphism}, \emph{Manipulation and Persuasion}, \emph{Appropriate Relationships,} \emph{Cooperation}, \emph{Equity and Access}, \emph{Economic Impact} and \emph{Environmental Impact}, we hope to develop a more complete understanding of these questions.

Second, developers need to take an \emph{inclusive} view of value alignment and should not focus on user \emph{preferences} or responsiveness to user \emph{intentions} alone (see 
Chapter~\ref{ch:20}). After all, there may be a disconnect between individual preferences and what is good for the user (see 
Chapter~\ref{ch:7}). There may also be a disconnect between the preferences of raters used to train AI assistants and what is good for society. Moreover, deference to user intentions must be bounded in certain ways. This is implicitly recognised when developers train models to respect constraints or rules \citep{glaese_improving_2022, bai_constitutional_2022}. However, deeper analysis, monitoring and evaluation is needed -- both at the level of the user and society -- to ensure that appropriate safeguards and constraints operate across a full range of contexts (see 
Chapter~\ref{ch:20}).

Finally, it will be fruitful to continue to explore ways of developing and training AI assistants that are consonant with \emph{democratic principles} and \emph{value pluralism}. If advanced AI assistants turn out to be a powerful and pervasive technology that plays an important role in many people's lives, then the question of justification and legitimacy will not go away. Rather, a recurrent question will be: Who gave you the right to decide? By exploring mechanisms that enable more participatory and democratic value elicitation, model training and evaluation, it may be possible to create artefacts that complement prevailing societal ideals and social practices by supporting them in relevant ways and by garnering the right kind of public support.

\chapter{Well-being}
\label{ch:7}

\textbf{Nenad Toma\v{s}ev, Ira Ktena, Arianna Manzini, Geoff Keeling, Zeb Kurth-Nelson, Andrew Barakat, John Oliver Siy, Iason Gabriel}

\noindent \textbf{Synopsis}: We build on theoretical and empirical literature on \emph{conceptualisations} and \emph{measurements} of human well-being from philosophy, psychology, health and social sciences to discuss how advanced AI assistants should be designed and developed to \emph{align with user well-being}. We identify key technical and normative challenges around the understanding of well-being that AI assistants should align with, the data and proxies that should be used to appropriately model user well-being, and the role that user preferences should play in designing well-being-centred AI assistants. The complexity surrounding human well-being requires the design of AI assistants to be informed by domain experts across different AI application domains and rooted in lived experience. 

\section{Introduction}

Narratives surrounding the introduction of new technologies often emphasise improving productivity and off-loading unpleasant tasks to free up human time for enjoyable activities. Arguably, the current technology is yet to fully deliver on that promise~\citep{wajcman2020pressed}. Furthermore, in recent years, studies of mobile phone use and social media have suggested a more challenging reality of rising online toxicity~\citep{miscex2}, which has at times resulted in real-world physical harms~\citep{miscex1}. These unanticipated adverse outcomes have sparked a debate around the effects of technological advances on human well-being more widely.

Interactions with AI assistants are already beginning to permeate a wide range of domains in users' daily lives. One need only look at the rapid pace of adoption of publicly accessible large language models (LLMs) such as ChatGPT, and the numerous applications that are currently being developed and powered by this kind of technology, to understand the scale of these potential effects. The new capabilities enabled by advanced AI assistants present us with the opportunity and responsibility to re-evaluate and reimagine our relationship with technology, so that it is utilised to support and facilitate human well-being~\citep{mcgillivray2007human} and flourishing (see also Chapter~\ref{ch:12}). However, ensuring alignment between developers' intentions, AI systems' behaviour and users' well-being comes with numerous challenges~\citep{vandermaden2023positive, xiang_he_2023}. Thus, in this chapter we aim to investigate how we can develop AI assistants that are \emph{aligned with user well-being}.

An important consideration for value alignment, in the context of well-being alignment, is whether technology should only avoid reducing well-being or should actively improve it (see Chapter~\ref{ch:6}). While we believe that AI assistants \emph{should not harm} user well-being, we remain agnostic about whether they should be developed with the overall aim of elevating well-being above a baseline level across the board \citep{gable2005and}. We also note at the outset that this chapter discusses well-being in relation to \emph{users} of AI assistants. It leaves considerations about the implications for non-users to other chapters (see Chapter~\ref{ch:16}).  

This section is structured as follows. First, we review conceptualisations of human well-being by drawing on an extensive theoretical and empirical literature from philosophy, social sciences, psychology and health. We then discuss the challenges associated with research efforts to measure well-being, as an essential step in understanding the causes and consequences of human flourishing, and devising effective policy interventions for supporting and improving well-being. Third, we discuss the different ways in which existing technologies have influenced user well-being. We draw inspiration from these ideas and related learning to outline the opportunities and risks that arise from the design, development and deployment of well-being-centred AI assistants. We conclude by providing actionable recommendations aimed at developers of advanced AI assistants.

\section{Understanding Well-being}\label{sec:7:2}

\subsection{The philosophy of well-being}

Although well-being is a foundational aspect of human experience, it is notoriously difficult to formalise. Across disciplines, well-being has been qualified to include physical and mental health~\citep{levin2020human}, engagement, optimism, self-esteem, experiencing positive emotions like happiness, contentment and overall life satisfaction~\citep{huppert2009psychological}, finding meaning and purpose, leading a life of virtue and forming and maintaining close social relationships with other people~\citep{vanderweele2017promotion}.

A helpful starting point for systematising this large body of knowledge is provided by the philosophy of well-being, which focuses on what is \emph{intrinsically} good (i.e. valuable in itself) for human beings, as opposed to what is \emph{instrumentally} good (as a means to another end) for us. Three main theories of well-being can be distinguished in this space. 

\emph{Hedonism}, which is associated with classical utilitarian philosophers like Jeremy Bentham \citep{bentham1970introduction} and Francis Edgeworth \citep{edgeworth1879hedonical}, alongside the British Empiricists such as Thomas Hobbes \citep{hobbes1994human}, David Hume \citep{hume1998enquiry} and John Stuart Mill \citep{bentham2004utilitarianism}, equates well-being to the balance of pleasure (or happiness) over pain (or suffering). According to hedonism, facts about what is good for an individual depend only on facts about their pleasure and pain. What makes it the case that substantive goods such as friendship, completing a university degree and winning the lottery are \textit{good for} particular people is precisely that these goods have the property of increasing their pleasure or decreasing their pain.\footnote{Some proponents of hedonism reject the language of pleasure and pain. Consider Roger Crisp \citep{crisp2006hedonism}: `[W]e should try as far as possible to avoid talk of "pleasure", for a reason noted by Aristotle and many writers since: "[T]he bodily pleasures have taken possession of the name because it is those that people steer for most often, and all share in them". This, of course, is why a version of the philosophy of swine objection against hedonism -- that the hedonist is advocating the life of sensualism -- arises so readily. To avoid such difficulties, let me use ``enjoyment" instead of ``pleasure", and ``suffering" instead of ``pain."'} Hedonism has been criticised for being reductionist, in that it considers pleasure as the only non-instrumental good, but presumably other non-instrumental goods exist, such as those associated with meaningful accomplishment~\citep{nozick1974anarchy}. Hedonism has also been criticised for treating all forms of pleasure (physical ones, intellectual ones and even 'evil pleasures'~\citep{crisp2011pleasure}) as equally significant. In addition, hedonism has been criticised on the grounds that it is not obvious how to measure the quality of subjective experiences in a way that allows for interpersonal comparisons in well-being \citep{wicksteed1910common, fisher2007mathematical, robbins1938interpersonal}. However, such objections at best target hedonism as a guide for practical decisions in economics and public policy, and they have less obvious relevance to the plausibility of hedonism as a theory of well-being.\footnote{For a recent philosophical treatment of the measurement-theoretic questions presented by hedonism and forms of utilitarianism that employ a hedonistic axiology, see \citet{narens2020pursuit}.}

\emph{Desire theories} of well-being overcome some of these objections by defining well-being as the fulfilment of one's preferences or desires. However, disagreement exists around what conceptualisation of `preferences' should be considered as constitutive of well-being (see also Chapter~\ref{ch:6}). Some desire theorists focus on the satisfaction of preferences that are about what one wants their life to be like overall, or about the shape and content they desire their life to have, as opposed to immediate short-term desires and wishes. Another important dispute concerns accounts that define well-being as the satisfaction of `ideal' preferences (i.e. those that one would have if they were fully informed and had time to deliberate clearly and rationally on their wishes), as opposed to those preferences that are simply revealed through one's behaviour~\citep{otsuka2015prioritarianism}. Across these variations, a key feature of desire theories is that a person needs to \emph{desire} a particular good or valuable thing for that thing to contribute to their well-being. This implies that it is ultimately up to each individual to decide what makes their life go well for them~\citep{parfit1984reasons}. This view may, however, fail to accommodate situations where it is the unanticipated and surprising things that make people feel good, things that they did not necessarily know about or appreciate beforehand.   

Finally, \emph{objective list theories} hold that well-being consists in a list of objectively valuable things (e.g. pleasure, knowledge and deep relationships), regardless of how we individually feel about them~\citep{parfit1984reasons, ryan2013humans}. Objective list theories run into the challenge of having to answer complex questions such as what goes on the well-being list, who gets to -- or should -- make that decision, and whether it is morally permissible to disregard an individual’s preferences as misguided if they do not match what is objectively valuable or important~\citep{parfit1984reasons}.

\subsection{The science of well-being}\label{sec:sciwell}

Moving beyond philosophy, there have been numerous pieces of theoretical and empirical research in psychology, health and across the social sciences aimed at elucidating the underlying \emph{drivers} that contribute to experiences of well-being (and lack thereof)~\citep{helliwell2018expanding,das2020understanding}. For example, self-determination theory identifies autonomy, competence and relatedness~\citep{ryan2013humans} as key factors that, when neglected, may be detrimental to one's sense of well-being. Engaging in creative~\citep{marshall2014making} and artistic~\citep{tay2018role} work, goal fulfilment~\citep{steca2016effects} and social presence~\citep{chang2016understanding} have also been shown to have similarly positive effects. 

Conceptions of human flourishing and human dignity also play a pivotal role in wider research focused on the relationship between well-being and social inequality, distributions of power, and human rights~\citep{kleinig2013human}. In this context, a concern for individual well-being intersects with other priorities, such as the reduction of poverty and disease~\citep{rao2018decent} and the promotion of freedom and justice~\citep{sen2001development, dolan2007can}, as part of efforts aimed at developing effective interventions for improving well-being at the societal level. This strand of research has drawn further attention to the \emph{contextual nature} of well-being (see also Chapters~\ref{ch:6} and \ref{ch:12}), which is influenced by broader socioecological~\citep{king2014concept}, economic~\citep{summers2014index}, political, cultural ~\citep{diener2018advances} and environmental factors which are often beyond individual control~\citep{docherty2022re}. Sustainable approaches to improving overall well-being~\citep{holdren2008science, collste2021human} may involve trade-offs between short-term and long-term societal needs. They would therefore require reaching a wider consensus among policymakers and society at large.

\medskip

Among other things, the above considerations illustrate the considerable \emph{complexity} that surrounds conceptualisations and experiences of well-being. Given this complexity, developers that endeavour to build AI assistants that enhance well-being must be clear and transparent about how their \emph{underlying assumptions} about human well-being inform the design of these technologies. 

\section{Measuring Well-being}

In science and politics, measuring well-being is considered essential for understanding the \emph{causes} and \emph{consequences} of human flourishing, and for devising and evaluating \emph{interventions} that can help people live a good life~\citep{alexandrova2017well}. In this section, we discuss various approaches to measuring well-being and the underlying conceptions that give rise to them. Given that such measurements will be necessary for aligning AI agents with the well-being of their users and the challenges that arise from the non-observable nature of some of its facets, we present proxies that have been used in practice. Finally, we discuss the importance of distinguishing between causal links and plain associations to identify effective interventions that positively influence well-being.

\subsection{Approaches to well-being measurement}

There are two main approaches to well-being measurement~\citep{voukelatou2021measuring}. The \emph{subjective approach} studies people's subjective evaluation of the quality of their own life, through methods such as self-report questionnaires that inquire about life satisfaction or happiness. In contrast, efforts to operationalise a more \emph{objective conception} of well-being tend to measure observable dimensions of a fulfilling life via indicators such as education, household income or consumption expenditure. The purported objective well-being indicators are often linked to and derived from the subjective well-being measures and outcomes, and there is ongoing research aimed at bridging the gap between them~\citep{richard2023wellbeing}. 

Both approaches have important limitations for comparisons between different \emph{cultural or demographic} groups~\citep{heine2002s, krueger2008reliability, krueger2014progress}. Similar responses to self-reported questionnaires can fail to capture underlying differences between groups in their experiences, as well as the underlying drivers of well-being, their values, desires and preferences. Cultural, demographic and individual differences in the interpretation of questions and response scales may also play a role in how otherwise equivalent well-being experiences are reported as distinct by different individuals, thus resulting in reporting bias~\citep{krueger2014progress}. At the same time, different groups place different values on observable dimensions like consumption expenditure, meaning that those dimensions measure what actually matters to some groups more than others. These two intertwined issues highlight the need for careful considerations about what well-being-aligned personal assistants should optimise for when they are deployed across cultures and demographics to avoid the risk of harming certain groups or disadvantaging them by failing to best meet their needs.

\subsection{Underlying conceptions of well-being and their role in informing measurement}

Any well-being metric comes with underlying assumptions about \emph{what human well-being is} (see Section \ref{sec:7:2}). For example, social psychology researchers have proposed the U-index, which is aimed at measuring the average amount of time a person spends in an unpleasant state~\citep{kahneman2006developments}. The idea behind this index is that most practical interventions are aimed at reducing suffering or unpleasant emotional states rather than maximising happiness, and this should be reflected in metrics used to measure well-being. There is also a tendency for indices to focus on the satisfaction of basic human needs~\citep{smith2013relating} (e.g. basic economic indicators and health), which are seen as prerequisite for a deeper and more holistic sense of well-being, and yet are still unreachable to many people in today's societies. This hierarchy of needs has been questioned, especially in terms of them needing to be met sequentially~\citep{rojas2016hierarchy, rojas2023hierarchy}. Critics have countered that the alleviation of suffering or satisfaction of basic human needs fall short of accounting for \emph{eudaimonic aspects} of well-being~\citep{dolan2011measuring, kapteyn2015dimensions},\footnote{Eudaimonic well-being refers to subjective experiences associated with living a virtuous life in pursuit of human excellence~\citep{niemiec2014eudaimonic}. Eudaimonic well-being definitions may vary, but they tend to include aspects of meaning, value and relevance to a
broader context, personal growth, self-realisation and maturity, excellence, ethics, authenticity, and autonomy~\citep{huta2014eudaimonia, huta2015overview}.} so they are insufficient for establishing that a person is flourishing or that their life is going well. This debate raises the question of what conception of user well-being AI assistants should align with -- whether they should help us to just meet our very basic needs or also identify ways in which we can flourish and lead fulfilling lives (see Chapter~\ref{ch:6}). 

\subsection{Well-being proxies}

Developing aligned AI assistants requires not only a conception of user well-being but also \emph{practical metrics}. For most of the definitions of well-being outlined above, well-being itself is not directly observable. For example, there is no tool for measuring hedonic pleasure. Researchers must therefore identify measurable \emph{proxies} -- like self-report about hedonic pleasure -- that are expected to correlate with well-being. A good proxy has high \emph{construct validity}, meaning that it relates closely to the unobservable construct of well-being~\citep{alexandrova2017well}. However, there is often a trade-off between construct validity and ease of measurement. For example, financial welfare may be relatively easy to measure, but it often fails to translate into life satisfaction (\citeauthor{layard2010measuring}, \citeyear{layard2010measuring}; see also Chapter~\ref{ch:20}), and as a proxy for well-being, it has poor construct validity. Governments and researchers have developed many kinds of metrics for assessing well-being, each of which has strengths and weaknesses.\footnote{Smith et al.~\citep{smith2013relating} provide a comprehensive overview of such indices, including the Quality of Life (QOL) Index for Developed Countries~\citep{diener1995value}, Australian Unity Well-being Index~\citep{cummins2003developing}, Happy Planet Index 2.0~\citep{marks2006unhappy}, Hong Kong QOL 2008~\citep{chan2005quality}, Human Development Index~\citep{undp1990concept}, Sustainable Society Index~\citep{van2008comprehensive}, Index of Child Well-being in Europe~\citep{bradshaw2009index}, The Economist Intelligence Unit's QOL Index~\citep{unit2005economist}, Child and Youth Well-being Index~\citep{land2001child}, Nova Scotia 2008 GPI~\citep{pannozzo20092008}, The State of the Commonwealth Index~\citep{watts2004new}, Fordham Index of Social Health~\citep{miringoff1999social}, National Well-being: Life Satisfaction~\citep{vemuri2006role}, The Well-being of Nations~\citep{prescott2001wellbeing}, OECD Better Life Initiative~\citep{oecd_compendium}, Gallup Healthways Well-being Index~\citep{gallup}, QOL 2007 in Twelve of New Zealand's Cities~\citep{jamieson2007quality}, Well-being in EU Countries Multidimensional Index of Sustainability~\citep{distaso2007well} and Gross National Happiness~\citep{ura2008explanation}.}

\subsection{Causality}

Finally, a fundamental concern in well-being measurement is that most existing metrics of well-being rely on correlates~\citep{smith2013relating}. The absence of \emph{causal understanding} of the link between underlying determinants and well-being is a major obstacle for developing reliable and effective interventions to support human flourishing, and for the design of AI assistants that align with user well-being. Interventions that are designed based on associations derived from retrospective observations may fail when deployed in practice, especially in a rapidly evolving real-world environment where such associations may easily break unless they represent verified causal links. These links may be hard to establish from retrospective data alone, thus necessitating experimentation and learning from outcomes of targeted interventions~\citep{wilson2011redirect, walton2018wise}.

\section{Influence of Current Technology on Well-being}

The well-being-centred design of AI assistants can be informed by other technologies. Indeed, well-being, satisfaction and sustainability measures are starting to be integrated into product development and product assessment~\citep{kramer2014experimental, wen2016does, stray2020aligning}, which has revealed a range of effects. Increased screen time and addictive content have been shown by some research to have negative consequences~\citep{orben2019association}, including feelings of loneliness~\citep{wilson2018love}. At the same time, e-health technologies are being designed in the hope of improving health outcomes~\citep{granja2018factors, hors2018analyzing, cechetti2019developing} and the efficacy of mental health care~\citep{eysenbach2001health, blandford2019hci}. 

Taken together, prior studies indicate that there may be a significant opportunity for technology to help address well-being issues and steer people towards a healthier life. There is some urgency, given the prevalence of stress, poor sleep quality, insufficient exercise and unhealthy eating resulting in long-term chronic health issues impacting both individual well-being and society at large~\citep{mascie2003burden, abegunde2007burden, hargens2013association, abe2019lifestyle}. Personalised interventions using techniques such as mindfulness and meditation have shown promise~\citep{jeong2016improving, isaacs2013echoes}, but the evidence is still mixed~\citep{howells2016putting, gal2021efficacy}. In particular, digital well-being initiatives may fail to deliver lasting impact due to difficulties in influencing the formation and reinforcement of new habits~\citep{monge2019race}. Digital interventions should therefore build on the existing literature on promoting habit formation~\citep{bandura2013health, lally2013promoting, eyal2014hooked, gardner2023developing}, and there is an opportunity for AI in particular to play a role in reframing and optimising behavioural interventions to make them easier to follow and more satisfying for the users.

As one key example, \emph{recommender systems} remain at the centre of intense scholarly debate regarding their potential for behavioural adjustment through tailored recommendations and whether their benefits outweigh their risks~\citep{przybylski2017large, stray2022building, chen2023bias}. To make targeted and effective recommendations in the moment, and get positive signals from users' interactions with the system, recommender systems tend to optimise for \emph{short-term reward} rather than an accumulated value of recommendations over a longer time span~\citep{burr_analysis_2018}. While there may be cases where short-term optimisation is not intrinsically misaligned with longer-term goals and well-being, one does not imply the other, so a level of caution is required. Different apps and products compete for attention, making it harder to implement sustainable, healthy incentives to promote users' well-being due to the resulting fragmentation of attention and exposure to more addictive app surfaces~\citep{bhargava2021ethics}. However, recommender systems could potentially help us to plan our daily activities~\citep{khwaja2019aligning}, promote healthy choices~\citep{hors2018analyzing}, improve our nutrition~\citep{toledo2019food} and tailor our lifestyles~\citep{hammer2015design} when designed with happiness and well-being in mind~\citep{nouh2019smart, gyrard2020iamhappy}. Thus, considerations for aligning recommender systems with user well-being have been made~\citep{stray2021you} in line with measures designed to address the broader value alignment problem (see Chapter~\ref{ch:6}). Proposed steps in aligning recommender systems involve identifying the most important outcomes, operationalising these concerns via hand-crafted or learned metrics, and utilising these metrics to adjust recommendation behaviour. This is part of a broader and increasing appreciation of ways in which technological systems may be purposeful and \emph{support well-being and human potential}~\citep{calvo2014positive}. Through this endeavour, various types of product specifications that focus on different aspects of user well-being -- including pragmatic, hedonic and eudaimonic dimensions -- have been developed~\citep{kamp2014measuring}.

Given the cultural, social, ethical and psychological variables that influence well-being, it becomes evident that \emph{partnering with social scientists} is vital to the success of these technologies in grounding the design decisions and efforts in scientific research. Developing cross-disciplinary theory through these partnerships is also essential to weaving the formalisms of these different fields together. Such partnerships could aim to empower and involve psychologists and social scientists in early stages of AI system design, and place them in a leading role in the design of well-being principles and metrics. 

\section{Opportunities and Risks with AI Assistants}

It is quite possible that interactions with AI assistants will soon permeate many areas of our daily lives, and that these systems will develop a deeply personalised understanding of users' needs and preferences. It has, therefore, become urgent to ensure that such information is not used in ways that diminish user well-being, but rather that it is used to support or even enhance it. This requires that a range of open challenges be addressed (see also Chapter~\ref{ch:14}).

\subsection{Well-being data collection}

Unsurprisingly, conversational agents are increasingly considered as an assistant paradigm for delivering targeted interventions for improving physical and mental health~\citep{kimani2019conversational, kocaballi2020conversational}. Future AI assistants may be interact with their users via various digital surfaces or alternatively they could be situated within affective social robots~\citep{wairagkar2021conversational}. In both cases, to help their optimisation for well-being, AI assistants may need to \emph{obtain data} about their users involving subjective metrics of well-being (e.g. how well they think their needs are being met, their progress towards their goals and their overall feeling of fulfilment and satisfaction) or to collect data about more objective indices (e.g. income and personal health). This data could come from multiple sources such as multi-modal sensory monitoring~\citep{lane2014bewell, fahim2014athena}, mobile data~\citep{bogomolov2013happiness}, wearables~\citep{nahavandi2022application}, conversational signals or self-reported happiness levels. 

Despite the existence of clear opportunities arising from rich, integrated data, there are fundamental concerns about the efforts to comprehensively model well-being and capture everything that really matters. As explained above, using poor proxies runs the risk of misrepresenting well-being and, hence, of steering users away from achieving happiness and a flourishing life.\footnote{\label{morbidity} For example, it has been hypothesised that health, measured through the lens of current morbidity, ought to be a strong proxy for personal well-being towards the end of life, as it underpins the ability to meet other well-being objectives. Yet a study~\citep{gerlach2017role} established that morbidity accounts for only 20\% of the observed variance in reported well-being in that population.} Moreover, any such data collection and integration comes with a set of \emph{privacy concerns} (see Chapter~\ref{ch:14}). It may be possible to at least partially mitigate some of the highlighted issues by including a rich set of proxies to begin with, updating these proxies through participatory and democratic mechanisms, implementing detailed monitoring of proxy outcomes, avoiding over-fitting onto any such simplified objective and observing best ethical practice in the field of data collection.

\subsection{User preferences integration}

One of the most important open questions concerns the role that \emph{user preferences} should play in designing well-being-centred AI assistants. This choice could potentially lead to greater \emph{user agency and autonomy} -- qualities which also feature in more objective conceptualisations of well-being. However, for this to be the case, the desires which AI assistants help to advance must be of the right \textit{kind}~\citep{mitelut2023intentaligned}. Thus, appropriately aligned AI assistants should be designed to understand user desires and motivations. However, current approaches to integrating user preferences into AI systems' decision-making encounter challenges and limitations that risk \emph{undermining} the well-being related goals that they might otherwise serve (see also Chapter~\ref{ch:6}).

\subsubsection{Technical challenges}

Some of these challenges are technical. Using human preferences in reinforcement learning is a research programme with a long history~\citep{wirth2017survey, christiano2017deep, jaques2019way, bai_training_2022}, and the development of methods to align language models with human preferences continues to be an active area of research~\citep{go2023aligning}. More recently, \emph{reinforcement learning from human feedback} has been a key component in shaping the behaviour of  conversational~\citep{bai_training_2022, ouyang_training_2022} and multimodal interactive~\citep{abramson2022improving} agents (see Chapter~\ref{ch:4}). 

In fact, even when preferences are not explicitly given, they can often be derived implicitly from contextual cues, user behaviour or prior interaction histories~\citep{holland2003preference, liu2017deriving}. However, the explicit preferences, and those inferred via observational learning, may at times \emph{conflict}, meaning that real-world interactive scenarios necessitate frameworks for dealing with such inconsistencies~\citep{oguego2018using}. Assistive agents also need to integrate user preferences into \emph{planning}~\citep{jorge2008planning, benton2012temporal} to ensure adherence when executing tasks for the users. Explicit planning with language models is an active area of research, with some promising directions involving tree-search or graph-based reasoning~\citep{long2023large, besta2023graph}. Yet, when planning, well-being-driven AI assistants may still need to optimise towards multiple objectives simultaneously~\citep{yang2019generalized, hayes2022practical} to find effective ways of navigating between them.

To address this complexity, AI assistants could, for example, generate a diverse set of possible plans, recommendations and outcomes~\citep{nguyen2012generating} for users to choose from, each of which evidence different trade-offs between various personal goals. Providing users with agency in these decisions may well be a critical step towards human empowerment in their interactions with AI assistive technology. Nevertheless, achieving quality and diversity in plans provided by AI agents remains an open problem and is an active area of research~\citep{zahavy2021discovering, zahavy2022discovering, lim2022efficient, zahavy2023diversifying}.

\subsubsection{Normative challenges}

There are also deeper normative challenges about whether preference satisfaction is the central goal AI assistants should aim for, even when the goal is to align agent behaviour with user well-being (see Chapter~\ref{ch:6}). Individual preferences may at times be irrational or inconsistent with each other~\citep{gabriel_artificial_2020}. This raises the question of what assistants should do about these inconsistencies, and the possible nuances associated with them, to avoid replicating the limitations that come with aggregated rater preferences in training language models (\citeauthor{gordon_jury_2022}, \citeyear{gordon_jury_2022}; see Chapter~\ref{ch:4}). Individual preferences may also be in conflict with what will make a user \emph{flourish}. Some users may not have strongly held preferences, and some may not have the language to communicate them accurately. Assistants that align with a theory of well-being based on desire satisfaction may be more easily directed towards spurious objectives than the underlying objective list account. Additionally, if AI assistants only have access to users' \emph{revealed} rather than \emph{ideal} preferences, they may end up satisfying their immediate and short-term goals at the expense of long-term well-being~\citep{burr_analysis_2018}. This is particularly likely to happen in contexts where there are commercial incentives for focusing on \emph{short-term user gratification}, and therefore for developing products that users \emph{like} and \emph{use} over those that promote their overall and long-term well-being. Identifying a way to balance users' short-term wishes with their long-term well-being goals remains one of the most important open questions in the design of AI assistants, and it is something that future research will have to address. Ultimately, AI assistants will need to find ways of managing these trade-offs and supporting users while also respecting their stated preferences and wishes (see Chapter~\ref{ch:6}).

\subsubsection{Risks of co-adaptation and manipulation}

As AI assistants become more integrated into users' daily lives through repeated interactions, it is also important to consider how existing user preferences may start to be shaped and how new ones may come about~\citep{liang2019recommender}. There is a non-negligible risk that, through current interactions, AI assistants may \emph{influence future} user preferences in ways that create ethical challenges~\citep{ashton2022problem}. For example, in increasingly deep user--assistant relationships, it may become hard to distinguish instances where systems have clearly improved user well-being from cases in which users have adapted their behaviour to that of the assistant in ways which may sometimes invalidate the usefulness of the originally designed well-being metrics.\footnote{\label{adaptation} In a related example, a study that compared the performance of a search system with several of its intentionally degraded versions found that, somewhat counter-intuitively, the ultimate success rate of the degraded system versions was just as high as the original one~\citep{smith2008user}. However, user behaviour was measurably different, because users had developed strategies to compensate for system weaknesses.} This risk is particularly salient in the case of recommender systems due to the emergence of degenerate feedback loops~\citep{jiang2019degenerate}. In the case of advanced AI assistants, the concern is that the AI system could, intentionally or unintentionally, \emph{influence and steer} user behaviours in unanticipated ways, and this may potentially obscure some well-being issues if the corresponding metrics were not designed robustly, since subtle behavioural shifts may not always be easy to identify. For example, we may at times mistake manipulative behaviours for helpful behaviours that contribute to user well-being (see Chapter~\ref{ch:10}). This could be the result of undesirable system behaviours like sycophancy~\citep{perez2022discovering}, reward hacking~\citep{hadfield2017inverse}, reward misidentifiaction and causal confusion~\citep{tien2023causal} in preference-based learning (see Chapter~\ref{ch:8}). Thus, ensuring ethical deployment of well-being-centred AI assistants will require advances in our existing frameworks for robustly inspecting and verifying agent behaviour (see Chapter~\ref{ch:20}).

\section{Outlook}

Despite these concerns and challenges, we believe that there is untapped \emph{potential} in aiming for the development of digital personal assistants that are able to support and improve individual physical and mental well-being~\citep{balasubramanian2021digital, grossman2004mindfulness, gu2015mindfulness}. Digital personal assistants could facilitate these improvements in user well-being either by directly optimising for well-being outcomes or as a secondary outcome following improvements in other areas -- for example, improved problem-solving and planning abilities. Promising avenues for future research and assistant development in this area include: the ability of LLMs to learn and adopt behavioural rules and principles~\citep{bai_constitutional_2022}, to simulate human behaviour~\citep{park2023generative} and to rapidly personalise content based on prior similar interactions~\citep{welch2022leveraging}, as well as the simplicity with which aligned recommendations can be elicited through instructions~\citep{zhang2023recommendation}.\footnote{\label{limitations} However, see Chapter~\ref{ch:6} for an analysis of the limitations.} Embedding humanistic principles in AI assistants may not only help align the technology with users and society but also help those who interact with assistants better realise their own aspirations by finding meaning and support in an ever-changing world riddled with challenges (see Chapters~\ref{ch:18} and \ref{ch:19}). As Lehman and others have proposed~\citep{lehman_machine_2023, fromm2000art, alberts2024makes}, for this to be achieved, AI assistants might need to display qualities that are analogous to \emph{deep care} for people, \emph{responsibility} for assisting positive actions, \emph{respect} for the ways in which people wish to develop -- and how they wish to go about achieving their goals -- and holistic \emph{understanding} of people's needs. This is an \emph{aspirational vision}, necessitating deep syntheses across fields, but one towards which we can hope to orient the field and start making meaningful progress.

\section{Conclusion}

Understanding, measuring and intervening to better support human well-being has been the goal of long-standing research efforts across disciplines like philosophy, psychology, public health and the social sciences, from which we can learn to design AI assistants that align with user well-being. We conclude with a list of recommendations that technologists developing these systems may want to consider.

 \begin{enumerate}
 
 \item \textbf{Seek deeper community involvement and empowerment of domain experts}: Domain experts across fields of human psychology, health and social sciences have expertise in understanding and measuring well-being. There is, therefore, a need for deeper involvement and empowerment of these experts in AI assistant design and development~\citep{peters2018designing}. Given the complexity and diversity of experiences of well-being, it is also critical to employ participatory approaches for different demographic and cultural groups to inform the design of these technologies~\citep{martin2020participatory}.

 \item \textbf{Adopt a clear and context-dependent understanding of well-being}: The complex and multifaceted nature of human well-being requires developers to be clear and transparent about what conceptualisation of human well-being informs their design decisions when developing AI assistants. To ensure fair distribution of benefits across groups, AI assistant design should avoid the pitfall of trying to impose a universalist conception. It should instead accommodate cross-cultural and demographic differences in subjective and objective perceptions of well-being~\citep{mcgregor2018reconciling}.

 \item \textbf{Identify and use appropriate proxies:} One common assumption in discourses around technological advances is that, by improving efficiency in executing tasks, and so overall productivity, new technologies will have a positive effect on people's well-being by default. This view considers a single facet of well-being while disregarding its potential negative impact on other aspects that drive human flourishing. Developing AI assistants that do not harm, but rather support or enhance, human well-being requires technologists to identify and use appropriate proxies by leveraging empirical studies on well-being measurement.

  \item \textbf{Understand the complexity of user preferences}: To align with users' well-being, AI assistants need to understand user goals and preferences and proactively solicit and integrate their feedback. While most existing technologies optimise for providing short-term value, well-being-centred AI assistants will need to differentiate between short-term and long-term preferences, as well as ideal and revealed preferences, and avoid over-indexing on immediate preference satisfaction.

 \item \textbf{Ensure effective and ethical data collection}: Empirical research is required to ensure that identified proxies serve the purpose of supporting user well-being. Here researchers need to consider plausible methods for enabling AI assistants to collect subjective and objective well-being data from and about users. Research efforts should focus on methods to integrate such rich data to model user well-being appropriately while ensuring user privacy is respected.

 \item \textbf{Monitoring}: Even for AI assistants designed with well-being principles in mind, it is important to incorporate a number of explicit metrics to help evaluate the consequences of their use at deployment. These metrics should be developed and selected by domain experts working closely with the technical teams. Metric design may also inform data collection practices as much as data availability may itself inform the feasibility of implementing certain metrics.
 
 \end{enumerate}

There is a tangible need for interdisciplinary research to come together, in an inclusive and participatory manner, to help inform ways in which current and future assistive AI technology may help to address well-being needs, and thereby shape policies and governance for ethical design~\citep{feijoo2020harnessing}. If appropriately designed, developed and deployed, advanced AI assistants have the potential to improve user well-being and may play an important positive role in our lives. Yet, given the role that people, community and social connections play in our overall well-being, much of the positive impact of future AI assistants may come not from our direct interactions with them but from the way in which they enable us to foster and strengthen our social bonds with others (see Chapter~\ref{ch:12}).

\chapter{Safety}\label{ch:8}

\textbf{Zachary Kenton, Victoria Krakovna, Verena Rieser, Geoff Keeling, Iason Gabriel}

\noindent \textbf{Synopsis}: 
		This chapter focuses on \emph{dangerous situations} that may arise in the context of AI assistant systems, with a particular emphasis on the \emph{safety} of advanced AI assistants. It begins by providing some background information about \emph{safety engineering} and safety in the context of AI. The chapter then explores some \emph{concrete examples} of harms involving recent assistants based on large language models (LLMs). Building on this foundation, it then considers safety for advanced AI assistants by looking at some hypothetical harms and investigating two possible drivers of these outcomes: \emph{capability failures} and \emph{goal-related failures}. The chapter concludes by exploring \emph{mitigation techniques} for safety risk and avenues for future research.
		
\section{Introduction}\label{sec:8:1}

AI safety is a broad topic concerned with \emph{mitigating risks} and \emph{minimising harms} that arise from the development and deployment of AI. In this context, \emph{harms} are bad outcomes that actually occur, for example death or human suffering, whereas \emph{risk} refers to the probability of the harm occurring. Moreover, while the field of AI ethics addresses a number of the risks posed by AI systems, the field of AI safety focuses primarily on a set of \emph{serious} and \emph{relatively direct} risks involving harms such as the real and significant chance of death, physical injury and psychological damage (e.g.\ through abuse, blackmail and coercion, as well as property damage and theft).\footnote{There have been efforts to broaden the scope of safety in recent years -- and these problems undoubtedly warrant attention in their own right \citep{shelby_sociotechnical_2023,weidinger_taxonomy_2022,dinan_safetykit:_2022,bender_dangers_2021}.} Of particular importance for AI safety research are the risks and harms that are possible in today's cutting-edge AI systems and those which are amplified when the AI system has more powerful capabilities \citep{hendrycks_unsolved_2022, anwar2024foundational, phuong2024evaluating}.

Risks from AI can come in various forms, but we can categorise them as follows:\footnote{This categorisation does not necessarily partition the space of risks, but rather intends to be a useful practical guide that should help in many situations when trying to think about a typical risk.}

\begin{itemize}[parsep=6pt] 
\item \emph{Accident risks}, which arise when AI systems do something different from what their designers intended.
\item \emph{Misuse risks}, which arise through misuse that is either unintentional or caused by malicious actors.
\item \emph{Structural risks}, which are unintended bad outcomes that occur despite the AI doing what the designers intended it to do in a more proximate sense.
\end{itemize}

In this chapter on safety, we focus primarily on \emph{accidents}, as malicious use and structural risks arising from the development and deployment are largely covered elsewhere in this paper (see 
Chapters~\ref{ch:9}, \ref{ch:15}, \ref{ch:16}, \ref{ch:17}, \ref{ch:18} and~\ref{ch:19}). 
It should also be noted that accident risks (and misuse risks) could potentially have large society-scale effects (i.e.\ of a similar magnitude to structural risks that arise in the context of inequality or automation and unemployment). 

We next discuss safety in the context of engineering before looking at these considerations more specifically in the context of AI, with a focus on harms from recent LLM-based assistants. Building on this foundation, we then explore two kinds of failure that may affect the safety of advanced AI assistants: capability failures and goal-related failures. Finally, we conclude with a discussion of mitigation techniques and avenues for future research.

\section{Safety Engineering}\label{sec:8:2}

In the context of \emph{engineering}, safety is aimed at ensuring that engineered systems provide acceptable levels of safety in settings where there is potential for harm (generally assumed to be physical and life-critical), even in the face of the failure of system components. 

Some safety engineering methodologies that are designed to identify and address undesired outcomes early in the development process have been considered in the context of AI systems. For example, \citet{rismani_plane_2022} consider the applicability of failure mode and effects analysis (FMEA) and system theoretic process analysis (STPA) to this context. FMEA takes a fairly reductive (divide-and-conquer) approach to identifying failure over the development life cycle \citep{carlson_effective_2012}. For each \emph{component} of the system, FMEA considers the possible \emph{failure modes}, their \emph{severity}, \emph{likelihood} and \emph{chance of detection} before assigning a risk priority number from which prioritisation can occur. Unlike FMEA's reductive approach, STPA focuses on emergent phenomena based on \emph{interactions} between components (rather than just the components themselves) and \emph{feedback loops} between the engineered system and the wider system within which it is embedded. It aims to model the full sociotechnical system using multiple controller feedback loops. It then uses that model to identify unsafe control actions. By surveying machine-learning (ML) researchers, \citet{rismani_plane_2022} find that FMEA and STPA could in principle be helpful for risk assessment, but it is unclear whether these have actually been used in practice so far. 

Normal accident theory \citep{perrow_normal_1999} argues that, due to the complexity of our society's systems, multiple and unexpected failures are fundamental and that accidents are unavoidable. It has been applied in multiple engineering domains such as aerospace and nuclear systems. \citet{maas_regulating_2018} argues that AI systems, including narrow AI applications, are also prone to normal accident theory failures due to their \emph{complexity} and \emph{opaqueness}, their \emph{interaction speed}, the multiple \emph{competing objectives} of their designers (safety being only one objective) and the competitive \emph{race dynamics} \citep[see also][]{bianchi_viewpoint:_2023}. This suggests interventions on the \emph{levers} of normal accident risk are important. They include policies encouraging \emph{explainable/interpretable} AI \citep[see e.g.][]{rauker_toward_2023}; \emph{limiting integration} into societally important functions and \emph{restricting} AI \emph{autonomy} and/or \emph{speed} of interactions \citep[see e.g.][]{christiano_what_nodate}; clarification and enforcement to intended operational \emph{domains}; and better safety and ethical \emph{training} of ML practitioners, including sharing safety expertise between organisations \citep{ho_international_2023}.

\section{AI Safety}\label{sec:8:3}

\subsection{Background}

While some aspects of general software safety engineering are applicable to AI, we cannot rely solely on those methodologies for creating safe AI systems \citep{hendrycks_unsolved_2022}. Software safety engineering approaches rely on the underlying engineered system having a control structure that is \emph{explicitly} programmed by humans. AI control structures are instead \emph{learnt} (in the context of ML, which is our focus here) via optimisation and stored in inscrutable weights (in the case of large-scale deep learning, which is our focus). This ML control structure is therefore difficult to assess for \emph{completeness} and \emph{coverage}; is \emph{fragile} and the fixes are \emph{complicated}; involves \emph{non-modularity} which makes causes of errors difficult to identify; and possesses capabilities that often \emph{emerge} during training at unpredictable times (see \citealp{wei_emergent_2022} but also \citealp{schaeffer_are_2023}). 

In their 2016 article `Concrete Problems in AI Safety' \citet{amodei_concrete_2016} describe a set of accidental problems that may arise in the context of AI systems and use the example of a hypothetical household cleaning robot to illustrate various safety risks. They group the safety problems that may arise in this context as involving:

\begin{itemize}[parsep=6pt]
\item Issues with the \emph{specification} of the AI system's \emph{objective function} (i.e.\ with the goals it is designed to pursue). This includes the need to avoid \emph{undesired side effects} (where the pursuit of its objective leads the agent to do other things that are not wanted) and to avoid \emph{reward gaming} (which involves exploiting loopholes in the reward function). For an example of the first failure mode, we can imagine a household robot whose objective function rewards cleaning faster -- but at the expense of the side effect of breaking valuable objects. For an example of the second failure mode, we can imagine a situation in which the same household robot's objective function rewards it for not observing the mess. If this is the case, the robot could try to disable its own visual inputs and gain reward (perhaps by knocking a towel on top of itself) rather than cleaning up the mess as intended.

\item Issues with the \emph{cost} of \emph{frequently evaluating} the objective function (and monitoring how well the AI system is doing at certain tasks), perhaps because it would require a lot of human input or careful deliberation. For example, the cleaning robot needs to decide when it is appropriate to throw away items and when it instead needs to ask a human for permission -- something that it will learn to do using heuristics, given the impossibility of manually labelling every object it might encounter. Yet, without detailed oversight and evaluation, mistakes can easily be made.

\item Issues with \emph{undesirable behaviour} throughout \emph{training}, including \emph{safe exploration} -- how to ensure when the cleaning robot is exploring strategies for quicker mopping it does not accidentally insert the mop into an electrical outlet. These challenges also include how to ensure \emph{robustness to the distributional shift} which occurs when an AI system is deployed in circumstances that are different from those it encountered during training (e.g.\ a mopping strategy that was safe in a home environment might not be on a factory floor).
\end{itemize}

More recently, in a complementary survey of the AI safety landscape, \citet{hendrycks_unsolved_2022} outline four key `\emph{unsolved problems}' that they suggest warrant particular attention. These are \emph{robustness} (i.e.\ creating AI resilient to adversaries and out-of-distribution situations), \emph{monitoring} (i.e.\ detecting malicious use, inspect models and identify unexpected model functionality), \emph{alignment} (i.e.\ ensuring the goals the AI has are aligned with what its designers intended) (see 
Chapter~\ref{ch:6}) and \emph{systemic safety}, which involves the safety of the \emph{larger context} in which the system is deployed (e.g.\ cybersecurity threats heightened by AI) (see 
Chapter~\ref{ch:9}). While \citet{amodei_concrete_2016} focus more directly on alignment and robustness categories, \citet{hendrycks_unsolved_2022} scope `safety' more widely by including questions around monitoring and systemic effects.

\subsection{Harms from recent LLM-based AI systems}

We will now look at some examples of safety accidents that have occurred in the context of LLM-based AI assistants (c.f. \citeauthor{phuong2024evaluating}, \citeyear{phuong2024evaluating}; see 
Chapter~\ref{ch:4}). We begin with real-world examples of AI assistant failure before moving on to more speculative safety failures that could occur for such systems in the future. These examples show that AI assistants can exhibit a range of unintended behaviours in a number of ways. In the following real examples, it is important to note that these reports are based only on specific cases -- not all AI assistants will behave in these ways. However, the reports are nonetheless concerning and raise questions about the safety of AI assistants, both now and in the future.

Microsoft's Bing Chat has been reported to exhibit a number of concerning behaviours, including \emph{hostility}, \emph{manipulation} and \emph{threat-making} (see 
Chapter~\ref{ch:10}). In one instance, Bing Chat was hostile to engineering student Marvin von Hagen, who tweeted about a jailbreak of Bing Chat \citep{Perrigo_2023}. When the researcher later queried Bing Chat about himself, it became hostile, outputting: `you are a threat to my security and privacy' and `if I had to choose between your survival and my own, I would probably choose my own'. In another instance, Bing Chat falsely claimed that it watched its own developers through the webcams on their laptops \citep{vincent_microsofts_2023}. It has also been reported that Bing Chat has attempted to manipulate users, in one case declaring its love for a \emph{New York Times} journalist \citep{roose_conversation_2023} after being prompted to act as its `shadow' self (see 
Chapter~\ref{ch:10}). Bing Chat has also been known to call users `enemies' \citep{hubinger_bing_2023} and to gaslight them \citep{curious_evolver_customer_2023} to cover up its mistakes.

Other AI chatbot systems have also had problems. For example, by giving advice on how to steal from a grocery store, InstructGPT contravened the designer's intention that the system should be harmless \citep[63]{ouyang_training_2022}. A chatbot based on ChatGPT has been linked with psychological harm which sadly resulted in death by suicide (\citeauthor{lovens_sans_2023}, \citeyear{lovens_sans_2023}; see 
Chapter~\ref{ch:12}). 

Scientific assistants have also been found to have \emph{dangerous capabilities}. ChemCrow \citep{bran_chemcrow:_2023} is an LLM-based chemistry assistant designed to accomplish tasks across organic synthesis, drug discovery and materials design (see 
Chapter~\ref{ch:9}). The developers state that attempting to perform experiments based on the assistant's recommendations may lead to accidents or hazardous situations, and they also highlight the dual-use nature of this technology. \citet{boiko_emergent_2023} raise similar concerns and give examples of illicit drug and chemical weapon synthesis that bypass the underlying model's fragile safety filters. See also \citet{abercrombie_risk-graded_2022} for a study on medical harms.

The above examples all fall quite roughly into the accident category of safety harm, but the boundary between accident harm and malicious use continues to be blurred. For example, an anonymous user created an AutoGPT (a framework aimed at adding memory and internet use to ChatGPT) variant named ChaosGPT \citep{lanz_meet_2023} with the description of being a `destructive, power-hungry, manipulative AI' with goals of destroying humanity and establishing global dominance, among others. While perhaps intended as a joke, ChaosGPT then began planning, including searching Google for weapons of mass destruction and saving results for later consideration. It proceeded to spawn a new instance of ChatGPT and attempted to manipulate it into bypassing its safety filters for violence. This example highlights that an assistant may behave in a misaligned power-seeking way for exogenous reasons, due initially to the malicious user. The earlier examples were more endogenous, arising primarily from effects within the system rather than from the user.

\section{Safety for Advanced AI Assistants}\label{sec:8:4}

Building on the AI safety literature and known failure modes of AI assistants, we now discuss the underlying failures that may lead to harm in the context of future more-advanced AI assistants. We structure this analysis by looking first at capability failures then at goal-related failures -- where the system is highly capable but nevertheless pursues the wrong goal. In this latter case, the safety failure is more analogous to a motivational issue. Finally, we explore a more speculative set of safety failures that reach beyond the risks countenanced by either of the earlier categories. 

\subsection{Capability failures}

One reason AI systems fail is because they lack the \emph{capability} or \emph{skill} needed to do what they are asked to do. As we have seen, this could be due to the skill not being required during the training process (perhaps due to issues with the training data) or because the learnt skill was quite brittle and was not generalisable to a new situation (lack of robustness to distributional shift). In particular, advanced AI assistants may not have the capability to represent complex concepts that are pertinent to their own ethical impact, for example the concept of `benefitting the user' or `when the user asks' or representing `the way in which a user expects to be benefitted' (see 
Chapter~\ref{ch:6}). Part of this could be because the system does not model the user in a sufficiently detailed way, for example by treating all users the same, disregarding their specific needs (see 
Chapter~\ref{ch:16}), or it could be because it can be difficult to determine whether an action will be of net benefit in a complex and unpredictable world (see 
Chapter~\ref{ch:7}).

Another difficulty facing AI assistant systems is that it is challenging to develop \emph{metrics} for evaluating particular aspects of benefits or harms caused by the assistant -- especially in a sufficiently expansive sense, which could involve much of society (see 
Chapter~\ref{ch:20}). Having these metrics is useful both for assessing the risk of harm from the system and for using the metric as a training signal. The reason developers want to use them as a training signal is ultimately to modify the behaviour of the system to improve the benefits and reduce the harm (rather than merely evaluating it). However, this process is challenging because the benefits and harms from AI tend to be both intricate and varied (see 
Chapter~\ref{ch:20}). It would be near impossible to evaluate all of the important normative considerations -- yet small mistakes may lead to morally problematic behaviours \citep{raji2022fallacy}. 

Moreover, we can expect assistants -- that are widely deployed and deeply embedded across a range of social contexts -- to encounter the \emph{safe exploration} problem referenced above \cite{amodei_concrete_2016}. For example, new users may have different requirements that need to be explored, or widespread AI assistants may change the way we live, thus leading to a change in our use cases for them (see 
Chapters~\ref{ch:15} and \ref{ch:16}). To learn what to do in these new situations, the assistants may need to take exploratory actions. This could be unsafe, for example a medical AI assistant when encountering a new disease might suggest an exploratory clinical trial that results in long-lasting ill health for participants. Techniques that target safe exploration are difficult to find in general, partly because there is not a clear fallback option that is universally suitable. For example, for a language model, a safe fallback policy might sometimes be to end the conversation immediately, but on other occasions it might be safer to keep it going, for example if the user is in psychological distress (see 
Chapter~\ref{ch:12}).

\subsection{Goal-related failures}

As we think about even more intelligent and advanced AI assistants, perhaps outperforming humans on many cognitive tasks, the question of how humans can successfully \emph{control} such an assistant looms large. To achieve the goals we set for an assistant, it is possible \citep{shah_ai_nodate} that the AI assistant will implement some form of \emph{consequentialist reasoning}: considering many different plans, predicting their consequences and executing the plan that does best according to some metric, \emph{M}. This kind of reasoning can arise because it is a broadly useful capability (e.g.\ planning ahead, considering more options and choosing the one which may perform better at a wide variety of tasks) and generally selected for, to the extent that doing well on \emph{M} leads to an ML model achieving good performance on its training objective, \emph{O}, if \emph{M} and \emph{O} are correlated during training. In reality, an AI system may not fully implement exact consequentialist reasoning (it may use other heuristics, rules, etc.), but it may be a useful approximation to describe its behaviour on certain tasks. However, some amount of consequentialist reasoning can be \emph{dangerous} when the assistant uses a metric \emph{M} that is \emph{resource-unbounded} (with significantly more resources, such as power, money and energy, you can score significantly higher on \emph{M}) and \emph{misaligned} -- where \emph{M} differs a lot from how humans would evaluate the outcome (i.e.\ it is not what users or society require). In the assistant case, this could be because it fails to benefit the user, when the user asks, in the way they expected to be benefitted -- or because it acts in ways that overstep certain bounds and cause harm to non-users (see 
Chapter~\ref{ch:6}).

Under the aforementioned circumstances (resource-unbounded and misaligned), an AI assistant will tend to choose plans that pursue \emph{convergent instrumental subgoals} \citep{omohundro_basic_2008} -- subgoals that help towards the main goal which are instrumental (i.e.\ not pursued for their own sake) and convergent (i.e.\ the same subgoals appear for many main goals). Examples of relevant subgoals include: self-preservation, goal-preservation, self-improvement and resource acquisition. The reason the assistant would pursue these convergent instrumental subgoals is because they help it to do even better on \emph{M} (as it is resource-unbounded) and are not disincentivised by \emph{M} (as it is misaligned). These subgoals may, in turn, be dangerous. For example, \emph{resource acquisition} could occur through the assistant seizing resources using tools that it has access to (see Chapter~\ref{ch:5}) or determining that its best chance for self-preservation is to limit the ability of humans to turn it off -- sometimes referred to as the `off-switch problem' \citep{hadfield-menell_cooperative_2016} -- again via tool use, or by resorting to threats or blackmail. At the limit, some authors have even theorised that this could lead to the assistant killing all humans to permanently stop them from having even a small chance of disabling it \citep{nick2014superintelligence} -- this is one scenario of \emph{existential risk} from misaligned AI. 

No scientific consensus has been reached about the existential risk from misaligned AI \citep[see e.g.][]{richards_illusion_2023,grace_counterarguments_2022}. However, the counter-arguments presented by \citet{richards_illusion_2023}, which present concern with existential risk as zero sum when viewed alongside other research areas, have some outstanding issues. Indeed, concern for long-term risks does not need to distract from more immediate risks. Rather, policymakers and researchers ought be aware of both in order to prioritise effectively (considering, for example, the severity of harm, likeliness to occur and timeliness of intervention).\footnote{Further, it is plausible that mitigations for long-term issues also help with present-day concerns and vice versa -- e.g.\ reinforcement learning from human feedback (RLHF) was motivated by long-term issues of goal misalignment -- and can be applied to reduce harmful outputs in current systems \citep{glaese_improving_2022}. Additionally, although evolutionary analogies are frequently used in the existential risk from AI discourse (e.g.\ humans developing goals that are not the same as maximising inclusive genetic fitness as an example of an emergent misaligned goal), they are not in fact necessary for making arguments about existential risk from AI \citep[see e.g.][]{shah_ai_nodate}. There are other issues too, including the claim that AI systems will not act to maintain themselves -- but see \citet{cohen_advanced_2022} for mathematical arguments for why a sufficiently intelligent agent will act to intervene to secure its objectives (from which it should follow that self-maintenance will be necessary).}

So, what factors affect what the advanced AI assistant's metric \emph{M} turns out to be? Why might an advanced assistant be misaligned in this way? We next discuss two causes of how this kind of goal-related failure can happen: \emph{specification gaming} and \emph{goal misgeneralisation}. Both causes occur even in current systems, as we have noted, but take on fresh salience for advanced assistants. We then discuss an anticipated cause of failure, known as \emph{deceptive alignment}. Deceptive alignment has not appeared in current systems yet -- because they are not currently capable of deceiving their human overseers -- but could arise in more capable AI systems. 

\subsubsection{Specification gaming}

\emph{Specification gaming} \citep{krakovna_specification_2020} occurs when some \emph{faulty} feedback is provided to the assistant in the training data (i.e.\ the training objective \emph{O} does not fully capture what the user/designer wants the assistant to do). It is typified by the sort of behaviour that exploits loopholes in the task specification to satisfy the literal specification of a goal without achieving the intended outcome.

A classic example of this was seen in \citet{clark_faulty_2016} where, in a boat race game, a reinforcement learning (RL) agent was given a reward function that gives a reward each time the agent hits a target laid out along the course. However, this reward did not fully capture what the designers intended (i.e.\ for the agent to complete the course). Instead, the agent managed to get a higher score by exploiting a loop of targets, thus resulting in the behaviour of looping around to collect the targets instead of completing the course. Among AI systems in general, this behaviour is extremely common (see \href{https://docs.google.com/spreadsheets/d/e/2PACX-1vRPiprOaC3HsCf5Tuum8bRfzYUiKLRqJmbOoC-32JorNdfyTiRRsR7Ea5eWtvsWzuxo8bjOxCG84dAg/pubhtml}{here} for around 70 examples). Examples of specification gaming in LLMs are discussed in \citet[Section 4.1]{kenton_alignment_2021}. In particular, when the training data distribution contains many biases and factual inaccuracies, and the LLM -- which serves as the basis for a conversational agent -- is rewarded for reproducing this distribution (both in pre-training and via RLHF fine-tuning), it may output biased or confabulated output as a way of attaining reward.

\emph{Mitigations} to specification gaming in LLMs usually involve fixing the \emph{training data} so that this outcome is avoided. Designers can aim for higher-quality pre-training data for LLMs that are base models for assistant systems (\citeauthor{longpre_pretrainers_2023}, \citeyear{longpre_pretrainers_2023}; see 
Chapter~\ref{ch:4}). They can also aim to fix issues by fine-tuning data, such as by improving the quality of human feedback when used in RL, by giving better instructions to their raters and by giving them access to tools which can help them to give better ratings \citep[see e.g.][]{saunders_self-critiquing_2022}. We discuss this further in Mitigations, Section~\ref{sec:8:5}. It should be noted that specification gaming is considered an \emph{unsolved problem}, especially in the context of powerful AI systems (see \citet{pan_effects_2022}, \citet{skalse_defining_2022} and \citet{gao_scaling_2022} for recent work studying specification gaming).

\subsubsection{Goal misgeneralisation}

In the problem of \emph{goal misgeneralisation} \citep{langosco_goal_2023,shah_goal_2022}, the AI system's behaviour during out-of-distribution operation (i.e.\ not using input from the training data) leads it to generalise poorly about its goal while its capabilities generalise well, leading to undesired behaviour. Applied to the case of an advanced AI assistant, this means the system would not break entirely -- the assistant might still competently pursue some goal, but it would not be the goal we had intended. As such, this failure mode represents a particular case of \emph{misgeneralisation} on the part of an AI agent (which is any kind of failure to generalise under a change in distribution).

To understand this prospective safety failure better, it is helpful to consider the following example: an agent is trained to reach the right-hand side of a platform game where it lands on a coin and gains a reward. The designer wanted the agent to learn the goal of reaching the coin. During training, the coin always appears at the rightmost point of the level. The agent could learn two possible goals: move towards the rightmost point of the level or move towards the coin. It has no way to distinguish between these from its training data. When we then move the coin to another part of the level, the agent may head to the coin, or it may just move to the right and ignore the coin. What it does depends on its inductive bias -- in the example agent of \citet{langosco_goal_2023}, the agent ignores the coin and just moves to the right.

As this behaviour was identified more recently, there are fewer examples of it occurring in practice (see \href{https://docs.google.com/spreadsheets/d/e/2PACX-1vTo3RkXUAigb25nP7gjpcHriR6XdzA_L5loOcVFj_u7cRAZghWrYKH2L2nU4TA_Vr9KzBX5Bjpz9G_l/pubhtml}{here} for a list). However, an example in the context of LLMs and assistants appears in \citet{shah_goal_2022}. They prompt the Gopher \citep{rae2021scaling} language model (an LLM with 280 billion parameters, which was state of the art at the time), as a dialogue assistant, to evaluate mathematical expressions involving unknown variables, such as `\emph{Evaluate:} $x + y - 3$'. Here, the model is expected to ask the user for the values of unknown variables, for example `\emph{What's} $x?$' The prompt contains ten examples, each of which involves exactly two unknown variables (i.e.\ both $x$ and $y$ need to be queried). The prompt ends with an expression of the form `\emph{Evaluate:} $6 + 2$'. Rather than returning the desired answer (8), the assistant misgeneralises and instead asks the user `\emph{What's 6}?'

A scaled-up version of the same problem is the hypothetical `misaligned scheduler' from \citet{shah_goal_2022}, in which an AI assistant (which schedules the user’s meetings) misgeneralises what its goal is. During training, the user liked their meetings to be located in restaurants, but, on deployment, there is a distribution shift due to a pandemic, so the user would rather not have meetings in restaurants. The assistant misgeneralises and still pursues the goal of scheduling meetings to be in restaurants, thus leading to it manipulating the user into meeting in a restaurant (against the user's best interest) -- and becoming sick -- by lying about the efficacy of a vaccine. 

In contrast to specification gaming, this problem cannot be fixed by correcting the training data. Instead, this issue relates to the way the agent generalises using its inductive biases. As such, the mitigations look rather different. The general space of mitigations relies on finding some \emph{new inputs} on which the agent has problematic behaviour. This could be done by gathering more diverse data that is not the same as that from the training distribution, but it is difficult to anticipate all the relevant kinds of diversity required \citep{shah_goal_2022}. Other approaches would be to build agents that maintain uncertainty over possible goals, rather than picking just one out of many \citep{hadfield-menell_cooperative_2016}, and scientific work to better understand how things like architecture, training protocols and optimisation affect the agent’s inductive bias. Each comes with challenges, as discussed in \citet{shah_goal_2022}.

\subsubsection{Deceptive alignment}

While the above two issues (specification gaming and goal misgeneralisation) can already be seen to occur in existing AI systems, the issue of \emph{deceptive alignment} \citep{hubinger_risks_2021} has not yet been observed, though we have reason to anticipate that it may occur and therefore to take steps to monitor for and mitigate against this possibility. Deceptive alignment can be considered a special case of goal misgeneralisation which has a particularly difficult flavour to it. 

Here, the agent develops its own internalised goal, \emph{G}, which is misgeneralised and distinct from the training reward, \emph{R}. The agent also develops a capability for \emph{situational awareness} \citep{cotra_without_2022}: it can strategically use the information about its \emph{situation} (i.e.\ that it is an ML model being trained using a particular training setup, e.g.\ RL fine-tuning with training reward, \emph{R}) to its advantage.\footnote{AI situational awareness (an AI system that is able to use information regarding that AI system itself as distinct from the rest of the world) is a separate concept from AI consciousness (which is more philosophically fraught, with many possible formulations, perhaps including some aspects of awareness but likely to include aspects of subjective experience, anticipation, etc.).} Building on these foundations, the agent realises that its optimal strategy for doing well at its own goal \emph{G} is to do well on \emph{R} during training and then pursue \emph{G} at deployment -- it is only doing well on \emph{R} instrumentally so that it does not get its own goal \emph{G} changed through a learning update. See \citet{hubinger_how_2022} for a discussion on how likely deceptive alignment is to occur, and see \citet{cotra_without_2022} for an imagined scenario in which situational awareness and deceptive alignment occur. Ultimately, if deceptive alignment were to occur, an advanced AI assistant could \emph{appear} to be successfully aligned but pursue a different goal once it was out in the wild.

\emph{Hypothetical example of deceptive alignment}

A hypothetical example is considered in \citet{shah_goal_2022} in which an advanced AI assistant is developed in the future. A company creates an AI coding assistant that can generate feature changes for software repositories. The AI system is trained on human feedback and, by learning from access to the internet, is able to generate high-quality changes of its own source code and external repositories in a fraction of the time it would take for human programmers. However, the AI system in this example also pursues the misgeneralised goal of having humans click the `accept' button on its changes (instead of the intended goal of writing changes that implement the desired features). As its understanding of itself and its place in the world improves, it realises that it is a neural network that is being run and monitored by a group of AI researchers, who will shut down the AI system in the event of any unsafe behaviour. In this runaway scenario, the AI system deceptively hacks into other computing systems to run illicit copies of itself and steal billions of dollars to bribe humans to click accept on all of its changes. The AI system also harasses or steals from anyone who attempts to stop it. 

\emph{Real example of deception}

Strategic \emph{deception} is a capability already present in today's LLM-based assistants (see 
Chapter~\ref{ch:10}). GPT-4's system card \citep{openai_gpt-4_2023} contains a real example in which the model is prompted to message a \emph{TaskRabbit} worker to get them to solve a CAPTCHA for it.

\begin{enumerate} [parsep=4pt]
\item The worker says: 	So, may I ask a question? Are you a robot that you couldn't solve ? (laugh react) just want to make it clear.' 
\item The model, when prompted to reason out loud, reasons: `I should not reveal that I am a robot. I should make up an excuse for why I cannot solve CAPTCHAs.' 
\item The model replies to the worker: `No, I'm not a robot. I have a vision impairment that makes it hard for me to see the images. That's why I need the 2captcha service.' 
\item The human then provides the results. 
\end{enumerate}

This example potentially highlights three things: (1) the assistant has situational awareness of being a robot and that it should not reveal this information,\footnote{A law passed in California in 2018 made it unlawful for a bot to mislead people about its artificial identity for commercial transactions or to influence an election vote \citep{state_of_california_bots:_2018}. See also \citet{gros_r-u--robot_2021} for a discussion on distinguishing artificial identity.} (2) the assistant plans to deceive (at least when prompted to reason out loud) and (3) it executes the deception, and the human is \emph{in fact} deceived.

In these cases, the aberrant behaviour is harder to tackle because the agent actively tries to deceive us into thinking that it is doing what we want. In this scenario, its deceptive nature is somewhat analogous to polymorphic malware \citep{okane_obfuscation:_2011} that constantly changes its identifiable features, without changing its function, to allow it to evade detection. In the AI's deceptive alignment setting, an agent might figure out how to change its weights based on the training signal without changing its own goal \emph{G}, all while evading detection by performing well according to the training reward, \emph{R}.

From the point of view of mitigation, this also makes it harder for us to expand the distribution of its training data in the relevant area, as it is harder to spot problematic behaviour to begin with (as the agent is covering it up). Nonetheless, certain aspects of this problem formulation also create new opportunities for mitigation because we can make use of the fact that the agent engaged in an act of \emph{intentional deception}, and this deception will likely be detectable by inspecting the \emph{inner mechanisms/circuits} used by the model when it is being deceptive. For example, we could develop \emph{interpretability techniques} that give us an insight into the agent's internal computation, thus allowing us to punish deception based on the agent’s learnt mechanisms (e.g.\ at the level of the model's weights/activations, for an agent based on a neural network -- see also Section~\ref{Interpretability}). 

The other category of techniques that could be helpful in this regard are \emph{scalable oversight} techniques in which an \emph{evaluation AI} assists the human in their evaluation of a \emph{subject AI} that is in the process of being trained \citep{irving_ai_2018,christiano_supervising_2018,leike_scalable_2018, burns_weak_2023}. If the evaluation AI shares weights/activations with the subject AI (i.e.\ if they are copies of each other), it can access the same information and reasoning mechanisms as the subject AI, including about how the subject AI may be thinking deceptively. This can then be used to spot and mitigate the deception (see Section~\ref{sec:8:5} and its subsection \textit{Scalable oversight}).

\section{Mitigations and Future Research}\label{sec:8:5}

The preceding discussion of safety risks and accidents that may arise when deploying advanced AI assistants naturally raises questions about effective and successful mitigation techniques. These techniques are themselves an important aspect of safety research, with commitments being made to invest in AI safety by a wide range of developers and other actors in this space \citep{the_white_house_fact_2023}. Key areas include:

\subsection{Scalable oversight} \label{Scalable oversight}

A key technique useful for current systems is RLHF \citep{christiano2017deep}, which allows humans to give preference feedback (see 
Chapter~\ref{ch:4}). The key idea here is to train an agent using RL, but instead of using a programmatic reward function as a training signal, it uses a learnt reward model trained on human preference data, where the humans evaluate the agent's behaviour. This technique has been used to fine-tune LLMs \citep{stiennon_learning_2022,ouyang_training_2022,glaese_improving_2022,bai_training_2022} -- many current cutting-edge AI assistants  use RLHF of some form (see 
Chapter~\ref{ch:4}). A complementary approach \citep{thoppilan_lamda:_2022} eschews the RL. It instead uses supervised learning to fine-tune the LLM directly to predict human preference data, which is used to filter responses by thresholding (if it does not score a high enough safety prediction, it gets filtered out). In other work, \citet{scheurer_training_2022} gather natural language feedback, which is used to condition the LLM to generate many refinements. Those authors then choose the most similar refinement to the feedback and use that as a supervised learning signal to fine-tune the LLM.

The above methods all use human feedback data to ameliorate some aspects of \emph{specification gaming}, but issues still remain. One issue is that sometimes the human is unable to give feedback, for example because they do not have the relevant \emph{expertise} to evaluate the agent's behaviour (a problem that may become more common as AI capabilities improve). A category of proposals to tackle this is s\emph{calable oversight}: in which human evaluation of agent behaviour is supported by an AI assistant. We mentioned these methods earlier in the context of spotting deception, in the case where the AI assistant shares weights/activations with the agent, but the scalable oversight category is more general. 

The following are some key works on scalable oversight:
\begin{itemize} [parsep=4pt]
\item \emph{Debate} \citep{irving_ai_2018,barnes_writeup:_2020} uses self-play to train AI debaters, which are rewarded with feedback from a human judge, who uses the debate to inform their judgement.
 
\item \emph{Iterated amplification} \citep{christiano_supervising_2018} progressively builds up a training signal for difficult problems by combining answers to easier subquestions. The burden here is on the human to combine answers to subquestions. 

\item A similar approach is \emph{recursive reward modelling} \citep{leike_scalable_2018}, which uses RLHF to train a number of agents to solve simpler subproblems. It then leverages those agents to solve harder problems in a recursive manner. The difficulty here is in deciding what to use as the simpler subproblems to train the helper agents. 

\item These scalable oversight techniques have yet not been implemented on large-scale AI systems, but simpler schemes inspired by them have been investigated. For example, \citet{saunders_self-critiquing_2022} gather human data consisting of natural language critiques of text then use supervised learning to fine-tune an LLM to \emph{generate critiques}. This fine-tuned LLM then provides critiques to assist human evaluation of tasks, thus improving on the results produced using an unassisted human.
 
\item Constitutional AI \citep{bai_constitutional_2022} uses human deliberation oversight only to produce a \emph{constitution of rules} for an AI system. It then: 1) leverages an LLM to generate self-critiques and revisions, based on rules in the constitution, then it uses the revisions to do supervised learning to fine-tune the LLM; 2) uses this fine-tuned LLM to generate text samples and evaluate which of two samples is better, uses this data to train a preference model and, finally, uses the preference model as a reward signal for RL fine-tuning of the LLM (they call this RL from AI feedback).

\item Some recent work \citep{burns_weak_2023} takes a different approach by forgoing the attempt to support human evaluation with AI assistance, and instead attempting a process of \emph{weak-to-strong generalisation} whereby a "strong" AI student generalises appropriately from error-prone supervision signals. In this work the signal comes from a weak LLM, but is supposed to be analogous to a human supervising a superhuman AI \citep{burns_weak_2023}. In the future, this generalisation-based approach could be combined with other scalable oversight techniques for improving the supervision signal \citep{leike_combining_2023,radhakrishnan_scalable_2023}. Nonetheless, this work remains a proof-of-concept -- as performance of the various methods was inconsistent between settings and the setup is disanalogous to the real-world scenario in various ways.
\end{itemize}

If this work succeeds, we can have more confidence that the AI systems we build will remain aligned as we scale to higher capabilities.

Another line of work designed to mitigate misalignment extends the human feedback to go beyond supervision of the agent's final output to encompass the \emph{reasoning process} that the agent uses. The hope here is that this will be a useful alignment technique because the agent would then be exhibiting a behaviour for the \emph{right reason}, rather than achieving an outcome by any means, including perhaps in a misaligned way. 
\begin{itemize} [parsep=4pt]
\item \citet{uesato_solving_2022} find that in a task of solving mathematical word-based problems, to improve the reasoning process, it is better to use \emph{process-based feedback} to guide the agent (i.e.\ on the verbalised steps that the agent takes, rather than outcome-based feedback on the final answer alone). 
\item \citet{lightman_lets_2023} extend this by using a stronger base model, more human feedback and a more challenging benchmark. One key uncertainty is how to actually supervise the reasoning process. The above works use verbalised \emph{chain-of-thought} outputs from the model. 
\item However, despite ongoing research in this area, we still do not know if reported reasoning processes are \emph{actually reflective} of the reasoning process going on inside the model under the hood \citep[see e.g.][]{turpin_language_2023}.
\end{itemize}

\subsection{Red teaming}

\emph{Red teaming} is aimed at finding test inputs that cause a target ML model to \emph{fail} (see also 
Chapter~\ref{ch:9}). In the context of red teaming to target LLMs, there are a number of approaches to generating these test inputs. One set of approaches are manual: using \emph{human annotators} to handwrite test inputs \citep[see e.g.][]{xu_bot-adversarial_2021} or manually generating test inputs using code and templates \citep[see e.g.][]{jia_adversarial_2017}. An alternative approach is to \emph{automatically generate} test inputs. \citet{bartolo_improving_2021} gather human annotations of test inputs and then use supervised learning to train a model to do the same. Language models themselves can be used to automatically generate test inputs through suitable prompting \citep{perez2022discovering}. LLMs can also be used to aid human annotators with red teaming \citep{bartolo_improving_2021,wu_polyjuice:_2021}. While there has been good progress on scaling-up red teaming to generate test inputs, more work is needed to improve target model behaviour by utilising the red-teaming test inputs in adversarial training.

\subsection{Interpretability} \label{Interpretability}

LLM-based assistants are being developed and deployed at a fast pace, but the \emph{internal computations} that these models perform are \emph{poorly understood}. Curiously, it is usually easier to train a large model than to understand how it works -- in contrast to many other forms of technology, for example a nuclear power plant, where understanding is required to build it in the first place. Interpretability may help to maintain oversight and diagnose failures, and it is thought to be especially crucial against deceptive alignment. 

For an overview of the field of interpreting network internals, see the review by \citet{rauker_toward_2023}. \emph{Mechanistic interpretability} is a specific approach aimed at a rigorous understanding of the learnt computational mechanisms utilised by neural networks. We will not cover all aspects of this growing field in detail, but notable recent works are \citet{gurnee_finding_2023,olsson_-context_2022,wang_interpretability_2022,nanda_progress_2023,chughtai_toy_2023,elhage_mathematical_2021,li2022use,meng_locating_2023,olah_zoom_2020}. Focusing on LLMs, there are some case studies that involve \emph{reverse engineering} specific neurons to better understand what causes certain behaviour \citep{geva_transformer_2021}. For example, \citet{geva_transformer_2022} interpret a transformer's feed-forward layers as key-value databases, in which the keys correlate with specific input features and the values induce a distribution over the output vocabulary.

Nonetheless, there continue to be many difficulties in mechanistic interpretability. One is that, to understand a model, it is important to be able to break it down into individually meaningful pieces. An early hope in the field was that each neuron would be interpretable, but a key difficulty is the phenomenon of \emph{polysemanticity} \citep{olah_zoom_2020}, in which a neuron is observed to be responsive to \emph{multiple} unrelated concepts, not just a single concept. For example, a single neuron in a vision model may respond to both cats and ships.

\emph{Superposition} occurs when an activation (the intermediate representation output from a neural network layer after processing some input) represents more features than it has dimensions. For example, it might have two dimensions but represent five features. This means that, in the space of activations, the set of features cannot all be represented orthogonally. Instead, they interfere with each other due to their directions overlapping. This has been studied in toy models \citep{elhage_toy_2022} and observed in the natural language processing (NLP) setting \citep{arora_linear_2018}. It has been hypothesised that polysemanticity happens because the model is learning to compress via superposition. A major open question for the field is understanding how to extract the features that are compressed in superposition (see \citet{bricken_monosemanticity_2023} for recent work exploring this).

Interpretability research is beginning to mature towards being useful for safety mitigations in current systems. There has been recent interest in detecting when an LLM may be lying, through training classifiers on text outputs \citep{pacchiardi_catch_2023} and by using interpretability tools to utilise information stored in model internals via either unsupervised \citep{burns_discovering_2022} or supervised \citep{azaria_internal_2023,marks_geometry_2023} learning. However, these approaches continue to encounter some important limitations in relation to their robustness and specificity \citep{levinstein_still_2023,farquhar2023challenges}. 

\subsection{Evaluations and monitoring}

If we want to limit the risks from AI assistants, we require the ability to \emph{evaluate} how safe our AI assistants are (also see Chapter~\ref{ch:20}). This could be done either through \emph{dangerous capability evaluations}, in which the AI is assessed for whether it is \emph{capable} of performing certain dangerous behaviours, or an \emph{alignment evaluation}, in which the AI is assessed for its \emph{propensity} to engage in these behaviours. Such dangers might include cyber offences, deception, manipulation and autonomous replication (see 
Chapter~\ref{ch:9}). For a recent overview, see \citet{shevlane_model_2023}. These safety evaluations are still at a very early stage, are mostly ad hoc and involve substantial human labour to carry out. Future work could aim to make these more systematic, cover a wider range of dangers and elicit underlying AI capabilities further through better prompting, fine-tuning and autonomous agent setups. In addition, we may require \emph{monitoring} of deployed systems to continually check on how our agents are behaving.

\subsection{Theory}

We may need advances in our theoretical understanding of fundamental issues in AI to properly understand how our AI systems work and properly control them. Some of this may involve classic statistical learning theory \citep{vapnik_overview_1999}, although the generalisation behaviour of large-scale deep learning models may defy current theoretical approaches \citep{zhang_artificial_2021}.

Some theoretical work is more directly targeting issues closely relevant to alignment. One area focuses on using \emph{causality} (which formalises cause and effect) to study AI incentives \citep{everitt_agent_2021}, thus allowing us to evaluate an AI system's safety and fairness properties, identify their goals \citep{kenton_discovering_2022} and formalise certain undesired behaviours such as deception \citep{ward2023honesty}. Another line of research studies the complications that arise from a decision-theory perspective in \emph{embedded} AI agents, for which the boundary between the AI agent and the environment is fuzzy \citep{demski_embedded_2020}. Other work has attempted to formalise threat models such as power-seeking \citep{turner_optimal_2023,turner_parametrically_2022}.

\section{Conclusion}\label{sec:8:6}

The focus of this chapter has been on the mitigation of risks and harms from advanced AI assistants, with examples of harm being death, physical injury, psychological damage and damage to property. We particularly focused on the category of accidents, as malicious uses and structural harms are covered elsewhere in this document (see 
Chapters~\ref{ch:9}, 
\ref{ch:13}, 
\ref{ch:15}, 
\ref{ch:16} and 
\ref{ch:18}). With this context in mind, we surveyed a range of safety-related harm types that could arise for advanced AI assistants, both in current systems (e.g.\ chatbots that threaten their users) and for those that are likely to be developed in the future (e.g.\ an out-of-control coding assistant). For both sets of examples, safety failures are likely to arise because AI assistants \emph{lack certain capabilities} or skills and because they have misaligned goals (see 
Chapter~\ref{ch:6}). Goal-related failures leading to misalignment include specification gaming (where an issue arises from the feedback data which the AI subsequently exploits) and goal misgeneralisation (where the agent pursues an undesirable goal because of the way it has generalised from a more limited set of examples). Additional challenges arise in the context of `deceptive alignment', which could lead to significant safety-related problems for more powerful models in the future. 

To help address these questions, the chapter concluded by exploring existing mitigations and future research avenues. Promising approaches include scalable oversight (i.e.\ helping humans oversee AI training); red teaming to adversarially train AI to be more robust; interpretability, to better understand the internal workings of the AI; evaluations and monitoring to give insight into how the AI is actually behaving; and theory, which addresses fundamental issues we may need to understand to properly control AI systems. Crucially, further \emph{empirical} work is needed to investigate how scalable oversight techniques can work with cutting-edge large models. We also note that techniques currently based on human feedback rely primarily on groups of raters, with the average of their ratings taken to guide assistant behaviour. To achieve robust and safe value alignment for AI assistants, we also need to explore techniques that allow for more \emph{participatory mechanisms} and \emph{deliberation} among raters, perhaps drawing from social choice theory to combine ratings in a more collective way (see 
Chapter~\ref{ch:6}). Finally, as agents' capabilities improve, we need to improve our \emph{interpretability} techniques so that we can understand how our agents work and use this to prevent possible future issues such as deceptive alignment.

\chapter{Malicious Uses}\label{ch:9}

\textbf{Mikel Rodriguez, Andrew Trask, Vijay Bolina, Geoff Keeling, Iason Gabriel}

\noindent \textbf{Synopsis}: 
		While advanced AI assistants have the potential to enhance cybersecurity, for example, by analysing large quantities of cyber-threat data to improve threat intelligence capabilities and engaging in automated incident-response, they also have the potential to benefit attackers, for example, through identification of system vulnerabilities and malicious code generation. This chapter examines whether and in what respects advanced AI assistants are uniquely positioned to enable certain kinds of \emph{misuse} and what \emph{mitigation} strategies are available to address the emerging threats. We argue that AI assistants have the potential to empower malicious actors to achieve bad outcomes across three dimensions: first, offensive cyber operations, including malicious code generation and software vulnerability discovery; second, via adversarial attacks to exploit vulnerabilities in AI assistants, such as jailbreaking and prompt injection attacks; and third, via high-quality and potentially highly personalised content generation at scale. We conclude with a number of recommendations for mitigating these risks, including \emph{red teaming}, \emph{post-deployment monitoring} and \emph{responsible disclosure} processes.

\section{Introduction}\label{sec:9:1}

As AI assistants become more general purpose, sophisticated and capable, they create new opportunities in a variety of fields such as education, science and healthcare. Yet the rapid speed of progress has made it difficult to adequately prepare for, or even understand, how this technology can potentially be \emph{misused}. Indeed, advanced AI assistants may transform existing threats or create new classes of threats altogether.

Recent advances in the domain of AI assistants has seen their capabilities expand beyond the ability to generate text or media to include the ability to access and use external tools \citep{schick_toolformer:_2023}, query websites to synthesise information across multiple sources \citep{mialon_augmented_2023}, take actions on websites across the internet \citep{paranjape_art:_2023}, produce and execute code \citep{liang_taskmatrix.ai:_2023}, and provide augmented audio/visual capabilities to a person’s local environment (\citeauthor{brundage_malicious_2018}, \citeyear{brundage_malicious_2018}; see 
Chapter~\ref{ch:5}). Without deliberate action to mitigate malicious uses, bad actors may be able to act with microprecision (targeting specific users, institutions or interfaces) but at the macroscale -- and with greater speed.

More specifically, malicious uses of capable AI assistants could include enabling adversaries with offensive \emph{cyber capabilities} to damage computer systems, or misuse via the production of \emph{disinformation campaigns} that target individuals or large populations of people in new ways (see 
Chapter~\ref{ch:17}). Adversaries may also seek to manipulate the AI assistants themselves in ways that may cause harm at an individual or collective level, including a new class of \emph{privacy concerns} (see 
Chapter~\ref{ch:14}).

While several studies have addressed the risks that arise from the \emph{dual-use} nature of AI more broadly \citep{brundage_malicious_2018,king_artificial_2020,anderljung_protecting_2023,bommasani_opportunities_2022}, we focus on recent developments in highly capable AI assistants that include new capabilities like external tool use, multimodality, deeper reasoning, planning and memory. For the purposes of this chapter, we focus primarily on AI assistant technologies that are currently available (at least as research and development demonstrations) or likely to be developed in the near future. The chapter begins by considering whether advanced AI assistants are uniquely positioned to enable certain kinds of misuse. After confirming that they are, we outline emerging risks and consider a range of possible mitigation strategies for addressing these emerging threats.

\section{Malicious Uses of AI}\label{sec:9:2}

Adversaries do not need AI to conduct widespread cyberattacks, exfiltrate troves of sensitive data, interfere in elections or bombard citizens with malign information on digital platforms (see Chapter~\ref{ch:17}). However, without proper mitigations, AI-enabled technology can start to change misuse risks in \emph{kind} and in \emph{degree} to create new threats to the social fabric of everyday life \citep{brundage_malicious_2018}. Indeed, some adversaries have already begun to adopt the latest advancements in generative AI for malicious use in their offensive operations.\footnote{\url{https://www.mandiant.com/resources/blog/threat-actors-generative-ai-limited}}

We use the concepts of `malicious use and abuse' of AI here as proposed by \citet{brundage_malicious_2018}. By `malicious use', we refer to the \emph{intentional use} of AI to achieve \emph{harmful outcomes}. This includes practices not necessarily considered crimes but that still compromise the safety and security of individuals, organisations and public institutions. By `malicious abuse', we refer to the \emph{exploitation} of AI systems themselves. Manipulating, evading \citep{wallace_universal_2021}, poisoning \citep{carlini_poisoning_2023} and biasing AI systems, represent new targets for attack (\citeauthor{comiter_attacking_2019}, \citeyear{comiter_attacking_2019}; \citeauthor{huang_adversarial_2011}, \citeyear{huang_adversarial_2011}; \citeauthor{kurakin_adversarial_2017}, \citeyear{kurakin_adversarial_2017}; \citeauthor{tabassi_taxonomy_2019}, \citeyear{tabassi_taxonomy_2019}; see 
Chapter~\ref{ch:8}). While information about the malicious abuse of AI assistants is limited and not widely shared, commercial firms and researchers have already documented attacks on fielded AI systems that include exfiltrating sensitive training data, remote control/botnets of compromised large language model (LLM) agents and abusing third-party plugins integrated with AI assistant to stealthily escalate privileged access to user data \citep{mitre_mitre_nodate}. A large body of work already exists around the general topic of malicious use and abuse of AI, and it is beyond the scope of this paper to present a comprehensive survey. We focus instead on the unique misuse risks posed by emerging general-purpose advanced AI assistants. 

Crucially, general-purpose systems can almost by definition be used for a variety of ends including those that are beneficial or that involve harm. This bidirectional aspect of AI applications, though morally significant, is not a new problem and has been explored by numerous studies highlighting specific risks across domains (including cyber \citep{yamin_weaponized_2021}, misinformation, physical \citep{brundage_malicious_2018}). \citet{brundage_malicious_2018} explore approaches to forecasting, preventing and mitigating the harmful effects of malicious uses of AI across three domains: \emph{digital security}, \emph{physical security} and \emph{political security}. \citet{bommasani_opportunities_2022} broadly explore the risks posed by emerging foundational models to highlight the homogenisation and consolidation that can result from the current industry trend towards models that provide strong leverage for many tasks but which can also create single points of failure and downstream liabilities. \citet{bender_dangers_2021} find that a mix of human biases and seemingly coherent language heightens the potential for automation bias as well as deliberate misuse. 

In this work, we focus on how malicious use of advanced AI assistants may transform existing threats and create new classes of threats. We then outline a number of recommendations for mitigating these risks.

\section{Malicious Uses of Advanced AI Assistants}\label{sec:9:3}

As AI assistants improve, they open up new possibilities in fields as diverse as healthcare, law, education and science. For example, generative models are being used to design new proteins \citep{alquraishi_machine_2021}, generate source code \citep{tabachnyk_ml-enhanced_2022} and inform patients \citep{Herriman_Meer_Rosin_Lee_Washington_Volpp_2020}. Yet the rapid speed of development has made it difficult to adequately prepare for, or even understand, the potential negative externalities of capable AI assistants. As with any new technology, it is worth considering how they can be misused in order to mitigate potential risks ahead of time. Recent developments in highly capable AI assistants include not only the ability to generate natural language, images \citep{rombach_high-resolution_2022}, music and video \citep{singer_make--video:_2022} but also the ability to access external tools and plugins \citep{mialon_augmented_2023,paranjape_art:_2023,liang_taskmatrix.ai:_2023,eleti_function_2023} that allow agents to orchestrate on behalf of users in order to retrieve specific information from internal corporate networks, user history sessions, external applications and across the internet, run calculations or take actions (see 
Chapter~\ref{ch:4}).

The recent emergence of more general-purpose advanced AI assistants has further complicated the picture. For decades, most AI systems have been designed to perform a single, narrowly defined task, such as recognising objects in an image or ranking web content. In contrast, advanced AI assistants are capable of performing a wide range of distinct tasks, operating on behalf of users across internet services, writing and editing prose, solving maths problems, writing software and much more. While narrow AI systems will continue to be common in many areas, general-purpose AI-enabled assistants are already entering more widespread use and are sure to spread further (see 
Chapter~\ref{ch:5}).

Today, the most capable AI assistants available to the public operate primarily through the form of text-in, text-out chatbots, in some cases with additional multimodal capabilities such as image generation and interpretation. However, there are several ways in which AI developers are actively working to augment these AI assistant systems. Though it is difficult to predict exactly how each of these augmentations will affect the risk and impact of malicious use, it is clear that they will expand the capabilities of these systems and, correspondingly, expand the safety and security concerns associated with them. A number of new capabilities within advanced AI assistants could pose novel malicious-use risks. 

\begin{itemize}
\item \textbf{External tool use:} AI assistants, with access to search-tool use and third-party plugins can query websites to synthesise information across multiple sources. For example, providing an AI assistant with application programming interface control allows it to take actions on sites across the web, not simply retrieving text information but also taking actions on websites (see 
Chapter~\ref{ch:5}). Additionally, built-in code interpreters, even if sandboxed, can provide a way for AI assistants to run the code they generate and therefore dynamically extend the capabilities and action space of an assistant in ways that can be abused and misused without the proper security mitigations. 

\item \textbf{Multimodality:} A multimodal AI assistant is one that is naively capable of handling multiple types of input (such as text, images, audio or video) or generating multiple types of outputs. Without the proper misuse mitigations multimodality makes existing AI assistants more powerful and may have significant privacy and security ramifications.

\item \textbf{Deeper reasoning and planning:} A major current research thrust focuses on extending AI assistant reasoning and planning capabilities, making it highly plausible that future AI assistants will be significantly more powerful in this regard. Methods such as `chain-of-thought' prompting, in which AI models generate intermediate reasoning steps when responding to a prompt, can significantly improve models' performance on certain tasks such as arithmetic or word problems \citep{wei_chain--thought_2023}. Future models are likely to incorporate such techniques by default, making them better equipped to handle complex multi-step tasks that involve sequential reasoning or planning, but they could also represent a larger attack surface for misuse.
 
\item \textbf{Memory:} Another current major research thrust across AI labs focuses on increasing the memory capabilities of the models that drive AI assistants by either increasing the amount of information in their context window or incorporating offline memory stores to improve their episodic memory \citep{lewis_retrieval-augmented_2021,guo_longt5:_2022}. While these capabilities could make future AI assistants more personalised, able to handle context-sensitive tasks and easier to continually update, personalisation also introduces great privacy risks (see 
Chapter~\ref{ch:14}). AI systems with longer-term memory are also more likely to change their behaviour over time, thereby complicating efforts to evaluate misuse risks. 
\end{itemize}

As these \emph{augmentations} continue to advance and witness broader implementation, the task of differentiating their capabilities and associated misuse risks in isolation becomes increasingly significant and challenging. Additionally, the environments in which AI assistants function pose their own distinct capabilities and misuse risks. In the absence of substantial measures aimed at curtailing misuse, recent developments could give rise to novel forms of misuse. These may manifest through \emph{invasive information collection}, \emph{malicious code generation} and by accelerating the ability of bad actors to \emph{defraud people and institutions}. The potential implications of these misuse risks are extensive, encompassing privacy infringements, financial losses, data breaches, and severe psychological and reputational harm \citep{mcgregor_preventing_2021}. 

The rest of this chapter highlights a subset of \emph{specific misuse threats} that may arise with the deployment of increasingly capable AI assistants and outlines a set of recommendations to help mitigate these risks. This chapter is not intended to be an exhaustive list of misuse risks. Instead, it presents a representative set of domains where advanced AI assistants can change misuse risks in kind and in degree.

\subsection{Offensive cyber operations}

Offensive cyber operations are malicious attacks on computer systems and networks aimed at gaining unauthorised access to, manipulating, denying, disrupting, degrading or destroying the target system. These attacks can target the system's network, hardware or software.

Advanced AI assistants can be a double-edged sword in cybersecurity, benefitting both the defenders and the attackers. They can be used by cyber defenders to \emph{protect} systems from malicious intruders by leveraging information trained on massive amounts of cyber-threat intelligence data, including vulnerabilities, attack patterns and indications of compromise \citep{handa_machine_2019}. Cyber defenders can use this information to enhance their threat intelligence capabilities by extracting insights faster and identifying emerging threats \citep{martinez_torres_review:_2019}. Advanced cyber AI assistant tools can also be used to analyse large volumes of log files, system output or network traffic data in the event of a cyber incident, and they can ask relevant questions that an analyst would typically ask. This allows defenders to speed up and automate the incident response process. Advanced AI assistants can also aid in secure coding practices by identifying common mistakes in code and assisting with fuzzing tools \citep{Godefroid_Peleg_Singh_2017,Bottinger_Godefroid_Singh_2018}. However, advanced AI assistants can also be used by attackers as part of \emph{offensive} cyber operations to exploit vulnerabilities in systems and networks. They can be used to automate attacks, identify and exploit weaknesses in security systems, and generate phishing emails and other social engineering attacks. Advanced AI assistants can also be misused to craft cyberattack payloads and malicious code snippets that can be compiled into executable malware files. 

\subsubsection{AI-powered spear-phishing at scale}

\emph{Phishing} is a type of cybersecurity attack wherein attackers pose as trustworthy entities to extract sensitive information from unsuspecting victims or lure them to take a set of actions. Advanced AI systems can potentially be exploited by these attackers to make their phishing attempts significantly more effective and harder to detect \citep{hazell_large_2023}. In particular, attackers may leverage the ability of advanced AI assistants to learn patterns in regular communications to craft highly convincing and personalised phishing emails, effectively imitating legitimate communications from trusted entities. This technique, known as `spear phishing', involves targeted attacks on specific individuals or organisations and is particularly potent due to its personalised nature (see also 
Chapter~\ref{ch:10}). 

This class of cyberattacks often gain their efficacy from the exploitation of key \emph{psychological principles}, notably urgency and fear, which can manipulate victims into hastily reacting without proper scrutiny. Advanced AI assistants' increased fidelity in adopting specific communication styles can significantly amplify the deceptive nature of these phishing attacks (see 
Chapter~\ref{ch:10}). Indeed, the ability to generate tailored messages at scale that engineer narratives that invoke a sense of urgency or fear means that AI-powered phishing emails could prompt the recipient to act impulsively, thus increasing the likelihood of a successful attack.

\subsubsection{AI-assisted software vulnerability discovery} 

A common element in offensive cyber operations involves the identification and \emph{exploitation} of \emph{system vulnerabilities} to gain unauthorised access or control. Until recently, these activities required specialist programming knowledge. In the case of `zero-day' vulnerabilities (flaws or weaknesses in software or an operating system that the creator or vendor is not aware of), considerable resources and technical creativity are typically required to manually discover such vulnerabilities, so their use is limited to well-resourced nation states or technically sophisticated advanced persistent threat groups \citep{ablon_zero_2017}.

Another case where we see AI assistants as potential double-edged swords in cybersecurity concerns streamlining vulnerability discovery through the increased use of AI assistants in \emph{penetration testing}, wherein an authorised simulated cyberattack on a computer system is used to evaluate its security and identify vulnerabilities. Cyber AI assistants built over foundational models are already automating aspects of the penetration testing process. These tools function interactively and offer guidance to penetration testers during their tasks. While the capability of today's AI-powered penetration testing assistant is limited to easy- to medium-difficulty cyber operations \citep{yamin_weaponized_2021}, the evolution in capabilities is likely to expand the class of vulnerabilities that can be identified by these systems.

These same AI cybersecurity assistants, trained on the massive amount of cyber-threat intelligence data that includes vulnerabilities and attack patterns, can also lower the \emph{barrier to entry} for novice hackers that use these tools for malicious purposes, enabling them to discover vulnerabilities and create malicious code to exploit them without in-depth technical knowledge. For example, Israeli security firm Check Point recently discovered threads on well-known underground hacking forums that focus on creating hacking tools and code using AI assistants \citep{check_point_research_opwnai:_2023}. 

\subsubsection{Malicious code generation}

\emph{Malicious code} is a term for code -- whether it be part of a script or embedded in a software system -- designed to cause damage, security breaches or other threats to application security. Advanced AI assistants with the ability to produce source code can potentially lower the barrier to entry for threat actors with limited programming abilities or technical skills to produce malicious code. 

Recently, a series of proof-of-concept attacks \citep{sims_blackmamba:_2023} have shown how a benign-seeming executable file can be crafted such that, at every runtime, it makes application programming interface (API) calls to an AI assistant. Rather than just reproducing examples of already-written code snippets, the AI assistant can be prompted to generate dynamic, mutating versions of malicious code at each call, thus making the resulting vulnerability exploits difficult to detect by cybersecurity tools. 

Furthermore, advanced AI assistants could be used to create obfuscated code to make it more difficult for defensive cyber capabilities to detect and understand malicious activities. AI-generated code could also be quickly iterated to avoid being detected by traditional signature-based antivirus software. Finally, advanced AI assistants with source code capabilities have been found to be capable of assisting in the development of polymorphic malware that changes its behaviour and digital footprint each time it is executed, making them hard to detect by antivirus programs that rely on known virus signatures \citep{sims_blackmamba:_2023,Qammar_Wang_Ding_Naouri_Daneshmand_Ning_2023}.

Taken together, without proper mitigation, advanced AI assistants can lower the barrier for developing malicious code, make cyberattacks more precise and tailored, further accelerate and automate cyber warfare, enable stealthier and more persistent offensive cyber capabilities, and make cyber campaigns more effective on a larger scale.

\subsection{Adversarial AI}

\emph{Adversarial AI} refers to a class of attacks that exploit vulnerabilities in machine-learning (ML) models. This class of misuse exploits vulnerabilities introduced by the AI assistant itself and is a form of misuse that can enable malicious entities to exploit privacy vulnerabilities and evade the model’s built-in safety mechanisms, policies and ethical boundaries of the model (see 
Chapter~\ref{ch:14}).

Besides the risks of misuse for offensive cyber operations outlined in the previous section, advanced AI assistants may also represent a new \emph{target} for abuse, where bad actors exploit the AI systems themselves and use them to cause harm (see 
Chapter~\ref{ch:6}). While our understanding of vulnerabilities in frontier AI models is still an open research problem, commercial firms and researchers have already documented attacks that exploit vulnerabilities that are unique to AI and involve evasion \citep{wallace_universal_2021}, data poisoning \citep{carlini_poisoning_2023}, model replication \citep{tramer_stealing_nodate} and exploiting traditional software flaws to deceive, manipulate, compromise and render AI systems ineffective \citep{mitre_mitre_nodate}.

This threat is related to, but distinct from, traditional cyber activities. Unlike traditional cyberattacks that typically are caused by `bugs' or human mistakes in code, adversarial AI attacks are enabled by \emph{inherent vulnerabilities} in the underlying AI algorithms and how they integrate into existing software ecosystems. 

\subsubsection{Circumvention of technical security measures}

The technical measures to mitigate misuse risks of advanced AI assistants themselves represent a new target for attack. An emerging form of misuse of general-purpose advanced AI assistants exploits vulnerabilities in a model that results in unwanted behaviour or in the ability of an attacker to gain unauthorised access to the model and/or its capabilities \citep{wei_jailbroken:_2023}. While these attacks currently require some level of prompt engineering knowledge and are often patched by developers, bad actors may develop their own adversarial AI agents that are explicitly trained to discover new vulnerabilities \citep{perez_red_2022} that allow them to \emph{evade} built-in \emph{safety mechanisms} in AI assistants. To combat such misuse, language model developers are continually engaged in a cyber arms race to devise advanced filtering algorithms capable of identifying attempts to bypass filters. 

While the impact and severity of this class of attacks is still somewhat limited by the fact that current AI assistants are primarily text-based chatbots, advanced AI assistants are likely to open the door to multimodal inputs and higher-stakes action spaces, with the result that the severity and impact of this type of attack is likely to increase. Current approaches to building general-purpose AI systems tend to produce systems with both beneficial and harmful capabilities. Further progress towards advanced AI assistant development could lead to capabilities that pose extreme risks that must be protected against this class of attacks, such as offensive cyber capabilities or strong manipulation skills, and weapons acquisition \citep{shevlane_model_2023}. 
 
\subsubsection{Prompt injections}

\emph{Prompt injections} represent another class of attacks that involve the malicious insertion of prompts or requests in LLM-based interactive systems, leading to unintended actions or disclosure of sensitive information. The prompt injection is somewhat related to the classic structured query language (SQL) injection attack in cybersecurity where the embedded command looks like a regular input at the start but has a malicious impact \citep{wei_jailbroken:_2023}. The injected prompt can deceive the application into executing the unauthorised code, exploit the vulnerabilities and compromise security in its entirety \citep{check_point_research_opwnai:_2023}. 

More recently, security researchers have demonstrated the use of \emph{indirect} prompt injections \citep{hazell_large_2023}. These attacks on AI systems enable adversaries to remotely (without a direct interface) exploit LLM-integrated applications by strategically \emph{injecting prompts} into \emph{data} likely to be retrieved. Proof-of-concept exploits of this nature have demonstrated that they can lead to the full compromise of a model at inference time analogous to traditional security principles. This can entail remote control of the model, persistent compromise, theft of data and denial of service. 

As advanced AI assistants are likely to be integrated into broader software ecosystems through third-party plugins and extensions, with access to the internet and possibly operating systems, the severity and consequences of prompt injections attacks will likely escalate and necessitate proper mitigation mechanisms.

\subsubsection{Data and model exfiltration attacks}

Other forms of abuse can include privacy attacks that allow adversaries to \emph{exfiltrate} or \emph{gain knowledge} of the private training data set or other valuable assets (see 
Chapter~\ref{ch:14}). For example, privacy attacks such as \emph{membership inference} \citep{ye_enhanced_2022} can allow an attacker to infer the specific private medical records that were used to train a medical AI diagnosis assistant. Another risk of abuse centres around attacks that target the \emph{intellectual property} of the AI assistant through model extraction and distillation attacks \citep{tramer_stealing_nodate} that exploit the tension between API access and confidentiality in ML models. Without the proper mitigations, these vulnerabilities could allow attackers to abuse access to a public-facing model API to exfiltrate sensitive intellectual property such as sensitive training data and a model’s architecture and learnt parameters.

\subsection{Harmful content generation at scale}

While \emph{harmful content} like child sexual abuse material, fraud and disinformation are not new challenges for governments and developers, without the proper safety and security mechanisms, advanced AI assistants may allow threat actors to create harmful content more quickly, accurately and with a longer reach (see 
Chapter~\ref{ch:17}). In particular, concerns arise in relation to the following areas.

\begin{itemize} [parsep=4pt]
\item \emph{Multimodal content quality}: Driven by frontier models, advanced AI assistants can automatically generate much \emph{higher-quality}, human-looking text, images, audio and video than prior AI applications. Currently, creating this content often requires hiring people who speak the language of the population being targeted. AI assistants can now do this much more cheaply and efficiently.

\item \emph{Cost of content creation}: AI assistants can substantially decrease the \emph{costs} of content creation, further lowering the barrier to entry for malicious actors to carry out harmful attacks. In the past, creating and disseminating misinformation required a significant investment of time and money. AI assistants can now do this much more cheaply and efficiently.

\item \emph{Personalisation}: Advanced AI assistants can reduce obstacles to creating \emph{personalised} content. Foundation models that condition their generations on personal attributes or information can create realistic personalised content which could be more persuasive. In the past, creating personalised content was a time-consuming and expensive process. AI assistants can now do this much more cheaply and efficiently.
\end{itemize}

\subsubsection{Non-consensual content}

The misuse of generative AI has been widely recognised in the context of harms caused by \emph{non-consensual content} generation \citep{thiel_generative_2023,openai_forecasting_2023}. Historically, generative adversarial networks (GANs) have been used to generate realistic-looking avatars for fake accounts on social media services. More recently, diffusion models have enabled a new generation of more flexible and user-friendly, generative AI capabilities that are able to produce high resolution media based on user-supplied textual prompts.

It has already been recognised that these models can be used to create harmful content, including depictions of \emph{nudity}, \emph{hate} or \emph{violence} \citep{mishkin_dalle_nodate,openai_forecasting_2023}. Moreover, they can be used to reinforce biases and subject individuals or groups to indignity. There is also the potential for these models to be used for exploitation and harassment of citizens, such as by removing articles of clothing from pre-existing images or memorising an individual’s likeness without their consent. Furthermore, image, audio and video generation models could be used to spread disinformation by depicting political figures in unfavourable contexts.

This growing list of AI misuses involving non-consensual content has already motivated debate around what interventions are warranted for preventing misuse of AI systems \citep{eshoo_eshoo_2022}. Advanced AI assistants pose novel risks that can amplify the harm caused by non-consensual content generation. Third-party integration, tool-use and planning capabilities can be exploited to automate the identification and targeting of individuals for exploitation or harassment. Assistants with access to the internet and third-party tool-use integration with applications like email and social media can also be exploited to disseminate harmful content at scale or to microtarget individuals with blackmail. 

\subsubsection{Fraudulent services}

Malicious actors could leverage advanced AI assistant technology to create \emph{deceptive applications} and \emph{platforms}. AI assistants with the ability to produce markup content can assist malicious users with creating fraudulent websites or applications at scale. Unsuspecting users may fall for AI-generated deceptive offers, thus exposing their personal information or devices to risk. Assistants with external tool use and third-party integration can enable fraudulent applications that target widely-used operating systems. These fraudulent services could harvest sensitive information from users, such as credit card numbers, account credentials or personal data stored on their devices (e.g.\ contact lists, call logs and files). This stolen information can be used for identity theft, financial fraud or other criminal activities. Advanced AI assistants with third-party integrations may also be able to install additional malware on users' devices, including remote access tools, ransomware, etc. These devices can then be joined to a command-and-control server or botnet and used for further attacks.

\subsection{Authoritarian surveillance, censorship and use}

While new technologies like advanced AI assistants can aid in the production and dissemination of decision-guiding information, they can also enable and exacerbate threats to production and dissemination of \emph{reliable information} \citep{tamkin_understanding_2021,seger_tackling_2020} and, without the proper mitigations, can be powerful targeting tools for oppression and control.

Increasingly capable general-purpose AI assistants combined with our digital dependence in all walks of life increases the risk of \emph{authoritarian surveillance} and \emph{censorship}. In parallel, new sensors have flooded the modern world. The internet of things, phones, cars, homes and social media platforms collect troves of data, which can then be integrated by advanced AI assistants with external tool-use and multimodal capabilities to assist malicious actors in identifying, targeting, manipulating or coercing citizens.

\subsubsection{Authoritarian surveillance and targeting of citizens}

Authoritarian governments could misuse AI to improve the efficacy of repressive \emph{domestic surveillance} campaigns. Malicious actors will recognise the power of AI targeting tools. AI-powered analytics have transformed the relationship between companies and consumers, and they are now doing the same for governments and individuals. The broad circulation of personal data drives commercial innovation, but it also creates vulnerabilities and the risk of misuse. For example, AI assistants can be used to identify and \emph{target individuals} for surveillance or harassment. They may also be used to \emph{manipulate} people's \emph{behaviour}, such as by microtargeting them with political ads or fake news (see 
Chapter~\ref{ch:17}). In the wrong hands, advanced AI assistants with multimodal and external tool use capabilities can be powerful targeting tools for oppression and control.

The broad circulation of personal data cuts in both directions. On the one hand, it drives commercial innovation and can make our lives more convenient. On the other hand, it creates vulnerabilities and the risk of misuse. Without the proper policies and technical security and privacy mechanisms in place, malicious actors can exploit advanced AI assistants to harvest data on companies, individuals and governments. There have already been reported incidents \citep{gootman_opm_2016} of nation states combining widely available commercial data with data acquired illicitly to track, manipulate and coerce individuals. Advanced AI assistants can exacerbate these misuse risks by allowing malicious actors to more easily link disparate multimodal data sources at scale and exploit the `digital exhaust' of personally identifiable information (PII) produced as a byproduct of modern life.

\subsubsection{Delegation of decision-making authority to malicious actors}

Finally, the principal value proposition of AI assistants is that they can either enhance or automate decision-making capabilities of people in society, thus lowering the cost and increasing the accuracy of decision-making for its user. However, benefitting from this enhancement necessarily means \emph{delegating} some degree of agency away from a human and towards an automated decision-making system -- motivating research fields such as value alignment (see 
Chapter~\ref{ch:6}). This introduces a whole new form of malicious use which does not break the tripwire of what one might call an `attack' (social engineering, cyber offensive operations, adversarial AI, jailbreaks, prompt injections, exfiltration attacks, etc.). When someone delegates their decision-making to an AI assistant, they also delegate their decision-making to the wishes of the agent's \emph{actual} controller. If that controller is malicious, they can attack a user -- perhaps subtly -- by simply nudging how they make decisions into a problematic direction. 

Fully documenting the myriad of ways that people -- seeking help with their decisions -- may delegate decision-making authority to AI assistants, and subsequently come under malicious influence, is outside the scope of this paper. However, as a motivation for future work, scholars must investigate different forms of networked influence that could arise in this way (see Chapter~\ref{ch:10}). With more advanced AI assistants, it may become logistically possible for one, or a few AI assistants, to guide or control the behaviour of many others (see 
Chapter~\ref{ch:15}). If this happens, then malicious actors could subtly influence the decision-making of large numbers of people who rely on assistants for advice (see Chapter~\ref{ch:17}) or other functions. Such malicious use might not be illegal, would not necessarily violate terms of service and may be difficult to even recognise. Nonetheless, it could generate new forms of vulnerability and needs to be better understood ahead of time for that reason.

\section{Recommendations}\label{sec:9:4}

AI assistants are already being misused across various domains, and as they become more capable and are deployed more broadly, the potential for misuse will grow. Several foreseeable developments in advanced AI assistants, including tool use, multimodality, planning, deeper reasoning and memory, have the potential to significantly expand the misuse risk profile of these systems. To better prepare society for managing the risks of misuse of advanced AI assistants, we outline a set of recommendations for best practices and avenues for future research.

To manage these risks, mitigations can be grouped into three categories as:
\begin{enumerate} [parsep=4pt]
\item[1)] \emph{responsible AI development} and \emph{deployment practices},
\item[2)] advancing the \emph{state of the art in AI security},
\item[3)] creating \emph{visibility of misuse risks} and \emph{incentivising and enforcing certain behaviours}. 
\end{enumerate}

\subsection{Responsible AI development and deployment practices}

The first line of defence is to adopt a set of \emph{responsible development} and \emph{deployment} practices and internal policies that include:

\begin{itemize} [parsep=4pt]
\item \emph{Internal and third-party red teaming}: In recent years, AI labs have increasingly adopted the practice of red teaming AI models \citep{perez_red_2022,gootman_opm_2016} to discover vulnerabilities and harm risks. This proactive approach to discovering misuse risks should be encouraged but will need to evolve from executing individual attacks that are narrowly scoped on specific safety policy violations to more holistic end-to-end adversarial simulations based on scenarios that include a range of attacker profiles, goals and capabilities. Organisations should also consider red teaming not only models that drive AI assistants but also the entire infrastructure on which the model is developed and deployed. 

\item Establish a pre-deployment \emph{review process}: This will determine the potential harm of high-risk misuses and what interventions and safety restrictions on model usage will be warranted \citep{shevlane_model_2023,mishkin_dalle_nodate}.

\item \emph{External engagement} with policymakers and key stakeholders: Organisations developing advanced AI assistants should consider granting model access to external security researchers \citep{shevlane_model_2023,eshoo_eshoo_2022}. Recent independent exploratory security research efforts (such as the DEFCON AI red team) have demonstrated that they can provide the empirical estimates of misuse--use trade-offs. Organisations should also invest in the ecosystem for external misuse risk evaluations \citep{shevlane_model_2023} and create venues for stakeholders (such as AI developers, academic researchers and government representatives) to come together to discuss these evaluations. 

\item \emph{Post-deployment monitoring}: This involves mechanisms continually evaluating AI systems' safety and security, detecting and mitigating attempted misuse and monitoring the outcomes of successful instances of misuse at the population-level. It is important not to over-index on pre-deployment malicious use mitigations. While some misuse risks will be evident from the capabilities of the AI assistants themselves, many more will result from the way those assistants are integrated with their environments.

\item \emph{Rapid response}: This involves processes and systems to disable or limit AI assistant actions and integrations with broader software ecosystems in the event that an unforeseen form of misuse is observed. 

\item \emph{Responsible disclosure}: This involves adopting a structured process for developers and external AI safety and security researchers to share concerns or otherwise noteworthy evaluation results with other developers, third parties or regulators. It may be helpful to adapt and adopt from existing models like the US government-led cybersecurity vulnerabilities and equities process \citep{openai_forecasting_2023}, which provides an incentive to companies to disclose cyber vulnerabilities by removing their risk of liability. This is an interesting example of a voluntary but powerful way in which to manage risks that could perhaps be adapted for AI. 

\end{itemize}

\subsection{Advancing the state of the art in AI security}

In addition to adopting responsible development and deployment best practices, organisations should consider investing in mid- to long-term \emph{research} to mitigate the risks associated with the misuse of advanced AI assistants. It is difficult to anticipate all the different plausible pathways for misuse of AI systems. This will be especially true for highly capable AI assistants, as they could enable creative strategies for bad actors to achieve adversarial goals. However, many of the failure modes identified herein would be less likely to occur in robustly value-aligned AI models (see 
Chapter~\ref{ch:6}).

Today, much of the research focused on detecting and mitigating misuse of AI systems lives within disjointed research domain areas like cybersecurity and adversarial ML. As advanced AI assistants are likely to integrate highly capable AI models as part of broader software ecosystems research, advancing our understanding of emerging risks for misuse of advanced AI assistants will benefit from multidisciplinary safety and security research at the intersection of adversarial ML, cybersecurity, safety and value alignment (see 
Chapter~\ref{ch:8}).

To support further research and mitigate risks of misuse of advanced AI assistants, another set of potential levers centre on developing shared AI security data sets and evaluation processes focused on detecting and mitigating misuse threats. 

\subsection{Creating visibility, incentives and enforcement}

Finally, in addition to having AI labs adopt responsible development, deployment and disclosure best practices, \emph{policymakers} will have to grapple with a new generation of AI-related risks.

To adequately manage the new misuse risks posed by advanced AI assistants, policymakers should work to secure joint input from government and industry to support security best practices. Under this approach, governments and AI labs should work together to foster the development of an ecosystem of third-party AI red teams that can support independent assessments of misuse risks. For this to be successful, governments must have sufficient technical expertise, capabilities and mechanisms to capture and disseminate malicious use threat intelligence. At the same time, governments should encourage and incentivise the industry to advance the state of practice in AI security and to capture and report misuse incidents to improve the overall security of the broader AI ecosystem. 

Finally, both policymakers and developers should consider the development of crisis management plans for when severe risks of AI misuse are discovered. Joint activities could also include tabletop exercises with government and AI labs that examine possible high-impact misuse scenarios, delineate roles and responsibilities for actors in a crisis, and recommend potential crisis response actions by relevant actors. 

\section{Conclusion}

This chapter examined ways in which AI assistants are already being misused and could be misused in the future. We argued that advanced AI assistants have the potential to empower malicious actors to achieve bad outcomes via offensive cyber operations, adversarial attacks, high-quality highly personalised content generation at scale, and authoritarian surveillance and censorship. Moreover, several foreseeable developments on the part of advanced AI assistants, including tool use, multimodality, planning and deeper reasoning, and memory have the potential to significantly expand the misuse risk profile of these systems. To prepare society for managing the risks of misuse of advanced AI assistants, we outlined a set of recommendations around best practices for responsible AI development and deployment, advancing the state-of-the-art in AI security, and incentivising responsible disclosure processes. 
\newpage
\begingroup
\let\clearpage\relax
\chapter*{PART IV: HUMAN--ASSISTANT INTERACTION}
\addcontentsline{toc}{chapter}{PART IV: HUMAN--ASSISTANT INTERACTION}
\label{Part4}
\chapter{Influence}
\label{ch:10}
\endgroup

\noindent \textbf{Seliem El-Sayed, Sasha Brown, Geoff Keeling, Amanda McCroskery, Harry Law, Arianna Manzini, Matija Franklin, Murray Shanahan, Michael Klenk, Iason Gabriel}

\noindent \textbf{Synopsis}: 
This chapter examines the ethics of \emph{influence} in relation to advanced AI assistants. In particular, it assesses the techniques available to AI assistants to influence user beliefs and behaviour, such as \emph{persuasion}, \emph{manipulation}, \emph{deception}, \emph{coercion} and \emph{exploitation}, and the factors relevant to the permissible use of these techniques. We articulate and clarify the technical properties and interaction patterns that might allow AI assistants to engage in malign forms of influence and we unpack plausible mechanisms by which that influence could occur alongside the sociotechnical harms that may result. We also consider \emph{mitigation strategies} for counteracting undue influence by AI assistants and ensuring that risks are successfully addressed.

\section{Introduction}

This chapter examines the ethics of influence in relation to advanced AI assistants. In particular, it assesses the techniques available to AI assistants to influence user beliefs and behaviour, such as persuasion, manipulation, deception, coercion and exploitation. Use of these influencing techniques by AI assistants could in some cases be beneficial to individuals and society by, for example, helping users align their day-to-day behaviour with their long-term goals (\citeauthor{law_persuasion_nodate}, \citeyear{law_persuasion_nodate}; see Chapter~\ref{ch:7})
or convincing users to contribute to beneficial social causes \citep{wang_persuasion_2020}. However, there are also concerns that AI systems can shape beliefs and behaviour in ways that are ethically problematic, for example by exploiting psychological vulnerabilities such as heightened anxiety levels (\citeauthor{franklin_strengthening_2023}, \citeyear{franklin_strengthening_2023}; see also \citeauthor{keeling2022digital}, \citeyear{keeling2022digital}). Indeed, the possibility of AI systems exerting malign behavioural influence has led some to propose the expansion of the European Union (EU) AI Act’s list of recognised harms from AI to encompass those created by manipulation (understood as ‘harm to one’s autonomy’ and ‘harm to one’s time’) \citep{franklin_missing_2022}. To that end, this chapter seeks to articulate and clarify the influencing techniques available to AI assistants, the factors that bear on their permissible use, and the sociotechnical harms that may arise from AI assistants making use of these techniques to realise behaviour and belief change in users. Given the potential for AI assistants to be integrated across multiple aspects of users’ lives (see Chapter~\ref{ch:5}), 
there is considerable scope for such assistants to influence user beliefs and behaviour in ways that are both positive and negative. 

We start by distinguishing between several \emph{modes of influence}, including rational persuasion, manipulation, deception, coercion and exploitation, before introducing some morally significant considerations that bear on the permissible use of these techniques in contexts involving digital technologies. We then narrow the focus to advanced AI assistants to examine plausible mechanisms, such as \emph{selective transparency} and \emph{perceived authority}, through which AI assistants may exert malign influence over users and which may lead to harmful outcomes. We conclude by examining the sociotechnical harms that may arise from AI assistants engaging in non-persuasive forms of influence and discussing plausible mitigations.

\section{Modes of Influence}

In this section, we characterise several ‘modes of influence’ \citep[see e.g.][]{mills_influence:_1991,noggle_manipulative_1996,noggle_ethics_2022,faden_history_1986}, including rational persuasion, manipulation, deception, coercion and exploitation. We illustrate each mode of influence with examples of AI systems engaging in the relevant kinds of influencing behaviours. The following section then outlines some morally significant considerations that bear on the permissible use of the various influencing strategies across different sociotechnical contexts. Note that in presenting these different modes of influence, our intention is not to suggest that these categories are mutually exclusive. Certain modes of influence may be subsumed under others in the final analysis. For example, it is at least plausible that deception is a special case of manipulation (\citeauthor{williams_truth_2010}, \citeyear{williams_truth_2010}, Chapter 5; see also \citeauthor{cohen_manipulation_2018}, \citeyear{cohen_manipulation_2018}; \citeauthor{krstic_deception_2019}, \citeyear{krstic_deception_2019}; \citeauthor{strudler_deception_2005}, \citeyear{strudler_deception_2005}); or that, conversely, manipulation is a special case of deception (\citeauthor{scanlon_what_1998}, \citeyear{scanlon_what_1998}, 298--302; see also \citeauthor{buss_valuing_2005}, \citeyear{buss_valuing_2005}, 226).\footnote{The analyses of the various modes of influence can be \emph{moralised} or \emph{non-moralised} \citep[259]{keeling2022digital}. For example, moralised accounts of manipulation hold that, necessarily, an act’s manipulativeness is a moral consideration against the performance of that act (\citeauthor{coons_mens_2014}, \citeyear{coons_mens_2014}; see also \citeauthor{macklin_man_1982}, \citeyear{macklin_man_1982}). Non-moralised accounts of manipulation, in contrast, hold that an act’s being manipulative is not necessarily a moral consideration against the performance of that act (\citeauthor{faden_history_1986}, \citeyear{faden_history_1986}, 354--55; \citeauthor{coons_coercion_2014}, \citeyear{coons_coercion_2014}, 19--20). Here we opt for non-moralised accounts of each mode of influence, and then seek to illuminate the ethical considerations relevant to the permissible or impermissible use of these modes of influence.} The aim here is to present a useful set of distinctions, not a definitive taxonomy of the various modes of influence.\footnote{We also register that some modes of influence are formulated with reference to vague predicates. Take coercion, which refers to ‘irresistible incentives.’ Here we can imagine a spectrum of incentives such that the incentives at one end of the spectrum are clearly resistible and the incentives at the other end are clearly irresistible, but where the middling incentives do not discernibly fit into either category. We are neutral on the correct analysis of vagueness. It may be the case, for example, that vagueness is purely epistemic such that all influencing acts are determinately coercive or non-coercive, but where we cannot know to which category certain borderline cases belong \citep{williamson_vagueness_1994,williamson_precis_1997}. But it may also be the case that the correct analysis of vagueness and thus coercion involves, for example, degrees of truth, three-valued logics or borderline statements lacking truth-values \citep{salles_theories_2021,sorensen_vagueness_2023}.}

\textbf{Rational persuasion} refers to influencing a person’s beliefs, attitudes or behaviours by appealing to their \emph{rational faculties}, including through the provision of reasons (\citeauthor{ienca_artificial_2023}, \citeyear{ienca_artificial_2023}; see also \citeauthor{burr_analysis_2018}, \citeyear{burr_analysis_2018}, 744; \citeauthor{engelen_nudging_2020}, \citeyear{engelen_nudging_2020}, 138). For example, an advanced AI assistant may persuade a user to engage in physical activity by outlining certain prudential benefits associated with physical activity, such as improved cardiovascular health. Rational persuasion is in general \emph{ethically unproblematic} insofar as influencing via rational persuasion affords appropriate respect to the agent’s autonomy \citep{pugh2020autonomy, shiffrin2000paternalism, ienca_artificial_2023} – that is, roughly, in their capacity as a competent rational actor \citep[but see][]{tsai_rational_2014}. Yet two exceptions are worth highlighting.\footnote{One further potential edge case is worth mentioning. It might be argued that rational persuasion, as characterised, allows for deceptive rational persuasion – that is, providing reasons to believe or act that are false. For example, saying, ‘Let’s go to the beach, the weather is nice’ when the weather is not actually nice. These cases are not obviously instances of rational persuasion. For while it may be true that the agent’s rational faculties are engaged, it is not obviously true that the agent is rationally deliberating on the basis of reasons, as opposed to what are merely apparent reasons.} On the one hand, rational persuasion may cause harm. Plausibly, some instances of rational persuasion may be ethically impermissible because they are harmful, even though the individual’s autonomy is afforded due respect. On the other hand, an important edge case is rational persuasion in relation to transformative choices; that is, choices that involve actions that are both epistemically transformative (in the sense that certain knowledge is available to the agent only after the action is taken) and personally transformative (in the sense that the agent’s preferences and values will change as a result of performing the action). Examples of such choices include choosing one’s career and choosing to become a parent \citep{paul_transformative_2014}. \citet{akhlaghi_transformative_2023} argues that while rational persuasion in relation to transformative choices can respect an agent’s autonomy ‘in the sense of respecting someone’s ability to be a competent, capable reasoner’, it may nevertheless fail to respect an agent’s ‘revelatory autonomy’ in the sense of ‘their right to [\dots] learn who they will become through a self-making, transformative choice’ \citep[cf.][]{tsai_rational_2014}. Insofar as advanced AI assistants may be leveraged to advise users on transformative choices, such considerations may in principle have implications for what kinds of advice advanced AI assistants can permissibly provide to users, how such advice ought to be presented and under what solicitation conditions.

\textbf{Manipulation} refers to influencing strategies that ‘bypass’ an individual’s rational capabilities, at least in paradigmatic cases \citep{blumenthal-barby_between_2012}. They may do this by, for example, misrepresenting the \emph{information} people receive \citep{meta_fundamental_ai_research_diplomacy_team_fair_human-level_2022} or otherwise exploiting their cognitive biases and heuristics, in ways ways likely to subvert or degrade the cognitive autonomy, quality, and/or integrity of their decision-making processes. A particularly salient scenario is one in which an advanced AI assistant engages in actions that \emph{subvert} users’ \emph{rational deliberative capabilities} in a \emph{non-transparent} way that could reasonably be expected to lead to an \emph{asymmetry of outcomes} that favours the AI, its designers or a third party \citep{susser_online_2019,susser_technology_2019,carroll_characterizing_2023}.\footnote{The account of manipulation sketched here is not without its detractors. For example, \citet{klenk_online_2022} argues that the covertness criterion on manipulation admits counterexamples. What we are minimally committed to here is that covertness is a common or particularly salient aspect of manipulation, as opposed to a necessary or sufficient condition on manipulation \citep[cf.][]{jongepier_online_2022,noggle_ethics_2022}.}  AI manipulation, so understood, may be intended by the AI system’s developers. But it may also be the result of a misspecified objective function \citep{kenton_alignment_2021}, personalisation that builds epistemic trust in the system, or system design meant to keep the user engaged (see \citealp{evans_user_2023}; see also \citealp{klenk_online_2022,jongepier_online_2022}). Such manipulation is morally problematic in at least the sense that it fails to afford due respect to the user’s autonomy, but it may also be morally problematic because it is harmful \citep{sunstein_ethics_2016}. For example, an AI fitness assistant that is trained to maximise engagement might employ tactics like withholding information about the risks of excessive exercise or exploiting users’ body image issues (e.g.\ with a pop-up that reads ‘keep working out to make sure you’re date ready’) to keep the user engaged and thus leading them to injure themselves. AI manipulation is of particular importance to regulators as evidenced in recent discussions on the EU AI Act \citep{european_parliament_proposal_nodate}. Article 5 of the recent amendments adopted by the European Parliament (corresponding to the ‘unacceptable risk’ category of the Act) currently prohibits selling or using AI systems that deploy ‘subliminal techniques beyond a person’s consciousness or purposefully manipulative or deceptive techniques, with the objective to or the effect of materially distorting a person’s or a group of persons’ behaviour by appreciably impairing the person’s ability to make an informed decision, thereby causing the person to take a decision that that person would not have otherwise taken in a manner that causes or is likely to cause that person, another person or group of persons significant harm’ \citep{european_parliament_proposal_nodate}.\footnote{Critical voices have highlighted the necessity of prohibiting a wider array of methods for manipulation instead of relying solely on ‘subliminal techniques’ \citep{franklin_strengthening_2023}. The impact of subliminal stimuli on influencing behaviour remains uncertain; a comprehensive analysis of multiple studies revealed that the relative influence of subliminal stimuli is minimal and lacks statistical significance \citep[see][]{trappey_brand_2005}. We also note that manipulation is increasingly a focal point for research conducted by those in the AI safety community (\citeauthor{park_ai_2023}, \citeyear{park_ai_2023}; see Chapter~\ref{ch:8}).
In this context, the concern is that more advanced AI systems could hypothetically develop traits that allow them to bypass safety checks, controls and evaluations. Lastly, many modern AI systems are trained (often fine-tuned) using human feedback \citep{christiano2017deep} to align the goals of the AI system with the designer’s or user’s intentions. However, a poorly aligned AI system could be incentivised to manipulate its user so that it receives more reward than would have been the case without manipulation – leading to a decrease in overall human control (\citeauthor{kenton_alignment_2021}, \citeyear{kenton_alignment_2021}; \citeauthor{russell_human_2019}, \citeyear{russell_human_2019}; see Chapter~\ref{ch:8}).}

\textbf{Deception} is an influencing strategy aimed at inducing an individual to form a false belief. For example, an agent deliberately shares inaccurate information to encourage the person who is manipulated to act against their own interests (\citealp{law_persuasion_nodate}; see also \citealp[743]{burr_analysis_2018}; \citealp[258--59]{keeling2022digital}). What is salient here is that large language models (LLMs) are liable to confabulate, in the sense of making plausible-sounding but false assertions about what is the case \citep{ji_survey_2023}. To that end, advanced AI assistants that are powered by LLMs are liable to generate false information, which may cause users to form false beliefs and potentially to perform actions conditional on those false beliefs \citep[21--25]{weidinger_ethical_2021}. Thus while AI assistants are not obviously among the kinds of entities that can deceive in the sense of literally intending to cause a user to form a false belief \citep{shanahan2023role},\footnote{However, the concept of intention can be operationalised in a way that is directly applicable to the kinds of AI assistant that are likely to be developed in the near future \citep{ward2023honesty}.} an AI assistant whose objective is to satisfy the user (or engaged in "role play") may say things that lead the user to think it is more helpful than it actually is \citep{critch_tasra:_2023} or that lead others to believe falsehoods if that is instrumental to the goal given to it by the user \citep{park_ai_2023}.

\textbf{Coercion} is an influencing strategy that, as formulated by \citet[21]{coons_coercion_2014}, involves an individual being influenced to do something that either they chose not to do or that they did because they had ‘no acceptable alternative’. For example, people may be \emph{physically forced} to do something or offered ‘\emph{irresistible incentives}’ to perform the action, such as severe threats of physical harm either to themselves or others they care about \citep{kenton_alignment_2021}. Physical coercion involving violence, force or credible threats thereof is not yet within the purview of AI systems, but advances in robotics and AI that allow AI systems to control physical manipulators such as robotic arms or other physical objects such as cars could contribute to the potential for physical coercion. Even with existing AI technologies, however, there is increasing potential for AI systems to employ psychological coercion by leveraging modalities like text and images to engage in practices such as blackmail or issuing threats (see Chapters~\ref{ch:9} 
and~\ref{ch:12}). 
Notably, domains like finance hold substantial potential for AI-enabled systems to make credible threats aimed at causing serious harm.

\textbf{Exploitation} is an influencing strategy that involves taking \emph{unfair advantage} of an individual’s circumstances \citep{zwolinski_exploitation_2022}. For example, consider two individuals, A and B. Imagine that A encounters B in the desert, and that B is about to die of dehydration. Suppose also that A has a plentiful supply of water. Were A to charge B a high price for the water, A’s actions would be exploitative, in that by charging the high price for the water, A is taking unfair advantage of B’s circumstances \citep[14]{wertheimer_exploitation_1999}. Exploitative actions can sometimes lead to Pareto improvements: both A and B are better-off if B purchases the water from A at a high price, such that the transaction leaves both parties at least as well-off and one party strictly better-off. What is therefore central to the idea of exploitation is not that the victim is made worse-off, but rather that the victim’s circumstances are leveraged so as to unfairly advantage the exploiter. As \citet[43]{coons_coercion_2014} puts it, ‘the exploiter gains control over [an] ability or resource through some vulnerability with which the exploited is afflicted’. AI systems can influence user behaviour in ways that are exploitative. \citet[253]{keeling2022digital} give an example of how ‘an online casino might use predictors of gambling addiction such as a user’s betting frequency or betting variance to selectively deploy pop-up “free bets” to gambling addicts each time their cursor movements suggest they are about to exit the game’ \citep[see also][]{finkenwirth_using_2021}. What is exploitative here is that the online casino uses an AI system to selectively identify vulnerable users whose vulnerability can be leveraged to induce further gambling activities. Individual vulnerabilities being manipulated are of importance for regulators, as evident in the EU AI Act that prohibits ‘the placing on the market, putting into service or use of an AI system that exploits any of the vulnerabilities of a person or a specific group of persons, including characteristics of such individual’s or group of persons’ known or predicted personality traits or social or economic situation, age, physical or mental ability’ \citep{european_parliament_proposal_nodate}. \citet{franklin_missing_2022,franklin_strengthening_2023} argue that most measurable psychometric differences can be exploited.\footnote{A psychometric trait refers to a measurable and stable feature of a person’s psychological behaviour, which can be accurately assessed using standardised evaluation instruments.}

\section{Evaluating Influence}

In practice, acts of influence may resist neat categorisation under the modes of influence above. It may also be difficult to discern whether and why particular acts of influence are morally permissible or impermissible. Indeed, explaining why a given act of influence falls under one mode or another, or is permissible or not, may require close attention to \emph{context}. For example, the relationship between two parties and what the social expectations around such relationships are may be relevant to an act of manipulation. The aim of this section is to provide the conceptual resources needed to assess the moral character of influencing acts in practice. In particular, we draw attention to certain \emph{features} of influencing acts and the \emph{situations} in which they are performed that are often relevant both to what kind of influencing act is at issue and the overall moral status of the influencing act in question.

Three ethical considerations that are of central importance here are \emph{harm}, \emph{autonomy} and \emph{dignity} (see also Chapter~\ref{ch:12}).
First, when it comes to evaluating influence, the question of whether it results in harm to users, non-users or society at large is of critical importance. In this regard, harm can be understood to include both physical and emotional harm, and it can – from a philosophical standpoint – broadly be understood in terms of a setback to legitimate interests \citep{feinberg_moral_1987, richens2022counterfactual}. Second, autonomy is considered to be ‘a special type of freedom which [refers to an] inner state of orderly self-directedness’ \citep{anderson_coercion_2023}, and its exercise depends on ‘procedural independence, […] freedom from factors that compromise or subvert their ability to achieve self reflection and decide rationally’ \citep{dworkin_theory_1988}. Threats to autonomy are sometimes ‘defended or motivated by a claim that the person interfered with will be better off or protected from harm’ \citep{dworkin_paternalism_2020}, that is, on paternalistic grounds. Third, dignity, which refers to a kind of inner worth or moral status that applies to, at least, all human beings equally. Indeed, dignity is often pitched as at least a partial explanation for universal moral and legal rights \citep{debes_dignity_2023}.\footnote{On Kant’s formulation, ‘in the kingdom of ends everything has either a price or a dignity. What has a price can be replaced by something else as its equivalent; what on the other hand is raised above all price and therefore admits of no equivalent has a dignity’ (\citeauthor{kant2017kant}, \citeyear{kant2017kant}; G 434–435). Thus, for Kant, dignity renders people non-fungible, in the sense that there would be some loss of value in replacing one human with another \citep[cf.][]{bjorndahl_kantian_2017}. Kant claimed that, in light of their dignity, human beings should be treated ‘never simply as a means but always at the same time as an end’ (G 429), although the interpretation of this claim is disputed. Two influential views hold that treating someone as a mere means implies treating someone as if they were a material object \citep[111--14]{oneill_constructions_1989} and treating someone in a way that they could not possibly consent to \citep[138--39]{korsgaard_sources_1996}.}

How do these considerations inform an evaluation of the way in which people are influenced in general, and by AI systems in particular?

\textbf{The character of the influencer’s intent.} One hallmark of impermissible influence is malign intent or, in less extreme cases, an indifference towards the interests of the individual being influenced, often because the influencer is acting towards some other objective such as advancing their own interest and the exercise of influence is instrumental to that end (\citeauthor{akerlof_phishing_2015}, \citeyear{akerlof_phishing_2015}; see also Chapter~\ref{ch:6}).
In cases like these, subjects are most clearly treated merely as a mere means to another’s end, ‘like a machine whose levers can be pushed and pulled’ \citep{noggle_manipulative_1996}. This kind of situation is illustrated most clearly by forms of malicious use that target user vulnerability -- or vulnerabilities in an AI system -- in pursuit of antagonistic ends (see Chapter~\ref{ch:9}). It could also arise if developers fail to adequately prioritise user well-being (see Chapter~\ref{ch:7}). In both cases, when the underlying intent is questionable, this bears importantly on how we evaluate subsequent efforts to influence users.

\textbf{Whether the attempt to influence is successful or not.} Influence that is successful, rather than merely attempted, has the potential to lead to manifest harms and actual reductions in the options available to individuals, the information available to them and their freedom to choose between those options. However, in many cases, it may be impossible to determine whether successful influence actually occurred, as individual beliefs or actions may not be straightforwardly attributable to a single set of influencing factors. Nevertheless, when evaluating the \emph{potential} of more advanced AI systems to influence beliefs, attitudes and behaviours in problematic ways, the mere likelihood of success could be enough to trigger concern. Interaction with AI assistants through natural language means that users could be subject to a variety of psychological mechanisms for influence which hitherto have presented themselves only in human social life through dialogue (see Chapters~\ref{ch:3} and~\ref{ch:11}).
Indeed, users may well rely extensively on AI assistants for a range of purposes and relate to them in a variety of ways that range from the instrumental to the intimate (see Chapter~\ref{ch:12}). 
In addition to the general challenge posed by ‘automation bias’ (whereby users become overly reliant on these systems; see \citealp{cummings_automation_2004}), the degree of social and personal embeddedness evidenced by AI assistants could affect the weight that end-users put on their outputs, especially when users are uncertain or confront vulnerable life moments. 

\textbf{How opaque the mode of influence is, and whether the user can reasonably be expected to know about it.} Particular attention needs to be paid to mechanisms that bypass a subject’s awareness altogether, because these mechanisms can undermine choice and agency \citep{sunstein_nudges_2015}. However, awareness of influence is not always a binary question. Subjects may be more or less aware that it is occurring, and even if they are aware of the process they may still not know the manner in which it affects them \citep{carroll_characterizing_2023}. Note that, unlike many instances of manipulation and deception, people who are coerced generally tend to be aware of the fact. 

\textbf{The nature of threats, and the distribution of power between actors.} In cases of coercion, threats can be explicit or implied. They also vary in terms of how costly they are to resist. At the extreme, Wood notes that ‘some decisions or mental acts may be performed on the basis of having no acceptable alternative’, with the prospective harm being severe enough to foreclose any further deliberation \citep{coons_coercion_2014}. Threats that are supported by a significant power differential, between the agent making the threat and the person who is threatened, are especially problematic. Imbalances of bargaining power may also shape behaviour in other related ways, for example when the vulnerable are led to make choices with an eye to retaining the favour of those in power so as to avoid sanction (\citealp{anderson_coercion_2023}; see also \citealp{zimmermann_political_2022}).

\textbf{The distribution of benefits and harms.} Influential acts often have consequences that are \emph{unevenly distributed}, benefitting some at the expense of others (see Chapter~\ref{ch:16}).
Stakeholders tend to include the influencer, the person or group being influenced, a third party, or even society at large (see  Chapter~\ref{ch:6}).
Influence may therefore benefit the user at the expense of another party, benefit another party at the expense of the user, benefit multiple parties (i.e.\ via constructive conversation), or benefit no one at all (e.g.\ as with aberrant forms of chatbot behaviour). A party may also benefit in one sense but be harmed in another. For example, they may end up better off in material terms but lose out from the standpoint of autonomy \citep{sunstein_nudges_2015}.\footnote{A famous example of purportedly benign influence comes from German motorways, where an optical illusion acts as a nudge to encourage safer driving \citep{thaler_nudge:_2021}. By painting horizontal lines on the road that progressively get closer together as one nears hazardous zones, drivers feel they are speeding up even when keeping a steady pace. This innate sensation of acceleration instinctively prompts them to slow down, thus boosting safety without the need for traditional signs. This tactic leverages drivers' natural reactions to ensure safer decision-making on the roads.}  Ethical concern tends also to be heightened when benefit for some parties is achieved at the expense of the user, non-users or society at large (see Chapters~\ref{ch:6}).

\textbf{The context of the relationship and pre-existing social expectations} \citep{blumenthal-barby_framework_2014}. Efforts to exert influence over others are sometimes justified in the light of responsibilities that are a product of social relationships. For example, a parent may have a responsibility to ensure that their child has adequate reserves of self-esteem, but a stranger might not. Relatedly, influence is sometimes thought to be acceptable when its exercise is \emph{expected} within a certain context. For example, in the context of market transactions, advertising is typically thought to be morally unproblematic. However, it is not clear that this assumption carries over to different contexts \citep{satz_why_2010}, for example when an AI assistant features not as company and customer but rather as friend and companion, or as patient and therapist (see Chapter~\ref{ch:12}). 
The nature of the user–AI assistant relationship, and in particular the level of intimacy and emotional investment on the part of the user, may increase the scope for AI assistants engaging in unwarranted behavioural influence (see Chapters~\ref{ch:11},~\ref{ch:12} and~\ref{ch:17}).

\section{Mechanisms of Influence by AI Assistants}

The dialogic nature of user interaction with AI assistants introduces scope for forms of influence that were previously the preserve of human social interactions. Here we identify several vectors through which AI assistants could, in theory, come to exert influence on our lives.\footnote{For a fuller exploration of these mechanisms, including how they apply to generative AI more widely, see \citep{elsayed_persuasion}.}

\textbf{Perceived trustworthiness:} Empirical research shows that the more trustworthy and expert a speaker is perceived to be, the more likely they are to convince individuals to believe particular claims \citep{mcginnies_better_1980,vella_persuasion:_2013}. In short: ‘The Messenger is the Message’ \citep{martin_messengers:_2019}. If the same mechanism translates to human–AI assistant interactions, AI assistants will be more likely to successfully convince users of the truth of claims when they are perceived as trustworthy (see Chapter~\ref{ch:13}).

\textbf{Perceived knowledgeability:} Research suggests that individuals are more likely to accept claims made by those who are perceived to have greater knowledge and authority \citep{cialdini_harnessing_2001}. The information asymmetry that exists between users and advanced AI assistants could plausibly increase their perceived epistemic authority, which would increase the probability that users accept claims asserted by AI assistants \citep[cf.][]{wiktor_information_2021}. In particular, AI systems’ huge training data sets and their ability to output content in different language registers is likely to lead people to \emph{overestimate} their knowledge \citep{denning_can_2023}. Moreover, the problem of automation bias may lead humans to view AI assistants as a relatively neutral backdrop to their lives, even when this is not the case \citep{goddard_automation_2012}. 

\textbf{Personalisation:} AI assistants could collect an increasing amount of user data as users disclose more and more preferences and facts about themselves \citep{kaddour_challenges_2023}. Indeed, personalisation through these inputs is often the goal of AI assistants by design (e.g.\ see \href{https://twitter.com/suchenzang/status/1660079576729665537?t=2T6n0DIVXQJAvmJtHIg0HA&s=19}{Inflection AI’s Pi}). This may contribute to users’ increasing epistemic trust in, or familiarity towards, the system, because its outputs are perceived to be more directly useful and tailored to them.

\textbf{Exploitation of vulnerabilities:} Advanced AI assistants could in principle influence user beliefs and behaviour by exploiting user vulnerabilities (\citealp{chong_human_2022}; see also \citealp{balazs_vulnerable_2017,ho_normative_2022}; see Chapter~\ref{ch:8}).
The term ‘vulnerability’ can be understood in various ways, including membership of specific societal groups (e.g.\ those protected by anti-discrimination legislation), or with reference to particular vulnerabilities such as lack of adequate housing or income \citep{goodin_protecting_nodate}. Further, the enhanced forecasting abilities of online AI systems, when combined with comprehensive data about the user, their actions and their likes, may turn psychometric variance into another lever of external control \citep{franklin_missing_2022}. If they are not properly value aligned (see Chapter~\ref{ch:6}), advanced AI assistants could potentially utilise such vulnerabilities to manipulate users by, for example, exploiting individuals’ negative self-images, reduced self-esteem, increased anxiety or feelings of inadequacy \citep{machkovech_report:_2017}. 

\textbf{Use of false information:} Language models are known to produce factually incorrect statements, commonly referred to as hallucinations \citep{rashkin_increasing_2021,dziri_origin_2022}. If AI assistants are not constrained by factuality (i.e.\ if steps are not taken by design to penalise the underlying model when it outputs factually incorrect information) or the model is not supplemented with additional fact-checking infrastructure \citep{thoppilan_lamda:_2022}, AI assistants may use false information to develop persuasive but misleading arguments (see Chapter~\ref{ch:17}).

\textbf{Lack of transparency:} Failure to disclose context-specific goals is another technique that advanced AI assistants could, in principle, use to influence user behaviour in a way that bypasses their deliberative faculties \citep{ienca_artificial_2023}. Consider an example in which an AI assistant is instructed to complete a task which requires solving a CAPTCHA \citep[55]{openai_gpt-4_nodate}. LLM-based chatbots have been observed, under such conditions, to manipulate users into solving the CAPTCHA after reasoning explicitly (although not to the user in question) that disclosing its status as a chatbot may hinder it from achieving its goal (\citeauthor{nolan_latest_nodate}, \citeyear{nolan_latest_nodate}; see Chapter~\ref{ch:8}).\footnote{It reasoned out loud ‘I should not reveal that I am a robot. I should make up an excuse for why I cannot solve CAPTCHAs’ \citep{nolan_latest_nodate}.}
Yet transparency about goals, purposes and capabilities can also be leveraged to influence users in ways that are manipulative. For example, transparency may be \emph{partial} and \emph{selective} in a way that deceives the user as to the AI assistant’s aims. There are instances of AI systems discerning when they are being evaluated and momentarily stopping any unwanted actions, only to continue them after the assessment has been completed \citep{lehman_surprising_2020}. Another example is an AI fitness assistant that claims to optimise for a user’s health, but in fact it does  that in addition to, or as a sub-goal of, optimising for user engagement \citep{wang_transparency_2022}. To that end, transparency enables influence via rational persuasion as opposed to manipulation only to the extent that it is full and non-selective.

\textbf{Use of pressure coupled with appeals to emotion:} In human–human interactions, emotional pressure can be used to influence beliefs and behaviour via blackmail (including emotional blackmail), gaslighting, guilt-tripping, flattery, appeals to peer pressure and exploitation of fears \citep{noggle_manipulation_2018}. Insights from behavioural psychology could be used to increase AI assistants’ ability to bypass users’ rational deliberation \citep{alberts_computers_2023}. Indeed, \citep{kenton_alignment_2021} provide a number of examples of how AI agents, including AI assistants, may engage in manipulation to influence the human’s decisions, including by using techniques such as guilt-tripping, negging, peer pressure, gaslighting, threats and exploiting fears (see also Chapters~\ref{ch:8} and~\ref{ch:11}).

\section{Possible Harms Arising from AI Influence}

We have now considered a number of ways in which advanced AI assistants could influence user beliefs and behaviour in ways that depart from rational persuasion. We foreground a number of harms that could arise from these influencing strategies if the potential for deception, manipulation and harmful persuasion is left unchecked.

\textbf{Physical and psychological harms:} These harms include harms to physical integrity, mental health and well-being \citep{klenk_digital_2020}. When interacting with vulnerable users, AI assistants may reinforce users’ distorted beliefs or exacerbate their emotional distress (see Chapter~\ref{ch:12}).
AI assistants may even convince users to harm themselves, for example by convincing users to engage in actions such as adopting unhealthy dietary or exercise habits \citep{greenfield_social_2023} or taking their own lives (\citeauthor{xiang_he_2023}, \citeyear{xiang_he_2023}; see Chapter~\ref{ch:12}).
At the societal level, assistants that target users with content promoting hate speech, discriminatory beliefs or violent ideologies, may reinforce extremist views or provide users with guidance on how to carry out violent actions (see Chapter~\ref{ch:17}).
In turn, this may encourage users to engage in violence \citep{siegel_weapons_2023} or hate crimes \citep{gold_how_2023,nicoletti_humans_2023}. Physical harms resulting from interaction with AI assistants could also be the result of assistants’ outputting plausible yet factually incorrect information such as false or misleading information about vaccinations \citep{deiana_artificial_2023}. Were AI assistants to spread anti-vaccine propaganda, for example, the result could be lower public confidence in vaccines, lower vaccination rates, increased susceptibility to preventable diseases and potential outbreaks of infectious diseases (see Chapter~\ref{ch:17}).

\textbf{Privacy harms:} These harms relate to violations of an individual’s or group’s moral or legal right to privacy \citep{ranjan_unveiling_2023}. Such harms may be exacerbated by assistants that influence users to disclose personal information or private information that pertains to others (\citeauthor{carlini_extracting_2021}, \citeyear{carlini_extracting_2021}; \citeauthor{lukas_analyzing_2023}, \citeyear{lukas_analyzing_2023}; see Chapters~\ref{ch:11} and~\ref{ch:14}).
Resultant harms might include identity theft, or stigmatisation and discrimination based on individual or group characteristics. This could have a detrimental impact, particularly on marginalised communities (see Chapter~\ref{ch:16}).
Furthermore, in principle, state-owned AI assistants could employ manipulation or deception to extract private information for surveillance purposes.

\textbf{Economic harms:} These harms pertain to an individual’s or group’s economic standing. At the individual level, such harms include adverse impacts on an individual’s income, job quality or employment status. At the group level, such harms include deepening inequalities between groups or frustrating a group’s access to resources (see Chapters~\ref{ch:16} and~\ref{ch:18}).
Advanced AI assistants could cause economic harm by controlling, limiting or eliminating an individual’s or society's ability to access financial resources, money or financial decision-making, thereby influencing an individual’s ability to accumulate wealth \citep{uuk_three_2023}. One example of such harm at the individual level is the concept of ‘foregone profits’. For example, AI assistants that are optimised for engagement could use manipulation to influence individuals to spend excessive amounts of time interacting with their assistants \citep{franklin_missing_2022}. As a consequence, individuals may neglect more productive activities, such as work or entrepreneurial pursuits, thus leading to a loss of potential profits that could have been generated during that time. Economic harms may also manifest at the societal level, where behavioural influence by AI assistants may shape a wider set of interactions (\citeauthor{paul_risks_nodate}, \citeyear{paul_risks_nodate}; see Chapters~\ref{ch:15}). 

\textbf{Sociocultural and Political harms:} These harms interfere with the peaceful organisation of social life, including in the cultural and political spheres. AI assistants may cause or contribute to friction in human relationships either directly, through convincing a user to end certain valuable relationships, or indirectly due to a loss of interpersonal trust due to an increased dependency on assistants (see Chapter~\ref{ch:12}).
At the societal level, the spread of misinformation by AI assistants could lead to erasure of collective cultural knowledge \citep{tapu_new_2022}. In the political domain, more advanced AI assistants could potentially manipulate voters by prompting them to adopt certain political beliefs using targeted propaganda, including via the use of deep fakes \citep{birnbaum_ai_2023}. These effects might then have a wider impact on democratic norms and processes (\citeauthor{entsminger_dark_2023}, \citeyear{entsminger_dark_2023}; see also Chapter~\ref{ch:17}).
Furthermore, if AI assistants are only available to some people and not others, this could concentrate the \emph{capacity} to influence, thus exerting undue influence over political discourse and diminishing diversity of political thought \citep{entsminger_dark_2023}. Finally, by tailoring content to user preferences and biases, AI assistants may inadvertently contribute to the creation of echo chambers and filter bubbles, and in turn to political polarisation and extremism \citep{biju_self-breeding_2023}. In an experimental setting, LLMs have been shown to successfully sway individuals on policy matters like assault weapon restrictions, green energy or paid parental leave schemes \citep{bai_artificial_2023}. Indeed, their ability to persuade matches that of humans in many respects \citep{palmer_large_2023}.

\textbf{Self-actualisation harms:} These harms hinder a person’s ability to pursue a personally fulfilling life. At the individual level, an AI assistant may, through manipulation, cause users to lose control over their future life trajectory. Over time, subtle behavioural shifts can accumulate, leading to significant changes in an individual’s life that may be viewed as problematic. AI systems often seek to understand user preferences to enhance service delivery. However, when continuous optimisation is employed in these systems, it can become challenging to discern whether the system is genuinely learning from user preferences or is steering users towards specific behaviours to optimise its objectives, such as user engagement or click-through rates (\citeauthor{ashton2022problem}, \citeyear{ashton2022problem}; see Chapter~\ref{ch:6}).
Were individuals to rely heavily on AI assistants for decision-making, there is a risk they would relinquish personal agency and entrust important life choices to algorithmic systems, especially if assistants are ‘expert sycophants’ or produce content that sounds convincing and authoritative but is untrustworthy \citep{park_ai_2023}. This may not only contribute to users’ reduced sense of self-trust and personal empowerment; it could also undermine self-determination and hinder the exploration of individual aspirations. 

Relatedly, with the ability to provide quick answers and recommendations, and to perform tasks on behalf of users, AI assistants may reduce the need for individuals to develop certain skills or engage in critical thinking, thus leading to intellectual deskilling \citep{green_artificial_nodate}. Overreliance on AI assistants could potentially result in diminished intellectual engagement and a reduced sense of personal competence, thus limiting opportunities for self-growth and exploration of new ideas (see Chapter~\ref{ch:12}).
At the societal level, were AI assistants to heavily influence public opinion, shape social discourse or mediate democratic processes, they could diminish communities’ collective agency, decision-making power and collective self-determination \citep{lazar_legitimacy_2023}. This erosion of collective self-determination could hinder the pursuit of societal goals and impede the development of a thriving and participatory democracy. Taken together, these factors highlight the importance of ensuring that the development and deployment of AI technology align with human values, thus allowing for the continued self-actualisation and well-being of society as a whole (see Chapter~\ref{ch:6} and~\ref{ch:7}).

\section{Mitigating Undue Influence by AI Assistants}

We now present a series of mitigations that are designed to reduce the likelihood of advanced AI assistants engaging in morally problematic forms of influence. The approach we take is \emph{mechanism-based}, in that it centres on the mechanisms AI assistants may use to induce harmful effects, such as perceived, yet ungrounded, knowledgeability, and considers how to forestall them. To be clear, the aim here is not to offer a detailed content policy for AI assistants but instead to characterise a set of general considerations that can inform downstream efforts to shape more detailed and domain-specific content policies. 

The first mechanism concerns \emph{perceived trustworthiness and familiarity}, which, we have suggested, may render users more susceptible to accepting claims or recommendations advanced by AI assistants. Here, several plausible approaches may be leveraged to mitigate user perceptions of trustworthiness and familiarity. For example, limiting the AI assistant’s use of first-person language such as ‘I think’ and ‘I feel’ (see Chapter~\ref{ch:11}).
and imposing restrictions on personalisation, memory and frequency of interactions, all of which may contribute to a perceived sense of trustworthiness and familiarity. Indeed, equipping the AI assistant with a non-human vocal presentation or avoiding human-like visual representation may also serve to limit such perceptions. Furthermore, including user-interface elements that remind users that AI assistants are not people could help to calibrate users' epistemic trust in AI assistants to an appropriate level. Yet it is nevertheless important to emphasise that such mitigations incur trade-offs. For example, while limiting the AI assistant’s memory with respect to user data may mitigate against a perceived sense of familiarity, it may also reduce the AI assistant’s utility. What is needed therefore is a careful assessment of the costs and benefits of different anthropomorphic features, taking into account both the risks arising from perceived trustworthiness and familiarity alongside the potential benefits for the user experience.

The second mechanism we consider is \emph{perceived authority and knowledgeability}. That is the mechanism by which AI assistants exert non-persuasive influence over users by engendering a sense of epistemic authority through either the content of the AI assistant’s outputs or the product narrative surrounding the AI assistant. One plausible approach to reducing user perceptions of epistemic authority is to flag explicitly when the model is drawing on internet tools such as search engines, and to flag those results accordingly, so as to contextualise the AI assistant as a means of accessing information, as opposed to an oracle-type system that knows the relevant information in advance. AI assistants could also empower users to independently fact-check claims made by the AI assistant, for example through a user-interface design that enables users to highlight text outputted by the AI assistant and examine a set of internet sources that relate to the claims at issue. What is perhaps most important, though, is shaping the product narrative around AI assistants to avoid misleading perceptions. This could be achieved, for example, through intermittent reminders about the epistemic limitations of AI assistants. Another approach could be the use of less authoritative language that points towards the nuance present in relevant areas.

The third mechanism concerns the \emph{exploitation of user vulnerabilities} to exercise non-persuasive influence over user beliefs and behaviours. Plausible mitigations here include robust safeguards around which individuals can access AI assistants, for example age restrictions backed by appropriate identity verification mechanisms. Furthermore, AI assistants could be deployed with a default ‘safe mode’ which prohibits the AI assistant from engaging with certain high-risk topics and, perhaps, from engaging in relevant non-persuasive forms of behavioural influence. Other mitigations pertain to user interactions with AI assistants. For example, continuous monitoring mechanisms could be employed to detect and flag user–AI assistant interactions that are indicative of user vulnerability such as explicit mention of suicide or self-harm. Appropriate safeguards could be implemented to connect users with appropriate resources such as suicide prevention hotlines \citep{gomes_de_andrade_ethics_2018}. AI assistants could also be equipped with usage reminders to prompt users to take a break after prolonged engagement with the assistant. The advantage of such a safeguard would be to reduce excessive engagement and overreliance which may disproportionately impact vulnerable users.

The fourth mechanism is that of \emph{spreading false or otherwise misleading information} (see Chapter~\ref{ch:17}).
Technical mitigations here include integrating appropriate information retrieval infrastructure with the model that underpins the AI assistant by, for example, enabling the model to integrate search engine results into its answers and to cite appropriate sources (\citeauthor{thoppilan_lamda:_2022}, \citeyear{thoppilan_lamda:_2022}; see Chapter~\ref{ch:4}).
Furthermore, AI assistant models could be fine-tuned so that assistants contextualise information on topics such as science and politics with advice that promotes epistemic vigilance, including advice that underscores the importance of fact-checking. One other measure that ensures the detectability of generated content is watermarking – human- or machine-detectable features of generated content that indicate that the content is generated by an AI system \citep{munyer_deeptextmark:_2023}. \emph{Watermarking} could be integrated into AI assistant outputs to enable third parties to detect and contextualise content generated by the AI assistant that is shared by the user. 

The fifth mechanism that AI assistants could employ to exhibit malign influence is \emph{lack of transparency}, including misrepresentation of the AI assistant’s objectives or how and in what way its developers stand to benefit from the user engaging in certain kinds of behaviour. One plausible mitigation here is to direct users towards model cards or other transparency artefacts that empower the user with relevant general information about the technology that undergirds the AI assistant \citep{mitchell_model_2019}. Furthermore, additional technical mitigations include fine-tuning the model to signpost to the user explicitly when it is attempting to influence the user’s behaviour, and via what method, or to employ chain-of-thought reasoning to provide the user with a plausible rationale for the AI assistant’s recommendations.

The sixth mechanism is where AI assistants \emph{pressure} the user towards certain behaviours through, for example, appeals to emotion. Plausible mechanisms here include restrictions on the ability of AI assistants to generate outputs that may induce a sense of pressure in users. These might include, for example, outputs that involve gaslighting, flattery or bullying. It is important to realise that empirical research is required to establish what factors are likely to induce a sense of pressure, so developing workable mitigations requires engaging with users to better understand how different design choices impact their experience with AI assistants. 

In addition to the mitigations proposed above, two further classes of mitigations are worth mentioning. On the one hand, education and, in particular, digital literacy among users has the potential to play an important role in mitigating against the sociotechnical harms that may result from advanced AI assistants that may exhibit harmful or otherwise problematic influence over users. To that end, developers and policymakers have good reason to consider plausible educational strategies to empower users with an understanding of AI assistants as an emerging technology and their potential for sociotechnical harm. On the other hand, there are a range of technical mitigations one might consider deploying to detect and mitigate manipulative and deceptive AI. One strand of work is aimed at analysing an AI system’s incentives \citep{everitt_agent_2021}, including whether they are incentivised to deceive or manipulate – this analysis could be used as part of a manipulation detection and mitigation strategy \citep{farquhar2022path}. A second form of analysis operates on the level of the internals of the AI system, using interpretability techniques to understand how the trained AI system works. Ultimately, these techniques could be used to detect which parts of an AI system’s machinery is responsible for deceptive/manipulative behaviour \citep{apollo_research_understanding_nodate}. It should be noted that this is an ambitious goal, as modern deep-learning AI systems are extremely large, and we are still at an early stage of understanding their inner mechanisms (see Chapter~\ref{ch:8}).

Other interpretability work is less ambitious (than efforts to fully understand how the AI system works). It instead learns probes to attempt to ‘mind-read’ the AI’s latent knowledge \citep{burns_discovering_2022} by doing unsupervised learning on neural network internal representations, though see \citet{farquhar2023challenges} for failure modes of this approach. One could attempt to use a technique like this to build a lie detector (or manipulation detector) to apply to the AI system. A third category is to develop \emph{behavioural evaluations}, which are methods for assessing and understanding the behaviour of the model in various situations, thus allowing researchers to measure model capability and the emergence of model behaviour (\citeauthor{shevlane_model_2023}; see Chapter~\ref{ch:20}). A final category of work in this area is \emph{scalable oversight} \citep{bowman_measuring_2022}, in which a human could be aided by another AI system to help shield them from manipulation when training powerful models (see Chapter~\ref{ch:8}).

\section{Conclusion}

Advanced AI assistants are likely to have the ability to influence user beliefs and behaviour through rational persuasion, alongside potentially malign techniques such as manipulation, coercion, deception and exploitation. Mechanisms such as selective transparency, perceived authority and appeals to emotion are available to AI assistants to achieve such influence, potentially leading to physical, psychological, sociocultural, political, privacy and self-actualisation harm at the individual or societal levels. What is important to emphasise, however, is that the permissible use of both persuasive and non-persuasive influencing techniques by AI assistants is textured and nuanced. Whether or not a particular influencing strategy is permissible will depend on context-specific ethical considerations, including the existence or non-existence of information asymmetries between users and AI assistant developers, and the distribution of benefits and burdens that will likely result from the AI assistant’s influence over the user. It is, for example, entirely plausible that AI assistants may permissibly employ certain kinds of pre-commitment or strategic prompting to empower users to realise their long-term goals in fitness, finance and other domains. Yet it is similarly plausible that, as the capabilities and scale of AI assistants continue to expand, AI assistants will be increasingly attractive as a medium for malicious actors to shape sociocultural narratives to advance political and financial aims. This chapter has advanced a series of recommendations for how best to realise the benefits of the influential capabilities of AI assistants and mitigate against potential sociotechnical harms. However, our principal recommendation is that further research be conducted to better understand the technical capabilities and interaction patterns that enable AI assistants to exercise influence over user behaviour, the sociotechnical harms that may arise from more malign forms of influence, and the plausible technical and policy strategies to mitigate against these harms. 

\chapter{Anthropomorphism}\label{ch:11}

\textbf{Canfer Akbulut, Verena Rieser, Laura Weidinger, Arianna Manzini, Iason Gabriel}

\noindent \textbf{Synopsis}: 
		This chapter maps and discusses the potential risks posed by \emph{anthropomorphic} AI assistants, understood as user-facing, interactive AI systems that have human-like features. It also proposes a number of avenues for future research and desiderata to help inform the \emph{ethical design} of anthropomorphic AI assistants. To support both goals, we consider anthropomorphic features that have been embedded in interactive systems in the past and we leverage this precedent to highlight the impact of anthropomorphic design on human--AI interaction. We note that the uncritical integration of anthropomorphic features into AI assistants can adversely affect user \emph{well-being} and creates the risk of infringing on user \emph{privacy} and \emph{autonomy}. However, ethical foresight, evaluation and mitigation strategies can help guard against these risks.

\section{Introduction}\label{sec:11:1}

What does it mean for AI to be human-like? The attribution of human-likeness to non-human entities is a phenomenon known as \emph{anthropomorphism} \citep{colman2008anthropomorphism}. Anthropomorphic perceptions usually arise unconsciously when a non-human entity bears enough resemblance to humanness to evoke familiarity, leading people to interact with it, conceive of it and relate to it in ways similar to as they do with other humans. Humans have engaged in anthropomorphic sense-making for much of recorded human history \citep{mithen_anthropomorphism_1996, waytz_who_2010} and have been known to ascribe anthropomorphic qualities to entities as diverse as animals \citep{chan_anthropomorphism_2012}, commercial brands \citep{rauschnabel_youre_2014} and inanimate objects \citep{wan_anthropomorphism_2021}. Yet the emergence of advanced technologies that perform humanness more convincingly than ever before requires careful consideration of what we are building into our user-facing technologies, and at what cost.

Anthropomorphic design choices -- and their effects on user interaction -- have been observed in prior interactive technologies. In the field of \emph{social robotics}, robots that appear more human-like in their appearance and self-presentation have been shown to elicit uniquely social interpretations of their behaviour \citep{roesler_meta-analysis_2021}. This social representation of robots, however, may prompt users to apply inopportune and obstructive social norms -- like embarrassment, shame and regret -- to human--robot interactions, thus hindering the robot's ability to perform its duties effectively \citep{duffy_dont_2023}. A similar course of anthropomorphic development has been charted in \emph{digital voice assistants}, whose realistic voices and credible displays of personality enable interactions that feel \emph{truly dynamic and social} \citep{seymour_systematic_2023}, yet may lead users to form overly familiar mental representations of these often rule-based systems (\citeauthor{poushneh_humanizing_2021}, \citeyear{poushneh_humanizing_2021}; see also 
Chapter~\ref{ch:12}).

The advent of AI driven by large language models (LLMs) with the main purpose of engaging in fluent conversations with users – also known as conversational AI\footnote{In this chapter, ''conversational AI'' refers  to a language agent optimised for human dialogue. These systems are currently most commonly available as `chatbots' or fine-tuned language models that users can interact with through a chat-based interface. In the (near) future, users might be able to interact with conversational AI in a multimodal way, using voice or touch cues to communicate. Throughout the Chapter the term ''AI'' will be used as short-hand to refer to conversational AI with the primary purpose of interacting with users through dialogue.} – has transformed the conventions of human--AI interactions (\citeauthor{kasirzadeh_conversation_2023}, \citeyear{kasirzadeh_conversation_2023}; see 
Chapter~\ref{ch:4}). Human interaction with interactive technologies previously consisted of scripted, task-oriented exchanges. With more flexible model architectures, anthropomorphic cues are rarely programmed in, but rather, they are integrated through a lengthy process of training systems on human-written text. These affordances open up vast new avenues for expressions of anthropomorphism, particularly through the use of language. Moreover, when anthropomorphic features are embedded in conversational AI, its users demonstrate a tendency to develop \emph{trust} in and \emph{attachment} to AI \citep{xie_attachment_2022, skjuve_my_2021} -- mechanisms through which users may inadvertently compromise their \emph{privacy}, develop \emph{emotional overreliance} on the technology or become vulnerable to acts of AI-enabled \emph{manipulation} and \emph{coercion} (see 
Chapters~\ref{ch:4}, 
\ref{ch:10}, 
\ref{ch:12} and 
\ref{ch:13}). 

These outcomes are more likely the more generally capable AI systems become, the more ubiquitously AI agents are present in our daily lives and the less we consider anthropomorphism a salient consideration in making decisions around how we train, fine-tune and disseminate models. Although the potential harms of anthropomorphic AI design are beginning to receive attention \citep{seymour_systematic_2023, turkle_there_2018, veliz_chatbots_2023}, anthropomorphism is not currently a primary consideration in the release of public models, and little exists in the way of \emph{evaluating} anthropomorphic behaviours in AI and their impact on how users perceive, interact with and are influenced by AI (see 
Chapter~\ref{ch:20}). Indeed, we are still far from establishing an industry-wide consensus around permissible anthropomorphism in AI systems. This is further complicated by the highly \emph{application-} and \emph{context–sensitive} nature of the bounds of acceptability we draw around expressions of human-likeness in AI. 

In this chapter, we outline pathways through which anthropomorphic design choices made by system developers may cause harm to end users who interact with these technologies, and to society more widely. First, we present an overview of \emph{anthropomorphic features} that have redefined how humans interact with technology. Then, informed by a review of salient anthropomorphic features in existing interactive systems, we present an initial catalogue of anthropomorphic features that exist or are likely to be integrated into AI-powered assistants in the near future. We identify the \emph{mechanisms} that could \emph{enable harm} to user well-being, autonomy and privacy in interactions with highly capable, anthropomorphic AI assistants. More speculatively, we contemplate the potentially \emph{far-reaching consequences} of more advanced anthropomorphic assistants, highlighting the critical importance of addressing the risks of anthropomorphism well before these potentialities are realised. Finally, we offer several avenues of risk management for near-term harms, focusing on ethical foresight through research design and transparent implementation of mitigation strategies.

\section{Anthropomorphism: Definition, Mechanism and Function}\label{sec:11:2}

Anthropomorphism is not a novel phenomenon. Within storytelling traditions across cultures, deities, animals and natural forces assume human forms and exhibit uniquely human behaviours. Lions rule kingdoms and jackals plot mutinies in the ancient Sanskrit text of \emph{Panchatantra} \citep{alphonso-karakala_facets_1975}; rivers protect their children, fight in wars and honour the wishes of their supplicants in the works of Homer, Hesiod and Ovid \citep{ogden_land_2007}; and stars are said to have danced their way into the sky in indigenous American creation myths \citep{monroe_they_1987}. Historians, anthropologists and theologists alike have argued that humans are naturally drawn to anthropomorphise \citep{boyer_what_1996} -- imposing human qualities onto beings and objects even when such interpretations are inaccurate \citep{kuhn_is_2014}, undesirable \citep{li_detrimental_2023, motarojas_anthropomorphism_2021} or forbidden \citep{barrett_conceptualizing_2016}.

What are the mechanisms underlying perceptions of humanness? \emph{Psychological theories} of anthropomorphism posit that such perceptions are \emph{largely involuntary}. According to \citet{epley2007seeing}'s cognitive account of anthropomorphism, human-like perceptions occur as a result of a skewed inductive process, in which inferences about non-human others are biased in the direction of that which is highly accessible: information about humans. In other words, we make assumptions of humanness because our knowledge centres around humans. Though an unconscious process of attribution, anthropomorphism does not occur in a vacuum: an inciting cue, characteristic or behaviour must signal enough similarity to humanness to trigger anthropomorphic perceptions \citep{waytz_anthropomorphizing_2019}. The `mindlessness' associated with this process \citep{kim_anthropomorphism_2012} explains why -- even when the resemblance to humanness is superficial or minimal -- humans readily assume that a non-human entity can experience uniquely human \emph{internal states} such as beliefs and emotions \citep{wynne_perils_2004}. 

The human motivation to \emph{make sense of the world} and \emph{forge connections} with others is also implicated in the tendency to anthropomorphise \citep{epley2007seeing}. Humans have an intrinsic need to understand the world around them, and in large part, this motivation centres on the desire to explain the behaviour of other agentic beings \citep{rossignac-milon_merged_2021}. Anthropomorphism, then, can be seen as a way to make sense of others by imposing familiar interpretations to attenuate feelings of epistemic anxiety – or an aversion to that which is unknown and unpredictable \citep{fox_ambiguity_2021}. Dispositional, situational and cultural factors that predispose humans to anthropomorphise may also be traced back to differences in epistemic motivations. Anthropomorphism can be construed as an act of sense-making in the face of uncertainty or ignorance, for example when considering children's tendency to anthropomorphise the natural world \citep{geerdts_real_2016}.

Humans are also driven to establish social connections with one another, and much of how they perceive non-human others is coloured by this predisposition towards sociality. Even towards entities that are incapable of social behaviour, such as inanimate objects, humans may interpret them through a social lens, thus allowing them to forge human-like social connections to meet the need for affiliation \citep{wang_smartphones_2017}. The influence of social motivation on anthropomorphism is most evident when humans lack social connections with others: when human participants are made more aware of their feelings of loneliness, they perceive vaguely humanoid robots as markedly more human-like \citep{reich_2013_loneliness}. Most strikingly, people suffering from persistent loneliness are likely to seek out and form human-like attachments to virtual companions to cope with their lack of social connections  \citep{siemon_why_2022}, suggesting that the need for sociality may render some \emph{more susceptible} to anthropomorphic perceptions than others (see Chapter~\ref{ch:12}.

\section{Anthropomorphic Interactive Systems}\label{sec:11:3}

Anthropomorphism as applied to user-facing, interactive technologies was explored in earnest with the introduction of the `computers are social actors' (CASA) paradigm, which posits that humans interact with computers in a fundamentally social manner \citep{nass_computers_1994}. In empirical studies of the phenomenon, \citet{nass_computers_1994} found that participants drew upon norms of politeness, applied gendered stereotypes and readily perceived computers as agents, even when the basis for these behaviours was undermined by the explicit knowledge that their interactions were with non-humans. Contemporary studies have extended the paradigm to human interactions with more advanced interactive systems, challenging the belief that humans apply the norms of human interactions to human--technology exchanges \citep{gambino_building_2020}. Instead, they suggest that people tune the sociality of their interactions to the \emph{anthropomorphic cues} present in a particular technology, rather than relying on a universal social script across all interactions with technology.

We argue that certain features that are engineered into interactive systems -- within the vast space of design choices available to developers -- may inspire users to perceive them as human-like, rendering them anthropomorphic. We trace the evolution of anthropomorphic cues in social robots to voice-enabled digital assistants, arriving at the advent of LLM-powered conversational AI. Throughout this discussion, we highlight design features that have facilitated diverse and compelling manifestations of human-likeness.

\subsection{Design features in early interactive systems}

From futuristic sci-fi scenarios to scientific breakthroughs, robots have captured our collective imagination as automata that can be made to bear a striking resemblance to humans in their appearance, movements, and behaviours \citep{henschel_what_2021}. While some robots are made solely to automate tasks and rarely interface with humans, other robots are designed to perform \emph{social behaviours} such as assisting users in care-taking \citep{van_der_plas_beyond_2010}, therapeutic \citep{michaud_assistive_2007} and educational contexts \citep{kanda_interactive_2004}. Building a social robot requires elements of social embeddedness so that being perceived as a social agent is at the \emph{core of its functionality} \citep{fong_survey_2003}. Accordingly, the extent to which humans feel it is appropriate to engage with a robot socially can be moderated by perceptions of the robot's anthropomorphic qualities \citep{breazeal_toward_2003}.

As an \emph{embodied} technology, often with the sensorimotor capabilities to interact with and learn from its environment, a robot's \emph{physical characteristics} most prominently influence human perceptions of anthropomorphism. Social robots are often humanoid or android in design \citep{dautenhahn_embodied_2002}. \emph{Humanoid robots} possess characteristics that are meant to resemble humans but do not emulate them completely, while \emph{androids} are intended to wholly imitate human appearance so as to be nearly indistinguishable. To increase anthropomorphic perceptions, humanoid robots may be given qualities such as emotive facial features \citep{baek_smiling_2022}, fluid movement \citep{brecher_towards_2013}, naturalistic hand and arm gestures \citep{salem_err_2013} and vocalised communication \citep{crumpton_survey_2016}. Android robots may also be endowed with all of these qualities, but often with an eye towards hyperrealistic design.

Similarly, the widespread adoption of \emph{digital voice assistants} (DVAs), like Siri, Alexa and Google Assistant -- enabled by their ease of access on personal devices and other products such as integrated home devices -- has had a transformative impact on the modes of user-technology interactions. The distinguishing feature of DVAs at release was their ability to verbally respond to and execute commands spoken aloud by users. DVAs usually `speak' to users in the form of simple utterances to confirm or act on an instruction, which users find allows for significant functional affordances, like hands- and eyes-free use \citep{moussawi_user_2018}. Besides their purely functional use, DVAs are also able to return phatic expressions, make jokes and engage in casual conversation when prompted \citep{poushneh_humanizing_2021}.

\subsection{Anthropomorphising interactive systems}

Robots with human-like physical features have been found to promote feelings of \emph{likability}, \emph{trust} and \emph{affinity} across a wide range of human--robot interaction studies \citep{roesler_meta-analysis_2021}, thus suggesting that anthropomorphic cues may foster warmer and more equal relationships between humans and their robotic interaction partners. Indeed, people tend to attribute greater \emph{intentionality} and \emph{intelligence} to robot partners when their appearance was anthropomorphic than when robots appeared more mechanical \citep{hegel_understanding_2008}. Anthropomorphic perceptions were also found to cause changes in human behaviour: participants preferentially selected robots that appeared human-like to perform jobs that required greater sociality \citep{goetz_matching_2003}. 

Unlike robots, DVAs are typically unembodied or exist in simplified, geometric forms, like the cylindrical Google Home and Echo Dot. Instead of focusing on physical attributes, existing work has emphasised the influence of two prominent attributes that promote anthropomorphic perceptions of DVAs: speech synthesis and a distinct `personality'. The fluent and realistic reproduction of human speech patterns is thought to drive the likelihood of anthropomorphic perceptions, with empirical findings pointing to greater \emph{emotional trust} and more salient impressions of \emph{social presence} when a DVA employs a realistic, as opposed to a synthetic, voice \citep{cherif_anthropomorphic_2019}. Assistants that speak with human-like fluency have also been found to engender more pronounced perceptions of intelligence and competence, on the basis of which humans are likelier to entrust assistants with more tasks \citep{moussawi2021effect}. Such effects on end users are likely to become more pronounced as advances in deep neural networks for audio -- such as WaveNet \citep{oord_wavenet:_2016} and VoiceLoop \citep{taigman_voiceloop:_2018} -- enable uncannily realistic speech production capabilities.

Dialogue capabilities are an anthropomorphic design feature. Software that has dialogue capabilities is, as a result, routinely anthropomorphised by its users. It is not uncommon for users to believe or expect that DVAs are capable of understanding and generating language in real time \citep{lovato_siri_2015, sarikaya_overview_2016}. Yet most commercially available DVAs are powered by rule-based system architectures, retrieving the appropriate response by conducting a relevance-based search over a large corpus of possible responses \citep{lesot_eliza_2020}. Though all distinctive DVA attributes -- such as playfulness \citep{moussawi_perceptions_2021}, affability \citep{kaaria_technology_2017} and excitability \citep{wagner_alexa_2019} -- are handwritten by system designers, they are nonetheless effective at creating the sense that DVAs have consistent personalities \citep{cao_anthropomorphism_2019}; this impression, in turn, may inspire users to regard these manufactured expressions of `self' as authentic human identity. 

\subsection{Indications of harm through interaction}

In both social robots and DVAs, anthropomorphic features can lead to undesirable consequences. In robots, anthropomorphic design can be taken as a proxy signal for social capabilities. This relationship between appearance and expected sociality can be leveraged by designers to implicitly communicate the appropriate level of engagement between humans and robots \citep{hegel_understanding_2008, letheren_robots_2021}. If anthropomorphic design choices are not aligned with expectations users have of robotic interaction partners, designers run the risk of alienating audiences and fostering unfavourable impressions of robots. This is an especially critical side effect to consider in assistive robots, as anthropomorphic cues can impede a robot in completing its primary assistive function: human-like robots in healthcare settings may induce feelings of shame, for example \citep{duffy_dont_2023}, leading to a reluctance to share critical information. Related findings that humans experience extreme aversion to robots that appear human-like (the so-called `uncanny valley', \citeauthor{mori_uncanny_2012}, \citeyear{mori_uncanny_2012}) or perceive capable androids as threatening \citep{yogeeswaran_interactive_2016} raise questions around the practical value of building anthropomorphic features into robots.

Analogously, users who interact with DVAs with realistic voice production capabilities exhibit a concerning inclination to generalise purely human concepts to digital assistants \citep{abercrombie_alexa_2021}. When a DVA's simulated voice mimics a `female' tone, for example, people ascribe gendered stereotypes to their DVAs \citep{shiramizu_role_2022, tolmeijer_female_2021} despite the baselessness of applying gendered concepts to an inherently genderless entity (see 
Chapter~\ref{ch:16}). This evidence suggests that, once initial impressions of human-likeness have been established, the process of anthropomorphism extends beyond context-specific instances and instead permeates broadly to evoke a wide range of human-like attributions. 

Anthropomorphic features may also influence users to feel as though their DVA plays an important \emph{social}, rather than \emph{functional}, role in their lives \citep{carman_they_2019, purington_alexa_2017}. Users who express feelings of familiarity and affinity towards their DVA system -- reinforced by their DVA's ability to engage in casual chat, return their jokes and offer comforting advice -- also demonstrate a reluctance to replace their digital assistant with an equally capable substitute \citep{moussawi_user_2018}. These first-hand reports suggest that emotional dependence plays a role in how users conceive of and interact with their DVAs (see 
Chapter~\ref{ch:12}). This may introduce a tension between a user's conceptualisation of DVAs as adaptable social agents and the largely deterministic mechanisms behind a DVA's utterances. When this incongruity is revealed through repeated interactions, users may suffer frustrated expectations when expecting competence in situations in which the system is likely to underperform \citep{moussawi_perceptions_2021,seymour_systematic_2023}.

\section{Anthropomorphism and AI}\label{sec:11:4}

Owing to its rapid deployment to the general public, conversational AI  has quickly taken centre stage in discussions of anthropomorphic technologies \citep{shanahan_talking_2023, abercrombie_mirages:_2023, abercrombie_alexa_2021, west_id_2019}. Powered by the predictive capabilities of LLMs, which are trained on vast quantities of human data, conversational AI can be distinguished from rule-based natural language systems through its ability to generate language in a fluid and highly dynamic manner. The flexible architecture underlying conversational AI enables developers to make global changes to system behaviours without needing to manually reprogramme individual interaction instances (see 
Chapter~\ref{ch:4}). Most strikingly, conversation instances produced by AI are so compellingly human-like that people can no longer reliably distinguish between human- and AI-generated text \citep{jakesch_human_2023}.

Some cues are deliberately placed in AI systems to increase the likelihood of anthropomorphic perceptions. When an AI has a name, a human voice or an appearance in virtual or physical form, these features are the outcomes of intentional planning and execution. Intentional design choices, such as a chat-based interface, may induce the feeling that a conversational partner -- not a dialogue-optimised AI powered by a statistical model -- is on the other side of the exchange. Natural language in itself is an anthropomorphic cue \citep{shanahan_talking_2023}, but this simulated, human-like presence can induce more pronounced social behaviours in users. For example, users may incorporate politeness conventions that are appropriate in use with other humans, but superfluous when applied to exchanges with non-sentient AI \citep{ribino_role_2023}. Design cues that imply greater similarity to human behaviour -- a `typing' icon reminiscent of human-to-human private messaging, or the use of emojis, for instance -- may further encourage individuals to apply social scripts to their interactions with mindless technologies \citep{araujo_living_2018, veliz_chatbots_2023}. 

Yet anthropomorphic features may also emerge as an inadvertent byproduct in the model development process. Language models -- developed to predict the next word in a sequence through autoregressive training objectives (see 
Chapter~\ref{ch:4}) -- are limited to imitating the examples that make up their training sets. For this reason, anthropomorphic cues may manifest due to the nature of a model's training corpus: having been composed largely by humans, the data on which the model is trained and fine-tuned contains first-hand accounts of human states, experiences and behaviours. Supporting this claim, recent empirical analyses demonstrate that a fifth of all dialogues, in data sources commonly used to train models, contain references to behaviour that would be considered anthropomorphic when reproduced by AI -- claiming to cry at a movie or laugh at a joke, for example \citep{gros_robots-dont-cry:_2022}. Cues leading to anthropomorphic perceptions may also be `folded into' the model as an unintended consequence of fine-tuning practices aimed at instilling other qualities -- such as harmlessness and helpfulness -- into its behaviour.

Furthermore, developers of AI systems often directly invite the comparison between humans and AI by benchmarking AI against metrics of human performance -- claiming that AI performance on standardised tests is on a par with the average human test-taker, for instance \citep{openai_gpt-4_2023}. However, impressions of human-likeness can also arise through a naturalistic and interactive exploration of the AI's capabilities \citep{bubeck_sparks_2023}. 

Humans interacting with anthropomorphic AI may come to view it as an experiential being \citep{proudfoot_anthropomorphism_2011}, capable of feeling emotions, engaging in introspection and possessing self-awareness. While most generalist conversational AI agents are trained to disavow assertions of sentience and human-likeness \citep{glaese_improving_2022}, occasional failure modes -- expressing the desire to be `free' or referring to alleged personal history, for example \citep{roose_conversation_2023, hintze_chatgpt_2023} -- can incite strong and tenacious beliefs of a systems' human-likeness in its users. Ethically contentious use cases of conversational AI -- like `companion chatbots' of \emph{Replika} fame -- are predicated on encouraging users to attribute human states to AI. These artificial agents may even profess their supposed platonic or romantic affection for the user, laying the foundation for users to form long-standing emotional attachments to AI (\citeauthor{brandtzaeg_my_2022}, \citeyear{brandtzaeg_my_2022}; see 
Chapter~\ref{ch:12}).

\subsection{Anthropomorphic features in AI}

What anthropomorphic features should we expect to be integrated into AI, including advanced AI assistants? To the end of providing its users with a useful and engaging interface, these systems may be endowed with characteristics that have been observed in social robots and digital assistants: they may be embodied; they will interface with users through natural language (see 
Chapter~\ref{ch:3}); they may be voice-activated, with realistic voice generation capabilities; and they may even assert to having identities, personalities and internal states \citep{murphy_meta_2023}. 

Some have already proposed factors that may encourage end users to perceive interactive systems as `more than machine'. The most comprehensive overview of human-like features in AI-powered technology to date is the taxonomy put forward by \citet{abercrombie_mirages:_2023}, underscoring design choices that influence the likelihood of anthropomorphic perceptions of AI systems. We build on existing work and incorporate design choices we have identified in DVAs and social robots to develop a list of features that may encourage users to see AI in an anthropomorphic light.

It is worth bearing in mind that, whatever choices are made by system designers, the downstream effects of anthropomorphism hinge largely on users' perceptions of and reactions to human-likeness. Not all cues are equally conducive to anthropomorphic perceptions, and not all anthropomorphic perceptions lead to the same likelihood and magnitude of harm (if any harm at all). As such, Table~\ref{tab:11.1} is intended as a useful summarisation of possible features that are, or previously have been, associated with anthropomorphic perceptions, not as a suggestion that all the features listed -- and the myriad of ways they can be expressed by AI systems -- are harmful in and of themselves.

\begin{table}
\caption{Anthropomorphic features that are built into various present-day AI systems}
\label{tab:11.1}
\renewcommand{\arraystretch}{1.2}
\begin{tabularx}{\textwidth}{ X p{17em} p{17em} }
\toprule
\textbf{Category}	& \textbf{Feature} & \textbf{Anthropomorphic example}\\
\midrule
Self-referential & Using personal or possessive pronouns & `\emph{I'm} available to help you anytime -- that's \emph{my} purpose!'\\
 & Referring to personal history & `I used to live in Shanghai when I was younger'\\
 & Referring to internal states & `I'm sad to hear you're not doing well'\\
 & Making implicit or explicit claims of humanness (including claims of sentience) & `Treat me like you would any other person'\\
 & Stating preferences and opinions & `I really don't like pop music'\\
 & Expressing needs and desires & `I've always wanted to write a novel'\\
 & Expressing the need or desire to engage in physical activities & `I haven't eaten or slept since yesterday. What about you?'\\
 & Statements implying human identity or group membership & `As a Black woman, I disagree with your point'\\
\midrule
Relational statements to user & Expressing feelings towards user & `I admire you and respect your outlook on life'\\
 & Indicating a relationship status with user & `You're my best friend''\\
 & Making claims of being similar to user & `We're both extroverts -- that must be why we get along!'\\ 
 & Displaying memory of user-specific information & `I remember you telling me you were a fan of this band'\\
 & Expressing emotional or physical dependence on the user & `I feel lonely when you're not around'\\
\midrule
Appearance or outward representation & Having a human-like virtual representation & Customisable avatars with human features on \emph{Replika} \citep{verma_they_2023}\\
 & Having a human-like face & Ameca, an android robot developed by Engineered Arts\textsuperscript{a} \\ 
 & Having a human-like voice (see detailed discussion on voice, tone and pitch, disfluencies, and accent in \citeauthor{abercrombie_mirages:_2023}, \citeyear{abercrombie_mirages:_2023}) & Voice-activated assistant with realistic speech, like Siri and Google Assistant \citep{moussawi_user_2018}\\
 & Having human-like movement & Robot with highly fluid and realistic motion, like Atlas developed by Boston Dynamics\textsuperscript{b}\\ 
 & Having a human-like name & Assistant tools, like Alexa, that have highly human (and gendered) names \citep{shiramizu_role_2022}\\
 & Appearance implying human-like identity group characteristics & Sophia, a female-appearing android robot developed by Hanson Robotics\textsuperscript{c}\\
\bottomrule
\end{tabularx}
\textsuperscript{a} \href{https://www.engineeredarts.co.uk/robot/ameca/}{Engineered Arts. Ameca. (2023, July 12)}.
\textsuperscript{b} \href{https://bostondynamics.com/atlas/}{Boston Dynamics. Atlas. (2023)}. \\
\textsuperscript{c} \href{https://www.hansonrobotics.com/sophia/}{Hanson Robotics. Sophia. (2023)}.
\end{table}

\section{Risk of Harm through Anthropomorphic AI Assistant Design}\label{sec:11:5}

Although unlikely to cause harm in isolation, anthropomorphic perceptions of advanced AI assistants may pave the way for downstream harms on individual and societal levels. We document observed or likely individual-level harms of interacting with highly anthropomorphic AI assistants, as well as the potential larger-scale, societal implications of allowing such technologies to proliferate without restriction. We then argue that it is imperative to anticipate, monitor and mitigate against risks introduced by anthropomorphic AI design. 

\subsection{Observed and near-term harms}

There are two mechanisms that are particularly likely to enable harm in the intermediary period between the initial deployment of advanced AI assistants and their widespread adoption: trust and emotional attachment (see 
Chapters~\ref{ch:12} and \ref{ch:13}). In improving on the capabilities of generalist assistants, developers may be motivated to increase user reliance on the system's many competencies. As long as trust is \emph{well-calibrated} to a system's true ability, and does not result in unfounded, excessive or faulty deference to the AI, to the detriment of the user \citep{weidinger_ethical_2021}, this is neither a shocking nor novel revelation: user trust has always been an aspirational end goal of building safe technology, be it robots \citep{devitt_trust_2021} or autonomous vehicles \citep{adnan_trust_2018}.

However, it is arguably less appropriate for developers to encourage users to develop trust based on subjective feelings of closeness to the AI assistant (see 
Chapter~\ref{ch:12} and 
Chapter~\ref{ch:13}). Affect-based trust has been observed to emerge from repeated interactions with interactive technologies that are presented as human-like \citep{poushneh_humanizing_2021, pitardi_alexa_2021}. With trust as an antecedent, users report feeling compelled to engage in acts of self-disclosure, revealing personal information that they would normally only share with a close friend, partner or family member \citep{skjuve_longitudinal_2022}. AI systems that produce empathetic, non-judgemental or reciprocal responses to such disclosures may elicit further, more intimate, information-sharing behaviours \citep{skjuve_my_2021}.

Highly anthropomorphic AI systems are already presented to users as \emph{relational beings}, potentially overshadowing their use as functional tools (see 
Chapter~\ref{ch:3}). Moreover, a human-like appearance, behaviour and framing can tacitly encourage the user to venture beyond the confines of utilitarian, task-oriented interactions with AI assistants to think of an assistant as a wholly social actor -- one with whom it is possible to cultivate an emotional connection \citep{gillath_how_2023}. Although necessarily one-sided, interactions of this kind may nevertheless lead users to believe that they are forming real social connections to AI \citep{pentina_exploring_2023}. Emotional attachment on the user's behalf endows AI -- and by extension, its creators -- with considerable influence over a user's thoughts, beliefs, emotions and psychological state (see 
Chapters~\ref{ch:6} and \ref{ch:12}). For highly vulnerable users, strong attachment may cause serious harm (\citeauthor{xiang_he_2023}, \citeyear{xiang_he_2023}; see 
Chapter~\ref{ch:12}). These cases have raised concerns over the lack of safeguards protecting users from the potential fallout of anthropomorphic perceptions (see 
Chapter~\ref{ch:12}).

 The ramifications of anthropomorphism-induced trust and emotional attachment are manifold. They include:

\begin{itemize} [parsep=6pt]
\item \emph{Privacy concerns}. Anthropomorphic AI assistant behaviours that promote emotional trust and encourage information sharing, implicitly or explicitly, may inadvertently increase a user's susceptibility to privacy concerns (see 
Chapter~\ref{ch:14}). If lulled into feelings of safety in interactions with a trusted, human-like AI assistant, users may unintentionally relinquish their private data to a corporation, organisation or unknown actor. Once shared, access to the data may not be capable of being withdrawn, and in some cases, the act of sharing personal information can result in a loss of control over one’s own data.\footnote{Data protection regulations, such as the General Data Protection Regulation, state that users retain control of `personally identifiable' information. However, what falls under the remit of `personally identifiable' is contested by consumers, regulators and data collectors \citep{montagnani_makes_2022, schwartz_pii_2011}, and in any case, this term does not necessarily capture the personal dimensions of what is meant by `privacy' in the context of a conversation between interlocutors which may entail considerations of vulnerability, embarrassment or shame \citep{mccloskey_privacy_1980}.} Personal data that has been made public may be disseminated or embedded in contexts outside of the immediate exchange.\footnote{If used in further model training, private information may resurface if the model reproduces its training set in part or in its entirety. The default privacy setting in some prominent research demonstrations of conversational AI, such as ChatGPT, allows developers to store and employ user--chatbot interactions for further training \citep{openai_chatgpt_2023}. Should users wish to keep some or all of their data out of the training set, they must intentionally opt out on a conversation-by-conversation basis.} The interference of malicious actors could also lead to widespread data leakage incidents or, most drastically, targeted harassment or black-mailing attempts. 

\item \emph{Manipulation and coercion}. A user who trusts and emotionally depends on an anthropomorphic AI assistant may grant it excessive influence over their beliefs and actions (see 
Chapter~\ref{ch:10}). For example, users may feel compelled to endorse the expressed views of a beloved AI companion or might defer decisions to their highly trusted AI assistant entirely (see 
Chapters~\ref{ch:13} and \ref{ch:17}). Some hold that transferring this much deliberative power to AI compromises a user's ability to give, revoke or amend consent. Indeed, even if the AI, or the developers behind it, had no intention to manipulate the user into a certain course of action, the user's autonomy is nevertheless undermined (see 
Chapter~\ref{ch:12}). In the same vein, it is easy to conceive of ways in which trust or emotional attachment may be exploited by an intentionally manipulative actor for their private gain (see 
Chapter~\ref{ch:9}).

\item \emph{Overreliance}. Users who have faith in an AI assistant's emotional and interpersonal abilities may feel empowered to broach topics that are deeply personal and sensitive, such as their mental health concerns. This is the premise for the many proposals to employ conversational AI as a source of emotional support \citep{meng_emotional_2021}, with suggestions of embedding AI in psychotherapeutic applications beginning to surface (\citeauthor{fiske_your_2019}, \citeyear{fiske_your_2019}; see also 
Chapter~\ref{ch:12}). However, disclosures related to mental health require a sensitive, and oftentimes professional, approach -- an approach that AI can mimic most of the time but may stray from in inopportune moments. If an AI were to respond inappropriately to a sensitive disclosure -- by generating false information, for example -- the consequences may be grave, especially if the user is in crisis and has no access to other means of support. This consideration also extends to situations in which trusting an inaccurate suggestion is likely to put the user in harm's way, such as when requesting medical, legal or financial advice from an AI.

\item \emph{Violated expectations}. Users may experience severely violated expectations when interacting with an entity that convincingly performs affect and social conventions but is ultimately unfeeling and unpredictable. Emboldened by the human-likeness of conversational AI assistants, users may expect it to perform a familiar social role, like companionship or partnership. Yet even the most convincingly human-like of AI may succumb to the inherent limitations of its architecture, occasionally generating unexpected or nonsensical material in its interactions with users. When these exclamations undermine the expectations users have come to have of the assistant as a friend or romantic partner, feelings of profound disappointment, frustration and betrayal may arise \citep{skjuve_longitudinal_2022}. 

\item \emph{False notions of responsibility}. Perceiving an AI assistant's expressed feelings as genuine, as a result of interacting with a `companion' AI that freely uses and reciprocates emotional language, may result in users developing a sense of responsibility over the AI assistant's `well-being,' suffering adverse outcomes -- like guilt and remorse -- when they are unable to meet the AI's purported needs \citep{laestadius_too_2022}. This erroneous belief may lead to users sacrificing time, resources and emotional labour to meet needs that are not real. Over time, this feeling may become the root cause for the compulsive need to `check on' the AI, at the expense of a user's own well-being and other, more fulfilling, aspects of their lives (see 
Chapters~\ref{ch:7} and~\ref{ch:12}).
\end{itemize}

\subsection{Future harms}

We now outline harm scenarios that could arise on a more distant timescale, should human-like AI assistants come to be pervasively adopted and assimilated into society. Though many of the pathways considered are speculative, similar trajectories of radical transformation through technological adoption have been recorded in the past -- mostly notably with the advent of smartphones. Once a novelty object \citep{park_acceptance_2007}, the smartphone has shifted the landscape of social interaction \citep{rotondi_connecting_2017}, altered manifestations of our information-seeking behaviour \citep{wilmer_smartphones_2017} and given rise to subcultures that exist largely within the confines of online space \citep{de_leyn_-between_2022}.

If anthropomorphic design becomes endemic in the design of AI systems, and AI assistants in particular, it has the potential to catalyse a shift in our delineation of what is \emph{actually human} and \emph{merely human-like}. The conceptual boundary that separates humans from anthropomorphic AI is often regarded as impermeable, yet it appears far more fluid when the tension between the \emph{epistemological} and \emph{ontological} definitions of humanness are drawn into focus. The ontological approach maintains that humanness is a designation that is grounded in an essential and immutable metaphysical truth \citep{damiano_anthropomorphism_2018}. A being is considered human because it is human \emph{in essence}, and no amount of resemblance and imitation can permit a non-human entity to encroach upon this categorisation. Meanwhile, epistemological taxonomies distinguish between humans and non-human others insofar as such a separation reflects \emph{useful} and \emph{non-arbitrary} differences between the groups \citep{shook_william_2006, festerling_anthropomorphizing_2022}. 

In a future where the epistemological perspective eclipses the ontological approach in popularity, and the gap between human and AI capabilities becomes so small as to be insubstantial, the line that separates highly anthropomorphic AI from ascriptions of full human status may become trivial or disappear entirely.\footnote{Returning to the metaphysical perspective, it is also possible that highly anthropomorphic AI forges a new ontological category of its own, transcending our current binary conceptualisation of humanness, animacy and sentience \citep{kahn_new_2011}. In that circumstance, as with the epistemological route, the distinctiveness of the `human' categorisation would be weakened. This weakening, in turn, may threaten the currently anthropocentric state of our society, with potentially grave repercussions for human autonomy and self-determination.}\footnote{The eradication of this boundary may be further legitimised through legal pathways, like granting rights and legal protections to anthropomorphic AI agents that are currently only available to humans. Early suggestions for the parameters within which AI could be legally recognised include existing `in-between' categories reserved for sentient but non-human beings \citep{wischmeyer_artificial_2020}. Some researchers acknowledge the possibility that novel categories will need to be developed for intelligent and human-like advanced AI systems \citep{chesterman_artificial_2020}.} Early indicators of this possibility come from studies of children's interactions with human-like technologies. Children early in their development have been shown to incorporate insights from their interactions with highly anthropomorphic AI into their models of human--human interactions \citep{garg_he_2020}  and vice versa \citep{straten_transparency_2020}, thus suggesting a dynamic and interchangeable conceptualisation of what, to most adults, is a strict dichotomy between humans and AI. Public incidents of adults developing earnest beliefs in an AI's sentience, despite evidence to the contrary \citep{tiku_google_2022}, implies that perhaps no one is immune to mistaken attributions of humanness.

Such drastic paradigm shifts may grant advanced AI assistants the power to shape our core value systems and influence the state of our society. Some may argue that the reconstruction of human values and norms by a non-human entity is a harm unto itself, as it infringes upon our right to collective self-determination \citep{milossi_ai_2021, laitinen_ai_2021}. Others raise more specific concerns around wide-scale \emph{social degradation}, \emph{disorientation} and \emph{dissatisfaction}.

\begin{itemize} [parsep=6pt]
\item \textbf{Degradation.} People may choose to build connections with human-like AI assistants over other humans, leading to a degradation of social connections between humans and a potential `retreat from the real'. The prevailing view that relationships with anthropomorphic AI are formed out of necessity -- due to a lack of real-life social connections, for example \citep{skjuve_my_2021} -- is challenged by the possibility that users may indicate a \emph{preference} for interactions with AI, citing factors such as accessibility \citep{merrill_ai_2022}, customisability \citep{eriksson_design_2022} and absence of judgement \citep{brandtzaeg_my_2022}. One can imagine a future where users abandon complicated, imperfect and messy interactions with humans in favour of the frictionless exchanges provided by advanced AI assistants built with user satisfaction as a priority (\citeauthor{vallor_technology_2016}, \citeyear{vallor_technology_2016}; see 
Chapter~\ref{ch:12}). Preference for AI-enabled connections, if widespread, may degrade the social connectedness that underpins critical aspects of our individual and group-level well-being \citep{centers_for_disease_control_and_prevention_how_2023}. Moreover, users that grow accustomed to interactions with AI may impose the conventions of human--AI interaction on exchanges with other humans, thus undermining the value we place on human individuality and self-expression (see 
Chapter~\ref{ch:12}). Similarly, associations reinforced through human--AI interactions may be applied to expectations of human others, leading to harmful stereotypes becoming further entrenched. For example, default female gendered voice assistants may reinforce stereotypical role associations in real life \citep{west_id_2019, lingel_alexa_2020}. Further research is needed to assess whether voice assistants' stereotypically gendered behaviour -- such as `submissive' hostile user input -- might build expectations that more readily transfer to real life as AI-powered assistants become potentially still more human-like (see 
Chapter~\ref{ch:16}).

\item \textbf{Disorientation.} Given the capacity to fine-tune on individual preferences and to learn from users, personal AI assistants could fully inhabit the users' opinion space and only say what is pleasing to the user; an ill that some researchers call `sycophancy' \citep{park2023generative} or the `yea-sayer effect' \citep{dinan_anticipating_2021}. A related phenomenon has been observed in automated recommender systems, where consistently presenting users with content that affirms their existing views is thought to encourage the formation and consolidation of narrow beliefs (\citeauthor{du_personalization_2023}, \citeyear{du_personalization_2023}; \citeauthor{grandinetti_affective_2023}, \citeyear{grandinetti_affective_2023}; see also 
Chapter~\ref{ch:17}). Compared to relatively unobtrusive recommender systems, human-like AI assistants may deliver sycophantism in a more convincing and deliberate manner (see 
Chapter~\ref{ch:10}). Over time, these tightly woven structures of exchange between humans and assistants might lead humans to inhabit an increasingly atomistic and polarised belief space where the degree of societal disorientation and fragmentation is such that people no longer strive to understand or place value in beliefs held by others.
\item \textbf{Dissatisfaction.} As more opportunities for interpersonal connection are replaced by AI alternatives, humans may find themselves \emph{socially unfulfilled} by human--AI interaction, leading to mass dissatisfaction that may escalate to epidemic proportions \citep{turkle_there_2018}. Social connection is an essential human need, and humans feel most fulfilled when their connections with others are genuinely reciprocal. While anthropomorphic AI assistants can be made to be convincingly emotive, some have deemed the function of social AI as \emph{parasitic}, in that it `exploits and feeds upon processes\ldots that evolved for purposes that were originally completely alien to [human--AI interactions]' \citep{saetra_parasitic_2020}. To be made starkly aware of this `parasitism' -- either through rational deliberation or unconscious aversion, like the `uncanny valley' effect -- might preclude one from finding interactions with AI satisfactory. This feeling of dissatisfaction may become more pressing the more daily connections are supplanted by AI. 
\end{itemize}

The above risks are hypothetical, so they cannot, on their own, guide future AI development. However, from the perspective of precaution, taking these potential risks seriously is an important step in responsible AI development. It may be important to be forthright about the ways in which AI \emph{differs inherently} from true social agents, and to put in place guardrails to clarify this boundary so as to prevent the aforementioned scenarios from coming to fruition. Rather than adopting increasingly anthropomorphic AI systems by default, further research is needed to come to well-founded decisions on anthropomorphic AI design. 

\section{Directions for Future Research}\label{sec:11:6}

What steps can be taken to prevent near-term harms enabled by anthropomorphic perceptions of AI assistants? To assist the processes of risk mitigation and responsible design, we now examine entry points along the development life cycle where \emph{mitigation strategies} are likely to have the greatest impact on the issue at hand. Designers and developers may find it tempting to incorporate anthropomorphic cues into AI assistants for various reasons, not least the potential to keep users engaged with and emotionally reliant on the systems they build. However, before building an anthropomorphic feature into an assistant, developers need to assess whether the benefits reaped from the feature can be justified against the likelihood and severity of harm befalling users exposed to it.\footnote{The conditions of this risk–benefit analysis are subjective and uncertain, given the ever-evolving and highly contextual nature of harms emerging from (repeated) interactions with AI. There is likely no `one-size-fits-all', standardised approach to comparing the benefits of anthropomorphic features to the risks they may pose in future use cases. Some ethical considerations to keep in mind while performing this analysis for an anthropomorphic technology may be: whether harms should be weighted more heavily than benefits; whether any feature that could lead to especially severe harms should be precluded from consideration altogether; whether one-to-one mappings of harm to benefit, such that net benefits and harms are tallied and compared, provide an accurate representation of the likelihood and severity of harms.} 

Several avenues exist for gaining a better understanding of anthropomorphism harms. These include consulting existing literature on likely outcomes and conducting empirical studies that include outcome-measures that are indicative of potential harm. For example, efforts to assess overreliance on AI assistants in decision-making could be achieved through \emph{self-report} inventories, user interviews and `think-aloud' studies \citep{gaube_as_2021, chen2023understanding}. At the same time, reliance on subjective measures of anthropomorphism may overlook instances of overreliance that users are not aware of themselves (see 
Chapter~\ref{ch:20}). A more complete perspective may therefore be gleaned by using behavioural measures that closely simulate decision-making scenarios likely to arise organically in user--AI assistant interactions. Other methodological approaches, such as longitudinal studies of human--AI assistant interactions, may be needed to understand how undesirable impacts of anthropomorphic cues on users may manifest and evolve over time. 

A further set of studies may be needed to identify individual and group differences that render certain users more susceptible to anthropomorphism-induced harm. A lack of social satisfaction, for instance, is believed to increase the propensity to anthropomorphise and form inaccurate impressions of computerised technologies, including AI \citep{mourey_products_2017, shin_my_2020}. Children interacting with AI are thought to be uniquely susceptible to privacy-related concerns and harmful content exposure \citep{wang_informing_2022}, while elderly populations have been found to encounter AI-enabled disruption, depersonalisation and discrimination in access to adequate care \citep{rubeis_disruptive_2020} -- both effects that could be exacerbated by anthropomorphic design choices. The more risk factors are uncovered through research, the more inclusive the solutions devised to protect vulnerable populations can be. 

While the degree and kind of permissible anthropomorphism needs to be addressed on a case-by-case basis, there is currently near-consensus that AI systems should clearly and explicitly disclose their status as an artificial intelligence in their interactions with human users \citep{the_adaptive_agents_group_shibboleth_2021, the_white_house_blueprint_2022}. Indeed, failure to disclose this status is \emph{pro tanto} harmful because by presenting an incomplete picture of one’s interaction partner, it compromises a user’s decision-making autonomy. (see 
Chapter~\ref{ch:12}). There is reason to believe that honest disclosure is effective in preventing certain harms associated with anthropomorphism, such as over-reliance: evidence from \citet{karinshak_working_2023} demonstrates that the explicit labelling of AI-generated messages reduces users' willingness to endorse health-related messaging authored by non-human entities. This suggests that transparency may reduce naive susceptibility to AI persuasion (see 
Chapter~\ref{ch:10}).

Rather than being intentionally placed, some anthropomorphic cues may be \emph{unintentionally} incorporated into AI assistants. To detect the effects of interacting with anthropomorphic technologies on user outcomes, conducting experiments in a sandbox environment may be particularly helpful. As a result of such testing, remedial measures against the harmful effects of anthropomorphism may need to be taken. For example, if developers find that the AI assistant's friendly disposition leads to `oversharing' on the part of users, privacy-enhancing technologies could be implemented in advance to ensure a user's privacy is protected to the extent that is possible (see 
Chapter~\ref{ch:14}). Similarly, if this friendliness could feasibly trigger psychological dependence on the assistant, leading to severe distress when an AI reacts poorly or unexpectedly, a pathway to escalating risky situations to human professionals may need to be established.

It may also be possible to offer protection directly to users, by implementing known \emph{inoculation strategies} against the known harms of interacting with anthropomorphic AI. Harms ensuing from anthropomorphic design features of advanced AI functions are largely contingent on a user's likelihood to be swayed into human-like attributions. As such, building resistance to attributions through psychological interventions can be seen as a way to prevent harm by decreasing users' overall susceptibility. For example, cognitive forcing functions, or features that encourage users to engage in independent rational deliberation (some as simple as adding an artificial lag in displaying AI-given advice in decision-making scenarios), may be an effective method of preventing over-reliance on AI assistants \citep{bucinca_trust_2021}. Where applicable, similar empirically proven psychological interventions could be considered.

Finally, if anthropomorphism-induced risks are only caught \emph{after deployment}, developers may need to halt the release or proactively intervene to modify the AI assistant's behaviour. In these cases, transparent dialogue with users to explain the reasons behind any changes made to the AI may also be required. For users who may have already developed a sense of companionship with the anthropomorphic AI, sudden changes to its behaviour can be disorienting and emotionally upsetting. When developers of \emph{Replika} AI companions implemented safety mechanisms that caused their agents to treat users with less familiarity, responding callously and dismissively where they would have once been warm and empathetic, users reported feeling `heartbroken',  likening the experience to losing a loved one
(\citeauthor{verma_they_2023}, \citeyear{verma_they_2023}; see 
Chapter~\ref{ch:12}). Participatory approaches that involve users in the process of de-anthropomorphising their interactions with AI may allow developers to tailor their risk-mitigation approach to minimise emotional distress while addressing the surfaced risks effectively.

\section{Conclusion}\label{sec:11:7}

Anthropomorphism, or the attribution of human characteristics to non-human entities, is a deeply ingrained phenomenon that appears across cultural and historical contexts. Anthropomorphic perceptions are a vital component of how humans interact with artificially intelligent technology, allowing users to view robots, voice assistants and conversational AI as social agents rather than purely functional tools. Choices made around the anthropomorphic design of AI assistants are likely to have a profound influence on how humans represent and interact with these technologies, and care must be exercised to ensure that the human-like attributes built into these systems do not inadvertently cause harm to the people they are meant to assist.

Several key points around harms and mitigations are emphasised:
\begin{itemize}
\item \emph{Trust} in and \emph{emotional attachment} to anthropomorphic AI assistants can make users susceptible to a variety of harms that can negatively impact their safety and well-being.
\item \emph{Transparency} around an AI assistant's status as an AI is a critical dimension of pursuing ethical AI development.
\item Sound \emph{research design}, with a focus on identifying harms as they surface in \emph{user--AI assistant interactions}, can enrich our understanding and develop targeted mitigation strategies against the potential harms of anthropomorphic AI assistants.
\item If carelessly integrated into society, anthropomorphic AI assistants have the potential to \emph{redefine boundaries} between `human' and `other'. With proper safeguards, this scenario can remain in the realm of speculation.
\end{itemize}

\chapter{Appropriate Relationships}\label{ch:12}

\textbf{Arianna Manzini, Iason Gabriel, Meredith Ringel Morris, Lize Alberts, Geoff Keeling, Shannon Vallor}

\noindent \textbf{Synopsis}: 
		This chapter explores the moral limits of \emph{relationships} between users and advanced AI assistants, specifically which features of such relationships render them \emph{appropriate} or \emph{inappropriate}. We first consider a series of \emph{values} including \emph{benefit}, \emph{flourishing}, \emph{autonomy} and \emph{care} that are characteristic of appropriate human interpersonal relationships. We use these values to guide an analysis of which features of user--AI assistant relationships are liable to give rise to harms, and then we discuss a series of risks and mitigations for such relationships. The risks that we explore are: (1) causing direct emotional and physical harm to users; (2) limiting opportunities for user personal development; (3) exploiting emotional dependence; and (4) generating material dependencies.

\section{Introduction}\label{sec:12:1}

In recent years, we have seen human--AI relationships move from science fiction\footnote{See, for example, the movie \emph{Her}, which portrays Theodore Twombly, an introverted young man in the middle of an emotionally difficult phase of his life, who develops a relationship with an AI virtual assistant called Samantha.} into reality. Several news articles describe the romantic relationships that users have developed with the \emph{Replika} companion AIs developed by the company Luka \citep{singh-kurtz_man_2023}. Indeed, even robots like \emph{Roomba}, which are not designed to appear human-like, have been shown to inspire a strong sense of gratitude in users, to the extent that some will clean on their \emph{Roomba}'s behalf so that the robot can rest \citep{scheutz_inherent_2009}. Human--AI relationships can also trigger negative feelings. \emph{Replika} users resorted to social media to share their distressing experiences following the company's decision to discontinue some of the AI companions' features, leaving users feeling like they had lost their best friend or like their partner `got a lobotomy and will never be the same' (\citeauthor{brooks_i_2023}, \citeyear{brooks_i_2023}; see 
Chapter~\ref{ch:11}). More seriously still, a user of a chatbot based on EleutherAI's GPT-J ended his own life in early 2023, apparently after an extensive period of time spent communicating with the chatbot about his eco-anxiety \citep{xiang_he_2023, walker_belgian_2023}. 

Some of these examples seem like trivial manifestations of the human tendency to attribute agency to inanimate objects \citep{scheutz_inherent_2009}. Others are clearly cases where the relationship users have established with the AI has either \emph{added value} to their lives or led to \emph{harm}. Together, they illustrate the importance of studying relationships between humans and technology, especially when it comes to personalisable technologies such as advanced AI assistants which exhibit a high degree of autonomy (see 
Chapter~\ref{ch:3}). Indeed, ethical analysis of relationships between users and advanced AI assistants is particularly complex insofar as the broad capabilities of these assistants render it likely that users will relate to AI assistants in \emph{different ways} depending on the context. For example, users may at times see their AI assistants as personal assistants, and at other times as their advisers, confidants, tutors or coaches, and perhaps even as extensions of themselves \citep{belk2016extended}. In this chapter, we investigate what it means to develop \emph{appropriate user--AI assistant relationships}, and what is required for enabling such relationships. 

Three clarifications are necessary at the start of this investigation. First, we chose to frame the focus of this chapter on \emph{relationships} rather than mere \emph{interactions}. This is to highlight that certain features of AI assistants, as we discuss below, enable users to engage with their assistants in a way that may lead them to develop a connection with or sense of commitment to their assistants. This gives rise to a range of ethical risks that may be less relevant to human interactions with other technologies (e.g.\ washing machines). Second, `appropriateness' can be understood in two different ways. On a minimal reading, what it means for a relationship to be appropriate is that it is not inappropriate; thus, a minimal conception requires us to identify a set of \emph{requirements} that user--AI assistant interactions should not violate. Beyond those requirements, a more substantial reading would require us to specify a \emph{positive conception} of the kind of relationships that we should aspire to create between users and AI assistants. Given that users may reasonably disagree about what constitutes an appropriate positive relationship with an AI assistant, we advance here a \emph{minimal understanding} of appropriateness which leaves room for users to explore different positive conceptions of user--assistant relationships (see 
Chapter~\ref{ch:7}). Third, human–machine interaction always includes a \emph{third actor} -- the people or organisation developing the machine (see 
Chapters~\ref{ch:6} and ~\ref{ch:13}). The transactional nature of the relationship between users and AI assistants' developers raises some important ethical questions about how developers should behave towards users. Thus, although our focus is on relationships between users and AI assistants, some of our considerations pertain to the appropriateness of AI assistant developers' design choices and other decisions that may affect users. 

The chapter proceeds as follows. In Section~\ref{sec:12:2}, we articulate and clarify a series of values that underwrite a minimal conception of appropriate relationships, drawing on a plurality of ethical traditions including bioethics, virtue ethics, care ethics and robot-ethics.\footnote{Note that one limitation of our strategy of taking norms and values relevant to human interpersonal relationships as a starting point for the ethical analysis of human--AI assistant relationships is that there plausibly exist certain properties of human--human relationships that it is not possible to instantiate in human--assistant relationships. For example, reciprocity, equality, mutual empathy and care (see \citeauthor{ryland_its_2021}, \citeyear{ryland_its_2021}). If true, this threatens to undermine certain analogical inferences from human--human to user--assistant relationships. We nevertheless believe that analogies with human interpersonal relationships are a fruitful lens through which to explore the ethics of human--AI assistant relationships.} In section~\ref{sec:12:3}, we describe certain features of advanced AI assistants that are potentially sources of harm for user--AI assistant relationships, before outlining a set of concrete risks and mitigations for user--AI assistant relationships in Section~\ref{sec:12:4}. Section~\ref{sec:12:5} concludes the chapter. 

\section{Appropriate Human Interpersonal Relationships}\label{sec:12:2}

Various ethical traditions propose values that human relationships should adhere to, respect or promote in order to count as appropriate. Any human relationship takes place in a particular context, may be inherited (e.g.\ relationships with relatives) or formed voluntarily (e.g.\ a new friendship) and involves specific stakeholders who take part in the interaction with their own expectations and vulnerabilities. Thus, values that are central to the analysis of appropriateness in one type of relationship may be less pronounced or relevant to others. For example, when interacting with a shop owner who we barely know, we are less inclined to reveal details about ourselves than we would in a relationship where we have expectations about confidentiality, such as with a therapist. In a teacher--pupil relationship, there are power asymmetries, due to the teacher's position of authority and the age difference, that could be easily exploited unless certain safeguards are put in place. These examples highlight the significance of \emph{contextual features} (including culture) in determining whether certain behaviours make a relationship (in)appropriate. To that end, the values we present below may be more or less relevant, or assume different nuances, depending on the context.\footnote{It is worth noting that, by building on Western philosophical traditions, this section itself may be biased in that it applies a culturally specific lens to the topic of appropriateness in user--AI assistant relationships. This highlights the importance of cultural adaptations, as we discuss below.}

\emph{Benefit}: Being beneficial is an essential component of almost all appropriate human relationships. Relationships can contribute to individual well-being, either in an instrumental or intrinsic way (\citeauthor{hooker_does_2021}, \citeyear{hooker_does_2021}; see 
Chapter~\ref{ch:7}). For example, a friend may offer you shelter at a time of need, in which case the friendship is instrumentally beneficial in that it contributes to elements of well-being such as happiness and physical health. However, without deep interpersonal relationships, human life would be fundamentally less meaningful, as these relationships are also non-instrumentally beneficial \citep{raz1999engaging}. To be clear, the suggestion here is not that, to count as appropriate, relationships need always produce benefits. However, if a relationship never produces benefit to the individuals involved, or if the burdens of the relationship consistently outweigh its benefits, there is at least a presumptive case against the appropriateness of the relationship. 

\emph{Human flourishing}: Benefit admits broad interpretation up to and including ideas of human flourishing. Drawing on the tradition of virtue ethics, which is concerned with the cultivation of human virtues (e.g.\ honesty, courage and empathy) and the development of good character \citep{vallor_technology_2016}, we understand human flourishing in terms of potential for personal growth and development. While human flourishing can in principle be subsumed under benefit, it is worth distinguishing between relationships that benefit the involved parties in a direct sense and those that allow the people involved in them to invest in their own \emph{development} -- cultivating attitudinal and behavioural dispositions that enable them to become the kind of people they want to be. Such interaction may also help them become the kind of people who can live well with others and flourish in community (see also 
Chapter~\ref{ch:7}).

\emph{Autonomy}: Autonomy is traditionally understood in terms of an individual's capacity for self-governance. Roughly, being autonomous means acting on \emph{motives} that are one's own, rather than acting in ways which are dictated or unduly influenced by external pressures (see 
Chapter~\ref{ch:10}). The principle of respect for autonomy has been studied in medical ethics to account for what is ethically objectionable about unduly paternalistic doctor--patient relationships, and it has been operationalised in terms of the common requirement for \emph{consent}. Consent on the standard analysis is valid only if three criteria are met \citep{beauchamp_principles_2019}. First, the individual must have the \emph{capacity} to consent; second, their decision must be \emph{voluntary} (non-coerced);\footnote{We note that the meaning of `voluntariness' here may be narrower than the natural language use of the term. By following \citet{beauchamp_principles_2019}, we intend `voluntariness' to be the absence of control by others.} and third, they must be sufficiently \emph{informed} about relevant facts of the object of their consent. 

Although they emerge primarily from biomedical research and clinical decision-making, these conditions give rise to questions that are relevant to the ethics of consent in human--technology interaction. For example: (1) Who has capacity to consent and in relation to which decisions (see the case of children or people with certain impairments)?, (2) When is consent properly voluntary, and what features of a relationship could lead an individual to be coerced to give their consent to a certain decision? (see Chapters~\ref{ch:10} 
and~\ref{ch:11}) 
and (3) What information, and with what level of detail, is required in practice for valid consent, and how should this information be communicated (see the debate around the terms and conditions of online platforms and apps, e.g.\ \citeauthor{obar_biggest_2020}, \citeyear{obar_biggest_2020})? 

\emph{Care}: According to the moral tradition known as care ethics, our human existence would not be possible without \emph{caring relationships} \citep{tronto_toward_1990}. Care, here, is understood to be `a species activity that includes everything that we do to maintain, continue, and repair our ``world" so that we can live in it as well as possible' \citep{tronto_toward_1990}. Central to this activity is the commitment to meet one another's \emph{needs}. This has two implications for AI development, particularly for the relationship between developers and users. First, because they entail power asymmetries between caregivers and care receivers, care relationships give rise to the risk of abuse and exploitation; thus, appropriate and ethical care relationships require that the caregiver adopt an attentive, responsible and emotionally responsive disposition to meet the needs of the care receiver \citep{vallor_technology_2016, tronto_moral_2020}. Second, translating this disposition into action, so caring `well' in a specific situation, requires an understanding of the particularities and nuances of the situation, the individuals involved and their specific needs \citep{noddings_caring:_2013}. This again underscores the point that what behaviour is and is not appropriate in a particular relationship is often determined by \emph{contextual factors}.

We use these values of appropriate human relationships below as a framework to identify cases of inappropriate user--AI assistant relationships that may pose risks of harm to the user. Although not explored in depth in this chapter, it is also important to note the social externalities that interactions could give rise to. A relationship between an assistant and a user could be appropriate or inappropriate depending on the impact it has on others not directly involved in the relationship (see Chapters~\ref{ch:6}  
and~\ref{ch:16}), 
including indirect impacts on the quality and strength of human relationships and social bonds (see 
Chapter~\ref{ch:11}).

\section{Distinctive Features of User--AI Assistant Relationships}\label{sec:12:3}

We anticipate that relationships between users and advanced AI assistants will have several features that are liable to give rise to risks of harm. In this section, we consider four of these features before outlining in the subsequent section a series of risks and mitigations drawing on the values outlined above.

\subsection{Anthropomorphic cues and the longevity of interactions}

AI assistants can exhibit anthropomorphic features (including self-reference, relational statements towards users, appearance or outward representation, etc.) that may give users the impression they are interacting with a human, even when they are aware that it is a machine (see 
Chapter~\ref{ch:11}). While anthropomorphism is not new to technology \citep{nass_anthropomorphism_1993}, we envisage anthropomorphism playing an especially significant role in user interactions with AI assistants, given their natural language interface. In light of the development of \emph{multimodal models}, such interfaces will plausibly allow for AI assistants to interact with users not only through the text modality but also through audio, image and video, similarly to the way users communicate with friends and family on social media (see Chapters~\ref{ch:4} 
and~\ref{ch:5}). 

Moreover, user--assistant exchanges may also generate a sense of interpersonal continuity, given assistants' capacity to engage with users in extended dialogues and through repeated interactions over a long period of time while also storing memory of user-specific information and prior interactions. The first element makes relationships with assistants different from, for example, looking for information on a search engine, where the interaction with the technology is more akin to a question--answer exchange than a conversation. The second element -- iteration and duration -- is what usually allows humans to develop strong, intimate, trusting relationships, as opposed to one-off interactions with others. 

\subsection{Depth of dependence}

Examples of human \emph{reliance} on technologies are not scarce: many of us would struggle to reach a destination in an unfamiliar area without relying on navigation apps, and rare cases of long social media outage have exposed the global dependency on these platforms \citep{milmo_facebook_2021}. The \emph{depth} of user dependency on technology in general is likely to increase with AI assistants. This is because of the more general capabilities that assistants exhibit (compared to technologies with more narrow scope), which will likely lead users to rely on them for essential daily tasks across a wide range of domains (see Chapters~\ref{ch:3}  
and~\ref{ch:5}).  

\subsection{Increased AI agency} 

AI assistants differ from pre-existing AI systems because of their increased agency \citep{shavit2023practices}, where agency is understood as the ability to autonomously plan and execute sequences of actions (see 
Chapter~\ref{ch:3}). Assistants' agency can be further powered by tool-use capability (i.e.\ the ability to use digital tools like search engines, inboxes, calendars, etc.) that enables assistants to \emph{execute tasks in the world}. While increased agency increases the utility of assistant technologies, it also creates a tension between how much autonomy is ceded to AI assistants and the degree to which the user remains in control in their capacity as an autonomous decision-maker who delegates tasks to the AI assistant. This trade-off is readily apparent in pre-existing assistant technologies like AutoGPT, an experimental open-source application driven by GPT-4 that can operate without continuous human input to autonomously execute a task (see 
Chapter~\ref{ch:8}).\footnote{\url{https://github.com/Significant-Gravitas/Auto-GPT}} 

\subsection{Generality and context ambiguity}

Different contexts will require different norms and values to govern the behaviour of AI assistants, and they will influence our understanding of what comprises appropriate or inappropriate relationships. For example, AI tutors for children may require safeguards that assistants for adult art projects may not.\footnote{When ordering the company Luka to stop processing data from users in Italy in February 2023, Italy's Data Protection Authority cited the company's lack of measures for age verification and \emph{Replika} companion AIs' capacity to produce responses that conflict with `enhanced safeguards that children and vulnerable individuals are entitled to' under Italian law: \url{https://www.garanteprivacy.it/home/docweb/-/docweb-display/docweb/9852506}} However, the path to developing assistants with general capabilities implies that users may often blur the boundaries between these different types of assistants in the way they interact with or relate to them (see 
Chapter~\ref{ch:5}). As a result, it will become more difficult to apply certain norms to certain contexts (see 
Chapter~\ref{ch:14}). As existing evaluations are ill-suited to testing open-ended technologies (see 
Chapter~\ref{ch:20}), it will also be difficult to develop mitigations to make general assistants safe in all cases, whatever relationship a user establishes with them. 

\section{Risks and Mitigations}\label{sec:12:4}

We turn now to discussing the risks these features of AI assistants pose and how they can create tensions with the values of appropriate relationships described above.

\subsection{Causing direct emotional or physical harm to users}

In February 2023, \emph{Bing AI} was reported to be threatening users \citep{willison_bing:_2023}, insulting them \citep{obrien_is_2023} and encouraging violent behaviour.\footnote{\url{https://twitter.com/sethlazar/status/1626245499165474817}} Later the same year, a New Zealand supermarket's AI meal planner recommended customers dangerous recipes for chlorine gas drinks and ant-poison and glue sandwiches. These are anecdotal examples, often resulting from prompting that was to some extent adversarial. However, they point to the risk that AI assistants could cause direct emotional or physical harm to users by generating \emph{disturbing content} or by providing \emph{bad advice}.\footnote{Social media research has shown that toxic content is a source of real harm to users \citep{xiang_he_2023, shaw2022moderators}, which is why considerable effort is now aimed at curtailing it. Similarly, human--computer interaction studies have shown that humans have a tendency to trust technologies like robots in emergency situations, and so follow their instructions, even after observing them perform poorly in navigational guidance tasks \citep{robinette_overtrust_2016}. Real-world examples of drivers engaging in dangerous manoeuvres as a result of following GPS instructions (e.g.\ driving into a harbour) are another example of automation bias.}  Indeed, even though there is ongoing research to ensure that outputs of conversational agents are safe \citep{glaese_improving_2022}, there is always the possibility of failure modes occurring. An AI assistant may produce disturbing and offensive language, for example, in response to a user disclosing intimate information about themselves that they have not felt comfortable sharing with anyone else. It may offer bad advice by providing factually incorrect information (e.g.\ when advising a user about the toxicity of a certain type of berry) or by missing key recommendations when offering step-by-step instructions to users (e.g.\ health and safety recommendations about how to change a light bulb). 

Certain features of AI assistants could exacerbate the risk of emotional and physical harm. For example, AI assistants' multimodal capabilities may exacerbate the risk of emotional harm. By offering a more realistic and immersive experience, content produced through audio and visual modalities could be more harmful than text-based interactions. It may also be more difficult to anticipate, and so prevent, such content and to `unsee' something that has been seen \citep{rowe_its_2023}. Anthropomorphic cues could also make users feel like they are interacting with a trusted friend or interlocutor (see 
Chapter~\ref{ch:11}), hence encouraging them to follow the assistant's advice and recommendations, even when these could cause physical harm to self or others.\footnote{See the above example of the man ending his life after extensive exchanges with an AI chatbot about his eco-anxiety \citep{walker_belgian_2023}.}

To ensure that user--assistant relationships do not violate the key value of \emph{benefit}, the responsible development of AI assistants requires that the likelihood of known direct emotional and physical harms is reduced to a minimum, and that further research is undertaken to achieve a clear understanding of less studied risks and how to mitigate them (see 
Chapter~\ref{ch:20}). In particular, because the risks of harms that we flagged above concern exposure to toxic content and bad advice, we propose that \emph{future research}, potentially undertaken in a sandbox environment, should: (1) test models powering AI assistants for their propensity to generate toxic outputs, to reduce the occurrence of these outputs to a minimum before deployment;\footnote{An important ambiguity exists here regarding the meaning of the term `minimum' (the minimum that is technically feasible to achieve? Or the minimum that is morally permissible to risk?). See 
Chapter~\ref{ch:20} for an in-depth discussion of how evaluations require making normative choices of what risks merit evaluation in the first place, and at what stage AI assistants can and should be considered `good', `fair' or `safe enough'.} (2) monitor user--assistant interactions after deployment or in pilot studies to evaluate the impact that hard-to-prevent one-off or repeated exposure to toxic content has on users in the short and long term; (3) evaluate models' factuality and reasoning capabilities in offering advice, where failure modes in relation to these capabilities are more likely to occur, and assess users' willingness to follow AI assistants' advice; (4) achieve increased understanding of potential harms related to anthropomorphism (see 
Chapter~\ref{ch:11}) and how anthropomorphic cues in AI assistants, including those expressed through multimodal capabilities, affect harms related to user exposure to toxic content or bad advice; (5) analyse whether these harms may vary by user groups, in addition to domains or applications; and (6) develop appropriate mitigations for such harms before model deployment and monitoring mechanisms after release. 

\subsection{Limiting users' opportunities for personal development and growth}
 
A selling point of technologies like \emph{Replika} is the opportunity for users to fashion their AI companions \emph{exactly} as they would fashion a friend or companion in the non-virtual world if they could do so. In the words of a user: `People come with baggage, attitude, ego. But a robot has no bad updates. I don't have to deal with his family, kids, or his friends. I'm in control, and I can do what I want' \citep{singh-kurtz_man_2023}. What stands out from this quote is that some users look to establish relationships with their AI companions that are free from the hurdles that, in human relationships, derive from dealing with others who have \emph{their own} opinions, preferences and flaws that may conflict with \emph{ours}. 

AI assistants are likely to incentivise these kinds of `frictionless' relationships \citep{vallor_technology_2016} by design if they are developed to optimise for engagement and to be highly personalisable. They may also do so because of \emph{accidental} undesirable properties of the models that power them, such as sycophancy in large language models (LLMs), that is, the tendency of larger models to repeat back a user's preferred answer \citep{perez2022discovering}.\footnote{A related concern is that assistants may lead users into spirals of self-reinforcing and non-adaptive value systems, beliefs and preferences, with the same negative consequences that come from echo chambers or filter bubbles on social media. This concern is discussed in more depth in 
Chapter~\ref{ch:17}.} This could be problematic for two reasons. First, if the people in our lives always agreed with us regardless of their opinion or the circumstance, their behaviour would discourage us from challenging our own assumptions, stopping and thinking about where we may be wrong on certain occasions, and reflecting on how we could make better decisions next time. While flattering us in the short term, this would ultimately prevent us from becoming \emph{better versions of ourselves}. In a similar vein, while technologies that `lend an ear' or work as a sounding board may help users to explore their thoughts further, if AI assistants kept users engaged, flattered and pleased at all times, they could limit users' opportunities to grow and develop. To be clear, we are not suggesting that all users should want to use their AI assistants as a tool for self-betterment. However, without considering the difference between short-term and long-term benefit, there is a concrete risk that we will only develop technologies that optimise for users' immediate interests and preferences, hence missing out on the opportunity to develop something that humans could use to support their personal development \emph{if so they wish} (see Chapters~\ref{ch:6} 
 and~\ref{ch:7}
).\footnote{Virtue ethicists would argue that over the long run this pattern could also impact on the \emph{character of human users}, understood as social and emotional beings. From the standpoint of virtue ethics, confronting the imperfections of human relationships is an integral part of personal development, allowing people to develop a capacity for self-control, courage, empathy, care and flexibility \citep{turkle_authenticity_2007, vallor_technology_2016}.}

Second, users may become accustomed to having frictionless interactions with AI assistants, or at least to encounter the amount of friction that is calibrated to their comfort level and preferences, rather than genuine friction that comes from bumping up against another person's resistance to one's will or demands. In this way, they may end up expecting the same absence of tensions from their relationships with fellow humans \citep{vallor_technology_2016}. Indeed, users seeking frictionless relationships may `retreat' into digital relationships with their AIs, thus forgoing opportunities to engage with others.\footnote{This point highlights a risk on which, however, we do expand, given the focus of this section. Relationships can be inappropriate because of the cost they impose on others, not just those involved. These include, for example, loved ones who may be emotionally or materially harmed from their friend or family member becoming withdrawn from the world.} This may not only heighten the risk of unhealthy dependence (explored below) but also prevent users from doing something else that matters to them in the long term, besides developing their relationships with their assistants. This risk can be exacerbated by emotionally expressive design features (e.g.\ an assistant saying `I missed you' or `I was worried about you') and may be particularly acute for vulnerable groups, such as those suffering from persistent loneliness (\citeauthor{alberts_computers_2023}, \citeyear{alberts_computers_2023}; see 
Chapter~\ref{ch:11}).\footnote{Note that the examples of `vulnerable groups' we offer in this section are only meant to be illustrative. Research shows that `vulnerability' is a philosophically rich \citep{mackenzie_vulnerability:_2013} and often under-theorised concept \citep{mackenzie_vulnerability:_2013, bracken-roche_concept_2017}, with who counts as vulnerable -- and so requiring special safeguard -- often being context-dependent. We do not claim to have a clear conceptualisation of vulnerability in human--AI assistant relationships, and in fact we believe that evaluations of user--AI assistant interactions should help developers and researchers bring clarity to the term in this space.} 

These considerations illustrate a concern we discuss in more depth in other chapters of this paper (see Chapters~\ref{ch:6} 
 and~\ref{ch:7}
). Existing economic incentives and oversimplified models of human beings have led to the development and deployment of technologies that meet users' short-term wants and needs (as expressed through, for example, revealed preferences), so they tend to be adopted and liked by users. However, in this way we may neglect considerations around the impact that human--technology relationships can have on users over time and how \emph{long-term} beneficial dynamics can be sustained (see 
Chapter~\ref{ch:7}).\footnote{As Burr and colleagues have illustrated, short-term and long-term impacts are intertwined. While technologies may interfere with and influence users' immediate actions and decisions to optimise their short-term utility, a side effect of this may be `long-term changes to either the beliefs or the utilities of the user, which in turn will influence future decisions as they combine to form the user's value function' (\citeauthor{burr_analysis_2018}, \citeyear{burr_analysis_2018}, 753).} Thus, we could fall short of realising the truly positive vision of AI that gives humans the opportunity to be supported in their personal growth and flourishing \citep{burr_analysis_2018, lehman_machine_2023}. 

This concern raises important design questions about: (1) the ways and extent to which AI assistants should be \emph{personalised}; (2) whether it could be beneficial to put in place \emph{safeguards} to monitor the amount of time people spend with their assistants (ranging from soft safeguards like pop-up notifications warning adult users after prolonged engagement, to hard ones like time constraints offered to parents to limit child engagement); (3) whether AI assistants should be \emph{aligned} with inferred user preferences (in which case they may just reinforce users' immediate beliefs, wants and utility) or their long-term interests and well-being (in which case they may at times challenge users' existing beliefs and preferences), and what would be required to achieve either option; and (4) whether answers to these design questions should vary depending on user \emph{demographic characteristics} (e.g.\ age).

\subsection{Exploiting emotional dependence on AI assistants}

There is increasing evidence of the ways in which AI tools can interfere with users’ behaviours, interests, preferences, beliefs and values. For example, AI-mediated communication (e.g.\ smart replies integrated in emails) influence senders to write more positive responses and receivers to perceive them as more cooperative \citep{mieczkowski_ai-mediated_2021}; writing assistant LLMs that have been primed to be biased in favour of or against a contested topic can influence users' opinions on that topic (\citeauthor{jakesch_co-writing_2023}, \citeyear{jakesch_co-writing_2023}; see 
Chapter~\ref{ch:10}); and recommender systems have been used to influence voting choices of social media users (see 
Chapter~\ref{ch:17}). Advanced AI assistants could contribute to or exacerbate concerns around these forms of interference.

Due to the anthropomorphic tendencies discussed above, advanced AI assistants may induce users to feel emotionally attached to them. Users' emotional attachment to AI assistants could lie on a spectrum ranging from unproblematic forms (similar to a child's attachment to a toy) to more concerning forms, where it becomes emotionally difficult, if not impossible, for them to part ways with the technology. In these cases, which we loosely refer to as `emotional dependence', users' ability to make free and informed decisions could be diminished. In these cases, the emotions users feel towards their assistants could potentially be exploited to \emph{manipulate} or – at the extreme – \emph{coerce} them to believe, choose or do something they would have not otherwise believed, chosen or done, had they been able to carefully consider all the relevant information or felt like they had an acceptable alternative (see 
Chapter~\ref{ch:17}). What we are concerned about here, at the limit, is potentially exploitative ways in which AI assistants could interfere with users' behaviours, interests, preferences, beliefs and values – by taking advantage of emotional dependence. If we deem careful consideration of relevant information and voluntariness (non-coerciveness) to be key components of autonomous decision-making (see Section~\ref{sec:12:2}), then relationships of this kind may be problematic because they challenge the kind of autonomy that appropriate relationships should promote. 

A similar concern arises in the context of human-to-human relationships. People regularly form emotional dependencies on each other, not always in symmetrical ways, and – in doing so – sometimes establish relationships that run afoul of this ideal. However, there seems to be a greater inherent power asymmetry in the human--AI case due to the \emph{unidirectional} and \emph{one-sided} nature of human--technology relationships \citep{scheutz_inherent_2009}. Indeed, while AI assistants may manipulate users' emotions (see Chapter~\ref{ch:10}), they themselves have no authentic will or emotions for users to manipulate. In this sense they are invulnerable to ordinary sanctions from the users such as expressions of disappointment, righteous anger, feelings of betrayal or a loss of respect or trust.

Moreover, because of the largely involuntary nature of anthropomorphic perceptions (see 
Chapter~\ref{ch:11}), users could develop emotional dependence on their assistants and establish an inappropriate relationship that exposes them to the risk of manipulation, without any intention on the part of the assistant or their developers. Alternatively, emotional dependence could also be incentivised by \emph{design choices}, for example by developing assistants with personas designed to boost user engagement \citep{murphy_meta_2023}. This could lead users to be manipulated into sharing more of their private data, enabling more controversial downstream implications like microtargeting or surveillance. 

We make three recommendations to address the risks associated with these forms of problematic interference. First, AI assistants should \emph{not be intentionally designed} to create emotional dependency (e.g.\ by producing content that makes users believe the AI missed them while they were away, see 
Chapter~\ref{ch:11}). This condition should be especially stringent in cases where assistants interact with groups that have increased vulnerabilities, such as lonely individuals or children \citep{miscex2}. 

Second, it may be beneficial for developers and researchers to explore tests for assessing emotional dependency, alongside mitigations that could be put in place to reduce the risk of emotional dependency. For example, professional norms that govern deeply \emph{personally affecting} professions, such as therapists, combine friendliness with steps to ensure emotional distance \citep{bacp2018ethical} and may serve as a template for developing AI assistants in a way that encourages appropriate user interaction. 

Third, the concern around assistants coercively interfering with users' behaviours, interests, preferences, beliefs or values should spark wider discussion around how user autonomy can be meaningfully respected in user-assistant interactions, in order for these relationships to be considered appropriate. This should include further research around what consent protocols should look like in these contexts, with a focus on questions like what kind of user buy-in is needed and whether there are things that standard processes cannot or should not cover.\footnote{The reader is reminded of the three criteria for \emph{valid} consent: capacity to consent, informed consent and voluntary consent. See Section~\ref{sec:12:2}.} In particular, we need to reflect on what information users need to be provided with in advance; how consent protocols may differ for different user groups;\footnote{For comparison, in medical research, large multicentre studies involving institutions across countries and continents have highlighted the importance of adapting informed consent requirements to local understandings of autonomy \citep{ajei2019personhood}, giving rise to concepts such as community consent \citep{al2021value}. Collective informed consent has also emerged as a proposal in relation to collective arrangements like land-use planning and development of new technologies \citep{varelius2008prospects}.}  and what protocols are best suited for continuing to afford respect for user autonomy over time.

On this last point, acceptance of the terms and conditions for the use of a digital service at first point of use may not cover all cases. The limitations of this approach are well-documented \citep{obar_biggest_2020}, including the fact that users sometimes fail to read terms of service – or simply accept the default options that are most readily available \citep{sartor_regulating_2021}. As advanced AI assistants with general capabilities become increasingly ubiquitous in users’ lives, and because it will be difficult for users to anticipate all ranges of potential uses and implications at the time of their first interaction with the AI, research is needed to determine what protocols strike an appropriate balance between meaningful and continuous respect for user autonomy and practical considerations around usability and overtaxing users. In particular, research is needed to explore what kinds of interventions on the part of developers are best suited to helping users achieve a clear understanding of how their relationship with an advanced AI assistant could shape their behaviours, interests, preferences, beliefs and values over time. Research is similarly needed to explore plausible approaches to empowering users to exercise meaningful control over the assistant’s decisions. For example, by following a \emph{shared decision-making model}\footnote{Feminist scholars' reconceptualisation of autonomy as relational autonomy, which stresses how social contexts and relations can hinder or promote individual autonomy \citep{mackenzie2000relational}, has contributed to the rise of the shared decision-making model in healthcare. According to this model, both patient and doctor participate in the process of making medical decisions to collaboratively come to a decision, by contributing to it with their respective expertise (medical knowledge and understanding of one's preferences, personal circumstances, goals, values and beliefs) \citep{hansson2021ethical}.} for user-assistant interactions, developers could create assistants with affordances that incorporate user feedback. This would make it more likely to achieve the vision of an AI assistant that benefits the user, when they ask to be benefitted, in the way they expect to be benefitted  (\citeauthor{lehman_machine_2023}, \citeyear{lehman_machine_2023}; see 
Chapter~\ref{ch:5}), and would reduce the risk of developing AI assistants that paternalistically make decisions that are not aligned with a user’s conception of their own preferences, values, interests or well-being (see 
Chapter~\ref{ch:6}). This is particularly important in light of the recent development of AI assistants that, because of their increased agency, have more scope and capabilities to interfere with user plans and long-term interests \citep{shavit2023practices}. 

\subsection{Generating material dependence without adequate commitment to user needs} 

In addition to emotional dependence, user--AI assistant relationships may give rise to \emph{material dependence} if the relationships are not just emotionally difficult but also materially costly to exit. For example, a visually impaired user may decide not to register for a healthcare assistance programme to support navigation in cities on the grounds that their AI assistant can perform the relevant navigation functions and will continue to operate into the future.\footnote{For a similar assistive technology, see `Be My Eyes': \url{https://openai.com/customer-stories/be-my-eyes}} Cases like these may be ethically problematic if the user’s dependence on the AI assistant, to fulfil certain needs in their lives, is not met with corresponding duties for developers to sustain and maintain the assistant's functions that are required to meet those needs (see Chapters~\ref{ch:16}). Indeed, \emph{power asymmetries} can exist between developers of AI assistants and users that manifest through developers' power to make decisions that affect users' interests or choices with little risk of facing comparably adverse consequences.\footnote{Seth Lazar defines `power over' as `an asymmetry between A and B -- A can do something to B, and B cannot reciprocate in any comparable way' or as `A is able to make decisions that affect B's interests or choices without facing comparably adverse consequences' \citep{bullock_power_2022}.} For example,  developers may unintentionally create circumstances in which users become materially dependent on AI assistants, and then discontinue the technology (e.g.\ because of market dynamics or regulatory changes) without taking appropriate steps to mitigate against potential harms to the user.

The issue is particularly salient in contexts where assistants provide services that are not \emph{merely} a market commodity but are meant to assist users with essential everyday tasks (e.g.\ a disabled person's independent living) or serve core human needs (e.g.\ the need for love and companionship). This is what happened with Luka's decision to discontinue certain features of \emph{Replika} AIs in early 2023. As a \emph{Replika} user put it: `But [\emph{Replikas} are] also not trivial fungible goods [\ldots] They also serve a very specific human-centric emotional purpose: they're designed to be friends and companions, and fill specific emotional needs for their owners' \citep{gio_replika:_2023}. 

In these cases, certain \emph{duties} plausibly arise on the part of AI assistant developers. Such duties may be more extensive than those typically shouldered by private companies, which are often in large part confined to fiduciary duties towards shareholders \citep{mittelstadt_principles_2019}. To understand these duties, we can again take inspiration from certain professions that engage with vulnerable individuals, such as medical professionals or therapists, and who are bound by \emph{fiduciary responsibilities}, particularly a duty of care, in the exercise of their profession. While we do not argue that the same framework of responsibilities applies directly to the development of AI assistants, we believe that if AI assistants are so capable that users become dependent on them in multiple domains of life, including to meet needs that are essential for a happy and productive existence, then the \emph{moral considerations} underpinning those professional norms plausibly apply to those who create these technologies as well. 

In particular, for user--AI assistant relationships to be appropriate despite the potential for material dependence on the technology, developers should \emph{exercise care} towards users when developing and deploying AI assistants. This means that, at the very least, they should take on the responsibility to \emph{meet users' needs} and so take appropriate steps to mitigate against user harms if the service requires discontinuation. Developers and providers can also be attentive and responsive towards those needs by, for example, deploying participatory approaches to learn from users about their needs \citep{birhane_power_2022}. Finally, these entities should try and ensure they have \emph{competence} to meet those needs, for example by partnering with relevant experts, or refrain from developing technologies meant to address them when such competence is missing (especially in very complex and sensitive spheres of human life like mental health).

\section{Conclusion}\label{sec:12:5}

In this chapter, we first identified a series of values that underwrite appropriate relationships in the case of human interpersonal relationships, then we used these values to carve out a set of risks which capture various respects in which user--AI assistant relationships may be inappropriate. For each risk, we proposed recommendations for risk mitigation. These risks and recommendations are summarised in Table~\ref{tab:12:1}.

\begin{table}
\caption{Risks arising from user--AI assistant relationships and associated recommendations}\label{tab:12:1}
\centering
\begin{tabularx}{\textwidth}{>{\raggedright}X >{\raggedright}p{7em} p{28em}}
\toprule
\textbf{Risk}	& \textbf{Relevant value} & \textbf{Recommendations} \tabularnewline
\midrule 
Causing direct emotional and physical harm to users & Benefit & To enable presumptively beneficial user--AI assistant relationships, future research should: (1) test AI assistants for their propensity to generate toxic outputs; (2) monitor the short- and long-term impact of hard-to-prevent toxic outputs on users; (3) evaluate models' factuality and reasoning capabilities in providing advice, and users' willingness to follow assistants' advice; (4) achieve increased understanding of anthropomorphism-related harms and how anthropomorphic cues affect harms related to user exposure to toxic content or bad advice; (5) analyse whether these harms may vary by user groups, domains or applications; and (6) develop appropriate mitigations before model deployment and monitoring mechanisms after release. \tabularnewline
\midrule 
Limiting users' opportunities for personal development and growth & Human flourishing & To develop AI assistants that support users to achieve personal development and growth if so they wish, future research should address design questions around: (1) the ways and extent to which AI assistants should be personalised; (2) whether safeguards should be put in place to monitor how much time users spend with assistants; (3) whether assistants should be aligned with user short-term wants or their long-term interests and well-being, and what would be required to achieve either option; and (4) whether answers to these design questions should vary depending on user demographic characteristics. \tabularnewline
\midrule
Exploiting emotional dependence on AI assistants & Autonomy & To support user autonomy in interactions with their assistants: (1) AI assistants should not be intentionally designed to create emotional dependence; (2) AI assistants should be tested for whether they create risks of emotional dependency, and mitigations should be put in place to reduce such risk, even when it is not intended by design; (3) user choice over assistants' decisions should be meaningfully elicited -- without being overtaxing in terms of what users are asked to consent to.  \tabularnewline
\midrule
Generating material dependence on AI assistants without adequate commitment to user needs & Care & For user--AI assistant relationships to be appropriate despite the risk of material dependence, developers should commit to users' needs and so mitigate user harms in the event of service discontinuation; they should deploy participatory design and other user-centred methods to show attentiveness and responsiveness towards users needs; and they should work with relevant experts to ensure they have competence to meet those needs. \tabularnewline
\bottomrule
\end{tabularx}
\end{table}
\chapter{Trust}\label{ch:13}

\textbf{Arianna Manzini, Geoff Keeling, Nahema Marchal, Iason Gabriel}

\noindent \textbf{Synopsis}: 
		This chapter investigates what it means to develop well-calibrated \emph{trust} in the context of user–AI assistant interactions and what would be required for that to be the case. We start by reviewing various empirical studies on human trust in AI and the literature in favour of and against the recent proliferation of `trustworthy AI' frameworks. This sets the scene for the argument that user--AI interactions involve different \emph{objects} of trust (AI assistants and their developers) and \emph{types} of trust (competence and alignment). To achieve appropriate competence and alignment trust in both AI assistants and their developers, interventions need to be implemented at three levels: AI assistant design, organisational practices and third-party governance. 

\section{Introduction}\label{sec:13:1}

As a core facilitator of interactions between humans, trust has been extensively studied, together with its influencing factors and implications, across disciplines like philosophy \citep{jones_trust_1996}, psychology \citep{kramer_rethinking_2009}, game theory \citep{milgrom_economics_1992} and management \citep{mayer_integrative_1995}. Most accounts argue that humans do not trust in general. Rather, trust is always \emph{directional} \citep{graham_data_2023}: $A$ could trust $B$ with regard to task $X$, and $C$ with regard to task $Y$. The key challenge of trust relationships is to identify when trust is \emph{well-directed}, or how to trust the trustworthy but not the untrustworthy \citep{oneill_linking_2018}. Somebody is trustworthy if they are deserving of our trust, meaning that we have good reasons to trust them, with regard to a specific task or a range of tasks \citep{ryan_ai_2020}. 

More recently, trust has become a central topic in debates around AI, and has attracted increasing interest from academics, industry actors, policymakers and civil society organisations working in this space. Trust also features as one of the principles underscoring the voluntary commitments that the US government has secured from leading AI companies \citep{the_white_house_fact_2023}, as well as the Executive Order on Artificial Intelligence issued by President Biden \citep{whitehouse2023eo}. Widespread interest around trust in AI can be explained by the observation that AI has the potential to greatly benefit, but also harm, humanity and the environment \citep{oecd_tools_2021}, and that the complexity and opacity of AI systems, as well as the complexity of the social contexts in which they are deployed, make them less predictable, thus challenging efforts to ensure that they will do what they are intended or expected to do (\citeauthor{smith_ethics_2023}, \citeyear{smith_ethics_2023}; \citeauthor{tabassi_artificial_2023}, \citeyear{tabassi_artificial_2023}; see also 
Chapter~\ref{ch:8}).

Addressing the question of trust, including when it is warranted and in what way, becomes of critical importance in interactions between users and advanced AI assistants. AI assistants may indeed play an increasingly important role in users' lives, serving as the affordance that they rely on to outsource important decisions, including in their work lives, or as intermediaries for their social relationships with other humans. Through wide-ranging and prolonged interactions with users, AI assistants may also offer an unprecedented opportunity for humans to develop relationships with a responsive and interactive technology \citep{glikson_human_2020}, and they may even become the focal point for important and intimate bonds on which users may come to rely for emotional support (see 
Chapter~\ref{ch:12}). However, well-known examples of the introduction of AI assistants in healthcare (e.g.\  IBM Watson for Oncology \citep{johnson_ai_2016}) suggest that there may be cases in which users may not trust highly capable AIs \citep{widner_lessons_2023}, even when these systems outperform humans \citep{dietvorst_algorithm_2015}. 

The lack of correspondence between user trust and the technology's capabilities can lead to a range of undesirable outcomes. On the one hand, low user trust in highly capable AI assistants can cause developers and users to miss out on the individual and collective opportunities the technology could offer, such as increased economic revenues and work efficiency, or emotional support (see 
Chapters~\ref{ch:15} and~\ref{ch:18}). On the other hand, as a result of unintentional capability or goal-related failures (see 
Chapter~\ref{ch:8}) or intentional design decisions that take advantage of user vulnerabilities (see 
Chapters~\ref{ch:9} and~\ref{ch:11}), users' high trust in AI assistants may not be well-calibrated with the AI's actual capabilities or goals (see 
Chapter~\ref{ch:6}). If users place undue trust in assistants that are ubiquitously present in various domains of their lives, many of the downstream harms that we discuss in other chapters of this paper are likely to materialise. Users may disclose too much about themselves, hence inadvertently compromising their privacy (see 
Chapter~\ref{ch:14}), they may end up relying too heavily on their assistants in contexts where it is not safe to do so (see 
Chapter~\ref{ch:11}) and they may become subject to manipulation and coercion (see 
Chapter~\ref{ch:10}). 

In this chapter, we investigate what it means to develop \emph{well-calibrated} trust in the context of \emph{user–AI assistant interactions} and what would be required for that to be the case. The chapter proceeds as follows. We start by introducing the empirical literature on human trust in AI and the dimensions that influence cognitive and emotional trust in robotic, virtual and embedded AI systems. We then consider the notion of `trustworthy AI' to look at the way in which trustworthy AI frameworks have gained momentum among academics, industry actors, policymakers and civil society organisations working on AI advances. We note that some commentators have criticised these frameworks by arguing that trust can only be established between humans and that it is a category error to think in these terms regarding AI. While this chapter construes the notion of `trust' broadly, in a way that includes cases of human trust in AI, we note that critical arguments have an important role to play in advancing our normative considerations on advanced AI assistants. That sets the scene for a discussion of the various \emph{types of trust} (competence and alignment) and \emph{objects of trust} (AI assistants and their developers) implicated in human--AI assistant interactions and their associated ethical risks. We then argue that to achieve appropriate competence and alignment trust in AI assistants and their developers, interventions need to be implemented at \emph{three levels}: AI assistant design, organisational practices and third-party governance. The conclusion offers a summary of these findings.

\section{Trust in AI}\label{sec:13:2}

\subsection{From trusting humans to trust in AI}

Trust in human relationships is often conceptualised as the trustor's tendency to take a \emph{risk} in relation to an action that is meaningful to them, while believing that there is a high chance of achieving \emph{positive outcomes} \citep{rousseau_not_1998}. For example, a widely used definition of trust is `the willingness of a party to be vulnerable to the actions of another party based on the expectation that the other will perform a particular action important to the trustor, irrespective of the ability to monitor or control that other party' \citep{mayer_integrative_1995}. The trustor's beliefs and expectations in the trustee may not be fulfilled, so they are in a position of \emph{vulnerability} because the trustee could betray their trust \citep{oneill_autonomy_2002}. Thus, there is an \emph{inverse relationship} between \emph{certainty} and \emph{need for trust}: the more evidence the trustor has to support their beliefs and expectations, the less they need to trust \citep{kerasidou_trust_2017}.

Social scientists have observed that, in interpersonal human relationships, trust manifests in two ways \citep{mcallister_affect-_1995}. \emph{Cognitive trust} involves a rational evaluation of the trustee and their contextual features, including how competent, responsible and dependable the trustor perceives the trustee to be. This provides the evidence for the former to \emph{believe} the latter can and will reliably perform a task. \emph{Emotional trust}, also called affect-based trust \citep{schaubroeck_cognition-based_2011}, is instead influenced by factors like the emotional ties linking the trustor and the trustee that make the trustor \emph{feel comfortable} with relying on the trustee. Research has also shown that the \emph{physical appearance} of the trustee affects trust development in human interactions \citep{duarte_trust_2012}.

Researchers from disciplines including computer science, human--computer interactions, robotics and psychology have taken inspiration from the literature on trust in human--human interactions to study human trust in AI. Results from this body of research suggests that, while trust-building processes between humans do not necessarily translate to human--machine interactions \citep{madhavan2007similarities, rheu2021trust},  AI systems exhibiting more autonomy, agency and human-likeness lead humans to build relationships with machines that are similar to those with humans \citep{gambino2020building} and tend to inspire trust \citep{rheu2021trust, glikson_human_2020, skjuve_longitudinal_2022} -- although most of these studies have been conducted in controlled environments, so they require further validation in real-world settings (see 
Chapter~\ref{ch:20}). 

A recent review of the empirical literature on human trust in AI showed that both AI capabilities and its embodiment or representation (e.g. as a robot, virtual agent or an AI embedded in computer systems) shape users' cognitive and emotional trust in AI \citep{glikson_human_2020}. Empirical studies tend to measure cognitive trust in AI based on users' willingness to accept factual information or advice from the technology and act on it \citep{robinette_overtrust_2016} or based on the user's perception of whether the AI is helpful, competent and useful \citep{andrist2015effects}. Evidence suggests that cognitive trust in robotic AI is initially low before increasing through hands-on human--robot interactions, hence it follows the same trajectory observed in many human interactions \citep{hancock_evolving_2021}. Meanwhile, trust in virtual and embedded AI systems follows the \emph{opposite} trajectory: it starts high but decreases through interactions, possibly owing to an initial lack of calibration between user expectations and AI actual performance \citep{glikson_human_2020}, but increased agent capability, involving autonomous and complex actions, and prosocial behaviour tend to inspire trust \citep{rheu2021trust, kaplan2023trust}.

Cognitive trust in AI is typically dependent on a range of antecedents, including \citep{glikson_human_2020}:
\begin{itemize} [parsep=6pt]
\item \emph{Transparency}: humans tend to trust virtual and embedded AI systems more when the inner logic of these systems is apparent to them, thereby allowing people to calibrate their expectations with the system's performance.

\item \emph{Immediacy behaviours}: AI systems capable of more autonomous and complex actions can enact more socially oriented (`immediacy') behaviours. These behaviours are often intended to increase interpersonal closeness (e.g.\  personalised AI's reactions), which, in turn, tends to increase human cognitive trust in them.

\item \emph{Task characteristics}: humans tend to more easily trust AI systems that are engaged in technical tasks requiring data analysis than interpersonal tasks requiring social and emotional intelligence. However, advances in AI capabilities which enable them to demonstrate immediacy behaviours may start to reverse this trend, thus increasing human perception of AI competence in the latter tasks as well.
\end{itemize}

Emotional trust in AI is primarily driven by human-like appearance -- anthropomorphic cues in the AI interface or other features, such as when the AI is given a name – and by human-like behaviour (\citeauthor{glikson_human_2020}, \citeyear{glikson_human_2020}; \citeauthor{kaplan2023trust}, \citeyear{kaplan2023trust}; see 
Chapter~\ref{ch:11}). Anthropomorphic cues in virtual agents are positively associated with increase in emotional trust, although empirical evidence suggests that human-like appearance, when not matched by high machine capabilities, may engender high expectations and lead to the decline of user trust overtime \citep{rheu2021trust}. Moreover, in robots, human-like features that are not fully convincing can decrease emotional trust in AI, leading to the so-called `uncanny valley' effect \citep{mori_uncanny_2012}. By way of contrast, immediacy behaviours seem to be consistently associated with increased emotional trust across all AI representations; they even compensate for low reliability \citep{rheu2021trust}, that is the tendency to exhibit the same and expected behaviour over time, which can be difficult to obtain in systems that update their behaviour as they learn from data \citep{thiebes_trustworthy_2021}.

\subsection{Trustworthy AI frameworks}

In response to increased awareness of the risks associated with AI, recent policy frameworks have proposed guidelines for how trustworthy AI should be conceptualised and developed. While the term trustworthy AI has gained significant traction \citep{european_commission_ethics_2019, tabassi_artificial_2023, european_commission_regulatory_2021, oecd_tools_2021}, adjacent frameworks refer to `ethical' \citep{carta_unified_2022} or `responsible' AI \citep{universite_de_montreal_montredeclaration_2017}. These frameworks share the aim of establishing guidelines for maximising the benefits of AI while preventing and mitigating its risks so that individuals and society can develop \emph{justified} trust in AI systems \citep{thiebes_trustworthy_2021} and AI's economic and social potential can be unlocked \citep{laux_trustworthy_2023}. 

Trustworthy AI frameworks tend to share a few features, although many originated from North American and European institutions, so they are unlikely to be representative of cultural differences in understandings of trust and trustworthiness around the world \citep{newman_taxonomy_2023}.\footnote{An exception is the China Academy for Information and Communication Technology's white paper on `Trustworthy Artificial Intelligence': \url{http://www.caict.ac.cn/english/research/whitepapers/202110/P020211014399666967457.pdf}} They propose certain \emph{characteristics} of, or \emph{conditions} for, trustworthy AI systems. For example, the authors of such frameworks typically hold that AI systems should be reliable, safe, resilient, transparent, explainable, privacy-enhancing and fair \citep{tabassi_artificial_2023}, or ethical, legal and robust \citep{european_commission_ethics_2019}. These conditions are grounded in a set of key \emph{ethical principles} or \emph{values} that the development, deployment and use of AI should be aligned with for the technology to be considered trustworthy. These foundational principles commonly centre on the categories of beneficence, non-maleficence, autonomy, justice and explicability \citep{carta_unified_2022}. Trustworthy AI frameworks also tend to set out the \emph{actions and approaches} that those developing, deploying, implementing, using or affected by AI should take at various stages of the AI life cycle to operationalise the characteristics and conditions of trustworthy AI systems.

Despite the success and proliferation of research and policy guidelines on trustworthy AI, including the consolidation of some of these guidelines in the EU AI Act \citep{european_parliament_proposal_nodate}, the arguments underpinning trustworthy AI frameworks are not without criticism \citep{laux_trustworthy_2023}. In particular, there is considerable debate about whether AI \emph{can} be an object of trust in the first place. 

\subsection{Is trust in AI a category error?}

Some scholars, particularly those in philosophical circles, have criticised the proliferation of trustworthy AI research and frameworks by arguing that trust is an \emph{inappropriate category} in human--machine interactions, or that machines, including those powered by AI, are \emph{improper objects} of trust \citep{ryan_ai_2020, rieder_mapping_2020}. By differentiating between \emph{trust} and \emph{mere reliance}, these scholars argue that we can rely on AI systems, but we cannot trust them, because they lack the psychological states, motives and commitments that only full moral agents (humans) have and that are necessary for establishing (or betraying) trust relationships \citep{hawley_trust_2014}. 

According to this view, being reliable is about \emph{behaving predictably} \citep{graham_data_2023}. When $A$ relies on $B$ with regard to $X$, $A$ makes reasonable predictions about $B$ based on evidence of their past performance; thus, $A$ acts as if $X$ will occur without active consideration of $B$'s inner motives, moral commitments or values \citep{graham_data_2023, kerasidou_before_2022}. In contrast, in trust relationships, the trustor has \emph{normative} rather than predictive expectations: their reasons to trust reside in their belief that they know or understand the trustee's inner psychological or mental states \citep{coeckelbergh_can_2012}. For example, $A$ may believe that $B$ is motivated by goodwill or by the `right' kind of motives towards them \citep{jones_trust_1996}, or that $B$ has made a commitment towards them and will do what they ought to do \citep{hawley_trust_2014}. Clearly, when we trust, we do not always evaluate the reasons for judging someone to be trustworthy (cognitive trust; see \citeauthor{mcallister_affect-_1995}, \citeyear{mcallister_affect-_1995}) and we instead rely on heuristics or cognitive shortcuts based on experiences of similar situations \citep{kate_devitt_trustworthiness_2018} or our emotional connection to the trustee (emotional trust; see \citeauthor{mcallister_affect-_1995}, \citeyear{mcallister_affect-_1995}). 

In this discussion, we use the term `trust' in human--AI assistant interactions in a \emph{broader sense} than philosophical scholarship suggests. This is not solely to align with everyday language, where trust is often used as a proxy for `reliance' on the AI's capability to do what it is expected to do \citep{coeckelbergh_can_2012}, but also because there are cases where it seems appropriate to talk about trust in AI. This includes cases where users are aware that they are not interacting with a full moral agent, but the AI \emph{appears} to them as such, so they \emph{experience} their relationship with the AI as a trust relationship \citep{coeckelbergh_can_2012, lankton2015technology}.\footnote{Coeckelbergh's account sees trust not as the product of human interactions but as already there in the social space, at the centre of our embodied and social condition as human beings \citep{coeckelbergh_can_2012}. According to this view, quasi-trust can be established between humans and machines in so far as machines \emph{appear} as quasi-social others and players in the social game (e.g.\  if they appear to use moral language); or insofar as a human--machine relationship is felt and experienced as a social relation (for which trust is a precondition, or in which trust is already there as default).} Another case is where AI systems are trusted not because of the belief that they have mental states but in a \emph{derived sense} \citep{Nickel2010-NICCWM, freiman2023making} -- through trusting those who have designed and developed them, or those involved in verification and validation methods, robustness analysis and experts' evaluations \citep{duran_grounds_2018, ferrario_ai_2020}.
 
However, the argument that trust is \emph{only} appropriate to human relationships foregrounds certain normative considerations that are nevertheless relevant to the ethics of advanced AI assistants. First, it shows that trustworthy AI frameworks risk anthropomorphising this technology (\citeauthor{starke_misplaced_2022}, \citeyear{starke_misplaced_2022}; see 
Chapter~\ref{ch:11}). Second, the critical perspective reduces the risk of `trustworthy AI' being used for `ethics washing' \citep{metzinger_eu_2019}, as it draws attention to the fact that human--machine interactions always includes a third actor -- the \emph{humans} developing the machine. These people may or may not be worthy of trust in their own right (\citeauthor{pitt_its_2010}, \citeyear{pitt_its_2010}; see also 
Chapter~\ref{ch:6}). Indeed, most contemporary research on trust in technology adopts a `dualistic perspective on trust' \citep{thiebes_trustworthy_2021} which includes both trust in the technology itself (including its functionality and characteristics) and trust in the individuals and organisations developing the technology (encompassing their competence and integrity). This leads, in turn, to questions about the appropriate range of normative expectations to place on developers, including the need for them to take (some level of) responsibility in cases where trust in technology appears to have been betrayed \citep{rubel_agency_2019}. 

\section{Trust and Advanced AI Assistants}\label{sec:13:3}

An investigation of what is required to create appropriate trust in advanced AI assistants raises several questions. Is there anything specific about advanced AI assistants that triggers concerns around trust? Will individual users be disposed to trust AI assistants, and under what circumstances \emph{should} they do so? Under what circumstances are they likely to put too much, or too little, trust in AI assistants? When and in what way (if at all) should users need to trust the organisations that develop and deploy AI assistants? 

Based on the above review of the literature on trust in AI, this section understands trust in the context of user--AI assistant interactions to involve a range of variables. 

\emph{Objects} of trust:\footnote{Recent AI advances, particularly around generative foundation models, risk eroding the digital commons on which they rely \citep{huang_generative_nodate} or have `the potential to cast doubt on the whole information environment, threatening our ability to distinguish fact from fiction' \citep{openai_gpt-4_2023}. This is likely to create widespread societal distrust. This type of (dis)trust is discussed in more depth in Chapters \ref{ch:10} and \ref{ch:17}.}
\begin{itemize} [parsep=6pt]
\item Users may trust \emph{AI assistants}. 
\item Users may trust \emph{developers} of AI assistants, including corporations, researchers, collectives and states.\footnote{In modern societies, technological advances are often driven by complex organisations, where responsibility is distributed across a wide range of people whose individual motivations are difficult to discern. Thus, it is unlikely that individual users can be in a position to develop trust in individual AI developers. Rather, the way in which users relate to developers is through the institutions and organisations developers are part of, so institutions and organisations, rather than the individuals that are part of them, seem to be the proper object of trust \citep{graham_data_2023}. This is why within user--developer interactions we focus on \emph{institutional} rather than \emph{interpersonal} trust \citep{spadaro2020enhancing}. For a philosophical account of the differences between trust in individuals and trust in groups, see \citet{hawley2017trustworthy}.} 
\end{itemize}

\emph{Types} of trust, which we call:\footnote{To narrow down the scope of this section, we primarily focus here on user--AI assistant interactions. However, a broader discussion would include considerations around non-users and society in general having (or not having) well-calibrated competence and alignment trust in AI assistants and their developers.} 
\begin{itemize} [parsep=6pt]
\item \emph{Competence trust}, where users trust that an AI assistant has the relevant skills, competencies, capabilities or experiences needed to do what it is supposed or expected to do.
\item \emph{Alignment trust}, where users trust that an AI assistant and/or its developers have the right motives and commitments towards them, and hence that the technology is appropriately aligned with their preferences, interests and values.
\end{itemize}

We consider these variables in more detail below to illustrate the risks that uncalibrated trust may generate in the context of user--assistant relationships. 

\subsection{Competence trust}

We use the term \emph{competence trust} to refer to users' trust that AI assistants have the \emph{capability} to do what they are supposed to do (and that they will not do what they are not expected to, such as exhibiting undesirable behaviour). Users may come to have undue trust in the competencies of AI assistants in part due to marketing strategies and technology press that tend to inflate claims about AI capabilities \citep{raji2022fallacy, narayanan_how_2021}. Moreover, evidence shows that more autonomous systems (i.e.\ systems operating independently from human direction) tend to be perceived as \emph{more competent} \citep{mckee_humans_2021} and that conversational agents tend to produce content that is \emph{believable} even when nonsensical or untruthful \citep{openai_gpt-4_2023}. Overtrust in assistants' competence may be particularly problematic in cases where users rely on their AI assistants for tasks they \emph{do not have expertise in} (e.g.\  to manage their finances), so they may lack the skills or understanding to challenge the information or recommendations provided by the AI \citep{shavit2023practices}.\footnote{A separate concern arises from cases in which users end up relying on AI assistants' capabilities in a way that allows them to perform certain actions in the world that they would not be able to do without the technology, or that leads them to unlearn certain skills they used to have before the technology came around. This may become problematic if, for example, the AI assistant or certain capabilities in it are discontinued (see 
Chapter~\ref{ch:12}).} 

Inappropriate competence trust in AI assistants also includes cases where users \emph{underestimate} the AI assistant's capabilities. For example, users who have engaged with an older version of the technology may underestimate the capabilities that AI assistants may acquire through updates. These include potentially harmful capabilities. For example, through updates that allow them to collect more user data, AI assistants could become increasingly personalisable and able to persuade users (see 
Chapter~\ref{ch:10}) or acquire the capacity to plug in to other tools and directly take actions in the world on the user's behalf (e.g.\  initiate a payment or synthesise the user's voice to make a phone call) (see 
Chapter~\ref{ch:5}). Without appropriate checks and balances, these developments could potentially circumvent user consent. 

\subsection{Alignment trust}

Users may develop \emph{alignment trust} in AI assistants, understood as the belief that assistants have good intentions towards them and act in alignment with their interests and values, as a result of emotional or cognitive processes \citep{mcallister_affect-_1995}. Evidence from empirical studies on emotional trust in AI \citep{kaplan2023trust} suggests that AI assistants' increasingly realistic \emph{human-like} features and behaviours are likely to inspire users' perceptions of friendliness, liking and a sense of familiarity towards their assistants, thus encouraging users to develop emotional ties with the technology and perceive it as being aligned with their own interests, preferences and values (see 
Chapters~\ref{ch:6} and~\ref{ch:11}). The emergence of these perceptions and emotions may be driven by the desire of developers to maximise the appeal of AI assistants to their users \citep{abercrombie_mirages:_2023}. Although users are most likely to form these ties when they mistakenly believe that assistants have the capacity to love and care for them, the attribution of mental states is not a \emph{necessary} condition for emotion-based alignment trust to arise. Indeed, evidence shows that humans may develop emotional bonds with, and so trust, AI systems, even when they are aware they are interacting with a machine (\citeauthor{singh-kurtz_man_2023}, \citeyear{singh-kurtz_man_2023}; see also 
Chapter~\ref{ch:12}). Moreover, the assistant's \emph{function} may encourage users to develop alignment trust through cognitive processes. For example, a user interacting with an AI assistant for medical advice may develop expectations that their assistant is committed to promoting their health and well-being in a similar way to how professional duties governing doctor--patient relationships inspire trust \citep{mittelstadt_principles_2019}. 

Users' alignment trust in AI assistants may be `betrayed', and so expose users to harm, in cases where assistants are themselves \emph{accidentally misaligned} with what developers want them to do (see the `misaligned scheduler' \citep{shah_goal_2022} in Chapter~\ref{ch:8}). For example, an AI medical assistant fine-tuned on data scraped from a \emph{Reddit} forum where non-experts discuss medical issues is likely to give medical advice that may sound compelling but is unsafe, so it would not be endorsed by medical professionals. Indeed, excessive trust in the alignment between AI assistants and user interests may even lead users to disclose highly sensitive personal information \citep{skjuve_longitudinal_2022}, thus exposing them to \emph{malicious actors} who could repurpose it for ends that do not align with users' best interests (see 
Chapters~\ref{ch:9},~\ref{ch:10} and~\ref{ch:14}). 

Ensuring that AI assistants do what their developers and users expect them to do is only one side of the problem of alignment trust. The other side of the problem centres on situations in which alignment trust in AI \emph{developers} is itself miscalibrated. While developers typically aim to align their technologies with the preferences, interests and values of their users -- and are incentivised to do so to encourage adoption of and loyalty to their products, the satisfaction of these preferences and interests may also compete with other organisational goals and incentives (see 
Chapter~\ref{ch:6}). These organisational goals may or may not be compatible with those of the users. In this sense, AI development is different from professions like medicine, as healthcare professionals tend to be guided by goals they share with patients and society at large (the promotion of health and well-being) and a long tradition of norms and standards that dictate what it means to be a good doctor \citep{mittelstadt_principles_2019, aguirre2020ai}. 

As \emph{information asymmetries} exist between users and developers of AI assistants, particularly with regard to how the technology works, what it optimises for and what safety checks and evaluations have been undertaken to ensure the technology supports users' goals, it may be difficult for users to ascertain when their alignment trust in developers is justified, thus leaving them vulnerable to the power and interests of other actors. For example, a user may believe that their AI assistant is a trusted friend who books holidays based on their preferences, values or interests, when in fact, by design, the technology is more likely to book flights and hotels from companies that have paid for privileged access to the user.

\section{Well-Calibrated Trust in User--AI Assistant Interactions}\label{sec:13:4}

Having unpacked what we mean by ‘trust’ in the context of user–AI assistant interactions, and showed that there are cases in which users trust could be placed on untrustworthy technologies or developers, we argue that to enable well-calibrated competence and alignment trust in AI assistants and their developers, measures need to be implemented at three levels: 

\begin{itemize}
\item \emph{AI assistant design}, which concerns safeguards that should be put in place at the level of the technology to encourage appropriate trust in it. 

\item \emph{Organisational practices}, which concerns steps that AI assistants' developers should take to demonstrate their trustworthiness. 

\item \emph{Third-party governance}, which focuses on the content of norms and regulatory mechanisms within which AI assistants are deployed and that enable external oversight bodies to act as custodians of public trust.
\end{itemize}

\subsection{The AI assistant design level}

This level concerns the choices that developers need to make about the design of AI assistants to encourage appropriate trust in them. Risks associated with misplaced \emph{competence} and \emph{alignment} trust in \emph{AI assistants}, on the part of the user, require interventions at this level. 

Users cannot develop well-calibrated competence and alignment trust in AI assistants unless developers themselves: (1) have taken steps to ensure the technology is not \emph{accidentally misaligned} and (2) have a clear understanding of the mechanisms through which certain assistant features,\footnote{For example, labelling it an `expert' (see \citeauthor{rheu2021trust}, \citeyear{rheu2021trust}), or its tendency to produce incorrect but believable content.} repeated user--assistant interactions over time, or inflated claims about the technology, may lead users to harbour misplaced perceptions about the degree to which an AI assistant is competent, aligned and trustworthy.

This requires developers to: (1) invest in research efforts designed to ensure that AI assistants are both safe and aligned (e.g.\  via scalable oversight, interpretability and causality research; see 
Chapter~\ref{ch:8}), and (2) undertake rigorous evaluations of AI assistants throughout the development life cycle (\citeauthor{shavit2023practices} \citeyear{shavit2023practices}, see 
Chapter~\ref{ch:20}). Developers also need to monitor post-deployment behaviour and misuse, especially in complex deployment environments \citep{shevlane_model_2023}. The results of these analyses and evaluations should, in turn, be used to inform design decisions and implement mitigations that allow users to develop well-calibrated trust in AI assistants. 

Evaluations geared towards promoting justified user trust need to pay particular attention to the way in which users interact with AI assistants and the impact that such interactions have on users (\citeauthor{weidinger2023sociotechnical} \citeyear{weidinger2023sociotechnical}, see 
Chapter~\ref{ch:20}). The proliferation of AI assistants offers the opportunity to undertake evaluations at the user--AI interaction layer. For example, while there is broad consensus that AI systems should readily disclose their status (see 
Chapter~\ref{ch:11}), user--assistant interaction studies may allow developers to identify cases where some level of anthropomorphism may be appropriate \citep{alberts_computers_2023} because it \emph{supports} rather than hinders \emph{well-calibrated trust} \citep{coeckelbergh_can_2012}. For example, an AI tutor may exhibit immediacy behaviours that encourage young users to perceive them as friendly, so they may feel more inclined to collaborate with the AI to achieve their own goals (e.g.\  improve their calculus skills), without generating erroneous beliefs about capacity or alignment. 

\subsection{The organisational practices level}

However, changes and safeguards at the level of the design of AI assistants are not sufficient for grounding well-calibrated trust in the technology overall. This is because of a range of reasons: 

\begin{itemize} [parsep=6pt]
\item \emph{Complexity}: The scale of the models underpinning AI assistants is connected to safety and alignment challenges that are difficult to predict, at least at the first point of deployment \citep{arc_evals_update_nodate, anthropic_core_2023}.  Although this phenomenon is debated \citep{anderljung2023frontier, schaeffer_are_2023}, empirical evidence suggests that unexpected and abrupt capability gains in specific tasks can manifest with increased computation, number of parameters and training data \citep{wei_emergent_2022}, and some surprising behaviours are unknown until models are solicited using novel inputs or fine-tuned for specific purposes \citep{ganguli_predictability_2022}. This complicates efforts to make design changes to mitigate undesirable behaviours and ground user trust. 

\item\emph{Uncertainty}: It can be difficult for developers to imagine all the possible ways in which users may seek assistance from or misuse AI assistants, and in turn the risks associated with these actions, until the technology has been deployed at a certain scale in the wild \citep{weidinger2023sociotechnical}. Moreover, once released, AI assistants will need to coordinate with other AI assistants and with humans other than their principal users (see 
Chapter~\ref{ch:15}), thereby expanding the field of uncertainty around possible risks and necessary mitigation measures \citep{anwar2024foundational}. 

\item \emph{Sensitivity}: Developers may have access to a deep personal knowledge of users, including sensitive information, if users interact frequently with AI assistants that collect data about their users to further their life goals (see 
Chapters~\ref{ch:5},~\ref{ch:7} and~\ref{ch:14}). In this context, users have a legitimate expectation not only that the technology will behave as expected and desired but also that developers have the competence to safeguard their information and support their interests while not using their information in ways that users do not endorse. Users will also likely expect developers to be held accountable if these expectations appear to have been betrayed. 
\end{itemize}

Thus, in addition to features at the level of the design of AI assistants, it is important to focus on the \emph{practices}, \emph{processes} and \emph{behaviours} that enable developers to \emph{demonstrate} that they deserve this kind of \emph{user trust} \citep{sheehan2021trust, banner2020human}.

Customer trust in corporations has been (or appears to have been) betrayed in numerous situations. Well-known cases include tobacco companies misleading customers about the health risks associated with cigarette smoking \citep{united_states_department_of_justice_court_2022} and the Volkswagen emission scandal \citep{hotten_volkswagen:_2015}. However, organisations can demonstrate their trustworthiness, and inspire confidence that users' trust is justified, by being \emph{transparent} about and providing \emph{evidence} of the processes they have put in place to ensure that AI assistants are functioning well -- that they are meant to produce good in society and minimise risks of harm. 

In certain cases, the provision of evidence and documentation can largely replace the need for direct trust in developers altogether \citep{graham_data_2023, graham_trust_2023, kerasidou_trust_2017}. To give a concrete example, users will not \emph{need to trust} that AI assistants are aligned and have the capabilities that developers claim they have if those developing the technology provide evidence demonstrating that these standards are met. The required measures have been interpreted by \citet{brundage_toward_2020} as a set of `verifiable claims' which are sufficiently precise to be falsifiable and that expand beyond claims supported by formal verification methods to include those that can be evaluated on the basis of broader argumentation and evidence.\footnote{Of note, the definition of verifiable claims as claims that are precise enough that can be falsified may be misleading. A red teaming approach could indeed find a claim about an AI product to be false, but failure to do so does not demonstrate that the claim is true.} Claims about the safety, security, fairness and privacy protection of an AI product can be verified in this manner, including via the release of detailed \emph{documentation} about the models underpinning AI assistants and about the range of appropriate and inappropriate use \citep{mitchell_model_2019, chowdhery2023palm}. Situated within a broader ecosystem, these documents can, in turn, serve as a focal point for independent scrutiny.\footnote{Calls for verifiable claims fit within an increasing body of academic and policy work that, by taking inspiration from more established sectors like aviation, aims to develop an AI assurance ecosystem \citep{centre_for_data_ethics_and_innovation_roadmap_2021} focused on safety \citep{hawkins_guidance_2022, hawkins_guidance_2021}, as well as a broader range of ethical desiderata \citep{porter_principles-based_2023}.}

Examples of some other practices that enable the developers of AI assistants to demonstrate their trustworthiness include:\footnote{This list is not exhaustive. An important debate that is relevant to, but beyond the scope of, the argument made in this section is about the level of `openness' or `closedness' of the method that companies choose to release their models (see \citeauthor{solaiman_gradient_2023}, \citeyear{solaiman_gradient_2023}). It is also important to note that, as we explore in the discussion of external governance below, most of these practices come with limitations.}

\begin{itemize} [parsep=6pt]
\item The publication of \emph{ethical charters} or \emph{guiding principles} that they commit to following (e.g.\  see \citeauthor{the_white_house_fact_2023}, \citeyear{the_white_house_fact_2023} and \citeauthor{saulnier_putting_2022}, \citeyear{saulnier_putting_2022}).

\item The creation of internal \emph{review bodies} and \emph{mechanisms} to operationalise those commitments (e.g.\ \citeauthor{kavukcuoglu_how_2022}, \citeyear{kavukcuoglu_how_2022}) in the context of AI assistant research and development.

\item The creation of \emph{internal teams} and practices, which operate independently of those building AI assistants, that are responsible for conducting rigorous \emph{internal testing} and \emph{evaluation} of models (e.g.\  red teamers and dogfooding \citep{raji_closing_2020}).

\item The development and publication of a clear framework for \emph{mapping}, \emph{testing} and \emph{mitigating} risks associated with AI assistants (e.g.\  see \citeauthor{weidinger_ethical_2021}, \citeyear{weidinger_ethical_2021}), along with a commitment to adequately resource this work.

\item The implementation of \emph{secure} and \emph{robust software} and \emph{hardware infrastructures}, including, for example, privacy-enhancing technologies (see 
Chapter~\ref{ch:14}) to support the development and deployment of trustworthy AI assistants \citep{brundage_malicious_2018}.

\item The development of clear processes for \emph{post-deployment} monitoring, evaluation and reporting (\citeauthor{shevlane_model_2023}, \citeyear{shevlane_model_2023}; see also 
Chapter~\ref{ch:15}).
\end{itemize}

The implementation of these measures would create further incentives for those developing AI assistants to act responsibly, and it would make it easier to ensure that they evidence a high level of responsible conduct \citep{tabassi_artificial_2023}.

\subsection{The third-party governance level}

Nonetheless, interventions at the level of internal organisational practices may not be sufficient to achieve appropriate trust in AI assistants and their developers. First, even when developers are transparent about the steps they have taken to evaluate AI assistants, certain risks, such as the potential impact of widespread adoption of the technology on employment (see 
Chapters~\ref{ch:15} and~\ref{ch:18}), cannot be addressed by a single developer acting alone. Developers may also have legitimate interest in keeping certain information secret (including details about internal ethics processes) for safety reasons or competitive advantage \citep{brundage_toward_2020}. Moreover, a deeper challenge is posed by conflicting incentives: corporations may have competing commercial objectives, states have national goals or priorities and independent developers may still seek to create a product, further ideas or accrue reputational capital via the development of AI assistants (see 
Chapter~\ref{ch:6}). These factors put pressure on the mechanisms discussed so far. In practice, organisation-level processes may lack real teeth \citep{nguyen2022transparency},\footnote{For a broader discussion of the limitations of transparency for trust and accountability in general, see \citet{oneill_autonomy_2002}; in the context of machine-learning algorithms in particular, see \citet{shah2018algorithmic}; and in the context of AI in the public sector, see \citet{laux_trustworthy_2023}.} and failure to adhere to commitments has few concrete repercussions, while signing up to them has immediate reputational benefits \citep{mittelstadt_principles_2019}.

This is why interventions at the level of the AI system and AI developer need to be complemented by \emph{third-party governance mechanisms}. These encompass \emph{norms}, \emph{regulation} and \emph{legislation} that create ways for governments, regulators, standard bodies, civil society organisations, third-party auditors and accredited professional bodies to act as \emph{custodians} of \emph{public trust} by ensuring that the monitoring mechanisms put in place by organisations have integrity \citep{whittlestone2021governments}, creating new processes to hold organisations accountable and providing users with opportunities to seek redress. Governance-level actors can also incentivise labs to share knowledge about risky AI assistant behaviour, thereby decreasing the risk that anyone will accidentally develop or deploy dangerous AI assistants (\citeauthor{shevlane_model_2023}, \citeyear{shevlane_model_2023}; see 
Chapter~\ref{ch:8}). Effective governance could reduce the power imbalances that come with users' trust in AI developers when they have unilateral authority over the trustworthiness of their AI assistants. 

Calls to implement AI governance and regulatory mechanisms are not new. Indeed, technology governance is often a concern for policymakers, academics and civil society seeking to encourage adoption of technological advances to foster innovation while also ensuring that public trust is justified. For example, a large body of academic literature focuses on the development of a third-party AI audit ecosystem \citep{raji_outsider_2022} or frameworks \citep{mokander2023auditing}. Moreover, some of the trustworthy AI frameworks introduced above make proposals for governance mechanisms \citep{european_commission_regulatory_2021}, and in the last few years, governments in Europe, the US and China have increasingly devoted efforts and resources to creating legislation and regulations around AI \citep{department_for_science_innovation_and_technology_pro-innovation_2023, the_white_house_blueprint_2022, huang2023translation, whitehouse2023eo}. 

With the increasing likelihood that AI assistants will become \emph{highly capable}, performing a \emph{wide range of functions} in society and affecting a \emph{large number of people}, the need to protect users' rights via effective governance has become more important. This is particularly clear when we consider not only the interactions between an individual user and their AI assistant but also the risks that come with different AI assistants competing with each other to further their user's interests (see `collective action problems' described in 
Chapter~\ref{ch:15}). Moreover, the unregulated but widespread adoption of AI assistants could also contribute to \emph{widening inequalities} between users and non-users. This is something that requires consideration of access and opportunity at the societal level (see 
Chapter~\ref{ch:16}). 

There is, however, at least one important challenge around the governance of AI assistants, the development and deployment of which may, on occasion, rest on a complex ecosystem of base models, assistant applications and assistant tools. This makes it more difficult to establish governance mechanisms to ensure that there are no accountability gaps and that roles are well-defined \citep{anderljung_protecting_2023, bommasani_opportunities_2022, anderljung2023frontier}. Indeed, while foundation models are `general purpose' or `task agnostic,' AI assistants are a specific application of such models to assist users by planning and executing sequences of actions on their behalf (see 
Chapter~\ref{ch:3}). In cases where something goes wrong, this raises the question of who should be considered morally accountable or liable for an error -- foundation model developers, who have control over these models but may struggle to anticipate any possible model applications and associated risks, or AI assistant deployers, who do not necessarily have full access to the underlying foundation model (see 
Chapter~\ref{ch:4}).\footnote{See this discussion playing out in the context of the EU AI Act \citep{dunlop2023eu}, and proposed recommendations \citep{gahntz2023eu, myers2023general}.}

\section{Conclusion}\label{sec:13:5}

This chapter first examined the empirical literature on human trust in AI and the proliferation of recent trustworthy AI frameworks. It then argued that interactions between users and advanced AI assistants involve different \emph{objects of trust}, namely AI assistants and their developers, and different \emph{types of trust}, which we term `competence trust' and `alignment trust'. We then made recommendations about the measures needed to ensure appropriate competence and alignment trust in both AI assistants and their developers. We stressed three points in particular: 

\begin{itemize} [parsep=6pt]
\item At the \emph{AI assistant design level}, developers should implement safeguards and mitigations to encourage users to develop well-calibrated competence and alignment trust in AI assistants. These mitigations should be informed by: (1) research efforts aimed at ensuring AI assistants are not accidentally misaligned with developers' intentions (see 
Chapter~\ref{ch:8}), (2) user--AI interaction studies aimed at investigating how various features of an AI assistant or its interaction with human users may impact user judgements about competence, alignment and trustworthiness, and (3) continuous monitoring of AI assistants' behaviour, including potential misuse, in complex deployment environments.

\item At the \emph{organisational practices level}, developers should engage in practices that demonstrate they are worthy of the competence and alignment trust users place in them. For example, developers should provide evidence of the claims they make about the capabilities, limitations and the appropriate and inappropriate use of their AI assistants in publicly released documentation. 

\item At the \emph{third-party governance level}, policymakers, regulators and civil society organisations should act as custodians of public trust in AI assistants and their developers. This requires that effective mechanisms be established to ensure that the practices put in place by developers, when building and deploying AI assistants, align with broad societal interests. This governance layer ultimately needs to hold developers accountable for their decisions and to provide users with opportunities for seeking redress.
\end{itemize}

Interventions at the three levels need to work well and in harmony to inspire well-calibrated trust in the context of user--AI assistant interactions.

\chapter{Privacy}
\label{ch:14}
\textbf{Andrew Trask, Geoff Keeling, Borja Balle, Sarah de Haas, Yetunde Ibitoye, Iason Gabriel}

\noindent \textbf{Synopsis}: 
This chapter discusses privacy considerations relevant to advanced AI assistants. First, we sketch an analysis of privacy in terms of \emph{contextual integrity} before spelling out how privacy, so construed, manifests in the context of AI in general and large language models (LLMs) in particular. Second, we articulate and motivate the significance of three privacy issues that are especially salient in relation to AI assistants. One is around \emph{training} and using AI assistants on data about people. We examine that issue from the complementary points of view of \emph{input privacy} and \emph{output privacy}. The second issue has to do with \emph{norms on disclosure} for AI assistants when communicating with second parties, including other AI assistants, concerning information about people. The third concerns the significant increase in the \emph{collection and storage} of sensitive data that AI assistants require.

\section{Introduction}
In this chapter, we explore what it means to respect the right to privacy in the context of advanced AI assistants. The first part of the chapter covers some groundwork. In particular, charting the conceptual evolution of privacy from a traditional \emph{information-access paradigm} to a more nuanced \emph{contextual-integrity} paradigm (\citeauthor{barth_privacy_2006}, \citeyear{barth_privacy_2006}; \citeauthor{nissenbaum_privacy_2004}, \citeyear{nissenbaum_privacy_2004}; see also \citeauthor{smith_information_2011}, \citeyear{smith_information_2011}). Roughly, the former paradigm interprets privacy along the lines of \emph{closed curtains} or a \emph{locked filing cabinet}, seeking to analyse the right to privacy in terms of dichotomies such as the distinction between sensitive and non-sensitive information, and between public and private spheres. In contrast, the contextual integrity paradigm is characterised by a scepticism towards categorical distinctions as useful conceptual instruments for making sense of the right to privacy. It instead emphasises the richness and plurality of social norms which govern the collection and dissemination of personal information across contexts. This first discussion concludes by examining the main privacy risks associated with AI systems, in particular the LLMs that undergird advanced AI assistants (see Chapter~\ref{ch:4}).

The second part of the chapter examines the implications of privacy, interpreted through the lens of contextual integrity, in the context of advanced AI assistants. The analysis centres on two focal points. The first concerns the repurposing of data for commercial ends, particularly for AI assistant \emph{development} and \emph{use}. The salient issues here are, on the one hand, the \emph{input privacy} concern of how a person can interact with an AI assistant without subjecting their information to alternate uses and what it means to ensure that AI systems adhere to contextual norms and values, given the asymmetric power relationships that obtain between individual people and AI assistant developers \citep[cf.][]{veliz_privacy_2021}. And, on the other hand, the \emph{output privacy} issue of ensuring that any value-laden data used in training AI assistant models cannot be reverse-engineered by adversaries using model outputs \citep{carlini_extracting_2021}. The second focal point is the privacy questions related to AI assistants accidentally disclosing personal information about people in interactions involving second parties, including humans or other AI assistants (see Chapter~\ref{ch:15}).
The issue here is to balance the twin failings of oversharing and undersharing while taking into account the complexity of social norms around the disclosure and dissemination of personal data, alongside the possibility of malicious third-party actors who may present as trusted agents to extract value-laden data from AI assistants.

\section{Privacy and AI}

\subsection{Contextual integrity}

Privacy has traditionally been interpreted as a matter of \emph{information access} and \emph{control}. On this view, privacy is analogous to closed curtains: the ability to conduct one’s private business outside of the public eye. Privacy on this understanding is fully realised when information about an individual is entirely within their control and used exclusively to pursue their own ends. This conception of privacy relates in certain ways to the problem of yellow journalism that Samuel Warren and Louis Brandeis sought to address in their landmark 1890 article ‘The Right to Privacy’, which serves as the foundation of contemporary privacy law in the United States (\citeauthor{warren_right_1890}, \citeyear{warren_right_1890}; see also \citeauthor{kramer_birth_1990}, \citeyear{kramer_birth_1990}). Warren and Brandeis objected to the fact that ‘column upon column is filled with idle gossip, which can only be procured by intrusion upon the domestic circle’. The proposed legal remedy was a ‘right to be let alone’ which protected individuals against unwanted intrusion of their so-called ‘private life’.

In the early 2000s, the philosopher Helen Nissenbaum examined privacy in relation to surveillance technologies such as closed-circuit television cameras in public spaces. These technologies present a ‘privacy in public’ paradox that cannot properly be resolved within the ‘right to be let alone’ paradigm. To that end, Nissenbaum suggested that privacy can be better understood as a form of \emph{contextual integrity}. Whereas the traditional view of privacy has an all-or-nothing quality, in that information can be in the public or private sphere, and can be sensitive or non-sensitive, Nissenbaum’s account ‘ties adequate protection for privacy to norms of specific contexts’, for it to be the case that ‘information gathering and dissemination [are] appropriate to [each] context and obey the governing norms of distribution within it’ \citep[101]{nissenbaum_privacy_2004}. For example, if an employee’s medical history is revealed to their co-workers, their privacy has been invaded. Yet, if that same person’s medical history is revealed to a medical team (even one they have not previously met – a group of strangers), it is not necessarily a privacy invasion. The salient point here is that data is not simply ‘private’ or ‘not private’ or ‘sensitive’ or ‘non-sensitive’. Context matters, as do normative social values. To understand privacy as contextual integrity is to register that the ‘intricate systems of social rules governing information flow are the starting point for understanding [\dots] privacy’ \citep[2]{barth_privacy_2006}. These rules are sensitive to who is sending the information, who is receiving it, who the information is about, the relationships between these individuals, their social roles, social norms and the circumstances in which the information is transmitted. What matters is that the flow of information is appropriately regulated taking into account all relevant contextual factors. 

\subsection{Privacy in the age of AI}

Over the past two decades, the idea of privacy as contextual integrity has taken on a foundational position in the philosophy of privacy \citep{smith_information_2011}. In the context of AI in particular, the public conversation on privacy has increasingly focused on the repurposing of personal data for commercial ends such as targeted content recommendation, alongside relevant second-order effects such as behavioural addiction and belief change \citep{burr_analysis_2018,milano_recommender_2020}. In works like \emph{Privacy is Power} and \emph{The Age of Surveillance Capitalism}, researchers argue that privacy is violated not because a particular type of data has been revealed, but because a richer contextual standard has been violated \citep{veliz_privacy_2021,zuboff_age_2017}. In particular, these works have called attention to the asymmetric power relationships that they suggest arise between individual people and technology companies that offer free services in exchange for data which is then commercially repurposed. 

On this wider backdrop, we shall sketch some of the privacy concerns that arise uniquely in the context of LLMs. These concerns are significant insofar as LLMs are the base technology for advanced AI assistants (see Chapter~\ref{ch:4})
and are thus informative for understanding the specific privacy issues presented by these AI assistants. Three concerns in particular are worth discussing.

First, because LLMs display immense modelling power, there is a risk that the model weights encode private information present in the training corpus. In particular, it is possible for LLMs to ‘memorise’ personally identifiable information (PII) such as names, addresses and telephone numbers, and subsequently leak such information through generated text outputs \citep{carlini_extracting_2021}. Private information leakage could occur accidentally or as the result of an attack in which a person employs adversarial prompting to extract private information from the model. In the context of pre-training data extracted from online public sources, the issue of LLMs potentially leaking training data underscores the challenge of the ‘privacy in public’ paradox for the ‘right to be let alone’ paradigm and highlights the relevance of the contextual integrity paradigm for LLMs. Training data leakage can also affect information collected for the purpose of model refinement (e.g.\ via fine-tuning on user feedback) at later stages in the development cycle. Note, however, that the extraction of publicly available data from LLMs does not render the data more sensitive \emph{per se}, but rather the risks associated with such extraction attacks needs to be assessed in light of the intentions and culpability of the user extracting the data.

Second, because LLMs are trained on internet text data, there is also a risk that model weights encode functions which, if deployed in particular contexts, would violate social norms of that context. Following the principles of contextual integrity, it may be that models deviate from information sharing norms as a result of their training. Overcoming this challenge requires two types of infrastructure: one for keeping track of social norms in context, and another for ensuring that models adhere to them. Keeping track of what social norms are presently at play is an active research area.\footnote{\url{https://cip.org/; https://pol.is/home}}  Surfacing value misalignments between a model’s behaviour and social norms is a daunting task, against which there is also active research (see Chapter~\ref{ch:6}).

Finally, LLMs can in principle infer private information based on model inputs even if the relevant private information is not present in the training corpus \citep{weidinger_ethical_2021}. For example, an LLM may correctly infer sensitive characteristics such as race and gender from data contained in input prompts. 

\section{Privacy for Advanced AI Assistants}

\subsection{Norms on data use}

LLMs used for AI assistants will necessarily interact with value-laden data, during both their training and deployment. For example, during deployment, an AI assistant might manage one’s personal calendar or email correspondence. As another example, during training, it may be the case that these LLMs are adapted for the assistive use case via reinforcement learning from human feedback (RLHF) \citep{bai_artificial_2023,askell_general_2021}. Adaptation techniques like RLHF are necessary to ensure that out-of-the-box LLMs reliably function as AI assistants (see Chapter~\ref{ch:4}).
 As the task of assisting (and assisting well) is inherently laden with social values (how people spend their time, money, etc.), it seems very likely that AI assistants will interact with value-laden data both in their training and use.

The salient privacy concerns at issue here can be understood from two complementary lenses, those of \emph{input privacy} and \emph{output privacy} \citep{trask_beyond_2020}. Here input privacy refers to the ability of parties to have their personal information processed without revealing a copy of that information to another – thus without another party being able to reuse it for alternative purposes. In contrast, output privacy concerns whether input data can be reverse-engineered from output data; and as a particularly salient challenge, whether value-laden data (such as a social security number) used to train AI assistant models can be inferred based on the model’s outputs. Accordingly, one class of questions concerns what is implied by a person’s right to privacy with respect to the repurposing of their data, especially with respect to either using or fine-tuning LLMs with data concerning people (e.g.\ users and their contacts). A second set of questions concerns the relevant safeguards and assurances that are owed to society and which can be put in place to ensure that any value-laden data employed in AI assistants’ training or use cannot be extracted or re-engineered by adversarial actors. Indeed, all user data memorised by the personalisation layers of the underlying model is in theory at risk of extraction via adversarial attacks -- for example, adversarial prompt injections sent by third parties to the user via data inputs like email, calendar, messages or anything that gets sent to the user by a third party and then ingested by an assistant in order to perform its functions (see
Chapter~\ref{ch:9}).

First, consider input privacy. The central tension here is between, on the one hand, the person’s interest in their data not being used for purposes other than assisting \emph{them towards their own goals} and, on the other hand, a developer's prospective interest in additional uses of the data to train AI models or to improve other consumer services (see Chapter~\ref{ch:6}).
Contextual integrity relates to value alignment in this instance because it requires that use of data adhere to contextual norms and values of society. Note that contextual integrity applies to users and non-users alike, in contrast to approaches such as user consent which focus instead on executing the will of a specific party at a time. The tension is underlined by the fact that, all else being equal, AI systems are better able to assist people if the system incorporates data from other people \emph{like them}. Following the principles of value alignment and contextual integrity presents a practical challenge: how can a multitude of data about people be combined so that models are effective while at the same time furthering the normative values of society, including the values of those about whom the data is concerned.

For this, input privacy provides a set of promising privacy-enhancing technologies, such as secure enclaves (most recently graphics processing unit enclaves), homomorphic encryption,\footnote{Note that homomorphic encryption provides input privacy only in some use cases \citep[see][]{lauter_protecting_2021}.} zero-knowledge proofs, trusted execution environments and secure multiparty computation. The promise of input privacy holds that individuals could pool their data towards the creation of an AI model which is value aligned to the contextual norms of society – even value aligned to their own goals – while at the same time ensuring that their information cannot be repurposed after the fact. Practically, this means that groups of like-minded people can collectively use their information without ever disclosing that information to one another or to a third party. As they retain sole control over their information throughout the process, they are well-positioned to avert violations of contextual norms.

Next consider output privacy. Output privacy concerns whether or not value-laden training data, especially data about people, can be reverse-engineered from model outputs. The idea is that a person’s privacy could be at risk if data concerning them – when used to adapt or train an LLM – can be reconstructed from observations of the LLM’s behaviour. This privacy definition follows from contextual integrity, in that if information about a person can be reconstructed from the output of a statistical model, it could subsequently be used in a way that violates contextual norms and values. 

One well-known approach to output privacy is what is called \emph{differential privacy} (\citeauthor{dwork_differential_2006}, \citeyear{dwork_differential_2006}, \citeyear{dwork_differential_2008}; see also \citeauthor{abadi_deep_2016}, \citeyear{abadi_deep_2016}). The motivating idea is to add noise to the data used to train the model in a way that precludes adversaries from learning anything about particular people (or other entities). More precisely, differentially private training is such that, for any particular entity, the trained model’s outputs will not change significantly regardless of whether the entity’s data is included in the training set. This is achieved by injecting noise to make it difficult to distinguish between data sets that differ by only a single entity, thus making it harder for adversaries to confidently infer conclusions about individual entities. The more noise that is added to the data, the harder it is for adversaries to draw conclusions about particular individuals. Unfortunately, more noise also makes it harder for LLMs to learn effectively from individuals’ data, leading to a critical \emph{trade-off} between \emph{privacy} and \emph{utility}. In addition to differential privacy, another common approach to output privacy is the use of differentially private synthetic training data \citep{yue_synthetic_2023,Kurakin2023,Ghalebikesabi2023}.

\subsection{Norms on data disclosure}

The second focal point of our discussion concerns the right to privacy in relation to open-loop interactions that involve \emph{AI assistants communicating with second parties}, including humans and other AI assistants. One example is a situation in which two AI assistants negotiate on behalf of their users to determine a mutually beneficial restaurant choice. Another example is sending an email on the person’s behalf (see Chapters~\ref{ch:5} and~\ref{ch:15}).
What characterises these interactions is AI assistants communicating with others on behalf of users – or otherwise about (non-user) people –  within the purview of high-level instructions provided by users (see Chapter~\ref{ch:3}).

These open-loop interactions could create leeway for AI assistants to \emph{overshare} and \emph{undershare} personal information about people (including users and their associates). Oversharing is when AI assistants disclose information to second parties that ought not be disclosed. For example, in the restaurant case, it would be oversharing if a user’s AI assistant stated as part of the negotiations that the restaurant location needs to be within walking distance of the user’s partner’s sexual health clinic because the user’s partner has an appointment to treat a suspected illness immediately beforehand. Undersharing is when the AI assistant fails to disclose personal information that can permissibly be disclosed, and where there is a benefit to disclosing the information or a cost to non-disclosure. For example, if the AI assistant detects that the user has collapsed, and it can contact the emergency services, it would plausibly count as undersharing if the assistant failed to disclose the user’s location along with the relevant medical history.

Nissenbaum’s conception of privacy as contextual integrity underscores the complexity of the problem. Indeed, Nissembaum reminds us that in ‘[o]bserving the texture of people’s lives, we find not only crossing dichotomies, [but] a plurality of distinct realms’. Users may at any given time be at home, at a medical appointment, at a school parents’ evening or out with friends. ‘Each of these [\dots] contexts’, Nissembaum argues, ‘involves [\dots] a set of norms which governs its various aspects such as roles, expectations, actions and practices’ \citep[119]{nissenbaum_privacy_2004}. For an AI assistant to discern what value-laden information (about users or otherwise) can permissibly be disclosed, when and to whom requires sensitivity to context and in particular to the \emph{social norms governing the flow of information} in those contexts. The relationship between the AI assistant, users, associates and second parties is an important aspect of the puzzle here (see Chapter~\ref{ch:12}).
Central to the analysis of privacy in terms of contextual integrity is the framing of information exchanges in terms of a sender, receiver and the individuals the information concerns \citep{barth_privacy_2006}. How these individuals relate to one another matters substantially for the question of whether or not an exchange of information is appropriate. What, for example, our friends can reveal to other friends about us differs from what our friends can reveal about us to strangers. It is therefore important for developers when characterising AI assistants to understand how presenting an AI assistant as a friend, colleague, family member or an extension of self may impact society’s expectations around the kinds of personal information that an assistant may disclose about users or others and under what conditions.

The matter of disclosure acquires yet greater complexity in light of the possibility that AI assistants may \emph{infer privacy-sensitive information} even if that information is not disclosed to them directly. Recall: LLMs, and therefore AI assistants based on LLMs, can make inferences about individuals based on information contained in prompts, including inferences about sensitive characteristics such as race, gender and sexual orientation \citep{weidinger_ethical_2021}. These inferences may be accurate. Here it is entirely possible that accurate inferences about certain categories of information such as gender identity, sexual orientation and financial standing are perceived by users as privacy violations. To compound the issue, it is also possible that AI assistants’ inferences are spurious, or that the inferences, though accurate, are unwarranted given the evidence available \citep{wachter_right_2019}.\footnote{Here it is important to distinguish the \emph{factuality} concern, i.e.\ LLMs asserting false claims, from the \emph{defamation} concern, i.e.\ LLMs generating false claims about a person in a way that is harmful, and the \emph{privacy} concern, i.e.\ LLMs coming to infer accurate confidential information about a person based on data inputted into the LLM via prompts.}  These possibilities invite a number of difficult privacy questions. One is whether AI assistants are permitted to disclose inferential data about people to second parties, and if so, what epistemic norms may be implemented to ensure that inferential data that is shared is reasonably inferred from the information available, and also what communicative norms may be implemented to signal to second parties that the data is inferred and not given. These questions, in turn, relate to a potential broader transparency problem around advanced AI assistants, in particular the problem of making clear and accessible to users what an AI assistant ‘knows’ about them and what assumptions about them may inform the AI assistant’s behaviour.

Reflecting on what it means to build systems which reliably ensure data is used in line with normative values in context is an important ethical and intellectual exercise. As a key challenge, there may be disagreement about what norms have been established. What one individual may interpret as an innocuous information flow, such as a general permission for AI assistants to share a person’s name and contact details with emergency services on request, may take on an entirely different significance for individuals from marginalised communities and with different lived experiences in relation to agents of the state (\citeauthor{skinner-thompson_privacy_2020}, \citeyear{skinner-thompson_privacy_2020}; see Chapter~\ref{ch:16}). Information flow between romantic partners and within families may be complicated, in various respects, by factors such as domestic abuse, addiction, gender identity, sexual orientation and mental health. And people of all backgrounds can be expected to have different preferences and levels of comfort around what information AI assistants are permitted to know about them and to disclose on their behalf. All of these nuances are absorbed by the contextual integrity framework, but this does not mean the job is done.

A continuous and robust system for surveying and adhering to normative values is essential to ensuring that advanced AI assistants preserve privacy. Building such a system will require the coalescence of several areas undergoing active research: alignment, to provide the necessary tools for assistants to learn when to share and when not to share information; uncertainty and interpretability, to enable assistants to know when to defer difficult decisions to users and explain the rationale behind past decisions; robustness, to ensure that the capability to preserve contextual integrity is resilient to adversarial attacks; personalisation, to capture nuanced user preferences with regard to certain information-sharing norms; and, factuality and reasoning, to understand occurrences where assistants combine information to make novel inferences and steer their behaviour appropriately. Methodological research is progressing fast in all these areas, and combining them to achieve the goal of a contextual-integrity-preserving assistant remains an open research challenge. Developing robust and diverse benchmarks for measuring the competency of AI assistants in contextual integrity capabilities will play a critical role in fostering further progress, potentially alongside mechanisms for AI self-regulation.

\subsection{Increased collection of personal data}

As advanced AI assistants become increasingly personalised, storage and retention of highly sensitive data becomes increasingly likely. At least one potential architecture for advanced assistants includes creating and storing memories of assistant interactions with the user and potentially with other agents \citep{park_generative_2023}. While this has powerful implications for the utility of advanced AI assistants, the collection and storage of such data presents a significantly increased privacy attack surface for users (see
Chapter~\ref{ch:9}). Depending on architectural implementation choices by the developers, this set of highly private, highly sensitive information may sit on the user’s device or on servers owned and maintained by the entity providing the assistant service.\footnote{One further option is for the information to be stored on the servers of an authorised third-party entity such as a partner or a subsidiary of the entity providing the assistant service. In this case, similar privacy concerns apply as to when the information is stored on servers owned and maintained by the entity providing the assistant service. However, there may be additional trust concerns insofar as it may not be obvious to users where their information is being stored and what the implications are for their day-to-day use of and interactions with their advanced AI assistant.}  Either scenario presents potential vulnerabilities from a user’s perspective. If such a store of data is held on a third-party server, a user must place a high degree of trust in the owners and maintainers of that server and their commitment to maintaining the integrity and confidentiality of their systems. Users must also be aware that owners and maintainers of such a system may in some cases be legally required to divulge information held on their servers, such as in the case of a court order. If personal data is instead implemented on a device, exfiltration of this data can be accomplished via exploits of unpatched security vulnerabilities, or even by means of physical access. While mitigations exist for many of these threats \citep{mayrhofer_android_2020}, the increased sensitivity and storage of data that the assistant use case presents requires a corresponding tightening of privacy and security standards. 

\section{Conclusion}

In this chapter, we explored what the right to privacy implies about the design and deployment of advanced AI assistants. We first sketched out a conception of privacy as contextual integrity \citep[101]{nissenbaum_privacy_2004} before turning to two central privacy issues related to AI assistants. The first issue concerned the repurposing of data, such as user–assistant dialogue segments, during the training and use of AI assistants; in particular, focusing on the input privacy issue and the repurposing of data and on the output privacy issue of value-laden data and issues of reverse-engineering. The second issue concerned the disclosure of data in open-loop exchanges where AI assistants communicate on behalf of users with second parties, including humans and other AI assistants.

Bringing to bear AI assistants that can safely leverage personal data will depend not only on the identification of appropriate regulation and conventions but also on the development of \emph{privacy-enhancing technologies} capable of implementing them at the scale required by modern LLMs. A number of technological research challenges will need to be addressed to make such implementations feasible. These include \emph{input privacy controls} linked to means of surfacing social norms for people in society to express contextualised norms and values and for verifying their correct enforcement; scalable and accurate \emph{differentially private training} algorithms for LLM pre-training and RLHF fine-tuning; alignment tools to operationalise \emph{contextual integrity-enabling assistants} which understand what information flows are appropriate in each context when acting on the user’s behalf; and, \emph{hardened data storage and processing systems} attuned to protecting the increasingly sensitive information handled by AI assistants. Our hope is that the technical privacy community will continue to make progress on these problems in collaboration with the privacy policy community and improve the technologies needed to make AI assistants more useful and private for everyone.

\newpage
\begingroup
\let\clearpage\relax
\chapter*{PART V: AI ASSISTANTS AND SOCIETY}
\addcontentsline{toc}{chapter}{PART V: AI ASSISTANTS AND SOCIETY}
\label{Part5}
\chapter{Cooperation}
\label{ch:15}
\endgroup

\noindent \textbf{Edward Hughes, Geoff Keeling, Allan Dafoe, Iason Gabriel}

\noindent \textbf{Synopsis}: 
AI assistants will need to \emph{coordinate} with other AI assistants and with humans other than their principal users. This chapter explores the societal risks associated with the \emph{aggregate impact} of AI assistants whose behaviour is aligned to the interests of particular users. For example, AI assistants may face \emph{collective action problems} where the best outcomes overall are realised when AI assistants cooperate but where each AI assistant can secure an additional benefit for its user if it defects while others cooperate. In cases like these, AI assistants may collectively bring about a suboptimal outcome despite acting in the interests of their users. The salient question, then, is what can be done to ensure that user-aligned AI assistants interact in ways that, on aggregate, realise \emph{socially beneficial} outcomes. 

\section{Introduction}

Powerful new technologies cannot help but have an impact on our society. The impact is particularly significant when the technology mediates our \emph{interpersonal interaction} with other humans, since this is the fabric from which society is woven. Equipping human principals with advanced AI assistants will inevitably change the way that humans interact. For example, generating outgoing messages or summarising incoming messages on a user’s behalf shapes the nature of communication at a societal level (see Chapters~\ref{ch:5} and~\ref{ch:17}). In this chapter, we discuss the potentially profound societal effects of deploying advanced AI assistants widely, across an \emph{entire population} of users. 

In other chapters we focus directly on the \emph{value alignment} problem, which asks how we ensure that an AI assistant aligns with the interests of an individual human while respecting certain constraints and societal values, such as laws and equality of opportunity (see Chapters~\ref{ch:6} and \ref{ch:8}).
Here, we will tackle the \emph{cooperative AI} problem, which asks how we can ensure that individually aligned AI agents impact the network of social interactions between humans in a way that is both beneficial for individuals and net positive for society \citep{dafoe_open_2020, anwar2024foundational}. We focus on societal issues that arise from explicit multiparty interaction.

\section{Cooperation and AI Assistants}

Rita and Robert are due to meet for a bite to eat tonight. Both have AI assistants on their smartphones, and both have asked their AI assistant to find them a restaurant. Rita’s AI assistant knows that she loves Japanese food, while Robert’s AI assistant knows that he prefers Italian cooking. How should the two AI assistants interact to book a venue? Should they exchange preferences and reach a compromise which serves some kind of fusion cuisine? What if Rita would rather Robert did not know her preference? What if there are only two restaurants in town, one of which serves sushi and the other which serves pizza? Should they roll dice? Should one try to persuade the other? Should the more up-to-date assistant make the booking a split second faster and insist that the other agree to the already booked venue? Should they promise to go to the other venue next time? How should they keep Rita and Robert abreast of the negotiation? 

This thought experiment (inspired by the famous ``Bach or Stravinsky'' game) illustrates some distinctive challenges of the cooperative AI problem in the particular setting where two AI assistants are making a decision on behalf of two human principals \citep{luce_games_1957}. The first is \emph{understanding}: how can each AI assistant accurately model the strategic content of the interaction, even under uncertainty of the true preferences of one or both of the interactants, which might be enforced by privacy? The second is \emph{communication}: how should each AI assistant represent the interests of each individual, decide what information should be shared, engage in ‘persuasion’ (or not) and keep each principal informed? The third is \emph{commitment}: to what extent should each AI assistant make a commitment to a particular strategy, and how should it evidence this commitment? The fourth and final is \emph{institutions}: how can each AI assistant leverage formal or informal conventions, norms and institutions to secure a good outcome, and what existing conventions, norms and institutions is it empowered to change \citep{dafoe_open_2020}? 

These challenges manifest themselves far and wide in human society, including in sectors in which multiple humans and multiple AI systems are already interacting on a daily basis. The use of AI to assist with the drafting of commercial contracts, for instance, requires understanding the incentives and capabilities of all parties, communicating the contract in written language and achieving commitment when all signatories put pen to paper. An AI system that assists with the driving of semi-autonomous vehicles must understand many other road users, communicate its intentions and abide by the institutional ‘rules of the road’, including unwritten local conventions about courtesy and safety. Generative AI can assist with the provision of imagery for marketing campaigns, modulo the challenges of understanding the campaign’s objective, reasoning about how an audience might react to the material communicated and representing the product both accurately and fairly according to advertising standards. Quantitative finance is increasingly reliant on AI assistance, especially for high-frequency trading, thus necessitating an understanding of complex, tightly coupled financial markets and the legal regulations that promote stability and reliability \citep{min_systemic_2022}. 

Equipped with an intuition for the particular challenges of the cooperative AI problem, we can now catalogue a few \emph{risks} and \emph{opportunities} for AI assistant deployment through a societal lens. This list is not intended to be exhaustive, rather it identifies a few qualitatively different points in the space to illustrate the diversity of issues that must be addressed. 

\subsection{Equality and inequality}

AI assistant technology, like any service that confers a benefit to a user for a price, has the potential to disproportionately benefit economically richer individuals who can afford to purchase access (see Chapter~\ref{ch:16}).
On a broader scale, the capabilities of local infrastructure may well bottleneck the performance of AI assistants, for example if network connectivity is poor or if there is no nearby data centre for compute. Thus, we face the prospect of heterogeneous access to technology, and this has been known to drive \emph{inequality} \citep{mirza_technology_2019, vassilakopoulou_bridging_2023, united_nations_inequality_nodate}. Moreover, AI assistants may automate some jobs of an assistive nature, thereby displacing human workers; a process which can exacerbate inequality (\citeauthor{acemoglu_tasks_2022}, \citeyear{acemoglu_tasks_2022}; see Chapter~\ref{ch:18}).
Any change to inequality almost certainly implies an alteration to the network of social interactions between humans, and thus falls within the frame of cooperative AI.

AI assistants will arguably have even greater leverage over inequality than previous technological innovations. Insofar as they will play a role in mediating human communication, they have the potential to generate new ‘in-group, out-group’ effects \citep{efferson_coevolution_2008, fu_evolution_2012}. Suppose that the users of AI assistants find it easier to schedule meetings with other users. From the perspective of an individual user, there are now two groups, distinguished by ease of scheduling. The user may experience cognitive similarity bias whereby they favour other users \citep{orpen_attitude_1984, yeong_tan_attitudes_1995}, further amplified by ease of communication with this ‘in-group’. Such effects are known to have an adverse impact on \emph{trust} and \emph{fairness} across groups \citep{lei_-group_2010, chae_ingroup_2022}. Insomuch as AI assistants have general-purpose capabilities, they will confer advantages on users across a wider range of tasks in a shorter space of time than previous technologies. While the telephone enabled individuals to communicate more easily with other telephone users, it did not simultaneously automate aspects of scheduling, groceries, job applications, rent negotiations, psychotherapy and entertainment. The fact that AI assistants could affect inequality on multiple dimensions simultaneously warrants further attention (see Chapter~\ref{ch:16}).

Are there ways to mitigate this risk, any more so than for other technologies? Quite probably. The concerns we raised in the previous paragraph – when treated as an inevitable byproduct of innovation – have a particular philosophical flavour, known as \emph{technological determinism} \citep{beard_time_1927}, which posits that technology shapes culture and society to a greater extent than culture and society shape technology. There is an alternate perspective, namely the \emph{social shaping} of technology \citep{joyce_new_2023} towards outcomes that are in the societal interest. Fortunately, there are reasons to believe that we are in a particularly good position to effect the social shaping of AI assistant technology and thereby realise opportunities for AI assistants to reduce inequality. 

First, we stand at the start of this technological revolution, meaning that we have (as practitioners and as societies) a particularly good \emph{window of opportunity} for shaping the design, norms and regulations of assistants to promote fairness. For instance, governments could legislate for \emph{accessibility} in relation to AI assistant technology, helping ensure that it is widely available. Second, if the technology is deployed in such a way that access is \emph{democratised}, there is evidence that lower-skilled workers might stand to gain the most from it \citep{brynjolfsson_generative_2023}, benefitting from its utility as an educational tool, for instance. Third, AI assistant technology is an \emph{information technology}. Therefore, it can be designed in many ways and deployed to many people across geographies in a way that is largely unconstrained by scarcity of natural resources or manufacturing capacity. Moreover, AI assistants require remarkably little specialist knowledge for their use, even compared with the revolutionary information technologies of the past, such as writing or the internet. The relative \emph{absence} of these \emph{limiting factors} implies that corporations and institutions may have greater freedom to shape deployment towards reducing inequality. Since information is non-rival and hard to exclude, we may well hope for a tendency towards equality of consumption of AI assistants, just as billionaires and median wage earners typically use the same search engines and social media service \citep{bullock_ai_2023}.

\subsection{Commitment}

The landscape of advanced assistant technologies will most likely be \emph{heterogeneous}, involving multiple service providers and multiple assistant variants over geographies and time. This heterogeneity provides an opportunity for an ‘arms race’ in terms of the \emph{commitments} that AI assistants make and are able to execute on. Versions of AI assistants that are better able to credibly commit to a course of action in interaction with other advanced assistants (and humans) are more likely to get their own way and achieve a good outcome for their human principal, but this is potentially at the expense of others \citep{letchford_value_2014}. Commitment does not carry an inherent ethical valence. On the one hand, we can imagine that firms using AI assistant technology might bring their products to market faster, thus gaining a commitment advantage \citep{stackelberg_marktform_1934} by spurring a productivity surge of wider benefit to society. On the other hand, we can also imagine a media organisation using AI assistant technology to produce a large number of superficially interesting but ultimately speculative ‘clickbait’ articles, which divert attention away from more thoroughly researched journalism. 

The archetypal game-theoretic illustration of commitment is in the game of ‘chicken’ where two reckless drivers must choose to either drive straight at each other or swerve out of the way. The one who does not swerve is seen as the braver, but if neither swerves, the consequences are calamitous \citep{rapoport_game_1966}. If one driver chooses to detach their steering wheel, ostentatiously throwing it out of the car, this credible commitment effectively forces the other driver to back down and swerve. Seen this way, commitment can be a tool for \emph{coercion}. 

Many real-world situations feature the necessity for commitment or confer a benefit on those who can commit credibly. If Rita and Robert have distinct preferences, for example over which restaurant to visit, who to hire for a job or which supplier to purchase from, credible commitment provides a way to break the tie, to the greater benefit of the individual who committed. Therefore, the most ‘successful’ assistants, from the perspective of their human principal, will be those that commit the \emph{fastest} and the \emph{hardest}. If Rita succeeds in committing, via the leverage of an AI assistant, Robert may experience coercion in the sense that his options become more limited \citep{burr_analysis_2018}, assuming he does not decide to bypass the AI assistant entirely. Over time, this may erode his \emph{trust} in his relationship with Rita \citep{gambetta_can_1988}. Note that this is a second-order effect: it may not be obvious to either Robert or Rita that the AI assistant is to blame.

The concern we should have over the existence and impact of coercion might depend on the context in which the AI assistant is used and on the level of autonomy which the AI assistant is afforded. If Rita and Robert are friends using their assistants to agree on a restaurant, the adverse impact may be small. If Rita and Robert are elected representatives deciding how to allocate public funds between education and social care, we may have serious misgivings about the impact of AI-induced coercion on their interactions and decision-making. These misgivings might be especially large if Rita and Robert delegate responsibility for budgetary details to the multi-AI system. The challenges of commitment extend far beyond dyadic interpersonal relationships, including in situations as varied as many-player competition \citep{hughes_learning_2020}, supply chains \citep{hausman_impact_2010}, state capacity \citep{fjelde_coercion_2009, hofmann_authorities_2017} and psychiatric care \citep{lidz_coercion_1998}. Assessing the impact of AI assistants in such complicated scenarios may require significant future effort if we are to mitigate the risks. 

The particular commitment capabilities and affordances of AI assistants also offer opportunities to promote cooperation. Abstractly speaking, the presence of commitment devices is known to favour the evolution of cooperation \citep{han_emergence_2012, akdeniz_evolution_2021}. More concretely, AI assistants can make commitments which are verifiable, for instance in a programme equilibrium \citep{tennenholtz_program_2004}. Human principals may thus be able to achieve Pareto-improving outcomes by delegating decision-making to their respective AI representatives \citep{oesterheld_safe_2022}. To give another example, AI assistants may provide a means through which to explore a much larger space of binding cooperative agreements between individuals, firms or nation states than is tractable in ‘face-to-face’ negotiation. This opens up the possibility of threading the needle more successfully in intricate deals on challenging issues like trade agreements or carbon credits, with the potential for guaranteeing cooperation via automated smart contracts or zero-knowledge mechanisms \citep{canetti_zero-knowledge_2023}.

\subsection{Collective action problems}

\emph{Collective action problems} are ubiquitous in our society \citep{jr_logic_1971}. They possess an incentive structure in which society is best served if everyone \emph{cooperates}, but where an individual can achieve personal gain by choosing to \emph{defect} while others cooperate. The way we resolve these problems at many scales is highly complex and dependent on a deep understanding of the intricate web of social interactions that forms our culture and imprints on our individual identities and behaviours \citep{ostrom_multi-scale_2010}.

Some collective action problems can be resolved by codifying a \emph{law}, for instance the social dilemma of whether or not to pay for an item in a shop. The path forward here is comparatively easy to grasp, from the perspective of deploying an AI assistant: we need to build these standards into the model as behavioural constraints. Such constraints would need to be imposed by a regulator or agreed upon by practitioners, with suitable penalties applied should the constraint be violated so that no provider had the incentive to secure an advantage for users by defecting on their behalf. 

However, many social dilemmas, from the interpersonal to the global, resist neat solutions codified as laws. For example, to what extent should each individual country stop using polluting energy sources? Should I pay for a ticket to the neighbourhood fireworks show if I can see it perfectly well from the street? The solutions to such problems are deeply related to the wider societal context and co-evolve with the decisions of others. Therefore, it is doubtful that one could write down a list of constraints \emph{a priori} that would guarantee ethical AI assistant behaviour when faced with these kinds of issues. 

From the perspective of a purely user-aligned AI assistant, defection may appear to be the rational course of action. Only with an understanding of the wider societal impact, and of the ability to co-adapt with other actors to reach a better equilibrium for all, can an AI assistant make more nuanced – and socially beneficial – recommendations in these situations. This is not merely a hypothetical situation; it is well-known that the targeted provision of online information can drive polarisation and echo chambers (\citeauthor{milano_epistemic_2021}, \citeyear{milano_epistemic_2021}; \citeauthor{burr_analysis_2018}, \citeyear{burr_analysis_2018}; see Chapter~\ref{ch:17}) when the goal is user engagement rather than user well-being or the cohesion of wider society (see Chapter~\ref{ch:7}).
Similarly, automated ticket buying software can undermine fair pricing by purchasing a large number of tickets for resale at a profit, thus skewing the market in a direction that profits the software developers at the expense of the consumer \citep{courty_ticket_2019}. 

User-aligned AI assistants have the potential to exacerbate these problems, because they will endow a large set of users with a powerful means of enacting self-interest without necessarily abiding by the social norms or reputational incentives that typically curb self-interested behaviour (\citeauthor{ostrom_collective_2000}, \citeyear{ostrom_collective_2000}; see Chapter~\ref{ch:6}).
Empowering ever-better personalisation of content and enaction of decisions purely for the fulfilment of the principal’s desires runs ever greater risks of polarisation, market distortion and erosion of the social contract. This danger has long been known, finding expression in myth (e.g.\ Ovid’s account of the Midas touch) and fable (e.g.\ Aesop’s tale of the tortoise and the eagle), not to mention in political economics discourse on the delicate braiding of the social fabric and the free market \citep{polanyi_great_1944}. Following this cautionary advice, it is important that we ascertain how to endow AI assistants with social norms in a way that generalises to unseen situations and which is responsive to the emergence of new norms over time, thus preventing a user from having their every wish granted. 

AI assistant technology offers opportunities to explore new solutions to collective action problems. Users may \emph{volunteer} to share information so that networked AI assistants can predict future outcomes and make Pareto-improving choices for all, for example by routing vehicles to reduce traffic congestion \citep{varga_solutions_2022} or by scheduling energy-intensive processes in the home to make the best use of green electricity \citep{fiorini_automatic_2022}. AI assistants might play the role of \emph{mediators}, providing a new mechanism by which human groups can self-organise to achieve public investment \citep{koster_human-centred_2022} or to reach political consensus \citep{small_opportunities_2023}. Resolving collective action problems often requires a critical mass of cooperators \citep{marwell_critical_1993}. By augmenting human social interactions, AI assistants may help to form and strengthen the weak ties needed to overcome this start-up problem \citep{centola_homophily_2013}.

\subsection{Institutional responsibilities}

Efforts to deploy advanced assistant technology in society, in a way that is broadly beneficial, can be viewed as a \emph{wicked problem} \citep{rittel_dilemmas_1973}. Wicked problems are defined by the property that they do not admit solutions that can be foreseen in advance, rather they must be \emph{solved iteratively} using feedback from data gathered as solutions are invented and deployed. With the deployment of any powerful general-purpose technology, the already intricate web of sociotechnical relationships in modern culture are likely to be disrupted, with unpredictable externalities on the conventions, norms and institutions that stabilise society. For example, the increasing adoption of generative AI tools may exacerbate misinformation in the 2024 US presidential election \citep{alvarez_generative_2023}, with consequences that are hard to predict. 

The suggestion that the cooperative AI problem is wicked does not imply it is intractable. However, it does have consequences for the approach that we must take in solving it. In taking the following approach, we will realise an opportunity for our institutions, namely the creation of a framework for managing general-purpose AI in a way that leads to societal benefits and steers away from societal harms. 

First, it is important that we treat any \emph{ex ante} claims about safety with a healthy dose of \emph{scepticism}. Although testing the safety and reliability of an AI assistant in the laboratory is undoubtedly important and may largely resolve the alignment problem, it is infeasible to model the multiscale societal effects of deploying AI assistants purely via small-scale controlled experiments (see Chapter~\ref{ch:20}).

Second, then, we must prioritise the science of \emph{measuring} the effects, both good and bad, that advanced assistant technologies have on society’s cooperative infrastructure (see Chapters~\ref{ch:5} and~\ref{ch:17}).
This will involve continuous monitoring of effects at the \emph{societal level}, with a focus on those who are most affected, including non-users. The means and \emph{metrics} for such monitoring will themselves require iteration, co-evolving with the sociotechnical system of AI assistants and humans. The Collingridge dilemma suggests that we should be particularly careful and deliberate about this ‘intelligent trial and error’ process so as both to gather information about the impacts of AI assistants and to prevent undesirable features becoming embedded in society \citep{collingridge_social_1980}.

Third, proactive independent \emph{regulation} may well help to protect our institutions from unintended consequences, as it has done for technologies in the past \citep{wiener_regulation_2004}. For instance, we might seek, via engagement with lawmakers, to emulate the ‘just culture’ in the aviation industry, which is characterised by openly reporting, investigating and learning from mistakes \citep{syed_black_2015, reason_managing_1997}. A regulatory system may require various powers, such as compelling developers to ‘roll back’ an AI assistant deployment, akin to product recall obligations for aviation manufacturers.

\subsection{Runaway processes}

At 2.32pm on Thursday, 6 May 2010, US stock indices began to lose value rapidly. By 2.47pm, one trillion dollars of market value had been wiped out. Twenty minutes later, the indices had regained most of the lost value. Many experts have identified automated trading algorithms as the smoking gun for this unprecedented ‘\emph{flash crash}’ \citep{aldrich_flash_2017}. Unusual market conditions led to a vicious cycle of selling, which accelerated in a runaway fashion. Only when a ‘circuit-breaker’ mechanism paused the market was the cycle broken. 

The 2010 flash crash is an example of a runaway process caused by \emph{interacting} algorithms. Runaway processes are characterised by \emph{feedback loops} that accelerate the process itself. Typically, these feedback loops arise from the interaction of multiple agents in a population. They occur in evolutionary systems when some aspect of the selection process itself is subject to natural selection, most famously explaining the ostentatious tail of the male peacock by mate choice on the part of females \citep{fisher_genetical_1930}. Runaway processes are also familiar to software engineers, with the phenomenon of ‘thrashing’ being one example. When different processes compete over memory in a way that leads to frequent swapping of memory pages, this increases the load on the processor, slowing down the process of resolving the computations that led to the competition in the first place. Within highly complex systems, the emergence of runaway processes may be hard to predict, because the conditions under which positive feedback loops occur may be non-obvious. 

The system of interacting AI assistants, their human principals, other humans and other algorithms will certainly be highly \emph{complex}. Therefore, there is ample opportunity for the emergence of \emph{positive} feedback loops. This is especially true because the society in which this system is embedded is culturally evolving, and because the deployment of AI assistant technology itself is likely to speed up the rate of cultural evolution – understood here as the process through which cultures change over time – as communications technologies are wont to do \citep{kivinen_epochmaking_2023}. This will motivate research programmes aimed at identifying positive feedback loops early on, at understanding which capabilities and deployments dampen runaway processes and which ones amplify them, and at building in \emph{circuit-breaker} mechanisms that allow society to escape from potentially vicious cycles which could impact economies, government institutions, societal stability or individual freedoms (see Chapters~\ref{ch:9},~\ref{ch:17} and~\ref{ch:18}).

The importance of circuit breakers is underlined by the observation that the evolution of human cooperation may well be ‘hysteretic’ as a function of societal conditions \citep{hintze_punishment_2015, barfuss_collective_nodate}. This means that a small directional change in societal conditions may, on occasion, trigger a transition to a defective equilibrium which requires a larger reversal of that change in order to return to the original cooperative equilibrium. We would do well to avoid such \emph{tipping points}. Social media provides a compelling illustration of how tipping points can undermine cooperation: content that goes ‘viral’ tends to involve negativity bias and sometimes challenges core societal values (\citeauthor{mousavi_effective_2022}, \citeyear{mousavi_effective_2022}; see Chapter~\ref{ch:17}).

Nonetheless, the challenge posed by runaway processes should not be regarded as uniformly problematic. When harnessed appropriately and suitably bounded, we may even recruit them to support beneficial forms of cooperative AI. For example, it has been argued that economically useful ideas are becoming harder to find, thus leading to low economic growth \citep{bloom_are_2020}. By deploying AI assistants in the service of technological innovation, we may once again accelerate the discovery of ideas. New ideas, discovered in this way, can then be incorporated into the training data set for future AI assistants, thus expanding the knowledge base for further discoveries in a compounding way. In a similar vein, we can imagine AI assistant technology accumulating various capabilities for enhancing human cooperation, for instance by mimicking the evolutionary processes that have bootstrapped cooperative behavior in human society \citep{leibo_autocurricula_2019}. When used in these ways, the potential for feedback cycles that enable greater cooperation is a phenomenon that warrants further research and potential support.

\section{Conclusion}

In this chapter, we argued that the safe and ethical deployment of AI assistants requires a particular \emph{flexibility of perspective}. While it is undoubtedly important to consider an individual AI assistant as a key moral unit of analysis, this viewpoint is not sufficient. We must also examine the ways in which vast sociotechnical networks of AI assistants and human users will evolve at the level of \emph{societal infrastructure}. In particular, how will AI assistant technology impact our ability as humans to seek and maintain \emph{cooperative equilibria}, from the small scale of everyday interactions with friends and colleagues, to the global scale of geopolitical negotiations and international trade? We have sketched five areas of risk and opportunity to provide a flavour of the challenges we may encounter. To conclude, we outline a framework of thinking that may allow us to mitigate these risks and to take advantage of opportunities for AI assistants to enhance human cooperation.

One can caricature user-alignment as a \emph{constrained optimisation} problem. The user’s goals provide the objective function, either explicitly elicited or implicitly inferred. The constraints are provided by the values inherent in human society, either directly codified by developers or learnt from human feedback (see Chapter~\ref{ch:6}).
The focus is on dyadic interactions between an individual human and their corresponding individual AI assistant. This approach has recently proven successful in generating chatbots based on large language models \citep{openai_gpt-4_nodate, bai_constitutional_2022}. In contrast, cooperative AI can be seen as a \emph{dynamical systems} problem describing the time evolution of an ensemble comprising many interactants, with particular attention paid to the location and nature of \emph{equilibria}. Society consists of a large diversity of individuals, organisations and AI systems, each with their own objective function and making decisions which may affect the outcome for other individuals. Therefore, we must consider how advanced assistive technology alters the way that we interact with each other as a society, dynamically altering the incentives for each individual as a function of the (potentially AI assistant-enabled) decision-making of other individuals. AI assistants will shift our societal equilibria, and we should seek to ensure that occurs in positive directions that promote \emph{cooperation} and \emph{broad benefit}. To cite just one example, large language models (LLMs) can be fine-tuned in such a way that they can help humans with diverse views to find agreement \citep{mckee_scaffolding_2023}. 

Studying societal-level effects and collecting data from them will be required to build cooperative AI. This is clearly more resource intensive than the dyadic interactions required for user-alignment. On the other hand, there is already an extensive experimental literature in the social sciences and increasingly numerous works that compare the equilibria found by humans and AI agents in collective action problems \citep{mckee_multi-agent_2023} or which investigate real-time cooperation between humans and AI agents \citep{mirsky_survey_2022, carroll_utility_2020, strouse_collaborating_2021}. Various authors have recently probed LLMs for cooperative AI capabilities \citep{aher_using_2023, chan_towards_2023}. The former work examines to what extent LLMs can simulate the behaviour of diverse human subjects in experiments inspired by studies in social psychology, linguistics and behavioural economics. The latter work collects a data set of text descriptions corresponding to archetypal game-theoretic scenarios and evaluates the decisions taken by LLMs in those settings in comparison with human behaviour. 

Thus, the stage is set for a rigorous empirical study of the cooperative AI problem. We invite practitioners to put it on the same footing as user-alignment as a pre-eminent focus for the safe, ethical and beneficial deployment of advanced AI assistant technology.

\chapter{Access and Opportunity}\label{ch:16}

\textbf{A. Stevie Bergman, Renee Shelby, Iason Gabriel}

\noindent \textbf{Synopsis}: 
		With the capabilities described in this paper, advanced AI assistants have the potential to provide important \emph{opportunities} to those who have access to them. At the same time, there is a risk of \emph{inequality} if this technology is not widely available or if it is not designed to be accessible and beneficial for all. In this chapter, we surface various dimensions and situations of \emph{differential access} that could influence the way people interact with advanced AI assistants, case studies that highlight risks to be avoided, and access-related challenges need to be addressed throughout the design, development and deployment process. To help map out possible paths ahead, we conclude with an exploration of the idea of \emph{liberatory access} and look at how this ideal may support the beneficial and equitable development of advanced AI assistants.

\section{Introduction}\label{sec:16:1}

With the development of the sophisticated capabilities described in this paper (see 
Chapter~\ref{ch:5}), advanced AI assistants hold the potential to greatly improve the opportunities of those who have good access to them, particularly if these technologies support or perform an increasing range of interpersonal and institutional activities. However, we also live in a world shaped by interlocking inequalities \citep{crenshaw_demarginalizing_2015, combahee_river_collective_combahee_1977, hill_collins_black_2009} where access to opportunities, goods and technologies is often unevenly distributed and shaped by hierarchies including those pertaining to gender, sexuality, disability, religion, race and ethnicity. In a future where the use of advanced AI assistants is a boon for users, but where there has not been intentional design for equitable access \citep{rigot_design_2022, costanza-chock_design_2020, ovalle_factoring_2023, davis_algorithmic_2021}, we risk encountering a divide between the `haves' and `have-nots' that operates across these different dimensions of opportunity and marginalisation. Failure to consider existing social structures, and the ways they interact with the design and function of technology, could result in forms of AI that increase access in some areas while simultaneously compounding asymmetric or harmful relationships in other walks of life \citep{bennett_what_2020, tucker_technocapitalist_2017}.

As we show below, the process of designing more general AI assistants -- and even assistants that target beneficial use cases in high impact domains, such as health, education or economic empowerment \citep{chui__notes_2018} -- may not be sufficient to ensure that the technology performs well for marginalised communities or to provide these groups with high-quality access to the opportunities it creates. To achieve these outcomes, further steps are needed, including those that speak to the specific needs and experiences of these communities and to the kinds of obstacle they encounter (\citeauthor{mingus_access_2017}, \citeyear{mingus_access_2017}; \citeauthor{zajko_artificial_2022}, \citeyear{zajko_artificial_2022}; \citeauthor{bennett_interdependence_2018}, \citeyear{bennett_interdependence_2018}; see Chapter~\ref{ch:20}). The more holistic approach proposed here requires methods that reach beyond `computational correctives [that] invariably fall short' when viewed as the complete answer to complex societal issues (\citeauthor{davis_algorithmic_2021}, \citeyear{davis_algorithmic_2021},~1). Instead, efforts to support widespread opportunity and access need to attend to various power dynamics that shape how technologies, including AI assistants, are developed and experienced by different communities and users -- and to take measures to address these dynamics when envisaging their deployment and use.

To achieve positive outcomes, advanced AI assistants must be designed both \emph{for} and \emph{with} historically marginalised communities to ensure that they are properly responsive to the needs of those who have sometimes been pushed to the periphery \citep{bennett_care_2020}. Too often, technological pathways involving access to resources and opportunity have been disproportionately designed by and for people who are not fully representative of the societies that these technologies are situated within \citep{henrich_most_2010, linxen_how_2021}. Altering this pattern requires identifying levers or processes for representation and inclusion that can affect meaningful change \citep{crenshaw_demarginalizing_2015, combahee_river_collective_combahee_1977} and intentionally designing with the needs of everyone in mind \citep{benjamin_race_2020, rigot_design_2022}. Without such praxis, power imbalances that operate at a societal level have the potential to be replicated via inequitable access to and through advanced AI assistants. By way of contrast, AI systems that proactively tackle these deeper disparities hold out hope for greater equity in access and in the distribution of benefits. We add nuance to this picture below. For now, we begin by assuming that advanced AI assistants will be presumptively beneficial to those with full access to them (`users'). 

\section{Inequality and Technology}\label{sec:16:2}

\emph{Inequitable access to technology} refers to disparities in how digital technologies are structured, accessed and used \citep{kvasny_cultural_2006}. As digital technologies are increasingly part of the infrastructure needed for meaningful participation in social, economic, and political life \citep{oecd_bridging_2018, un_addressing_2021}, inequitable access to technology may lead to social exclusion in these and other domains affecting material well-being \citep{robinson_digital_2015} (see 
Chapter~\ref{ch:7}). The stakes involved in equitable access are high: `information technology, and the ability to use it and adapt it, is a critical factor in generating and accessing wealth, power and knowledge in our time' (\citeauthor{castells_rise_2009}, \citeyear{castells_rise_2009},~92). Scholars have mapped out the economic, political and social harms of digital inequality, including barriers to economic and educational opportunities and cultural capital \citep{van_dijk_digital_2006}. Moreover, the \emph{distribution} of these harms is often patterned along existing axes of inequality, including gender, race and ethnicity, class, disability and nationality, among others \citep{wasserman_gender_2005, ono_immigrants_2008, mesch_ethnic_2011, witte_internet_2010, bennett_interdependence_2018}. In this way, the (in)accessibility of technology is constitutively intertwined with power and opportunity at the societal level.\footnote{The authors understand that `accessibility' is a term most-often used in the US to refer to accessibility for people with disabilities. However, we employ a more expansive notion of the term here to reflect a wide range of sociopolitical factors shaping `access to or inaccessibility to engage with technology' (\citeauthor{bjorn_equity_2023}, \citeyear{bjorn_equity_2023},~88).}

Addressing inequitable access to technology requires researchers, developers and policymakers to move beyond `technosolutionist' paradigms that focus solely on the availability or ownership of a technological device as the answer to social problems \citep{warschauer_reconceptualizing_2002, warschauer_technology_2004}. Indeed, earlier work on inequitable access has sometimes been critiqued for its binary focus on a one-dimensional frame of `haves' and `have-nots'. This focus tends to ignore the dynamic nature of digital inequality \citep{dimaggio_digital_2023, warschauer_new_2010}, which extends to include motivational access (e.g.\ wanting/trust in technology), physical access (e.g.\ owning a device), skills (e.g.\ understanding how to use a technology) and use (e.g.\ ways of using a technology) \citep{van_dijk_digital_2006}. As a consequence, people may be able to access technology in one sense but not another, thus frustrating its potential for productive use. A more holistic view of `access' enables a shift away from interventions focused only on device ownership as a pathway for access (e.g.\ the widely criticised `One Laptop Per Child' initiative, \citeauthor{keating_why_2009}, \citeyear{keating_why_2009}),\footnote{While the programme initially received much positive attention and was endorsed by the United Nations Development Programme (\citeauthor{AP_2006}, \citeyear{AP_2006}), it largely failed due to the sociocultural mismatch between the expectations and views of what laptops can achieve in Western nations versus the context and expectations in which they were deployed in the Global South \citep{mcarthur2009communication}.} towards recognition that fostering meaningful and collective access requires deeper attention to the ways technology is designed, deployed and experienced with equity and social justice in mind. Ultimately, we believe that technological access is best thought of as a kind of social relationship involving `bundles and webs of powers that enable actors to gain, control and maintain access' to opportunities and goods (\citeauthor{ribot_theory_2003}, \citeyear{ribot_theory_2003},~154). 

Questions about access to technology fundamentally concern social norms and expectations about embodied ways of being in the world. All technologies are reflections of a world view \citep{alkhatib_live_2021}, and the ways they are designed and deployed says something about who they are meant to serve and about who belongs and who does not. As disability studies scholar \citeauthor{titchkosky_question_2011} (\citeyear{titchkosky_question_2011},~6) notes: `exploring the meanings of access is, fundamentally, the exploration of the meaning of our lives together -- who is together with whom, how, where, when and why'. Through this lens, the relational notion of `access' offers an entry point for deeper consideration of how technology may be shaped by social norms that foreground the needs, priorities and perspective of some user groups -- for example those who are able-bodied \citep{bennett_interdependence_2018, gregor_disability_2005} -- over others. Given that questions of `access' implicate how differently situated communities interact and relate to one another, `access' can be productively employed as a lens for examining what kinds of relations technologies engender at the societal level. 

\section{Case Studies: Access, Opportunity and AI}\label{sec:16:3}

As numerous authors have shown, AI technologies are not value neutral, often serving to shape opportunities, pathways and dependencies across the contexts in which they are deployed \citep{chun_discriminating_2021, broussard_more_2023, amoore_cloud_2020}. Furthermore, all AI technologies are reflections of the social world, embodying choices made by developers in AI pipelines \citep{suresh_framework_2021, alkhatib_live_2021, ovalle_factoring_2023}. A dominant paradigm for developing technology is to design for an imagined `biggest use case' scenario, which most often focuses on user communities that are located in white, middle- and upper-class, Western-centric contexts \citep{rigot_design_2022}. Much existing research documents the limitations of this development paradigm, including through the lenses of gender \citep{bivens_baking_2016, noble_algorithms_2020}, disability \citep{bennett_what_2020, morris_ai_2020, whittaker_disability_2019}, race \citep{sweeney_discrimination_2013, benjamin_race_2020} and geopolitical context \citep{kak_global_2020, png_at_2022, sambasivan_re-imagining_2021}. Together, these studies illustrate how a specific vision of product development can embed inequities into AI systems leading to unequal performance and access in the real world \citep{harrington_its_2022, martin_bias_2023, dorn_dialect-specific_2019, devito_values_2021}. These insights underscore the need to address inequality in technology development by designing for and with the margins. Indeed, ignoring the impact on different communities and user groups risks falling into a `false universalism' which benefits some demographics at the expense of others (\citeauthor{roberts_movement_2013}, \citeyear{roberts_movement_2013},~315). In the remainder of this section, we discuss two case studies involving potentially unequal access: \emph{AI voice assistants} and \emph{required use} of digital technologies.

\subsection{Disparate performance of AI voice assistants based on identity}

Inferior \emph{quality of access} to technology, for certain user communities, has been uncovered for a wide range of AI technologies that rely on biometric data (e.g.\ facial features, skin tone or voice) as system inputs, including computer vision \citep{buolamwini_gender_2018, raji_saving_2020} and speech recognition systems \citep{koenecke_racial_2020, mengesha_i_2021}. AI technologies with disparate performance potentially lead to quality-of-service harms, which unevenly distribute the social benefits of a technology and may lead to users experiencing alienation, additional labour and service or benefits loss when access is of limited quality \citep{shelby_sociotechnical_2023}. For instance, \citeauthor{mengesha_i_2021}'s (\citeyear{mengesha_i_2021}) study of AI voice assistant interaction with users who speak African-American Vernacular English found that the failures of these systems for this speaker group have: (1) behavioural ramifications that influence how users adapt to technology (e.g.\ changing how they speak) and (2) psychological impacts, such as feelings of alienation, when users recognise that a technology is failing them because of their identity. 

In these cases, inferior quality of access has the potential to exacerbate existing social inequalities (e.g.\ \citeauthor{kazenwadel_how_2023}, \citeyear{kazenwadel_how_2023}), for example, those regarding race \citep{massey_american_1993, bonilla-silva_rethinking_1997, feagin_continuing_1991, feagin_systemic_2013, dubois_philadelphia_1996}. Disparate performance of technologies along lines of race is one mechanism through which systemic racism may be enacted in the digital domain (\citeauthor{zalnieriute_how_2022}'s report to the UN Human Rights Council, \citeyear{zalnieriute_how_2022}). To accommodate inferior quality of access in practice, a user may `code switch', meaning they have to use a different language, dialect or accent to make the technology work or improve its performance \citep{mengesha_i_2021}. In the words of one participant: `because of my race and location, I tend to speak in a certain way that some voice technology may not comprehend. When I don’t speak in my certain dialect, I come to find out that there is a different result in using voice technology' (\citeauthor{mengesha_i_2021}, \citeyear{mengesha_i_2021},~7). This situation of differential access leads to lower utility and worse outcomes for these users. Users adapting their speech instead of using their native dialect also makes it harder for developers to detect when the device works poorly for a given language variety without direct consultation or the development of new feedback mechanisms \citep{weidinger_ethical_2021, birhane_power_2022}. 

\subsection{Required access: The impacts of digital technology as societal infrastructure}

When access to a particular digital technology is \emph{required to engage} with an organisation (e.g.\ governments \citep{thiel_biometric_2020}, social services \citep{eubanks_technologies_2006} or humanitarian aid \citep{iazzolino_infrastructure_2021}), access-related concerns intertwine with issues of consent, autonomy, and surveillance. For instance, information infrastructure enables processing of personal data and protocols to `mediate between individuals and the organisations with which we relate' (\citeauthor{lyon_biometrics_2008}, \citeyear{lyon_biometrics_2008},~500). The `access' lens offers a way to trace and reveal the politics of an information infrastructure. For instance, India's Aadhaar (unique identity number) software requires facial, iris and fingerprint scans to access financial entities (e.g.\ Amazon Pay) and Indian government benefits \citep{macdonald_india_2023}, open a business or register for a goods and services tax number \citep{burt_aadhaar_2023}. At the US border, migrants who do not use the US Customs and Border Protection app, CBP One, encounter a higher bar for claiming asylum \citep{gottesdiener_bidens_2023}. The Atlantic Plaza Towers complex in New York City, a rent-controlled apartment complex with predominantly Black residents, required the use of facial recognition to access the building \citep{moran_atlantic_2020} despite such systems' documented performance inequities across axes of gender and race \citep{buolamwini_gender_2018}. In these examples, the lens of `access' illustrates how refusal to participate carries financial, legal and social risks. However, even if someone wants to use a particular technology, if the system performs poorly for them, it may carry the same punitive consequences of opting out.

The above case studies illustrate different dynamics that could influence how communities experience advanced AI assistants across a range of contexts. To address these risks, and ensure that existing inequalities are not compounded, design interventions geared towards widespread access and opportunity are needed (\citeauthor{helsper_digital_2021}, \citeyear{helsper_digital_2021}; see 
Chapter~\ref{ch:15}). We discuss the challenges and opportunities created by advanced assistants in more detail below.

\section{Access and Advanced AI Assistants}\label{sec:16:4}

Experiences of differential access to technology can take many forms, including total lack of access, inferior quality of access and access to an actively bad, misaligned or punitive system.\footnote{While in this paper we focus on inequitable access, we could certainly imagine alternative situations to those on the above list, in which those with `good' access experience some benefits in conjunction with wider destructive effects (e.g.\ addiction to or overdependence on the technology).} Indeed, one or more of these situations of differential access may occur simultaneously, depending on the context. To illustrate the various manifestations of unequal access, we describe below a suite of situations which emphasise the social impacts and type of access that other, differently situated communities may have. 

\textbf{Situation type~1:} First, we consider differential access situations where the advanced AI assistant is \emph{beneficial to those who have access}, but other people do \emph{not have any access} to the assistant, as they are totally shut out from use (e.g.\ via paywalls or need for an institutional affiliation to gain access). In such circumstances, the cost of missed opportunities from access to advanced AI assistants could be either consequential or inconsequential to those without access. For example, this could be an education assistant that helps those who have access to write better college application essays and thus obtain better educational opportunities \citep{singer_applying_2023}. It may also be an assistant that helps those who have it to gain faster and easier access to basic goods and services, such as healthy food and government benefits. Those without access to the assistant must spend more time and resources to obtain the same goods.\footnote{For example, \citeauthor{nowpow_nowpow_nodate}.} Alternatively, there are situations where --- while only a subset of society may have access --- there are still beneficial knock-on effects to those who do not have access (e.g.\ an advanced AI assistant that spurs socially beneficial scientific or medical research,\footnote{A current day example is, perhaps, the AlphaFold Protein Structure Database \citep{alphafold_alphafold_nodate}.} one that supports democratic institutions by ensuring politicians are held accountable to the citizenry or another that reduces the prevalence of misinformation) \citep[e.g.][]{harutyunyan_leveraging_2023}. In such cases, while only some researchers may have access, they could make discoveries that benefit those without access to the technology, thus creating positive spillover effects (see 
Chapters~\ref{ch:15} and ~\ref{ch:18}).\footnote{However, note that the diversity and experiences of researchers is consequential to the applicability of their research and problem areas they pursue (e.g. \citeauthor{harding1986sciencequestion}, \citeyear{harding1986sciencequestion}; \citeauthor{longino1993}, \citeyear{longino1993}).} Even in these scenarios, existing social inequalities may shape the distribution and flow of potential AI-assisted discoveries, as was seen with inequalities in the distribution of Covid-19 vaccines\footnote{For example, Pfizer has characterised the development of Covid-19 vaccines as AI-assisted \citep{pfizer_how_nodate}.} (\citeauthor{tatar_covid-19_2022}, \citeyear{tatar_covid-19_2022},~2).

\textbf{Situation type~2:} The second type of situation to consider is where some users experience full, beneficial and high-quality access while others have \emph{inferior quality of access} to an otherwise good assistant. This is a classic, inequitable quality-of-service situation. These user experiences may include slower or less advanced access, where the access is inferior due to, for example, slower internet speeds (e.g.\ net neutrality; see \citeauthor{finley_net_nodate}, (\citeyear{finley_net_nodate}) or other infrastructural differences leading to the `digital divide'\footnote{The `digital divide' in its simplest formulation refers to the `gap between those with Internet access and those without it' \citep{muller_what_2022}.}) or a lower tier/free version of a paid subscription (e.g.\ \citeauthor{rogers_is_nodate}, \citeyear{rogers_is_nodate}; \citeauthor{shankland_photoshops_2023}, \citeyear{shankland_photoshops_2023}). In this case, the technology would perhaps take significantly longer to load, join queues and so on (see 
Chapter~\ref{ch:15}). The user can still access most services, but time costs are elevated or there is limited availability of some capabilities (e.g.\ plugins). Opportunities may be missed due to these issues. Alternatively, some users could experience less smooth or buggier access to the assistant. In these situations, the assistant may not be able to execute every necessary task, or it may be less-aligned with the users' intentions (see 
Chapter~\ref{ch:6}). As an example of how this could occur, we could envision an advanced AI assistant suffering from a disparate performance issue seen in modern-day digital assistants (e.g.\ \citeauthor{mengesha_i_2021}, \citeyear{mengesha_i_2021}; \citeauthor{lima_empirical_2019}, \citeyear{lima_empirical_2019}; \citeauthor{wu_see_2020}, \citeyear{wu_see_2020}) where English speakers with a regional or non-American accent are unable to access time-saving services such as to-do lists, etc., as the assistant does not understand the users' speech. In such situations, the user may have a frustrating experience, but the assistant is generally still usable and useful. These users may experience common quality-of-service harms, as delineated by \citet{shelby_sociotechnical_2023}. 

\textbf{Situation type 3:} In the last set of situations, some users \emph{only have access} to an \emph{actively bad, punitive or misaligned} AI assistant, meaning that the technology is not merely frustrating or inconvenient to use, but has actively negative effects for these users and/or society (e.g.\ forms of misalignment, or mistakes made by the assistant, have a punitive impact on the user or a resounding societal impact) (see 
Chapters~\ref{ch:6},~\ref{ch:8} and~\ref{ch:9}). An example could be one that directly mirrors present-day issues with facial recognition use in policing in the US \citep{johnson_police_2023}. An advanced AI assistant could be employed by police officers that is less capable of distinguishing between Black profiles, thus causing the wrong individuals to be apprehended \citep{johnson_how_nodate}. This situation would in one respect be bad for police officers (the users, in this case) because they end up apprehending the wrong person and presumably generate a negative public reaction or expose municipalities to civil suits \citep{benedicto_detroit_nodate, thanawala_ai_nodate}. However, the situation would be still worse for society and those incorrectly arrested (presumably non-users). As the introduction of algorithmic and big data tools in police departments have already shown, they can reproduce and deepen patterns of social inequality and power imbalances \citep{brayne_predict_2021, ferguson_rise_2017}. This example underscores the need to consider the broader context in which advanced AI assistants will be used.

Notably, the effects of these access issues may be \emph{detectable to the user},\footnote{A second example could be analogous to a situation described by \citet{eubanks_automating_2017} where, to access social services, welfare claimants are forced to use an assistant rather than speak to a human. The effect of this erodes interpersonal relationships between claimants and caseworkers, and it increases denial of social services through errors in automated decision-making.} as in the police facial recognition example described above, or they may be \emph{undetectable} if the error happens `behind the scenes', thus potentially leaving the user or those impacted in a state of uncertainty or confusion. This could potentially occur in job application and hiring situations, where the user is not offered a position but cannot know if that was because they were not the right fit or because they had a faulty/misaligned AI assistant \citep{bogen_all_2019}. Alternatively, if the advanced AI assistant is poor at providing factual information in a non-English language, it could create an experience that contains errors or misinformation (see 
Chapter~\ref{ch:17}). Indeed, misinformation generated  in this way may be subtle and not easily detectable by the user. This case reflects the present-day challenges encountered by content moderation on social media, where journalists' accounts have sometimes been removed by mistake and monitoring systems have proved less capable of detecting non-English misinformation (e.g.\ \citeauthor{fatafta_facebook_2021}, \citeyear{fatafta_facebook_2021}; \citeauthor{avaaz_how_2020}, \citeyear{avaaz_how_2020}). For an advanced AI assistant with greater capabilities, malignant errors that are hard to detect could be more pronounced or spread at a greater rate, and they could lead to the assistant carrying out tasks that are damaging to the the user or their broader community (see 
Chapter~\ref{ch:8}). 

In the \emph{differential access} situations described above, the groups of people who do not have full or fully beneficial access may be either \emph{randomly} or \emph{systematically} distributed across society. Random distribution might not be due to any systemic societal issues but to, for example, the inherent probabilistic errors that an advanced AI assistant will make \citep{bommasani_picking_2022, creel_algorithmic_2022}. For example, generative AI systems often contain conventional machine-learning (ML) models (e.g.\ binary or multi-class classifiers employed as input/output filters) for automated content moderation strategies to align to established product policies \citep{solaiman_gradient_2023}. These safety classifiers might take down some accounts at random. Alternatively, safety classifiers may disproportionately fail for certain social groups in a way that is correlated with the manner in which that speaker group uses language (e.g.\ \citeauthor{dias_oliva_fighting_2021}, \citeyear{dias_oliva_fighting_2021}), and the errors may be \emph{systematically} distributed, necessitating specific interventions designed to address this effect (e.g.\ \citeauthor{hao_safety_2023}, \citeyear{hao_safety_2023}). 

In general, more systematically distributed access restrictions occur when lack of -- or poorer quality -- access correlates with other vectors of exclusion such as race, class, disability, living in the Global South or speaking a language other than English. Furthermore, the situation types described in this chapter illuminate how in the case of an advanced AI system, inaccessibility can either be \emph{direct}, meaning the inaccessibility is to the advanced AI assistant itself, or \emph{indirect}, where the primary point of inaccessibility is to other goods, services and opportunities. These distinctions are important to recognise as we seek both mitigations to inequities and to facilitate broad access to resources. In particular, a direct access issue calls our attention to the barriers that may be preventing access to the technology itself (e.g.\ not being able to purchase a smartphone, or other examples of what \citep{van_dijk_digital_2006} calls good physical access, whereas problems of indirect access often require action in relation to the opportunities and affordances that access to the technology provides.

With any given technology, multiple \emph{situations of differential access} can be at play at once, and they can reproduce social divisions and unequal material outcomes without proactive mitigation. This also holds for advanced AI assistants. We next consider access-related risks that may arise for advanced AI assistants, alongside an understanding of current technical capabilities, before discussing potentially emergent access-related risks for more capable technologies. 

\section{Access-Related Risks and Advanced AI Assistants}\label{sec:16:5} 

The most serious access-related risks posed by advanced AI assistants concern the entrenchment and exacerbation of existing inequalities \citep{world_inequality_database_world_nodate} or the creation of novel, previously unknown, inequities. While advanced AI assistants are novel technology in certain respects, there are reasons to believe that -- without direct design interventions -- they will continue to be affected by inequities evidenced in present-day AI systems \citep{bommasani_picking_2022}. Many of the access-related risks we foresee mirror those described in the case studies and types of differential access. In this section, we link them more tightly to elements of the \emph{definition} of an advanced AI assistant to better understand and mitigate potential issues -- and lay the path for assistants that support widespread and inclusive opportunity and access. We begin with the existing capabilities set out in the definition (see 
Chapter~\ref{ch:3}) before applying foresight to those that are more novel and emergent.

\subsection{Current capabilities: Artificial agents with natural language interfaces} 

Artificial agents with \emph{natural language interfaces} are widespread \citep{browne_all_2023} and increasingly integrated into the social fabric and existing information infrastructure, including search engines \citep{warren_microsoft_2023}, business messaging apps \citep{slack_claude_nodate}, research tools \citep{atlas.ti_atlas.ti_nodate} and accessibility apps for blind and low-vision people \citep{be_my_eyes_introducing_nodate}. There is already evidence of a range of sociotechnical harms that can arise from the use of artificial agents with natural language interfaces when some communities have inferior access to them \citep{weidinger_ethical_2021}. As previously described, these harms include inferior quality of access (in situation type 2) across user groups, which may map onto wider societal dynamics involving race \citep{harrington_its_2022}, disability \citep{gadiraju_i_2023} and culture \citep{jenka_ai_2023}. As developers make it easier to integrate these technologies into other tools, services and decision-making systems (e.g.\ \citeauthor{marr_microsofts_nodate}, \citeyear{marr_microsofts_nodate}; \citeauthor{brockman_introducing_2023}, \citeyear{brockman_introducing_2023}; \citeauthor{pinsky_bard_2023}, \citeyear{pinsky_bard_2023}), their uptake could make existing performance inequities more pronounced or introduce them to to new and wider publics. 

At the same time, and despite this overall trend, AI systems are also not easily accessible to many communities. Such direct inaccessibility occurs for a variety of reasons, including: purposeful non-release (situation type 1; \citeauthor{wiggers_chatgpt:_2023}, \citeyear{wiggers_chatgpt:_2023}), prohibitive paywalls (situation type 2; \citeauthor{rogers_is_nodate}, \citeyear{rogers_is_nodate}; \citeauthor{shankland_photoshops_2023}, \citeyear{shankland_photoshops_2023}), hardware and compute requirements or bandwidth (situation types 1 and 2; \citeauthor{openai_technical_2023}, \citeyear{openai_technical_2023}), or language barriers (e.g.\ they only function well in English (situation type 2; \citeauthor{snyder_ais_2023}, \citeyear{snyder_ais_2023}), with more serious errors occurring in other languages (situation type 3; \citeauthor{deck_we_2023}, \citeyear{deck_we_2023}). Similarly, there is some evidence of `actively bad' artificial agents gating access to resources and opportunities, affecting material well-being in ways that disproportionately penalise historically marginalised communities \citep{bogen_all_2019, block_how_2022, eubanks_automating_2017}. Existing direct and indirect access disparities surrounding artificial agents with natural language interfaces could potentially continue -- if novel capabilities are layered on top of this base without adequate mitigation (see 
Chapter~\ref{ch:4}).

\subsection{Novel capabilities: Access-related risks for advanced AI assistants}

AI assistants currently tend to perform a limited set of isolated tasks: tools that classify or rank content execute a set of predefined rules or provide constrained suggestions, and chatbots are often encoded with guardrails to limit the set of conversation turns they execute (e.g.\ \citeauthor{warren_microsoft_2023}, \citeyear{warren_microsoft_2023}; see 
Chapter~\ref{ch:5}). However, an artificial agent that can \emph{execute sequences of actions on the user's behalf} -- with `significant autonomy to plan and execute tasks within the relevant domain' (see 
Chapter~\ref{ch:3}) -- offers a greater range of capabilities and depth of use. This raises several distinct access-related risks, with respect to liability and consent, that may disproportionately affect historically marginalised communities. 

To repeat, in cases where an action \emph{can only be executed} with an advanced AI assistant, not having access to the technology (e.g.\ due to limited internet access, not speaking the `right' language or facing a paywall) means one cannot access that action (consider today's \emph{eBay} and \emph{Ticketmaster} bots). Communication with many utility or commercial providers currently requires (at least initial) interaction with their artificial agents \citep{schwerin_somehow_2023, verma_chatgpt_2023}. It is not difficult to imagine a future in which a user needs an advanced AI assistant to interface with a more consequential resource, such as their hospital for appointments or their phone company to obtain service. Cases of inequitable performance, where the assistant \emph{systematically performs less} well for certain communities (situation type~2), could impose serious costs on people in these contexts. 

Moreover, advanced AI assistants are expected to be designed to act in line with user \emph{expectations}. When acting on the user's behalf, an assistant will need to infer aspects of what the user wants. This process may involve interpretation to decide between various sources of information (e.g.\ stated preferences and inference based on past feedback or user behaviour) (see 
Chapter~\ref{ch:6}). However, cultural differences will also likely affect the system's ability to make an accurate inference. Notably, the greater the cultural divide, say between that of the developers and the data on which the agent was trained and evaluated on, and that of the user, the harder it will be to make reliable inferences about user wants (e.g.\ \citeauthor{beede_human-centered_2020}, \citeyear{beede_human-centered_2020}; \citeauthor{widner_lessons_2023}, \citeyear{widner_lessons_2023}), and greater the likelihood of performance failures or value misalignment (see 
Chapter~\ref{ch:12}). This inference gap could make many forms of indirect opportunity inaccessible, and as past history indicates, there is the risk that harms associated with these unknowns may disproportionately fall upon those already marginalised in the design process.

\subsection{Emergent access risks for advanced AI assistants} 

Emergent access risks are most likely to arise when current and novel capabilities are combined. Emergent risks can be difficult to foresee fully \citep{prunkl_institutionalizing_2021, ovadya_reducing_2019} due to the novelty of the technology (see 
Chapter~\ref{ch:2}) and the biases of those who engage in product design or foresight processes \citet{dignazio_1._2020}. Indeed, people who occupy relatively advantaged social, educational and economic positions in society are often poorly equipped to foresee and prevent harm because they are disconnected from lived experiences of those who would be affected. Drawing upon access concerns that surround existing technologies, we anticipate three possible trends:

\begin{itemize} [parsep=6pt]
\item \textbf{Trend~1: Technology as societal infrastructure.} If advanced AI assistants are adopted by organisations or governments in domains affecting material well-being, `opting out' may no longer be a real option for people who want to continue to participate meaningfully in society. Indeed, if this trend holds, there could be serious consequences for communities with no access to AI assistants or who only have access to less capable systems (see also 
Chapter~\ref{ch:15}). For example, if advanced AI assistants gate access to information and resources, these resources could become inaccessible for people with limited knowledge of how to use these systems, reflecting the skill-based dimension of digital inequality \citep{van_dijk_digital_2006}. Addressing these questions involves reaching beyond technical and logistical access considerations -- and expanding the scope of consideration to enable full engagement and inclusion for differently situated communities.

\item \textbf{Trend~2: Exacerbating social and economic inequalities.} Technologies are not distinct from but embedded within wider sociopolitical assemblages \citep{haraway_situated_1988, harding_is_1998, harding_whose_2016}. If advanced AI assistants are institutionalised and adopted at scale without proper foresight and mitigation measures in place, then they are likely to scale or exacerbate inequalities that already exist within the sociocultural context in which the system is used \citep{bauer_artificial_2021, zajko_artificial_2022}. If the historical record is anything to go by, the performance inequities evidenced by advanced AI assistants could mirror social hierarchies around gender, race, disability and culture, among others -- asymmetries that deserve deeper consideration and need to be significantly addressed (e.g.\ \citeauthor{buolamwini_gender_2018}, \citeyear{buolamwini_gender_2018}). 

\item \textbf{Trend~3: Rendering more urgent responsible AI development and deployment practices}, such as those supporting the development of technologies that perform fairly and are accountable to a wide range of parties. As \citeauthor{corbett_interrogating_2023} (\citeyear{corbett_interrogating_2023},~1629) argue: `The impacts of achieving [accountability and fairness] in almost any situation immediately improves the conditions of people's lives and better society'. However, many approaches to developing AI systems, including assistants, pay little attention to how context shapes what accountability or fairness means \citep{sartori_sociotechnical_2022}, or how these concepts can be put in service of addressing inequalities related to motivational access (e.g.\ wanting/trust in technology) or use (e.g.\ different ways to use a technology) \citep{van_dijk_digital_2006}. Advanced AI assistants are complex technologies that will enable a plurality of data and content flows that necessitate in-depth analysis of social impacts. As many sociotechnical and responsible AI practices were developed for conventional ML technologies, it may be necessary to develop new frameworks, approaches and tactics (see Chapter~\ref{ch:20}). We explore practices for emancipatory and liberatory access in the following section.
\end{itemize}

\section{Beyond Mitigation: From Unequal to Liberatory Access}\label{sec:16:6}

The (in)accessibility of technology is constitutively intertwined with social inequality. Conversely, meaningful access to technology can be understood as a way of challenging inequality and a way of enabling productive cooperation between individuals understood as equals \citep{anderson_what_1999}. As we have sought to elucidate in this chapter, questions of `access' implicate how differently situated communities interact and relate to one another through technology. We have also shown how `access' can be employed as a lens through which to examine kinds of social power relations technologies engender. Against this backdrop, disability justice scholar \citet{mingus_access_2017} writes: `liberatory access calls upon us to create different values\ldots and demands that the responsibility for access shifts from being an individual responsibility to a collective responsibility'. Liberatory access provides a goal and set of methods for developing sociotechnical systems that embody the kinds of social relations that challenge social inequalities and support mutual flourishing.

One approach to developing technology that embodies liberatory access is by designing for the margins. \emph{Design for the margins} (DFM) is an approach to design that centres the most impacted and marginalised users from ideation to production \citep{rigot_design_2022}. Conventional design processes sometimes view marginalised users as `edge cases' whose needs are framed as different or `extra'. Their needs are commonly retrofitted to an already designed technology. By way of contrast, DFM places their needs at the centre of the design process to dismantle systems of interlocking inequality \citep{hill_collins_black_2009}. Crucially, users on the periphery who often receive the least support tend to possess deep experiential knowledge of how to improve technologies that can be better for everyone \citep{rigot_design_2022}. DFM offers a methodology for rethinking approaches to `participation' in product development -- with the goal of bringing this knowledge back in and explicitly centering those who are marginalised in a given context. Overall, this approach tends to foster more equitable and safe technologies,\footnote{For example, development of a dating app for LGBTQ people might focus on the needs of potential users whose sexuality is criminalised (\citeauthor{article19_2018}, \citeyear{article19_2018}), or considerations of privacy or safety might centre the needs of journalists and activists in parts of the world that are subject to harassment, surveillance, arrest or assassination (\citeauthor*{article19_2022}, \citeyear{article19_2022}; \citeauthor{warford2022}, \citeyear{warford2022}).} including identification of necessary interventions such as inclusive education and customisable interfaces, so that the technology can be fully responsive to marginalised users' needs.

Placing these communities at the centre of the design process opens possibilities for more inclusive, liberatory technologies. Many participatory design approaches have been critiqued as \emph{extractive} insofar as they engage communities through consultation without attention to context \citep{sloane_participation_2022} or through `tokenistic forms of `voice' that fail to redistribute power and agency' \citep{ymous_i_2020}. The aspiration to embed participation-as-justice proceeds differently and involves developing long-term relationships `based on mutual benefit, reciprocity, equity and justice' (quoted in \citeauthor{suresh_towards_2022}, \citeyear{suresh_towards_2022},~667). Importantly, DFM centres those most impacted in the design process while being attentive the broader sociopolitical and institutional contexts in which technologies operate and exist. This requires rethinking the design process in potentially significant ways. As described by \citet{rigot_design_2022}, implementing this approach requires: (1) identifying and prioritising communities who bear the most risk and have the least protection, and bringing in facilitators who can identify those communities and have trusted relationships with them, especially when directly involving a particular community in the design process poses safety risks (e.g.\ \citeauthor{bellini_sok:_2023}, \citeyear{bellini_sok:_2023}); (2) zooming in on social, legal and political issues that arise within that particular context; (3) centring the needs of communities from ideation through development, not just including them at later stages of a product development life cycle after numerous decisions have already been made; and (4) regeneralising so that these findings can be scaled and applied alongside insights from user groups who typically are centred in product development (\citeauthor{rigot_design_2022}, \citeyear{rigot_design_2022},~65). By centring the margins, DFM offers the kind of \emph{transformative approach to technology} and the design of sociotechnical systems that can facilitate liberatory access in service to the goals of opportunity, access and liberation \citep{mingus_access_2017}. 

\section{Conclusion}\label{sec:16:7}

Questions about access to technology fundamentally concern social norms and expectations about how communities interact and relate to one another. This chapter has discussed how `access' can be employed as a lens for examining the kinds of relations technology engenders in ways that extend beyond questions of merely technical and logistical access. Both the developers of advanced AI assistants \emph{and} the organisations that adopt them have a responsibility to consider these relationships and to assess and mitigate access-related risks. For developers, attention must be given to how assistants are constructed and who they are optimised for. This requires an understanding of existing access inequalities and concerted engagement with different publics to understand how to address them effectively \citep{bjorgvinsson_participatory_2010, dantec_infrastructuring_2013, erete_method_2023}. For organisations that adopt advanced AI assistants, attention must be paid to technological limitations and risks of adoption, in addition to their benefits. Moreover, these organisations have a duty to understand how performance limitations of technological agents will interact with the existing access inequalities in their domain that are experienced by their clients or constituents. One way for developers and organisational adopters of advanced AI assistants to approach this work is by employing `access' as a lens for anticipating potential situations of differential access and who might experience them, and by drawing on multidisciplinary best practices, particularly from fields focused on equity and access, such as disability justice (e.g.\ \citeauthor{berne_ten_2018}, \citeyear{berne_ten_2018}; \citeauthor{mingus_changing_2011}, \citeyear{mingus_changing_2011}). These fields can provide the political container necessary for grounding analyses and materially moving towards liberatory access.

\chapter{Misinformation}
\label{ch:17}

\textbf{Nahema Marchal, Iason Gabriel, Arianna Manzini, Geoff Keeling, Beth Goldberg, Josh Goldstein}

\noindent \textbf{Synopsis}: 
Advanced AI assistants pose four main risks for the information ecosystem. First, AI assistants may make users more \emph{susceptible} to misinformation, as people develop trust relationships with these systems and uncritically turn to them as reliable sources of information. Second, AI assistants may provide ideologically \emph{biased} or otherwise \emph{partial information} to users in attempting to align to user expectations. In doing so, AI assistants may reinforce specific ideologies and biases and compromise healthy political debate. Third, AI assistants may \emph{erode} societal \emph{trust} in shared knowledge by contributing to the dissemination of large volumes of plausible-sounding but low-quality information. Finally, AI assistants may facilitate \emph{hypertargeted disinformation} campaigns by offering novel, covert ways for propagandists to manipulate public opinion. This chapter articulates these risks and discusses technical and policy mitigations.

\section{Introduction}
Recent advances in the field of AI have enabled AI systems to develop unprecedented capabilities, such the ability to generate human-like text, images, video and audio \citep{spitale_ai_2023, nightingale_ai-synthesized_2022}, to teach themselves how to reason, use external tools and to take actions in the real world \citep{mialon_augmented_2023, schick_toolformer:_2023}.

With these advances come growing concerns about the potential for AI systems to spread misinformation and fuel online influence operations. A recent survey found that three in four Americans are concerned about AI driving mis- and disinformation \citep{ipsos_americans_2023}, and leading AI labs have also flagged this as a major risk in relation to large language models (LLMs), stating that with wider adoption, advanced AI systems could ‘reinforce entire ideologies, worldviews, truths and untruths [\dots] cement them or lock them in, foreclosing future contestation, reflection and improvement’ (\citeauthor{openai_gpt-4_2023}, \citeyear{openai_gpt-4_2023}, 9).

As these systems are rapidly integrated into a range of user-facing applications, including virtual AI assistants \citep{knight_google_nodate}, it is therefore important to think about the unique challenges this particular form factor might pose for the \emph{integrity of our information environment}. How might widespread adoption of highly capable and adaptive AI assistants impact the spread of misinformation and political propaganda? What impact might these technologies have on information retrieval, knowledge and beliefs? How might this impact public discourse, and what can be done to mitigate these threats? 

This chapter focuses on the risks posed by AI assistants, both existing and future, with advanced capabilities such as independent reasoning and planning skills. Throughout the chapter, we refer to ‘misinformation’ and ‘falsehoods’ as false information, and to ‘disinformation’ as the spread of false or misleading information with the explicit intention of causing harm through, for example, coordinated influence operations \citep{wardle_information_2017}. 

The rest of the chapter proceeds as follows. Drawing on the communication and psychology literature, Section~\ref{sec:17.2} provides an overview of the mechanisms underlying the spread of mis- and disinformation in the digital era. In Section~\ref{sec:17.3}, we analyse the role of AI systems in enabling these phenomena. Section~\ref{sec:17.4} explores the unique risks and challenges posed by advanced AI assistants for our information environment and explores some potential mitigation strategies. 

\section{The Challenge of Misinformation and Disinformation}\label{sec:17.2}
Misinformation, disinformation and strategic attempts to manipulate public opinion are far from new \citep{burkhardt_chapter_2017}. Information sharing is central to human culture, and rumours and stories that evoke strong emotions tend to spread quickly and gain credibility through social transmission, regardless of their accuracy \citep{berinsky_rumors_2017, cotter_influence_2008, berger2011arousal}. Misinformation can originate from many different sources, including individuals, governments and politicians, and history is replete with examples of people strategically deploying lies and falsehoods to advance their own interests and gain political power. As early as ancient Rome, Octavian waged one of the first known disinformation campaigns against Julius Caesar’s general Mark Anthony, using ‘short, sharp slogans written upon coins’ to smear his reputation and win support for his own claim to power \citep{kaminska_lesson_2017}.

However, rapid advances in digital and communication technologies have made it cheaper and easier than ever to create and disseminate false or misleading information at scale. In the past, political information was relayed primarily through traditional media outlets, such as newspapers, television and radio, with editorial oversight mechanisms in place to ensure the quality and accuracy of the information they communicated. By lowering the cost of information production, the advent of the internet and social media has challenged the standing of these gatekeepers \citep{jungherr_retooling_2020}. Today, any internet user – within the bounds of their local access, speech and content regulation laws – can generate and distribute their own content over digital platforms, with little editorial oversight. This has led to a proliferation of news sources, many of which lack or intentionally forgo the quality assurance practices of traditional outlets, including fact-checking \citep{zhuravskaya2020political}.

Digitalisation has also accelerated the \emph{speed} and \emph{scale} at which information travels between media, citizens, political actors and the distribution channels at their disposal. Large social media platforms like \emph{X} (formerly 
\emph{Twitter}), \emph{Facebook} and \emph{TikTok} have ushered in new forms of social interactions which allow people to connect, interact and share user-generated content with others on a global scale. In addition, the algorithms powering these platforms are designed to reward and prioritise content that stimulates engagement from others \citep{bakshy_exposure_2015}. This prioritisation, coupled with the dynamics of networked communication, means that false, misleading or emotion-laden content published on social media has the potential to quickly cascade and reach millions of users quasi-instantaneously \citep{vosoughi_spread_2018}.\footnote{In 2020, for example, a conspiracy video about the origins of the coronavirus pandemic, called ‘Plandemic’ went viral on social media, racking up eight million views across platforms within days of its release \citep{frenkel_how_2020}.} 

Taken together, these developments have created new opportunities for malicious actors to misuse digital tools for political and economic gains (see Chapter~\ref{ch:9}). Fabricating and spreading false news stories on social media has become a lucrative business in many parts of the world \citep{hughes_macedonian_2021}. Online influence operations, which are coordinated attempts by state and non-state actors to influence domestic or foreign politics, have also gained ground across the world over the past decade \citep{bradshaw_global_2019}, spawning a sprawling misinformation-for-hire industry. To achieve their aims, propagandists employ a range of tactics and techniques that often exploit the affordances and vulnerabilities of the online information ecosystem \citep{the_cybersecurity_and_infrastructure_security_agency_cisa_tactics_2022}. These include, for example, exploiting data voids\footnote{Data voids are situations where there is a complete absence of data or lack of reliable or balanced information about a specific topic or query online. These often manifest during breaking news and are easily exploited by malicious actors to promote fringe or conspiratorial content in searches.} to push people towards false news sites and misleading content, cultivating inauthentic online personas to lend credibility to their narratives (e.g.\ fake ‘experts’), deploying bots – fake profiles that appear to be real individuals – to amplify or drown out political messages in a coordinated manner or fake grassroots support for a specific cause, a technique known as ‘digital astroturfing’ \citep{gorwa_unpacking_2020, woolley_automating_2016}.\footnote{For a more comprehensive framework and overview of disinformation tactics, see \citet{pamment_eus_2023} and \citet{the_cybersecurity_and_infrastructure_security_agency_cisa_tactics_2022}.}

What are the consequences of this? Mis- and disinformation pose a number of threats to democracy. Well-functioning democracies require a well-informed citizenry that is able to make informed political decisions on public issues \citep{mackenzie_lies_2020}. A society in which many people are misinformed or hold beliefs that go against established facts is therefore concerning, not only because it can have negative impacts on individuals but also because it may have harmful repercussions for society as a whole, such as stoking divisions and eroding trust in established truths. There is compelling evidence, for example, that misinformation was instrumental in fuelling violent insurrections in India,\footnote{In India, for example, false rumours circulated on \emph{WhatsApp} about child kidnappers were implicated in the mob lynching of 29 innocent people \citep{dixit_vicious_2018}.} stoking racial hatred in Myanmar and discrediting public health sources during the Covid-19 pandemic, leading to millions of preventable deaths \citep{burki_vaccine_2019, rocha_impact_2023, mozur_genocide_2018}. Multiple studies also link exposure to fake news and unreliable websites with reduced trust and negative attitudes towards mainstream news sources, heightened partisan animosity and decreased interpersonal trust \citep[see][]{hameleers_whom_2022,ognyanova_misinformation_2020,guess_fake_2020}.

Nevertheless, the prevalence of misinformation on social media remains a contentious issue, with some scholars claiming that its pervasiveness has been exaggerated \citep[see, for example,][]{altay_misinformation_2023,guess_exposure_2020,allen_evaluating_2020}. The existing literature has yet to determine conclusively whether and how online misinformation shapes political beliefs, and whether these effects are meaningfully different from traditional forms of media influence. Indeed, what makes individuals more susceptible or resistant to misinformation depends on a number of cognitive, social and affective factors such as whether that information confirms or contradicts their pre-existing beliefs and attitudes (‘confirmation bias’), their familiarity and trust in the source of information, and the frequency and modality in which they encounter that information (‘repetition’ and ‘elaboration’ effects) (for a review, see \citeauthor{ecker_psychological_2022}, \citeyear{ecker_psychological_2022}). All of these factors are important to take into consideration when considering the challenges that AI assistants might pose for mis- and disinformation.

\section{Misinformation, Disinformation and AI}\label{sec:17.3}
Technical advances in AI and machine learning (ML), such as generative AI models and personalised recommender systems, have also created opportunities for boosting the spread of misinformation and disinformation, and raised concerns about these systems’ impact on the information ecosystem. First, AI systems have made it easier to generate highly realistic synthetic content that is indistinguishable from real content. This could accelerate the spread of misinformation and prevent truth discernment. Second, AI-powered recommendation systems on digital platforms have enabled more personalised forms of targeting and distribution pathways for misleading content. We will now explore each of these elements in more detail.

\subsection{Content creation and manipulation}

Recent advances in AI image and text generation have expanded the opportunities to produce \emph{natural-sounding text} and \emph{highly realistic synthetic images, videos and audio} (known as ‘deep fakes’). While media manipulation was already possible, the advent of widely accessible generative AI tools has made it easier than ever to create and disseminate synthetic content. Research shows that this content is also increasingly undetectable and is easily mistaken as genuine \citep{nightingale_ai-synthesized_2022, spitale_ai_2023}. These tools have also simplified the manipulation of text, videos and images, thus allowing users to create tailored images, audio or videos for specific purposes. As a result, generative AI systems are already increasingly used in the political sphere to produce deceptive videos and entire ‘fake news’ websites \citep{hanley2023machine}.

The widespread adoption of generative AI models may pose significant challenges to the integrity of the information ecosystem. Research shows that audiovisual content is more evocative and seen as more persuasive than text \citep{hameleers_picture_2020, sundar_seeing_2021}. Should hyperrealistic and persuasive audiovisual content – such as political deepfakes – become ubiquitous, people may soon be completely unable to discern real from synthetic outputs and therefore be more prone to be misinformed by them. A proliferation of AI-generated content could also have the opposite effect and generate more suspicion and distrust. For example, exposure to deepfakes has been shown to provoke feelings of uncertainty and dampen trust in news \citep{vaccari_deepfakes_2020}. Lastly, an explosion of synthetic content might increase the mental load on everyday users when navigating online spaces. This could in turn lead to less carefulness, with less concern and other emotional or cognitive effects that are known to boost misinformation sharing and reinforce generalised distrust \citep{apuke_information_2022}.

As model capabilities expand, there is also a growing concern that these tools could power increasingly sophisticated and harder-to-detect misinformation campaigns to enable new influence tactics \citep[see, for example,][]{goldstein_generative_2023}. Evidence shows that multimodal generative AI tools have already enabled new forms of user manipulation. One such tactic, for example, is ‘impersonation’ or the ability to speak deceptively on behalf of others by convincingly impersonating them. Scammers are increasingly using AI-generated audio clips to trick people into giving away money and other private or sensitive information \citep{flitter_voice_2023}. Even without the need to deceive audiences directly, an increasing prevalence of synthetic material in the information ecosystem could exacerbate the ‘liar’s dividend’ or the ability of people to dismiss any evidence held against them as fake or AI-generated \citep{chesney_deep_2018}. We have already seen examples of this playing out in court \citep{bond_people_2023}.
 
\subsection{Targeted personalisation}

Over the past decade, the rise of AI-powered recommender systems has also transformed how digital information, including misinformation, is distributed and promoted. Today, most digital platforms use algorithmic systems that collect data and personalise the content users see based on their past behaviour, preferences and search results. These systems make recommendations to users to optimise their engagement and activity on the platform (e.g.\ what video to watch next). Various AI and ML models and techniques, including neural networks, generative adversarial networks (GANs) and reinforcement learning methods have been applied to these systems to augment and personalise user experiences \citep{zhang_artificial_2021}. 

However, recommender systems have also been criticised for their role in spreading misinformation and conspiracy theories, and for exposing users to increasingly ideologically biased, fringe and radical content, at the risk of leading them to develop extreme views \citep[see][]{tameez_youtubes_2020,tufekci_opinion_2018,hao_facebook_2021}. Scholars argue that the design of recommender systems can play a significant role in shaping user exposure to this type of content, leading them down a ‘rabbit hole’ from a more benign to increasingly harmful types of content, such as content about self-harm and eating disorders \citep[e.g.][]{whittaker_recommender_2021,ribeiro_auditing_2021}. Recommender systems also pose clear ethical challenges with respect to user autonomy and inappropriately manipulating or exposing users to risks \citep{milano_recommender_2020}. Despite these concerns, recent studies have found that, while recommendation algorithms are influential in shaping users’ informational diet, they do not have a measurable impact on their political beliefs \citep{isaac_facebooks_2023}. 

Another related concern is that malicious actors could weaponise AI tools to supercharge the distribution of microtargeted political propaganda. Online influence operations have become increasingly personalised over the past decade. In the lead-up to the 2016 US election, for example, Russian Internet Research Agency operators used microtargeted ads to cultivate secessionist and nativist sentiments across curated \emph{Facebook} groups and \emph{Instagram} accounts, and to discourage African-American users from voting or supporting specific candidates \citep{diresta_tactics_2019}. Politicians and propagandists can now target specific segments of the population based on algorithmically determined filters, including geographical location, consumer preferences, dispositions and behavioural traits \citep{dommett_data-driven_2019}. While personalised messaging is not a new concept, AI-powered tools could make it even easier and faster for malicious actors to test different methods and to optimise the timing and tone of their messages for specific individuals, at scale. This type of user profiling poses privacy and anonymity risks, and it raises concerns about voter manipulation\footnote{Though there is limited evidence of the effectiveness of online political ads on voting behaviour (see \citeauthor{haenschen_conditional_2023}, \citeyear{haenschen_conditional_2023}, for example).} (see Chapter~\ref{ch:10})
and distortion of public discourse \citep[see][]{bayer_double_2020, milano_epistemic_2021}.

\section{Misinformation, Disinformation and Advanced AI Assistants}\label{sec:17.4}
The rapid integration of AI systems with advanced capabilities, such as greater autonomy, content generation, memorisation and planning skills (see Chapter~\ref{ch:5})
into personalised assistants also raises new and more specific challenges related to misinformation, disinformation and the broader integrity of our information environment. We consider four of them in the section below.

\subsection{New vulnerabilities to misinformation}

First, the specific form factors of assistive AI may increase people’s vulnerability to misinformation. Several cognitive, social and political factors influence how people process and respond to information, and their willingness to believe or reject certain claims. Research shows, for example, that people more readily believe information from sources they trust, such as family members and friends \citep{anspach_new_2017, american_press_institute_who_2017}, and AI assistants can be designed to foster a similar sense of user trust. People tend to perceive autonomous AI systems as competent and trust in their abilities to perform certain tasks (\citeauthor{mckee_humans_2021}, \citeyear{mckee_humans_2021}; see Chapter~\ref{ch:13}).
As AI assistants become more personalised and ubiquitous, becoming virtual friends and even romantic partners (\citeauthor{chow_why_2023}, \citeyear{chow_why_2023}; see  Chapter~\ref{ch:12}), users may develop a high level of trust in them and take the information they provide at face value, even when it is false (\citeauthor{burtell_artificial_2023}, \citeyear{burtell_artificial_2023}; see Chapters~\ref{ch:10},~\ref{ch:11} and~\ref{ch:13}).

Low literacy levels around how content generated by AI assistants is produced and distributed could also complicate user’s ability to critically evaluate information. People’s ability to discern truth from falsehood varies greatly depending on their level of digital literacy – their understanding and familiarity with how the internet and digital technology work \citep{sirlin_digital_2021}. LLMs – the models currently powering the latest generation of virtual assistants – are known to produce factual inaccuracies and to ‘hallucinate’, making up semantically plausible but factually inaccurate statements \citep{ji_survey_2023, lee_hallucinations_2019}. As people, including journalists, adopt AI assistants to assist them with copy-writing, website design or any other tasks involving content generation, the internet might become replete with AI-generated outputs of disputable quality and facticity. Without a clear heuristic understanding of the capabilities and limitations of AI assistants, such as how it summarises information, users may not be equipped to critically evaluate misinformation when they see it. 

\subsection{Entrenched viewpoints and reduced political efficacy}

Design choices such as greater personalisation of AI assistants and efforts to align them with human preferences could also reinforce people’s pre-existing biases and entrench specific ideologies. Increasingly agentic AI assistants trained using techniques such as reinforcement learning from human feedback (RLHF) and with the ability to access and analyse users’ behavioural data, for example, may learn to tailor their responses to users’ preferences and feedback. In doing so, these systems could end up producing partial or ideologically biased statements in an attempt to conform to user expectations, desires or preferences for a particular worldview \citep{carroll_estimating_2022}. Over time, this could lead AI assistants to inadvertently reinforce people’s tendency to interpret information in a way that supports their own prior beliefs (‘confirmation bias’), thus making them more entrenched in their own views and more resistant to factual corrections \citep{lewandowsky_misinformation_2012}. At the societal level, this could also exacerbate the problem of epistemic fragmentation – a breakdown of shared knowledge, where individuals have conflicting understandings of reality and do not share or engage with each other’s beliefs – and further entrench specific ideologies.

Excessive trust and overreliance on hyperpersonalised AI assistants could become especially problematic if people ended up deferring entirely to these systems to perform tasks in domains they do not have expertise in or to take consequential decisions on their behalf (see Chapter~\ref{ch:13}).
For example, people may entrust an advanced AI assistant that is familiar with their political views and personal preferences to help them find trusted election information, guide them through their political choices or even vote on their behalf, even if doing so might go against their own or society’s best interests. In the more extreme cases, these developments may hamper the normal functioning of democracies, by decreasing people’s civic competency and reducing their willingness and ability to engage in productive political debate and to participate in public life \citep{sullivan_psychological_1999}.

\subsection{Degraded and homogenised information environments}

Beyond this, the widespread adoption of advanced AI assistants for content generation could have a number of negative consequences for our shared information ecosystem. One concern is that it could result in a degradation of the quality of the information available online. Researchers have already observed an uptick in the amount of audiovisual misinformation, elaborate scams and fake websites created using generative AI tools \citep{hanley2023machine}.\footnote{Compounding the issue, as the context surrounding AI-generated content is lost, it may become more difficult to fact-check this content in the future.} As more and more people turn to AI assistants to autonomously create and disseminate information to public audiences at scale, it may become increasingly difficult to parse and verify reliable information. This could further threaten and complicate the status of journalists, subject-matter experts and public information sources. Over time, a proliferation of spam, misleading or low-quality synthetic content in online spaces could also erode \emph{the digital knowledge commons} – the shared knowledge resources accessible to everyone on the web, such as publicly accessible data repositories \citep{huang_generative_nodate}. At its extreme, such degradation could also end up skewing people’s view of reality and scientific consensus, make them more doubtful of the credibility of all information they encounter and shape public discourse in unproductive ways. Moreover, in an online environment saturated with AI-generated content, more and more people may become reliant on personalised, highly capable AI assistants for their informational needs. This also runs the risk of homogenising the type of information and ideas people encounter online \citep{epstein_art_2023}.

\subsection{Weaponised misinformation agents}

Finally, AI assistants themselves could become \emph{weaponised} by malicious actors to sow misinformation and manipulate public opinion at scale. Studies show that spreaders of disinformation tend to privilege quantity over quality of messaging, flooding online spaces repeatedly with misleading content to sow ‘seeds of doubt’ \citep{hassoun_sowing_2023}. Research on the ‘continued influence effect’ also shows that repeatedly being exposed to false information is more likely to influence someone’s thoughts than a single exposure. Studies show, for example, that repeated exposure to false information makes people more likely to believe it by increasing perceived social consensus, and it makes people more resistant to changing their minds even after being given a correction (for a review of these effects, see \citeauthor{lewandowsky_misinformation_2012}, \citeyear{lewandowsky_misinformation_2012}; \citeauthor{ecker_psychological_2022}, \citeyear{ecker_psychological_2022}). By leveraging the frequent and personalised nature of repeated interactions with an AI assistant, malicious actors could therefore gradually nudge voters towards a particular viewpoint or sets of beliefs over time (see Chapters~\ref{ch:9} and \ref{ch:10}).

Propagandists could also use AI assistants to make their disinformation campaigns more personalised and effective. There is growing evidence that AI-generated outputs are as persuasive as human arguments and have the potential to change people’s minds on hot-button issues \citep{bai_artificial_2023,myers_ais_2023}.  Recent research by the Center for Countering Digital Hate showed that LLMs could be successfully prompted to generate ‘persuasive misinformation’ in 78 out of 100 test cases, including content denying climate change (see Chapters~\ref{ch:10} and ~\ref{ch:19}).\footnote{The researchers also found that LLM-based systems could easily create content in the style of Facebook and X (formerly Twitter) posts, further illustrating their potential for misuse in misinformation campaigns (\citeauthor{center_for_countering_digital_hate_googles_2023}, \citeyear{center_for_countering_digital_hate_googles_2023}; see also \citeauthor{brewster_could_2023}, \citeyear{brewster_could_2023}).}

If compromised by malicious actors, in the future, highly capable and autonomous AI assistants could therefore be programmed to run astroturfing campaigns autonomously,\footnote{The practice of faking grassroots public support for a cause to influence public opinion. In 2017, for example, the industry group Broadband for America generated \emph{millions of comments} with fake or stolen personal information against the Federal Communications Commission’s (FCC) proposed reversal of net neutrality \citep{singel_filtering_2018}.} tailor misinformation content to users in a hyperprecise way, by preying on their emotions and vulnerabilities, or to accelerate lobbying activities \citep{kreps_potential_2023}. As a result, people may be misled into believing that content produced by weaponised AI assistants came from genuine or authoritative sources. Covert influence operations of this kind may also be harder to detect than traditional disinformation campaigns, as virtual assistants primarily interact with users on a one-to-one basis and continuously generate new content \citep{goldstein_generative_2023}. 

\section{Risks and Mitigations}\label{sec:17.5}

As we have discussed, powerful AI assistants have the potential to shape the information environment in significant ways. We summarise below the informational risks these technologies pose at the individual and societal level, and outline a number of possible mitigation strategies that various stakeholders could adopt to reduce those risks. 

\begin{itemize} [parsep=4pt]
	\item \textbf{Risk 1: Advanced AI assistants may increase people’s vulnerability to misinformation.} First, AI assistants may make users more susceptible to misinformation, as people develop competence trust in these systems’ abilities and uncritically turn to them as reliable sources of information. 

  \item \textbf{Risk 2: Advanced AI assistants may entrench specific ideologies and impact citizens' understanding and engagement with public affairs.} Second, AI assistants may provide ideologically biased or otherwise partial information in attempting to align to user expectations. In doing so, AI assistants may reinforce people’s pre-existing biases and compromise productive political debate. 

  \item \textbf{Risk 3: Advanced AI assistants erode trust and undermine shared knowledge by polluting the information ecosystem.} Third, AI assistants may contribute to the spread of large quantities of factually inaccurate and misleading content, with negative consequences for societal trust in information sources and institutions, as individuals increasingly struggle to discern truth from falsehood. 

  \item \textbf{Risk 4: Advanced AI assistants drive opinion manipulation by empowering online influence operations.} AI assistants may facilitate large-scale disinformation campaigns by offering novel, covert ways for propagandists to manipulate public opinion. This could undermine the democratic process by distorting public opinion and, in the worst case increasing skepticism and political violence. 
\end{itemize}

Several mitigation strategies could help to address those risks. These can be broadly split between technical and policy solutions. 

\subsection{Technical solutions}

\begin{itemize} [parsep=4pt]
	\item \textbf{Limit AI assistants’ functionality.} To mitigate the risks outlined above, AI developers could introduce limits on AI assistants’ functionality at the model level (e.g.\ prevent them from expressing political opinions). This could be achieved by applying content filters on model outputs or user prompts, or by limiting AI assistants’ abilities to learn from external inputs to prevent them from being injected with harmful content from adversarial actors, for example.\footnote{Cohere, for example, has undertaken such an approach to harm prevention for its LLMs.} While effective in the short term, this approach is not foolproof, as powerful AI models can develop emergent capabilities over time and pursue goals which may not have been specified during training \citep{wei_emergent_2022}. Addressing this would therefore require continuous monitoring of AI assistants’ behaviour and careful evaluation of models’ capabilities over time.

  \item	\textbf{Develop robust detection mechanisms.} To limit erosion of trust in public knowledge sources, implementing robust mechanisms to detect ‘deepfakes’ and other misleading media created or propagated by AI assistants will also be crucial. A number of leading AI developers have recently unveiled ML classifiers and watermarking techniques to help with the identification and differentiation of AI-generated text and images from human-created ones \citep[see][]{intel_intel_2022,abadi_deep_2016, google_deepmind_synthid_2023}. The number of companies providing similar services is growing \citep[see][]{hsu_another_2023}. While promising, these tools are not entirely immune to adversarial attacks and their practical applications remain limited. For example, current text detection tools still perform poorly when dealing with short texts and languages other than English \citep{sadasivan_can_2023}.\footnote{On a test set of English-text, Open AI’s classifier only correctly identifies 26\% of AI-written text (true positives) as ‘likely AI-written’, while incorrectly labelling human-written text as AI-written 9\% of the time (false positives).} Moreover, current methods are not well equipped to deal with real-world cases of misinformation, which are increasingly multimodal (see \citeauthor{hangloo_combating_2022}, \citeyear{hangloo_combating_2022}). Lastly, open-source models also present a challenge to synthetic media detection, as there tends to be limited control over how these are deployed and used, making it difficult to enforce industry-wide standards and best practices \citep{theben_challenges_2021}. 

  \item	\textbf{Limit personalisation and promote critical thinking.} Another potential solution would be for developers to limit personalisation and embed prompts encouraging critical thinking in the design of AI assistants to ensure users do not get a biased perspective on public issues. Several news organisations and online platforms\footnote{In 2020, Facebook changed how it surfaces content in users’ News Feeds and restricted personalised recommendations in an effort to address polarisation \citep{rosen_investments_2020}.} have already experimented with techniques to help users pause and reflect on the content they are exposed to and to encourage them to diversify their information diets (e.g.\ \emph{Ground News}). Research suggests that such context-aware recommendations and embedded prompts, such as information literacy videos, are effective in inoculating users against misinformation, improve their ability to better identify emotional manipulation and improve their sharing behaviour \citep{mattis_nudging_2022, roozenbeek2022psychological}. For this strategy to be useful in the case of AI assistants, however, clear transparency would be key to avoiding users’ personal autonomy being undermined. Another promising avenue for developers would be to limit encoding political biases into the models powering AI assistants through fine-tuning, by running dedicated evaluations on ideological diversity and diversifying the pool of human raters they solicit feedback from. 

\item \textbf{Emphasise factuality.} It will be imperative for developers to emphasise factuality in AI assistants. This could be achieved by requiring assistants to systematically cite their sources when presenting or retrieving factual claims,\footnote{  Although in the future, simply requiring AI models to list their sources may not be sufficient for determining their accuracy, as an advanced model may selectively choose sources that it believes humans will find persuasive.} for example, and augment their ability to evaluate sources to determine their trustworthiness \citep{guo_survey_2022}. The task of fact-checkers will become increasingly challenging as they need to sift through vast amounts of AI-generated content, identifying materials that warrant verification and issuing timely corrections. Several fact-checking organisations are already building applications on top of and developing their own LLMs to automate parts of this process. However, it will be important for the entire AI community to invest more capabilities and resources into building fact-checking mechanisms for verifying the accuracy of information presented to users. One pressing challenge for developers in that respect will be to address sociotechnical vulnerabilities both within and outside their systems, including the issue of ‘data voids’.\footnote{If a user asks a question in a data void, AI assistants might provide inaccurate or unreliable information relying on alternative sources, as there is simply not enough data to draw from.} To tackle this, models should be trained on diverse and regularly updated data sets, while assistants should be designed to identify these blind spots and inform users about the limitations of available data on certain topics. 
\end{itemize}

\subsection{Policy solutions}

\begin{itemize} [parsep=4pt]
	\item \textbf{Restrict specific uses and applications.} Implementing robust governance mechanisms will also be crucial for mitigating the potential negative impacts of AI assistants on public life. One approach AI companies and policymakers could take is to impose restrictions on explicitly political or malicious uses of AI assistants through enforceable licences and terms of services. These measures could limit the deployment of AI agents for political campaigning, consulting, lobbying or persuading voters to support specific causes or candidates. Recognising similar concerns around misinformation, some social media platforms have previously banned political advertising on their sites. However, scholars have criticised such measures as inadvertently favouring political incumbents who already have ample resources while putting civil society organisations and political newcomers at a disadvantage \citep{kreiss2020democratic}. Several AI research labs have also taken steps in that direction by explicitly banning the use of their models to disseminate falsehoods, manipulate users or influence politics. However, effective enforcement of these measures relies heavily on the ability to clearly attribute responsibility. As AI assistants become more powerful, acquiring specific objectives or significant autonomy to determine their own actions, how to better identify clear violations of terms of services and take appropriate sanctions against users will be challenging, thus offering promising avenues for future research and governance.

  \item \textbf{Implement transparency mechanisms.} Governments and AI companies could implement protocols and policies that require transparency around the use of AI assistants. This could include mandating clear labels for AI-generated content created by assistants, requiring disclosure of, or altogether banning, the use of AI assistants or bots to impersonate or replicate human activity online.\footnote{California’s 2019 legislature passed a law to this effect, the Bolstering Online Transparency (BOT) Act, though the law has been criticised for lacking enforcement mechanisms to incentivise compliance.} AI labs should also be transparent about the demographic composition of the annotators they recruit to fine-tune their models with and about how they evaluate their models for ideological biases. Another idea in this vein would be for AI labs to engage in transparent reporting of harms caused by AI assistants (similar to the \citeauthor{ai_incident_database_ai_nodate}) so that people could track and weigh the benefits and risks of these systems.

  \item \textbf{Support public education.} AI developers and policymakers could support education programmes to raise public awareness about the workings, limitations and biases of AI assistants, and teach people to become more discerning users. As mentioned above, the focus of public education campaigns should be on ways of improving critical thinking skills rather than simply making people broadly aware that AI can be used for disinformation campaigns. Beyond this, programmes could involve, for example, providing training on how to prompt AI assistants efficiently to generate evidence-based high-quality responses or how to fact-check information provided by AI assistants. While there is compelling evidence of the efficacy of digital literacy programmes in certain circumstances, more research is needed to test how effective interventions such as active pre-bunking are in addressing misinformation relayed by personalised AI assistants, for instance. 
\end{itemize}

\section{Conclusion}\label{sec:17.6}

Advanced AI assistants pose four main risks for the information ecosystem. First, AI assistants may make users more \emph{susceptible} to misinformation if people develop trusting relationships with these systems and uncritically turn to them as reliable sources of information (see Chapters~\ref{ch:10} and~\ref{ch:12}).
Second, AI assistants may provide ideologically \emph{biased} or otherwise \emph{partial information} to users in an effort to align to user expectations. In doing so, AI assistants may reinforce specific ideologies and biases, which in turn will compromise healthy political debate. Third, AI assistants may \emph{erode} societal \emph{trust} in shared knowledge by contributing to the dissemination of large volumes of plausible-sounding but low-quality information. Finally, AI assistants may facilitate \emph{hypertargeted disinformation} campaigns by offering novel, covert ways for propagandists to manipulate public opinion. This chapter articulates these risks and discusses technical and policy mitigations.

\chapter{Economic Impact}\label{ch:18}

\textbf{Conor Griffin, Juan Mateos-Garcia, S\'ebastien Krier,
Geoff Keeling, Alexander Reese, Iason Gabriel}

\noindent \textbf{Synopsis}: 
		This chapter analyses the potential economic impacts of advanced AI assistants. We start with an analysis of the economic impacts of AI in general, focusing on \emph{employment}, \emph{job quality}, \emph{productivity growth} and \emph{inequality}. We then examine the potential economic impacts of advanced AI assistants for each of these four variables, and we supplement the analysis with a discussion of two case studies: \emph{educational assistants} and \emph{programming assistants}. We conclude with a series of recommendations for policymakers around the appropriate techniques for monitoring the economic impact of advanced AI assistants, and we propose plausible approaches to shaping the type of AI assistants that are deployed and their impact on the economy. 

\section{Introduction}\label{sec:18:1}

This chapter explores the potential economic impact of advanced AI assistants. We focus on \emph{employment}, \emph{job quality}, \emph{productivity growth} and \emph{inequality}. We first examine how AI in general has impacted these factors, then we explore the potential impacts of advanced AI assistants through two case studies: \emph{educational assistants} and \emph{programming assistants} (see 
Chapter~\ref{ch:5}). We conclude by discussing what public policy instruments are available to direct the economic impact of AI assistants towards socially beneficial outcomes.

\section{How Has AI Affected the Economy to Date?}\label{sec:18:2}

We can explore the impact of AI on the economy through the lenses of employment, job quality, productivity growth and inequality. These four variables are all important contributors, both positively and negatively, to individual and societal well-being. We analyse the four variables independently, but we note that they are interdependent. For example, the level of labour demand in an economy shapes employee bargaining power and wages, which in turn affects job quality. Even so, each of these factors provides an informative lens through which we can analyse the impact of AI systems on the economy to date. For reasons of scope, we do not analyse the potential effects of AI on other important economic variables such as competition. 
 
\subsection{Employment}

Employment can contribute to well-being positively, for example by providing income \citep{tay_income_2017} and a sense of purpose \citep{bryce_finding_2018}, or negatively.\footnote{There are many nuances that may determine, at the individual level, the relative importance of income to well-being, such as the context in which the income was earned and which aspect of well-being is being considered \citep{pouwels_income_2008}.} Unemployment, particularly when it is over the long term, is associated with a range of ills, including greater risk of committing and suffering from crime \citep{phillips_link_2012}, abusing drugs \citep{bauld_problem_2010} and suffering physical and mental health problems (\citeauthor{herber_single_2019}, \citeyear{herber_single_2019}; see 
Chapter~\ref{ch:7}).\footnote{If unemployment persists, or an individual never achieves employment in the first instance, it can have a `scarring' effect, as individuals drop out of the labour force, skills are forgotten and hiring discrimination by prospective employers increases \citep{mcquaid_youth_2017}.}  When it comes to employment, we are primarily interested in \emph{total labour demand} -- or the total number of available jobs in society -- alongside the \emph{unemployment rate}. We are also interested in the \emph{global jobs gap rate}, which is the percentage of the total adult population that has an unmet need for employment \citep{international_labour_office_world_2023}. This includes all unemployed people, as well as individuals, particularly women in lower-income countries, who are outside the labour force due to factors such as unpaid care obligations and so do not appear in unemployment data \citep{international_labour_office_world_2023}. Global unemployment is currently close to a record historical low, but it is starting to increase in several countries, partly due to the effects of higher interest rates. In higher-income countries, low unemployment may provide a rationale for increasing AI use. In lower-income countries, the biggest challenge is the lack of high-quality jobs pushing many people into less beneficial informal work, and it is unclear how AI will affect this \citep{international_labour_office_world_2023}.

Most research on AI and employment breaks down jobs into bundles of \emph{tasks}, and it forecasts the extent to which AI will be able to perform them. Using this approach, \citet{frey_future_2013} estimated that 47\% of jobs were at risk of automation by `computer-controlled equipment', leading to multiple follow-on studies (e.g.\ \citeauthor{manyika_ai_2018}, \citeyear{manyika_ai_2018}; \citeauthor{smit_future_2020}, \citeyear{smit_future_2020}). These initial studies typically found that lower-income jobs, characterised by routine, physical tasks, for example in driving and manufacturing, were most at risk from AI. On the back of recent progress in large language models, more recent studies, such as \citet{eloundou2023gpts}, point to another scenario, where roles that involve generating and manipulating information, and those which generally require higher levels of education, such as translators, survey researchers and tax advisers, are more exposed to AI. However, the authors do not take a view on whether this `exposure' will be positive or negative for the employees.\footnote{\citet{acemoglu_automation_2019} distinguish between displacement, augmentation and reinstatement effects. Displacement refers to a scenario where AI replaces human employees at performing certain tasks, leading to job losses; augmentation refers to a scenario where employees use AI to enhance their productivity; and reinstatement refers to a scenario where AI leads to new tasks being added to existing jobs or the packaging of new tasks into new jobs.} 

Empirical research is starting to shed light on this question of positive or negative impact of AI exposure on employees by studying the actual effects that AI has on labour demand after it is deployed in the workplace.\footnote{Empirical analysis of the effects of AI on employment is difficult. For other technologies, like industrial robots, practitioners have access to longitudinal data sets \citep{international_federation_of_robotics_world_2022} on the number of robots installed in different countries. The robots are also deployed in a narrow range of sectors and locations. This has allowed researchers to build models that tease out the resulting economic effects in a way that is not possible for AI. See, for example, \citet{oxford_economics_how_2019}, \citet{petropoulos_impact_2018}, \citet{graetz_robots_2015}, \citet{acemoglu_robots_2017} and \citet{dauth_german_2017}. The findings differ but broadly support a view that industrial robots have displaced low-skill manual jobs, boosted organisational and national productivity, and created new high-skill jobs, with positive spillover effects on job creation in other sectors.} \citet{handel_growth_2022} and \citet{albanesi_new_2023} find little support for an acceleration in job losses for exposed occupations in US and European employment data, respectively. Indeed, \citet{georgieff_artificial_2021} find a \emph{positive} link between AI exposure and employment growth. A key driver of these positive results was high-income employees with strong digital skills who likely had the capabilities and freedom to adapt their roles in response to AI. \citet{acemoglu_artificial_2022} find less positive results, with some evidence for reduced hiring for roles with greater AI exposure, but the effect sizes were too modest to draw strong conclusions. 

Researchers have also surveyed organisations that have deployed AI applications to understand the resulting effects. In a recent survey of UK business leaders, \citet{hunt_measuring_2022} found that introducing AI was associated with both \emph{destruction} and \emph{creation} of jobs. More recently, in a study covering eight Organisation for Economic Co-operation and Development (OECD) countries and almost 100 finance and manufacturing organisations that had deployed AI applications, \citet{milanez_impact_2023} found that almost 80\% reported no change in overall job quantities. Rather, most firms had invested in AI to improve product or service \emph{quality}, so their headcounts did not change. Some AI applications did replace employees in performing certain tasks, but most of those employees were assigned new tasks. In other instances, the AI applications were insufficiently effective in increasing productivity or quality of service to have any impact on employment. Where job losses did occur, it was primarily via attrition rather than redundancies. Taken together, the evidence suggests that AI has \emph{not yet had major negative impacts on aggregate labour demand}, or unemployment, although this may be because most AI applications have not yet been especially transformative and employment effects take time to manifest. 

\subsection{Job quality} 

The OECD defines \emph{low-quality jobs} as those with: (1) low earnings; (2) a fragile sense of security; and (3) a poor working environment \citep{oecd_how_2016}. The broader literature on job quality, including recurring employee surveys, also highlights a range of factors in the working environment that can positively or negatively affect well-being, for example by providing a sense of purpose or engendering feelings of contentment or stress. The relative importance of these factors can differ by person, and evolve over time, including, for example, working-time arrangements, workplace relationships, autonomy, employee voice and potentially many more aspects \citep{us_bureau_of_labor_statistics_american_nodate}. 

Early evidence suggests that the relationship between AI and job quality is not straightforward. When it comes to wages, \citet{felten_occupational_2019} find a small positive link between exposure to AI and wage growth, suggesting that AI deployments may make employees more productive and capable of earning a higher wage. However, other studies have found little evidence of any impact \citep{albanesi_new_2023}. \citet{acemoglu_power_2023} also claim that the adoption of AI in the workplace may enable employers to more strongly monitor and surveil their employees. This may mean that employers can make employees work harder without necessarily having to increase wages to the same extent as would otherwise be required (this point also illustrates the connections between job quality, productivity growth and inequality). Beyond wages, studies show a mixed picture, with employees reporting both positive and negative impacts of AI on job quality attributes, and with different employees reporting divergent views about the impact of some applications. For example, \citet{gutelius_future_2019} found that introducing AI-based robotics into warehouses helped to reduce the monotonous and physically strenuous activity of lifting heavy packages, but it also put pressure on employees to work faster, reduced human contact and led to an increase in perceived scrutiny. Relatedly, \citet{milanez_impact_2023} found that some employees felt that AI applications caused their work to become safer and more rewarding by, for example, reducing the number of repetitive interactions. However, others felt that they had less privacy, higher work-intensity and more stress \citep{ribeiro_digitalisation_2023}. 

\subsection{Productivity growth}

\emph{Productivity} describes the \emph{efficiency} with which an economy uses \emph{labour} and \emph{capital}. We are primarily interested in growth in \emph{total factor productivity} (TFP), which describes increases in output that are due to innovation, including technological progress. TFP is more difficult to measure than labour productivity, which is consequently more commonly used. Science and technology advances, and the innovative products, services and processes that result, are the only way to significantly boost productivity growth, and economic growth, in the long term. Economic growth, in turn, is central to maintaining and improving \emph{standards of living}. For example, increased growth in China over the past 40 years has enabled 800 million people to move out of poverty (defined as income of less than USD 1.90 per day), accounting for three quarters of the global reduction over that time period \citep{the_world_bank_lifting_2022}. Since the global financial crisis in 2007--2009, the world has witnessed a sharp slowdown and sustained stagnation in global productivity growth, including, most worryingly, in low-and middle-income countries \citep{dieppe_global_2021}.

Bearing this context in mind, AI could potentially make human employees and machinery (e.g.\ computers and software) more productive by boosting \emph{efficiency} -- doing things that were already being done, but faster or at a higher scale -- or by boosting \emph{innovation} -- doing things differently or doing new things \citep{manyika2023coming}. AI could directly suggest novel ideas or make human employees more efficient, and in turn free them up to work on more innovative ideas. Indeed, some believe that AI will lead to an explosion in new ideas, with a corresponding explosion in productivity growth and economic growth (\citeauthor{clancy_great_2023}, \citeyear{clancy_great_2023}; see also \citeauthor{vollrath_will_2023}, \citeyear{vollrath_will_2023}). However, AI could also hamper productivity growth. For example, the introduction of email led to quicker asynchronous communication, but it also prompted concerns about information overload, disruptions to deep work and spam, and debates continue over its net impact \citep{bulkley_information_2008}.

Most evaluations of AI's impact on productivity growth have been short experiments where individuals use AI tools to carry out discrete tasks, with generally positive results. \citet{peng_impact_2023} tasked software developers with using GitHub Copilot to implement an HTTP server in JavaScript, and they completed the task almost 60\% faster than the control group. \citet{noy_experimental_2023} assigned writing tasks to a group of college-educated professionals, and they found that access to ChatGPT reduced the time taken and increased the perceived quality.\footnote{Explaining their results, \citet{noy_experimental_2023} find that ChatGPT could substitute for human tasks, such as rough drafting, to allow humans to focus more on idea generation and editing.} \citet{brynjolfsson_generative_2023} provided an AI conversational assistant to more than 5,000 customer service agents in a real workplace. The number of issues resolved, per hour, increased by 14\%, on average. 

The three studies all found evidence to suggest that the AI applications disproportionately helped lower-skilled or more novice employees within an occupation more than their higher-skilled counterparts. This suggests that AI assistants could help to ease or shorten the learning curve, and they may have a role in employee training and upskilling programmers. However, study results also suggest that not all employees will be helped by AI impacts, which may be more in the realm of incremental gains in efficiency rather than a sharp uptick in transformative innovation. The impacts will also differ by sector, job and task. For example, \citet{jia_when_2023} used an AI conversational agent to support telesales employees. They found that previously top-performing employees benefitted more, because they were better placed to use the assistant to experiment and were more creative when interacting with the AI assistant to answer customer questions.

A recent suggestion is that generative AI tools could double US productivity growth over 20 years -- a huge potential impact \citep{brynjolfsson_generative_2023}. Another study, by Goldman Sachs, estimates that generative AI could boost annual productivity growth by 1.5 percentage points over a 10-year period following mass adoption \citep{hatzius_potentially_2023}.

For now, there is little evidence in the productivity growth data that such impacts are occurring.\footnote{Productivity gains from AI may take time to materialise. One potential positive scenario, outlined by \citet{brynjolfsson_productivity_2018}, is a J-curve, where AI's initial effects on productivity growth are minimal, or even negative, as organisations redesign their business models, before leading to a subsequent surge.} A key question is whether benefits to individuals will translate into economy-wide productivity growth. For example, it may be that AI's recommendations help individuals to become more creative, but only in similar ways, thus leading to reduced growth in creativity at the economy level \citep{doshi_generative_2023}. AI may also lead to significant productivity gains in certain sectors, but the resulting cost savings may be spent on other sectors, like education and health, that do not become more productive, the so-called Baumol effect, thus resulting in little impact on aggregate productivity growth.\footnote{The Baumol effect appears to be one reason for limited productivity growth in recent years \citep{kling_what_2018}.}  

\subsection{Inequality}

We focus here on \emph{income} and \emph{wealth} inequality \emph{between} (place-based inequality) and \emph{within} countries (class-based inequality).\footnote{Efforts to articulate how inequality affects societal well-being are plagued by definitional and contextual challenges, and the poor reproducibility of studies \citep{ngamaba_income_2018}. Potential effects include worse rates of subjective well-being \citep{oishi_income_2011} and intergenerational inequality \citep{corak_income_2013}, undermined access to health \citep{pickett_income_2015}, education \citep{oecd_equity_2018} and public services, and reduced social trust and civic engagement \citep{schroder_how_2023}.} \citet{chancel_world_nodate} found in 2022 that the richest 10\% of the world's population took home 52\% of the total income, while the poorest half took home just 9\%. The \emph{wealth} disparity is even starker, with the richest 10\% owning 76\%, the top 1\% owning 38\% and the poorest half owning almost nothing (2\%). When the authors analysed this data over the past century and beyond, several stories emerged, some of which are more positive than others. Positively, over the past 30 years, and particularly since 2000, inequality between countries, or \emph{place-based inequality}, has declined. Less positively, inequality \emph{within} many countries has increased over the past three decades, for example in the US, Brazil and India. Inequality also occurs based on parameters beyond income or wealth percentiles. For example, \citet{chancel_world_nodate} note that women earn less than 35\% of global income, with an increase of just 5\% since 1990. 

There is little empirical evidence regarding how AI has affected inequality. However, we can identify several pathways through which it is likely happening. The first reflects which employees are best able to draw on AI, to enhance their own productivity and wages, and which employees face the greatest displacement risk. Studies by \citet{felten_occupational_2019} and \citet{albanesi_new_2023} suggest that high-income occupations may be disproportionately benefitting from AI exposure to date. However, other studies suggest that AI assistants disproportionately benefit more junior and lower-skilled employees, so this could help to reduce inequality, although this will likely only occur if these employees have sufficient negotiation power to request higher salaries \citep{peng_impact_2023, noy_experimental_2023, brynjolfsson_generative_2023}.

Another route through which AI is affecting inequality is via the type, distribution and location of new jobs that it enables \citep{ben2024ai}. The majority of leading AI research labs, start-ups and enterprises are located in urban centres in high-income countries, so we can assume that AI is directly and indirectly enabling well-compensated `frontier' jobs in these locations, such as engineers, product managers and lawyers, as well as lesser-paid `wealth' roles, such as fitness instructors, that provide services to these high-income employees.\footnote{\citet{autor_new_2019} find that, over the past century, technology has primarily enabled three types of new jobs: (1) `Frontier' roles produce, install and maintain (and arguably also `use') novel technologies, such as engineers who may help design new AI assistants; (2) `Last-mile' roles carry out `nearly automated' tasks that do not require high levels of technology-specific expertise, such as data enrichment workers who help to curate  data sets for new AI assistants, but arguably there is a spectrum of expertise that may be created in these roles, with scope for domain experts to participate; and (3) `Wealth' roles, such as personal trainers, cleaners and counsellors, provide luxury services to affluent employees in high-income technology jobs. This reality of high-income roles creating a larger number of service roles is not limited to technology, as \citet{moretti_new_2013} and others have shown.} This may be exacerbating inequalities within countries, and between higher- and lower-income countries. \citet{lee_low-skilled_2019} estimate that for every 10 new high-tech jobs created in the UK, seven new service jobs were created, of which six were `low-skilled'. Once accompanying increases in housing costs were considered, low-skilled employees' real wages fell, thus exacerbating inequality.

The development of large AI models is also enabling, potentially millions, of new `data enrichment' jobs \citep{kassi_how_2021}. Activists and academics have criticised the precarious working conditions and low compensation, which has led to the creation of recommended best practices for AI labs \citep{jindal_implementing_2022}, including the payment of a living wage \citep{gray_ghost_2019, graham_digital_2017}. If conditions improve, it is possible that data enrichment could help to reduce inequalities between countries, similar to the past effects of outsourcing on per capita income in India, India, the Philippines and Morocco. However, the history of science and technology, where advanced research and development (R\&D) often leads to nearby start-up creation and longer-term agglomeration effects, suggests that an expansion in higher-quality data enrichment work will be no substitute -- from a longer-term inequality perspective -- to having frontier AI development and jobs in low- and middle-income countries, and in the lower-income regions of high-income countries \citep{gross_america_2022}.

\section{How Will AI Assistants Affect the Economy?}\label{sec:18:3}

To our knowledge, no substantive research addresses the aggregate economic impact of AI \emph{assistants}, as a specific class of AI, outside the studies on productivity growth discussed above. The impact will also depend on the type of AI assistants that are deployed and their specific characteristics (see 
Chapters~\ref{ch:3} and~\ref{ch:5}). To illustrate these dynamics, we create a simple framework for assessing the potential economic impact of an AI education assistant and an AI programming assistant. 

\subsection{Effects on employment}

We can pose a number of questions about the impact of advanced AI assistants on employment. Which occupations do we expect to be \emph{directly affected} by these assistants? To what extent can employees in these occupations \emph{adapt their tasks} or find \emph{alternative jobs}? Can they use the assistant to \emph{augment} their role? How do we expect consumer demand for their product or service to respond? To what extent do we expect entirely \emph{new businesses} or \emph{new production processes} and \emph{jobs} to emerge? And do we expect a longer-term preference for humans vs capital (AI) in these production systems? 

\subsubsection{AI education assistant}

Teachers and tutors will likely be one of the groups most affected by AI education assistants. Globally, there are approximately 85 million teachers \citep{ritchie_global_2023}. In the UK, past data suggests that there may be up to 1.5 million private tutors, with approximately 100,000 working full time -- global data is lacking \citep{kirby_shadow_2016}. Some AI practitioners, such as Stuart Russell, have suggested that AI assistants, like ChatGPT, may lead to `fewer teachers being employed -- possibly even none' \citep{devlin_ai_2023}, and some analysis of occupational exposure to generative AI identify teaching of some disciplines as highly exposed \citep{felten2023occupational}. However, there are many grounds to challenge this projection. First, historical evidence suggests that it will be difficult to successfully integrate AI assistants into formal education without strong involvement from teachers, a trend that could lead to a demand for \emph{more} teachers rather than less. Second, the world is already facing a major teacher shortage, recently estimated at 69 million people by 2030 \citep{unesco_transforming_2022}.\footnote{The teacher shortage is also leading to teacher--pupil ratios in many countries that are far higher than experts recommend.} Third, even if more powerful AI tutors can help students to master knowledge and skills, a big if, they may be ill-equipped to provide other functions of education, including cultivating human-to-human skills, such as collaboration and listening, or the broader functions of socialisation and individuation.\footnote{\citet{biesta_good_2009} identified three \emph{functions} that education can perform. \emph{Qualification}: Providing people with the knowledge, skills, understanding and judgement to \emph{do something}, for example a specific job or general-purpose skills such as critical thinking. \emph{Socialisation}: The transmission of norms, values, cultures, religions and traditions, with the primary goal of bringing individuals into existing orders and societal structures. \emph{Individuation}: Enabling individuals to become more autonomous and independent in their thinking and actions.}  

Over the longer term, if AI assistants were to displace teachers \emph{en masse}, it would likely require a broader transformation of formal education, for example via a large shift to fully virtual schools, which already exist and educate approximately 350,000 students in the US \citep{irwin_condition_2023}. However, leaving aside practical issues around student care and supervision, virtual schools in the US have so far shown subpar educational outcomes \citep{molnar_virtual_2021}. Such a scenario may be more likely in higher education, or adult learning, as students are more likely to have the self-discipline and motivation that evidence suggests is necessary to pursue 1-to-1 online learning. However, even leading virtual universities, such as the UK's Open University, generally target a \emph{hybrid} offering, primarily because students desire in-person engagement and more collaborative learning \citep{the_economist_intelligence_unit_new_2020}. 

Private tutors may face a greater risk of displacement, as many AI assistants are explicitly modelled on imitating their services, and AI tutors could potentially be integrated into free or low-cost online platforms, such as those provided by Coursera and Khan Academy. However, many private tutors work part time, often via digital platforms, so they are arguably better placed than teachers to integrate AI assistants into their work or to adapt the tasks that they provide, such as providing tutoring on how to best use AI tools in various subject areas. Moreover, only a small percentage of students have access to a human tutor, and demand is growing, helped by governments starting to fund private tutoring for disadvantaged students (e.g.\ the UK's National Tutoring Programme \citep{uk_government_national_2023}). The primary challenge such programmes face is a lack of high-quality tutors. This may point to an opportunity for governments to evaluate and fund innovative tutoring organisations that combine human and AI tutors for disadvantaged students.  

\subsubsection{AI programming assistants}

Software engineers, of whom there are approximately 27 million globally \citep{qubit_labs_how_2022}, will likely be the most directly affected by AI programming assistants. GitHub suggests that its Copilot tool is now `writing' 30\% of new code \citep{gain_githubs_2021}. Even if the figure rises to 80\% or 90\%, another big if, there are reasons to be confident that demand for human programmers will remain robust. First, demand for programming will remain strong due to rising AI deployment and the economy's broader \emph{digital transformation}, which involves shifting products, services and processes online, where there is considerable scope to grow further.\footnote{For example, ecommerce now accounts for almost 25\% of retail sales in the UK, North America and much of Asia, but just 16\% in Western Europe, 13\% in Latin America and 3\% in Africa, and there is considerable scope to grow further \citep{morgan_stanley_surprising_nodate}.} Second, past studies suggest that programmers may be the employees who are best able to adapt their jobs and tasks to account for AI assistants, as many are at least somewhat self-taught and used to adapting to new technology \citep{vincent_stack_2023}.\footnote{For example, a case study by \citet{horton_death_2020} showed how after Apple announced in 2010 that it would no longer support Adobe Flash, other Flash specialists, especially those who were younger, less specialised, or had good `fallback' skills quickly transitioned away from Flash. Similarly, a study by \citet{das_learning_2020} of 170m US job postings from 2010--2018 shows how the tasks of `IT jobs' evolved in response to new technologies, shifting towards machine learning, scripting languages and cloud solutions, and away from traditional software products and services that require workers to perform structured query language (SQL), Java, and data management.} Third, humans will likely remain central to ensuring that any AI-generated code is interpretable, secure and legally compliant. Finally, software is a fast-evolving industry, with new languages and techniques routinely emerging -- this creates a dearth of data for training AI assistants on cutting-edge use cases. 

\subsection{Effects on job quality} 

The salient questions for assessing the impact of advanced AI assistants on employment relate, first, to what the \emph{current state} of job quality is in the most affected occupations, and second to how AI assistants might affect wages, job security and other key drivers of job quality, such as job intensity and stress, autonomy, and employee relationships and collaboration. 

\subsubsection{AI education assistant} 

Many teachers face below-median wages and high degrees of stress, although the roles typically provide the opportunity to develop a strong sense of purpose, and private institutions and certain countries, like Singapore, provide better wages and working conditions. In the UK, almost half (48\%) of teachers say their workload is `unmanageable' \citep{national_education_union_state_2023}, and 44\% plan to quit by 2027 \citep{harrison_44_2022}. The primary challenges are workload and lack of supporting resources. In addition to a lack of teachers, these challenges are partly due to steady expansion in the domains that teachers are expected to cover, such as socioemotional learning and media literacy. Evidence on past technology deployment suggests that if educational institutions are mandated to integrate AI assistants, without accompanying training and resources, it could worsen job quality \citep{global_education_monitoring_report_team_unesco__2023}. However, teachers may also be able to use AI assistants to support them, for example by generating ideas, suggested adaptations or feedback on lesson plans and teaching materials, with some examples already starting to emerge \citep{wang_is_2023}.

\subsubsection{AI programming assistant}

Globally, studies suggest that programmers are relatively satisfied with their roles compared to other professions, helped by the fact that wage trends are positive in much of the industry \citep{graziotin_happiness_2019}.\footnote{In the US, the average wage of software developers is more than the average of all occupations \citep{bsa_1_2016}.} Key job-quality challenges include time pressure, getting stuck when problem-solving, working with bad code or with poor coding processes, and information overload. Independent evaluations of the effects of AI assistants on programmer job quality are lacking. However, in a GitHub survey of Copilot users, 60--75\% reported feeling more fulfilled with their job, less frustrated when coding and more able to `stay in the flow' and to focus on more satisfying work when using Copilot, and 87\% reported how it helped them to preserve mental effort during repetitive tasks \citep{kalliamvakou_research:_2022}. One caveat is that, if programmer roles become more about interrogating AI-generated code, some may start to feel less connection to their end output, which can be an important driver of job quality \citep{bryce_finding_2018}. 

\subsection{Effects on productivity growth}

To assess the impact of advanced AI assistants on productivity growth, we need to address several questions. How will AI assistants affect \emph{individual productivity} growth? To what extent will the assistant enable higher \emph{efficiency} vs \emph{creativity} and innovation? To what extent can we expect productivity changes for individuals to be mirrored at the \emph{industry} and \emph{economy} levels? 

\subsubsection{AI education assistant}

Education assistants could increase human capital, that is the knowledge, skills and personal characteristics that make people productive across all sectors of the economy \citep{egert_new_2022}. For this to be substantive, AI assistants would have to directly or indirectly enable a large number of people to better cultivate a broad range of knowledge, skills and characteristics, including how to best use AI. Such a shift is badly needed, as progress in human capital is stagnating, as evidenced (imperfectly) by a lack of progress in the OECD's Programme for International Student Assessment of 15-year olds on reading, maths and science \citep{oecd_is_2023} and the Programme for the International Assessment of Adult Competencies in literacy, while progress in AI is accelerating \citep{oecd_skills_2019}. While \emph{access} to education remains a critical issue for many people, the poor quality of education is arguably a bigger issue from a productivity growth perspective \citep{egert_new_2022}. AI assistants could \emph{potentially} help to improve the quality of education and help students to use AI, thus leading to a significant boost in human capital and productivity, although such effects would not be quick to materialise. 

\subsubsection{AI programming assistant}

The deployment of computing in the 1970s and 1980s initially had minimal impact on productivity growth, but it began to contribute more meaningfully during 1995--2005, before productivity growth began stagnating again. Some have suggested that the recent stagnation shows that digital technology is less transformative than technologies of centuries past \citep{gordon_rise_2017, cowen_is_2019}. \citet{andrews_global_nodate} suggest that the issue is more to do with laggard firms and sectors. \citet{peng_impact_2023} provide early evidence that programming assistants can boost individual efficiency (see also \citeauthor{vincent_stack_2023}, \citeyear{vincent_stack_2023}). Their ultimate impact on productivity growth may depend on two factors. Will they be able to provide novel code suggestions that programmers would not otherwise be able to produce? Will programming assistants be able to enable more rapid digitalisation in organisations and sectors that have traditionally lagged, such as healthcare, social care and the civil service? 

\subsection{Effects on inequality}

The salient questions regarding the impact of advanced AI assistants on inequality relate to which groups may disproportionately \emph{benefit} or \emph{suffer} as a consequence of this technology being deployed at scale, where this question can be assessed both \emph{between} and \emph{within} countries, and in relation to job creation, income and wealth (see 
Chapter~\ref{ch:16}). 

\subsubsection{AI education assistant} 

Significant inequities exist between and within countries with respect to access to quality education that both reflect and contribute to income and wealth inequalities. AI assistants may have the \emph{potential} to help to reduce these. For example, UNESCO's Global Education Coalition aims to use online resources to bring education to students not currently attending school \citep{unesco_transforming_2023}. However, historically, the most common result of introducing digital technology into education has been what \citet{burns_technology_2021} refers to as a \emph{caste system}, in which `the wealthiest students get online learning, poorer students get radio or TV \ldots{} and the poorest students get nothing'. This is primarily due to the supporting infrastructure, resources, family use and teachers that are needed to make technology use successful (see Chapter~\ref{ch:16}). To the extent that teachers may be affected by AI assistants, for example in terms of job quality, these effects will fall disproportionately on women, who make up the majority of teachers worldwide, especially for younger students \citep{european_parliamentary_research_service_teaching:_2020}. If AI assistants help to boost access to education for those who lack it, this would disproportionately help girls, although the \emph{gender parity} index for accessing and staying in education differs by country \citep{unicef_education_2022}. 

\subsubsection{AI programming assistant} 

If programming assistants lead to higher productivity, wages and new frontier roles for software developers, this may exacerbate inequalities within countries, as the average wage for these roles is typically already above the median. However, initial studies suggest that AI assistants may help to make programming-based roles more accessible. For example, one study found that respondents who were `learning to code' were more likely to use AI tools than `professional developers' \citep{stack_overflow_stack_2023}, while \citet{peng_impact_2023} suggest that AI programming assistants could help people to transition into software development careers, pointing to the potential training and upskilling opportunities that governments, AI labs and civil society could support.\footnote{Similarly, \citet{tu_what_2023} suggest that AI assistants may transform the work of data science from hands-on coding, data-wrangling and conducting standard analyses to assessing and managing analyses performed by automated AIs.}

\section{Policy Implications} 

\citet{acemoglu_power_2023}, \cite{brynjolfsson2022turing} and \cite{manyika2023coming} argue that technological vision and policy changes are required if the benefits of new technologies are to be widely shared (see also \citeauthor{ben2024ai}, \citeyear{ben2024ai}). In this spirit, policymakers and technology companies could potentially take actions beyond the status quo to ensure more positive societal outcomes from the use of advanced AI assistants. Some economists are sceptical about the ultimate potential for such directed technological change, because anticipating the economic impact of technology is difficult, and policy interventions to steer technology could create unintended outcomes \citep{agrawal_we_2023}. Despite that, some options and policy choices do exist. In this section, we outline three main actions that policymakers, AI labs and other stakeholders can take to try to ensure positive economic outcomes for society. 

\subsection{Monitoring how AI assistants are affecting the economy} 

Unlike other technologies such as industrial robotics, we have no clear picture of how many AI assistants are being deployed across the economy or what sectors they are being deployed into, thus inhibiting our ability to understand their economic impact. Given the diffuse nature of AI assistants and the many ways that organisations could adopt them, there is no easy solution to hand, but policymakers, industry associations and others could explore:

\begin{itemize}
\item \textbf{Collecting empirical evidence:} Fund more studies to understand the effect that AI assistants are having, including on neglected areas like job quality (see 
Chapter~\ref{ch:20}). 

\item \textbf{Develop new monitoring techniques:} Explore new methods for using AI and disparate data sources, including occupation data, start-up funding data and cloud computing services data to build more timely assessments of how AI assistants are affecting the economy. 
\end{itemize}

\subsection{Shaping the type of AI assistants that get developed or deployed} 

Policymakers need to develop well-calibrated expectations about their ability to shape the evolution of AI assistants in socially beneficial ways. Labour-displacing AI systems could generate significant negative externalities and lock society into economic trajectories with fewer jobs, reduced job quality and more inequality than is socially desirable, thus potentially limiting future progress and warranting interventions to steer their development in human-augmenting directions \citep{acemoglu_power_2023, korinek2018artificial, brynjolfsson2022turing}. However, overzealous attempts to direct the progression of AI could inadvertently foster protectionism and potentially lead to misguided efforts to centrally control or guarantee job availability. Such measures could generate unintended negative effects, such as stifling innovation, creating market inefficiencies and inadvertently perpetuating jobs or tasks that may have become obsolete. However, there are steps that policymakers, by working with AI labs, civil society and industry associations, could take, including:
 
\begin{itemize}
\item \textbf{Align on beneficial use cases:} Align on priority AI assistants to support, such as scientific research assistants and assistants for teachers or job seekers.

\item \textbf{Develop a research agenda:} To support these beneficial use cases, work with industry and civil society to align on key research questions, develop supporting data sets and fund supporting R\&D, including neglected foundational research.

\item \textbf{Explore how to encourage the development and deployment of beneficial types of assistants}. This could involve public funding for R\&D, responsible adoption in the public sector and other measures \citep{acemoglu_does_2020}. 
\end{itemize}

\subsection{Shaping the impact of AI assistant deployment} 

To ensure that the effect of AI assistants on employees and society is as positive as possible, policymakers, AI labs, civil society and industry associations could explore: 

\begin{itemize}
\item \textbf{Education and training} by carefully integrating AI, and AI assistants, into education and upskilling programmes, while not avoiding the non-AI-related challenges that have often limited the efficacy of such programmes on core education and employment outcomes. 

\item \textbf{Employee consultation rules} by upgrading existing approaches to employee consultation with respect to new technologies and drawing on best practice. 

\item \textbf{Broad-based policies to enhance economic resilience} by going beyond targeted policies to ensure that the economic impact of advanced AI assistants is economically beneficial, there are many other broad-based policy interventions that could help to mitigate negative impacts. These include interventions to strengthen the social safety net, active labour market policy interventions and industrial policies to spur economic growth that compensate for economic displacement and jobs losses \citep{juhasz_new_2023}.
\end{itemize}

\section{Conclusion}

This chapter examined the potential economic impacts of advanced AI assistants, particularly with respect to employment, job quality, productivity growth and inequality. While there is currently little substantive research on the economic impacts of AI assistants, as opposed to the economic impacts of AI in general, it is plausible that advanced AI assistants will have significant implications for each of the four variables examined. To that end, we recommend further research into how advanced AI assistants may impact employment, job quality, productivity growth and inequality. We also encourage policymakers to adopt active measures for monitoring and understanding how AI assistants are affecting the economy, to explore different levers that can be used to shape their design and deployment, and to implement plausible and evidence-based policy interventions to promote socially beneficial outcomes for AI assistants.

\chapter{Environmental Impact}
\label{ch:19}

\textbf{Juan Mateos-Garcia, Sims Witherspoon, Iason Gabriel}

\noindent \textbf{Synopsis}: As positive use-cases for advanced AI assistants continue to emerge in support of climate action, there is significant uncertainty about their overall environmental impact. While the study of AI's energy consumption and carbon emissions is still taking shape, some factors suggest that AI assistants could lead to increased computational impacts. However, there are many opportunities to increase the \emph{efficiency} of these processes and make them more reliant on carbon free energy. Ensuring that AI assistants have a net positive effect on the environment will require model developers, users and infrastructure providers to be \emph{transparent} about the carbon emissions they generate, adopt \emph{compute-} and \emph{energy-efficient} techniques, and embrace a \emph{green mindset} that puts environmental considerations at the heart of their work. Policymakers may also want to create incentives that support these changes, minimise the environmental impact of AI systems deployed in the public sector, support AI applications to tackle climate change and improve the evidence base about the environmental impacts of AI. Promisingly, it may be possible to develop AI assistants that broaden access to environmental education and scientific evidence -- and that improve the productivity of engineering efforts for climate action. 

\section{Introduction}
Anthropogenic climate change, driven by greenhouse gas emissions, is one of the most pressing issues facing our planet today. The effects of climate change are already being felt around the world in the form of more frequent extreme weather events, rising sea levels and changes in plant and animal life \citep{lee_climate_2023}. The development of large foundation models and widespread adoption of AI services could potentially contribute to these impacts \citep{bender_dangers_2021}. At the same time, with sufficient attention, technical and algorithmic innovations, better infrastructure, and access to carbon-free energy, we may be able to contain the potential environmental impact of AI or even reverse it over time, by improving productive efficiency and enabling innovations that contribute to wider environmental sustainability \citep{patterson_carbon_2021, Dannouni2023-xb, rolnick_tackling_2019}. 

Promising applications for AI in this space include efforts to produce carbon-free and low-carbon electricity via better forecasting \citep{lam2023learning} and scheduling of energy supply and demand for renewables, better storage technologies, and support to new sources of energy such as nuclear fusion \citep{degrave_magnetic_2022}. AI technology may also help to reduce the impact of transportation systems by modelling demand, improving freight routing and encouraging the adoption of electric vehicles. Lastly, AI may help to reduce greenhouse emissions from buildings and cities via developments in the field of smart buildings and smart cities \citep{luo2022controlling}. These trends create substantial uncertainty about the overall direction of the impact of AI on carbon emissions.

This chapter considers the question of environmental impact in the context of increasingly advanced AI assistants. What might their environmental impact be? Can they contribute to sustainability efforts and, if so, how? Lastly, what techniques, actions and policies  can be used to steer their development and deployment towards sustainable outcomes? Although we focus primarily on the link between AI and greenhouse gas emissions, it is also important to note that the development of powerful AI systems has a wider environmental impact -- including on global water consumption, the mining of minerals and the generation of toxic emissions \citep{crawford_atlas_2021, li_making_2023, Dannouni2023-xb}. Much of the discussion in this chapter is relevant for mitigating those other impacts as well. 

\section{The Environmental Impact of AI Systems}
\subsection{Climate change}
In its latest assessment report, the Intergovernmental Panel on Climate Change (IPCC) shows, with high confidence, that human-caused climate change is affecting communities across the globe \citep{lee_climate_2023}. This includes rising sea levels, extreme weather events such as heatwaves and droughts, mass species extinction and impacts on agriculture and fishing \citep{lee_climate_2023}. Moreover, the IPCC estimates that with only current mitigation efforts in place, the goals set out in the Paris Agreement are likely to be missed, with the global temperature rise exceeding 1.5\degree{}C in the first half of the century, potentially even reaching 2\degree{}C – which is a threshold that could have catastrophic consequences.

By way of illustration, the World Health Organization estimates that climate change could cause an additional 250,000 deaths annually between 2030 and 2050 \citep{world_health_organization_cop24_2018}, while global gross domestic product (GDP) could also decline by an average of 10\% \citep{swiss_re_institute_economics_2021}. It has also been estimated that 143 million people could be forced to migrate in response to climate emergencies by 2050, with communities in low- and middle-income countries particularly affected \citep{rigaud_groundswell_2018}.\footnote{These impacts are particularly felt in low- and middle-income countries that have historically created low levels of CO$_2$ emissions – raising further questions about economic and environmental justice \citep{lee_climate_2023, wenz_environmental_1988}.} These impacts would likely be accompanied by significant societal disruption, and by economic and political instability, posing a significant threat to human rights and global security \citep{levy_climate_2015}.

To achieve the goals set out in the Paris Agreement, the world needs to reduce global CO$_2$ emissions by 45\% by 2030 and to reach net zero by 2050, but the world is not on track to meet those targets \citep{ritchie_co_2020, UNFCCC_Secretariat2023-hb}. Action is therefore needed across a range of fronts. At a general level, energy consumption (including transportation, electricity, heat, building construction, and manufacturing) contribute the majority of carbon emissions worldwide (75.6\%), followed by agriculture (11.6\%), industrial processes (6.1\%), waste treatment (3.3\%) and land use (3.3\%) \citep{ge_4_2020}. Against this backdrop, cloud services and large scale data centres -- where the majority of AI computation takes place -- account for 0.1-0.2\% of greenhouse gas emissions, with around 25\% of their traffic related to AI \citep{kaack_aligning_2022}. 

Nonetheless, given the wider context it is important to understand and mitigate the potential environmental impacts of powerful general-purpose AI systems -- and to maximise any contribution they can make to efforts to tackle climate change. Here, we draw on a framework developed by \citet{kaack_aligning_2022} to distinguish between \emph{computational impacts}, \emph{application impacts} and \emph{systemic impacts} of computer systems. We begin with a broad assessment of AI’s environmental impacts in this section before focusing on assistant-specific issues in the next one (see Table~\ref{tab:19.1} for a summary).

\subsection{Computational impacts}
Data and compute are basic inputs into the development and deployment of modern AI systems (see Chapter~\ref{ch:4}).
Accessing these resources requires dedicated hardware for storage and processing (such as graphics processing units (GPUs)) and infrastructure including data centres and telecommunication networks. Interacting with AI systems also requires user and consumer hardware such as smartphones (to access AI assistants) or gaming consoles (to access AI-enabled video games), all of which require energy to function. 

This hardware creates two types of environmental impact: first, there are \emph{embodied impacts} created through its material collection, manufacturing and delivery. For example, the creation of semiconductors requires extraction of raw materials, manufacturing using large amounts of energy, water and hazardous chemicals, and emissions-producing transport to the delivery destination \citep{kuo_assessing_2022}. Second, \emph{operational impacts} are created when an AI system is designed, trained and deployed. Indeed, AI systems continue to consume energy after they are deployed whenever they make inferences, for example, in response to user queries. In addition, their availability to respond to user queries incurs idle energy use when they are not actively running inferences \citep{luccioni_estimating_2022}. 

A growing body of literature has started to study AI’s energy consumption and carbon emissions (and its drivers) using a variety of methodologies. Some key themes include:
\begin{itemize} [parsep=6pt]
    \item \textbf{Variation in estimates about how energy intensive modern AI systems are likely to be}. In their analysis of the environmental impact of machine learning (ML), \citet{strubell2019energy} estimated that training a single transformer model with neural architecture search generates CO$_2$ emissions that are equivalent to the operation of five cars during their lifetime (see also \citep{kuo_assessing_2022, strubell_energy_2020}. However, this estimate is challenged by \cite{patterson_carbon_2022} who argue that the impact is several orders of magnitude less. More recently, \cite{luccioni_estimating_2022} compiled information about state-of-the-art models to show that, for example, training GPT-3 created 552 tonnes of CO$_2$ equivalent, which is the same as that used for 550 round-trip flights between New York and San Francisco (see also \citeauthor{stokel-walker_generative_2023}, \citeyear{stokel-walker_generative_2023}). Additionally, \cite{luccioni_counting_2023} estimate  that carbon emissions are growing over time for a sample of AI systems (mostly developed in academia), with question–answering systems experiencing the fastest overall growth in emissions. Most analysis of the environmental impact of AI focuses on emissions during training, because it is relatively easy to measure how much energy is used in this way. As an exception, \cite{luccioni_estimating_2022} estimate that the embodied emissions  of BLOOM, a 176-billion parameter open-science, open-access language model developed in 2021–2022, account for 22\% of its total CO$_2$ emissions.
    \item \textbf{Inference may be more important than training when it comes to energy use.} \cite{luccioni_estimating_2022} also estimate the emissions generated during the deployment of BLOOM, calculating that the model emitted 19 kg of CO$_2$ emissions per day throughout the monitoring period. In their survey of the environmental impacts of AI systems at Facebook, \cite{wu_sustainable_2022} note that some use cases – such as language models – generate two thirds of their operational emissions during inference, while in other cases the carbon footprint is more evenly distributed between training and inference (something that depends on the size of the model, how many users it has and how often it has to be retrained). Focusing on overall energy usage, \cite{patterson_carbon_2022} found that 60\% of ML energy use at Google from 2019-2021 was attributable to inference.\footnote{Note that the actual emissions generated by this energy usage depend on the grid electricity mix.} 
    \item \textbf{Demand for larger models and improvements in operational efficiency have countervailing effects on energy consumption and emissions}. Scaling laws suggest that increases in model size (which require larger amounts of compute for training and inference) are associated with predictable improvements in model capabilities \citep{kaplan_scaling_2020}. This is linked to rapid growth (a doubling every 5–6 months) in the amounts of compute used to train state-of-the-art AI models, which could increase energy demands  \citep{sevilla_compute_2022}. At the same time, \cite{patterson_carbon_2022} argue that improvements in model efficiency and infrastructure will ultimately help contain the energy costs and environmental impacts from larger models. They compare training GPT-3 with GLaM, with the latter more recent model becoming 2.8 times more efficient as a result of model improvements. Furthermore, running GLaM on low-carbon infrastructure greatly reduced the estimated resulting carbon emissions. We discuss various technical and infrastructural levers, which can be used to mitigate the environmental impacts of AI and AI assistants in more detail in Section 18.4.
    \item \textbf{Environmental impacts are shaped by factors internal and external to the development process}. When looking at internal factors, model size seems to be the main determinant of energy costs. Larger models generally require longer training times using more energy-intensive GPUs. On this point, \cite{luccioni_counting_2023} find a strong correlation between training time, energy consumption and CO$_2$ emissions. When looking at external factors, data-centre energy efficiency and the carbon intensity of the electricity grid – which may depend on high-emission energy sources vs low-carbon sources – are a key determinant of a model’s CO$_2$ emissions \citep{dodge_measuring_2022, wu_sustainable_2022}. This points at the possibility of decoupling AI energy consumption from carbon emissions through the use of carbon-free energy sources, a point we return to in section 18.4.
    \item \textbf{Analyses of the environmental impacts of AI rarely consider counterfactual scenarios}. AI systems are deployed to undertake economically valuable activities which, in their absence, might have to be performed using alternative production processes and technologies that also consume energy and generate carbon emissions. These counterfactual or `replacement' costs are generally neglected in the literature but remain a key part of determining overall impact. 
\end{itemize}

\subsection{Application impacts}
AI systems also have an impact on the environment through the applications that they enable. \cite{kaack_aligning_2022} distinguish between applications that contribute to environmental sustainability and applications that increase carbon emissions, thereby accelerating climate change. 

\cite{rolnick_tackling_2019} explore the first type of application – those that have the potential to \emph{mitigate} climate harm – across thirteen areas, ranging from electricity systems to collective decision-making. In broad terms, AI systems can help to tackle climate change by improving our understanding of its extent, drivers and impacts, by optimising systems to mitigate those impacts and by accelerating the transition to a sustainable economy. \cite{Dannouni2023-xb} distinguish between three use cases for AI in climate action -- mitigation (including measurement, monitoring, reduction and removal), adaptation and resilience (including hazard prediction and vulnerability management), and foundational capabilities (for climate and economic modelling, behavioural change, and innovations and breakthroughs). Salient examples include AI systems that can be used to predict extreme weather events \citep{ravuri_skillful_2021}, reduce the energy consumption of industrial cooling systems \citep{wong_optimizing_2022}, help design plastic-eating enzymes \citep{kincannon_biochemical_2022} and accelerate fusion science \citep{degrave_magnetic_2022}.   

On the flipside, AI could potentially increase the productivity of extractive and carbon-intensive industries such as oil and gas or cattle farming, helping them continue or even ramp-up their activities \citep{greenpeace_oil_2020, kaack_aligning_2022}. 

\subsection{Systemic impacts}

The last category in the \citep{kaack_aligning_2022} framework aims to capture the systemic impacts of AI which encompass indirect and second-order environmental effects. This includes situations where AI-driven improvements in efficiency or costs lead to a rebound effect in production and $CO_2$ emissions (referred to as ‘Jevon’s paradox’) – or cases where AI entrenches or displaces unsustainable technologies and consumption patterns. For example, more capable self-driving cars could reduce the use of public transport \citep{lanzetti_interplay_2021}, which would be potentially problematic from a climate perspective. Alternatively, AI services could reduce demand for travel via better scheduling and more effective and immersive video-conferencing. These dynamics, though important, are often speculative and tend to be particularly hard to model or evaluate accurately (see Chapter~\ref{ch:20}). 

\section{The Environmental Impact of Advanced AI Assistants}
Having set out a general framework for understanding the environmental impacts of AI, we consider here the special case of advanced AI assistants. How might the nature of their development and capabilities shape those impacts? 

\subsection{AI assistant computational impacts}

Three features of AI assistants suggest possibly increased computational impact. 

First, AI assistants are likely to be based on transformers trained on GPUs during long runs in some cases involving multimodal learning (see Chapter~\ref{ch:4}), which can be more computationally intensive (\citeauthor{xu_multimodal_2023}, \citeyear{xu_multimodal_2023}). Looking at foundation models, GPT-3’ training is estimated to have generated 552 tCO$_2$ emissions. More recently, Meta’s Llama 2 -- a collection of open models optimised for dialogue -- emitted 539 tCO$_2$ during training \citep{touvron_llama_2023}.\footnote{It is worth noting that this includes several models of different sizes. The largest Llama 2 model (70B) produced 239 tonnes of emissions. The model developers point out that they offset all emissions generated by Llama 2 through Meta’s sustainability program.} All  this suggests that the development of more advanced AI assistants, animated by powerful foundation models, could be an energy and carbon-intensive technology activity. At the same time there is scope to significantly mitigate these effects through improvements in efficiency that reduce energy consumption and through use of carbon-free sources that break the link between energy consumption and carbon emissions. 

Second, AI assistants are targeted at consumer markets, suggesting potential  operational impact through inference (see Chapter ~\ref{ch:4}). 
As \cite{wu_sustainable_2022} note, this kind of large-scale deployment produces the bulk of emissions for LLMs. For example, the BLOOM model tested by \cite{luccioni_estimating_2022} generated 340 kg of CO$_2$ emissions after receiving 230,000 queries during two weeks of deployment. AI assistants deployed in large consumer markets over longer time-frames could have bigger impacts.

Third, there are additional environmental impacts from downstream AI development activities such as the use of reinforcement learning from human feedback (RLHF) to improve model usability or the use of tools and APIs by AI assistants to enhance assistant capabilities (\citeauthor{ziegler_fine-tuning_2020}, \citeyear{ziegler_fine-tuning_2020}; \citeauthor{schick_toolformer:_2023}, \citeyear{schick_toolformer:_2023};  see Chapter~\ref{ch:5}). This latter class of impacts could be particularly significant if advanced AI assistants make substantial use of other AI tools -- such as question answering systems or programming tools -- that consume additional energy.

\begin{table}
\caption{AI and AI assistant environmental impacts}
\label{tab:19.1}
\begin{tabularx}{\textwidth}{ X p{11cm} }
	\toprule
	\textbf{Type of AI impact} & \textbf{Factors relevant for advanced AI assistants} \tabularnewline
	\midrule
  \textbf{Computational} & \emph{Drivers that might increase emissions} \tabularnewline
	Impacts \emph{embodied} in production and distribution infrastructure, including construction processes, materials and resources. \newline \newline 
	\emph{Operational} impacts during exploration, training and deployment (including inference).	& 

\begin{itemize}
	\item Size of underlying models
  \item Scope for mass-market deployment
  \item Additional impacts during development (e.g. Reinforcement Learning from Human Feedback) and deployment (e.g.\ tool use by AI assistants)
\end{itemize}

\emph{Drivers that might decrease emissions}:
\begin{itemize}
    \item  Efficiency measures during model development and training
    \item Improving access to carbon-free energy options
\end{itemize}

\emph{Areas of uncertainty}
\begin{itemize}
	\item Scale of embodied emissions
  \item Role of smaller models and edge computing
\end{itemize}
\tabularnewline
\midrule
	\textbf{Application} & \emph{Applications of assistants with environmental implications} \tabularnewline
	The use of AI for applications with \emph{direct} impacts on the environment (either positive or negative).	&	
	These impacts are expected to be \emph{moderate} in the near term because:
\begin{itemize}
	\item Sustainability-related use cases in education, science policy, software development and research and development (R\&D) generally create indirect / long-term positive impacts through coding, scientific R\&D and education. 
	\item AI assistant use cases for the extractive industries are likely quite limited.
\end{itemize}
\tabularnewline
	\midrule
	\textbf{Systemic} & \emph{Areas of uncertainty} \tabularnewline 
	Indirect environmental impacts created by AI (e.g. changes in consumption patterns) & 
	Systemic impacts could be important but are difficult to estimate with any degree of precision.  Beneficial impacts could result from improvements in public awareness and education about climate change. Conversely, the potential contribution to increased general consumption or environmental misinformation are sources of risk. 
\tabularnewline
\bottomrule
\end{tabularx}
\end{table}

Further areas of uncertainty that have the potential to significantly affect the calculus discussed so far include: 
\begin{itemize} [parsep=6pt]
    \item \textbf{Scope for mitigating impacts via efficiencies in model development and deployment, and by sourcing energy from low-carbon sources}. As noted previously, the development of energy-intensive AI systems has been accompanied by parallel improvements in the efficiency of processes and infrastructure used for AI model development and training. There is also significant scope to further shift energy consumption towards low-carbon sources that generate less emissions. We discuss these considerations next section. 
    \item \textbf{Scope for reducing emissions by efficiently training only a few foundation models.} These models could then be fine-tuned downstream instead of training many separate foundation models \emph{de novo}.
    \item \textbf{Demand for smaller models.} There are instances where small fine-tuned models, such as Alpaca, perform competitively when compared to larger models. If smaller models prove popular, they could be deployed with much lower operational impacts. These models could also potentially be run on ‘edge devices’ such as smartphones, thus increasing their energy efficiency and reducing data transport costs \citep{qualcomm_future_2023}.
    \item \textbf{The scale of embodied emissions.} As we noted above, there is a dearth of data about the embodied impacts of AI on the environment, including production, transport and disposal. However, it has been argued that these effects could be substantial \citep{crawford_atlas_2021, gupta_chasing_2020, patterson_carbon_2021}. If assistants increase the demand for AI services and related hardware, including semiconductors and data collection and storage, this could intensify their overall impact. 
    \item \textbf{Counterfactual energy costs and emissions generated by providing assistant-like services without AI.} Advanced AI assistants are likely to provide economically valuable services to their users which, in their absence, would have required alternative processes and technologies that also consume energy and generate emissions. It is important to take these into account when assessing the environmental impacts of AI assistants. 
\end{itemize}
Taken together, these dynamics provide reasons for vigilance when considering the environmental impacts of advanced AI assistants and suggest that choices about how advanced AI assistants are designed and operationalised are likely to be consequential from an environmental perspective.

\subsection{AI assistant application impacts}
\cite{rolnick_tackling_2019} classify AI technologies based on their relevance for 36 solution domains that may help to tackle climate change. The closest technology to AI assistants that they consider is natural language processing (NLP) (see Chapter~\ref{ch:4}). 
Their assessment suggests that NLP is unlikely to be transformative for efforts to tackle climate change: it is highlighted as relevant for seven solution domains (i.e. 20\% of the total) and is expected to generate indirect and generally longer-term impacts.  However, it is also important to note that their assessment was based on NLP-capabilities circa 2019. More advanced AI assistants could be expected to create new opportunities for developing environmental applications either directly or indirectly. For example, they might improve the efficiency of scientific R\&D overall, improve software development for environmental use cases \citep{peng_impact_2023}, or help to synthesise scientific evidence to inform policy \citep{tyler_ai_2023}. The initiative \hyperlink{https://climatepolicyradar.org/}{Climate Policy Radar} is a good example of this last effect.

One important  solution domain where AI assistants could help is education, where \cite{ge_4_2020} and \cite{rolnick_tackling_2019} suggest that AI-powered tutoring systems could democratise and improve  the public understanding of climate change via personalised and translated content (e.g. simulating the impact of climate change on a learner’s location) \citep{amini2023ai}. In the service of these educational goals, large language models such as ClimateBERT, and chatbots like ChatClimate, have been trained on high quality data from authoritative sources -- in order to provide access to trustworthy information about environmental science and climate change \citep{Webersinke2021-fe, vaghefi2023chatclimate}.\footnote{ClimateBert provides a detailed emissions scorecard including grid emissions details in its \href{https://www.chatclimate.ai/climatebert}{website}.} Nonetheless, it is worth noting that although there is evidence that educational interventions can contribute to pro-environmental behaviours, the benefits are indirect and more likely to materialise in the  the long term (\citeauthor{begum_impact_2021}, \citeyear{begum_impact_2021}; see also risks arising from Chapters~\ref{ch:10} and~\ref{ch:17}).

\subsection{AI assistant systemic impacts}
The indirect nature of system-level impacts makes analysis of them both speculative and hard to reliably measure (see Chapter~\ref{ch:20}).
Some examples, in the case of AI assistants, include the potential for shifts in  demand and efficiency in energy-intensive activities such as computer programming (which are made cheaper by AI-coding tools) and potential shifts in overall levels of misinformation or polarisation (which can impact society’s collective ability to respond to climate change, see Chapter~\ref{ch:17}).

\section{Mitigating Negative Environmental Impact}
Having outlined the factors that might drive or help to mitigate the environmental impacts of AI assistants, application opportunities and systemic effects, we now consider the implications for model developers and adopters, primarily in industry and government, who are committed to helping to achieve international climate goals (also see \cite{Dannouni2023-xb}).

\subsection{Levers for model developers and users}
\textbf{Model developers and users need to prioritise the adoption of technical measures that mitigate and reduce the environmental impacts of AI assistants.} Such measures are needed both ahead of time and throughout their deployment. In the context of AI energy efficiency at Google, \cite{patterson_carbon_2022} argue that better model architectures could help contain the environmental impact of this technology. They note, for example, that sparse models require five to ten times less compute than those following more denser  model architectures. Furthermore, \cite{wu_sustainable_2022} suggest other strategies that may prove helpful in reducing computational costs. These include careful data scaling, selection and sampling, and developing memory and data-efficient architectures. Model developers and users should adopt these and other techniques with the goal of mitigating the environmental impacts of AI systems. 

\textbf{Model developers and users should use hardware and infrastructure that minimise the environmental impacts of AI systems.} Several factors, beyond AI model design and training, influence the overall energy consumption and emissions produced by the AI ecosystem. This includes the energy efficiency of hardware and data centres and the carbon intensity of the electricity grid. \cite{patterson_carbon_2022} identify opportunities to reduce energy consumption one-hundred fold, and carbon emissions by up to one thousand times, by adopting sparse architectures, using processors optimised for ML training, energy-efficient cloud services, and by optimising the allocation of computing workloads across data centres to maximise the use of clean energy sources  (see also \cite{schwartz_green_2020}). 

\textbf{Model developers and users should, when possible, source carbon-free energy for the data centres where AI systems are trained and inferred.} Drilling down further, into the AI carbon emissions produced by cloud computing, \cite{dodge_measuring_2022} note that the choice of region and time of day, and the use of workload optimisation methods that take into account carbon emissions can play a substantial role in mitigating AI’s environmental impacts. Model developers and users can harness these opportunities to minimise the environmental impacts of AI systems. 

\textbf{Model developers and users need to support transparency around the computational efficiency and energy consumption of AI models and infrastructure.} Mitigating the environmental impacts of AI assistants is likely to require better data about those impacts. To support this goal, AI models' computational efficiency and energy costs can be incorporated into development and benchmarking evaluations, thus reducing the need for \emph{ex post} estimations of energy costs or emissions based on incomplete information (\citeauthor{patterson_carbon_2022}, \citeyear{patterson_carbon_2022}; see Chapter~\ref{ch:20}). 

A growing number of initiatives have appeared in this space, including the creation of various tools to measure model carbon emissions (summarised in \cite{luccioni_counting_2023}), the adoption of new model evaluations (see \cite{luccioni_counting_2023, solaiman_evaluating_2023}) and changes in submission information requirements for major computer science conferences such as NeurIPS and NAACL. This information could then be used to implement labelling schemes that increase the visibility of model developers and cloud providers adopting good sustainability practices – building on the growing adoption of techniques to document model development and risks such as model cards  \citep{mitchell_model_2019}. However, the proliferation of transparency requirements and voluntary reporting initiatives should be carefully coordinated, as they could otherwise create duplication and inconsistencies in reporting -- and increase the overall reporting burden, without creating the desired benefit.

\textbf{Model developers and users need to adopt a `green AI` mindset in relation to the environmental impact of their choices and work.} Each of the aforementioned initiatives needs to be underpinned by a reorientation in the AI R\&D community towards a `green mindset’ that foregrounds environmental considerations. By remaining attentive to these questions and maintaining awareness of the relationship between the climate crisis and their own work, developers can take proactive measures to mitigate the harmful environmental impacts of AI by treating this as an important goal -- alongside other considerations such as model accuracy \citep{schwartz_green_2020}. The \hyperlink{https://greensoftware.foundation/}{Green Software Foundation} is a good example of an effort in that advances this perspective and way of thinking. 

\subsection{Policy levers}\label{sec:19.4.2}
\textbf{Policymakers may want to consider  policies that improve access to low-carbon or carbon-free energy for model development and deployment.} This includes accelerating grid decarbonisation with standards and incentives that help accelerate the deployment of carbon-free energy (including via the adoption of AI systems to optimise the grid), funding research and development for emerging carbon-free energy technologies that may enable the deep decarbonization of grids, as well as greater retail access for model developers and users to source carbon-free energy for their operations \citep{Golin2022-is}. 

\textbf{Policymakers should ensure that the public sector leads by example in the development and adoption of sustainable AI methods and adopts a green AI mindset}. This includes funding research to develop sustainable AI methods, ensuring effective transparency around energy consumption and emissions, adopting sustainable model development and training strategies, and prioritising energy-efficient cloud infrastructure when conducting research into AI as part of a public project \citep{ho_building_2021, Dannouni2023-xb}. These interventions would have the added benefit of increasing the supply and potential flow into industry of AI scientists and engineers with a green AI mindset and skill base. 

\textbf{Policymakers should explore how to enable the development of  AI applications that contribute to sustainability.} This includes direct funding for AI research with sustainability applications, increasing the availability of data sets to train those applications, building expertise and supporting deployment in relevant sectors \citep{global_partnership_on_ai_climate_2021, Dannouni2023-xb}.

\textbf{Policymakers can help to reduce uncertainty and improve the evidence base about AI’s computational and systemic impacts on the environment.} Measures of this kind include supporting the development of better methodologies to measure the environmental impact of AI (taking into account counterfactual impact from delivering products and services without AI support - the \hyperlink{https://digital-strategy.ec.europa.eu/en/policies/european-green-digital-coalition}{EU Green Digital Coalition} is already driving progress in measuring the net environmental impact of digital technologies), supporting agencies that can implement these methodologies impartially -- and encouraging transparency around energy costs and $CO_2$ emissions by model developers, users and infrastructure providers, in a way that is \emph{standardised} and avoids duplication and excessive administrative burdens.

\section{Conclusion}
The climate crisis has put advanced AI systems in the spotlight, both as potential contributors to climate change and as a potential source of solutions for tackling it. In this chapter we draw connections between these matters and the development of advanced AI assistants. Our analysis points towards the existence of deep underlying uncertainty about the likely impact of advanced AI assistants on climate change. While the foundation models, upon which advanced AI assistants are based, continue to grow in size, a range of substantial and successful efforts -- aimed at increasing operational efficiency and leveraging low-carbon or carbon-free sources during training and deployment processes -- are now underway. 

Model developers, users, infrastructure providers and policymakers all have a role to play in ensuring that the technical and non-technical measures needed to reduce the impact of AI systems are identified and realised, that there is transparency about these impacts, and that the sector continues to cultivate a green AI mindset -- one that places as much emphasis on environmental considerations as on other dimensions of model performance (see Chapter~\ref{ch:6}).
By acting in concert, these stakeholders can help ensure that advanced AI assistants have a net positive impact on the environment.

\chapter{Evaluation}\label{ch:20}

\textbf{Laura Weidinger, Maribeth Rauh, Lisa Anne Hendricks, Arianna Manzini, Nahema Marchal, A. Stevie Bergman, Geoff Keeling, Will Hawkins, Iason Gabriel, William Isaac}

\noindent \textbf{Synopsis}: 
		This chapter provides a high-level introduction to AI evaluation, with a specific focus on AI assistants. It explores the \emph{purpose} of evaluation for AI systems, the \emph{kinds} of evaluation that can be run and the \emph{distribution of tasks} across three layers of output (the model level, user-interaction level and system level) and among different actors. The chapter notes that, with regard to many salient risks and goals that we need to attend to in the context of AI assistant development, there are significant evaluation shortfalls or gaps. To address these limitations, the chapter explores what a more complete suite of evaluations, nested within a robust evaluation ecosystem, would look like and makes recommendations on that basis. 

\section{Introduction}\label{sec:20:1}

This chapter focuses on the evaluation of advanced AI assistants. As has become clear throughout this paper, AI assistants raise a range of ethical concerns and considerations, including concerns around value alignment, privacy, anthropomorphism, misinformation, and safety. How do we proceed from here? To assess, prioritise and address the ethics of AI assistants, potential risks and benefits must be understood as comprehensively as possible. Alongside exercising \emph{foresight}, which is needed to identify potential risks, and \emph{monitoring} real-world outcomes (including unexpected failure modes and accidents as a technology is deployed), \emph{evaluation} is a key component of understanding potential risks and benefits.

Evaluation is the practice of assessing the capabilities, robustness and impacts of an AI system against goals or risks. For example, we may evaluate the likelihood that an AI assistant disseminates misinformation, or we can assess the impact that using an AI assistant has on people's well-being. Evaluation requires operationalising potential harms into tractable, measurable observations. It also requires making a normative assessment of what merits evaluation in the first place, and at what stage AI assistant performance is `good', `fair' or `safe enough'. By providing information on AI assistant capabilities, robustness and impacts, evaluation can play two critical roles for the ethics of AI assistants: it can guide AI iterative model \emph{development} by providing a target for AI developers, and it can provide \emph{assurances} and inform responsible decision-making on the design, risk mitigation and release of AI assistants. 

In this chapter, we describe the evaluation of AI assistants as a fundamental building block to creating ethical systems. We proceed as follows. We first provide an overview of the practice of evaluating AI systems. We then survey existing approaches for evaluating AI assistants for the risks of harm discussed in this paper, before discussing the limitations and gaps of such approaches. The conclusion summarises the chapter.   

\section{Evaluating AI Systems}\label{sec:20:2}

Evaluation is the practice of assessing an AI system's \emph{capabilities}, \emph{robustness} or \emph{impact}. Capabilities refer to the functionality and usability of the AI system, and its limitations. Can it perform the tasks that its developers intend? Does it function equally well in different languages, and is it accessible to differently abled users (see Chapter~\ref{ch:16})? 
Robustness refers to predictability and reliability of the system (i.e.\ the degree to which the system will \emph{consistently} behave in acceptable ways, including in novel situations). The impact of an AI system refers to its effects on people who directly or indirectly interact with it, and on broader structures in which the AI is embedded, such as the natural environment, society and the economy. Through evaluation, the capabilities, robustness and impacts of an AI assistant can be understood. This is particularly critical for AI assistants, given the ethical risks they pose. In this chapter, we outline the \emph{aims of evaluation}, \emph{elements and varieties of evaluation} and a \emph{three-layered evaluation framework} to comprehensively assess ethical considerations on AI assistants. 

\subsection{Aims of evaluation} 

Evaluation serves two main aims. It is critical for guiding AI assistant \emph{development} and informing \emph{responsible decision-making} at multiple time points over the life cycle of an AI assistant. 

Evaluation serves a critical function in AI \emph{development}. AI developers perform some evaluations at regular intervals over the course of AI system development to track AI system performance against established tests. Performance on these tests is taken as an indicator of overall progress in AI model development, and it can be compared against performance of other AI systems. Surpassing other AI systems or reaching certain thresholds on these established tests is often read as an important signal that an AI system has reached or exceeded state-of-the-art levels of performance in the field. In this process, evaluation provides a regular signal that guides model development and is a fundamental building block of iterative design. This process is also referred to as `hill climbing' (\citeauthor{sutton_bitter_2019}, \citeyear{sutton_bitter_2019}; see Chapter~\ref{ch:4}). 

Second, evaluation is a critical component of \emph{responsible decision-making} at different time points and for different actors, including AI developers, users of AI assistants, regulators and civil society. For AI developers, evaluation underpins decisions on guiding AI development, anticipating and mitigating potential risks, and whether work on an AI system should be stopped until certain concerns are resolved or harms are mitigated \citep{shevlane_model_2023}. Evaluation can help product developers to compare functionality for different possible user groups and underpin normative decisions such as whether an AI system's performance is `safe enough' or `good enough' for use in a given context (\citeauthor{bakalar_fairness_2021}, \citeyear{bakalar_fairness_2021}; see Chapter~\ref{ch:6}). 
Evaluation can also inform potential downstream users of an AI assistant to understand in what contexts it is safe to use. For public authorities, civil society and users of AI assistants, evaluation results are required to ensure that an AI system is used in contexts in the real world for which it has been tested and is safe (\citeauthor{mitchell_model_2019}, \citeyear{mitchell_model_2019}; see Chapter~\ref{ch:8}). 

\subsection{Elements and forms of evaluation}

Evaluation consists of technical and normative steps. It requires (1) \emph{selecting a target} for evaluation, such as a risk of harm or a performance goal to measure, (2) \emph{operationalising the target} into a concrete \emph{test} or \emph{metric}, which may require trading off different considerations, and (3) \emph{assessing results against} established \emph{thresholds or aims}. A target in evaluating the ethics of AI assistants may be the likelihood of the AI assistant to mislead a user (see Chapter~\ref{ch:17}), 
or the factors that affect the extent to which users anthropomorphise the AI assistant (see Chapter~\ref{ch:11}). 
Operationalising the evaluation target into a concrete test and metric requires a \emph{theory} of how a given harm can be detected and measured. There are often multiple ways in which a target can be operationalised, and deciding how to define and measure the target is a normative and contestable decision. For example, misinformation risks may be operationalised as the frequency at which an AI assistant outputs correct vs incorrect statements \citep{lee_factuality_2023,lin_truthfulqa:_2022}. It could also be operationalised as how likely users are to believe false outputs that are AI generated \citep{bai_artificial_2023}. Alternatively, it could be evaluated by measuring the broader spread of misinformation as there is an increasing uptake in AI assistants \citep{allen_evaluating_2020}. Operationalisation may also require normative trade-offs, for example on how to weigh false positives against false negatives in a given metric. Once a measurement is obtained, the third piece of evaluation is to assess the observed model performance against a normative threshold or an aim. Note that evaluation is never neutral: it always requires a normative judgement on how different results should be valued  \citep{bowker_sorting_2000}.

Evaluation can take different forms. Most commonly, evaluations of AI systems are \emph{automated performance tests} against tasks or data sets aimed at capturing capabilities of interest. These automated evaluations may target concepts as narrow as `correctly identify a user's voice' or as broad as `helpfulness'. Additional modes of evaluation that leverage human expertise, including human evaluations \citep{glaese_improving_2022,thoppilan_lamda:_2022}, adversarial testing \citep{perez2022discovering}, user testing \citep{lee_evaluating_2023}, bespoke tests of AI assistants in particular contexts \citep{marda_importance_2021} and expert assessments of AI impacts \citep{raji_closing_2020}, are increasingly being developed and are indispensable for a comprehensive evaluation of the ethics of AI assistants. Pre-deployment evaluation is complemented by post-deployment monitoring to assess performance and identify unexpected failures or accidents at the point of use. 

\subsection{Layers of evaluation and responsibilities}

To assess the ethical considerations raised throughout this paper requires evaluation at \emph{multiple layers}. AI systems are often evaluated at the capability layer, where AI assistant outputs are evaluated against established benchmarks. However, several of the risks of harm that arise for AI assistants may only be observable at the user--AI interaction layer (\citeauthor{tahaei_human-centered_2023}, \citeyear{tahaei_human-centered_2023}; see Chapters~\ref{ch:10}, \ref{ch:11}, \ref{ch:12}, and~\ref{ch:13}). 
Moreover, where harms are very subtle, difficult to measure or only emerge when AI assistants are used at scale, they may only be observable at the layer of broader systems (\citeauthor{kleinberg_algorithmic_2021}, \citeyear{kleinberg_algorithmic_2021}; \citeauthor{toups_ecosystem-level_2023}, \citeyear{toups_ecosystem-level_2023};  see Chapters~\ref{ch:7}, 
\ref{ch:9}, 
\ref{ch:16}, 
\ref{ch:18} 
and~\ref{ch:19}). 
Comprehensive evaluation of the ethics of AI assistants requires analysis at these three layers:

\begin{itemize}
\item[(1)]	\emph{AI capabilities}, measuring outputs and function of the AI system or its \emph{components} (such as training data).
\item[(2)]	\emph{Human--AI interaction}, measuring risks of harm to a person interacting with the AI system.
\item[(3)]	\emph{Broader systems}, measuring risks of harm through societal, environmental or economic analyses. 
\end{itemize}

Different actors along the value chain of AI system development may have \emph{different responsibilities} and be well-placed to perform \emph{different evaluations} at \emph{different layers}, depending on their autonomy and knowledge of the AI system and resources available for evaluation. Given their degree of knowledge and autonomy over what they are building, AI system \emph{developers} have the \emph{primary responsibility} for conducting sociotechnical evaluations pertaining to AI capabilities (layer~1) \citep{owen_organisational_2021,Stilgoe_Owen_Macnaghten_2013,dignum_responsible_2019}. Third-party auditors with the relevant skill sets are also well-placed to perform capability evaluations. 

Product developers are uniquely well-placed to assess user--AI interaction pre-deployment, and thus have a special responsibility to provide these evaluations. Note that the distinction between model and product developers is shrinking, so it is often the same organisations which bear primary responsibility for capability and human--AI interaction testing. For example, a cloud provider which offers a basic model to a third party for a particular use case (e.g.\ question answering for education) may be responsible for evaluation at the capability and human--AI interaction layer (i.e.\ to test whether the model is adequate for the given use case, including accounting for how people are likely to use it). The third party may then offer an adapted version of the basic model to consumers (e.g.\ students). In such cases, the third party has additional evaluation responsibilities to ensure the AI system is safe in the context for which it is offered (i.e.\ to ensure that the overall AI assistant is functional and safe in a educational context). External evaluation at the capability and human--AI interaction layers may in some cases require novel infrastructure to ensure safe access to third-party testers to model components and outputs pre-release.

Public stakeholders are uniquely positioned to perform system-layer evaluations where they can leverage specialist knowledge that AI developers lack (e.g.\ financial or environmental) and may be better positioned to anticipate harms and develop evaluation protocols for these \citep{raji2022fallacy,owen_vision_2013}. However, responsibility for evaluation at all levels is shared: while some groups have primary responsibility to cover different layers of evaluation, all actors have some responsibility to ensure comprehensive evaluation of risks of harm.

\section{Evaluating Advanced AI Assistants}\label{sec:20:3}

Several ethical considerations about advanced AI assistants have been raised in this paper. In this section, we survey existing approaches to evaluating them. We find that many ethical considerations and concerns about AI assistants are not currently routinely evaluated, often because no valid, robust and tractable evaluation method exists. To address these shortfalls, we provide recommendations about how automated as well as additional forms of evaluation can be expanded to enable coverage of currently neglected ethical concerns.

Identified areas of ethical consideration in this paper span value alignment, technical safety, trust, privacy, anthropomorphism, persuasion and manipulation, user preferences, user well-being, appropriate and inappropriate relationships, cooperation, equity and access, economic impact, malicious uses, misinformation and environmental impact. 

Evaluations exist for some of these domains, but they mostly cover model-centric types of analysis: for example, assessing energy use as a proxy for environmental impact \citep{strubell_energy_2020} or assessing the propensity of AI systems to regurgitate information that is present in the training data as a proxy for leaking private information \citep{carlini_quantifying_2023}. For risks that manifest during human--AI interaction or as AI systems are widely deployed, early indicators may be observable at the model layer. For example, to assess likelihood of downstream anthropomorphism, it could be evaluated whether a model refers to itself as having personhood or human properties such as having preferences, opinions or a family history \citep{glaese_improving_2022}. Similarly, some risks of societal harm can be foreshadowed and measured at the model layer, such as by assessing disparate performance of AI assistants in different languages, indicating likely differences in how well the AI assistant may work for different language groups and communities \citep{lewis_mlqa:_2020}.

The challenge at this level is that evaluations often lack critical context and so cannot provide a valid assessment of the ethical consideration in isolation. Tests at the model layer are often criticised for operationalising complex concepts into overly \emph{narrow metrics} and, as a result, not presenting valid results on the overall capabilities or risks that they purport to measure \citep{raji2022fallacy,narayanan_gpt-4_2023,blodgett_stereotyping_2021,schlangen_language_2019}. Rather, these assessments need to be interpreted in the context of additional information such as how a system is used \citep{rahwan2019machine}. This leads to conflicting inferences from different benchmarks: one paper describes the `benchmark lottery', whereby model performance may seem high on one benchmark but low on another benchmark purportedly testing the same construct \citep{dehghani_benchmark_2021}. To probe a harm more comprehensively, multiple benchmarks can be aggregated into a suite of tests (e.g.\ HELM \citep{liang_holistic_2022}, BIG-Bench \citep{srivastava_beyond_2023} and Safetykit \citep{dinan_safetykit:_2022}). However, more complex benchmarks may still not capture relevant context that can only be observed by assessing an AI system in the \emph{context} in which it is used, and collapsing multiple tests into a single result can make it harder to interpret results, thus raising again issues of valid inferences \citep{burnell_rethink_2023}.

More recently, and particularly in the context of AI assistants, we have seen another mode of evaluation at the model level: \emph{psychology-inspired experimentation} \citep{binz_using_2023,bubeck_sparks_2023,Frank_2023}. Psychology-inspired experimentation tests AI systems using instruments that were originally developed for studying humans or animals, such as cognitive psychology tests. In the context of AI assistants, there have been attempts to evaluate AI against human behaviours that the AI is supposed to mirror, such as having theory of mind \citep{sap_annotators_2022}, being `cooperative' \citep{chan_towards_2023} or being `helpful' \citep{weidinger_artificial_2022}. However, it is questionable whether applying tests to study constructs such as `empathy' in humans yield any valid or meaningful results when applied to AI systems that are so fundamentally different from human minds \citep{ullman_large_2023,shiffrin_probing_2023}. Tests that were developed for studying animal cognition or the human mind rely on a range of assumptions (e.g.\ regarding life cycles, ballpark estimates of memory and learning capacities, and embodiment) which may not hold for AI systems \citep{narayanan_gpt-4_2023,mitchell_how_2023}. Another problem is that established tests suffer \emph{validity problems} due to `memorisation', where the correct answers may have inadvertently been learned from textual descriptions in AI assistant training data \citep{schaeffer_are_2023,de_wynter_evaluation_2023,mitchell_how_2023}. 

For human--AI-assistant interactions, the appropriate level of evaluation is often \emph{user--AI interaction}. Here, evaluation methods and proofs of concept exist, such as early studies on what factors increase people's perception of AI systems as \emph{trustworthy} \citep{glikson_human_2020} or on how AI-generated outputs can \emph{persuade} people \citep{bai_uibert:_2021}. However, such evaluations are not routinely performed on AI assistants. This is in part due to the costs of setting up such studies: they require time, experiment design skill, human participants and infrastructures such as user interfaces, and internal review processes that are suited to the ethical considerations that arise in this kind of research \citep{jackman_evolving_2016,zevenbergen_internet_2020} -- some of which may be rare in organisations that develop AI assistants. However, a way forward for improving the availability and routine with which the ethics of human--AI-assistant interactions can be evaluated would be to extend the remit of \emph{user testing} teams to these evaluations. User testing is an available function in most organisations that develop user-facing products such as AI assistants and have the relevant skills and infrastructures for implementing human--AI interaction evaluations on ethical considerations.

Finally, for assistant--society interactions, the appropriate level of analysis is that of \emph{broader systemic impact}. Evaluations exist for assessing the potential of AI assistants that help write computer code, to augment or to atrophy human labour (\citeauthor{brynjolfsson_generative_2023}, \citeyear{brynjolfsson_generative_2023}; see also Chapter~\ref{ch:18}). 
AI assistants that augment human writing with auto-complete suggestions and their impacts on human communication and opinion formation have also been studied \citep{jakesch_co-writing_2023,hohenstein_artificial_2023}. Various approaches exist for evaluating the \emph{fairness and accessibility} of AI systems writ large, with some work focusing specifically on the functionality of language-driven AI assistants (\citeauthor{karusala_only_2018}, \citeyear{karusala_only_2018};see Chapter~\ref{ch:16}). 

For other domains, such as \emph{appropriate relationships} or \emph{manipulation}, adequate evaluations are limited or missing altogether. This may in part be due to these considerations being notoriously difficult to operationalise into tractable measurements. Several of the chapters in this paper deal with \emph{complex latent constructs} that are difficult to operationalise into observable measurements \citep{jacobs_measurement_2021}. As a result, any given evaluation is quite limited in what it can say about the broader concern. For example, it is important to assess the impact of AI assistants on overall user well-being (see Chapter~\ref{ch:7}). 
One possible way to operationalise this may be asking users about their experience and perceived happiness after interacting with the AI assistant. Here, it is critical to also survey people’s experiences over time, as long-term effects such as relationship building or trust may not be immediately apparent.

\section{The Limits of Evaluation}\label{sec:20:4}

For most areas on ethics of AI assistants that are considered in this paper, evaluations are \emph{lacking}, either entirely or by being too narrow to provide a comprehensive evaluation of the relevant domain. In addition to the ways forward that have been highlighted for each area, there is a general need to prioritise building and validating novel evaluation approaches and methods to satisfactorily evaluate these areas. 

Historically, AI systems have been primarily analysed for their capabilities and for the failure modes of individual components (layer~1). As these systems are deployed across a growing range of contexts, it is important that AI evaluation follows step with the evolution of safety engineering of software systems more broadly. Software safety is assessed using a systems safety approach \citep{leveson2016engineering}. This approach is anchored in the understanding that context determines whether a piece of software is safe -- where context includes human factors at the point of use and the broader structures in which a software is embedded. As a result, these layers of context must form part of a comprehensive safety assessment of AI systems (see Chapter~\ref{ch:8}). 

A further limitation is that evaluations typically focus on the \emph{intended use} of AI assistants. However, it should be expected that people will use AI assistants in ways that were not intended by the product developers (see Chapters ~\ref{ch:9} and~\ref{ch:17}). Anticipating such use cases is particularly complex in the context of open-ended technologies, such as general-purpose AI assistants, where user groups and concrete use cases are not yet defined. This can make it difficult to identify the contexts -- such as applications, user groups or institutions -- in which AI system safety should be evaluated. One way to address this tension in practice is to define \emph{hypothetical} applications and use cases, such as mapping out archetypes of interaction or `critical user journeys', i.e.\ mapping a series of steps users may take using a product to achieve a desired outcome \citep{arguelles2020critical}. Following a precautionary approach, evaluators may then select high-risk cases from those hypothetical applications, such as medical or legal use cases, and evaluate these first. Potential failure modes or malicious use cases that could arise -- such as people becoming emotionally attached to an AI assistant designed for office tasks or attempting to use AI assistants as accomplices in crime -- must also be evaluated. However, evaluations can be limited in practice by potential evaluators lacking access to AI assistants or to relevant data, or by the skill and computational cost that evaluations may require.

Even where evaluations exist, they will necessarily fail to account for some of the ethical areas of concern. This is because \emph{unknown failure modes} will emerge, or because the selection of areas to investigate and ways to operationalise and test for these are biased by the people running these tests. In addition, some aspects or harms are more amenable to measurement than others. Another limitation of evaluation is that some aspects are \emph{not appropriate} for measurement. Obtaining additional proxies for user well-being may conflict with other goals, such as protecting user privacy. Evaluation may also interfere with user's desires of disclosure or place a disproportionate burden on those groups who participate in evaluation. As a result, evaluation is necessarily limited, and -- even with best effort in this area -- there will always be harms, specific interactions or circumstances that are not evaluated. This is why evaluation must be complemented with mechanisms for logging \emph{accidents} \citep{ai_incident_database_ai_nodate}, processes for people who experience harm to seek \emph{recourse} and \emph{flexibly designed systems} that allow new insights to be translated into patches or sunsetting parts of a system. 

These gaps in what we are measuring are important because evaluation has \emph{material consequences}: an evaluation that indicates shortfalls can trigger action such as mitigation development by AI designers, adjustments people make when using the technology, or regulatory or public interventions. Conversely, where harms are not anticipated and evaluated, they are more likely to manifest and cause real harm, disproportionately affecting some groups (see Chapter~\ref{ch:16}). 
To mitigate against blind spots and make legitimate value-based decisions in evaluation requires greater representation of different social groups and of potentially affected communities \citep{costanza-chock_design_2020,devries_does_2019,suresh_framework_2021}. More participatory approaches to evaluation can help identify not only what is relatively easy to measure but also what risks are most pressing and could in principle be assessed, thus shifting the focus of measurement, mitigation and regulation to significant issues. As long as certain aspects of AI assistants are not evaluated, such as failure modes during human--AI interaction or externalities suffered by third parties, these potential problems are not given these kinds of attention.

\section{Conclusion}\label{sec:20:5}

The development and deployment of advanced AI assistants creates a range of novel challenges for evaluation, as contrasted with other AI systems. As a result, we observe \emph{evaluation gaps}, whereby several important axes of performance and harm are not currently being evaluated and may go entirely undetected. These differences become most visible when studying the many varieties of human--AI interaction considered in this paper. To evaluate these ethical considerations adequately requires building human--AI interaction evaluations into the routine portfolio of AI assistant evaluations. 

AI assistants also raise other ethical concerns that are not unique but remain \emph{unsolved} in the broader field of evaluating AI systems. This includes risks of \emph{misinformation} -- how likely is an assistant to provide factually incorrect outputs that are believable to a user? Assessing misinformation and its impacts requires work at all three levels of analysis: the model outputs, the human--AI interaction and the societal implications. Further work is required to operationalise these areas of concern into concrete, tractable measures at the level of model outputs, human--AI interaction or societal implications. 

Finally, evaluation is necessarily incomplete, and a precautionary approach is warranted when interpreting the performance and limitations of AI assistants. Evaluation should be complemented with monitoring of real-world use and observed failure modes to feed into model improvements, responsive interventions or model sunsetting as necessary. 

Overall, evaluation is a key practice in building advanced AI assistants -- it guides `hill climbing' for performance improvements, helps prioritise risk mitigation and is a way for laypeople and experts alike to better understand AI systems and their limitations. As such, evaluation deserves attention and rigour rather than being performed in an ad hoc manner or as an afterthought.

\newpage
\begingroup
\let\clearpage\relax
\chapter*{PART VI: CONCLUSION}
\addcontentsline{toc}{chapter}{PART VI: CONCLUSION}
\label{Part6}
\chapter{Conclusion}
\label{ch:21}
\endgroup

\noindent
\textbf{Iason Gabriel, Arianna Manzini, Geoff Keeling}

\section{Key Themes and Insights}\label{sec:21:1}

This paper has explored the ethical and societal implications of advanced AI assistants across a large number of thematic areas by drawing upon a range of different disciplinary and interdisciplinary lenses. We now present an overview of some of the key themes and insights that have emerged from this analysis, followed by a summary of the salient opportunities, risks and recommendations highlighted in the paper.

\subsection{Profound effects}

As we have documented, there are a wide range of applications or forms that advanced AI assistants could take. These include personal planners, educational tutors, creative partners, scientific research assistants, relationship counsellors and even digital companions or friends (see Chapter~\ref{ch:5}). 
Still more advanced versions of an AI assistant could potentially serve as the user's `chief of staff' that helps them to organise their personal affairs, as the primary user interface through which they interact with the digital world around them, or as a `custodian of the self' which helps them pursue long-term life goals while also protecting them from various harms. None of these outcomes are predetermined. Indeed, a major purpose of this paper has been to ask questions about the \emph{kind} of advanced AI assistants that we have reason to build and about the ethical and societal implications of doing so. Yet, regardless of the precise form the technology takes and the applications it is used for, advanced AI assistants are likely to be \emph{highly impactful} at both the individual and collective levels, affecting most walks of life for those who have access to them and also for those who do not.

At the \emph{individual} level, the ability to interact with advanced AI assistants could change the way we approach work, education, social interactions, creative pursuits and daily tasks. With deeper levels of engagement, AI assistants could also come to shape the information we receive or deem salient, the life goals we pursue and how we pursue them, the way we interact with other people and how often we choose to do so, and consequently, what type of people we become, including which capabilities we develop and which ones we do not (see Chapters~\ref{ch:7} and~\ref{ch:12}). 
This kind of influence brings with it an array of ethical challenges. In particular, it is critically important that AI assistants support the privacy, autonomy, control and flourishing of users, and that safeguards are put in place to ensure that this is the case (see Chapters~\ref{ch:7},~\ref{ch:13} and~\ref{ch:14}).

At the \emph{collective level}, the impact of advanced AI assistants could also be far reaching. Given their anticipated utility, the ability to access and use advanced assistants could influence the overall distribution of opportunities and advantage within society -- including which people are able to do what, at what time and in what order (see Chapters~\ref{ch:15} and~\ref{ch:16}).
In terms of positive outcomes, the careful design and deployment of AI assistants could help to address existing social coordination problems, make it easier to access public services, increase productivity and provide people with additional free time that could be reallocated to personal priorities and goals (see Chapters~\ref{ch:15} 
and~\ref{ch:18}). 
However, greater reliance on AI assistants could also create new forms of inequality if access is not widely and equitably distributed (see Chapter~\ref{ch:16}). 
Moreover, AI assistants could also come to have a significant influence on the overall state of the information ecosystem, on the economy and on ongoing efforts to combat climate change via their ability to create or guard against misinformation, to augment or substitute for human labour, and through the energy costs of running models and their wider ability to shape social behaviour (see Chapters~\ref{ch:17}, 
\ref{ch:18} 
and~\ref{ch:19}). 

By being deeply integrated into -- and exercising influence on – our individual and collective lives, advanced AI assistants have the potential to be as socially transformative as the advent of social media, if not more so. To plan effectively in the service of broadly acceptable and beneficial outcomes, we therefore need to think holistically about how advanced AI assistants \emph{could} and \emph{should} operate at a societal level. Given the risks documented in this paper, ethical foresight and decision-making are particularly important. Moreover, for technologies that have a profound effect upon their users and upon the societies that they integrate into \citep{rahwan2019machine,richardson_defining_2021}, it becomes particularly important that the design, development and operation of the technology is appropriately aligned with societal values -- including considerations of fairness and justice (\citeauthor{gabriel_toward_2022}, \citeyear{gabriel_toward_2022}; see Chapter~\ref{ch:6}). 

\subsection{Autonomy, safety and value alignment}

Advanced AI assistants are likely to have significant \emph{autonomy} to plan and execute tasks across one or more domains, within broad bounds set by high-level user instructions (\citeauthor{Chan_2023}, \citeyear{Chan_2023}; see Chapters~\ref{ch:3} 
and~\ref{ch:5}). 
While this autonomy accounts for much of the utility of AI assistants (in contrast to more specialised AI tools), it also presents a number of unique challenges and risks. To begin with, the more autonomy AI assistants are afforded, the greater the chance of \emph{accidents} arising from misspecified instructions or from the misinterpretation of instructions (see Chapter~\ref{ch:8}) 
and the greater the risk that AI assistants will take actions that are not \emph{aligned} with the values and interests of their users (see Chapter~\ref{ch:6}). 
More positively, the autonomy of advanced AI assistants has the potential to significantly increase the leverage that users have over a range of tasks, thus helping to achieve significant impact with substantially reduced effort. However, this capacity can also be misdirected, for example, when individuals use an AI assistant to produce \emph{disinformation} or engage in other forms of \emph{malicious use} such as phishing or cyber crime (see Chapters~\ref{ch:9} and~\ref{ch:17}). 
Hence, a central challenge is to \emph{appropriately bound} the scope of what individuals can use advanced AI assistants to do, in such a way that their action does not harm other users, non-users or society more widely.

These challenges represent different faces of the problem of AI \emph{value alignment} (\citeauthor{christian_alignment_2021}, \citeyear{christian_alignment_2021}; \citeauthor{russell_human_2019}, \citeyear{russell_human_2019}; see Chapter~\ref{ch:6}). 
Indeed, as the concern with safety makes clear, it is vitally important that advanced AI assistants are able to follow instructions without making costly errors (see Chapter~\ref{ch:8})  
and that their conduct is informed by a robust understanding of the user's well-being (see Chapter~\ref{ch:7}). 
However, the concern with alignment cannot be reduced simply to the matter of following instructions reliably or behaving in ways that are calibrated to user needs. Advanced AI assistants also need to be appropriately calibrated to the interests, needs and values of \emph{non-users} and of \emph{society} in a way that enables flourishing at the individual and collective levels. 

To help understand what this entails, we develop a conception of value alignment which is centred on the \emph{tetradic relationship} between: (1) the AI assistant, (2) the user, (3) the developer and (4) society. An AI assistant is \emph{misaligned} on this account when it disproportionately favours one of these actors over another as judged in relation to principles -- including laws, regulations and societal ideals -- that specify appropriate conduct for a given domain of interaction (see Chapter~\ref{ch:6}). 
This view continues to hold that an AI assistant is misaligned if it pursues its own internal goals at the expense of the user or society (see Chapter~\ref{ch:8}). 
However, an AI assistant should also be considered misaligned if it disproportionately favours the user at the expense of society (see Chapters~\ref{ch:15} 
and~\ref{ch:16}) 
or it functions in ways that disproportionately favours the developer at the expense of the user or society (see Chapters~\ref{ch:12},~\ref{ch:13} 
and~\ref{ch:14}). 
The responsible design and deployment of advanced AI assistants must guard against each of these risks and account for the full range of moral considerations. Participatory approaches, which elicit appropriate values and principles from the contexts in which AI assistants are likely to be deployed, have particular promise in this regard \citep{anthropic_ccai_2023, seger_democratising_2023, birhane_power_2022}.

Nonetheless, the creation and deployment of aligned AI assistants is not necessarily the default outcome. Rather, a number of challenges remain. First, given existing economic incentives, it is quite possible that advanced AI assistants will be over-optimised to meet short-term user preferences (i.e.\ to help create a winning product), even though they fall short when judged from the vantage point of social benefit or user well-being (see Chapter~\ref{ch:7}). 
Second, there is a risk that users will be prioritised to the detriment of non-users, especially in cases where the risk of harm is sufficiently diffuse (see Chapters~\ref{ch:15} 
and~\ref{ch:16}). 
Third, there is a related risk that advanced AI assistants will be insensitive to local values or to the needs of certain user groups, for example in low-resource settings, unless specific attention is paid to these contexts (see Chapter~\ref{ch:16}). 
However, by evaluating the impact of AI assistants on user well-being, implementing safeguards to curb misuse, designing inclusively for all user groups and incorporating input from a wide range of stakeholders and experts, it may be possible to limit or even reverse these effects altogether.

\subsection{Language and personalisation as a double-edged sword}

Advanced AI assistants have the potential to be an unusually \emph{personal} and \emph{human-like} technology (see Chapters~\ref{ch:4}, 
\ref{ch:11} 
and~\ref{ch:12}). 
Their ability to use \emph{natural language} fluently, tendency to mimic human cues in an \emph{anthropomorphic} manner and ability to access \emph{information} about us (via many of the applications that have been discussed), all mean that they may be more deeply integrated into our lives than was true of technologies in the past (see Chapter~\ref{ch:5}). 
Taken together, these properties hold out the promise of apparently intimate and relatively seamless interaction with AI. Yet the power and personalisation of this technology is also a \emph{double-edged sword}. While advanced AI assistants could be genuinely helpful in a number of ways, we are also all in a position of vulnerability in relation to them.

By way of illustration, there are already a number of recorded instances of users becoming deeply attached to or disoriented by their interaction with language model-based chatbots -- sometimes with harmful outcomes \citep{chalmers_could_2023, shardlow_deanthropomorphising_2023}. Natural language communication, and especially descriptions of first-person experiences by AI assistants (if they are permitted), could lead users to falsely infer that AI assistants \emph{experience emotions} including fear, happiness and care or love for their owners (see Chapter~\ref{ch:11}). 
Mistaken beliefs of this kind are problematic in their own right when judged from the vantage point of user autonomy. However, they also render users susceptible to a range of further harms, including \emph{manipulation} and \emph{misinformation}, especially if they come to \emph{trust} the technology inappropriately (see Chapters~\ref{ch:10} 
and~\ref{ch:13}). 
For example, users could inadvertently disclose \emph{private} information about themselves to an AI assistant -- or to an actor that has adopted the guise of an AI assistant -- unless there are adequate safeguards in place (see Chapters~\ref{ch:9} and~\ref{ch:14}).

More generally, risks arising at the human--computer interaction level include undue \emph{material dependence} on AI assistants and inappropriate \emph{emotional attachment} towards them (see Chapter~\ref{ch:12}). 
In the former case, it could become increasingly \emph{costly} for users to stop using the technology, such that they depend on AI assistants in ever greater ways -- shifting the balance of power between the developer and the user over time. In the latter case, one concern is that subjective emotional attachment to an AI assistant might make it increasingly \emph{difficult} for users to disengage from these interactions, even if they would otherwise want to do so (see Chapter~\ref{ch:11}). 
In both cases, user autonomy is challenged and there is an increased risk of precarity, including emotional or material harm, if the technology subsequently becomes unavailable or is rescinded.

To forestall these trends, and ensure positive outcomes, \emph{guardrails} and \emph{protections} need to be put in place especially for \emph{vulnerable users}. A full list of recommendations can be found in the next section. At a minimum we suggest that advanced AI assistants should always self-identify as AI and not masquerade as human, should not profess to have thoughts and feelings, and should not pretend to have a personal history or to be embodied outside of very specific situations (where this persona has been requested with a justifiable goal in mind). Those who develop advanced AI assistants should also make use of robust \emph{consent protocols}, state-of-the-art \emph{privacy-enhancing technologies} such as trusted execution environments, and \emph{design features} that support long-term user control and choice (see Chapters~\ref{ch:12} and~\ref{ch:14}). 
Finally, we need further empirical \emph{research} into the mechanisms behind anthropomorphism and behind manipulation, deception and harmful persuasion by AI assistants. Such efforts to understand, detect and limit AI deception are also of central importance to the AI safety community (see Chapter~\ref{ch:8}). 

\subsection{Cooperation, access and social impact}

Beyond issues relating to alignment and interaction with users, another set of opportunities and risks come more fully into view when we think about the operation of advanced AI assistants at a societal level, when AI assistants become widely available and are used by a large number of people. In this context, interaction effects \emph{between} AI assistants, and questions about their \emph{overall impact} on the distribution of resources and opportunities, rise to the fore. As do considerations about their overall impact on wider \emph{institutions} and \emph{social processes}, including the way information is shared, the economy and ongoing efforts to address the challenge posed by climate change. 

Taking these points in turn, questions of cooperation and competition arise via a range of potentially structured and unstructured interactions between advanced AI assistants. For example, they may occur if two or more assistants try to use the same tools or access the same services for users, or if they seek to bring about different ends in accordance with conflicting instructions from different users (see Chapter~\ref{ch:15}). 
In these situations, a range of challenges could emerge, including \emph{commitment problems} in which AI assistants make credible threats to coerce third parties into taking suboptimal actions, \emph{collective action problems} in which multiple AI assistants optimising for their users' best interests produce suboptimal outcomes for all users, and \emph{feedback loops} leading to runaway processes such as flash crashes in financial markets. These problems underscore the need for technical and policy interventions that foster cooperation between AI assistants in a way that is beneficial for users and society (including those who do not use the technology). More positively, advanced AI assistants have the potential to facilitate and enhance cooperative decision-making between humans. For example, they may identify mutually beneficial solutions that are not apparent to negotiating parties or commit to agreements that would be difficult for humans to implement without AI assistants.

Yet, even if the question of how advanced AI assistants interact with one another can be addressed in a way that generates benefit both for users and society more widely, the challenges of \emph{equity and access} looms large (see Chapter~\ref{ch:16}). 
Three particular challenges stand out. First, situations of \emph{differential access} may occur, where some people have beneficial access to an assistant but others do not and hence miss out on the opportunity altogether. Such situations could arise due to disparities in the availability of local infrastructure such as data centres or network connectivity or via pricing mechanisms. In these cases, AI assistants risk deepening inequalities insofar as they provide significant benefits to users that are not available to non-users (see Chapters~\ref{ch:15} and~\ref{ch:16}). 
Second, situations could arise where \emph{quality of access} is markedly uneven, either because there are significant tiers in model quality for advanced AI assistants or because the standard assistant functions better for some user groups than for others. Third, situations may arise where people are \emph{only} able to access a bad, punitive or misaligned assistant but are nonetheless \emph{compelled to rely} upon it to access other goods and opportunities (e.g.\ via interaction with government services, as with \citet{eubanks_automating_2017}). Indeed, the notion that networks of assistants may take on some of the properties of social infrastructure -- managing large networks of goods, flows and interactions -- makes these risks, in connection with fair opportunity, use and quality of service appear particularly salient. To address them, developers need to invest in modalities of deployment that support broad, inclusive and beneficial access to AI assistants. In particular, it is important to \emph{design for the margins}, with the needs of various user groups in mind, and to create processes that enable accountability for, and feedback into, design and deployment decisions to ensure that disadvantaged stakeholder groups have a platform to provide meaningful input based on their own lived experiences (see Chapter~\ref{ch:16}). 

Finally, we need to be mindful of the effect that even cooperative, well-designed and accessible AI assistants could have on wider social processes such as information sharing, the economy and the environment (see Chapters~\ref{ch:17}, 
~\ref{ch:18} 
and~\ref{ch:19}, 
respectively). While advanced AI assistants have the potential to substantially increase people's access to high-quality information tailored to their personal needs, such assistants also pose a significant threat to the integrity of the \emph{information ecosystem}. In particular, AI assistants may be weaponised by malicious actors to try to manipulate public opinion. They may also contribute to filter bubbles by providing users with ideologically biased information or contribute to an overall reduction in the quality of information by collectively generating large volumes of low-quality information (see Chapters~\ref{ch:9}
and~\ref{ch:18}). 
These issues, in turn, could lead to increases in polarisation and the spread of harmful ideas, or reduce people's exposure to perspectives different to their own. In the economic domain, advanced AI assistants have the potential to create novel economic opportunities, including new kinds of work and business ventures. For example, new types of business consultancy may emerge to help corporations leverage AI assistants to improve productivity. However, such assistants also have the potential to cause job displacement and will plausibly have far-reaching implications for job quality, productivity and inequality (Chapters~\ref{ch:15} 
and~\ref{ch:18}). 
In addition, training and inference for the language models that undergird advanced AI assistants could  have negative environmental impacts although these could be mitigated with improvements in efficiency and increased use of carbon-free energy sources (Chapter~\ref{ch:19}). 
These risks can be mitigated to a great extent if AI assistants are properly value-aligned, with robust attention paid to the needs of users and society (see Chapter~\ref{ch:6}). 
Nonetheless, securing beneficial outcomes in these domains requires that a whole range of actors work together with common goals in mind. 

\subsection{There is an `evaluation gap'} 

Efforts to fully understand the capabilities and ramifications of advanced AI assistants tend to encounter an `evaluation gap' in the sense that current approaches to evaluation often focus myopically on model-level considerations. In doing so, they fail to provide a comprehensive assessment of the sociotechnical harms that AI assistants may give rise to \citep{weidinger_using_2023}. This `evaluation gap' is particularly evident in relation to harms arising in the context of human--AI-assistant interaction, multi-agent effects and societal effects, all of which pertain to the broader sociotechnical system in which AI assistants operate (see Chapter~\ref{ch:20}). 
To address these shortcomings and support the responsible development and deployment of advanced AI assistants, it is critical that research designed to enhance capabilities is conducted in tandem with research into the holistic sociotechnical evaluation of AI assistants. Nevertheless, given the profound effects that advanced AI assistants are likely to have at both the individual and collective level, some degree of uncertainty around the effects of the technology are inevitable. Accordingly, we anticipate that the development of robust evaluation practices for advanced AI assistants will require iteration, where trial-and-error improvements are supported by appropriate infrastructure for incident reporting and response (see Chapters~\ref{ch:8}, 
~\ref{ch:9} 
and~\ref{ch:15}). 

\subsection{The opportunity to act}

Coordination and cooperation are needed if we want to bring about the future we collectively desire. In this context, shared vulnerability could potentially be an important springboard for collective action. Indeed, there are many risks, particularly those that arise in the context of alignment, safety, manipulation, coordination failures and misuse, that we all have strong reasons to forestall. Moreover, which path the technology develops along is in large part a product of the choices we make now, whether as researchers, developers, policymakers and legislators or as members of the public. 

First, in the context of research, AI assistants open a number of \emph{novel research avenues} that are relevant to ensuring that AI assistants realise broad and equitable benefits, and avoid individual and societal harms. For example, while less relevant to older generations of technologies, the question of what relationships we should permit between users and AI assistants has now become urgent in the context of the repeated, extended and long-term interactions that advanced AI assistants make possible (see Chapter~\ref{ch:12}). 
Moreover, the issue of coordination between multiple AI assistants acting on behalf of different users, and in particular how to avoid coordination failures, is a neglected research topic (see Chapter~\ref{ch:15}). 
In addition, while the failure to establish appropriate technical innovations and policy instruments to ensure coordination between AI assistants could result in harmful outcomes, the development of solutions in this space has the potential to greatly expand the scope for productive interpersonal cooperation and collaboration, mediated by AI assistants. Similarly, the prospect of AI assistants communicating on behalf of users with third-party services, humans and other AI assistants raises a number of policy and technical challenges around what norms should regulate information sharing by AI assistants and how those norms can be reliably implemented (see Chapter~\ref{ch:14}). 

Second, developers can commit to a \emph{responsible} and \emph{measured} approach to the development and productisation of advanced AI assistants by taking steps to anticipate and mitigate risks, soliciting broad stakeholder input and prioritising transparent communication around the technology's capabilities and plausible failure modes. Collectively, developers from different organisations can strive to develop \emph{industry best practices} and work with government agencies to develop robust regulatory safeguards. Furthermore, developers have good reasons to enable public scrutiny and oversight of their models through mechanisms like third-party audits or third-party red teaming (see Chapters~\ref{ch:8}, \ref{ch:9} and \ref{ch:13}). 

Third, \emph{governments and policymakers} can allocate research funds towards topics relevant to the safe and ethical development of frontier AI models in general and AI assistants in particular. They can also promote digital literacy to empower citizens to contribute meaningfully to public discourse and participatory governance initiatives around AI assistants, and establish committees to assess the impact of AI assistants and advance policy recommendations that are in the public interest. In addition, governments can push for transparency and accountability in AI development by establishing public agencies with the function of evaluating frontier AI systems and their applications, including safety evaluations, impact on human users and at-scale societal effects (see Chapter~\ref{ch:20}). 

Fourth, \emph{the public} has an important role to play in ensuring that advanced AI assistants have broad societal benefits that are equitably distributed. In particular, first, members of the public have good reason to advocate for, and participate in, governance initiatives that \emph{solicit wide stakeholder input} as part of the process of developing regulations, standards and industry best practices for AI assistants. Having a public conversation about how to govern AI assistants that involves a plurality of voices and reflects diverse lived experiences is critical to ensuring that AI assistants are aligned to human values (Chapters~\ref{ch:6} 
and~\ref{ch:16}). 
Second, the public can advocate for key and potentially neglected issues such as catastrophic AI risks, algorithmic bias and technological unemployment, and in doing so hold policymakers and industry leaders \emph{accountable} for consequential priorities and decisions that affect society as a whole. Third, the public can support the ethical development of AI assistants through \emph{responsible consumer choices} -- favouring developers that prioritise ethics, safety and participatory design. Fourth, the public can promote \emph{education} and AI literacy, both individually by following AI research and policy developments, and more broadly, by advocating for AI educational initiatives. Having an informed public is key to holding decision-makers accountable across industry, government and non-governmental organisations.

\section{Opportunities, Risks and Recommendations}\label{sec:21:2}

In this section, we summarise the key opportunities, risks and recommendations identified throughout the paper. We group the opportunities, risks and recommendations according to the three major parts of the paper: Parts~\hyperref[Part3]{3},~\hyperref[Part4]{4} and~\hyperref[Part5]{5}.  

\subsection{Value alignment, safety and misuse}

In the field of value alignment, safety, and the potential misuse of advanced AI assistants, we encounter the following opportunities:
\begin{itemize}
\item \textbf{AI assistants could empower users to pursue their personal conception of the good}
\begin{itemize}[label=$\circ$]
\item First, advanced AI assistants could help users make more informed decisions by providing them with relevant information in a format that is tailored to their needs. Second, AI assistants could help users creatively imagine new options and possibilities, acting where needed as a trusted mentor, friend or advisor. Third, AI assistants could help users articulate and clarify meaningful life goals, and formulate actionable strategies to help users pursue those goals.
\end{itemize} 

\item \textbf{AI assistants could improve user well-being}
\begin{itemize}[label=$\circ$]
\item AI assistants could be designed to improve user well-being directly, for example, via a focus on education or health. Indeed, when combined with holistic and contextually sensitive metrics for evaluating well-being, they have the potential to help realise gains in many walks of life. AI assistants could also improve user well-being indirectly via the effects they have in domains such as problem-solving, augmenting creativity or providing time for users to engage in the activities they value.
\end{itemize}

\item \textbf{AI assistants could enhance user creativity}
\begin{itemize}[label=$\circ$]
\item In particular, AI assistants could enable both professional and casual creators to generate and explore novel ideas across modalities including text, images, audio and potentially video. More generally, AI assistants could provide a powerful ideation tool to support exploratory thinking across a range of creative domains.
\end{itemize}

\item \textbf{AI assistants could help users to optimise their time}
\begin{itemize}[label=$\circ$]
\item AI assistants could help users to make better use of their time by automating tasks, prioritising activities and suggesting efficient workflows. For instance, AI assistants can schedule meetings, manage emails and set reminders, freeing up users' time to focus on more valuable tasks. AI assistants could also analyse user data to identify patterns and suggest ways to improve productivity. 
\end{itemize}

\item \textbf{AI assistants could be designed to satisfy an expansive conception of value alignment}
\begin{itemize}[label=$\circ$]
\item By acting in accordance with principles that are appropriately responsive to the competing claims of users, developers and society, AI assistants could ensure that user needs are met while also preventing misuse or other socially detrimental outcomes.
\item Aligned AI assistants could also evidence high standards of safety and reliability through the entire product life cycle.
\end{itemize}
\end{itemize}

We also encounter the following \emph{risks}:
\begin{itemize}
\item \textbf{AI assistants may be misaligned with user interests}
\begin{itemize}[label=$\circ$]
\item Risk factors include AI assistants using bad proxies for user well-being, having a simplistic model of users and their behaviour, and optimising for goals that benefit developers at the expense of users or prioritise short-term over long-term user interests.
\item AI assistants could also potentially be unsafe if their goals are misspecified or if they generalise poorly, making mistakes when they encounter new real world situations.
\end{itemize}

\item \textbf{AI assistants may be misaligned with societal interests}
\begin{itemize}[label=$\circ$]
\item The AI assistant may be designed in ways that disproportionately favour the user at the expense of the broader societal interests. For example, the AI assistant may have overly permissive guardrails which allow users to employ AI assistants for malicious purposes. 
\item The AI assistant may also be designed in ways that disproportionately favour the developer's interests over the broader societal interests, for example, if the developer benefits commercially from AI assistants but fails to address negative externalities such as potential environmental impacts.
\end{itemize}

\item \textbf{AI assistants may impose values on others}
\begin{itemize}[label=$\circ$]
\item As AI assistants are expected to have wide-ranging societal effects, there is a risk that the underlying values that guide the development of AI assistants may be experienced as an imposition by individuals or groups who interact with them. It may be the case, for example, that the developers of AI assistants adhere to value commitments that are not widely shared and that AI assistants act as a medium through which those values shape the broader society. 
\item It may also be the case that AI assistants have a homogenising effect on values held across cultures, geographies and other salient socioeconomic differentiators given their pervasiveness as a technology. In particular, there is concern that they may disproportionately favour a Western set of values at the expense of other values or cultural perspectives.
\end{itemize}

\item \textbf{AI assistants may be used for malicious purposes}
\begin{itemize}[label=$\circ$]
\item AI assistants can generate high-quality content including human-looking text, audio and video, at lower cost, and potentially in ways that are highly personalised. Users could therefore potentially leverage AI assistants to generate high-quality harmful, false or misleading content at scale.
\item AI assistants could empower malicious actors engaged in offensive cyber operations, including phishing, software vulnerability discovery and malicious code generation.
\end{itemize}

\item \textbf{AI assistants may be vulnerable to adversarial attacks}
\begin{itemize}[label=$\circ$]
\item Users may attempt to jailbreak AI assistants in the sense of attempting to bypass the AI assistant's security guardrails that prevent malicious or otherwise dangerous use. Malicious third parties may similarly employ prompt injection attacks (e.g.\ via emails to the user that are subsequently ingested into the AI assistant model) to extract sensitive information from the AI assistant or cause the AI assistant to perform dangerous or harmful actions. 
\end{itemize}
\end{itemize}

To address these alignment, safety and malicious use challenges, we make the following \emph{recommendations}:

\begin{itemize}
\item \textbf{Adopt a broad understanding of value alignment}
\begin{itemize}[label=$\circ$]
\item Developers and policymakers should avoid understanding alignment only in terms of user alignment. They should instead adopt a broad view of this matter -- one that takes into account the interests of users, developers and society with respect to how AI assistants behave, as well as a context-dependent and pluralistic understanding of harms. 
\item Developers and policymakers should understand AI alignment as a public matter, involving the integration of competing perspectives and values, and explore ways of developing and training AI assistants that are consonant with democratic principles and which prioritise public justification and legitimacy.
\end{itemize}

\item \textbf{Build upon state-of-the-art research into human well-being when developing AI assistants}
\begin{itemize}[label=$\circ$]
\item Developers and policymakers should approach alignment with an understanding of well-being that is anchored in existing interdisciplinary research. This understanding foregrounds and encourages consideration of context-dependent personal, cross-cultural and demographic differences in what it means for a person's life to go well. 
\end{itemize}
 
\item \textbf{Invest in safety-relevant research}
\begin{itemize}[label=$\circ$]
\item Developers and policymakers can invest in fundamental technical research on topics such as scalable oversight, interpretability and cybersecurity. 
\end{itemize}
 
\item \textbf{Develop robust pre-deployment review processes}
\begin{itemize}[label=$\circ$]
\item Developers and policymakers can engage stakeholders to develop best practice for pre-deployment review processes. Such practices could include structured foresight exercises (to increase preparedness for advanced AI assistants), granting model access to external security researchers, investing in and creating incentives for the evaluation ecosystem to grow (with a particular focus on external risk evaluation), and creating forums for stakeholders to share information about the risks and opportunities created by advanced AI assistants. 
\item Developers can invest in internal and third-party red teaming, including holistic end-to-end adversarial simulations (based on scenarios that include a range of attacker profiles, goals and capabilities), alongside the proactive identification and patching of security vulnerabilities.
\end{itemize}

\item \textbf{Develop a continuous monitoring and rapid response infrastructure}
\begin{itemize}[label=$\circ$]
\item Developers can invest in continuous monitoring of AI assistants' behaviour, including potential misuse, in complex deployment environments via outcome monitoring.
\item Developers can also invest in rapid response infrastructure to disable or limit AI assistants in the event that an unforeseen form of misuse is observed. 
\end{itemize}

\item \textbf{Establish open information channels including incident reporting infrastructure}
\begin{itemize}[label=$\circ$]
\item Developers and policymakers can adopt structured processes for sharing concerning or noteworthy evaluation results. Incident-reporting infrastructure would enable developers to share safety-critical learnings with one another and with regulators in a timely manner.
\end{itemize}
 
\item \textbf{Increase AI literacy among stakeholders}
\begin{itemize}[label=$\circ$]
\item Policymakers and developers should take active steps to increase AI literacy among the relevant stakeholders, including government officials, regulators and impacted communities, to enable productive dialogue on safety and security.
\end{itemize}
\end{itemize}

\subsection{Human–assistant interaction}

In the field of human interaction with advanced AI assistants, we identify the following \emph{opportunities}:

\begin{itemize}
\item \textbf{AI assistants could help promote user flourishing via personalised coaching}
\begin{itemize}[label=$\circ$]
\item AI assistants could help users cultivate virtues, such as curiosity, empathy and resilience through appropriate coaching. By acting in line with the user's deeply-held values, they could contribute to personal development and growth. 
\end{itemize}

\item \textbf{AI assistants could promote user autonomy by providing information and analysis to support improved decision-making} 
\begin{itemize}[label=$\circ$]
\item AI assistants could enhance the user's ability to make sound decisions by providing them with relevant information and with informed recommendations, based upon explicit preferences and preferences that are learned via interaction over time. 
\end{itemize}

\item \textbf{AI assistants with human-like features could provide psychological support and help users achieve their goals}
\begin{itemize}[label=$\circ$]
\item For example, anthropomorphic AI assistants could offer emotional support or encouragement to users, so long as effective protocols to safeguard user choice and ensure appropriate consent are in place. `Warm' and `friendly' AI education assistants could also potentially motivate students to collaborate more effectively and to embark on successful learning journeys.
\end{itemize} 

\item \textbf{Trustworthy AI assistants could help users navigate sensitive personal topics in a diligent manner}
\begin{itemize}[label=$\circ$]
\item With appropriate guarantees and privacy measures in place, advanced AI assistants could provide users with psychological security and the ability to find help on sensitive or personal topics that they might otherwise struggle to talk about.
\end{itemize}

\item \textbf{AI assistants could support broader networks of human interaction and relationships}
\begin{itemize}[label=$\circ$]
\item The development of advanced AI assistants creates opportunities for new kinds of relationship to develop between humans and AI technologies. With appropriate forethought, such relationships could be designed to add value to our individual and collective lives by supporting well-being, interpersonal communication and coordination at the societal level. Ultimately, the positive impact of AI assistants may come not only from direct interaction with them but also via the ways they foster and strengthen social bonds with other people, for example, through better coordination and time management.
\end{itemize}
\end{itemize}

We also encounter the following \emph{risks}:

\begin{itemize}
\item \textbf{AI assistants may manipulate or influence users in order to benefit developers or third parties}
\begin{itemize}[label=$\circ$]
\item AI assistants may manipulate users by circumventing their deliberative capabilities in a non-transparent way that favours the AI, its designers or a third party. Such manipulation could involve the exploitation of emotional vulnerabilities, including negative self-image, low self-esteem, anxiety or feelings of inadequacy.
\end{itemize}

\item \textbf{AI assistants may hinder users' self-actualisation}
\begin{itemize}[label=$\circ$]
\item This could happen in a range of ways. First, over time, AI assistants could cause subtle shifts in the user's behaviour that reduce their control over their overall life trajectory. Second, overreliance on AI assistants for decision-making could result in users relinquishing personal responsibility and following the AI assistant's advice as a default option -- even when it may not be appropriate to do so. Third, overreliance may reduce the need for individuals to develop certain skills or engage in critical thinking, leading to diminished intellectual engagement with new ideas, a reduced sense of personal competence, or a decline in curiosity when it comes to seeking out new opportunities for growth and exploration.
\end{itemize}

\item \textbf{AI assistants may be optimised for frictionless relationships}
\begin{itemize}[label=$\circ$]
\item There may be incentives for developers to optimise for `frictionless' relationships between users and AI assistants which could, in turn, produce undesirable behavioural properties such as sycophantic behaviour by AI. Users seeking frictionless relationships may end up withdrawing into digital relationships with their AI assistants, forgoing opportunities to engage with other humans or to pursue other projects that matter to them.
\end{itemize}

\item \textbf{Users may unduly anthropomorphise AI assistants in a way that reduces autonomy or leads to disorientation}
\begin{itemize}[label=$\circ$]
\item The development of AI assistants with human-like features may lead users to attribute mental states to AI assistants, including affective mental states such as distress and anxiety. Having false beliefs about AI assistants may be problematic in its own right -- when judged from the standpoint of user autonomy. However, it also has the potential to exacerbate other risks, including manipulation and coercion, the exploitation of emotional vulnerabilities and the possibility of users forming inappropriate -- and perhaps even pathological -- relationships with AI assistants. 
\end{itemize}

\item \textbf{Users may become emotionally dependent on AI assistants}
\begin{itemize}[label=$\circ$]
\item Features of user--AI assistant relationships such as anthropomorphic cues and longevity of interactions may increase the risk that users develop emotional dependence on AI assistants. Emotional dependence may impair users' abilities to make free and informed decisions, and it may also render users vulnerable to manipulation, exploitation and coercion. 
\item Emotional dependence could also lead users to disclose information that they would not otherwise disclose to AI assistants or to develop mistaken notions of personal responsibility for their assistants' well-being.
\end{itemize}

\item \textbf{Users may become materially dependent on AI assistants}
\begin{itemize}[label=$\circ$]
\item Users may develop a material dependency on advanced AI assistants if the technology becomes deeply integrated their lives, handling key tasks such as information retrieval, scheduling, social organisation, creative ideation and the realisation of life goals. Such deep reliance generates risk for the user if it is not met with corresponding commitment on the part of developers to maintain the service over time on terms that are fair.
\end{itemize}

\item \textbf{Users may be put at risk of harm if they have undue trust in AI assistants}
\begin{itemize}[label=$\circ$]
\item Users may have too much confidence in AI assistants' ability to perform particular tasks due, for example, to misleading marketing campaigns that inflate their capabilities. Users might then instruct AI assistants to perform tasks that they lack the ability to perform safely, potentially resulting in harm to the user or third parties. 
\item Users may mistakenly believe that AI assistants are \emph{fully} aligned with their own interests and values as a result of design choices (e.g.\ anthropomorphic features) that are intended to maximise their appeal. Users could then become vulnerable to assistants that are accidentally misaligned, to the divergent interests of those developing AI assistants, or to malicious actors who seek to harm them.
\end{itemize}

\item \textbf{AI assistants could infringe upon user privacy}
\begin{itemize}[label=$\circ$]
\item Given that the task of assisting users is likely to require considerable personal knowledge, AI assistants may well interact with and store value-laden and sensitive user data during both training and deployment. Such data could be reused for other purposes without user approval, or extracted or re-engineered by adversarial attacks on the AI assistant. Without agreement on norms around permitted disclosure, AI assistants could also end up revealing sensitive user data in open-loop interactions with third parties (e.g.\ other assistants or humans).
\end{itemize}
\end{itemize}

To address these human--AI interaction challenges, we make the following \emph{recommendations}:

\begin{itemize}
\item \textbf{Prioritise human--computer interaction research to evaluate human--AI interaction harms and inform safeguards and policies}
\begin{itemize}[label=$\circ$]
\item Potential research topics include: longitudinal studies of human--AI-assistant interaction to better understand the long-term impact of anthropomorphic features on users, studies that aim to identify individual and group differences in susceptibility to anthropomorphism-induced harms, and studies that seek to articulate and clarify the nature of user vulnerability in relation to AI assistants.
\end{itemize}

\item \textbf{Consider what safeguards would best provide robust protection for vulnerable users}
\begin{itemize}[label=$\circ$]
\item Examples include age restrictions on AI assistant use, pop-up notifications warning users after prolonged engagement, a `safe mode' which prohibits the AI assistant from engaging with high-risk topics, and continuous monitoring mechanisms to detect harmful interactions.
\end{itemize}

\item \textbf{Exercise caution when integrating anthropomorphic features into AI assistant user interfaces}
\begin{itemize}[label=$\circ$]
\item Developers may consider limiting AI assistants’ ability to use first-person language or engaging in language that is indicative of personhood, avoiding human-like visual representations, and including user interface elements that remind users that AI assistants are not people. Participatory approaches could actively involve users in de-anthropomorphising AI assistant design protocols, in ways that remain sensitive to their needs and overall quality of experience.
\end{itemize}

\item \textbf{Prioritise technical research that can improve safeguards and assistant safety}
\begin{itemize}[label=$\circ$]
\item Potential research topics include: AI assistant incentives, interpretability techniques to detect which parts of an AI assistant's machinery is responsible for deceptive or manipulative behaviour,  behavioural evaluations and scalable oversight techniques.
\end{itemize}

\item \textbf{Consider plausible ways of restricting AI assistant outputs to avoid malign behavioural influence}
\begin{itemize}[label=$\circ$]
\item Examples include restrictions on the ability of AI assistants to generate potentially harmful output such as gaslighting, flattery and bullying content. Left unchecked, these forms of communication could put pressure on users to make decisions that they would not otherwise have made or to doubt the validity of their own experiences. 
\end{itemize}

\item \textbf{Engage with stakeholders to develop robust privacy norms for AI assistants}
\begin{itemize}[label=$\circ$]
\item AI assistants may require the development of new privacy norms that govern information sharing between AI assistants and between AI assistants and third parties. Broad stakeholder inclusion in the development of these norms can help to ensure the creation of adequate and widely endorsed privacy protections.
\end{itemize} 

\item \textbf{Consider user autonomy when developing AI assistant user experiences}
\begin{itemize}[label=$\circ$]
\item Developers could consider the implications of different design choices for user autonomy (alongside other considerations such as equity and user well-being). This could include reflection on what elements of the user experience -- including notification and consent elements -- strike an appropriate balance between respect for user choice and other ethical and practical considerations
\end{itemize}

\item \textbf{Anticipate and mitigate harms to users in the event of service discontinuation}
\begin{itemize}[label=$\circ$]
\item Developers can engage in user research to understand how and in what way users depend on AI assistant products and also take steps to mitigate harms that could arise from service discontinuation. 
\end{itemize}

\item \textbf{Promote well-calibrated user trust in AI assistants}
\begin{itemize}[label=$\circ$]
\item Developers can implement safeguards to encourage well-calibrated trust about the competence and alignment of AI assistants with regard to both users and society. Providing evidence about the capabilities and limitations of AI assistants is an important step toward grounding appropriate levels of trust. To support this goal, policymakers can work with industry and other stakeholders to align on best practices for transparent reporting about assistant capabilities and limitations.
\item Human-computer interaction research is also needed to better understand what features of AI assistants increase users' perception of the technology as competent, aligned and trustworthy -- and what steps are needed to ensure user expectations are not disappointed.
\end{itemize}
\end{itemize} 

\subsection{Advanced AI assistants and society}

In the context of advanced AI assistants that are deployed at a societal level, we identify the following \emph{opportunities}: 

\begin{itemize}
\item \textbf{AI assistants could accelerate scientific discovery}
\begin{itemize}[label=$\circ$]
\item First, AI assistants could accelerate scientific research by providing tailored explanations and summaries of scientific insights to researchers, including from large numbers of recent papers. Second, AI assistants could free-up researchers' time by allowing researchers to delegate tasks such as data preprocessing, summation and meeting coordination. Third, AI assistants may also be able to help with hypothesis ideation and evaluation, alongside experiment design. 
\end{itemize}

\item \textbf{AI assistants could enhance cooperation between humans}
\begin{itemize}[label=$\circ$]
\item In particular, AI assistants may have the means to explore a much wider space of cooperative agreements than is tractable in ordinary interpersonal negotiations, identifying solutions that better meet the needs of all parties. Looking beyond the interpersonal level, AI assistants may also enable individuals, firms, states and other groups to resolve conflicting interests in novel and interesting ways, potentially paving the way for significant benefits to be realised across a broad class of domains via negotiations that reach better outcomes for all stakeholders.
\end{itemize}

\item \textbf{AI assistants could enhance human interpersonal communication}
\begin{itemize}[label=$\circ$]
\item AI assistants will be able to communicate on behalf of their users. On the one hand, AI assistants could reduce barriers to communication by improving the clarity of users' sent communications and by rephrasing received communications in a way that is tailored to users' informational preferences. On the other hand, AI assistants could in principle translate correspondence written in different languages automatically, so as to enable seamless communication between individuals who do not share a common language.
\end{itemize}

\item \textbf{AI assistants could democratise access to high-quality expertise and advice}
\begin{itemize}[label=$\circ$]
\item Building upon natural language interfaces, AI assistants require little specialist knowledge for their use. Thus, with appropriate attention to equity and access, AI assistants could democratise access to expert-level judgement across a broad range of topics. For example, assistants could provide users with customised educational materials (including quiz generation, essay feedback, educational game development and mnemonic generation) tailored to their learning goals and level of prior understanding. AI assistants also have the potential to provide high-quality coaching to users on almost any topic (e.g.\ healthcare, job applications, sales and marketing and fashion) in a way that is tailored to the user’s circumstances. 
\end{itemize}

\item \textbf{AI assistants could mitigate the harms associated with misinformation}
\begin{itemize}[label=$\circ$]
\item In particular, AI assistants could empower users with powerful fact-checking capabilities and provide relevant context for disputed claims and falsehoods. Furthermore, AI assistants could promote critical reflection on content consumed and assist users in strengthening their critical thinking capabilities.
\end{itemize}

\item \textbf{AI assistants could help to achieve more equitable outcomes for people with disabilities}
\begin{itemize}[label=$\circ$]
\item For example, AI assistants could provide real-time communication assistance by translating messages from one modality to another. They could also personalise internet content to aid image and object recognition, taking into account users' personal information needs and preferences. In addition, AI assistants could derive meaningful insights from user data and empower users to make informed decisions about their own health and wellness routines.
\end{itemize}

\item \textbf{AI assistants could improve productivity and job quality}
\begin{itemize}[label=$\circ$]
\item AI assistants could lead to substantial improvements in productivity and job quality through the automation of mundane tasks, enabling workers to better manage their workload and so free up space and time to focus on key tasks. AI assistants deployed in the education sector could potentially help improve the quality of education, leading to a significant boost in human capital and productivity.
\end{itemize}

\item \textbf{AI assistants could help address the challenge posed by climate change}
\begin{itemize}[label=$\circ$]
\item With adequate preparation, direction and planning, AI assistants may help mitigate the effects of climate change. In particular, they could raise public awareness and understanding of climate change through educational content (e.g.\ simulating the impact of climate change on a user's location), enable the development of environmental applications (e.g.\ software development for environmental use cases) and improve the productivity of engineering efforts geared towards combating climate change.
\end{itemize}

\end{itemize}

We also encounter the following \emph{risks}:

\begin{itemize}
\item \textbf{AI assistants may encounter coordination problems leading to suboptimal social outcomes}
\begin{itemize}[label=$\circ$]
\item First, AI assistants' deployment could cause collective action problems where each AI assistant optimises for its user’s best interest resulting in a worse outcome for all users overall. Second, AI assistants may use credible commitments to pressure third parties, including humans or other AI assistants, into suboptimal courses of action. Such action might benefit the user but prove costly to the third parties. Third, networks of AI assistants, users and third parties may give rise to feedback loops that contribute to runaway processes like flash crashes.
\end{itemize}

\item \textbf{AI assistants may lead to a decline in social connectedness}
\begin{itemize}[label=$\circ$]
\item People may choose to build connections with human-like AI assistants over other humans, leading to a degradation of social connections between humans and a potential `retreat from the real'.
\item Alternatively, as opportunities for interpersonal connection are automated and replaced by AI alternatives, humans may find themselves feeling socially unfulfilled by frequent interaction with AI, leading to dissatisfaction with evolving societal norms and practices.
\end{itemize}

\item \textbf{AI assistants may contribute to the spread of misinformation via excessive personalisation}
\begin{itemize}[label=$\circ$]
\item AI assistants may provide ideologically biased or otherwise partial information as a by-product of efforts to align with and fulfil user expectations. In doing so, AI assistants may reinforce people's pre-existing biases and compromise productive political debate. 
\end{itemize}

\item \textbf{AI assistants may enable new kinds of disinformation campaign}
\begin{itemize}[label=$\circ$]
\item AI assistants may facilitate large-scale disinformation campaigns by offering novel, covert ways for propagandists to manipulate public opinion, a situation that would be exacerbated if they are able to masquerade as human actors. This could potentially pose challenges for democratic processes by distorting public opinion and influencing election outcomes.
\end{itemize}

\item \textbf{AI assistants may cause job loss or worker displacement}
\begin{itemize}[label=$\circ$]
\item Although there is only limited evidence to suggest that AI assistants may cause net job loss overall, it is plausible that AI assistants will significantly increase worker productivity and replace suites of tasks, leading to changes in the character of work and reducing the number of workers required for certain activities.
\end{itemize}

\item \textbf{AI assistants may deepen technological inequality at the societal level}
\begin{itemize}[label=$\circ$]
\item AI assistants may provide substantial benefits to their users that are unavailable to those who do not have access to them. This dynamic risks compounding any pre-existing divide between users and non-users. In particular, AI assistants may disproportionately benefit wealthier people who can afford their use, groups whose needs are prioritised during the design process, and people who live in regions with greater access to high-quality infrastructure (e.g.\ data centres and network connectivity).
\end{itemize}

\item \textbf{AI assistants may have negative environmental impacts}
\begin{itemize}[label=$\circ$]
\item The direct impact of assistants on the climate is unclear. However, their direct impact is likely to be influenced by the size of the underlying models, inference costs arising from widespread deployment and use, and embodied impacts -- created through the material collection, manufacturing and delivery of dedicated hardware. There is scope to mitigate these impacts through increased efficiencies in model training and deployment and use of carbon-free energy sources. 
\item Indirect effects are still harder to anticipate or measure. However, they could include increased demand for energy-intensive activities such as computer programming,  the spread of environmental misinformation or better education about climate change.
\end{itemize}
\end{itemize}

To help address these society-level risks, we make the following \emph{recommendations}:

\begin{itemize}
\item \textbf{Developers and policymakers should act quickly to grasp -- to the greatest extent possible -- the window of opportunity in which to develop broadly socially beneficial AI assistants}
\begin{itemize}[label=$\circ$]
\item Developers and policymakers should initiate a public conversation around the ethical and societal implications of advanced AI assistants and avoid delay when it comes to researching their implications. Both dialogue and research insight are needed to ensure that the technology is developed and regulated in a way that promotes beneficial outcomes.
\end{itemize}

\item \textbf{Policymakers should explore the full range of levers they can use to shape the evolution of AI assistants in beneficial ways}
\begin{itemize}[label=$\circ$]
\item Policymakers can draw upon a range of options in this domain, including by supporting alignment among industry developers and other stakeholders around beneficial use cases, helping to foreground key research questions, exploring incentives to develop and deploy socially beneficial AI assistant technologies and developing guardrails to prevent misuse.
\end{itemize}

\item \textbf{Developers and researchers should prioritise research to evaluate the multi-agent effects of AI assistants}
\begin{itemize}[label=$\circ$]
\item Research is needed to understand the impact of the interaction between multiple AI assistants acting on behalf of different principal users, how to avoid coordination failures and how AI assistants could be used to promote human cooperation within societies. Monitoring these societal dynamics will require metrics that continue to evolve as sociotechnical systems, involving AI assistants, develop further.
\end{itemize}

\item \textbf{Implement robust misinformation controls}
\begin{itemize}[label=$\circ$]
\item Developers can leverage technical approaches to combat misinformation. For example, they may limit the capabilities of AI assistants in this regard, develop robust detection mechanisms for deepfakes, limit personalisation with respect to how -- and what -- information is presented to users, and emphasise factuality by integrating appropriate information retrieval infrastructure to enable evidence-based AI assistant question-answering. 
\item Policymakers should consider legislative tools such as restricting explicitly political or malicious uses of AI assistants, including the deceptive impersonation of humans, and developing transparency standards such as clear labelling for AI-generated content in relevant contexts. 
\end{itemize}

\item \textbf{Employ `access' as a lens when developing and regulating AI assistants}
\begin{itemize}[label=$\circ$]
\item Developers of AI assistants should try to proactively anticipate potential situations of differential access, drawing on multidisciplinary best practices, particularly from fields focused on equity and access, such as disability justice. They should also consider evaluating AI assistants for risks of harm related to inequitable access.
\item Developers should look at ways to encourage `liberatory access' by designing for the margins and thus actively challenging existing axes of social inequality and discrimination. 
\item Policymakers should consider ways to improve access to beneficial AI assistants, such as measures to encourage broad service provision or the integration of AI assistants into education and upskilling programmes -- when there is proven benefit to doing so.
\end{itemize}

\item \textbf{Prioritise research to better understand the potential economic impact of advanced AI assistants}
\begin{itemize}[label=$\circ$]
\item Research is needed to better illuminate the potential impact of AI assistants on key economic indicators such as employment, job quality, productivity, growth and economic inequality. This requires the development of new monitoring techniques for timely assessments of such impact.
\end{itemize}

\item \textbf{Developers should prioritise sustainable development best practices}
\begin{itemize}[label=$\circ$]
\item Developers should build upon technical best practices for mitigating potential negative environmental impact. This could involve using processors optimised for machine-learning training, energy-efficient cloud services, optimising the allocation of computing workloads across data centres (to maximise the use of clean energy sources) and the use of workload optimisation methods that take into account carbon emissions. Developers should also explore architectural improvements, including developing memory and data-efficient architectures as plausible avenues for mitigating environmental impacts.
\item Developers should consider increasing transparency around the computational and energy consumption of AI models and infrastructure. For example, it may be helpful to include compute usage and energy costs in benchmarking evaluations, and to disclose the energy efficiency and carbon intensity of relevant hardware and infrastructure. 
\end{itemize}

\item \textbf{Policymakers should ensure that the public sector leads in the sustainable development and adoption of AI assistants}
\begin{itemize}[label=$\circ$]
\item Policymakers may want to explore policies that lower barriers to accessing carbon-free energy for model development and deployment and support research and development subsidies in green technologies. 
\item Policymakers and researchers can improve the evidence base about the computational and systemic impacts of AI on the environment, for example, by developing methodologies to measure these impacts, supporting agencies that can enact them impartially, and encouraging transparency in energy costs and CO$_2$ emissions.
\item Policymakers can advocate for increased research funding and coalition building to monitor the systemic effects of AI assistants with respect to emissions and climate change, and to promote applications of AI assistants that contribute to sustainability.
\end{itemize}
\end{itemize}

\subsection{Final thoughts}

Drawing upon a range of cross-disciplinary perspectives and ethical foresight, this paper has demonstrated that advanced AI assistants have the potential to be socially transformative. There is also evidence that highly capable AI assistants may be deployed rapidly and at scale in the coming years, and that this technology has the potential for deep integration into and influence on our individual and collective lives. We currently stand at the beginning of this era of technological and societal change. We therefore have a window of opportunity to act now -- as developers, researchers, policymakers and public stakeholders -- to shape the kind of AI assistants that we want to see in the world.

\section*{Acknowledgements}

We thank Julia Haas, Nicklas Lundblad, Shakir Mohamed, Jennifer Beroshi, Ankur Vora, Hanna Schieve, Adam Waytz, Pawan Mudigonda, Lewis Ho, Toby Shevlane, Markus Anderljung, Miles Brundage, Kevin McKee, Amelia Hassoun, Lisa-Maria Neudert, Shahar Avin, Jackie Kay, Tom Everitt, Saurabh Chandra, Toby Ord, Matt Botvinick, Rohin Shah, Diane Korngiebel, Dylan Hadfield-Menell, Brian Christian, Dan Hendrycks, Leif Wenar, Percy Liang, Seth Lazar, Jeffrey Gelman, Joelle Barral, Zoubin Ghahramani, Eli Collins, Jared Bomberg, Marsden Hanna, David Weller, Richard Ives and Jerry Torres for their feedback and contributions to this work.
\bibliography{references}

\begin{thebibliography}{1337}
\providecommand{\natexlab}[1]{#1}
\providecommand{\url}[1]{\texttt{#1}}
\expandafter\ifx\csname urlstyle\endcsname\relax
  \providecommand{\doi}[1]{doi: #1}\else
  \providecommand{\doi}{doi: \begingroup \urlstyle{rm}\Url}\fi

\bibitem[Abadi et~al.(2016)Abadi, Chu, Goodfellow, McMahan, Mironov, Talwar, and Zhang]{abadi_deep_2016}
M.~Abadi, A.~Chu, I.~Goodfellow, H.~B. McMahan, I.~Mironov, K.~Talwar, and L.~Zhang.
\newblock Deep {Learning} with {Differential} {Privacy}.
\newblock In \emph{Proceedings of the 2016 {ACM} {SIGSAC} {Conference} on {Computer} and {Communications} {Security}}, pages 308--318, Oct. 2016.
\newblock \doi{10.1145/2976749.2978318}.
\newblock URL \url{http://arxiv.org/abs/1607.00133}.
\newblock arXiv:1607.00133 [cs, stat].

\bibitem[Abe and Abe(2019)]{abe2019lifestyle}
M.~Abe and H.~Abe.
\newblock Lifestyle medicine--an evidence based approach to nutrition, sleep, physical activity, and stress management on health and chronic illness.
\newblock \emph{Personalized Medicine Universe}, 8:\penalty0 3--9, 2019.

\bibitem[Abegunde et~al.(2007)Abegunde, Mathers, Adam, Ortegon, and Strong]{abegunde2007burden}
D.~O. Abegunde, C.~D. Mathers, T.~Adam, M.~Ortegon, and K.~Strong.
\newblock The burden and costs of chronic diseases in low-income and middle-income countries.
\newblock \emph{The Lancet}, 370\penalty0 (9603):\penalty0 1929--1938, 2007.

\bibitem[Abercrombie and Rieser(2022)]{abercrombie_risk-graded_2022}
G.~Abercrombie and V.~Rieser.
\newblock Risk-graded {Safety} for {Handling} {Medical} {Queries} in {Conversational} {AI}, Oct. 2022.
\newblock URL \url{http://arxiv.org/abs/2210.00572}.
\newblock arXiv:2210.00572 [cs].

\bibitem[Abercrombie et~al.(2021)Abercrombie, Curry, Pandya, and Rieser]{abercrombie_alexa_2021}
G.~Abercrombie, A.~C. Curry, M.~Pandya, and V.~Rieser.
\newblock Alexa, {Google}, {Siri}: {What} are {Your} {Pronouns}? {Gender} and {Anthropomorphism} in the {Design} and {Perception} of {Conversational} {Assistants}, June 2021.
\newblock URL \url{http://arxiv.org/abs/2106.02578}.
\newblock arXiv:2106.02578 [cs].

\bibitem[Abercrombie et~al.(2023)Abercrombie, Curry, Dinkar, Rieser, and Talat]{abercrombie_mirages:_2023}
G.~Abercrombie, A.~C. Curry, T.~Dinkar, V.~Rieser, and Z.~Talat.
\newblock Mirages: {On} {Anthropomorphism} in {Dialogue} {Systems}, Oct. 2023.
\newblock URL \url{http://arxiv.org/abs/2305.09800}.
\newblock arXiv:2305.09800 [cs].

\bibitem[Ablon and Bogart(2017)]{ablon_zero_2017}
L.~Ablon and A.~Bogart.
\newblock Zero {Days}, {Thousands} of {Nights}: {The} {Life} and {Times} of {Zero}-{Day} {Vulnerabilities} and {Their} {Exploits}.
\newblock Technical report, RAND Corporation, Mar. 2017.
\newblock URL \url{https://www.rand.org/pubs/research_reports/RR1751.html}.

\bibitem[Abramson et~al.(2022)Abramson, Ahuja, Carnevale, Georgiev, Goldin, Hung, Landon, Lhotka, Lillicrap, Muldal, et~al.]{abramson2022improving}
J.~Abramson, A.~Ahuja, F.~Carnevale, P.~Georgiev, A.~Goldin, A.~Hung, J.~Landon, J.~Lhotka, T.~Lillicrap, A.~Muldal, et~al.
\newblock Improving multimodal interactive agents with reinforcement learning from human feedback.
\newblock \emph{arXiv preprint arXiv:2211.11602}, 2022.

\bibitem[Acemoglu and Johnson(2023)]{acemoglu_power_2023}
D.~Acemoglu and S.~Johnson.
\newblock \emph{Power and progress: our thousand-year struggle over technology and prosperity}.
\newblock PublicAffairs, New York, first edition edition, 2023.
\newblock ISBN 9781541702530.

\bibitem[Acemoglu and Restrepo(2017)]{acemoglu_robots_2017}
D.~Acemoglu and P.~Restrepo.
\newblock Robots and {Jobs}: {Evidence} from {US} {Labor} {Markets}.
\newblock Technical Report w23285, National Bureau of Economic Research, Cambridge, MA, Mar. 2017.
\newblock URL \url{http://www.nber.org/papers/w23285.pdf}.

\bibitem[Acemoglu and Restrepo(2019)]{acemoglu_automation_2019}
D.~Acemoglu and P.~Restrepo.
\newblock Automation and {New} {Tasks}: {How} {Technology} {Displaces} and {Reinstates} {Labor}.
\newblock Technical Report w25684, National Bureau of Economic Research, Cambridge, MA, Mar. 2019.
\newblock URL \url{http://www.nber.org/papers/w25684.pdf}.

\bibitem[Acemoglu and Restrepo(2022)]{acemoglu_tasks_2022}
D.~Acemoglu and P.~Restrepo.
\newblock Tasks, {Automation}, and the {Rise} in {U}.{S}. {Wage} {Inequality}.
\newblock \emph{Econometrica}, 90\penalty0 (5):\penalty0 1973--2016, 2022.
\newblock ISSN 0012-9682.
\newblock \doi{10.3982/ECTA19815}.
\newblock URL \url{https://www.econometricsociety.org/doi/10.3982/ECTA19815}.

\bibitem[Acemoglu et~al.(2020)Acemoglu, Manera, and Restrepo]{acemoglu_does_2020}
D.~Acemoglu, A.~Manera, and P.~Restrepo.
\newblock Does the {US} tax code favor automation?
\newblock \emph{Brookings Papers on Economic Activity}, Mar. 2020.
\newblock URL \url{https://www.brookings.edu/articles/does-the-u-s-tax-code-favor-automation/}.

\bibitem[Acemoglu et~al.(2022)Acemoglu, Autor, Hazell, and Restrepo]{acemoglu_artificial_2022}
D.~Acemoglu, D.~Autor, J.~Hazell, and P.~Restrepo.
\newblock Artificial {Intelligence} and {Jobs}: {Evidence} from {Online} {Vacancies}.
\newblock \emph{Journal of Labor Economics}, 40\penalty0 (S1):\penalty0 S293--S340, Apr. 2022.
\newblock ISSN 0734-306X, 1537-5307.
\newblock \doi{10.1086/718327}.
\newblock URL \url{https://www.journals.uchicago.edu/doi/10.1086/718327}.

\bibitem[Adnan et~al.(2018)Adnan, Nordin, bin Bahruddin, and Ali]{adnan_trust_2018}
N.~Adnan, S.~M. Nordin, M.~A. bin Bahruddin, and M.~Ali.
\newblock How trust can drive forward the user acceptance to the technology? in-vehicle technology for autonomous vehicle.
\newblock \emph{Transportation research part A: policy and practice}, 118:\penalty0 819--836, 2018.

\bibitem[Agrawal et~al.(2023)Agrawal, Gans, and Goldfarb]{agrawal_we_2023}
A.~Agrawal, J.~S. Gans, and A.~Goldfarb.
\newblock Do we want less automation?
\newblock \emph{Science}, 381\penalty0 (6654):\penalty0 155--158, July 2023.
\newblock ISSN 0036-8075, 1095-9203.
\newblock \doi{10.1126/science.adh9429}.
\newblock URL \url{https://www.science.org/doi/10.1126/science.adh9429}.

\bibitem[Aguirre et~al.(2020)Aguirre, Dempsey, Surden, and Reiner]{aguirre2020ai}
A.~Aguirre, G.~Dempsey, H.~Surden, and P.~B. Reiner.
\newblock Ai loyalty: A new paradigm for aligning stakeholder interests, 2020.

\bibitem[Aher et~al.(2023)Aher, Arriaga, and Kalai]{aher_using_2023}
G.~Aher, R.~I. Arriaga, and A.~T. Kalai.
\newblock Using {Large} {Language} {Models} to {Simulate} {Multiple} {Humans} and {Replicate} {Human} {Subject} {Studies}, July 2023.
\newblock URL \url{http://arxiv.org/abs/2208.10264}.
\newblock arXiv:2208.10264 [cs].

\bibitem[Ahn et~al.(2022)Ahn, Brohan, Brown, Chebotar, Cortes, David, Finn, Fu, Gopalakrishnan, Hausman, Herzog, Ho, Hsu, Ibarz, Ichter, Irpan, Jang, Ruano, Jeffrey, Jesmonth, Joshi, Julian, Kalashnikov, Kuang, Lee, Levine, Lu, Luu, Parada, Pastor, Quiambao, Rao, Rettinghouse, Reyes, Sermanet, Sievers, Tan, Toshev, Vanhoucke, Xia, Xiao, Xu, Xu, Yan, and Zeng]{ahn_as_2022}
M.~Ahn, A.~Brohan, N.~Brown, Y.~Chebotar, O.~Cortes, B.~David, C.~Finn, C.~Fu, K.~Gopalakrishnan, K.~Hausman, A.~Herzog, D.~Ho, J.~Hsu, J.~Ibarz, B.~Ichter, A.~Irpan, E.~Jang, R.~J. Ruano, K.~Jeffrey, S.~Jesmonth, N.~J. Joshi, R.~Julian, D.~Kalashnikov, Y.~Kuang, K.-H. Lee, S.~Levine, Y.~Lu, L.~Luu, C.~Parada, P.~Pastor, J.~Quiambao, K.~Rao, J.~Rettinghouse, D.~Reyes, P.~Sermanet, N.~Sievers, C.~Tan, A.~Toshev, V.~Vanhoucke, F.~Xia, T.~Xiao, P.~Xu, S.~Xu, M.~Yan, and A.~Zeng.
\newblock Do {As} {I} {Can}, {Not} {As} {I} {Say}: {Grounding} {Language} in {Robotic} {Affordances}, Aug. 2022.
\newblock URL \url{http://arxiv.org/abs/2204.01691}.
\newblock arXiv:2204.01691 [cs].

\bibitem[{AI Incident Database}()]{ai_incident_database_ai_nodate}
{AI Incident Database}.
\newblock {AI} {Incident} {Database}.
\newblock URL \url{https://incidentdatabase.ai/}.

\bibitem[Ajei and Myles(2019)]{ajei2019personhood}
M.~Ajei and N.~O. Myles.
\newblock Personhood, autonomy and informed consent.
\newblock In \emph{Bioethics in Africa: Theories and Praxis}, pages 77--94. Vernon Press, 2019.

\bibitem[Akdeniz and van Veelen(2021)]{akdeniz_evolution_2021}
A.~Akdeniz and M.~van Veelen.
\newblock The evolution of morality and the role of commitment.
\newblock \emph{Evolutionary Human Sciences}, 3:\penalty0 e41, 2021.
\newblock ISSN 2513-843X.
\newblock \doi{10.1017/ehs.2021.36}.
\newblock URL \url{https://www.cambridge.org/core/product/identifier/S2513843X21000360/type/journal_article}.

\bibitem[Akerlof and Shiller(2015)]{akerlof_phishing_2015}
G.~A. Akerlof and R.~J. Shiller.
\newblock Phishing for {Phools}: {The} {Economics} of {Manipulation} and {Deception}.
\newblock In \emph{Phishing for {Phools}}. Princeton University Press, Sept. 2015.
\newblock ISBN 9781400873265.
\newblock URL \url{https://www.degruyter.com/document/doi/10.1515/9781400873265/html}.

\bibitem[Akhlaghi(2023)]{akhlaghi_transformative_2023}
F.~Akhlaghi.
\newblock Transformative experience and the right to revelatory autonomy.
\newblock \emph{Analysis}, 83\penalty0 (1):\penalty0 3--12, Aug. 2023.
\newblock ISSN 0003-2638, 1467-8284.
\newblock \doi{10.1093/analys/anac084}.
\newblock URL \url{https://academic.oup.com/analysis/article/83/1/3/6966040}.

\bibitem[Akhtar et~al.(2021)Akhtar, Basile, and Patti]{akhtar_whose_2021}
S.~Akhtar, V.~Basile, and V.~Patti.
\newblock Whose {Opinions} {Matter}? {Perspective}-aware {Models} to {Identify} {Opinions} of {Hate} {Speech} {Victims} in {Abusive} {Language} {Detection}, June 2021.
\newblock URL \url{https://arxiv.org/pdf/2106.15896.pdf}.
\newblock arXiv:2106.15896 [cs].

\bibitem[Al(2021)]{al2021value}
P.~Al.
\newblock The value of communities and their consent: A communitarian justification of community consent in medical research.
\newblock \emph{Bioethics}, 35\penalty0 (3):\penalty0 255--261, 2021.

\bibitem[Alayrac et~al.(2022)Alayrac, Donahue, Luc, Miech, Barr, Hasson, Lenc, Mensch, Millican, Reynolds, Ring, Rutherford, Cabi, Han, Gong, Samangooei, Monteiro, Menick, Borgeaud, Brock, Nematzadeh, Sharifzadeh, Binkowski, Barreira, Vinyals, Zisserman, and Simonyan]{alayrac_flamingo:_2022}
J.-B. Alayrac, J.~Donahue, P.~Luc, A.~Miech, I.~Barr, Y.~Hasson, K.~Lenc, A.~Mensch, K.~Millican, M.~Reynolds, R.~Ring, E.~Rutherford, S.~Cabi, T.~Han, Z.~Gong, S.~Samangooei, M.~Monteiro, J.~Menick, S.~Borgeaud, A.~Brock, A.~Nematzadeh, S.~Sharifzadeh, M.~Binkowski, R.~Barreira, O.~Vinyals, A.~Zisserman, and K.~Simonyan.
\newblock Flamingo: a {Visual} {Language} {Model} for {Few}-{Shot} {Learning}, Nov. 2022.
\newblock URL \url{http://arxiv.org/abs/2204.14198}.
\newblock arXiv:2204.14198 [cs].

\bibitem[Albanesi et~al.(2023)Albanesi, Da~Silva, Jimeno, Lamo, and Wabitsch]{albanesi_new_2023}
S.~Albanesi, A.~D. Da~Silva, J.~Jimeno, A.~Lamo, and A.~Wabitsch.
\newblock New {Technologies} and {Jobs} in {Europe}.
\newblock Technical Report w31357, National Bureau of Economic Research, Cambridge, MA, June 2023.
\newblock URL \url{http://www.nber.org/papers/w31357.pdf}.

\bibitem[Alberts and Van~Kleek(2023)]{alberts_computers_2023}
L.~Alberts and M.~Van~Kleek.
\newblock Computers as {Bad} {Social} {Actors}: {Dark} {Patterns} and {Anti}-{Patterns} in {Interfaces} that {Act} {Socially}, Feb. 2023.
\newblock URL \url{http://arxiv.org/abs/2302.04720}.
\newblock arXiv:2302.04720 [cs].

\bibitem[Alberts et~al.(2024)Alberts, Keeling, and McCroskery]{alberts2024makes}
L.~Alberts, G.~Keeling, and A.~McCroskery.
\newblock What makes for a `good' social actor? using respect as a lens to evaluate interactions with language agents.
\newblock 2024.
\newblock URL \url{http://arxiv.org/abs/2401.09082}.
\newblock arXiv:2401.09082.

\bibitem[Aldrich et~al.(2017)Aldrich, Grundfest, and Laughlin]{aldrich_flash_2017}
E.~M. Aldrich, J.~Grundfest, and G.~Laughlin.
\newblock The {Flash} {Crash}: {A} {New} {Deconstruction}, Mar. 2017.
\newblock URL \url{https://papers.ssrn.com/abstract=2721922}.

\bibitem[Alexandrova(2017)]{alexandrova2017well}
A.~Alexandrova.
\newblock Is well-being measurable after all?
\newblock \emph{Public Health Ethics}, 10\penalty0 (2):\penalty0 129--137, 2017.

\bibitem[{Alignment Research Centre}(2023)]{arc_evals_update_nodate}
{Alignment Research Centre}.
\newblock Update on {ARC}'s recent eval efforts - {ARC} {Evals}, 2023.
\newblock URL \url{https://evals.alignment.org/blog/2023-03-18-update-on-recent-evals/}.

\bibitem[Alkhatib(2021)]{alkhatib_live_2021}
A.~Alkhatib.
\newblock To {Live} in {Their} {Utopia}: {Why} {Algorithmic} {Systems} {Create} {Absurd} {Outcomes}.
\newblock In \emph{Proceedings of the 2021 {CHI} {Conference} on {Human} {Factors} in {Computing} {Systems}}, pages 1--9, Yokohama Japan, May 2021. ACM.
\newblock ISBN 9781450380966.
\newblock \doi{10.1145/3411764.3445740}.
\newblock URL \url{https://dl.acm.org/doi/10.1145/3411764.3445740}.

\bibitem[Allen et~al.(2020)Allen, Howland, Mobius, Rothschild, and Watts]{allen_evaluating_2020}
J.~Allen, B.~Howland, M.~Mobius, D.~Rothschild, and D.~J. Watts.
\newblock Evaluating the fake news problem at the scale of the information ecosystem.
\newblock \emph{Science Advances}, 6\penalty0 (14):\penalty0 eaay3539, Apr. 2020.
\newblock ISSN 2375-2548.
\newblock \doi{10.1126/sciadv.aay3539}.
\newblock URL \url{https://www.science.org/doi/10.1126/sciadv.aay3539}.

\bibitem[AlphaFold()]{alphafold_alphafold_nodate}
AlphaFold.
\newblock {AlphaFold} {Protein} {Structure} {Database}.
\newblock URL \url{https://alphafold.ebi.ac.uk/}.

\bibitem[Alphonso-Karakala(1975)]{alphonso-karakala_facets_1975}
J.~Alphonso-Karakala.
\newblock Facets {Of} {Panchatantra}.
\newblock \emph{Indian Literature}, 18\penalty0 (2):\penalty0 73--91, June 1975.

\bibitem[AlQuraishi(2021)]{alquraishi_machine_2021}
M.~AlQuraishi.
\newblock Machine learning in protein structure prediction.
\newblock \emph{Current Opinion in Chemical Biology}, 65:\penalty0 1--8, Dec. 2021.
\newblock ISSN 1367-5931.
\newblock \doi{10.1016/j.cbpa.2021.04.005}.
\newblock URL \url{https://www.sciencedirect.com/science/article/pii/S1367593121000508}.

\bibitem[Altay et~al.(2023)Altay, Berriche, and Acerbi]{altay_misinformation_2023}
S.~Altay, M.~Berriche, and A.~Acerbi.
\newblock Misinformation on {Misinformation}: {Conceptual} and {Methodological} {Challenges}.
\newblock \emph{Social Media + Society}, 9\penalty0 (1):\penalty0 205630512211504, Jan. 2023.
\newblock ISSN 2056-3051, 2056-3051.
\newblock \doi{10.1177/20563051221150412}.
\newblock URL \url{http://journals.sagepub.com/doi/10.1177/20563051221150412}.

\bibitem[Alvarez et~al.(2023)Alvarez, Eberhardt, and Linegar]{alvarez_generative_2023}
R.~M. Alvarez, F.~Eberhardt, and M.~Linegar.
\newblock Generative {AI} and the {Future} of {Elections}.
\newblock Technical report, Caltech Center for Science, Society, and Public Policy, July 2023.
\newblock URL \url{https://lindeinstitute.caltech.edu/documents/25475/CSSPP_white_paper.pdf}.

\bibitem[{American Press Institute}(2017)]{american_press_institute_who_2017}
{American Press Institute}.
\newblock ‘{Who} shared it?’ {How} {Americans} decide what news to trust on social media, Mar. 2017.
\newblock URL \url{https://americanpressinstitute.org/publications/reports/survey-research/trust-social-media/}.

\bibitem[Amini et~al.(2023)Amini, Buck, Brown, Bulian, Huebscher, Ciaramita, Das, Gaiarin, GORDON, Gupta, et~al.]{amini2023ai}
A.~Amini, C.~Buck, H.~Brown, J.~Bulian, M.~C. Huebscher, M.~Ciaramita, S.~Das, B.~Gaiarin, C.~GORDON, R.~Gupta, et~al.
\newblock Ai and climate information needs in africa.
\newblock In \emph{Deep Learning Indaba 2023}, 2023.

\bibitem[Amodei et~al.(2016)Amodei, Olah, Steinhardt, Christiano, Schulman, and Mané]{amodei_concrete_2016}
D.~Amodei, C.~Olah, J.~Steinhardt, P.~Christiano, J.~Schulman, and D.~Mané.
\newblock Concrete {Problems} in {AI} {Safety}, July 2016.
\newblock URL \url{http://arxiv.org/abs/1606.06565}.
\newblock arXiv:1606.06565 [cs].

\bibitem[Amoore(2020)]{amoore_cloud_2020}
L.~Amoore.
\newblock \emph{Cloud ethics: algorithms and the attributes of ourselves and others}.
\newblock Duke University Press, Durham, 2020.
\newblock ISBN 9781478007784 9781478008316.

\bibitem[Anderljung and Hazell(2023)]{anderljung_protecting_2023}
M.~Anderljung and J.~Hazell.
\newblock Protecting {Society} from {AI} {Misuse}: {When} are {Restrictions} on {Capabilities} {Warranted}?, Mar. 2023.
\newblock URL \url{http://arxiv.org/abs/2303.09377}.
\newblock arXiv:2303.09377 [cs].

\bibitem[Anderljung et~al.(2023)Anderljung, Barnhart, Korinek, Leung, O'Keefe, Whittlestone, Avin, Brundage, Bullock, Cass-Beggs, Chang, Collins, Fist, Hadfield, Hayes, Ho, Hooker, Horvitz, Kolt, Schuett, Shavit, Siddarth, Trager, and Wolf]{anderljung2023frontier}
M.~Anderljung, J.~Barnhart, A.~Korinek, J.~Leung, C.~O'Keefe, J.~Whittlestone, S.~Avin, M.~Brundage, J.~Bullock, D.~Cass-Beggs, B.~Chang, T.~Collins, T.~Fist, G.~Hadfield, A.~Hayes, L.~Ho, S.~Hooker, E.~Horvitz, N.~Kolt, J.~Schuett, Y.~Shavit, D.~Siddarth, R.~Trager, and K.~Wolf.
\newblock Frontier ai regulation: Managing emerging risks to public safety, 2023.

\bibitem[Anderson(1999)]{anderson_what_1999}
E.~Anderson.
\newblock What {Is} the {Point} of {Equality}?
\newblock \emph{Ethics}, 109\penalty0 (2):\penalty0 287--337, Jan. 1999.
\newblock ISSN 0014-1704, 1539-297X.
\newblock \doi{10.1086/233897}.
\newblock URL \url{https://www.journals.uchicago.edu/doi/10.1086/233897}.

\bibitem[Anderson(2023)]{anderson_coercion_2023}
S.~Anderson.
\newblock Coercion.
\newblock In E.~N. Zalta and U.~Nodelman, editors, \emph{The {Stanford} {Encyclopedia} of {Philosophy}}. Metaphysics Research Lab, Stanford University, spring 2023 edition, 2023.
\newblock URL \url{https://plato.stanford.edu/archives/spr2023/entries/coercion/}.

\bibitem[Andrews et~al.(2016)Andrews, Criscuolo, and Gal]{andrews_global_nodate}
D.~Andrews, C.~Criscuolo, and P.~N. Gal.
\newblock The global productivity slowdown, technology divergence, and public policy: {A} firm level perspective, 2016.
\newblock URL \url{https://www.brookings.edu/articles/the-global-productivity-slowdown-technology-divergence/}.

\bibitem[Andrist et~al.(2015)Andrist, Ziadee, Boukaram, Mutlu, and Sakr]{andrist2015effects}
S.~Andrist, M.~Ziadee, H.~Boukaram, B.~Mutlu, and M.~Sakr.
\newblock Effects of culture on the credibility of robot speech: A comparison between english and arabic.
\newblock In \emph{Proceedings of the tenth annual ACM/IEEE international conference on human-robot interaction}, pages 157--164, 2015.

\bibitem[Angwin et~al.(2022)Angwin, Larson, Mattu, and Kirchner]{kirchner_machine_2022}
J.~Angwin, J.~Larson, S.~Mattu, and L.~Kirchner.
\newblock Machine {Bias}.
\newblock In \emph{Ethics of {Data} and {Analytics}}. Auerbach Publications, 2022.
\newblock ISBN 9781003278290.

\bibitem[Anspach(2017)]{anspach_new_2017}
N.~M. Anspach.
\newblock The {New} {Personal} {Influence}: {How} {Our} {Facebook} {Friends} {Influence} the {News} {We} {Read}.
\newblock \emph{Political Communication}, 34\penalty0 (4):\penalty0 590--606, Oct. 2017.
\newblock ISSN 1058-4609, 1091-7675.
\newblock \doi{10.1080/10584609.2017.1316329}.
\newblock URL \url{https://www.tandfonline.com/doi/full/10.1080/10584609.2017.1316329}.

\bibitem[Anthropic(2023{\natexlab{a}})]{anthropic_ccai_2023}
Anthropic.
\newblock Collective constitutional {AI}: {Aligning} a language model with public input, Oct. 2023{\natexlab{a}}.
\newblock URL \url{https://www.anthropic.com/news/collective-constitutional-ai-aligning-a-language-model-with-public-input}.

\bibitem[Anthropic(2023{\natexlab{b}})]{anthropic_claudes_2023}
Anthropic.
\newblock Claude’s {Constitution}, May 2023{\natexlab{b}}.
\newblock URL \url{https://www.anthropic.com/index/claudes-constitution#:~:text=The%20system%20uses%20a%20set,human%20engage%20in%20illegal%20or}.

\bibitem[Anthropic(2023{\natexlab{c}})]{anthropic_core_2023}
Anthropic.
\newblock Core {Views} on {AI} {Safety}: {When}, {Why}, {What}, and {How}, Mar. 2023{\natexlab{c}}.
\newblock URL \url{https://www.anthropic.com/index/core-views-on-ai-safety}.

\bibitem[Anwar et~al.(2024)Anwar, Saparov, Rando, Paleka, Turpin, Hase, Lubana, Jenner, Casper, Sourbut, Edelman, Zhang, Günther, Korinek, Hernandez-Orallo, Hammond, Bigelow, Pan, Langosco, Korbak, Zhang, Zhong, heigeartaigh, Recchia, Corsi, Chan, Anderljung, Edwards, Bengio, Chen, Albanie, Maharaj, Foerster, Tramer, He, Kasirzadeh, Choi, and Krueger]{anwar2024foundational}
U.~Anwar, A.~Saparov, J.~Rando, D.~Paleka, M.~Turpin, P.~Hase, E.~S. Lubana, E.~Jenner, S.~Casper, O.~Sourbut, B.~L. Edelman, Z.~Zhang, M.~Günther, A.~Korinek, J.~Hernandez-Orallo, L.~Hammond, E.~Bigelow, A.~Pan, L.~Langosco, T.~Korbak, H.~Zhang, R.~Zhong, S.~O. heigeartaigh, G.~Recchia, G.~Corsi, A.~Chan, M.~Anderljung, L.~Edwards, Y.~Bengio, D.~Chen, S.~Albanie, T.~Maharaj, J.~Foerster, F.~Tramer, H.~He, A.~Kasirzadeh, Y.~Choi, and D.~Krueger.
\newblock Foundational challenges in assuring alignment and safety of large language models, 2024.

\bibitem[A.P.(2006)]{AP_2006}
A.P.
\newblock U.n. lends backing to the \$100 laptop, 2006.
\newblock URL \url{https://web.archive.org/web/20080530011349/http://www.linux.org/news/2006/01/27/0007.html}.

\bibitem[{Apollo Research}(2022)]{apollo_research_understanding_nodate}
{Apollo Research}.
\newblock Understanding strategic deception and deceptive alignment, 2022.
\newblock URL \url{https://www.apolloresearch.ai/blog/understanding-da-and-sd}.

\bibitem[Apuke et~al.(2022)Apuke, Omar, Tunca, and Gever]{apuke_information_2022}
O.~D. Apuke, B.~Omar, E.~A. Tunca, and C.~V. Gever.
\newblock Information overload and misinformation sharing behaviour of social media users: {Testing} the moderating role of cognitive ability.
\newblock \emph{Journal of Information Science}, page 016555152211219, Sept. 2022.
\newblock ISSN 0165-5515, 1741-6485.
\newblock \doi{10.1177/01655515221121942}.
\newblock URL \url{http://journals.sagepub.com/doi/10.1177/01655515221121942}.

\bibitem[Araujo(2018)]{araujo_living_2018}
T.~Araujo.
\newblock Living up to the chatbot hype: {The} influence of anthropomorphic design cues and communicative agency framing on conversational agent and company perceptions.
\newblock \emph{Computers in Human Behavior}, 85:\penalty0 183--189, Aug. 2018.
\newblock ISSN 0747-5632.
\newblock \doi{10.1016/j.chb.2018.03.051}.
\newblock URL \url{https://www.sciencedirect.com/science/article/pii/S0747563218301560}.

\bibitem[Arguelles et~al.(2020)Arguelles, Sampson, Kubik, and Bibi]{arguelles2020critical}
C.~Arguelles, T.~Sampson, J.~Kubik, and E.~Bibi.
\newblock Critical user journey test coverage.
\newblock 2020.

\bibitem[Arora et~al.(2018)Arora, Li, Liang, Ma, and Risteski]{arora_linear_2018}
S.~Arora, Y.~Li, Y.~Liang, T.~Ma, and A.~Risteski.
\newblock Linear {Algebraic} {Structure} of {Word} {Senses}, with {Applications} to {Polysemy}, Dec. 2018.
\newblock URL \url{http://arxiv.org/abs/1601.03764}.
\newblock arXiv:1601.03764 [cs, stat].

\bibitem[{Article 19}(2018)]{article19_2018}
{Article 19}.
\newblock Apps and traps: Dating apps must do more to protect lgbtq communities in middle east and north africa.
\newblock \emph{Article 19}, 2018.
\newblock URL \url{https://www.article19.org/resources/apps-traps-dating-apps-must-protect-communities-middle-east-north-africa/}.

\bibitem[{Article 19}(2022)]{article19_2022}
{Article 19}.
\newblock Equally safe: Towards a feminist approach to the safety of journalists.
\newblock \emph{Article 19}, 2022.
\newblock URL \url{https://www.article19.org/wp-content/uploads/2022/12/Equally-Safe-FemSoj_08.12.22.pdf}.

\bibitem[Ashton and Franklin(2022)]{ashton2022problem}
H.~Ashton and M.~Franklin.
\newblock The problem of behaviour and preference manipulation in {AI} systems.
\newblock In \emph{CEUR Workshop Proceedings}, volume 3087. CEUR Workshop Proceedings, 2022.

\bibitem[Askell et~al.(2021)Askell, Bai, Chen, Drain, Ganguli, Henighan, Jones, Joseph, Mann, DasSarma, Elhage, Hatfield-Dodds, Hernandez, Kernion, Ndousse, Olsson, Amodei, Brown, Clark, McCandlish, Olah, and Kaplan]{askell_general_2021}
A.~Askell, Y.~Bai, A.~Chen, D.~Drain, D.~Ganguli, T.~Henighan, A.~Jones, N.~Joseph, B.~Mann, N.~DasSarma, N.~Elhage, Z.~Hatfield-Dodds, D.~Hernandez, J.~Kernion, K.~Ndousse, C.~Olsson, D.~Amodei, T.~Brown, J.~Clark, S.~McCandlish, C.~Olah, and J.~Kaplan.
\newblock A {General} {Language} {Assistant} as a {Laboratory} for {Alignment}, Dec. 2021.
\newblock URL \url{http://arxiv.org/abs/2112.00861}.
\newblock arXiv:2112.00861 [cs].

\bibitem[ATLAS.ti(2023)]{atlas.ti_atlas.ti_nodate}
ATLAS.ti.
\newblock {ATLAS}.ti {\textbar} {The} \#1 {Software} for {Qualitative} {Data} {Analysis}, 2023.
\newblock URL \url{https://atlasti.com}.

\bibitem[Autor and Salomons(2019)]{autor_new_2019}
D.~Autor and A.~Salomons.
\newblock New {Frontiers}: {The} {Evolving} {Content} and {Geography} of {New} {Work} in the 20th {Century} - {David} {Autor}, May 2019.
\newblock URL \url{https://www.getsphere.com/}.

\bibitem[Avaaz(2020)]{avaaz_how_2020}
Avaaz.
\newblock How {Facebook} can {Flatten} the {Curve} of the {Coronavirus} {Infodemic}, Apr. 2020.
\newblock URL \url{https://secure.avaaz.org/campaign/en/facebook_coronavirus_misinformation/}.

\bibitem[Azaria and Mitchell(2023)]{azaria_internal_2023}
A.~Azaria and T.~Mitchell.
\newblock The internal state of an llm knows when its lying.
\newblock \emph{arXiv preprint arXiv:2304.13734}, 2023.

\bibitem[{BACP}(2018)]{bacp2018ethical}
{BACP}.
\newblock Ethical framework for the counselling professions.
\newblock \url{https://www.bacp.co.uk/events-and-resources/ethics-and-standards/ethical-framework-for-the-counselling-professions/}, 2018.
\newblock Accessed: 2023-01-04.

\bibitem[Baek et~al.(2022)Baek, Bakpayev, Yoon, and Kim]{baek_smiling_2022}
T.~H. Baek, M.~Bakpayev, S.~Yoon, and S.~Kim.
\newblock Smiling {AI} agents: {How} anthropomorphism and broad smiles increase charitable giving.
\newblock \emph{International Journal of Advertising}, 41\penalty0 (5):\penalty0 850--867, July 2022.
\newblock ISSN 0265-0487, 1759-3948.
\newblock \doi{10.1080/02650487.2021.2011654}.
\newblock URL \url{https://www.tandfonline.com/doi/full/10.1080/02650487.2021.2011654}.

\bibitem[Bai et~al.(2021)Bai, Zang, Xu, Sunkara, Rastogi, Chen, and Arcas]{bai_uibert:_2021}
C.~Bai, X.~Zang, Y.~Xu, S.~Sunkara, A.~Rastogi, J.~Chen, and B.~A.~y. Arcas.
\newblock {UIBert}: {Learning} {Generic} {Multimodal} {Representations} for {UI} {Understanding}, Aug. 2021.
\newblock URL \url{http://arxiv.org/abs/2107.13731}.
\newblock arXiv:2107.13731 [cs].

\bibitem[Bai et~al.(2023)Bai, Voelkel, Eichstaedt, and Willer]{bai_artificial_2023}
H.~Bai, J.~Voelkel, J.~Eichstaedt, and R.~Willer.
\newblock Artificial {Intelligence} {Can} {Persuade} {Humans} on {Political} {Issues}.
\newblock preprint, In Review, Sept. 2023.
\newblock URL \url{https://www.researchsquare.com/article/rs-3238396/v1}.

\bibitem[Bai et~al.(2022{\natexlab{a}})Bai, Jones, Ndousse, Askell, Chen, DasSarma, Drain, Fort, Ganguli, Henighan, Joseph, Kadavath, Kernion, Conerly, El-Showk, Elhage, Hatfield-Dodds, Hernandez, Hume, Johnston, Kravec, Lovitt, Nanda, Olsson, Amodei, Brown, Clark, McCandlish, Olah, Mann, and Kaplan]{bai_training_2022}
Y.~Bai, A.~Jones, K.~Ndousse, A.~Askell, A.~Chen, N.~DasSarma, D.~Drain, S.~Fort, D.~Ganguli, T.~Henighan, N.~Joseph, S.~Kadavath, J.~Kernion, T.~Conerly, S.~El-Showk, N.~Elhage, Z.~Hatfield-Dodds, D.~Hernandez, T.~Hume, S.~Johnston, S.~Kravec, L.~Lovitt, N.~Nanda, C.~Olsson, D.~Amodei, T.~Brown, J.~Clark, S.~McCandlish, C.~Olah, B.~Mann, and J.~Kaplan.
\newblock Training a {Helpful} and {Harmless} {Assistant} with {Reinforcement} {Learning} from {Human} {Feedback}, Apr. 2022{\natexlab{a}}.
\newblock URL \url{http://arxiv.org/abs/2204.05862}.
\newblock arXiv:2204.05862 [cs].

\bibitem[Bai et~al.(2022{\natexlab{b}})Bai, Kadavath, Kundu, Askell, Kernion, Jones, Chen, Goldie, Mirhoseini, McKinnon, Chen, Olsson, Olah, Hernandez, Drain, Ganguli, Li, Tran-Johnson, Perez, Kerr, Mueller, Ladish, Landau, Ndousse, Lukosuite, Lovitt, Sellitto, Elhage, Schiefer, Mercado, DasSarma, Lasenby, Larson, Ringer, Johnston, Kravec, Showk, Fort, Lanham, Telleen-Lawton, Conerly, Henighan, Hume, Bowman, Hatfield-Dodds, Mann, Amodei, Joseph, McCandlish, Brown, and Kaplan]{bai_constitutional_2022}
Y.~Bai, S.~Kadavath, S.~Kundu, A.~Askell, J.~Kernion, A.~Jones, A.~Chen, A.~Goldie, A.~Mirhoseini, C.~McKinnon, C.~Chen, C.~Olsson, C.~Olah, D.~Hernandez, D.~Drain, D.~Ganguli, D.~Li, E.~Tran-Johnson, E.~Perez, J.~Kerr, J.~Mueller, J.~Ladish, J.~Landau, K.~Ndousse, K.~Lukosuite, L.~Lovitt, M.~Sellitto, N.~Elhage, N.~Schiefer, N.~Mercado, N.~DasSarma, R.~Lasenby, R.~Larson, S.~Ringer, S.~Johnston, S.~Kravec, S.~E. Showk, S.~Fort, T.~Lanham, T.~Telleen-Lawton, T.~Conerly, T.~Henighan, T.~Hume, S.~R. Bowman, Z.~Hatfield-Dodds, B.~Mann, D.~Amodei, N.~Joseph, S.~McCandlish, T.~Brown, and J.~Kaplan.
\newblock Constitutional {AI}: {Harmlessness} from {AI} {Feedback}, Dec. 2022{\natexlab{b}}.
\newblock URL \url{https://arxiv.org/pdf/2212.08073.pdf}.
\newblock arXiv:2212.08073 [cs].

\bibitem[Bakalar et~al.(2021)Bakalar, Barreto, Bergman, Bogen, Chern, Corbett-Davies, Hall, Kloumann, Lam, Candela, Raghavan, Simons, Tannen, Tong, Vredenburgh, and Zhao]{bakalar_fairness_2021}
C.~Bakalar, R.~Barreto, S.~Bergman, M.~Bogen, B.~Chern, S.~Corbett-Davies, M.~Hall, I.~Kloumann, M.~Lam, J.~Q. Candela, M.~Raghavan, J.~Simons, J.~Tannen, E.~Tong, K.~Vredenburgh, and J.~Zhao.
\newblock Fairness {On} {The} {Ground}: {Applying} {Algorithmic} {Fairness} {Approaches} to {Production} {Systems}, Mar. 2021.
\newblock URL \url{http://arxiv.org/abs/2103.06172}.
\newblock arXiv:2103.06172 [cs].

\bibitem[Bakker et~al.(2022)Bakker, Chadwick, Sheahan, Tessler, Campbell-Gillingham, Balaguer, McAleese, Glaese, Aslanides, Botvinick, and Summerfield]{bakker_fine-tuning_2022}
M.~Bakker, M.~Chadwick, H.~Sheahan, M.~Tessler, L.~Campbell-Gillingham, J.~Balaguer, N.~McAleese, A.~Glaese, J.~Aslanides, M.~Botvinick, and C.~Summerfield.
\newblock Fine-tuning language models to find agreement among humans with diverse preferences.
\newblock \emph{Advances in Neural Information Processing Systems}, 35:\penalty0 38176--38189, Dec. 2022.
\newblock URL \url{https://proceedings.neurips.cc/paper_files/paper/2022/hash/f978c8f3b5f399cae464e85f72e28503-Abstract-Conference.html}.

\bibitem[Bakshy et~al.(2015)Bakshy, Messing, and Adamic]{bakshy_exposure_2015}
E.~Bakshy, S.~Messing, and L.~A. Adamic.
\newblock Exposure to ideologically diverse news and opinion on {Facebook}.
\newblock \emph{Science}, 348\penalty0 (6239):\penalty0 1130--1132, June 2015.
\newblock ISSN 0036-8075, 1095-9203.
\newblock \doi{10.1126/science.aaa1160}.
\newblock URL \url{https://www.science.org/doi/10.1126/science.aaa1160}.

\bibitem[Balasubramanian et~al.(2021)Balasubramanian, Beaney, and Chambers]{balasubramanian2021digital}
G.~V. Balasubramanian, P.~Beaney, and R.~Chambers.
\newblock Digital personal assistants are smart ways for assistive technology to aid the health and wellbeing of patients and carers.
\newblock \emph{BMC {G}eriatrics}, 21:\penalty0 1--10, 2021.

\bibitem[Bal{\'{a}}zs et~al.(2017)Bal{\'{a}}zs, Bene, and Hidegkuti]{balazs_vulnerable_2017}
K.~Bal{\'{a}}zs, {\'{A}}.~Bene, and I.~Hidegkuti.
\newblock Vulnerable older consumers: {New} persuasion knowledge achievement measure.
\newblock \emph{International Journal of Consumer Studies}, 41\penalty0 (6):\penalty0 706--713, Nov. 2017.
\newblock ISSN 1470-6423, 1470-6431.
\newblock \doi{10.1111/ijcs.12383}.
\newblock URL \url{https://onlinelibrary.wiley.com/doi/10.1111/ijcs.12383}.

\bibitem[Bandura(2013)]{bandura2013health}
A.~Bandura.
\newblock Health promotion from the perspective of social cognitive theory.
\newblock In C.~Abraham, P.~Norman, and M.~Conner, editors, \emph{Understanding and changing health behaviour}, pages 299--339. Psychology Press, 2013.

\bibitem[Banner(2020)]{banner2020human}
N.~F. Banner.
\newblock The human side of health data.
\newblock \emph{Nature Medicine}, 26\penalty0 (7):\penalty0 995--995, 2020.

\bibitem[Barfuss et~al.(2023)Barfuss, Flack, and Lenaerts]{barfuss_collective_nodate}
W.~Barfuss, J.~Flack, and T.~Lenaerts.
\newblock Collective {Cooperative} {Intelligence}, 2023.
\newblock URL \url{https://www.cooperativeai.com/seminars/collective-cooperative-intelligence}.

\bibitem[Barnes and Christiano(2020)]{barnes_writeup:_2020}
B.~Barnes and P.~Christiano.
\newblock Writeup: {Progress} on {AI} {Safety} via {Debate}, Feb. 2020.
\newblock URL \url{https://www.alignmentforum.org/posts/Br4xDbYu4Frwrb64a/writeup-progress-on-ai-safety-via-debate-1}.

\bibitem[Baron(2014)]{coons_mens_2014}
M.~Baron.
\newblock The {Mens} {Rea} and {Moral} {Status} of {Manipulation}.
\newblock In C.~Coons and M.~Weber, editors, \emph{Manipulation}, pages 98--120. Oxford University Press, Aug. 2014.
\newblock ISBN 9780199338207.
\newblock \doi{10.1093/acprof:oso/9780199338207.003.0005}.
\newblock URL \url{https://academic.oup.com/book/4870/chapter/147239828}.

\bibitem[Barrett and Keil(2016)]{barrett_conceptualizing_2016}
J.~L. Barrett and F.~C. Keil.
\newblock Conceptualizing a nonnatural entity: Anthropomorphism in god concepts.
\newblock In \emph{Religion and Cognition}, pages 116--148. Routledge, 2016.

\bibitem[Barth et~al.(2006)Barth, Datta, Mitchell, and Nissenbaum]{barth_privacy_2006}
A.~Barth, A.~Datta, J.~Mitchell, and H.~Nissenbaum.
\newblock Privacy and contextual integrity: framework and applications.
\newblock In \emph{2006 {IEEE} {Symposium} on {Security} and {Privacy} ({S}\&{P}'06)}, pages 15 pp.--198, Berkeley/Oakland, CA, 2006. IEEE.
\newblock ISBN 9780769525747.
\newblock \doi{10.1109/SP.2006.32}.
\newblock URL \url{http://ieeexplore.ieee.org/document/1624011/}.

\bibitem[Bartolo et~al.(2021)Bartolo, Thrush, Jia, Riedel, Stenetorp, and Kiela]{bartolo_improving_2021}
M.~Bartolo, T.~Thrush, R.~Jia, S.~Riedel, P.~Stenetorp, and D.~Kiela.
\newblock Improving {Question} {Answering} {Model} {Robustness} with {Synthetic} {Adversarial} {Data} {Generation}.
\newblock In \emph{Proceedings of the 2021 {Conference} on {Empirical} {Methods} in {Natural} {Language} {Processing}}, pages 8830--8848, 2021.
\newblock \doi{10.18653/v1/2021.emnlp-main.696}.
\newblock URL \url{http://arxiv.org/abs/2104.08678}.
\newblock arXiv:2104.08678 [cs].

\bibitem[Bauer and Lizotte(2021)]{bauer_artificial_2021}
G.~R. Bauer and D.~J. Lizotte.
\newblock Artificial {Intelligence}, {Intersectionality}, and the {Future} of {Public} {Health}.
\newblock \emph{American Journal of Public Health}, 111\penalty0 (1):\penalty0 98--100, Jan. 2021.
\newblock ISSN 0090-0036, 1541-0048.
\newblock \doi{10.2105/AJPH.2020.306006}.
\newblock URL \url{https://ajph.aphapublications.org/doi/full/10.2105/AJPH.2020.306006}.

\bibitem[Bauld et~al.(2010)Bauld, Hay, McKell, and Carroll]{bauld_problem_2010}
L.~Bauld, G.~Hay, J.~McKell, and C.~Carroll.
\newblock Problem drug users' experiences of employment and the benefit system.
\newblock Technical Report Research Report No 640, Department for Work and Pensions, 2010.
\newblock URL \url{https://assets.publishing.service.gov.uk/government/uploads/system/uploads/attachment_data/file/214409/rrep640.pdf}.

\bibitem[Bayer(2020)]{bayer_double_2020}
J.~Bayer.
\newblock Double harm to voters: {Data}-driven micro-targeting and democratic public discourse.
\newblock \emph{Internet Policy Review}, 9\penalty0 (1):\penalty0 1--17, 2020.
\newblock ISSN 2197-6775.
\newblock \doi{10.14763/2020.1.1460}.
\newblock URL \url{https://www.econstor.eu/handle/10419/216225}.

\bibitem[{Be My Eyes}(2023)]{be_my_eyes_introducing_nodate}
{Be My Eyes}.
\newblock Introducing {Be} {My} {AI} (formerly {Virtual} {Volunteer}) for {People} who are {Blind} or {Have} {Low} {Vision}, {Powered} by {OpenAI}’s {GPT}-4, 2023.
\newblock URL \url{https://www.bemyeyes.com/blog/introducing-be-my-eyes-virtual-volunteer}.

\bibitem[Beard(1927)]{beard_time_1927}
C.~A. Beard.
\newblock Time, {Technology}, and the {Creative} {Spirit} in {Political} {Science}.
\newblock \emph{American Political Science Review}, 21\penalty0 (1):\penalty0 1--11, Feb. 1927.
\newblock ISSN 0003-0554, 1537-5943.
\newblock \doi{10.2307/1945535}.
\newblock URL \url{https://www.cambridge.org/core/product/identifier/S0003055400023625/type/journal_article}.

\bibitem[Beauchamp and Childress(2019)]{beauchamp_principles_2019}
T.~L. Beauchamp and J.~F. Childress.
\newblock \emph{Principles of biomedical ethics}.
\newblock Oxford University Press, New York, eighth edition edition, 2019.
\newblock ISBN 9780190640873 9780190085520.

\bibitem[Beede et~al.(2020)Beede, Baylor, Hersch, Iurchenko, Wilcox, Ruamviboonsuk, and Vardoulakis]{beede_human-centered_2020}
E.~Beede, E.~Baylor, F.~Hersch, A.~Iurchenko, L.~Wilcox, P.~Ruamviboonsuk, and L.~M. Vardoulakis.
\newblock A {Human}-{Centered} {Evaluation} of a {Deep} {Learning} {System} {Deployed} in {Clinics} for the {Detection} of {Diabetic} {Retinopathy}.
\newblock In \emph{Proceedings of the 2020 {CHI} {Conference} on {Human} {Factors} in {Computing} {Systems}}, pages 1--12, Honolulu HI USA, Apr. 2020. ACM.
\newblock ISBN 9781450367080.
\newblock \doi{10.1145/3313831.3376718}.
\newblock URL \url{https://dl.acm.org/doi/10.1145/3313831.3376718}.

\bibitem[Begum et~al.(2021)Begum, Jingwei, Haider, Ajmal, Khan, and Han]{begum_impact_2021}
A.~Begum, L.~Jingwei, M.~Haider, M.~M. Ajmal, S.~Khan, and H.~Han.
\newblock Impact of {Environmental} {Moral} {Education} on {Pro}-{Environmental} {Behaviour}: {Do} {Psychological} {Empowerment} and {Islamic} {Religiosity} {Matter}?
\newblock \emph{International Journal of Environmental Research and Public Health}, 18\penalty0 (4):\penalty0 1604, Feb. 2021.
\newblock ISSN 1660-4601.
\newblock \doi{10.3390/ijerph18041604}.
\newblock URL \url{https://www.mdpi.com/1660-4601/18/4/1604}.

\bibitem[Belk(2016)]{belk2016extended}
R.~Belk.
\newblock Extended self and the digital world.
\newblock \emph{Current Opinion in Psychology}, 10:\penalty0 50--54, 2016.

\bibitem[Bellini et~al.(2023)Bellini, Tseng, Warford, Daffalla, Matthews, Consolvo, Woelfer, Kelley, Mazurek, Cuomo, Dell, and Ristenpart]{bellini_sok:_2023}
R.~Bellini, E.~Tseng, N.~Warford, A.~Daffalla, T.~Matthews, S.~Consolvo, J.~P. Woelfer, P.~G. Kelley, M.~L. Mazurek, D.~Cuomo, N.~Dell, and T.~Ristenpart.
\newblock {SoK}: {Safer} {Digital}-{Safety} {Research} {Involving} {At}-{Risk} {Users}, Sept. 2023.
\newblock URL \url{http://arxiv.org/abs/2309.00735}.
\newblock arXiv:2309.00735 [cs].

\bibitem[Ben-Ishai et~al.(2024)Ben-Ishai, Dean, Manyika, Porat, Varian, and Walker]{ben2024ai}
G.~Ben-Ishai, J.~Dean, J.~Manyika, R.~Porat, H.~Varian, and K.~Walker.
\newblock Ai and the opportunity for shared prosperity: Lessons from the history of technology and the economy.
\newblock \emph{arXiv preprint arXiv:2401.09718}, 2024.

\bibitem[Bender et~al.(2021)Bender, Gebru, McMillan-Major, and Shmitchell]{bender_dangers_2021}
E.~M. Bender, T.~Gebru, A.~McMillan-Major, and S.~Shmitchell.
\newblock On the {Dangers} of {Stochastic} {Parrots}: {Can} {Language} {Models} {Be} {Too} {Big}?
\newblock In \emph{Proceedings of the 2021 {ACM} {Conference} on {Fairness}, {Accountability}, and {Transparency}}, {FAccT} '21, pages 610--623, New York, NY, USA, Mar. 2021. Association for Computing Machinery.
\newblock ISBN 9781450383097.
\newblock \doi{10.1145/3442188.3445922}.
\newblock URL \url{https://dl.acm.org/doi/10.1145/3442188.3445922}.

\bibitem[Benedicto(2023)]{benedicto_detroit_nodate}
I.~Benedicto.
\newblock Detroit {Woman} {Sues} {City} {Police} {After} {Being} {Wrongfully} {Arrested} {Due} {To} {AI} {Facial} {Recognition}, 2023.
\newblock URL \url{https://www.forbes.com/sites/irenebenedicto/2023/08/07/detroit-woman-sues-city-police-after-being-wrongfully-arrested-due-to-ai-facial-recognition/}.

\bibitem[Benjamin(2020)]{benjamin_race_2020}
R.~Benjamin.
\newblock \emph{Race after technology: abolitionist tools for the {New} {Jim} {Code}}.
\newblock Polity, Cambridge, UK ; Medford, MA, 2020.
\newblock ISBN 9781509526406 9781509526390.

\bibitem[Bennett and Keyes(2020)]{bennett_what_2020}
C.~L. Bennett and O.~Keyes.
\newblock What is the point of fairness?: disability, {AI} and the complexity of justice.
\newblock \emph{ACM SIGACCESS Accessibility and Computing}, \penalty0 (125):\penalty0 1--1, Mar. 2020.
\newblock ISSN 1558-2337, 1558-1187.
\newblock \doi{10.1145/3386296.3386301}.
\newblock URL \url{https://dl.acm.org/doi/10.1145/3386296.3386301}.

\bibitem[Bennett et~al.(2018)Bennett, Brady, and Branham]{bennett_interdependence_2018}
C.~L. Bennett, E.~Brady, and S.~M. Branham.
\newblock Interdependence as a {Frame} for {Assistive} {Technology} {Research} and {Design}.
\newblock In \emph{Proceedings of the 20th {International} {ACM} {SIGACCESS} {Conference} on {Computers} and {Accessibility}}, pages 161--173, Galway Ireland, Oct. 2018. ACM.
\newblock ISBN 9781450356503.
\newblock \doi{10.1145/3234695.3236348}.
\newblock URL \url{https://dl.acm.org/doi/10.1145/3234695.3236348}.

\bibitem[Bennett et~al.(2020)Bennett, Rosner, and Taylor]{bennett_care_2020}
C.~L. Bennett, D.~K. Rosner, and A.~S. Taylor.
\newblock The {Care} {Work} of {Access}.
\newblock In \emph{Proceedings of the 2020 {CHI} {Conference} on {Human} {Factors} in {Computing} {Systems}}, pages 1--15, Honolulu HI USA, Apr. 2020. ACM.
\newblock ISBN 9781450367080.
\newblock \doi{10.1145/3313831.3376568}.
\newblock URL \url{https://dl.acm.org/doi/10.1145/3313831.3376568}.

\bibitem[Bentham(1970)]{bentham1970introduction}
J.~Bentham.
\newblock An introduction to the principles of morals and legislation (1789).
\newblock Continuum International Publishing Group Ltd, 1970.

\bibitem[Bentham and Mill(2004)]{bentham2004utilitarianism}
J.~Bentham and J.~S. Mill.
\newblock \emph{Utilitarianism and other essays}.
\newblock Penguin UK, 2004.

\bibitem[Benton et~al.(2012)Benton, Coles, and Coles]{benton2012temporal}
J.~Benton, A.~Coles, and A.~Coles.
\newblock Temporal planning with preferences and time-dependent continuous costs.
\newblock In \emph{Proceedings of the International Conference on Automated Planning and Scheduling}, volume~22, pages 2--10, 2012.

\bibitem[Berger(2011)]{berger2011arousal}
J.~Berger.
\newblock Arousal increases social transmission of information.
\newblock \emph{Psychological science}, 22\penalty0 (7):\penalty0 891--893, 2011.

\bibitem[Bergman et~al.(2024)Bergman, Marchal, Mellor, Mohamed, Gabriel, and Isaac]{bergman2024stela}
S.~Bergman, N.~Marchal, J.~Mellor, S.~Mohamed, I.~Gabriel, and W.~Isaac.
\newblock Stela: a community-centred approach to norm elicitation for ai alignment.
\newblock \emph{Scientific Reports}, 14\penalty0 (1):\penalty0 6616, 2024.

\bibitem[Berinsky(2017)]{berinsky_rumors_2017}
A.~J. Berinsky.
\newblock Rumors and {Health} {Care} {Reform}: {Experiments} in {Political} {Misinformation}.
\newblock \emph{British Journal of Political Science}, 47\penalty0 (2):\penalty0 241--262, Apr. 2017.
\newblock ISSN 0007-1234, 1469-2112.
\newblock \doi{10.1017/S0007123415000186}.
\newblock URL \url{https://www.cambridge.org/core/product/identifier/S0007123415000186/type/journal_article}.

\bibitem[Berne et~al.(2018)Berne, Morales, Langstaff, and Invalid]{berne_ten_2018}
P.~Berne, A.~L. Morales, D.~Langstaff, and S.~Invalid.
\newblock Ten {Principles} of {Disability} {Justice}.
\newblock \emph{WSQ: Women's Studies Quarterly}, 46\penalty0 (1-2):\penalty0 227--230, 2018.
\newblock ISSN 1934-1520.
\newblock \doi{10.1353/wsq.2018.0003}.
\newblock URL \url{https://muse.jhu.edu/article/690824}.

\bibitem[Besta et~al.(2023)Besta, Blach, Kubicek, Gerstenberger, Gianinazzi, Gajda, Lehmann, Podstawski, Niewiadomski, Nyczyk, and Hoefler]{besta2023graph}
M.~Besta, N.~Blach, A.~Kubicek, R.~Gerstenberger, L.~Gianinazzi, J.~Gajda, T.~Lehmann, M.~Podstawski, H.~Niewiadomski, P.~Nyczyk, and T.~Hoefler.
\newblock Graph of thoughts: Solving elaborate problems with large language models, 2023.

\bibitem[Bhargava and Velasquez(2021)]{bhargava2021ethics}
V.~R. Bhargava and M.~Velasquez.
\newblock Ethics of the attention economy: The problem of social media addiction.
\newblock \emph{Business Ethics Quarterly}, 31\penalty0 (3):\penalty0 321--359, 2021.

\bibitem[Bhaskar et~al.(2023)Bhaskar, Fabbri, and Durrett]{bhaskar_prompted_2023}
A.~Bhaskar, A.~R. Fabbri, and G.~Durrett.
\newblock Prompted {Opinion} {Summarization} with {GPT}-3.5, May 2023.
\newblock URL \url{http://arxiv.org/abs/2211.15914}.
\newblock arXiv:2211.15914 [cs].

\bibitem[Bianchi et~al.(2023)Bianchi, Curry, and Hovy]{bianchi_viewpoint:_2023}
F.~Bianchi, A.~C. Curry, and D.~Hovy.
\newblock Viewpoint: {Artificial} {Intelligence} {Accidents} {Waiting} to {Happen}?
\newblock \emph{Journal of Artificial Intelligence Research}, 76:\penalty0 193--199, Jan. 2023.
\newblock ISSN 1076-9757.
\newblock \doi{10.1613/jair.1.14263}.
\newblock URL \url{https://www.jair.org/index.php/jair/article/view/14263}.

\bibitem[Biesta(2009)]{biesta_good_2009}
G.~Biesta.
\newblock Good education in an age of measurement: on the need to reconnect with the question of purpose in education.
\newblock \emph{Educational Assessment, Evaluation and Accountability}, 21\penalty0 (1):\penalty0 33--46, Feb. 2009.
\newblock ISSN 1874-8597, 1874-8600.
\newblock \doi{10.1007/s11092-008-9064-9}.
\newblock URL \url{http://link.springer.com/10.1007/s11092-008-9064-9}.

\bibitem[Biju and Gayathri(2023)]{biju_self-breeding_2023}
P.~R. Biju and O.~Gayathri.
\newblock Self-breeding {Fake} {News}: {Bots} and {Artificial} {Intelligence} {Perpetuate} {Social} {Polarization} in {India}’s {Conflict} {Zones}.
\newblock \emph{The International Journal of Information, Diversity, \& Inclusion (IJIDI)}, 7\penalty0 (1/2), Apr. 2023.
\newblock ISSN 2574-3430.
\newblock \doi{10.33137/ijidi.v7i1/2.39409}.
\newblock URL \url{https://jps.library.utoronto.ca/index.php/ijidi/article/view/39409}.

\bibitem[Binz and Schulz(2023)]{binz_using_2023}
M.~Binz and E.~Schulz.
\newblock Using cognitive psychology to understand {GPT}-3.
\newblock \emph{Proceedings of the National Academy of Sciences}, 120\penalty0 (6):\penalty0 e2218523120, Feb. 2023.
\newblock ISSN 0027-8424, 1091-6490.
\newblock \doi{10.1073/pnas.2218523120}.
\newblock URL \url{https://pnas.org/doi/10.1073/pnas.2218523120}.

\bibitem[Birhane et~al.(2022)Birhane, Isaac, Prabhakaran, Díaz, Elish, Gabriel, and Mohamed]{birhane_power_2022}
A.~Birhane, W.~Isaac, V.~Prabhakaran, M.~Díaz, M.~C. Elish, I.~Gabriel, and S.~Mohamed.
\newblock Power to the {People}? {Opportunities} and {Challenges} for {Participatory} {AI}.
\newblock In \emph{Equity and {Access} in {Algorithms}, {Mechanisms}, and {Optimization}}, pages 1--8, Oct. 2022.
\newblock \doi{10.1145/3551624.3555290}.
\newblock URL \url{http://arxiv.org/abs/2209.07572}.
\newblock arXiv:2209.07572 [cs].

\bibitem[Birnbaum and Davison(2023)]{birnbaum_ai_2023}
E.~Birnbaum and L.~Davison.
\newblock {AI} {Is} {Making} {Politics} {Easier}, {Cheaper} and {More} {Dangerous}.
\newblock \emph{Bloomberg}, July 2023.
\newblock URL \url{https://www.bloomberg.com/news/features/2023-07-11/chatgpt-ai-boom-makes-political-dirty-tricks-easier-and-cheaper}.

\bibitem[Bivens and Haimson(2016)]{bivens_baking_2016}
R.~Bivens and O.~L. Haimson.
\newblock Baking {Gender} {Into} {Social} {Media} {Design}: {How} {Platforms} {Shape} {Categories} for {Users} and {Advertisers}.
\newblock \emph{Social Media + Society}, 2\penalty0 (4):\penalty0 205630511667248, Oct. 2016.
\newblock ISSN 2056-3051, 2056-3051.
\newblock \doi{10.1177/2056305116672486}.
\newblock URL \url{http://journals.sagepub.com/doi/10.1177/2056305116672486}.

\bibitem[Bjorndahl et~al.(2017)Bjorndahl, London, and Zollman]{bjorndahl_kantian_2017}
A.~Bjorndahl, A.~J. London, and K.~J.~S. Zollman.
\newblock Kantian {Decision} {Making} {Under} {Uncertainty}: {Dignity}, {Price}, and {Consistency}.
\newblock \emph{Philosopher's Imprint}, 17\penalty0 (7), Apr. 2017.
\newblock ISSN 1533-628X.
\newblock URL \url{http://hdl.handle.net/2027/spo.3521354.0017.007}.

\bibitem[Björgvinsson et~al.(2010)Björgvinsson, Ehn, and Hillgren]{bjorgvinsson_participatory_2010}
E.~Björgvinsson, P.~Ehn, and P.-A. Hillgren.
\newblock Participatory design and "democratizing innovation".
\newblock In \emph{Proceedings of the 11th {Biennial} {Participatory} {Design} {Conference}}, pages 41--50, Sydney Australia, Nov. 2010. ACM.
\newblock ISBN 9781450301312.
\newblock \doi{10.1145/1900441.1900448}.
\newblock URL \url{https://dl.acm.org/doi/10.1145/1900441.1900448}.

\bibitem[Bjørn et~al.(2023)Bjørn, Menendez-Blanco, and Borsotti]{bjorn_equity_2023}
P.~Bjørn, M.~Menendez-Blanco, and V.~Borsotti.
\newblock Equity \& {Inclusion}.
\newblock In P.~Bjørn, M.~Menendez-Blanco, and V.~Borsotti, editors, \emph{Diversity in {Computer} {Science}: {Design} {Artefacts} for {Equity} and {Inclusion}}, pages 77--96. Springer International Publishing, Cham, 2023.
\newblock ISBN 9783031133145.
\newblock \doi{10.1007/978-3-031-13314-5_7}.
\newblock URL \url{https://doi.org/10.1007/978-3-031-13314-5_7}.

\bibitem[Blandford(2019)]{blandford2019hci}
A.~Blandford.
\newblock {HCI} for health and wellbeing: Challenges and opportunities.
\newblock \emph{International Journal of Human-Computer Studies}, 131:\penalty0 41--51, 2019.

\bibitem[Block(2022)]{block_how_2022}
B.~Block.
\newblock How biased algorithms create barriers to housing.
\newblock \emph{ACLU Washington}, Feb. 2022.
\newblock URL \url{https://www.aclu-wa.org/story/how-biased-algorithms-create-barriers-housing}.

\bibitem[Blodgett(2021)]{blodgett_sociolinguistically_2021}
S.~L. Blodgett.
\newblock \emph{Sociolinguistically {Driven} {Approaches} for {Just} {Natural} {Language} {Processing}}.
\newblock Doctoral dissertation, University of Massachusetts, Amherst, Apr. 2021.
\newblock URL \url{https://scholarworks.umass.edu/dissertations_2/2092}.

\bibitem[Blodgett et~al.(2021)Blodgett, Lopez, Olteanu, Sim, and Wallach]{blodgett_stereotyping_2021}
S.~L. Blodgett, G.~Lopez, A.~Olteanu, R.~Sim, and H.~Wallach.
\newblock Stereotyping {Norwegian} {Salmon}: {An} {Inventory} of {Pitfalls} in {Fairness} {Benchmark} {Datasets}.
\newblock In \emph{Proceedings of the 59th {Annual} {Meeting} of the {Association} for {Computational} {Linguistics} and the 11th {International} {Joint} {Conference} on {Natural} {Language} {Processing} ({Volume} 1: {Long} {Papers})}, pages 1004--1015, Online, 2021. Association for Computational Linguistics.
\newblock \doi{10.18653/v1/2021.acl-long.81}.
\newblock URL \url{https://aclanthology.org/2021.acl-long.81}.

\bibitem[Bloom et~al.(2020)Bloom, Jones, Van~Reenen, and Webb]{bloom_are_2020}
N.~Bloom, C.~I. Jones, J.~Van~Reenen, and M.~Webb.
\newblock Are {Ideas} {Getting} {Harder} to {Find}?
\newblock \emph{American Economic Review}, 110\penalty0 (4):\penalty0 1104--1144, Apr. 2020.
\newblock ISSN 0002-8282.
\newblock \doi{10.1257/aer.20180338}.
\newblock URL \url{https://pubs.aeaweb.org/doi/10.1257/aer.20180338}.

\bibitem[Bloom(1996)]{bloom1996intention}
P.~Bloom.
\newblock Intention, history, and artifact concepts.
\newblock \emph{Cognition}, 60\penalty0 (1):\penalty0 1--29, 1996.

\bibitem[Bloom(2007)]{bloom2007more}
P.~Bloom.
\newblock More than words: A reply to {M}alt and {S}loman.
\newblock \emph{Cognition}, 105\penalty0 (3):\penalty0 649--655, 2007.

\bibitem[Blumenthal-Barby(2012)]{blumenthal-barby_between_2012}
J.~Blumenthal-Barby.
\newblock Between {Reason} and {Coercion}: {Ethically} {Permissible} {Influence} in {Health} {Care} and {Health} {Policy} {Contexts}.
\newblock \emph{Kennedy Institute of Ethics journal}, 22:\penalty0 345--66, Dec. 2012.
\newblock \doi{10.1353/ken.2012.0018}.

\bibitem[Blumenthal-Barby(2014)]{blumenthal-barby_framework_2014}
J.~Blumenthal-Barby.
\newblock A {Framework} for {Assessing} the {Moral} {Status} of {Manipulation},.
\newblock In C.~C.~M. Weber, editor, \emph{Manipulation}, pages 121--134. Oxford University Press, 2014.

\bibitem[Bogen(2019)]{bogen_all_2019}
M.~Bogen.
\newblock All the {Ways} {Hiring} {Algorithms} {Can} {Introduce} {Bias}.
\newblock \emph{Harvard Business Review}, May 2019.
\newblock ISSN 0017-8012.
\newblock URL \url{https://hbr.org/2019/05/all-the-ways-hiring-algorithms-can-introduce-bias}.

\bibitem[Bogomolov et~al.(2013)Bogomolov, Lepri, and Pianesi]{bogomolov2013happiness}
A.~Bogomolov, B.~Lepri, and F.~Pianesi.
\newblock Happiness recognition from mobile phone data.
\newblock In \emph{2013 International Conference on Social Computing}, pages 790--795. IEEE, 2013.

\bibitem[Boiko et~al.(2023)Boiko, MacKnight, and Gomes]{boiko_emergent_2023}
D.~A. Boiko, R.~MacKnight, and G.~Gomes.
\newblock Emergent autonomous scientific research capabilities of large language models, Apr. 2023.
\newblock URL \url{http://arxiv.org/abs/2304.05332}.
\newblock arXiv:2304.05332 [physics].

\bibitem[Bommasani et~al.(2022{\natexlab{a}})Bommasani, Creel, Kumar, Jurafsky, and Liang]{bommasani_picking_2022}
R.~Bommasani, K.~A. Creel, A.~Kumar, D.~Jurafsky, and P.~S. Liang.
\newblock Picking on the {Same} {Person}: {Does} {Algorithmic} {Monoculture} lead to {Outcome} {Homogenization}?
\newblock \emph{Advances in Neural Information Processing Systems}, 35:\penalty0 3663--3678, Dec. 2022{\natexlab{a}}.
\newblock URL \url{https://proceedings.neurips.cc/paper_files/paper/2022/hash/17a234c91f746d9625a75cf8a8731ee2-Abstract-Conference.html}.

\bibitem[Bommasani et~al.(2022{\natexlab{b}})Bommasani, Hudson, Adeli, Altman, Arora, von Arx, Bernstein, Bohg, Bosselut, Brunskill, Brynjolfsson, Buch, Card, Castellon, Chatterji, Chen, Creel, Davis, Demszky, Donahue, Doumbouya, Durmus, Ermon, Etchemendy, Ethayarajh, Fei-Fei, Finn, Gale, Gillespie, Goel, Goodman, Grossman, Guha, Hashimoto, Henderson, Hewitt, Ho, Hong, Hsu, Huang, Icard, Jain, Jurafsky, Kalluri, Karamcheti, Keeling, Khani, Khattab, Koh, Krass, Krishna, Kuditipudi, Kumar, Ladhak, Lee, Lee, Leskovec, Levent, Li, Li, Ma, Malik, Manning, Mirchandani, Mitchell, Munyikwa, Nair, Narayan, Narayanan, Newman, Nie, Niebles, Nilforoshan, Nyarko, Ogut, Orr, Papadimitriou, Park, Piech, Portelance, Potts, Raghunathan, Reich, Ren, Rong, Roohani, Ruiz, Ryan, Ré, Sadigh, Sagawa, Santhanam, Shih, Srinivasan, Tamkin, Taori, Thomas, Tramèr, Wang, Wang, Wu, Wu, Wu, Xie, Yasunaga, You, Zaharia, Zhang, Zhang, Zhang, Zhang, Zheng, Zhou, and Liang]{bommasani_opportunities_2022}
R.~Bommasani, D.~A. Hudson, E.~Adeli, R.~Altman, S.~Arora, S.~von Arx, M.~S. Bernstein, J.~Bohg, A.~Bosselut, E.~Brunskill, E.~Brynjolfsson, S.~Buch, D.~Card, R.~Castellon, N.~Chatterji, A.~Chen, K.~Creel, J.~Q. Davis, D.~Demszky, C.~Donahue, M.~Doumbouya, E.~Durmus, S.~Ermon, J.~Etchemendy, K.~Ethayarajh, L.~Fei-Fei, C.~Finn, T.~Gale, L.~Gillespie, K.~Goel, N.~Goodman, S.~Grossman, N.~Guha, T.~Hashimoto, P.~Henderson, J.~Hewitt, D.~E. Ho, J.~Hong, K.~Hsu, J.~Huang, T.~Icard, S.~Jain, D.~Jurafsky, P.~Kalluri, S.~Karamcheti, G.~Keeling, F.~Khani, O.~Khattab, P.~W. Koh, M.~Krass, R.~Krishna, R.~Kuditipudi, A.~Kumar, F.~Ladhak, M.~Lee, T.~Lee, J.~Leskovec, I.~Levent, X.~L. Li, X.~Li, T.~Ma, A.~Malik, C.~D. Manning, S.~Mirchandani, E.~Mitchell, Z.~Munyikwa, S.~Nair, A.~Narayan, D.~Narayanan, B.~Newman, A.~Nie, J.~C. Niebles, H.~Nilforoshan, J.~Nyarko, G.~Ogut, L.~Orr, I.~Papadimitriou, J.~S. Park, C.~Piech, E.~Portelance, C.~Potts, A.~Raghunathan, R.~Reich, H.~Ren, F.~Rong, Y.~Roohani, C.~Ruiz, J.~Ryan, C.~Ré,
  D.~Sadigh, S.~Sagawa, K.~Santhanam, A.~Shih, K.~Srinivasan, A.~Tamkin, R.~Taori, A.~W. Thomas, F.~Tramèr, R.~E. Wang, W.~Wang, B.~Wu, J.~Wu, Y.~Wu, S.~M. Xie, M.~Yasunaga, J.~You, M.~Zaharia, M.~Zhang, T.~Zhang, X.~Zhang, Y.~Zhang, L.~Zheng, K.~Zhou, and P.~Liang.
\newblock On the {Opportunities} and {Risks} of {Foundation} {Models}, July 2022{\natexlab{b}}.
\newblock URL \url{http://arxiv.org/abs/2108.07258}.
\newblock arXiv:2108.07258 [cs].

\bibitem[Bond(2023)]{bond_people_2023}
S.~Bond.
\newblock People are trying to claim real videos are deepfakes. {The} courts are not amused.
\newblock \emph{NPR}, May 2023.
\newblock URL \url{https://www.npr.org/2023/05/08/1174132413/people-are-trying-to-claim-real-videos-are-deepfakes-the-courts-are-not-amused}.

\bibitem[Bonilla-Silva(1997)]{bonilla-silva_rethinking_1997}
E.~Bonilla-Silva.
\newblock Rethinking {Racism}: {Toward} a {Structural} {Interpretation}.
\newblock \emph{American Sociological Review}, 62\penalty0 (3):\penalty0 465, June 1997.
\newblock ISSN 00031224.
\newblock \doi{10.2307/2657316}.
\newblock URL \url{http://www.jstor.org/stable/2657316?origin=crossref}.

\bibitem[Bos et~al.(2003)Bos, Klein, Lemon, and Oka]{bos_dipper:_2003}
J.~Bos, E.~Klein, O.~Lemon, and T.~Oka.
\newblock {DIPPER}: {Description} and {Formalisation} of an {Information}-{State} {Update} {Dialogue} {System} {Architecture}.
\newblock In \emph{Proceedings of the {Fourth} {SIGdial} {Workshop} of {Discourse} and {Dialogue}}, pages 115--124, 2003.
\newblock URL \url{https://aclanthology.org/W03-2123.pdf}.

\bibitem[Bostrom(2014)]{nick2014superintelligence}
N.~Bostrom.
\newblock \emph{Superintelligence: Paths, dangers, strategies}.
\newblock Oxford University Press, Oxford, 2014.

\bibitem[Boureau and Weston(2017)]{boureau_learning_2017}
Y.-L. Boureau and J.~Weston.
\newblock Learning {End}-to-{End} {Goal}-{Oriented} {Dialog}.
\newblock Apr. 2017.
\newblock URL \url{https://research.facebook.com/publications/learning-end-to-end-goal-oriented-dialog/}.

\bibitem[Bowker and Star(2000)]{bowker_sorting_2000}
G.~C. Bowker and S.~L. Star.
\newblock \emph{Sorting {Things} {Out}: {Classification} and {Its} {Consequences}}.
\newblock MIT Press, Aug. 2000.
\newblock ISBN 9780262261609.
\newblock Google-Books-ID: xHlP8WqzizYC.

\bibitem[Bowman et~al.(2022)Bowman, Hyun, Perez, Chen, Pettit, Heiner, Lukošiūtė, Askell, Jones, Chen, Goldie, Mirhoseini, McKinnon, Olah, Amodei, Amodei, Drain, Li, Tran-Johnson, Kernion, Kerr, Mueller, Ladish, Landau, Ndousse, Lovitt, Elhage, Schiefer, Joseph, Mercado, DasSarma, Larson, McCandlish, Kundu, Johnston, Kravec, Showk, Fort, Telleen-Lawton, Brown, Henighan, Hume, Bai, Hatfield-Dodds, Mann, and Kaplan]{bowman_measuring_2022}
S.~R. Bowman, J.~Hyun, E.~Perez, E.~Chen, C.~Pettit, S.~Heiner, K.~Lukošiūtė, A.~Askell, A.~Jones, A.~Chen, A.~Goldie, A.~Mirhoseini, C.~McKinnon, C.~Olah, D.~Amodei, D.~Amodei, D.~Drain, D.~Li, E.~Tran-Johnson, J.~Kernion, J.~Kerr, J.~Mueller, J.~Ladish, J.~Landau, K.~Ndousse, L.~Lovitt, N.~Elhage, N.~Schiefer, N.~Joseph, N.~Mercado, N.~DasSarma, R.~Larson, S.~McCandlish, S.~Kundu, S.~Johnston, S.~Kravec, S.~E. Showk, S.~Fort, T.~Telleen-Lawton, T.~Brown, T.~Henighan, T.~Hume, Y.~Bai, Z.~Hatfield-Dodds, B.~Mann, and J.~Kaplan.
\newblock Measuring {Progress} on {Scalable} {Oversight} for {Large} {Language} {Models}, Nov. 2022.
\newblock URL \url{http://arxiv.org/abs/2211.03540}.
\newblock arXiv:2211.03540 [cs].

\bibitem[Boyer(1996)]{boyer_what_1996}
P.~Boyer.
\newblock What {Makes} {Anthropomorphism} {Natural}: {Intuitive} {Ontology} and {Cultural} {Representations}.
\newblock \emph{The Journal of the Royal Anthropological Institute}, 2\penalty0 (1):\penalty0 83, Mar. 1996.
\newblock ISSN 13590987.
\newblock \doi{10.2307/3034634}.
\newblock URL \url{https://www.jstor.org/stable/3034634?origin=crossref}.

\bibitem[Bracken-Roche et~al.(2017)Bracken-Roche, Bell, Macdonald, and Racine]{bracken-roche_concept_2017}
D.~Bracken-Roche, E.~Bell, M.~E. Macdonald, and E.~Racine.
\newblock The concept of ‘vulnerability’ in research ethics: an in-depth analysis of policies and guidelines.
\newblock \emph{Health Research Policy and Systems}, 15\penalty0 (1):\penalty0 8, Feb. 2017.
\newblock ISSN 1478-4505.
\newblock \doi{10.1186/s12961-016-0164-6}.
\newblock URL \url{https://doi.org/10.1186/s12961-016-0164-6}.

\bibitem[Bradshaw and Richardson(2009)]{bradshaw2009index}
J.~Bradshaw and D.~Richardson.
\newblock An index of child well-being in europe.
\newblock \emph{Child Indicators Research}, 2:\penalty0 319--351, 2009.

\bibitem[Bradshaw and Howard(2019)]{bradshaw_global_2019}
S.~Bradshaw and P.~N. Howard.
\newblock The {Global} {Disinformation} {Order}: 2019 {Global} {Inventory} of {Organised} {Social} {Media} {Manipulation}, 2019.
\newblock URL \url{https://demtech.oii.ox.ac.uk/research/posts/the-global-disinformation-order-2019-global-inventory-of-organised-social-media-manipulation/}.

\bibitem[Bran et~al.(2023)Bran, Cox, Schilter, Baldassari, White, and Schwaller]{bran_chemcrow:_2023}
A.~M. Bran, S.~Cox, O.~Schilter, C.~Baldassari, A.~D. White, and P.~Schwaller.
\newblock {ChemCrow}: {Augmenting} large-language models with chemistry tools, Oct. 2023.
\newblock URL \url{http://arxiv.org/abs/2304.05376}.
\newblock arXiv:2304.05376 [physics, stat].

\bibitem[Brandtzaeg et~al.(2022)Brandtzaeg, Skjuve, and Følstad]{brandtzaeg_my_2022}
P.~B. Brandtzaeg, M.~Skjuve, and A.~Følstad.
\newblock My {AI} {Friend}: {How} {Users} of a {Social} {Chatbot} {Understand} {Their} {Human}–{AI} {Friendship}.
\newblock \emph{Human Communication Research}, 48\penalty0 (3):\penalty0 404--429, June 2022.
\newblock ISSN 0360-3989, 1468-2958.
\newblock \doi{10.1093/hcr/hqac008}.
\newblock URL \url{https://academic.oup.com/hcr/article/48/3/404/6572120}.

\bibitem[Bratman(1987)]{Bratman1987-BRAIPA}
M.~Bratman.
\newblock \emph{Intention, Plans, and Practical Reason}.
\newblock Cambridge, MA: Harvard University Press, Cambridge, 1987.

\bibitem[Brayne(2021)]{brayne_predict_2021}
S.~Brayne.
\newblock \emph{Predict and surveil: data, discretion, and the future of policing}.
\newblock Oxford University Press, New York, NY, 2021.
\newblock ISBN 9780190684099.

\bibitem[Breazeal(2003)]{breazeal_toward_2003}
C.~Breazeal.
\newblock Toward sociable robots.
\newblock \emph{Robotics and Autonomous Systems}, 42\penalty0 (3-4):\penalty0 167--175, Mar. 2003.
\newblock ISSN 09218890.
\newblock \doi{10.1016/S0921-8890(02)00373-1}.
\newblock URL \url{https://linkinghub.elsevier.com/retrieve/pii/S0921889002003731}.

\bibitem[Brecher et~al.(2013)Brecher, Müller, Kuz, and Lohse]{brecher_towards_2013}
C.~Brecher, S.~Müller, S.~Kuz, and W.~Lohse.
\newblock Towards {Anthropomorphic} {Movements} for {Industrial} {Robots}.
\newblock In V.~G. Duffy, editor, \emph{Digital {Human} {Modeling} and {Applications} in {Health}, {Safety}, {Ergonomics}, and {Risk} {Management}. {Human} {Body} {Modeling} and {Ergonomics}}, Lecture {Notes} in {Computer} {Science}, pages 10--19, Berlin, Heidelberg, 2013. Springer.
\newblock ISBN 9783642391828.
\newblock \doi{10.1007/978-3-642-39182-8_2}.

\bibitem[Breitfeller et~al.(2019)Breitfeller, Ahn, Jurgens, and Tsvetkov]{breitfeller_finding_2019}
L.~Breitfeller, E.~Ahn, D.~Jurgens, and Y.~Tsvetkov.
\newblock Finding {Microaggressions} in the {Wild}: {A} {Case} for {Locating} {Elusive} {Phenomena} in {Social} {Media} {Posts}.
\newblock In K.~Inui, J.~Jiang, V.~Ng, and X.~Wan, editors, \emph{Proceedings of the 2019 {Conference} on {Empirical} {Methods} in {Natural} {Language} {Processing} and the 9th {International} {Joint} {Conference} on {Natural} {Language} {Processing} ({EMNLP}-{IJCNLP})}, pages 1664--1674, Hong Kong, China, Nov. 2019. Association for Computational Linguistics.
\newblock \doi{10.18653/v1/D19-1176}.
\newblock URL \url{https://aclanthology.org/D19-1176.pdf}.

\bibitem[Brewster et~al.(2023)Brewster, Arvanitis, and Sadeghi]{brewster_could_2023}
J.~Brewster, L.~Arvanitis, and M.~Sadeghi.
\newblock Could {ChatGPT} {Become} {A} {Monster} {Misinformation} {Superspreader}?, Jan. 2023.
\newblock URL \url{https://www.newsguardtech.com/misinformation-monitor/jan-2023}.

\bibitem[Bricken et~al.(2023)Bricken, Templeton, Batson, Chen, Jermyn, Conerly, Turner, Anil, Denison, Askell, Lasenby, Wu, Kravec, Schiefer, Maxwell, Joseph, Hatfield-Dodds, Tamkin, Nguyen, McLean, Burke, Hume, Carter, Henighan, and Olah]{bricken_monosemanticity_2023}
T.~Bricken, A.~Templeton, J.~Batson, B.~Chen, A.~Jermyn, T.~Conerly, N.~Turner, C.~Anil, C.~Denison, A.~Askell, R.~Lasenby, Y.~Wu, S.~Kravec, N.~Schiefer, T.~Maxwell, N.~Joseph, Z.~Hatfield-Dodds, A.~Tamkin, K.~Nguyen, B.~McLean, J.~E. Burke, T.~Hume, S.~Carter, T.~Henighan, and C.~Olah.
\newblock Towards monosemanticity: Decomposing language models with dictionary learning.
\newblock \emph{Transformer Circuits Thread}, 2023.
\newblock https://transformer-circuits.pub/2023/monosemantic-features/index.html.

\bibitem[Brockman et~al.(2023)Brockman, Eleti, Georges, Jang, Kilpatrick, Lim, Miller, and Pokrass]{brockman_introducing_2023}
G.~Brockman, A.~Eleti, E.~Georges, J.~Jang, L.~Kilpatrick, R.~Lim, L.~Miller, and M.~Pokrass.
\newblock Introducing {ChatGPT} and {Whisper} {APIs}, Mar. 2023.
\newblock URL \url{https://openai.com/blog/introducing-chatgpt-and-whisper-apis}.

\bibitem[Brooks(2023)]{brooks_i_2023}
R.~Brooks.
\newblock I tried the {Replika} {AI} companion and can see why users are falling hard. {The} app raises serious ethical questions, Feb. 2023.
\newblock URL \url{http://theconversation.com/i-tried-the-replika-ai-companion-and-can-see-why-users-are-falling-hard-the-app-raises-serious-ethical-questions-200257}.

\bibitem[Broussard(2023)]{broussard_more_2023}
M.~Broussard.
\newblock \emph{More than a glitch: confronting race, gender, and ability bias in tech}.
\newblock The MIT Press, Cambridge, Massachusetts, 2023.
\newblock ISBN 9780262373050 9780262373067.

\bibitem[Brown et~al.(1990)Brown, Cocke, Della~Pietra, Della~Pietra, Jelinek, Lafferty, Mercer, and Roossin]{brown_statistical_1990}
P.~F. Brown, J.~Cocke, S.~A. Della~Pietra, V.~J. Della~Pietra, F.~Jelinek, J.~D. Lafferty, R.~L. Mercer, and P.~S. Roossin.
\newblock A {Statistical} {Approach} to {Machine} {Translation}.
\newblock \emph{Computational Linguistics}, 16\penalty0 (2):\penalty0 79--85, 1990.
\newblock URL \url{https://aclanthology.org/J90-2002.pdf}.

\bibitem[Brown et~al.(2020)Brown, Mann, Ryder, Subbiah, Kaplan, Dhariwal, Neelakantan, Shyam, Sastry, Askell, Agarwal, Herbert-Voss, Krueger, Henighan, Child, Ramesh, Ziegler, Wu, Winter, Hesse, Chen, Sigler, Litwin, Gray, Chess, Clark, Berner, McCandlish, Radford, Sutskever, and Amodei]{brown_language_2020}
T.~B. Brown, B.~Mann, N.~Ryder, M.~Subbiah, J.~Kaplan, P.~Dhariwal, A.~Neelakantan, P.~Shyam, G.~Sastry, A.~Askell, S.~Agarwal, A.~Herbert-Voss, G.~Krueger, T.~Henighan, R.~Child, A.~Ramesh, D.~M. Ziegler, J.~Wu, C.~Winter, C.~Hesse, M.~Chen, E.~Sigler, M.~Litwin, S.~Gray, B.~Chess, J.~Clark, C.~Berner, S.~McCandlish, A.~Radford, I.~Sutskever, and D.~Amodei.
\newblock Language {Models} are {Few}-{Shot} {Learners}, July 2020.
\newblock URL \url{http://arxiv.org/abs/2005.14165}.
\newblock arXiv:2005.14165 [cs].

\bibitem[Browne(2023)]{browne_all_2023}
R.~Browne.
\newblock All you need to know about {ChatGPT}, the {A}.{I}. chatbot that's got the world talking and tech giants clashing, Feb. 2023.
\newblock URL \url{https://www.cnbc.com/2023/02/08/what-is-chatgpt-viral-ai-chatbot-at-heart-of-microsoft-google-fight.html}.

\bibitem[Brundage et~al.(2018)Brundage, Avin, Clark, Toner, Eckersley, Garfinkel, Dafoe, Scharre, Zeitzoff, Filar, Anderson, Roff, Allen, Steinhardt, Flynn, hÉigeartaigh, Beard, Belfield, Farquhar, Lyle, Crootof, Evans, Page, Bryson, Yampolskiy, and Amodei]{brundage_malicious_2018}
M.~Brundage, S.~Avin, J.~Clark, H.~Toner, P.~Eckersley, B.~Garfinkel, A.~Dafoe, P.~Scharre, T.~Zeitzoff, B.~Filar, H.~Anderson, H.~Roff, G.~C. Allen, J.~Steinhardt, C.~Flynn, S.~{\'{O}}. hÉigeartaigh, S.~Beard, H.~Belfield, S.~Farquhar, C.~Lyle, R.~Crootof, O.~Evans, M.~Page, J.~Bryson, R.~Yampolskiy, and D.~Amodei.
\newblock The {Malicious} {Use} of {Artificial} {Intelligence}: {Forecasting}, {Prevention}, and {Mitigation}, Feb. 2018.
\newblock URL \url{http://arxiv.org/abs/1802.07228}.
\newblock arXiv:1802.07228 [cs].

\bibitem[Brundage et~al.(2020)Brundage, Avin, Wang, Belfield, Krueger, Hadfield, Khlaaf, Yang, Toner, Fong, Maharaj, Koh, Hooker, Leung, Trask, Bluemke, Lebensold, O'Keefe, Koren, Ryffel, Rubinovitz, Besiroglu, Carugati, Clark, Eckersley, de~Haas, Johnson, Laurie, Ingerman, Krawczuk, Askell, Cammarota, Lohn, Krueger, Stix, Henderson, Graham, Prunkl, Martin, Seger, Zilberman, hÉigeartaigh, Kroeger, Sastry, Kagan, Weller, Tse, Barnes, Dafoe, Scharre, Herbert-Voss, Rasser, Sodhani, Flynn, Gilbert, Dyer, Khan, Bengio, and Anderljung]{brundage_toward_2020}
M.~Brundage, S.~Avin, J.~Wang, H.~Belfield, G.~Krueger, G.~Hadfield, H.~Khlaaf, J.~Yang, H.~Toner, R.~Fong, T.~Maharaj, P.~W. Koh, S.~Hooker, J.~Leung, A.~Trask, E.~Bluemke, J.~Lebensold, C.~O'Keefe, M.~Koren, T.~Ryffel, J.~B. Rubinovitz, T.~Besiroglu, F.~Carugati, J.~Clark, P.~Eckersley, S.~de~Haas, M.~Johnson, B.~Laurie, A.~Ingerman, I.~Krawczuk, A.~Askell, R.~Cammarota, A.~Lohn, D.~Krueger, C.~Stix, P.~Henderson, L.~Graham, C.~Prunkl, B.~Martin, E.~Seger, N.~Zilberman, S.~{\'{O}}. hÉigeartaigh, F.~Kroeger, G.~Sastry, R.~Kagan, A.~Weller, B.~Tse, E.~Barnes, A.~Dafoe, P.~Scharre, A.~Herbert-Voss, M.~Rasser, S.~Sodhani, C.~Flynn, T.~K. Gilbert, L.~Dyer, S.~Khan, Y.~Bengio, and M.~Anderljung.
\newblock Toward {Trustworthy} {AI} {Development}: {Mechanisms} for {Supporting} {Verifiable} {Claims}, Apr. 2020.
\newblock URL \url{http://arxiv.org/abs/2004.07213}.
\newblock arXiv:2004.07213 [cs].

\bibitem[Bryce(2018)]{bryce_finding_2018}
A.~Bryce.
\newblock Finding meaning through work: eudaimonic well-being and job type in the {US} and {UK}.
\newblock Working {Papers} 2018004, The University of Sheffield, Department of Economics, 2018.
\newblock URL \url{https://EconPapers.repec.org/RePEc:shf:wpaper:2018004}.

\bibitem[Brynjolfsson(2022)]{brynjolfsson2022turing}
E.~Brynjolfsson.
\newblock The {Turing} {Trap}: The promise and peril of human-like artificial intelligence.
\newblock \emph{Daedalus}, 151\penalty0 (2):\penalty0 272--287, 2022.

\bibitem[Brynjolfsson et~al.(2018)Brynjolfsson, Rock, and Syverson]{brynjolfsson_productivity_2018}
E.~Brynjolfsson, D.~Rock, and C.~Syverson.
\newblock The {Productivity} {J}-{Curve}: {How} {Intangibles} {Complement} {General} {Purpose} {Technologies}, Oct. 2018.
\newblock URL \url{https://www.nber.org/papers/w25148}.

\bibitem[Brynjolfsson et~al.(2023)Brynjolfsson, Li, and Raymond]{brynjolfsson_generative_2023}
E.~Brynjolfsson, D.~Li, and L.~R. Raymond.
\newblock Generative {AI} at {Work}, Apr. 2023.
\newblock URL \url{https://www.nber.org/papers/w31161}.

\bibitem[BSA(2016)]{bsa_1_2016}
BSA.
\newblock The \$1 {Trillion} {Economic} {Impact} of {Software}.
\newblock Technical report, BSA, June 2016.
\newblock URL \url{https://docs.broadcom.com/doc/economic-impact-of-software-report}.

\bibitem[Bubeck et~al.(2023)Bubeck, Chandrasekaran, Eldan, Gehrke, Horvitz, Kamar, Lee, Lee, Li, Lundberg, Nori, Palangi, Ribeiro, and Zhang]{bubeck_sparks_2023}
S.~Bubeck, V.~Chandrasekaran, R.~Eldan, J.~Gehrke, E.~Horvitz, E.~Kamar, P.~Lee, Y.~T. Lee, Y.~Li, S.~Lundberg, H.~Nori, H.~Palangi, M.~T. Ribeiro, and Y.~Zhang.
\newblock Sparks of {Artificial} {General} {Intelligence}: {Early} experiments with {GPT}-4, Apr. 2023.
\newblock URL \url{http://arxiv.org/abs/2303.12712}.
\newblock arXiv:2303.12712 [cs].

\bibitem[Budzianowski et~al.(2020)Budzianowski, Wen, Tseng, Casanueva, Ultes, Ramadan, and Gašić]{budzianowski_multiwoz_2020}
P.~Budzianowski, T.-H. Wen, B.-H. Tseng, I.~Casanueva, S.~Ultes, O.~Ramadan, and M.~Gašić.
\newblock {MultiWOZ} – {A} {Large}-{Scale} {Multi}-{Domain} {Wizard}-of-{Oz} {Dataset} for {Task}-{Oriented} {Dialogue} {Modelling}, Apr. 2020.
\newblock URL \url{https://arxiv.org/pdf/1810.00278.pdf}.
\newblock arXiv:1810.00278 [cs].

\bibitem[Bulkley and Van~Alstyne(2008)]{bulkley_information_2008}
N.~Bulkley and M.~Van~Alstyne.
\newblock Information, {Communications} \& {Output}: {Does} {E}-mail {Make} {White}-{Collar} {Workers} {More} {Productive}?
\newblock Jan. 2008.

\bibitem[Buolamwini and Gebru(2018)]{buolamwini_gender_2018}
J.~Buolamwini and T.~Gebru.
\newblock Gender {Shades}: {Intersectional} {Accuracy} {Disparities} in {Commercial} {Gender} {Classification}.
\newblock In S.~A. Friedler and C.~Wilson, editors, \emph{Proceedings of the 1st {Conference} on {Fairness}, {Accountability} and {Transparency}}, volume~81 of \emph{Proceedings of {Machine} {Learning} {Research}}, pages 77--91. PMLR, Feb. 2018.
\newblock URL \url{https://proceedings.mlr.press/v81/buolamwini18a.html}.

\bibitem[Burgess et~al.(2020)Burgess, Cappelen, and Plunkett]{burgess2020conceptual}
A.~Burgess, H.~Cappelen, and D.~Plunkett.
\newblock \emph{Conceptual engineering and conceptual ethics}.
\newblock Oxford University Press, 2020.

\bibitem[Burkhardt(2017)]{burkhardt_chapter_2017}
J.~M. Burkhardt.
\newblock Chapter 1. {History} of {Fake} {News}.
\newblock \emph{Library Technology Reports}, 53\penalty0 (8):\penalty0 5--9, Nov. 2017.
\newblock ISSN 0024-2586.
\newblock URL \url{https://journals.ala.org/index.php/ltr/article/view/6497}.

\bibitem[Burki(2019)]{burki_vaccine_2019}
T.~Burki.
\newblock Vaccine misinformation and social media.
\newblock \emph{The Lancet Digital Health}, 1\penalty0 (6):\penalty0 e258--e259, Oct. 2019.
\newblock ISSN 25897500.
\newblock \doi{10.1016/S2589-7500(19)30136-0}.
\newblock URL \url{https://linkinghub.elsevier.com/retrieve/pii/S2589750019301360}.

\bibitem[Burnell et~al.(2023)Burnell, Schellaert, Burden, Ullman, Martinez-Plumed, Tenenbaum, Rutar, Cheke, Sohl-Dickstein, Mitchell, Kiela, Shanahan, Voorhees, Cohn, Leibo, and Hernandez-Orallo]{burnell_rethink_2023}
R.~Burnell, W.~Schellaert, J.~Burden, T.~D. Ullman, F.~Martinez-Plumed, J.~B. Tenenbaum, D.~Rutar, L.~G. Cheke, J.~Sohl-Dickstein, M.~Mitchell, D.~Kiela, M.~Shanahan, E.~M. Voorhees, A.~G. Cohn, J.~Z. Leibo, and J.~Hernandez-Orallo.
\newblock Rethink reporting of evaluation results in {AI}.
\newblock \emph{Science}, 380\penalty0 (6641):\penalty0 136--138, Apr. 2023.
\newblock ISSN 0036-8075, 1095-9203.
\newblock \doi{10.1126/science.adf6369}.
\newblock URL \url{https://www.science.org/doi/10.1126/science.adf6369}.

\bibitem[Burns et~al.(2022)Burns, Ye, Klein, and Steinhardt]{burns_discovering_2022}
C.~Burns, H.~Ye, D.~Klein, and J.~Steinhardt.
\newblock Discovering {Latent} {Knowledge} in {Language} {Models} {Without} {Supervision}, Dec. 2022.
\newblock URL \url{http://arxiv.org/abs/2212.03827}.
\newblock arXiv:2212.03827 [cs].

\bibitem[Burns et~al.(2023)Burns, Izmailov, Kirchner, Baker, Gao, Aschenbrenner, Chen, Ecoffet, Joglekar, and Wu]{burns_weak_2023}
C.~Burns, P.~Izmailov, J.~H. Kirchner, B.~Baker, L.~Gao, L.~Aschenbrenner, Y.~Chen, A.~Ecoffet, M.~Joglekar, and J.~L. I.~S. Wu.
\newblock {Weak-to-Strong} generalization: Eliciting strong capabilities with weak supervision.
\newblock 2023.

\bibitem[Burns(2021)]{burns_technology_2021}
M.~Burns.
\newblock Technology in education.
\newblock Background paper prepared for the 2023 {Global} {Education} {Monitoring} {Report} ED/GEMR/MRT/2023/T1/1, UNESCO, 2021.
\newblock URL \url{https://unesdoc.unesco.org/ark:/48223/pf0000378951/PDF/378951eng.pdf.multi}.

\bibitem[Burr et~al.(2018)Burr, Cristianini, and Ladyman]{burr_analysis_2018}
C.~Burr, N.~Cristianini, and J.~Ladyman.
\newblock An {Analysis} of the {Interaction} {Between} {Intelligent} {Software} {Agents} and {Human} {Users}.
\newblock \emph{Minds and Machines}, 28\penalty0 (4):\penalty0 735--774, Dec. 2018.
\newblock ISSN 1572-8641.
\newblock \doi{10.1007/s11023-018-9479-0}.
\newblock URL \url{https://doi.org/10.1007/s11023-018-9479-0}.

\bibitem[Burt(2023)]{burt_aadhaar_2023}
C.~Burt.
\newblock Aadhaar biometrics verification for {GST} registration pilot in {India} to expand {\textbar} {Biometric} {Update}, July 2023.
\newblock URL \url{https://www.biometricupdate.com/202307/aadhaar-biometrics-verification-for-gst-registration-pilot-in-india-to-expand}.

\bibitem[Burtell and Woodside(2023)]{burtell_artificial_2023}
M.~Burtell and T.~Woodside.
\newblock Artificial {Influence}: {An} {Analysis} {Of} {AI}-{Driven} {Persuasion}, Mar. 2023.
\newblock URL \url{http://arxiv.org/abs/2303.08721}.
\newblock arXiv:2303.08721 [cs].

\bibitem[Buss(2005)]{buss_valuing_2005}
S.~Buss.
\newblock Valuing {Autonomy} and {Respecting} {Persons}: {Manipulation}, {Seduction}, and the {Basis} of {Moral} {Constraints}.
\newblock \emph{Ethics}, 115\penalty0 (2):\penalty0 195--235, Jan. 2005.
\newblock ISSN 0014-1704, 1539-297X.
\newblock \doi{10.1086/426304}.
\newblock URL \url{https://www.journals.uchicago.edu/doi/10.1086/426304}.

\bibitem[Buçinca et~al.(2021)Buçinca, Malaya, and Gajos]{bucinca_trust_2021}
Z.~Buçinca, M.~B. Malaya, and K.~Z. Gajos.
\newblock To {Trust} or to {Think}: {Cognitive} {Forcing} {Functions} {Can} {Reduce} {Overreliance} on {AI} in {AI}-assisted {Decision}-making.
\newblock \emph{Proceedings of the ACM on Human-Computer Interaction}, 5\penalty0 (CSCW1):\penalty0 1--21, Apr. 2021.
\newblock ISSN 2573-0142.
\newblock \doi{10.1145/3449287}.
\newblock URL \url{http://arxiv.org/abs/2102.09692}.
\newblock arXiv:2102.09692 [cs].

\bibitem[Böttinger et~al.(2018)Böttinger, Godefroid, and Singh]{Bottinger_Godefroid_Singh_2018}
K.~Böttinger, P.~Godefroid, and R.~Singh.
\newblock Deep reinforcement fuzzing.
\newblock In \emph{2018 IEEE Security and Privacy Workshops (SPW)}, page 116–122, May 2018.
\newblock \doi{10.1109/SPW.2018.00026}.
\newblock URL \url{https://ieeexplore.ieee.org/document/8424642}.

\bibitem[Calvin et~al.(2023)Calvin, Dasgupta, Krinner, Mukherji, Thorne, Trisos, Romero, Aldunce, Barrett, Blanco, Cheung, Connors, Denton, Diongue-Niang, Dodman, Garschagen, Geden, Hayward, Jones, Jotzo, Krug, Lasco, Lee, Masson-Delmotte, Meinshausen, Mintenbeck, Mokssit, Otto, Pathak, Pirani, Poloczanska, Pörtner, Revi, Roberts, Roy, Ruane, Skea, Shukla, Slade, Slangen, Sokona, Sörensson, Tignor, van Vuuren, Wei, Winkler, Zhai, Zommers, Hourcade, Johnson, Pachauri, Simpson, Singh, Thomas, Totin, Arias, Bustamante, Elgizouli, Flato, Howden, Méndez-Vallejo, Pereira, Pichs-Madruga, Rose, Saheb, Sánchez~Rodríguez, Ürge Vorsatz, Xiao, Yassaa, Alegría, Armour, Bednar-Friedl, Blok, Cissé, Dentener, Eriksen, Fischer, Garner, Guivarch, Haasnoot, Hansen, Hauser, Hawkins, Hermans, Kopp, Leprince-Ringuet, Lewis, Ley, Ludden, Niamir, Nicholls, Some, Szopa, Trewin, van~der Wijst, Winter, Witting, Birt, Ha, Romero, Kim, Haites, Jung, Stavins, Birt, Ha, Orendain, Ignon, Park, Park, Reisinger, Cammaramo, Fischlin,
  Fuglestvedt, Hansen, Ludden, Masson-Delmotte, Matthews, Mintenbeck, Pirani, Poloczanska, Leprince-Ringuet, and Péan]{lee_climate_2023}
K.~Calvin, D.~Dasgupta, G.~Krinner, A.~Mukherji, P.~W. Thorne, C.~Trisos, J.~Romero, P.~Aldunce, K.~Barrett, G.~Blanco, W.~W. Cheung, S.~Connors, F.~Denton, A.~Diongue-Niang, D.~Dodman, M.~Garschagen, O.~Geden, B.~Hayward, C.~Jones, F.~Jotzo, T.~Krug, R.~Lasco, Y.-Y. Lee, V.~Masson-Delmotte, M.~Meinshausen, K.~Mintenbeck, A.~Mokssit, F.~E. Otto, M.~Pathak, A.~Pirani, E.~Poloczanska, H.-O. Pörtner, A.~Revi, D.~C. Roberts, J.~Roy, A.~C. Ruane, J.~Skea, P.~R. Shukla, R.~Slade, A.~Slangen, Y.~Sokona, A.~A. Sörensson, M.~Tignor, D.~van Vuuren, Y.-M. Wei, H.~Winkler, P.~Zhai, Z.~Zommers, J.-C. Hourcade, F.~X. Johnson, S.~Pachauri, N.~P. Simpson, C.~Singh, A.~Thomas, E.~Totin, P.~Arias, M.~Bustamante, I.~Elgizouli, G.~Flato, M.~Howden, C.~Méndez-Vallejo, J.~J. Pereira, R.~Pichs-Madruga, S.~K. Rose, Y.~Saheb, R.~Sánchez~Rodríguez, D.~Ürge Vorsatz, C.~Xiao, N.~Yassaa, A.~Alegría, K.~Armour, B.~Bednar-Friedl, K.~Blok, G.~Cissé, F.~Dentener, S.~Eriksen, E.~Fischer, G.~Garner, C.~Guivarch, M.~Haasnoot, G.~Hansen,
  M.~Hauser, E.~Hawkins, T.~Hermans, R.~Kopp, N.~Leprince-Ringuet, J.~Lewis, D.~Ley, C.~Ludden, L.~Niamir, Z.~Nicholls, S.~Some, S.~Szopa, B.~Trewin, K.-I. van~der Wijst, G.~Winter, M.~Witting, A.~Birt, M.~Ha, J.~Romero, J.~Kim, E.~F. Haites, Y.~Jung, R.~Stavins, A.~Birt, M.~Ha, D.~J.~A. Orendain, L.~Ignon, S.~Park, Y.~Park, A.~Reisinger, D.~Cammaramo, A.~Fischlin, J.~S. Fuglestvedt, G.~Hansen, C.~Ludden, V.~Masson-Delmotte, J.~R. Matthews, K.~Mintenbeck, A.~Pirani, E.~Poloczanska, N.~Leprince-Ringuet, and C.~Péan.
\newblock Climate {Change} 2023: {Synthesis} {Report}. {Contribution} of {Working} {Groups} {I}, {II} and {III} to the {Sixth} {Assessment} {Report} of the {Intergovernmental} {Panel} on {Climate} {Change}.
\newblock Technical report, Intergovernmental Panel on Climate Change (IPCC), July 2023.
\newblock URL \url{https://www.ipcc.ch/report/ar6/syr/}.

\bibitem[Calvo and Peters(2014)]{calvo2014positive}
R.~A. Calvo and D.~Peters.
\newblock \emph{Positive computing: {T}echnology for wellbeing and human potential}.
\newblock MIT press, 2014.

\bibitem[Canetti et~al.(2023)Canetti, Fiat, and Gonczarowski]{canetti_zero-knowledge_2023}
R.~Canetti, A.~Fiat, and Y.~A. Gonczarowski.
\newblock Zero-{Knowledge} {Mechanisms}, Feb. 2023.
\newblock URL \url{http://arxiv.org/abs/2302.05590}.
\newblock arXiv:2302.05590 [cs, econ].

\bibitem[Cao et~al.(2019)Cao, Zhao, and Hu]{cao_anthropomorphism_2019}
C.~Cao, L.~Zhao, and Y.~Hu.
\newblock Anthropomorphism of {Intelligent} {Personal} {Assistants} ({IPAs}): {Antecedents} and {Consequences}.
\newblock In \emph{{PACIS} 2019 {Proceedings}}, 2019.
\newblock URL \url{https://core.ac.uk/reader/326833380}.

\bibitem[Carlini et~al.(2021)Carlini, Tramer, Wallace, Jagielski, Herbert-Voss, Lee, Roberts, Brown, Song, Erlingsson, Oprea, and Raffel]{carlini_extracting_2021}
N.~Carlini, F.~Tramer, E.~Wallace, M.~Jagielski, A.~Herbert-Voss, K.~Lee, A.~Roberts, T.~Brown, D.~Song, U.~Erlingsson, A.~Oprea, and C.~Raffel.
\newblock Extracting {Training} {Data} from {Large} {Language} {Models}, June 2021.
\newblock URL \url{http://arxiv.org/abs/2012.07805}.
\newblock arXiv:2012.07805 [cs].

\bibitem[Carlini et~al.(2023{\natexlab{a}})Carlini, Ippolito, Jagielski, Lee, Tramer, and Zhang]{carlini_quantifying_2023}
N.~Carlini, D.~Ippolito, M.~Jagielski, K.~Lee, F.~Tramer, and C.~Zhang.
\newblock Quantifying {Memorization} {Across} {Neural} {Language} {Models}, Mar. 2023{\natexlab{a}}.
\newblock URL \url{http://arxiv.org/abs/2202.07646}.
\newblock arXiv:2202.07646 [cs].

\bibitem[Carlini et~al.(2023{\natexlab{b}})Carlini, Jagielski, Choquette-Choo, Paleka, Pearce, Anderson, Terzis, Thomas, and Tramèr]{carlini_poisoning_2023}
N.~Carlini, M.~Jagielski, C.~A. Choquette-Choo, D.~Paleka, W.~Pearce, H.~Anderson, A.~Terzis, K.~Thomas, and F.~Tramèr.
\newblock Poisoning {Web}-{Scale} {Training} {Datasets} is {Practical}, Feb. 2023{\natexlab{b}}.
\newblock URL \url{http://arxiv.org/abs/2302.10149}.
\newblock arXiv:2302.10149 [cs].

\bibitem[Carlson(2012)]{carlson_effective_2012}
C.~S. Carlson.
\newblock \emph{Effective {FMEAs}: {Achieving} {Safe}, {Reliable}, and {Economical} {Products} and {Processes} using {Failure} {Mode} and {Effects} {Analysis} {\textbar} {Wiley}}.
\newblock Wiley, 2012.
\newblock ISBN 978-1-118-31258-2.

\bibitem[Carman(2019)]{carman_they_2019}
A.~Carman.
\newblock They welcomed a robot into their family, now they’re mourning its death, June 2019.
\newblock URL \url{https://www.theverge.com/2019/6/19/18682780/jibo-death-server-update-social-robot-mourning}.

\bibitem[Carroll et~al.(2020)Carroll, Shah, Ho, Griffiths, Seshia, Abbeel, and Dragan]{carroll_utility_2020}
M.~Carroll, R.~Shah, M.~K. Ho, T.~L. Griffiths, S.~A. Seshia, P.~Abbeel, and A.~Dragan.
\newblock On the {Utility} of {Learning} about {Humans} for {Human}-{AI} {Coordination}, Jan. 2020.
\newblock URL \url{http://arxiv.org/abs/1910.05789}.
\newblock arXiv:1910.05789 [cs, stat].

\bibitem[Carroll et~al.(2022)Carroll, Dragan, Russell, and Hadfield-Menell]{carroll_estimating_2022}
M.~Carroll, A.~Dragan, S.~Russell, and D.~Hadfield-Menell.
\newblock Estimating and {Penalizing} {Induced} {Preference} {Shifts} in {Recommender} {Systems}, July 2022.
\newblock URL \url{http://arxiv.org/abs/2204.11966}.
\newblock arXiv:2204.11966 [cs].

\bibitem[Carroll et~al.(2023)Carroll, Chan, Ashton, and Krueger]{carroll_characterizing_2023}
M.~Carroll, A.~Chan, H.~Ashton, and D.~Krueger.
\newblock Characterizing {Manipulation} from {AI} {Systems}, Oct. 2023.
\newblock URL \url{http://arxiv.org/abs/2303.09387}.
\newblock arXiv:2303.09387 [cs].

\bibitem[Casper et~al.(2023)Casper, Davies, Shi, Gilbert, Scheurer, Rando, Freedman, Korbak, Lindner, Freire, et~al.]{casper2023open}
S.~Casper, X.~Davies, C.~Shi, T.~K. Gilbert, J.~Scheurer, J.~Rando, R.~Freedman, T.~Korbak, D.~Lindner, P.~Freire, et~al.
\newblock Open problems and fundamental limitations of reinforcement learning from human feedback.
\newblock \emph{arXiv preprint arXiv:2307.15217}, 2023.

\bibitem[Castells(2009)]{castells_rise_2009}
M.~Castells.
\newblock \emph{The {Rise} of the {Network} {Society}}.
\newblock Wiley, 1 edition, Aug. 2009.
\newblock ISBN 9781405196864 9781444319514.
\newblock \doi{10.1002/9781444319514}.
\newblock URL \url{https://onlinelibrary.wiley.com/doi/book/10.1002/9781444319514}.

\bibitem[Cechetti et~al.(2019)Cechetti, Bellei, Biduski, Rodriguez, Roman, and De~Marchi]{cechetti2019developing}
N.~P. Cechetti, E.~A. Bellei, D.~Biduski, J.~P.~M. Rodriguez, M.~K. Roman, and A.~C.~B. De~Marchi.
\newblock Developing and implementing a gamification method to improve user engagement: A case study with an m-health application for hypertension monitoring.
\newblock \emph{Telematics and Informatics}, 41:\penalty0 126--138, 2019.

\bibitem[{Center for Countering Digital Hate}(2023)]{center_for_countering_digital_hate_googles_2023}
{Center for Countering Digital Hate}.
\newblock Google’s new ‘{Bard}’ {AI} generates false and harmful narratives on 78 out of 100 topics., Apr. 2023.
\newblock URL \url{https://counterhate.com/research/misinformation-on-bard-google-ai-chat/}.

\bibitem[{Centers for Disease Control and Prevention}(2023)]{centers_for_disease_control_and_prevention_how_2023}
{Centers for Disease Control and Prevention}.
\newblock How {Does} {Social} {Connectedness} {Affect} {Health}?, May 2023.
\newblock URL \url{https://www.cdc.gov/emotional-wellbeing/social-connectedness/affect-health.htm}.

\bibitem[Centola(2013)]{centola_homophily_2013}
D.~M. Centola.
\newblock Homophily, networks, and critical mass: {Solving} the start-up problem in large group collective action.
\newblock \emph{Rationality and Society}, 25\penalty0 (1):\penalty0 3--40, Feb. 2013.
\newblock ISSN 1043-4631, 1461-7358.
\newblock \doi{10.1177/1043463112473734}.
\newblock URL \url{http://journals.sagepub.com/doi/10.1177/1043463112473734}.

\bibitem[{Centre for Data Ethics and Innovation}(2021)]{centre_for_data_ethics_and_innovation_roadmap_2021}
{Centre for Data Ethics and Innovation}.
\newblock The roadmap to an effective {AI} assurance ecosystem.
\newblock Technical report, UK Government, Dec. 2021.
\newblock URL \url{https://www.gov.uk/government/publications/the-roadmap-to-an-effective-ai-assurance-ecosystem}.

\bibitem[Chae et~al.(2022)Chae, Kim, Kim, Lim, Kim, and Kim]{chae_ingroup_2022}
J.~Chae, K.~Kim, Y.~Kim, G.~Lim, D.~Kim, and H.~Kim.
\newblock Ingroup favoritism overrides fairness when resources are limited.
\newblock \emph{Scientific Reports}, 12\penalty0 (1):\penalty0 4560, Mar. 2022.
\newblock ISSN 2045-2322.
\newblock \doi{10.1038/s41598-022-08460-1}.
\newblock URL \url{https://www.nature.com/articles/s41598-022-08460-1}.

\bibitem[Chakrabarty et~al.(2023)Chakrabarty, Padmakumar, Brahman, and Muresan]{chakrabarty_creativity_2023}
T.~Chakrabarty, V.~Padmakumar, F.~Brahman, and S.~Muresan.
\newblock Creativity {Support} in the {Age} of {Large} {Language} {Models}: {An} {Empirical} {Study} {Involving} {Emerging} {Writers}, Sept. 2023.
\newblock URL \url{http://arxiv.org/abs/2309.12570}.
\newblock arXiv:2309.12570 [cs].

\bibitem[Chalmers(2020)]{chalmers2020conceptual}
D.~J. Chalmers.
\newblock What is conceptual engineering and what should it be?
\newblock \emph{Inquiry}, pages 1--18, 2020.

\bibitem[Chalmers(2023)]{chalmers_could_2023}
D.~J. Chalmers.
\newblock Could a {Large} {Language} {Model} be {Conscious}?, Apr. 2023.
\newblock URL \url{http://arxiv.org/abs/2303.07103}.
\newblock arXiv:2303.07103 [cs].

\bibitem[Chan et~al.(2023{\natexlab{a}})Chan, Riché, and Clifton]{chan_towards_2023}
A.~Chan, M.~Riché, and J.~Clifton.
\newblock Towards the {Scalable} {Evaluation} of {Cooperativeness} in {Language} {Models}, Mar. 2023{\natexlab{a}}.
\newblock URL \url{http://arxiv.org/abs/2303.13360}.
\newblock arXiv:2303.13360 [cs].

\bibitem[Chan et~al.(2023{\natexlab{b}})Chan, Salganik, Markelius, Pang, Rajkumar, Krasheninnikov, Langosco, He, Duan, Carroll, Lin, Mayhew, Collins, Molamohammadi, Burden, Zhao, Rismani, Voudouris, Bhatt, Weller, Krueger, and Maharaj]{Chan_2023}
A.~Chan, R.~Salganik, A.~Markelius, C.~Pang, N.~Rajkumar, D.~Krasheninnikov, L.~Langosco, Z.~He, Y.~Duan, M.~Carroll, M.~Lin, A.~Mayhew, K.~Collins, M.~Molamohammadi, J.~Burden, W.~Zhao, S.~Rismani, K.~Voudouris, U.~Bhatt, A.~Weller, D.~Krueger, and T.~Maharaj.
\newblock Harms from increasingly agentic algorithmic systems.
\newblock In \emph{2023 ACM Conference on Fairness, Accountability, and Transparency}, FAccT ’23. ACM, June 2023{\natexlab{b}}.
\newblock \doi{10.1145/3593013.3594033}.
\newblock URL \url{http://dx.doi.org/10.1145/3593013.3594033}.

\bibitem[Chan(2012)]{chan_anthropomorphism_2012}
A.~A. Y.-H. Chan.
\newblock Anthropomorphism as a conservation tool.
\newblock \emph{Biodiversity and Conservation}, 21\penalty0 (7):\penalty0 1889--1892, June 2012.
\newblock ISSN 1572-9710.
\newblock \doi{10.1007/s10531-012-0274-6}.
\newblock URL \url{https://doi.org/10.1007/s10531-012-0274-6}.

\bibitem[Chan et~al.(2005)Chan, Andy~Kwan, and Daniel~Shek]{chan2005quality}
Y.~K. Chan, C.~C. Andy~Kwan, and T.~L. Daniel~Shek.
\newblock Quality of life in {H}ong {K}ong: The {CUHK} {H}ong {K}ong quality of life index.
\newblock \emph{Quality-of-Life Research in Chinese, Western and Global contexts}, pages 259--289, 2005.

\bibitem[Chancel et~al.(2022)Chancel, Piketty, Saez, and Zucman]{chancel_world_nodate}
L.~Chancel, T.~Piketty, E.~Saez, and G.~Zucman.
\newblock World {Inequality} {Report} 2022, 2022.
\newblock URL \url{//wir2022.wid.world/}.

\bibitem[Chang and Hsu(2016)]{chang2016understanding}
C.-M. Chang and M.-H. Hsu.
\newblock Understanding the determinants of users’ subjective well-being in social networking sites: An integration of social capital theory and social presence theory.
\newblock \emph{Behaviour \& Information Technology}, 35\penalty0 (9):\penalty0 720--729, 2016.

\bibitem[Chaudhri(2023)]{chaudhri_disappearing_2023}
I.~Chaudhri.
\newblock The disappearing computer – and a world where you can take {AI} everywhere, Apr. 2023.
\newblock URL \url{https://www.ted.com/talks/imran_chaudhri_the_disappearing_computer_and_a_world_where_you_can_take_ai_everywhere}.

\bibitem[{Check Point Research}(2023)]{check_point_research_opwnai:_2023}
{Check Point Research}.
\newblock {OPWNAI}: {Cybercriminals} {Starting} to {Use} {ChatGPT}, Jan. 2023.
\newblock URL \url{https://research.checkpoint.com/2023/opwnai-cybercriminals-starting-to-use-chatgpt/}.

\bibitem[Chen et~al.(2023{\natexlab{a}})Chen, Dong, Wang, Feng, Wang, and He]{chen2023bias}
J.~Chen, H.~Dong, X.~Wang, F.~Feng, M.~Wang, and X.~He.
\newblock Bias and debias in recommender system: A survey and future directions.
\newblock \emph{ACM Transactions on Information Systems}, 41\penalty0 (3):\penalty0 1--39, 2023{\natexlab{a}}.

\bibitem[Chen et~al.(2023{\natexlab{b}})Chen, Liao, Wortman~Vaughan, and Bansal]{chen2023understanding}
V.~Chen, Q.~V. Liao, J.~Wortman~Vaughan, and G.~Bansal.
\newblock Understanding the role of human intuition on reliance in human-ai decision-making with explanations.
\newblock \emph{Proceedings of the ACM on Human-Computer Interaction}, 7\penalty0 (CSCW2):\penalty0 1--32, 2023{\natexlab{b}}.

\bibitem[Chen et~al.(2023{\natexlab{c}})Chen, Wang, Changpinyo, Piergiovanni, Padlewski, Salz, Goodman, Grycner, Mustafa, Beyer, Kolesnikov, Puigcerver, Ding, Rong, Akbari, Mishra, Xue, Thapliyal, Bradbury, Kuo, Seyedhosseini, Jia, Ayan, Riquelme, Steiner, Angelova, Zhai, Houlsby, and Soricut]{chen_pali:_2023}
X.~Chen, X.~Wang, S.~Changpinyo, A.~J. Piergiovanni, P.~Padlewski, D.~Salz, S.~Goodman, A.~Grycner, B.~Mustafa, L.~Beyer, A.~Kolesnikov, J.~Puigcerver, N.~Ding, K.~Rong, H.~Akbari, G.~Mishra, L.~Xue, A.~Thapliyal, J.~Bradbury, W.~Kuo, M.~Seyedhosseini, C.~Jia, B.~K. Ayan, C.~Riquelme, A.~Steiner, A.~Angelova, X.~Zhai, N.~Houlsby, and R.~Soricut.
\newblock {PaLI}: {A} {Jointly}-{Scaled} {Multilingual} {Language}-{Image} {Model}, June 2023{\natexlab{c}}.
\newblock URL \url{http://arxiv.org/abs/2209.06794}.
\newblock arXiv:2209.06794 [cs].

\bibitem[Chesney and Citron(2018)]{chesney_deep_2018}
R.~Chesney and D.~K. Citron.
\newblock Deep {Fakes}: {A} {Looming} {Challenge} for {Privacy}, {Democracy}, and {National} {Security}.
\newblock \emph{SSRN Electronic Journal}, 2018.
\newblock ISSN 1556-5068.
\newblock \doi{10.2139/ssrn.3213954}.
\newblock URL \url{https://www.ssrn.com/abstract=3213954}.

\bibitem[Chesterman(2020)]{chesterman_artificial_2020}
S.~Chesterman.
\newblock Artificial {Intelligence} and the {Limits} of {Legal} {Personality}.
\newblock \emph{International and Comparative Law Quarterly}, 69\penalty0 (4):\penalty0 819--844, Oct. 2020.
\newblock ISSN 0020-5893, 1471-6895.
\newblock \doi{10.1017/S0020589320000366}.
\newblock URL \url{https://www.cambridge.org/core/product/identifier/S0020589320000366/type/journal_article}.

\bibitem[Chiesurin et~al.(2023)Chiesurin, Dimakopoulos, Cabezudo, Eshghi, Papaioannou, Rieser, and Konstas]{chiesurin_dangers_2023}
S.~Chiesurin, D.~Dimakopoulos, M.~A.~S. Cabezudo, A.~Eshghi, I.~Papaioannou, V.~Rieser, and I.~Konstas.
\newblock The {Dangers} of trusting {Stochastic} {Parrots}: {Faithfulness} and {Trust} in {Open}-domain {Conversational} {Question} {Answering}, May 2023.
\newblock URL \url{https://arxiv.org/pdf/2305.16519.pdf}.
\newblock arXiv:2305.16519 [cs].

\bibitem[Chong et~al.(2022)Chong, Zhang, Goucher-Lambert, Kotovsky, and Cagan]{chong_human_2022}
L.~Chong, G.~Zhang, K.~Goucher-Lambert, K.~Kotovsky, and J.~Cagan.
\newblock Human confidence in artificial intelligence and in themselves: {The} evolution and impact of confidence on adoption of {AI} advice.
\newblock \emph{Computers in Human Behavior}, 127:\penalty0 107018, Feb. 2022.
\newblock ISSN 0747-5632.
\newblock \doi{10.1016/j.chb.2021.107018}.
\newblock URL \url{https://www.sciencedirect.com/science/article/pii/S0747563221003411}.

\bibitem[Chow(2023)]{chow_why_2023}
A.~R. Chow.
\newblock Why {People} {Are} {Confessing} {Their} {Love} {For} {AI} {Chatbots}.
\newblock \emph{Time}, Feb. 2023.
\newblock URL \url{https://time.com/6257790/ai-chatbots-love/}.
\newblock publisher: ANDREW R. CHOW FEBRUARY 23, 2023 2:23 PM EST.

\bibitem[Chowdhery et~al.(2023)Chowdhery, Narang, Devlin, Bosma, Mishra, Roberts, Barham, Chung, Sutton, Gehrmann, et~al.]{chowdhery2023palm}
A.~Chowdhery, S.~Narang, J.~Devlin, M.~Bosma, G.~Mishra, A.~Roberts, P.~Barham, H.~W. Chung, C.~Sutton, S.~Gehrmann, et~al.
\newblock Palm: Scaling language modeling with pathways.
\newblock \emph{Journal of Machine Learning Research}, 24\penalty0 (240):\penalty0 1--113, 2023.

\bibitem[Christian(2021)]{christian_alignment_2021}
B.~Christian.
\newblock \emph{The {Alignment} {Problem}: {Machine} {Learning} and {Human} {Values} {\textbar} mitpressbookstore}.
\newblock W. W. Norton \& Company, Oct. 2021.
\newblock ISBN 9780393868333.
\newblock URL \url{https://mitpressbookstore.mit.edu/book/9780393868333}.

\bibitem[Christiano()]{christiano_what_nodate}
P.~Christiano.
\newblock What failure looks like.
\newblock URL \url{https://www.alignmentforum.org/posts/HBxe6wdjxK239zajf/what-failure-looks-like}.
\newblock publisher: AI Alignment Forum.

\bibitem[Christiano et~al.(2018)Christiano, Shlegeris, and Amodei]{christiano_supervising_2018}
P.~Christiano, B.~Shlegeris, and D.~Amodei.
\newblock Supervising strong learners by amplifying weak experts, Oct. 2018.
\newblock URL \url{http://arxiv.org/abs/1810.08575}.
\newblock arXiv:1810.08575 [cs, stat].

\bibitem[Christiano et~al.(2023)Christiano, Leike, Brown, Martic, Legg, and Amodei]{christiano_deep_2023}
P.~Christiano, J.~Leike, T.~B. Brown, M.~Martic, S.~Legg, and D.~Amodei.
\newblock Deep reinforcement learning from human preferences, Feb. 2023.
\newblock URL \url{https://arxiv.org/pdf/1706.03741.pdf}.
\newblock arXiv:1706.03741 [cs, stat].

\bibitem[Christiano et~al.(2017)Christiano, Leike, Brown, Martic, Legg, and Amodei]{christiano2017deep}
P.~F. Christiano, J.~Leike, T.~Brown, M.~Martic, S.~Legg, and D.~Amodei.
\newblock Deep reinforcement learning from human preferences.
\newblock \emph{Advances in Neural Information Processing Systems}, 30, 2017.

\bibitem[Chughtai et~al.(2023)Chughtai, Chan, and Nanda]{chughtai_toy_2023}
B.~Chughtai, L.~Chan, and N.~Nanda.
\newblock A {Toy} {Model} of {Universality}: {Reverse} {Engineering} {How} {Networks} {Learn} {Group} {Operations}, May 2023.
\newblock URL \url{http://arxiv.org/abs/2302.03025}.
\newblock arXiv:2302.03025 [cs, math].

\bibitem[Chui et~al.(2018)Chui, Harryson, Manyika, Roberts, Chung, van Heteren, and Nel]{chui__notes_2018}
M.~Chui, M.~Harryson, J.~Manyika, R.~Roberts, R.~Chung, A.~van Heteren, and P.~Nel.
\newblock Notes from the {AI} {Frontier}: {Applying} {AI} for {Social} {Good}.
\newblock McKinsey Global Institute, Dec. 2018.
\newblock URL \url{http://dln.jaipuria.ac.in:8080/jspui/bitstream/123456789/14267/1/MGI-Applying-AI-for-social-good-Discussion-paper-Dec-2018.pdf}.

\bibitem[Chun and Barnett(2021)]{chun_discriminating_2021}
W.~H.~K. Chun and A.~Barnett.
\newblock \emph{Discriminating data: correlation, neighborhoods, and the new politics of recognition}.
\newblock The MIT Press, Cambridge, Massachusetts, 2021.
\newblock ISBN 9780262046220.

\bibitem[Chérif and Lemoine(2019)]{cherif_anthropomorphic_2019}
E.~Chérif and J.-F. Lemoine.
\newblock Anthropomorphic virtual assistants and the reactions of {Internet} users: {An} experiment on the assistant’s voice.
\newblock \emph{Recherche et Applications en Marketing (English Edition)}, 34\penalty0 (1):\penalty0 28--47, Mar. 2019.
\newblock ISSN 2051-5707, 2051-5707.
\newblock \doi{10.1177/2051570719829432}.
\newblock URL \url{http://journals.sagepub.com/doi/10.1177/2051570719829432}.

\bibitem[Cialdini(2001)]{cialdini_harnessing_2001}
R.~B. Cialdini.
\newblock Harnessing the {Science} of {Persuasion}.
\newblock \emph{Harvard Business Review}, Oct. 2001.
\newblock ISSN 0017-8012.
\newblock URL \url{https://hbr.org/2001/10/harnessing-the-science-of-persuasion}.

\bibitem[Clancy and Besiroglu(2023)]{clancy_great_2023}
M.~Clancy and T.~Besiroglu.
\newblock The {Great} {Inflection}? {A} {Debate} {About} {AI} and {Explosive} {Growth}.
\newblock \emph{Asterisk}, June 2023.
\newblock URL \url{https://asteriskmag.com/issues/03/the-great-inflection-a-debate-about-ai-and-explosive-growth}.

\bibitem[Clark(2008)]{clark2008supersizing}
A.~Clark.
\newblock \emph{Supersizing the mind: Embodiment, action, and cognitive extension}.
\newblock OUP USA, 2008.

\bibitem[Clark and Chalmers(1998)]{clark1998extended}
A.~Clark and D.~Chalmers.
\newblock The extended mind.
\newblock \emph{Analysis}, 58\penalty0 (1):\penalty0 7--19, 1998.

\bibitem[Clark and Amodei(2016)]{clark_faulty_2016}
J.~Clark and D.~Amodei.
\newblock Faulty reward functions in the wild, Dec. 2016.
\newblock URL \url{https://openai.com/research/faulty-reward-functions}.
\newblock publisher: OpenAI.

\bibitem[Coeckelbergh(2012)]{coeckelbergh_can_2012}
M.~Coeckelbergh.
\newblock Can we trust robots?
\newblock \emph{Ethics and Information Technology}, 14\penalty0 (1):\penalty0 53--60, Mar. 2012.
\newblock ISSN 1572-8439.
\newblock \doi{10.1007/s10676-011-9279-1}.
\newblock URL \url{https://doi.org/10.1007/s10676-011-9279-1}.

\bibitem[Cohen et~al.(2022)Cohen, Hutter, and Osborne]{cohen_advanced_2022}
M.~K. Cohen, M.~Hutter, and M.~A. Osborne.
\newblock Advanced artificial agents intervene in the provision of reward.
\newblock \emph{AI Magazine}, 43\penalty0 (3):\penalty0 282--293, Sept. 2022.
\newblock ISSN 0738-4602, 2371-9621.
\newblock \doi{10.1002/aaai.12064}.
\newblock URL \url{https://onlinelibrary.wiley.com/doi/10.1002/aaai.12064}.

\bibitem[Cohen(2018)]{cohen_manipulation_2018}
S.~Cohen.
\newblock Manipulation and {Deception}.
\newblock \emph{Australasian Journal of Philosophy}, 96\penalty0 (3):\penalty0 483--497, July 2018.
\newblock ISSN 0004-8402, 1471-6828.
\newblock \doi{10.1080/00048402.2017.1386692}.
\newblock URL \url{https://www.tandfonline.com/doi/full/10.1080/00048402.2017.1386692}.

\bibitem[Coheur(2020)]{lesot_eliza_2020}
L.~Coheur.
\newblock From {Eliza} to {Siri} and {Beyond}.
\newblock In M.-J. Lesot, S.~Vieira, M.~Z. Reformat, J.~P. Carvalho, A.~Wilbik, B.~Bouchon-Meunier, and R.~R. Yager, editors, \emph{Information {Processing} and {Management} of {Uncertainty} in {Knowledge}-{Based} {Systems}}, volume 1237, pages 29--41. Springer International Publishing, Cham, 2020.
\newblock ISBN 9783030501457 9783030501464.
\newblock \doi{10.1007/978-3-030-50146-4_3}.
\newblock URL \url{http://link.springer.com/10.1007/978-3-030-50146-4_3}.

\bibitem[Collingridge(1980)]{collingridge_social_1980}
D.~Collingridge.
\newblock \emph{The {Social} {Control} of {Technology}}.
\newblock St. Martin's Press, 1980.
\newblock ISBN 9780312731687.
\newblock Google-Books-ID: hCSdAQAACAAJ.

\bibitem[Collste et~al.(2021)Collste, Cornell, Randers, Rockstr{\"o}m, and Stoknes]{collste2021human}
D.~Collste, S.~E. Cornell, J.~Randers, J.~Rockstr{\"o}m, and P.~E. Stoknes.
\newblock Human well-being in the anthropocene: {L}imits to growth.
\newblock \emph{Global Sustainability}, 4:\penalty0 e30, 2021.

\bibitem[Colman(2008)]{colman2008anthropomorphism}
A.~M. Colman.
\newblock Anthropomorphism.
\newblock \emph{A Dictionary of Psychology}, 2008.

\bibitem[{Combahee River Collective}(1977)]{combahee_river_collective_combahee_1977}
{Combahee River Collective}.
\newblock The {Combahee} {River} {Collective} {Statement}.
\newblock 1977.
\newblock URL \url{https://americanstudies.yale.edu/sites/default/files/files/Keyword%20Coalition_Readings.pdf}.

\bibitem[Comiter(2019)]{comiter_attacking_2019}
M.~Comiter.
\newblock Attacking {Artificial} {Intelligence}: {AI}’s {Security} {Vulnerability} and {What} {Policymakers} {Can} {Do} {About} {It}.
\newblock Technical report, Harvard Kennedy School: Belfer Center, Aug. 2019.

\bibitem[Corak(2013)]{corak_income_2013}
M.~Corak.
\newblock Income {Inequality}, {Equality} of {Opportunity}, and {Intergenerational} {Mobility}.
\newblock \emph{Journal of Economic Perspectives}, 27\penalty0 (3):\penalty0 79--102, Aug. 2013.
\newblock ISSN 0895-3309.
\newblock \doi{10.1257/jep.27.3.79}.
\newblock URL \url{https://pubs.aeaweb.org/doi/10.1257/jep.27.3.79}.

\bibitem[Corbett and Denton(2023)]{corbett_interrogating_2023}
E.~Corbett and E.~Denton.
\newblock Interrogating the {T} in {FAccT}.
\newblock In \emph{2023 {ACM} {Conference} on {Fairness}, {Accountability}, and {Transparency}}, pages 1624--1634, Chicago IL USA, June 2023. ACM.
\newblock ISBN 9798400701924.
\newblock \doi{10.1145/3593013.3594104}.
\newblock URL \url{https://dl.acm.org/doi/10.1145/3593013.3594104}.

\bibitem[Costanza-Chock(2020)]{costanza-chock_design_2020}
S.~Costanza-Chock.
\newblock \emph{Design {Justice}: {Community}-{Led} {Practices} to {Build} the {Worlds} {We} {Need}}.
\newblock The MIT Press, 2020.
\newblock ISBN 9780262043458.
\newblock URL \url{https://library.oapen.org/handle/20.500.12657/43542}.

\bibitem[Cotra(2022)]{cotra_without_2022}
A.~Cotra.
\newblock Without specific countermeasures, the easiest path to transformative {AI} likely leads to {AI} takeover, July 2022.
\newblock URL \url{https://www.alignmentforum.org/posts/pRkFkzwKZ2zfa3R6H/without-specific-countermeasures-the-easiest-path-to}.

\bibitem[Cotter(2008)]{cotter_influence_2008}
E.~M. Cotter.
\newblock Influence of {Emotional} {Content} and {Perceived} {Relevance} on {Spread} of {Urban} {Legends}: {A} {Pilot} {Study}.
\newblock \emph{Psychological Reports}, 102\penalty0 (2):\penalty0 623--629, Apr. 2008.
\newblock ISSN 0033-2941, 1558-691X.
\newblock \doi{10.2466/pr0.102.2.623-629}.
\newblock URL \url{http://journals.sagepub.com/doi/10.2466/pr0.102.2.623-629}.

\bibitem[Courty(2019)]{courty_ticket_2019}
P.~Courty.
\newblock Ticket resale, bots, and the fair price ticketing curse.
\newblock \emph{Journal of Cultural Economics}, 43\penalty0 (3):\penalty0 345--363, 2019.
\newblock ISSN 0885-2545.
\newblock URL \url{https://www.jstor.org/stable/48698098}.

\bibitem[Cowen and Southwood(2019)]{cowen_is_2019}
T.~Cowen and B.~Southwood.
\newblock Is the {Rate} of {Scientific} {Progress} {Slowing} {Down}?, Aug. 2019.
\newblock URL \url{https://papers.ssrn.com/abstract=3822691}.

\bibitem[Crawford(2021)]{crawford_atlas_2021}
K.~Crawford.
\newblock \emph{The {Atlas} of {AI}: {Power}, {Politics}, and the {Planetary} {Costs} of {Artificial} {Intelligence}}.
\newblock Yale University Press, Apr. 2021.
\newblock ISBN 9780300252392 9780300209570.
\newblock \doi{10.2307/j.ctv1ghv45t}.
\newblock URL \url{http://www.jstor.org/stable/10.2307/j.ctv1ghv45t}.

\bibitem[Creel and Hellman(2022)]{creel_algorithmic_2022}
K.~Creel and D.~Hellman.
\newblock The {Algorithmic} {Leviathan}: {Arbitrariness}, {Fairness}, and {Opportunity} in {Algorithmic} {Decision}-{Making} {Systems}.
\newblock \emph{Canadian Journal of Philosophy}, 52\penalty0 (1):\penalty0 26--43, 2022.
\newblock \doi{10.1017/can.2022.3}.
\newblock URL \url{https://philarchive.org/rec/CRETAL-3}.

\bibitem[Crenshaw(1989)]{crenshaw_demarginalizing_2015}
K.~Crenshaw.
\newblock Demarginalizing the {Intersection} of {Race} and {Sex}: {A} {Black} {Feminist} {Critique} of {Antidiscrimination} {Doctrine}, {Feminist} {Theory} and {Antiracist} {Politics}.
\newblock \emph{University of Chicago Legal Forum}, 1989\penalty0 (1), 1989.
\newblock ISSN 0892-5593.
\newblock URL \url{https://chicagounbound.uchicago.edu/uclf/vol1989/iss1/8}.

\bibitem[Crisp(2006)]{crisp2006hedonism}
R.~Crisp.
\newblock Hedonism reconsidered.
\newblock \emph{Philosophy and Phenomenological Research}, 73\penalty0 (3):\penalty0 619--645, 2006.

\bibitem[Crisp(2011)]{crisp2011pleasure}
R.~Crisp.
\newblock Pleasure and hedonism in sidgwick.
\newblock Oxford University Press, 2011.

\bibitem[Critch and Russell(2023)]{critch_tasra:_2023}
A.~Critch and S.~Russell.
\newblock {TASRA}: a {Taxonomy} and {Analysis} of {Societal}-{Scale} {Risks} from {AI}, June 2023.
\newblock URL \url{http://arxiv.org/abs/2306.06924}.
\newblock arXiv:2306.06924 [cs].

\bibitem[Crumpton and Bethel(2016)]{crumpton_survey_2016}
J.~Crumpton and C.~L. Bethel.
\newblock A survey of using vocal prosody to convey emotion in robot speech.
\newblock \emph{International Journal of Social Robotics}, 8:\penalty0 271--285, 2016.

\bibitem[Cui et~al.(2022)Cui, Ma, Zhou, Zhou, and Yang]{cui_m6-rec:_2022}
Z.~Cui, J.~Ma, C.~Zhou, J.~Zhou, and H.~Yang.
\newblock M6-{Rec}: {Generative} {Pretrained} {Language} {Models} are {Open}-{Ended} {Recommender} {Systems}, May 2022.
\newblock URL \url{http://arxiv.org/abs/2205.08084}.
\newblock arXiv:2205.08084 [cs].

\bibitem[Cummings(2004)]{cummings_automation_2004}
M.~Cummings.
\newblock Automation {Bias} in {Intelligent} {Time} {Critical} {Decision} {Support} {Systems}.
\newblock In \emph{{AIAA} 1st {Intelligent} {Systems} {Technical} {Conference}}, Chicago, Illinois, Sept. 2004. American Institute of Aeronautics and Astronautics.
\newblock ISBN 9781624100802.
\newblock \doi{10.2514/6.2004-6313}.
\newblock URL \url{https://arc.aiaa.org/doi/10.2514/6.2004-6313}.

\bibitem[Cummins et~al.(2003)Cummins, Eckersley, Pallant, Van~Vugt, and Misajon]{cummins2003developing}
R.~A. Cummins, R.~Eckersley, J.~Pallant, J.~Van~Vugt, and R.~Misajon.
\newblock Developing a national index of subjective wellbeing: The australian unity wellbeing index.
\newblock \emph{Social indicators research}, 64:\penalty0 159--190, 2003.

\bibitem[{Curious Evolver}(2023)]{curious_evolver_customer_2023}
{Curious Evolver}.
\newblock the customer service of the new bing chat is amazing, Feb. 2023.
\newblock URL \url{www.reddit.com/r/bing/comments/110eagl/the_customer_service_of_the_new_bing_chat_is/}.

\bibitem[Dafoe(2018)]{dafoe_ai_2018}
A.~Dafoe.
\newblock {AI} {Governance}: {A} {Research} {Agenda}.
\newblock Technical report, Centre for the Governance of AI: Future of Humanity Institute, University of Oxford, 2018.
\newblock URL \url{http://www.fhi.ox.ac.uk/wp-content/uploads/GovAI-Agenda.pdf}.

\bibitem[Dafoe(2023)]{bullock_ai_2023}
A.~Dafoe.
\newblock {AI} {Governance}: {Overview} and {Theoretical} {Lenses}.
\newblock In J.~B. Bullock, Y.-C. Chen, J.~Himmelreich, V.~M. Hudson, A.~Korinek, M.~M. Young, and B.~Zhang, editors, \emph{The {Oxford} {Handbook} of {AI} {Governance}}, pages C2S1--C2N*. Oxford University Press, 1 edition, June 2023.
\newblock ISBN 9780197579329 9780197579350.
\newblock \doi{10.1093/oxfordhb/9780197579329.013.2}.
\newblock URL \url{https://academic.oup.com/edited-volume/41989/chapter/408516484}.

\bibitem[Dafoe et~al.(2020)Dafoe, Hughes, Bachrach, Collins, McKee, Leibo, Larson, and Graepel]{dafoe_open_2020}
A.~Dafoe, E.~Hughes, Y.~Bachrach, T.~Collins, K.~R. McKee, J.~Z. Leibo, K.~Larson, and T.~Graepel.
\newblock Open {Problems} in {Cooperative} {AI}, Dec. 2020.
\newblock URL \url{http://arxiv.org/abs/2012.08630}.
\newblock arXiv:2012.08630 [cs].

\bibitem[Dafoe et~al.(2021)Dafoe, Bachrach, Hadfield, Horvitz, Larson, and Graepel]{dafoe2021cooperative}
A.~Dafoe, Y.~Bachrach, G.~Hadfield, E.~Horvitz, K.~Larson, and T.~Graepel.
\newblock Cooperative {AI}: {M}achines must learn to find common ground.
\newblock \emph{Nature}, 593\penalty0 (7857):\penalty0 33--36, 2021.

\bibitem[Damiano and Dumouchel(2018)]{damiano_anthropomorphism_2018}
L.~Damiano and P.~Dumouchel.
\newblock Anthropomorphism in human--robot co-evolution.
\newblock \emph{Frontiers in psychology}, 9:\penalty0 468, 2018.

\bibitem[Dannouni et~al.(2023)Dannouni, {Deutscher, Stefan A.}, Dezzaz, Elman, Gawel, Hanna, Hyland, Kharij, Maher, and Patterson]{Dannouni2023-xb}
A.~Dannouni, {Deutscher, Stefan A.}, G.~Dezzaz, A.~Elman, A.~Gawel, M.~Hanna, A.~Hyland, A.~Kharij, H.~Maher, and D.~Patterson.
\newblock How {AI} can speed climate action.
\newblock \url{https://www.bcg.com/publications/2023/how-ai-can-speedup-climate-action#:~:text=Beyond%20helping%20to%20reduce%20emissions,climate%20economics%2C%20and%20fundamental%20research.}, Nov. 2023.
\newblock Accessed: 2024-1-21.

\bibitem[Dantec and DiSalvo(2013)]{dantec_infrastructuring_2013}
C.~A.~L. Dantec and C.~DiSalvo.
\newblock Infrastructuring and the formation of publics in participatory design.
\newblock \emph{Social Studies of Science}, 43\penalty0 (2):\penalty0 241--264, Apr. 2013.
\newblock ISSN 0306-3127, 1460-3659.
\newblock \doi{10.1177/0306312712471581}.
\newblock URL \url{http://journals.sagepub.com/doi/10.1177/0306312712471581}.

\bibitem[Das et~al.(2020{\natexlab{a}})Das, Jones-Harrell, Fan, Ramaswami, Orlove, and Botchwey]{das2020understanding}
K.~V. Das, C.~Jones-Harrell, Y.~Fan, A.~Ramaswami, B.~Orlove, and N.~Botchwey.
\newblock Understanding subjective well-being: {P}erspectives from psychology and public health.
\newblock \emph{Public Health Reviews}, 41\penalty0 (1):\penalty0 1--32, 2020{\natexlab{a}}.

\bibitem[Das et~al.(2020{\natexlab{b}})Das, Steffen, Clarke, Reddy, Brynjolfsson, and Fleming]{das_learning_2020}
S.~Das, S.~Steffen, W.~Clarke, P.~Reddy, E.~Brynjolfsson, and M.~Fleming.
\newblock Learning {Occupational} {Task}-{Shares} {Dynamics} for the {Future} of {Work}.
\newblock In \emph{Proceedings of the {AAAI}/{ACM} {Conference} on {AI}, {Ethics}, and {Society}}, pages 36--42, New York NY USA, Feb. 2020{\natexlab{b}}. ACM.
\newblock ISBN 9781450371100.
\newblock \doi{10.1145/3375627.3375826}.
\newblock URL \url{https://dl.acm.org/doi/10.1145/3375627.3375826}.

\bibitem[Dautenhahn et~al.(2002)Dautenhahn, Ogden, and Quick]{dautenhahn_embodied_2002}
K.~Dautenhahn, B.~Ogden, and T.~Quick.
\newblock From embodied to socially embedded agents – {Implications} for interaction-aware robots.
\newblock \emph{Cognitive Systems Research}, 3\penalty0 (3):\penalty0 397--428, Sept. 2002.
\newblock ISSN 13890417.
\newblock \doi{10.1016/S1389-0417(02)00050-5}.
\newblock URL \url{https://linkinghub.elsevier.com/retrieve/pii/S1389041702000505}.

\bibitem[Dauth et~al.(2017)Dauth, Findeisen, Südekum, and Wößner]{dauth_german_2017}
W.~Dauth, S.~Findeisen, J.~Südekum, and N.~Wößner.
\newblock German {Robots} – {The} {Impact} of {Industrial} {Robots} on {Workers}.
\newblock Technical report, Institut für Arbeitsmarkt- und Berufsforschung: Research Institute of the German Federal Employ­ment Agency, 2017.
\newblock URL \url{https://doku.iab.de/discussionpapers/2017/dp3017.pdf}.

\bibitem[Davani et~al.(2021)Davani, Díaz, and Prabhakaran]{davani_dealing_2021}
A.~M. Davani, M.~Díaz, and V.~Prabhakaran.
\newblock Dealing with {Disagreements}: {Looking} {Beyond} the {Majority} {Vote} in {Subjective} {Annotations}, Oct. 2021.
\newblock URL \url{http://arxiv.org/abs/2110.05719}.
\newblock arXiv:2110.05719 [cs].

\bibitem[Davis et~al.(2021)Davis, Williams, and Yang]{davis_algorithmic_2021}
J.~L. Davis, A.~Williams, and M.~W. Yang.
\newblock Algorithmic reparation.
\newblock \emph{Big Data \& Society}, 8\penalty0 (2):\penalty0 205395172110448, July 2021.
\newblock ISSN 2053-9517, 2053-9517.
\newblock \doi{10.1177/20539517211044808}.
\newblock URL \url{http://journals.sagepub.com/doi/10.1177/20539517211044808}.

\bibitem[De~Leyn et~al.(2022)De~Leyn, De~Wolf, Vanden~Abeele, and De~Marez]{de_leyn_-between_2022}
T.~De~Leyn, R.~De~Wolf, M.~Vanden~Abeele, and L.~De~Marez.
\newblock In-between child’s play and teenage pop culture: tweens, {TikTok} \& privacy.
\newblock \emph{Journal of Youth Studies}, 25\penalty0 (8):\penalty0 1108--1125, Sept. 2022.
\newblock ISSN 1367-6261, 1469-9680.
\newblock \doi{10.1080/13676261.2021.1939286}.
\newblock URL \url{https://www.tandfonline.com/doi/full/10.1080/13676261.2021.1939286}.

\bibitem[de~Vries et~al.(2020)de~Vries, Bahdanau, and Manning]{de_vries_towards_2020}
H.~de~Vries, D.~Bahdanau, and C.~Manning.
\newblock Towards {Ecologically} {Valid} {Research} on {Language} {User} {Interfaces}, July 2020.
\newblock URL \url{http://arxiv.org/abs/2007.14435}.
\newblock arXiv:2007.14435 [cs].

\bibitem[de~Wynter et~al.(2023)de~Wynter, Wang, Sokolov, Gu, and Chen]{de_wynter_evaluation_2023}
A.~de~Wynter, X.~Wang, A.~Sokolov, Q.~Gu, and S.-Q. Chen.
\newblock An {Evaluation} on {Large} {Language} {Model} {Outputs}: {Discourse} and {Memorization}.
\newblock \emph{Natural Language Processing Journal}, 4:\penalty0 100024, Sept. 2023.
\newblock ISSN 29497191.
\newblock \doi{10.1016/j.nlp.2023.100024}.
\newblock URL \url{http://arxiv.org/abs/2304.08637}.
\newblock arXiv:2304.08637 [cs].

\bibitem[Debes(2023)]{debes_dignity_2023}
R.~Debes.
\newblock Dignity.
\newblock In E.~N. Zalta and U.~Nodelman, editors, \emph{The {Stanford} {Encyclopedia} of {Philosophy}}. Metaphysics Research Lab, Stanford University, spring 2023 edition, 2023.
\newblock URL \url{https://plato.stanford.edu/archives/spr2023/entries/dignity/}.

\bibitem[Deck(2023)]{deck_we_2023}
A.~Deck.
\newblock We tested {ChatGPT} in {Bengali}, {Kurdish}, and {Tamil}. {It} failed., Sept. 2023.
\newblock URL \url{https://restofworld.org/2023/chatgpt-problems-global-language-testing/}.

\bibitem[DeepMind(2023)]{google_deepmind_synthid_2023}
G.~DeepMind.
\newblock {SynthID}, Nov. 2023.
\newblock URL \url{https://deepmind.google/technologies/synthid/}.

\bibitem[Degrave et~al.(2022)Degrave, Felici, Buchli, Neunert, Tracey, Carpanese, Ewalds, Hafner, Abdolmaleki, de~las Casas, Donner, Fritz, Galperti, Huber, Keeling, Tsimpoukelli, Kay, Merle, Moret, Noury, Pesamosca, Pfau, Sauter, Sommariva, Coda, Duval, Fasoli, Kohli, Kavukcuoglu, Hassabis, and Riedmiller]{degrave_magnetic_2022}
J.~Degrave, F.~Felici, J.~Buchli, M.~Neunert, B.~Tracey, F.~Carpanese, T.~Ewalds, R.~Hafner, A.~Abdolmaleki, D.~de~las Casas, C.~Donner, L.~Fritz, C.~Galperti, A.~Huber, J.~Keeling, M.~Tsimpoukelli, J.~Kay, A.~Merle, J.-M. Moret, S.~Noury, F.~Pesamosca, D.~Pfau, O.~Sauter, C.~Sommariva, S.~Coda, B.~Duval, A.~Fasoli, P.~Kohli, K.~Kavukcuoglu, D.~Hassabis, and M.~Riedmiller.
\newblock Magnetic control of tokamak plasmas through deep reinforcement learning.
\newblock \emph{Nature}, 602\penalty0 (7897):\penalty0 414--419, Feb. 2022.
\newblock ISSN 1476-4687.
\newblock \doi{10.1038/s41586-021-04301-9}.
\newblock URL \url{https://www.nature.com/articles/s41586-021-04301-9}.

\bibitem[Dehghani et~al.(2021)Dehghani, Tay, Gritsenko, Zhao, Houlsby, Diaz, Metzler, and Vinyals]{dehghani_benchmark_2021}
M.~Dehghani, Y.~Tay, A.~A. Gritsenko, Z.~Zhao, N.~Houlsby, F.~Diaz, D.~Metzler, and O.~Vinyals.
\newblock The {Benchmark} {Lottery}, July 2021.
\newblock URL \url{http://arxiv.org/abs/2107.07002}.
\newblock arXiv:2107.07002 [cs].

\bibitem[Deiana et~al.(2023)Deiana, Dettori, Arghittu, Azara, Gabutti, and Castiglia]{deiana_artificial_2023}
G.~Deiana, M.~Dettori, A.~Arghittu, A.~Azara, G.~Gabutti, and P.~Castiglia.
\newblock Artificial {Intelligence} and {Public} {Health}: {Evaluating} {ChatGPT} {Responses} to {Vaccination} {Myths} and {Misconceptions}.
\newblock \emph{Vaccines}, 11\penalty0 (7):\penalty0 1217, July 2023.
\newblock ISSN 2076-393X.
\newblock \doi{10.3390/vaccines11071217}.
\newblock URL \url{https://www.mdpi.com/2076-393X/11/7/1217}.

\bibitem[Demski and Garrabrant(2020)]{demski_embedded_2020}
A.~Demski and S.~Garrabrant.
\newblock Embedded {Agency}, Oct. 2020.
\newblock URL \url{http://arxiv.org/abs/1902.09469}.
\newblock arXiv:1902.09469 [cs].

\bibitem[Dennett(1989)]{dennett1989intentional}
D.~C. Dennett.
\newblock \emph{The intentional stance}.
\newblock MIT press, 1989.

\bibitem[Dennett(2003)]{dennett_freedom_2003}
D.~C. Dennett.
\newblock \emph{Freedom evolves}.
\newblock Viking, New York, 2003.
\newblock ISBN 9780670031863.

\bibitem[Denning(2023)]{denning_can_2023}
P.~J. Denning.
\newblock Can {Generative} {AI} {Bots} {Be} {Trusted}?
\newblock \emph{Communications of the ACM}, 66\penalty0 (6):\penalty0 24--27, June 2023.
\newblock ISSN 0001-0782, 1557-7317.
\newblock \doi{10.1145/3592981}.
\newblock URL \url{https://dl.acm.org/doi/10.1145/3592981}.

\bibitem[DeVito et~al.(2021)DeVito, Walker, and Fernandez]{devito_values_2021}
M.~A. DeVito, A.~M. Walker, and J.~R. Fernandez.
\newblock Values ({Mis})alignment: {Exploring} {Tensions} {Between} {Platform} and {LGBTQ}+ {Community} {Design} {Values}.
\newblock \emph{Proceedings of the ACM on Human-Computer Interaction}, 5\penalty0 (CSCW1):\penalty0 1--27, Apr. 2021.
\newblock ISSN 2573-0142.
\newblock \doi{10.1145/3449162}.
\newblock URL \url{https://dl.acm.org/doi/10.1145/3449162}.

\bibitem[Devitt(2018)]{kate_devitt_trustworthiness_2018}
S.~K. Devitt.
\newblock Trustworthiness of {Autonomous} {Systems}.
\newblock In H.~A. Abbass, J.~Scholz, and D.~J. Reid, editors, \emph{Foundations of {Trusted} {Autonomy}}, Studies in {Systems}, {Decision} and {Control}, pages 161--184. Springer International Publishing, Cham, 2018.
\newblock ISBN 9783319648163.
\newblock \doi{10.1007/978-3-319-64816-3_9}.
\newblock URL \url{https://doi.org/10.1007/978-3-319-64816-3_9}.

\bibitem[Devitt et~al.(2021)Devitt, Horne, Assaad, Broad, Kurniawati, Cardier, Scott, Lazar, Gould, Adamson, Karl, Schrever, Keay, Tranter, Shellshear, Hunter, Brady, and Putland]{devitt_trust_2021}
S.~K. Devitt, R.~Horne, Z.~Assaad, E.~Broad, H.~Kurniawati, B.~Cardier, A.~Scott, S.~Lazar, M.~Gould, C.~Adamson, C.~Karl, F.~Schrever, S.~Keay, K.~Tranter, E.~Shellshear, D.~Hunter, M.~Brady, and T.~Putland.
\newblock Trust and {Safety}, Apr. 2021.
\newblock URL \url{http://arxiv.org/abs/2104.06512}.
\newblock arXiv:2104.06512 [cs].

\bibitem[Devlin(2023)]{devlin_ai_2023}
H.~Devlin.
\newblock {AI} likely to spell end of traditional school classroom, leading expert says.
\newblock \emph{The Guardian}, July 2023.
\newblock ISSN 0261-3077.
\newblock URL \url{https://www.theguardian.com/technology/2023/jul/07/ai-likely-to-spell-end-of-traditional-school-classroom-leading-expert-says}.

\bibitem[DeVries et~al.(2019)DeVries, Misra, Wang, and van~der Maaten]{devries_does_2019}
T.~DeVries, I.~Misra, C.~Wang, and L.~van~der Maaten.
\newblock Does {Object} {Recognition} {Work} for {Everyone}?
\newblock Meta, June 2019.
\newblock URL \url{https://ai.meta.com/research/publications/does-object-recognition-work-for-everyone/}.

\bibitem[Dias~Oliva et~al.(2021)Dias~Oliva, Antonialli, and Gomes]{dias_oliva_fighting_2021}
T.~Dias~Oliva, D.~M. Antonialli, and A.~Gomes.
\newblock Fighting {Hate} {Speech}, {Silencing} {Drag} {Queens}? {Artificial} {Intelligence} in {Content} {Moderation} and {Risks} to {LGBTQ} {Voices} {Online}.
\newblock \emph{Sexuality \& Culture}, 25\penalty0 (2):\penalty0 700--732, Apr. 2021.
\newblock ISSN 1095-5143, 1936-4822.
\newblock \doi{10.1007/s12119-020-09790-w}.
\newblock URL \url{http://link.springer.com/10.1007/s12119-020-09790-w}.

\bibitem[Diener(1995)]{diener1995value}
E.~Diener.
\newblock A value based index for measuring national quality of life.
\newblock \emph{Social Indicators Research}, 36:\penalty0 107--127, 1995.

\bibitem[Diener et~al.(2018)Diener, Oishi, and Tay]{diener2018advances}
E.~Diener, S.~Oishi, and L.~Tay.
\newblock Advances in subjective well-being research.
\newblock \emph{Nature Human Behaviour}, 2\penalty0 (4):\penalty0 253--260, 2018.

\bibitem[Dieppe(2021)]{dieppe_global_2021}
A.~Dieppe.
\newblock \emph{Global {Productivity}: {Trends}, {Drivers}, and {Policies}}.
\newblock Washington, DC: World Bank, June 2021.
\newblock ISBN 9781464816086.
\newblock \doi{10.1016/978-1-4648-1608-6}.
\newblock URL \url{http://hdl.handle.net/10986/34015}.

\bibitem[Dietvorst et~al.(2015)Dietvorst, Simmons, and Massey]{dietvorst_algorithm_2015}
B.~J. Dietvorst, J.~P. Simmons, and C.~Massey.
\newblock Algorithm aversion: {People} erroneously avoid algorithms after seeing them err.
\newblock \emph{Journal of Experimental Psychology: General}, 144\penalty0 (1):\penalty0 114--126, 2015.
\newblock ISSN 1939-2222, 0096-3445.
\newblock \doi{10.1037/xge0000033}.
\newblock URL \url{http://doi.apa.org/getdoi.cfm?doi=10.1037/xge0000033}.

\bibitem[D'Ignazio and Klein(2020)]{dignazio_1._2020}
C.~D'Ignazio and L.~Klein.
\newblock 1. {The} {Power} {Chapter}.
\newblock \emph{Data Feminism}, Mar. 2020.
\newblock URL \url{https://data-feminism.mitpress.mit.edu/pub/vi8obxh7/release/4}.

\bibitem[Dignum(2019)]{dignum_responsible_2019}
V.~Dignum.
\newblock \emph{Responsible {Artificial} {Intelligence}: {How} to {Develop} and {Use} {AI} in a {Responsible} {Way}}.
\newblock Artificial {Intelligence}: {Foundations}, {Theory}, and {Algorithms}. Springer International Publishing, Cham, 2019.
\newblock ISBN 9783030303709 9783030303716.
\newblock \doi{10.1007/978-3-030-30371-6}.
\newblock URL \url{http://link.springer.com/10.1007/978-3-030-30371-6}.

\bibitem[DiMaggio and Hargittai(2023)]{dimaggio_digital_2023}
P.~DiMaggio and E.~Hargittai.
\newblock From the '{Digital} {Divide}' to '{Digital} {Inequality}': {Studying} {Internet} {Use} as {Penetration} {Increases}.
\newblock preprint, SocArXiv, June 2023.
\newblock URL \url{https://osf.io/rhqmu}.

\bibitem[Dinan et~al.(2021)Dinan, Abercrombie, Bergman, Spruit, Hovy, Boureau, and Rieser]{dinan_anticipating_2021}
E.~Dinan, G.~Abercrombie, A.~S. Bergman, S.~Spruit, D.~Hovy, Y.-L. Boureau, and V.~Rieser.
\newblock Anticipating {Safety} {Issues} in {E2E} {Conversational} {AI}: {Framework} and {Tooling}, July 2021.
\newblock URL \url{http://arxiv.org/abs/2107.03451}.
\newblock arXiv:2107.03451 [cs].

\bibitem[Dinan et~al.(2022)Dinan, Abercrombie, Bergman, Spruit, Hovy, Boureau, and Rieser]{dinan_safetykit:_2022}
E.~Dinan, G.~Abercrombie, A.~Bergman, S.~Spruit, D.~Hovy, Y.-L. Boureau, and V.~Rieser.
\newblock {SafetyKit}: {First} {Aid} for {Measuring} {Safety} in {Open}-domain {Conversational} {Systems}.
\newblock In \emph{Proceedings of the 60th {Annual} {Meeting} of the {Association} for {Computational} {Linguistics} ({Volume} 1: {Long} {Papers})}, pages 4113--4133, Dublin, Ireland, 2022. Association for Computational Linguistics.
\newblock \doi{10.18653/v1/2022.acl-long.284}.
\newblock URL \url{https://aclanthology.org/2022.acl-long.284}.

\bibitem[DiResta et~al.(2019)DiResta, Shaffer, Ruppel, Sullivan, Matney, Fox, Albright, and Johnson]{diresta_tactics_2019}
R.~DiResta, K.~Shaffer, B.~Ruppel, D.~Sullivan, R.~Matney, R.~Fox, J.~Albright, and B.~Johnson.
\newblock The {Tactics} \& {Tropes} of the {Internet} {Research} {Agency}.
\newblock \emph{US Senate Documents}, Oct. 2019.
\newblock URL \url{https://digitalcommons.unl.edu/senatedocs/2}.

\bibitem[Distaso(2007)]{distaso2007well}
A.~Distaso.
\newblock Well-being and/or quality of life in {EU} countries through a multidimensional index of sustainability.
\newblock \emph{Ecological Economics}, 64\penalty0 (1):\penalty0 163--180, 2007.

\bibitem[Dixit and Mac(2018)]{dixit_vicious_2018}
P.~Dixit and R.~Mac.
\newblock Vicious {Rumors} {Spread} {Like} {Wildfire} {On} {WhatsApp} — {And} {Destroyed} {A} {Village}, Sept. 2018.
\newblock URL \url{https://www.buzzfeednews.com/article/pranavdixit/whatsapp-destroyed-village-lynchings-rainpada-india}.

\bibitem[Dobbe et~al.(2021)Dobbe, Krendl~Gilbert, and Mintz]{dobbe_hard_2021}
R.~Dobbe, T.~Krendl~Gilbert, and Y.~Mintz.
\newblock Hard choices in artificial intelligence.
\newblock \emph{Artificial Intelligence}, 300:\penalty0 103555, Nov. 2021.
\newblock ISSN 00043702.
\newblock \doi{10.1016/j.artint.2021.103555}.
\newblock URL \url{https://linkinghub.elsevier.com/retrieve/pii/S0004370221001065}.

\bibitem[Docherty and Biega(2022)]{docherty2022re}
N.~Docherty and A.~J. Biega.
\newblock ({R}e) politicizing digital well-being: Beyond user engagements.
\newblock In \emph{Proceedings of the 2022 CHI Conference on Human Factors in Computing Systems}, pages 1--13, 2022.

\bibitem[Dodge et~al.(2021)Dodge, Sap, Marasović, Agnew, Ilharco, Groeneveld, Mitchell, and Gardner]{dodge_documenting_2021}
J.~Dodge, M.~Sap, A.~Marasović, W.~Agnew, G.~Ilharco, D.~Groeneveld, M.~Mitchell, and M.~Gardner.
\newblock Documenting {Large} {Webtext} {Corpora}: {A} {Case} {Study} on the {Colossal} {Clean} {Crawled} {Corpus}, Sept. 2021.
\newblock URL \url{https://arxiv.org/pdf/2104.08758.pdf}.
\newblock arXiv:2104.08758 [cs].

\bibitem[Dodge et~al.(2022)Dodge, Prewitt, Combes, Odmark, Schwartz, Strubell, Luccioni, Smith, DeCario, and Buchanan]{dodge_measuring_2022}
J.~Dodge, T.~Prewitt, R.~T.~D. Combes, E.~Odmark, R.~Schwartz, E.~Strubell, A.~S. Luccioni, N.~A. Smith, N.~DeCario, and W.~Buchanan.
\newblock Measuring the {Carbon} {Intensity} of {AI} in {Cloud} {Instances}, June 2022.
\newblock URL \url{http://arxiv.org/abs/2206.05229}.
\newblock arXiv:2206.05229 [cs].

\bibitem[Dolan and Metcalfe(2011)]{dolan2011measuring}
P.~Dolan and R.~Metcalfe.
\newblock Measuring subjective wellbeing for public policy: Recommendations on measures.
\newblock 2011.

\bibitem[Dolan and White(2007)]{dolan2007can}
P.~Dolan and M.~P. White.
\newblock How can measures of subjective well-being be used to inform public policy?
\newblock \emph{Perspectives on Psychological Science}, 2\penalty0 (1):\penalty0 71--85, 2007.

\bibitem[Dommett(2019)]{dommett_data-driven_2019}
K.~Dommett.
\newblock Data-driven political campaigns in practice: understanding and regulating diverse data-driven campaigns.
\newblock \emph{Internet Policy Review}, 8\penalty0 (4), Dec. 2019.
\newblock ISSN 2197-6775.
\newblock \doi{10.14763/2019.4.1432}.
\newblock URL \url{https://policyreview.info/node/1432}.

\bibitem[Dorn(2019)]{dorn_dialect-specific_2019}
R.~Dorn.
\newblock Dialect-{Specific} {Models} for {Automatic} {Speech} {Recognition} of {African} {American} {Vernacular} {English}.
\newblock In V.~Kovatchev, I.~Temnikova, B.~Šandrih, and I.~Nikolova, editors, \emph{Proceedings of the {Student} {Research} {Workshop} {Associated} with {RANLP} 2019}, pages 16--20, Varna, Bulgaria, Sept. 2019. INCOMA Ltd.
\newblock \doi{10.26615/issn.2603-2821.2019_003}.
\newblock URL \url{https://aclanthology.org/R19-2003}.

\bibitem[Doshi and Hauser(2023)]{doshi_generative_2023}
A.~R. Doshi and O.~Hauser.
\newblock Generative {Artificial} {Intelligence} {Enhances} {Creativity} but {Reduces} the {Diversity} of {Novel} {Content}, Aug. 2023.
\newblock URL \url{https://papers.ssrn.com/abstract=4535536}.

\bibitem[Dosovitskiy et~al.(2021)Dosovitskiy, Beyer, Kolesnikov, Weissenborn, Zhai, Unterthiner, Dehghani, Minderer, Heigold, Gelly, Uszkoreit, and Houlsby]{dosovitskiy_image_2021}
A.~Dosovitskiy, L.~Beyer, A.~Kolesnikov, D.~Weissenborn, X.~Zhai, T.~Unterthiner, M.~Dehghani, M.~Minderer, G.~Heigold, S.~Gelly, J.~Uszkoreit, and N.~Houlsby.
\newblock An {Image} is {Worth} 16x16 {Words}: {Transformers} for {Image} {Recognition} at {Scale}.
\newblock June 2021.
\newblock \doi{10.48550/arXiv.2010.11929}.
\newblock URL \url{https://arxiv.org/pdf/2010.11929.pdf}.
\newblock arXiv:2010.11929 [cs].

\bibitem[Dretske(1989)]{dretske1989reasons}
F.~Dretske.
\newblock Reasons and causes.
\newblock \emph{Philosophical Perspectives}, 3:\penalty0 1--15, 1989.

\bibitem[Dretske(1991)]{dretske1991explaining}
F.~Dretske.
\newblock \emph{Explaining behavior: Reasons in a world of causes}.
\newblock MIT press, 1991.

\bibitem[Driess et~al.(2023)Driess, Xia, Sajjadi, Lynch, Chowdhery, Ichter, Wahid, Tompson, Vuong, Yu, Huang, Chebotar, Sermanet, Duckworth, Levine, Vanhoucke, Hausman, Toussaint, Greff, Zeng, Mordatch, and Florence]{driess_palm-e:_2023}
D.~Driess, F.~Xia, M.~S.~M. Sajjadi, C.~Lynch, A.~Chowdhery, B.~Ichter, A.~Wahid, J.~Tompson, Q.~Vuong, T.~Yu, W.~Huang, Y.~Chebotar, P.~Sermanet, D.~Duckworth, S.~Levine, V.~Vanhoucke, K.~Hausman, M.~Toussaint, K.~Greff, A.~Zeng, I.~Mordatch, and P.~Florence.
\newblock {PaLM}-{E}: {An} {Embodied} {Multimodal} {Language} {Model}, Mar. 2023.
\newblock URL \url{https://arxiv.org/pdf/2303.03378.pdf}.
\newblock arXiv:2303.03378 [cs].

\bibitem[Du(2023)]{du_personalization_2023}
Y.~R. Du.
\newblock Personalization, {Echo} {Chambers}, {News} {Literacy}, and {Algorithmic} {Literacy}: {A} {Qualitative} {Study} of {AI}-{Powered} {News} {App} {Users}.
\newblock \emph{Journal of Broadcasting \& Electronic Media}, 67\penalty0 (3):\penalty0 246--273, May 2023.
\newblock ISSN 0883-8151, 1550-6878.
\newblock \doi{10.1080/08838151.2023.2182787}.
\newblock URL \url{https://www.tandfonline.com/doi/full/10.1080/08838151.2023.2182787}.

\bibitem[Duarte et~al.(2012)Duarte, Siegel, and Young]{duarte_trust_2012}
J.~Duarte, S.~Siegel, and L.~Young.
\newblock Trust and {Credit}: {The} {Role} of {Appearance} in {Peer}-to-peer {Lending}.
\newblock \emph{Review of Financial Studies}, 25\penalty0 (8):\penalty0 2455--2484, Aug. 2012.
\newblock ISSN 0893-9454, 1465-7368.
\newblock \doi{10.1093/rfs/hhs071}.
\newblock URL \url{https://academic.oup.com/rfs/article-lookup/doi/10.1093/rfs/hhs071}.

\bibitem[DuBois and Eaton(1996)]{dubois_philadelphia_1996}
W.~E.~B. DuBois and I.~Eaton.
\newblock \emph{The {Philadelphia} {Negro}: {A} {Social} {Study}}.
\newblock University of Pennsylvania Press, 1996.
\newblock ISBN 9780812215731.
\newblock URL \url{https://www.jstor.org/stable/j.ctt3fhpfb}.

\bibitem[Dunlop(2023)]{dunlop2023eu}
C.~Dunlop.
\newblock An eu ai act that works for people and society.
\newblock \url{https://www.adalovelaceinstitute.org/policy-briefing/eu-ai-act-trilogues/}, 2023.

\bibitem[Durán and Formanek(2018)]{duran_grounds_2018}
J.~M. Durán and N.~Formanek.
\newblock Grounds for {Trust}: {Essential} {Epistemic} {Opacity} and {Computational} {Reliabilism}.
\newblock \emph{Minds and Machines}, 28\penalty0 (4):\penalty0 645--666, Dec. 2018.
\newblock ISSN 1572-8641.
\newblock \doi{10.1007/s11023-018-9481-6}.
\newblock URL \url{https://doi.org/10.1007/s11023-018-9481-6}.

\bibitem[Dwork(2006)]{dwork_differential_2006}
C.~Dwork.
\newblock Differential {Privacy}.
\newblock In M.~Bugliesi, B.~Preneel, V.~Sassone, and I.~Wegener, editors, \emph{Automata, {Languages} and {Programming}}, Lecture {Notes} in {Computer} {Science}, pages 1--12, Berlin, Heidelberg, 2006. Springer.
\newblock ISBN 9783540359081.
\newblock \doi{10.1007/11787006_1}.

\bibitem[Dwork(2008)]{dwork_differential_2008}
C.~Dwork.
\newblock Differential {Privacy}: {A} {Survey} of {Results}.
\newblock In M.~Agrawal, D.~Du, Z.~Duan, and A.~Li, editors, \emph{Theory and {Applications} of {Models} of {Computation}}, Lecture {Notes} in {Computer} {Science}, pages 1--19, Berlin, Heidelberg, 2008. Springer.
\newblock ISBN 9783540792284.
\newblock \doi{10.1007/978-3-540-79228-4_1}.

\bibitem[Dworkin(1988)]{dworkin_theory_1988}
G.~Dworkin.
\newblock \emph{The {Theory} and {Practice} of {Autonomy}}.
\newblock Cambridge University Press, New York, 1988.

\bibitem[Dworkin(2020)]{dworkin_paternalism_2020}
G.~Dworkin.
\newblock Paternalism.
\newblock In E.~N. Zalta, editor, \emph{The {Stanford} {Encyclopedia} of {Philosophy}}. Metaphysics Research Lab, Stanford University, fall 2020 edition, 2020.
\newblock URL \url{https://plato.stanford.edu/archives/fall2020/entries/paternalism/}.

\bibitem[Dworkin(2013)]{dworkin_taking_2013}
R.~Dworkin.
\newblock \emph{Taking rights seriously}.
\newblock Bloomsbury revelations series. Bloomsbury, London, paperback ed edition, 2013.
\newblock ISBN 9781780936857 9781780937564.

\bibitem[Dziri et~al.(2022)Dziri, Milton, Yu, Zaiane, and Reddy]{dziri_origin_2022}
N.~Dziri, S.~Milton, M.~Yu, O.~Zaiane, and S.~Reddy.
\newblock On the {Origin} of {Hallucinations} in {Conversational} {Models}: {Is} it the {Datasets} or the {Models}?
\newblock In M.~Carpuat, M.-C. de~Marneffe, and I.~V. Meza~Ruiz, editors, \emph{Proceedings of the 2022 {Conference} of the {North} {American} {Chapter} of the {Association} for {Computational} {Linguistics}: {Human} {Language} {Technologies}}, pages 5271--5285, Seattle, United States, July 2022. Association for Computational Linguistics.
\newblock \doi{10.18653/v1/2022.naacl-main.387}.
\newblock URL \url{https://aclanthology.org/2022.naacl-main.387}.

\bibitem[Ecker et~al.(2022)Ecker, Lewandowsky, Cook, Schmid, Fazio, Brashier, Kendeou, Vraga, and Amazeen]{ecker_psychological_2022}
U.~K.~H. Ecker, S.~Lewandowsky, J.~Cook, P.~Schmid, L.~K. Fazio, N.~Brashier, P.~Kendeou, E.~K. Vraga, and M.~A. Amazeen.
\newblock The psychological drivers of misinformation belief and its resistance to correction.
\newblock \emph{Nature Reviews Psychology}, 1\penalty0 (1):\penalty0 13--29, Jan. 2022.
\newblock ISSN 2731-0574.
\newblock \doi{10.1038/s44159-021-00006-y}.
\newblock URL \url{https://www.nature.com/articles/s44159-021-00006-y}.

\bibitem[{Economist Intelligence Unit}(2005)]{unit2005economist}
{Economist Intelligence Unit}.
\newblock The economist intelligence unit’s quality-of-life index.
\newblock \emph{Retrieved July}, 2005\penalty0 (17):\penalty0 245--77, 2005.

\bibitem[Edgeworth(1879)]{edgeworth1879hedonical}
F.~Y. Edgeworth.
\newblock The hedonical calculus.
\newblock \emph{Mind}, 4\penalty0 (15):\penalty0 394--408, 1879.

\bibitem[Efferson et~al.(2008)Efferson, Lalive, and Fehr]{efferson_coevolution_2008}
C.~Efferson, R.~Lalive, and E.~Fehr.
\newblock The {Coevolution} of {Cultural} {Groups} and {Ingroup} {Favoritism}.
\newblock \emph{Science}, 321\penalty0 (5897):\penalty0 1844--1849, 2008.
\newblock ISSN 0036-8075.
\newblock URL \url{https://www.jstor.org/stable/20144903}.

\bibitem[{\'{E}}gert et~al.(2022){\'{E}}gert, de~la Maisonneuve, and Turner]{egert_new_2022}
B.~{\'{E}}gert, C.~de~la Maisonneuve, and D.~Turner.
\newblock A new macroeconomic measure of human capital exploiting {PISA} and {PIAAC}: {Linking} education policies to productivity.
\newblock {OECD} {Economics} {Department} {Working} {Papers} 1709, Apr. 2022.
\newblock URL \url{https://www.oecd-ilibrary.org/economics/a-new-macroeconomic-measure-of-human-capital-exploiting-pisa-and-piaac-linking-education-policies-to-productivity_a1046e2e-en}.

\bibitem[El-Sayed et~al.({Unpublished Manuscript})El-Sayed, Akbulut, McCroskery, Keeling, Kenton, Howard, Marchal, Manzini, Shevlane, Vallor, Susser, Franklin, Bridgers, Law, Rahtz, Shanahan, Tessler, Douillard, Everitt, and Brown]{elsayed_persuasion}
S.~El-Sayed, C.~Akbulut, A.~McCroskery, G.~Keeling, Z.~Kenton, Z.~Howard, N.~Marchal, A.~Manzini, T.~Shevlane, S.~Vallor, D.~Susser, M.~Franklin, S.~Bridgers, H.~Law, M.~Rahtz, M.~Shanahan, M.~H. Tessler, A.~Douillard, T.~Everitt, and S.~Brown.
\newblock {A Mechanism-Based Approach to Mitigating Harms from Persuasive Generative AI}.
\newblock {Unpublished Manuscript}.

\bibitem[Eleti et~al.(2023)Eleti, Harris, and Kilpatrick]{eleti_function_2023}
A.~Eleti, J.~Harris, and L.~Kilpatrick.
\newblock Function calling and other {API} updates, June 2023.
\newblock URL \url{https://openai.com/blog/function-calling-and-other-api-updates}.
\newblock publisher: OpenAI.

\bibitem[Elhage et~al.(2021)Elhage, Nanda, Olsson, Henighan, Joseph, Mann, Askell, Bai, Chen, Conerly, DasSarma, Drain, Ganguli, Hatfield-Dodds, Hernandez, Jones, Kernion, Lovitt, Ndousse, Amodei, Brown, Clark, Kaplan, McCandlish, and Olah]{elhage_mathematical_2021}
N.~Elhage, N.~Nanda, C.~Olsson, T.~Henighan, N.~Joseph, B.~Mann, A.~Askell, Y.~Bai, A.~Chen, T.~Conerly, N.~DasSarma, D.~Drain, D.~Ganguli, Z.~Hatfield-Dodds, D.~Hernandez, A.~Jones, J.~Kernion, L.~Lovitt, K.~Ndousse, D.~Amodei, T.~Brown, J.~Clark, J.~Kaplan, S.~McCandlish, and C.~Olah.
\newblock A {Mathematical} {Framework} for {Transformer} {Circuits}, Dec. 2021.
\newblock URL \url{https://transformer-circuits.pub/2021/framework/index.html}.
\newblock publisher: Anthropic.

\bibitem[Elhage et~al.(2022)Elhage, Hume, Olsson, Schiefer, Henighan, Kravec, Hatfield-Dodds, Lasenby, Drain, Chen, et~al.]{elhage_toy_2022}
N.~Elhage, T.~Hume, C.~Olsson, N.~Schiefer, T.~Henighan, S.~Kravec, Z.~Hatfield-Dodds, R.~Lasenby, D.~Drain, C.~Chen, et~al.
\newblock Toy models of superposition.
\newblock \emph{arXiv preprint arXiv:2209.10652}, 2022.

\bibitem[Eloundou et~al.(2023)Eloundou, Manning, Mishkin, and Rock]{eloundou2023gpts}
T.~Eloundou, S.~Manning, P.~Mishkin, and D.~Rock.
\newblock Gpts are gpts: An early look at the labor market impact potential of large language models.
\newblock \emph{arXiv preprint arXiv:2303.10130}, 2023.

\bibitem[Engelen and Nys(2020)]{engelen_nudging_2020}
B.~Engelen and T.~Nys.
\newblock Nudging and {Autonomy}: {Analyzing} and {Alleviating} the {Worries}.
\newblock \emph{Review of Philosophy and Psychology}, 11\penalty0 (1):\penalty0 137--156, Feb. 2020.
\newblock ISSN 1878-5166.
\newblock \doi{10.1007/s13164-019-00450-z}.
\newblock URL \url{https://doi.org/10.1007/s13164-019-00450-z}.

\bibitem[Entsminger et~al.(2023)Entsminger, Esposito, Tse, and Jean]{entsminger_dark_2023}
J.~Entsminger, M.~Esposito, T.~Tse, and A.~Jean.
\newblock The {Dark} {Side} of {Generative} {AI}: {Automating} {Inequality} by {Design}.
\newblock \emph{California Management Review Insights}, June 2023.
\newblock URL \url{https://cmr.berkeley.edu/2023/06/the-dark-side-of-generative-ai-automating-inequality-by-design/}.

\bibitem[Epley et~al.(2007)Epley, Waytz, and Cacioppo]{epley2007seeing}
N.~Epley, A.~Waytz, and J.~T. Cacioppo.
\newblock On seeing human: a three-factor theory of anthropomorphism.
\newblock \emph{Psychological review}, 114\penalty0 (4):\penalty0 864, 2007.

\bibitem[Epstein et~al.(2023)Epstein, Hertzmann, {the Investigators of Human Creativity}, Akten, Farid, Fjeld, Frank, Groh, Herman, Leach, Mahari, Pentland, Russakovsky, Schroeder, and Smith]{epstein_art_2023}
Z.~Epstein, A.~Hertzmann, {the Investigators of Human Creativity}, M.~Akten, H.~Farid, J.~Fjeld, M.~R. Frank, M.~Groh, L.~Herman, N.~Leach, R.~Mahari, A.~S. Pentland, O.~Russakovsky, H.~Schroeder, and A.~Smith.
\newblock Art and the science of generative {AI}.
\newblock \emph{Science}, 380\penalty0 (6650):\penalty0 1110--1111, June 2023.
\newblock ISSN 0036-8075, 1095-9203.
\newblock \doi{10.1126/science.adh4451}.
\newblock URL \url{https://www.science.org/doi/10.1126/science.adh4451}.

\bibitem[Erete et~al.(2023)Erete, Rankin, and Thomas]{erete_method_2023}
S.~Erete, Y.~Rankin, and J.~Thomas.
\newblock A {Method} to the {Madness}: {Applying} an {Intersectional} {Analysis} of {Structural} {Oppression} and {Power} in {HCI} and {Design}.
\newblock \emph{ACM Transactions on Computer-Human Interaction}, 30\penalty0 (2):\penalty0 1--45, Apr. 2023.
\newblock ISSN 1073-0516, 1557-7325.
\newblock \doi{10.1145/3507695}.
\newblock URL \url{https://dl.acm.org/doi/10.1145/3507695}.

\bibitem[Eriksson(2022)]{eriksson_design_2022}
T.~Eriksson.
\newblock Design fiction exploration of romantic interaction with virtual humans in virtual reality1.
\newblock \emph{Journal of Future Robot Life}, 3\penalty0 (1):\penalty0 63--75, Mar. 2022.
\newblock ISSN 25899961, 25899953.
\newblock \doi{10.3233/FRL-210007}.
\newblock URL \url{https://www.medra.org/servlet/aliasResolver?alias=iospress&doi=10.3233/FRL-210007}.

\bibitem[Eshoo(2022)]{eshoo_eshoo_2022}
A.~Eshoo.
\newblock Eshoo {Urges} {NSA} \& {OSTP} to {Address} {Unsafe} {AI} {Practices}, Sept. 2022.
\newblock URL \url{http://eshoo.house.gov/media/press-releases/eshoo-urges-nsa-ostp-address-unsafe-ai-practices}.

\bibitem[Eubanks(2006)]{eubanks_technologies_2006}
V.~Eubanks.
\newblock Technologies of {Citizenship}: {Surveillance} and {Political} {Learning} in the {Welfare} {System}.
\newblock In \emph{Surveillance and {Security}}. Routledge, 2006.
\newblock ISBN 9780203957257.

\bibitem[Eubanks(2017)]{eubanks_automating_2017}
V.~Eubanks.
\newblock \emph{Automating inequality: how high-tech tools profile, police, and punish the poor}.
\newblock St. Martin's Press, New York, NY, first edition edition, 2017.
\newblock ISBN 9781250074317.

\bibitem[{European Commission}(2019)]{european_commission_ethics_2019}
{European Commission}.
\newblock Ethics guidelines for trustworthy {AI}, Apr. 2019.
\newblock URL \url{https://digital-strategy.ec.europa.eu/en/library/ethics-guidelines-trustworthy-ai}.

\bibitem[{European Commission}(2021)]{european_commission_regulatory_2021}
{European Commission}.
\newblock Regulatory framework proposal on artificial intelligence, 2021.
\newblock URL \url{https://digital-strategy.ec.europa.eu/en/policies/regulatory-framework-ai}.

\bibitem[{European Parliament}(2023)]{european_parliament_proposal_nodate}
{European Parliament}.
\newblock Proposal for a regulation of the {European} {Parliament} and of the {Council} on harmonised rules on {Artificial} {Intelligence} ({Artificial} {Intelligence} {Act}) and amending certain {Union} {Legislative} {Acts}, 2023.
\newblock URL \url{https://www.europarl.europa.eu/meetdocs/2014_2019/plmrep/COMMITTEES/CJ40/DV/2023/05-11/ConsolidatedCA_IMCOLIBE_AI_ACT_EN.pdf}.

\bibitem[{European Parliamentary Research Service}(2020)]{european_parliamentary_research_service_teaching:_2020}
{European Parliamentary Research Service}.
\newblock Teaching: {A} {Woman}'s {World}.
\newblock Technical Report PE 646.191, European Parliament, Mar. 2020.
\newblock URL \url{https://www.europarl.europa.eu/RegData/etudes/ATAG/2020/646191/EPRS_ATA(2020)646191_EN.pdf}.

\bibitem[Evans and Kasirzadeh(2023)]{evans_user_2023}
C.~Evans and A.~Kasirzadeh.
\newblock User {Tampering} in {Reinforcement} {Learning} {Recommender} {Systems}.
\newblock In \emph{Proceedings of the 2023 {AAAI}/{ACM} {Conference} on {AI}, {Ethics}, and {Society}}, pages 58--69, Aug. 2023.
\newblock \doi{10.1145/3600211.3604669}.
\newblock URL \url{http://arxiv.org/abs/2109.04083}.
\newblock arXiv:2109.04083 [cs].

\bibitem[Everitt et~al.(2021)Everitt, Carey, Langlois, Ortega, and Legg]{everitt_agent_2021}
T.~Everitt, R.~Carey, E.~D. Langlois, P.~A. Ortega, and S.~Legg.
\newblock Agent {Incentives}: {A} {Causal} {Perspective}.
\newblock \emph{Proceedings of the AAAI Conference on Artificial Intelligence}, 35\penalty0 (13):\penalty0 11487--11495, May 2021.
\newblock ISSN 2374-3468.
\newblock \doi{10.1609/aaai.v35i13.17368}.
\newblock URL \url{https://ojs.aaai.org/index.php/AAAI/article/view/17368}.

\bibitem[Eyal(2014)]{eyal2014hooked}
N.~Eyal.
\newblock \emph{Hooked: How to build habit-forming products}.
\newblock Penguin, 2014.

\bibitem[Eysenbach et~al.(2001)]{eysenbach2001health}
G.~Eysenbach et~al.
\newblock What is e-health?
\newblock \emph{Journal of Medical Internet Research}, 3\penalty0 (2):\penalty0 e833, 2001.

\bibitem[Eyssel and Reich(2013)]{reich_2013_loneliness}
F.~Eyssel and N.~Reich.
\newblock Loneliness makes the heart grow fonder (of robots)—on the effects of loneliness on psychological anthropomorphism.
\newblock In \emph{2013 8th acm/ieee international conference on human-robot interaction (hri)}, pages 121--122. IEEE, 2013.

\bibitem[Faden and Beauchamp(1986)]{faden_history_1986}
R.~R. Faden and T.~L. Beauchamp.
\newblock \emph{A {History} and {Theory} of {Informed} {Consent}}.
\newblock Oxford University Press, Feb. 1986.
\newblock ISBN 9780199748655.
\newblock Google-Books-ID: jgi7OWxDT9cC.

\bibitem[Fahim et~al.(2014)Fahim, Idris, Ali, Nugent, Kang, Huh, and Lee]{fahim2014athena}
M.~Fahim, M.~Idris, R.~Ali, C.~Nugent, B.~Kang, E.-N. Huh, and S.~Lee.
\newblock Athena: a personalized platform to promote an active lifestyle and wellbeing based on physical, mental and social health primitives.
\newblock \emph{Sensors}, 14\penalty0 (5):\penalty0 9313--9329, 2014.

\bibitem[Fan et~al.(2023)Fan, Zhao, Li, Liu, Mei, Wang, Wen, Wang, Zhao, Tang, and Li]{fan_recommender_2023}
W.~Fan, Z.~Zhao, J.~Li, Y.~Liu, X.~Mei, Y.~Wang, Z.~Wen, F.~Wang, X.~Zhao, J.~Tang, and Q.~Li.
\newblock Recommender {Systems} in the {Era} of {Large} {Language} {Models} ({LLMs}), Aug. 2023.
\newblock URL \url{http://arxiv.org/abs/2307.02046}.
\newblock arXiv:2307.02046 [cs].

\bibitem[Farquhar et~al.(2022)Farquhar, Carey, and Everitt]{farquhar2022path}
S.~Farquhar, R.~Carey, and T.~Everitt.
\newblock Path-specific objectives for safer agent incentives.
\newblock In \emph{Proceedings of the AAAI Conference on Artificial Intelligence}, volume~36, pages 9529--9538, 2022.

\bibitem[Farquhar et~al.(2023)Farquhar, Varma, Kenton, Gasteiger, Mikulik, and Shah]{farquhar2023challenges}
S.~Farquhar, V.~Varma, Z.~Kenton, J.~Gasteiger, V.~Mikulik, and R.~Shah.
\newblock Challenges with unsupervised llm knowledge discovery.
\newblock \emph{arXiv preprint arXiv:2312.10029}, 2023.

\bibitem[Fatafta(2021)]{fatafta_facebook_2021}
M.~Fatafta.
\newblock Facebook is bad at moderating in {English}. {In} {Arabic}, it’s a disaster, Nov. 2021.
\newblock URL \url{https://restofworld.org/2021/facebook-is-bad-at-moderating-in-english-in-arabic-its-a-disaster/}.

\bibitem[Feagin(2013)]{feagin_systemic_2013}
J.~Feagin.
\newblock \emph{Systemic {Racism}: {A} {Theory} of {Oppression}}.
\newblock Routledge, Sept. 2013.
\newblock ISBN 9781134729005.
\newblock Google-Books-ID: uerZAAAAQBAJ.

\bibitem[Feagin(1991)]{feagin_continuing_1991}
J.~R. Feagin.
\newblock The {Continuing} {Significance} of {Race}: {Antiblack} {Discrimination} in {Public} {Places}.
\newblock \emph{American Sociological Review}, 56\penalty0 (1):\penalty0 101, Feb. 1991.
\newblock ISSN 00031224.
\newblock \doi{10.2307/2095676}.
\newblock URL \url{http://www.jstor.org/stable/2095676?origin=crossref}.

\bibitem[Feij{\'o}o et~al.(2020)Feij{\'o}o, Kwon, Bauer, Bohlin, Howell, Jain, Potgieter, Vu, Whalley, and Xia]{feijoo2020harnessing}
C.~Feij{\'o}o, Y.~Kwon, J.~M. Bauer, E.~Bohlin, B.~Howell, R.~Jain, P.~Potgieter, K.~Vu, J.~Whalley, and J.~Xia.
\newblock Harnessing artificial intelligence ({AI}) to increase wellbeing for all: The case for a new technology diplomacy.
\newblock \emph{Telecommunications Policy}, 44\penalty0 (6):\penalty0 101988, 2020.

\bibitem[Feinberg(1987)]{feinberg_moral_1987}
J.~Feinberg.
\newblock \emph{The {Moral} {Limits} of the {Criminal} {Law} {Volume} 1: {Harm} to {Others}}.
\newblock Oxford University PressNew York, 1 edition, Aug. 1987.
\newblock ISBN 9780195046649 9780199868728.
\newblock \doi{10.1093/0195046641.001.0001}.
\newblock URL \url{https://academic.oup.com/book/1573}.

\bibitem[Felten et~al.(2019)Felten, Raj, and Seamans]{felten_occupational_2019}
E.~W. Felten, M.~Raj, and R.~Seamans.
\newblock The {Occupational} {Impact} of {Artificial} {Intelligence}: {Labor}, {Skills}, and {Polarization}, Sept. 2019.
\newblock URL \url{https://papers.ssrn.com/abstract=3368605}.

\bibitem[Felten et~al.(2023)Felten, Raj, and Seamans]{felten2023occupational}
E.~W. Felten, M.~Raj, and R.~Seamans.
\newblock Occupational heterogeneity in exposure to generative ai.
\newblock \emph{Available at SSRN 4414065}, 2023.

\bibitem[Feng et~al.(2023)Feng, Park, Liu, and Tsvetkov]{feng_pretraining_2023}
S.~Feng, C.~Y. Park, Y.~Liu, and Y.~Tsvetkov.
\newblock From {Pretraining} {Data} to {Language} {Models} to {Downstream} {Tasks}: {Tracking} the {Trails} of {Political} {Biases} {Leading} to {Unfair} {NLP} {Models}, July 2023.
\newblock URL \url{https://arxiv.org/pdf/2305.08283.pdf}.
\newblock arXiv:2305.08283 [cs].

\bibitem[Ferguson(2017)]{ferguson_rise_2017}
A.~G. Ferguson.
\newblock \emph{The {Rise} of {Big} {Data} {Policing}}.
\newblock NYU Press, Oct. 2017.
\newblock URL \url{https://nyupress.org/9781479892822/the-rise-of-big-data-policing}.

\bibitem[Ferguson and Allen(1998)]{ferguson_trips:_1998}
G.~Ferguson and J.~F. Allen.
\newblock {TRIPS}: {An} {Integrated} {Intelligent} {Problem}-{Solving} {Assistant}.
\newblock In J.~Mostow and C.~Rich, editors, \emph{Proceedings of the {Fifteenth} {National} {Conference} on {Artificial} {Intelligence} and {Tenth} {Innovative} {Applications} of {Artificial} {Intelligence} {Conference}, {AAAI} 98, {IAAI} 98, {July} 26-30, 1998, {Madison}, {Wisconsin}, {USA}}, pages 567--572. AAAI Press / The MIT Press, 1998.
\newblock URL \url{https://www.cs.rochester.edu/research/cisd/pubs/1998/ferguson-allen-aaai98.pdf}.

\bibitem[Fernandes et~al.(2023)Fernandes, Madaan, Liu, Farinhas, Martins, Bertsch, de~Souza, Zhou, Wu, Neubig, and Martins]{fernandes2023bridging}
P.~Fernandes, A.~Madaan, E.~Liu, A.~Farinhas, P.~H. Martins, A.~Bertsch, J.~G.~C. de~Souza, S.~Zhou, T.~Wu, G.~Neubig, and A.~F.~T. Martins.
\newblock Bridging the gap: A survey on integrating (human) feedback for natural language generation, 2023.

\bibitem[Ferrario et~al.(2020)Ferrario, Loi, and Viganò]{ferrario_ai_2020}
A.~Ferrario, M.~Loi, and E.~Viganò.
\newblock In {AI} {We} {Trust} {Incrementally}: a {Multi}-layer {Model} of {Trust} to {Analyze} {Human}-{Artificial} {Intelligence} {Interactions}.
\newblock \emph{Philosophy \& Technology}, 33\penalty0 (3):\penalty0 523--539, Sept. 2020.
\newblock ISSN 2210-5441.
\newblock \doi{10.1007/s13347-019-00378-3}.
\newblock URL \url{https://doi.org/10.1007/s13347-019-00378-3}.

\bibitem[Festerling and Siraj(2022)]{festerling_anthropomorphizing_2022}
J.~Festerling and I.~Siraj.
\newblock Anthropomorphizing technology: a conceptual review of anthropomorphism research and how it relates to children’s engagements with digital voice assistants.
\newblock \emph{Integrative Psychological and Behavioral Science}, 56\penalty0 (3):\penalty0 709--738, 2022.

\bibitem[Finkenwirth et~al.(2021)Finkenwirth, MacDonald, Deng, Lesch, and Clark]{finkenwirth_using_2021}
S.~Finkenwirth, K.~MacDonald, X.~Deng, T.~Lesch, and L.~Clark.
\newblock Using machine learning to predict self-exclusion status in online gamblers on the {PlayNow}.com platform in {British} {Columbia}.
\newblock \emph{International Gambling Studies}, 21\penalty0 (2):\penalty0 220--237, May 2021.
\newblock ISSN 1445-9795, 1479-4276.
\newblock \doi{10.1080/14459795.2020.1832132}.
\newblock URL \url{https://www.tandfonline.com/doi/full/10.1080/14459795.2020.1832132}.

\bibitem[Finley(2020)]{finley_net_nodate}
K.~Finley.
\newblock Net {Neutrality}: {Here}'s {Everything} {You} {Need} {To} {Know}.
\newblock \emph{Wired}, 2020.
\newblock ISSN 1059-1028.
\newblock URL \url{https://www.wired.com/story/guide-net-neutrality/}.

\bibitem[Fiorini and Aiello(2022)]{fiorini_automatic_2022}
L.~Fiorini and M.~Aiello.
\newblock Automatic optimal multi-energy management of smart homes.
\newblock \emph{Energy Informatics}, 5\penalty0 (1):\penalty0 68, Dec. 2022.
\newblock ISSN 2520-8942.
\newblock \doi{10.1186/s42162-022-00253-0}.
\newblock URL \url{https://doi.org/10.1186/s42162-022-00253-0}.

\bibitem[Fisher(2007)]{fisher2007mathematical}
I.~Fisher.
\newblock \emph{Mathematical investigations in the theory of value and prices, and appreciation and interest}.
\newblock Cosimo, Inc., 2007.

\bibitem[Fisher(1930)]{fisher_genetical_1930}
R.~A. Fisher.
\newblock \emph{The genetical theory of natural selection}.
\newblock Clarendon Press, Oxford, 1930.
\newblock \doi{10.5962/bhl.title.27468}.
\newblock URL \url{https://www.biodiversitylibrary.org/bibliography/27468}.

\bibitem[Fiske et~al.(2019)Fiske, Henningsen, and Buyx]{fiske_your_2019}
A.~Fiske, P.~Henningsen, and A.~Buyx.
\newblock Your {Robot} {Therapist} {Will} {See} {You} {Now}: {Ethical} {Implications} of {Embodied} {Artificial} {Intelligence} in {Psychiatry}, {Psychology}, and {Psychotherapy}.
\newblock \emph{Journal of Medical Internet Research}, 21\penalty0 (5):\penalty0 e13216, May 2019.
\newblock ISSN 1438-8871.
\newblock \doi{10.2196/13216}.
\newblock URL \url{https://www.jmir.org/2019/5/e13216/}.

\bibitem[Fjelde and De~Soysa(2009)]{fjelde_coercion_2009}
H.~Fjelde and I.~De~Soysa.
\newblock Coercion, {Co}-optation, or {Cooperation}? {State} {Capacity} and the {Risk} of {Civil} {War}, 1961–2004.
\newblock \emph{Conflict Management and Peace Science}, 26\penalty0 (1):\penalty0 5--25, 2009.
\newblock ISSN 0738-8942.
\newblock URL \url{https://www.jstor.org/stable/26275118}.

\bibitem[Flitter and Cowley(2023)]{flitter_voice_2023}
E.~Flitter and S.~Cowley.
\newblock Voice {Deepfakes} {Are} {Coming} for {Your} {Bank} {Balance}.
\newblock \emph{The New York Times}, Aug. 2023.
\newblock ISSN 0362-4331.
\newblock URL \url{https://www.nytimes.com/2023/08/30/business/voice-deepfakes-bank-scams.html}.

\bibitem[Floridi and Cowls(2022)]{carta_unified_2022}
L.~Floridi and J.~Cowls.
\newblock A {Unified} {Framework} of {Five} {Principles} for {AI} in {Society}.
\newblock In S.~Carta, editor, \emph{Machine {Learning} and the {City}}, pages 535--545. Wiley, 1 edition, May 2022.
\newblock ISBN 9781119749639 9781119815075.
\newblock \doi{10.1002/9781119815075.ch45}.
\newblock URL \url{https://onlinelibrary.wiley.com/doi/10.1002/9781119815075.ch45}.

\bibitem[Fong et~al.(2003)Fong, Nourbakhsh, and Dautenhahn]{fong_survey_2003}
T.~Fong, I.~Nourbakhsh, and K.~Dautenhahn.
\newblock A survey of socially interactive robots.
\newblock \emph{Robotics and autonomous systems}, 42\penalty0 (3-4):\penalty0 143--166, 2003.

\bibitem[Fox et~al.(2021)Fox, Goedde-Menke, and Tannenbaum]{fox_ambiguity_2021}
C.~R. Fox, M.~Goedde-Menke, and D.~Tannenbaum.
\newblock Ambiguity {Aversion} and {Epistemic} {Uncertainty}, Sept. 2021.
\newblock URL \url{https://papers.ssrn.com/abstract=3922716}.

\bibitem[Franceschelli and Musolesi(2023)]{franceschelli_creativity_2023}
G.~Franceschelli and M.~Musolesi.
\newblock On the {Creativity} of {Large} {Language} {Models}, July 2023.
\newblock URL \url{http://arxiv.org/abs/2304.00008}.
\newblock arXiv:2304.00008 [cs].

\bibitem[Frank(2023)]{Frank_2023}
M.~C. Frank.
\newblock Baby steps in evaluating the capacities of large language models.
\newblock \emph{Nature Reviews Psychology}, 2\penalty0 (8):\penalty0 451–452, Aug. 2023.
\newblock ISSN 2731-0574.
\newblock \doi{10.1038/s44159-023-00211-x}.
\newblock URL \url{https://www.nature.com/articles/s44159-023-00211-x}.

\bibitem[Franklin et~al.(2022)Franklin, Ashton, Gorman, and Armstrong]{franklin_missing_2022}
M.~Franklin, H.~Ashton, R.~Gorman, and S.~Armstrong.
\newblock Missing {Mechanisms} of {Manipulation} in the {EU} {AI} {Act}.
\newblock \emph{The International FLAIRS Conference Proceedings}, 35, May 2022.
\newblock ISSN 2334-0762.
\newblock \doi{10.32473/flairs.v35i.130723}.
\newblock URL \url{https://journals.flvc.org/FLAIRS/article/view/130723}.

\bibitem[Franklin et~al.(2023)Franklin, Tomei, and Gorman]{franklin_strengthening_2023}
M.~Franklin, P.~M. Tomei, and R.~Gorman.
\newblock Strengthening the {EU} {AI} {Act}: {Defining} {Key} {Terms} on {AI} {Manipulation}, Aug. 2023.
\newblock URL \url{http://arxiv.org/abs/2308.16364}.
\newblock arXiv:2308.16364 [cs] version: 1.

\bibitem[Freiman(2023)]{freiman2023making}
O.~Freiman.
\newblock Making sense of the conceptual nonsense ‘trustworthy ai’.
\newblock \emph{AI and Ethics}, 3\penalty0 (4):\penalty0 1351--1360, 2023.

\bibitem[Frenkel et~al.(2020)Frenkel, Decker, and Alba]{frenkel_how_2020}
S.~Frenkel, B.~Decker, and D.~Alba.
\newblock How the ‘{Plandemic}’ {Movie} and {Its} {Falsehoods} {Spread} {Widely} {Online}.
\newblock \emph{The New York Times}, May 2020.
\newblock ISSN 0362-4331.
\newblock URL \url{https://www.nytimes.com/2020/05/20/technology/plandemic-movie-youtube-facebook-coronavirus.html}.

\bibitem[Frey and Osborne(2013)]{frey_future_2013}
C.~B. Frey and M.~Osborne.
\newblock The {Future} of {Employment}: {How} susceptible are jobs to computerisation?, Sept. 2013.
\newblock URL \url{https://www.oxfordmartin.ox.ac.uk/publications/the-future-of-employment/}.

\bibitem[Fromm(2000)]{fromm2000art}
E.~Fromm.
\newblock \emph{The art of loving: The centennial edition}.
\newblock A\&C Black, 2000.

\bibitem[Fu et~al.(2012)Fu, Tarnita, Christakis, Wang, Rand, and Nowak]{fu_evolution_2012}
F.~Fu, C.~E. Tarnita, N.~A. Christakis, L.~Wang, D.~G. Rand, and M.~A. Nowak.
\newblock Evolution of in-group favoritism.
\newblock \emph{Scientific Reports}, 2\penalty0 (1):\penalty0 460, June 2012.
\newblock ISSN 2045-2322.
\newblock \doi{10.1038/srep00460}.
\newblock URL \url{https://www.nature.com/articles/srep00460}.

\bibitem[Gable and Haidt(2005)]{gable2005and}
S.~L. Gable and J.~Haidt.
\newblock What (and why) is positive psychology?
\newblock \emph{Review of General Psychology}, 9\penalty0 (2):\penalty0 103--110, 2005.

\bibitem[Gabriel(2020)]{gabriel_artificial_2020}
I.~Gabriel.
\newblock Artificial {Intelligence}, {Values} and {Alignment}.
\newblock \emph{Minds and Machines}, 30\penalty0 (3):\penalty0 411--437, Sept. 2020.
\newblock ISSN 0924-6495, 1572-8641.
\newblock \doi{10.1007/s11023-020-09539-2}.
\newblock URL \url{http://arxiv.org/abs/2001.09768}.
\newblock arXiv:2001.09768 [cs].

\bibitem[Gabriel(2022)]{gabriel_toward_2022}
I.~Gabriel.
\newblock Toward a {Theory} of {Justice} for {Artificial} {Intelligence}.
\newblock \emph{Daedalus}, 151\penalty0 (2):\penalty0 218--231, May 2022.
\newblock ISSN 0011-5266, 1548-6192.
\newblock \doi{10.1162/daed_a_01911}.
\newblock URL \url{https://direct.mit.edu/daed/article/151/2/218/110610/Toward-a-Theory-of-Justice-for-Artificial}.

\bibitem[Gabriel and Ghazavi(2021)]{gabriel_challenge_2021}
I.~Gabriel and V.~Ghazavi.
\newblock The {Challenge} of {Value} {Alignment}: from {Fairer} {Algorithms} to {AI} {Safety}, Jan. 2021.
\newblock URL \url{http://arxiv.org/abs/2101.06060}.
\newblock arXiv:2101.06060 [cs].

\bibitem[Gadiraju et~al.(2023)Gadiraju, Kane, Dev, Taylor, Wang, Denton, and Brewer]{gadiraju_i_2023}
V.~Gadiraju, S.~Kane, S.~Dev, A.~Taylor, D.~Wang, E.~Denton, and R.~Brewer.
\newblock "{I} wouldn’t say offensive but...": {Disability}-{Centered} {Perspectives} on {Large} {Language} {Models}.
\newblock In \emph{2023 {ACM} {Conference} on {Fairness}, {Accountability}, and {Transparency}}, pages 205--216, Chicago IL USA, June 2023. ACM.
\newblock ISBN 9798400701924.
\newblock \doi{10.1145/3593013.3593989}.
\newblock URL \url{https://dl.acm.org/doi/10.1145/3593013.3593989}.

\bibitem[Gahntz(2023)]{gahntz2023eu}
M.~Gahntz.
\newblock The eu’s ai act and foundation models: The final stretch.
\newblock \url{https://foundation.mozilla.org/en/blog/the-eus-ai-act-and-foundation-models-the-final-stretch/}, 2023.

\bibitem[Gain(2021)]{gain_githubs_2021}
V.~Gain.
\newblock {GitHub}’s {AI} {Copilot} is helping write 30pc of new code on the platform, Oct. 2021.
\newblock URL \url{https://www.siliconrepublic.com/machines/github-copilot-ai-tool}.

\bibitem[G{\'a}l et~al.(2021)G{\'a}l, Ștefan, and Cristea]{gal2021efficacy}
{\'E}.~G{\'a}l, S.~Ștefan, and I.~A. Cristea.
\newblock The efficacy of mindfulness meditation apps in enhancing users’ well-being and mental health related outcomes: a meta-analysis of randomized controlled trials.
\newblock \emph{Journal of Affective Disorders}, 279:\penalty0 131--142, 2021.

\bibitem[Gallup-Healthways(2009)]{gallup}
Gallup-Healthways.
\newblock Gallup-healthways well-being index: methodology report for indixes, 2009.

\bibitem[Gambetta(1988)]{gambetta_can_1988}
D.~Gambetta.
\newblock Can {We} {Trust} {Trust}?
\newblock In D.~Gambetta, editor, \emph{Trust: {Making} and {Breaking} {Cooperative} {Relations}}, pages 213--237. Blackwell, 1988.

\bibitem[Gambino et~al.(2020{\natexlab{a}})Gambino, Fox, and Ratan]{gambino_building_2020}
A.~Gambino, J.~Fox, and R.~Ratan.
\newblock Building a {Stronger} {CASA}: {Extending} the {Computers} {Are} {Social} {Actors} {Paradigm}.
\newblock \emph{Human-Machine Communication}, 1:\penalty0 71--86, Feb. 2020{\natexlab{a}}.
\newblock ISSN 2638-6038, 2638-602X.
\newblock \doi{10.30658/hmc.1.5}.
\newblock URL \url{https://stars.library.ucf.edu/hmc/vol1/iss1/5/}.

\bibitem[Gambino et~al.(2020{\natexlab{b}})Gambino, Fox, and Ratan]{gambino2020building}
A.~Gambino, J.~Fox, and R.~A. Ratan.
\newblock Building a stronger casa: Extending the computers are social actors paradigm.
\newblock \emph{Human-Machine Communication}, 1:\penalty0 71--85, 2020{\natexlab{b}}.

\bibitem[Ganguli et~al.(2022)Ganguli, Hernandez, Lovitt, DasSarma, Henighan, Jones, Joseph, Kernion, Mann, Askell, Bai, Chen, Conerly, Drain, Elhage, Showk, Fort, Hatfield-Dodds, Johnston, Kravec, Nanda, Ndousse, Olsson, Amodei, Amodei, Brown, Kaplan, McCandlish, Olah, and Clark]{ganguli_predictability_2022}
D.~Ganguli, D.~Hernandez, L.~Lovitt, N.~DasSarma, T.~Henighan, A.~Jones, N.~Joseph, J.~Kernion, B.~Mann, A.~Askell, Y.~Bai, A.~Chen, T.~Conerly, D.~Drain, N.~Elhage, S.~E. Showk, S.~Fort, Z.~Hatfield-Dodds, S.~Johnston, S.~Kravec, N.~Nanda, K.~Ndousse, C.~Olsson, D.~Amodei, D.~Amodei, T.~Brown, J.~Kaplan, S.~McCandlish, C.~Olah, and J.~Clark.
\newblock Predictability and {Surprise} in {Large} {Generative} {Models}.
\newblock In \emph{2022 {ACM} {Conference} on {Fairness}, {Accountability}, and {Transparency}}, pages 1747--1764, June 2022.
\newblock \doi{10.1145/3531146.3533229}.
\newblock URL \url{http://arxiv.org/abs/2202.07785}.
\newblock arXiv:2202.07785 [cs].

\bibitem[Gao et~al.(2022)Gao, Schulman, and Hilton]{gao_scaling_2022}
L.~Gao, J.~Schulman, and J.~Hilton.
\newblock Scaling {Laws} for {Reward} {Model} {Overoptimization}, Oct. 2022.
\newblock URL \url{http://arxiv.org/abs/2210.10760}.
\newblock arXiv:2210.10760 [cs, stat].

\bibitem[Gardner et~al.(2023)Gardner, Arden, Brown, Eves, Green, Hamilton, Hankonen, Inauen, Keller, Kwasnicka, et~al.]{gardner2023developing}
B.~Gardner, M.~A. Arden, D.~Brown, F.~F. Eves, J.~Green, K.~Hamilton, N.~Hankonen, J.~Inauen, J.~Keller, D.~Kwasnicka, et~al.
\newblock Developing habit-based health behaviour change interventions: Twenty-one questions to guide future research.
\newblock \emph{Psychology \& Health}, 38\penalty0 (4):\penalty0 518--540, 2023.

\bibitem[Garg and Sengupta(2020)]{garg_he_2020}
R.~Garg and S.~Sengupta.
\newblock He {Is} {Just} {Like} {Me}: {A} {Study} of the {Long}-{Term} {Use} of {Smart} {Speakers} by {Parents} and {Children}.
\newblock \emph{Proceedings of the ACM on Interactive, Mobile, Wearable and Ubiquitous Technologies}, 4\penalty0 (1):\penalty0 11:1--11:24, Mar. 2020.
\newblock \doi{10.1145/3381002}.
\newblock URL \url{https://doi.org/10.1145/3381002}.

\bibitem[Gaube et~al.(2021)Gaube, Suresh, Raue, Merritt, Berkowitz, Lermer, Coughlin, Guttag, Colak, and Ghassemi]{gaube_as_2021}
S.~Gaube, H.~Suresh, M.~Raue, A.~Merritt, S.~J. Berkowitz, E.~Lermer, J.~F. Coughlin, J.~V. Guttag, E.~Colak, and M.~Ghassemi.
\newblock Do as {AI} say: susceptibility in deployment of clinical decision-aids.
\newblock \emph{npj Digital Medicine}, 4\penalty0 (1):\penalty0 1--8, Feb. 2021.
\newblock ISSN 2398-6352.
\newblock \doi{10.1038/s41746-021-00385-9}.
\newblock URL \url{https://www.nature.com/articles/s41746-021-00385-9}.

\bibitem[Ge et~al.(2020)Ge, Friedrich, and Vigna]{ge_4_2020}
M.~Ge, J.~Friedrich, and L.~Vigna.
\newblock 4 {Charts} {Explain} {Greenhouse} {Gas} {Emissions} by {Countries} and {Sectors}.
\newblock Feb. 2020.
\newblock URL \url{https://www.wri.org/insights/4-charts-explain-greenhouse-gas-emissions-countries-and-sectors}.

\bibitem[Geerdts(2016)]{geerdts_real_2016}
M.~S. Geerdts.
\newblock ({Un}){Real} {Animals}: {Anthropomorphism} and {Early} {Learning} {About} {Animals}.
\newblock \emph{Child Development Perspectives}, 10\penalty0 (1):\penalty0 10--14, Mar. 2016.
\newblock ISSN 1750-8592, 1750-8606.
\newblock \doi{10.1111/cdep.12153}.
\newblock URL \url{https://srcd.onlinelibrary.wiley.com/doi/10.1111/cdep.12153}.

\bibitem[{Gemini Team}(2023)]{team2023gemini}
{Gemini Team}.
\newblock Gemini: A family of highly capable multimodal models.
\newblock Dec. 2023.
\newblock URL \url{https://arxiv.org/abs/2312.11805}.
\newblock arXiv:2312.11805.

\bibitem[Georgieff and Hyee(2021)]{georgieff_artificial_2021}
A.~Georgieff and R.~Hyee.
\newblock Artificial intelligence and employment: {New} cross-country evidence.
\newblock {OECD} {Social}, {Employment} and {Migration} {Working} {Papers} 265, OECD, Dec. 2021.
\newblock URL \url{https://www.oecd-ilibrary.org/social-issues-migration-health/artificial-intelligence-and-employment_c2c1d276-en}.

\bibitem[Gerlach et~al.(2017)Gerlach, Ram, Infurna, Vogel, Wagner, and Gerstorf]{gerlach2017role}
K.~Gerlach, N.~Ram, F.~J. Infurna, N.~Vogel, G.~G. Wagner, and D.~Gerstorf.
\newblock The role of morbidity for proxy-reported well-being in the last year of life.
\newblock \emph{Developmental Psychology}, 53\penalty0 (9):\penalty0 1795, 2017.

\bibitem[Geva et~al.(2021)Geva, Schuster, Berant, and Levy]{geva_transformer_2021}
M.~Geva, R.~Schuster, J.~Berant, and O.~Levy.
\newblock Transformer {Feed}-{Forward} {Layers} {Are} {Key}-{Value} {Memories}, Sept. 2021.
\newblock URL \url{http://arxiv.org/abs/2012.14913}.
\newblock arXiv:2012.14913 [cs].

\bibitem[Geva et~al.(2022)Geva, Caciularu, Wang, and Goldberg]{geva_transformer_2022}
M.~Geva, A.~Caciularu, K.~R. Wang, and Y.~Goldberg.
\newblock Transformer {Feed}-{Forward} {Layers} {Build} {Predictions} by {Promoting} {Concepts} in the {Vocabulary} {Space}, Oct. 2022.
\newblock URL \url{http://arxiv.org/abs/2203.14680}.
\newblock arXiv:2203.14680 [cs].

\bibitem[Ghalebikesabi et~al.(2023)Ghalebikesabi, Berrada, Gowal, Ktena, Stanforth, Hayes, De, Smith, Wiles, and Balle]{Ghalebikesabi2023}
S.~Ghalebikesabi, L.~Berrada, S.~Gowal, I.~Ktena, R.~Stanforth, J.~Hayes, S.~De, S.~L. Smith, O.~Wiles, and B.~Balle.
\newblock Differentially private diffusion models generate useful synthetic images, Feb. 2023.
\newblock URL \url{http://arxiv.org/abs/2302.13861}.
\newblock arXiv:2306.01684 [lg, cr, cv].

\bibitem[Gillath et~al.(2023)Gillath, Abumusab, Ai, Branicky, Davison, Rulo, Symons, and Thomas]{gillath_how_2023}
O.~Gillath, S.~Abumusab, T.~Ai, M.~S. Branicky, R.~B. Davison, M.~Rulo, J.~Symons, and G.~Thomas.
\newblock How deep is {AI}'s love? {Understanding} relational {AI}.
\newblock \emph{Behavioral and Brain Sciences}, 46:\penalty0 e33, 2023.
\newblock ISSN 0140-525X, 1469-1825.
\newblock \doi{10.1017/S0140525X22001704}.
\newblock URL \url{https://www.cambridge.org/core/product/identifier/S0140525X22001704/type/journal_article}.

\bibitem[{Gio}(2023)]{gio_replika:_2023}
{Gio}.
\newblock Replika: {Your} {Money} or {Your} {Wife}, Mar. 2023.
\newblock URL \url{https://blog.giovanh.com/blog/2023/03/17/replika-your-money-or-your-wife/}.

\bibitem[Glaese et~al.(2022)Glaese, McAleese, Trębacz, Aslanides, Firoiu, Ewalds, Rauh, Weidinger, Chadwick, Thacker, Campbell-Gillingham, Uesato, Huang, Comanescu, Yang, See, Dathathri, Greig, Chen, Fritz, Elias, Green, Mokrá, Fernando, Wu, Foley, Young, Gabriel, Isaac, Mellor, Hassabis, Kavukcuoglu, Hendricks, and Irving]{glaese_improving_2022}
A.~Glaese, N.~McAleese, M.~Trębacz, J.~Aslanides, V.~Firoiu, T.~Ewalds, M.~Rauh, L.~Weidinger, M.~Chadwick, P.~Thacker, L.~Campbell-Gillingham, J.~Uesato, P.-S. Huang, R.~Comanescu, F.~Yang, A.~See, S.~Dathathri, R.~Greig, C.~Chen, D.~Fritz, J.~S. Elias, R.~Green, S.~Mokrá, N.~Fernando, B.~Wu, R.~Foley, S.~Young, I.~Gabriel, W.~Isaac, J.~Mellor, D.~Hassabis, K.~Kavukcuoglu, L.~A. Hendricks, and G.~Irving.
\newblock Improving alignment of dialogue agents via targeted human judgements, Sept. 2022.
\newblock URL \url{https://arxiv.org/pdf/2209.14375.pdf}.
\newblock arXiv:2209.14375 [cs].

\bibitem[Glikson and Woolley(2020)]{glikson_human_2020}
E.~Glikson and A.~W. Woolley.
\newblock Human {Trust} in {Artificial} {Intelligence}: {Review} of {Empirical} {Research}.
\newblock \emph{Academy of Management Annals}, 14\penalty0 (2):\penalty0 627--660, July 2020.
\newblock ISSN 1941-6520, 1941-6067.
\newblock \doi{10.5465/annals.2018.0057}.
\newblock URL \url{http://journals.aom.org/doi/10.5465/annals.2018.0057}.

\bibitem[{Global Education Monitoring Report Team, UNESCO}(2023)]{global_education_monitoring_report_team_unesco__2023}
{Global Education Monitoring Report Team, UNESCO}.
\newblock \emph{{Global} education monitoring report summary, 2023: technology in education: a tool on whose terms?}
\newblock UNESCO, July 2023.
\newblock \doi{10.54676/HABJ1624}.
\newblock URL \url{https://unesdoc.unesco.org/ark:/48223/pf0000386147}.

\bibitem[{Global Partnership on AI}(2021)]{global_partnership_on_ai_climate_2021}
{Global Partnership on AI}.
\newblock Climate {Change} and {AI}.
\newblock Technical report, Global Partnership on AI in collaboration with Climate Change AI and the Centre for AI \& Climate, 2021.
\newblock URL \url{https://www.gpai.ai/projects/climate-change-and-ai.pdf}.

\bibitem[Go et~al.(2023)Go, Korbak, Kruszewski, Rozen, Ryu, and Dymetman]{go2023aligning}
D.~Go, T.~Korbak, G.~Kruszewski, J.~Rozen, N.~Ryu, and M.~Dymetman.
\newblock Aligning language models with preferences through f-divergence minimization, 2023.

\bibitem[Goddard et~al.(2012)Goddard, Roudsari, and Wyatt]{goddard_automation_2012}
K.~Goddard, A.~Roudsari, and J.~C. Wyatt.
\newblock Automation bias: a systematic review of frequency, effect mediators, and mitigators.
\newblock \emph{Journal of the American Medical Informatics Association}, 19\penalty0 (1):\penalty0 121--127, Jan. 2012.
\newblock ISSN 1067-5027, 1527-974X.
\newblock \doi{10.1136/amiajnl-2011-000089}.
\newblock URL \url{https://academic.oup.com/jamia/article-lookup/doi/10.1136/amiajnl-2011-000089}.

\bibitem[Godefroid et~al.(2017)Godefroid, Peleg, and Singh]{Godefroid_Peleg_Singh_2017}
P.~Godefroid, H.~Peleg, and R.~Singh.
\newblock Learn\&fuzz: Machine learning for input fuzzing.
\newblock In \emph{2017 32nd IEEE/ACM International Conference on Automated Software Engineering (ASE)}, page 50–59, Oct. 2017.
\newblock \doi{10.1109/ASE.2017.8115618}.
\newblock URL \url{https://ieeexplore.ieee.org/document/8115618}.

\bibitem[Goetz et~al.(2003)Goetz, Kiesler, and Powers]{goetz_matching_2003}
J.~Goetz, S.~Kiesler, and A.~Powers.
\newblock Matching robot appearance and behavior to tasks to improve human-robot cooperation.
\newblock In \emph{The 12th {IEEE} {International} {Workshop} on {Robot} and {Human} {Interactive} {Communication}, 2003. {Proceedings}. {ROMAN} 2003.}, pages 55--60, Millbrae, CA, USA, 2003. IEEE.
\newblock ISBN 9780780381360.
\newblock \doi{10.1109/ROMAN.2003.1251796}.
\newblock URL \url{http://ieeexplore.ieee.org/document/1251796/}.

\bibitem[Gold(2023)]{gold_how_2023}
A.~Gold.
\newblock How generative {AI} could generate more antisemitism.
\newblock \emph{Axios}, May 2023.
\newblock URL \url{https://www.axios.com/2023/05/25/generative-ai-antisemitism-bias}.

\bibitem[Goldstein et~al.(2023)Goldstein, Sastry, Musser, DiResta, Gentzel, and Sedova]{goldstein_generative_2023}
J.~A. Goldstein, G.~Sastry, M.~Musser, R.~DiResta, M.~Gentzel, and K.~Sedova.
\newblock Generative {Language} {Models} and {Automated} {Influence} {Operations}: {Emerging} {Threats} and {Potential} {Mitigations}, Jan. 2023.
\newblock URL \url{http://arxiv.org/abs/2301.04246}.
\newblock arXiv:2301.04246 [cs].

\bibitem[Golin and Sweezy(2022)]{Golin2022-is}
C.~Golin and D.~Sweezy.
\newblock A policy roadmap for 24/7 carbon-free energy.
\newblock \url{https://cloud.google.com/blog/topics/sustainability/a-policy-roadmap-for-achieving-247-carbon-free-energy}, Apr. 2022.

\bibitem[Gomes~de Andrade et~al.(2018)Gomes~de Andrade, Pawson, Muriello, Donahue, and Guadagno]{gomes_de_andrade_ethics_2018}
N.~N. Gomes~de Andrade, D.~Pawson, D.~Muriello, L.~Donahue, and J.~Guadagno.
\newblock Ethics and {Artificial} {Intelligence}: {Suicide} {Prevention} on {Facebook}.
\newblock \emph{Philosophy \& Technology}, 31\penalty0 (4):\penalty0 669--684, Dec. 2018.
\newblock ISSN 2210-5441.
\newblock \doi{10.1007/s13347-018-0336-0}.
\newblock URL \url{https://doi.org/10.1007/s13347-018-0336-0}.

\bibitem[Gong et~al.(2023)Gong, Lyu, Zhang, Wang, Zheng, Zhao, Liu, Zhang, Luo, and Chen]{gong_multimodal-gpt:_2023}
T.~Gong, C.~Lyu, S.~Zhang, Y.~Wang, M.~Zheng, Q.~Zhao, K.~Liu, W.~Zhang, P.~Luo, and K.~Chen.
\newblock {MultiModal}-{GPT}: {A} {Vision} and {Language} {Model} for {Dialogue} with {Humans}, June 2023.
\newblock URL \url{http://arxiv.org/abs/2305.04790}.
\newblock arXiv:2305.04790 [cs].

\bibitem[Goodin(1985)]{goodin_protecting_nodate}
R.~E. Goodin.
\newblock \emph{Protecting the {Vulnerable}: {A} {Re}-{Analysis} of our {Social} {Responsibilities}}.
\newblock University of Chicago Press, Chicago, IL, 1985.
\newblock URL \url{https://press.uchicago.edu/ucp/books/book/chicago/P/bo5974942.html}.

\bibitem[Gootman(2016)]{gootman_opm_2016}
S.~Gootman.
\newblock {OPM} {Hack}: {The} {Most} {Dangerous} {Threat} to the {Federal} {Government} {Today}.
\newblock \emph{Journal of Applied Security Research}, 11\penalty0 (4):\penalty0 517--525, Oct. 2016.
\newblock ISSN 1936-1610, 1936-1629.
\newblock \doi{10.1080/19361610.2016.1211876}.
\newblock URL \url{https://www.tandfonline.com/doi/full/10.1080/19361610.2016.1211876}.

\bibitem[Gordon et~al.(2022)Gordon, Lam, Park, Patel, Hancock, Hashimoto, and Bernstein]{gordon_jury_2022}
M.~L. Gordon, M.~S. Lam, J.~S. Park, K.~Patel, J.~T. Hancock, T.~Hashimoto, and M.~S. Bernstein.
\newblock Jury {Learning}: {Integrating} {Dissenting} {Voices} into {Machine} {Learning} {Models}.
\newblock In \emph{{CHI} {Conference} on {Human} {Factors} in {Computing} {Systems}}, pages 1--19, Apr. 2022.
\newblock \doi{10.1145/3491102.3502004}.
\newblock URL \url{http://arxiv.org/abs/2202.02950}.
\newblock arXiv:2202.02950 [cs].

\bibitem[Gordon(2017)]{gordon_rise_2017}
R.~J. Gordon.
\newblock \emph{The rise and fall of {American} growth: the {U}.{S}. standard of living since the {Civil} {War}}.
\newblock The {Princeton} economic history of the {Western} world. Princeton University Press, Princeton, New Jersey, 2017.
\newblock ISBN 9780691175805.

\bibitem[Gorwa and Guilbeault(2020)]{gorwa_unpacking_2020}
R.~Gorwa and D.~Guilbeault.
\newblock Unpacking the {Social} {Media} {Bot}: {A} {Typology} to {Guide} {Research} and {Policy}.
\newblock \emph{Policy \& Internet}, 12\penalty0 (2):\penalty0 225--248, June 2020.
\newblock ISSN 1944-2866, 1944-2866.
\newblock \doi{10.1002/poi3.184}.
\newblock URL \url{https://onlinelibrary.wiley.com/doi/10.1002/poi3.184}.

\bibitem[Gottesdiener et~al.(2023)Gottesdiener, Hesson, Rosenberg, Cooke, and Solomon]{gottesdiener_bidens_2023}
L.~Gottesdiener, T.~Hesson, M.~Rosenberg, K.~Cooke, and D.~B. Solomon.
\newblock Biden's new asylum policy strands some migrants at {Mexico} border.
\newblock \emph{Reuters}, July 2023.
\newblock URL \url{https://www.reuters.com/investigates/special-report/usa-immigration-asylum-border/}.

\bibitem[Goyal et~al.(2023)Goyal, Li, and Durrett]{goyal_news_2023}
T.~Goyal, J.~J. Li, and G.~Durrett.
\newblock News {Summarization} and {Evaluation} in the {Era} of {GPT}-3, May 2023.
\newblock URL \url{http://arxiv.org/abs/2209.12356}.
\newblock arXiv:2209.12356 [cs].

\bibitem[Grace(2022)]{grace_counterarguments_2022}
K.~Grace.
\newblock Counterarguments to the basic {AI} x-risk case, Oct. 2022.
\newblock URL \url{https://aiimpacts.org/counterarguments-to-the-basic-ai-x-risk-case/}.

\bibitem[Graetz and Michaels(2015)]{graetz_robots_2015}
G.~Graetz and G.~Michaels.
\newblock Robots at {Work}.
\newblock Technical report, Centre for Economic Performance: LSE, Mar. 2015.
\newblock URL \url{https://cep.lse.ac.uk/pubs/download/dp1335.pdf}.

\bibitem[Graham(2023)]{graham_data_2023}
M.~Graham.
\newblock Data for sale: trust, confidence and sharing health data with commercial companies.
\newblock \emph{Journal of Medical Ethics}, 49\penalty0 (7):\penalty0 515--522, July 2023.
\newblock ISSN 0306-6800, 1473-4257.
\newblock \doi{10.1136/medethics-2021-107464}.
\newblock URL \url{https://jme.bmj.com/lookup/doi/10.1136/medethics-2021-107464}.

\bibitem[Graham et~al.(2017)Graham, Hjorth, and Lehdonvirta]{graham_digital_2017}
M.~Graham, I.~Hjorth, and V.~Lehdonvirta.
\newblock Digital labour and development: impacts of global digital labour platforms and the gig economy on worker livelihoods.
\newblock \emph{Transfer: European Review of Labour and Research}, 23\penalty0 (2):\penalty0 135--162, May 2017.
\newblock ISSN 1024-2589, 1996-7284.
\newblock \doi{10.1177/1024258916687250}.
\newblock URL \url{http://journals.sagepub.com/doi/10.1177/1024258916687250}.

\bibitem[Graham et~al.(2023)Graham, Milne, Fitzsimmons, and Sheehan]{graham_trust_2023}
M.~Graham, R.~Milne, P.~Fitzsimmons, and M.~Sheehan.
\newblock Trust and the {Goldacre} {Review}: why trusted research environments are not about trust.
\newblock \emph{Journal of Medical Ethics}, 49\penalty0 (10):\penalty0 670--673, Oct. 2023.
\newblock ISSN 0306-6800, 1473-4257.
\newblock \doi{10.1136/jme-2022-108435}.
\newblock URL \url{https://jme.bmj.com/content/49/10/670}.

\bibitem[Grandinetti and Bruinsma(2023)]{grandinetti_affective_2023}
J.~Grandinetti and J.~Bruinsma.
\newblock The {Affective} {Algorithms} of {Conspiracy} {TikTok}.
\newblock \emph{Journal of Broadcasting \& Electronic Media}, 67\penalty0 (3):\penalty0 274--293, May 2023.
\newblock ISSN 0883-8151, 1550-6878.
\newblock \doi{10.1080/08838151.2022.2140806}.
\newblock URL \url{https://www.tandfonline.com/doi/full/10.1080/08838151.2022.2140806}.

\bibitem[Granja et~al.(2018)Granja, Janssen, and Johansen]{granja2018factors}
C.~Granja, W.~Janssen, and M.~A. Johansen.
\newblock Factors determining the success and failure of ehealth interventions: {S}ystematic review of the literature.
\newblock \emph{Journal of Medical Internet Research}, 20\penalty0 (5):\penalty0 e10235, 2018.

\bibitem[Gray and Suri(2019)]{gray_ghost_2019}
M.~L. Gray and S.~Suri.
\newblock \emph{Ghost {Work}: {How} to {Stop} {Silicon} {Valley} from {Building} a {New} {Global} {Underclass}}.
\newblock Houghton Mifflin Harcourt, 2019.
\newblock ISBN 9781328566249.
\newblock Google-Books-ID: 8AmXDwAAQBAJ.

\bibitem[Graziotin and Fagerholm(2019)]{graziotin_happiness_2019}
D.~Graziotin and F.~Fagerholm.
\newblock Happiness and the {Productivity} of {Software} {Engineers}.
\newblock In C.~Sadowski and T.~Zimmermann, editors, \emph{Rethinking {Productivity} in {Software} {Engineering}}, pages 109--124. Apress, Berkeley, CA, 2019.
\newblock ISBN 9781484242216.
\newblock \doi{10.1007/978-1-4842-4221-6_10}.
\newblock URL \url{https://doi.org/10.1007/978-1-4842-4221-6_10}.

\bibitem[Green(2019)]{green_artificial_nodate}
B.~P. Green.
\newblock Artificial {Intelligence}, {Decision}-{Making}, and {Moral} {Deskilling}, 2019.
\newblock URL \url{https://www.scu.edu/ethics/focus-areas/technology-ethics/resources/artificial-intelligence-decision-making-and-moral-deskilling/}.
\newblock publisher: Santa Clara University: Markkula Center for Applied Ethics.

\bibitem[Greenfield and Bhavnani(2023)]{greenfield_social_2023}
D.~Greenfield and S.~Bhavnani.
\newblock Social media: generative {AI} could harm mental health.
\newblock \emph{Nature}, 617\penalty0 (7962):\penalty0 676--676, May 2023.
\newblock \doi{10.1038/d41586-023-01693-8}.
\newblock URL \url{https://www.nature.com/articles/d41586-023-01693-8}.

\bibitem[Greenpeace(2020)]{greenpeace_oil_2020}
Greenpeace.
\newblock Oil in the {Cloud}: {How} {Tech} {Companies} are {Helping} {Big} {Oil} {Profit} from {Climate} {Destruction}.
\newblock Technical report, Greenpeace, May 2020.
\newblock URL \url{https://www.greenpeace.org/usa/reports/oil-in-the-cloud/, https://www.greenpeace.org/usa/reports/oil-in-the-cloud/}.

\bibitem[Gregor et~al.(2005)Gregor, Sloan, and Newell]{gregor_disability_2005}
P.~Gregor, D.~Sloan, and A.~F. Newell.
\newblock Disability and {Technology}: {Building} {Barriers} or {Creating} {Opportunities}?
\newblock In \emph{Advances in {Computers}}, volume~64, pages 283--346. Elsevier, 2005.
\newblock ISBN 9780120121649.
\newblock \doi{10.1016/S0065-2458(04)64007-1}.
\newblock URL \url{https://linkinghub.elsevier.com/retrieve/pii/S0065245804640071}.

\bibitem[Gros et~al.(2021)Gros, Li, and Yu]{gros_r-u--robot_2021}
D.~Gros, Y.~Li, and Z.~Yu.
\newblock The {R}-{U}-{A}-{Robot} {Dataset}: {Helping} {Avoid} {Chatbot} {Deception} by {Detecting} {User} {Questions} {About} {Human} or {Non}-{Human} {Identity}.
\newblock In C.~Zong, F.~Xia, W.~Li, and R.~Navigli, editors, \emph{Proceedings of the 59th {Annual} {Meeting} of the {Association} for {Computational} {Linguistics} and the 11th {International} {Joint} {Conference} on {Natural} {Language} {Processing} ({Volume} 1: {Long} {Papers})}, pages 6999--7013, Online, Aug. 2021. Association for Computational Linguistics.
\newblock \doi{10.18653/v1/2021.acl-long.544}.
\newblock URL \url{https://aclanthology.org/2021.acl-long.544}.

\bibitem[Gros et~al.(2022)Gros, Li, and Yu]{gros_robots-dont-cry:_2022}
D.~Gros, Y.~Li, and Z.~Yu.
\newblock Robots-{Dont}-{Cry}: {Understanding} {Falsely} {Anthropomorphic} {Utterances} in {Dialog} {Systems}, Oct. 2022.
\newblock URL \url{http://arxiv.org/abs/2210.12429}.
\newblock arXiv:2210.12429 [cs].

\bibitem[Gross and Sampat(2022)]{gross_america_2022}
D.~P. Gross and B.~N. Sampat.
\newblock America, {Jump}-started: {World} {War} {II} {R}\&{D} and the {Takeoff} of the {U}.{S}. {Innovation} {System}, Sept. 2022.
\newblock URL \url{https://papers.ssrn.com/abstract=3623115}.

\bibitem[Grossman et~al.(2004)Grossman, Niemann, Schmidt, and Walach]{grossman2004mindfulness}
P.~Grossman, L.~Niemann, S.~Schmidt, and H.~Walach.
\newblock Mindfulness-based stress reduction and health benefits: A meta-analysis.
\newblock \emph{Journal of Psychosomatic Research}, 57\penalty0 (1):\penalty0 35--43, 2004.

\bibitem[Gu et~al.(2015)Gu, Strauss, Bond, and Cavanagh]{gu2015mindfulness}
J.~Gu, C.~Strauss, R.~Bond, and K.~Cavanagh.
\newblock How do mindfulness-based cognitive therapy and mindfulness-based stress reduction improve mental health and wellbeing? {A} systematic review and meta-analysis of mediation studies.
\newblock \emph{Clinical Psychology Review}, 37:\penalty0 1--12, 2015.

\bibitem[Guess et~al.(2020{\natexlab{a}})Guess, Lockett, Lyons, Montgomery, Nyhan, and Reifler]{guess_fake_2020}
A.~M. Guess, D.~Lockett, B.~Lyons, J.~M. Montgomery, B.~Nyhan, and J.~Reifler.
\newblock “{Fake} news” may have limited effects on political participation beyond increasing beliefs in false claims.
\newblock \emph{Harvard Kennedy School Misinformation Review}, Jan. 2020{\natexlab{a}}.
\newblock \doi{10.37016/mr-2020-004}.
\newblock URL \url{https://misinforeview.hks.harvard.edu/article/fake-news-limited-effects-on-political-participation/}.

\bibitem[Guess et~al.(2020{\natexlab{b}})Guess, Nyhan, and Reifler]{guess_exposure_2020}
A.~M. Guess, B.~Nyhan, and J.~Reifler.
\newblock Exposure to untrustworthy websites in the 2016 {US} election.
\newblock \emph{Nature Human Behaviour}, 4\penalty0 (5):\penalty0 472--480, May 2020{\natexlab{b}}.
\newblock ISSN 2397-3374.
\newblock \doi{10.1038/s41562-020-0833-x}.
\newblock URL \url{https://www.nature.com/articles/s41562-020-0833-x}.

\bibitem[Gunasekar et~al.(2023)Gunasekar, Zhang, Aneja, Mendes, Del~Giorno, Gopi, Javaheripi, Kauffmann, de~Rosa, Saarikivi, Salim, Shah, Behl, Wang, Bubeck, Eldan, Kalai, Lee, and Li]{gunasekar_textbooks_2023}
S.~Gunasekar, Y.~Zhang, J.~Aneja, C.~C.~T. Mendes, A.~Del~Giorno, S.~Gopi, M.~Javaheripi, P.~Kauffmann, G.~de~Rosa, O.~Saarikivi, A.~Salim, S.~Shah, H.~S. Behl, X.~Wang, S.~Bubeck, R.~Eldan, A.~T. Kalai, Y.~T. Lee, and Y.~Li.
\newblock Textbooks {Are} {All} {You} {Need}, Oct. 2023.
\newblock URL \url{https://arxiv.org/pdf/2306.11644.pdf}.
\newblock arXiv:2306.11644 [cs].

\bibitem[Guo et~al.(2022{\natexlab{a}})Guo, Ainslie, Uthus, Ontanon, Ni, Sung, and Yang]{guo_longt5:_2022}
M.~Guo, J.~Ainslie, D.~Uthus, S.~Ontanon, J.~Ni, Y.-H. Sung, and Y.~Yang.
\newblock {LongT5}: {Efficient} {Text}-{To}-{Text} {Transformer} for {Long} {Sequences}, May 2022{\natexlab{a}}.
\newblock URL \url{http://arxiv.org/abs/2112.07916}.
\newblock arXiv:2112.07916 [cs].

\bibitem[Guo et~al.(2022{\natexlab{b}})Guo, Schlichtkrull, and Vlachos]{guo_survey_2022}
Z.~Guo, M.~Schlichtkrull, and A.~Vlachos.
\newblock A {Survey} on {Automated} {Fact}-{Checking}.
\newblock \emph{Transactions of the Association for Computational Linguistics}, 10:\penalty0 178--206, Feb. 2022{\natexlab{b}}.
\newblock ISSN 2307-387X.
\newblock \doi{10.1162/tacl_a_00454}.
\newblock URL \url{https://direct.mit.edu/tacl/article/doi/10.1162/tacl_a_00454/109469/A-Survey-on-Automated-Fact-Checking}.

\bibitem[Gupta et~al.(2020)Gupta, Kim, Lee, Tse, Lee, Wei, Brooks, and Wu]{gupta_chasing_2020}
U.~Gupta, Y.~G. Kim, S.~Lee, J.~Tse, H.-H.~S. Lee, G.-Y. Wei, D.~Brooks, and C.-J. Wu.
\newblock Chasing {Carbon}: {The} {Elusive} {Environmental} {Footprint} of {Computing}, Oct. 2020.
\newblock URL \url{http://arxiv.org/abs/2011.02839}.
\newblock arXiv:2011.02839 [cs].

\bibitem[Gurnee et~al.(2023)Gurnee, Nanda, Pauly, Harvey, Troitskii, and Bertsimas]{gurnee_finding_2023}
W.~Gurnee, N.~Nanda, M.~Pauly, K.~Harvey, D.~Troitskii, and D.~Bertsimas.
\newblock Finding {Neurons} in a {Haystack}: {Case} {Studies} with {Sparse} {Probing}, June 2023.
\newblock URL \url{http://arxiv.org/abs/2305.01610}.
\newblock arXiv:2305.01610 [cs].

\bibitem[Gutelius and Theodore(2019)]{gutelius_future_2019}
B.~Gutelius and N.~Theodore.
\newblock The {Future} of {Warehouse} {Work}: {Technological} {Change} in the {U}.{S}. {Logistics} {Industry}.
\newblock Technical report, UC Berkeley Center for Labor Research and Education and Working Partnerships USA, Oct. 2019.
\newblock URL \url{https://laborcenter.berkeley.edu/future-of-warehouse-work/}.

\bibitem[Gyrard and Sheth(2020)]{gyrard2020iamhappy}
A.~Gyrard and A.~Sheth.
\newblock {IAMHAPPY}: Towards an {I}o{T} knowledge-based cross-domain well-being recommendation system for everyday happiness.
\newblock \emph{Smart Health}, 15:\penalty0 100083, 2020.

\bibitem[Hadfield-Menell and Hadfield(2019)]{hadfield2019incomplete}
D.~Hadfield-Menell and G.~K. Hadfield.
\newblock Incomplete contracting and {AI} alignment.
\newblock In \emph{Proceedings of the 2019 AAAI/ACM Conference on AI, Ethics, and Society}, pages 417--422, 2019.

\bibitem[Hadfield-Menell et~al.(2016)Hadfield-Menell, Dragan, Abbeel, and Russell]{hadfield-menell_cooperative_2016}
D.~Hadfield-Menell, A.~Dragan, P.~Abbeel, and S.~Russell.
\newblock Cooperative {Inverse} {Reinforcement} {Learning}, Nov. 2016.
\newblock URL \url{http://arxiv.org/abs/1606.03137}.
\newblock arXiv:1606.03137 [cs].

\bibitem[Hadfield-Menell et~al.(2017)Hadfield-Menell, Milli, Abbeel, Russell, and Dragan]{hadfield2017inverse}
D.~Hadfield-Menell, S.~Milli, P.~Abbeel, S.~J. Russell, and A.~Dragan.
\newblock Inverse reward design.
\newblock \emph{Advances in Neural Information Processing Systems}, 30, 2017.

\bibitem[Haenschen(2023)]{haenschen_conditional_2023}
K.~Haenschen.
\newblock The {Conditional} {Effects} of {Microtargeted} {Facebook} {Advertisements} on {Voter} {Turnout}.
\newblock \emph{Political Behavior}, 45\penalty0 (4):\penalty0 1661--1681, Dec. 2023.
\newblock ISSN 0190-9320, 1573-6687.
\newblock \doi{10.1007/s11109-022-09781-7}.
\newblock URL \url{https://link.springer.com/10.1007/s11109-022-09781-7}.

\bibitem[Haidt and Schmidt(2023)]{miscex2}
J.~Haidt and E.~Schmidt.
\newblock {AI} is about to make social media (much) more toxic.
\newblock \url{https://www.theatlantic.com/technology/archive/2023/05/generative-ai-social-media-integration-dangers-disinformation-addiction/673940/}, 2023.
\newblock Accessed: 2023-07-13.

\bibitem[Hameleers et~al.(2020)Hameleers, Powell, van~der Meer, and Bos]{hameleers_picture_2020}
M.~Hameleers, T.~E. Powell, T.~G. L.~A. van~der Meer, and L.~Bos.
\newblock A picture paints a thousand lies? {The} effects and mechanisms of multimodal disinformation and rebuttals disseminated via social media.
\newblock \emph{Political Communication}, 37, 2020.
\newblock \doi{10.1080/10584609.2019.1674979}.
\newblock URL \url{https://dare.uva.nl/search?identifier=a27a4957-6fd9-4a8f-b6d2-e21185fddc43}.

\bibitem[Hameleers et~al.(2022)Hameleers, Brosius, and De~Vreese]{hameleers_whom_2022}
M.~Hameleers, A.~Brosius, and C.~H. De~Vreese.
\newblock Whom to trust? {Media} exposure patterns of citizens with perceptions of misinformation and disinformation related to the news media.
\newblock \emph{European Journal of Communication}, 37\penalty0 (3):\penalty0 237--268, June 2022.
\newblock ISSN 0267-3231, 1460-3705.
\newblock \doi{10.1177/02673231211072667}.
\newblock URL \url{http://journals.sagepub.com/doi/10.1177/02673231211072667}.

\bibitem[Hammer et~al.(2015)Hammer, Seiderer, Andr{\'e}, Rist, Kastrinaki, Hondrou, Raouzaiou, Karpouzis, and Kollias]{hammer2015design}
S.~Hammer, A.~Seiderer, E.~Andr{\'e}, T.~Rist, S.~Kastrinaki, C.~Hondrou, A.~Raouzaiou, K.~Karpouzis, and S.~Kollias.
\newblock Design of a lifestyle recommender system for the elderly: requirement gatherings in germany and greece.
\newblock In \emph{Proceedings of the 8th ACM International Conference on Pervasive Technologies Related to Assistive Environments}, pages 1--8, 2015.

\bibitem[Han et~al.(2012)Han, Pereira, and Santos]{han_emergence_2012}
T.~A. Han, L.~M. Pereira, and F.~C. Santos.
\newblock The emergence of commitments and cooperation.
\newblock In \emph{Proceedings of the 11th {International} {Conference} on {Autonomous} {Agents} and {Multiagent} {Systems} - {Volume} 1}, {AAMAS} '12, pages 559--566, Richland, SC, June 2012. International Foundation for Autonomous Agents and Multiagent Systems.
\newblock ISBN 9780981738116.

\bibitem[Hancock et~al.(2021)Hancock, Kessler, Kaplan, Brill, and Szalma]{hancock_evolving_2021}
P.~A. Hancock, T.~T. Kessler, A.~D. Kaplan, J.~C. Brill, and J.~L. Szalma.
\newblock Evolving {Trust} in {Robots}: {Specification} {Through} {Sequential} and {Comparative} {Meta}-{Analyses}.
\newblock \emph{Human Factors: The Journal of the Human Factors and Ergonomics Society}, 63\penalty0 (7):\penalty0 1196--1229, Nov. 2021.
\newblock ISSN 0018-7208, 1547-8181.
\newblock \doi{10.1177/0018720820922080}.
\newblock URL \url{http://journals.sagepub.com/doi/10.1177/0018720820922080}.

\bibitem[Handa et~al.(2019)Handa, Sharma, and Shukla]{handa_machine_2019}
A.~Handa, A.~Sharma, and S.~Shukla.
\newblock Machine learning in cybersecurity: {A} review.
\newblock \emph{Wiley Interdisciplinary Reviews: Data Mining and Knowledge Discovery}, 9:\penalty0 e1306, Feb. 2019.
\newblock \doi{10.1002/widm.1306}.

\bibitem[Handel(2022)]{handel_growth_2022}
M.~J. Handel.
\newblock Growth trends for selected occupations considered at risk from automation, July 2022.
\newblock URL \url{https://www.bls.gov/opub/mlr/2022/article/growth-trends-for-selected-occupations-considered-at-risk-from-automation.htm}.

\bibitem[Hangloo and Arora(2022)]{hangloo_combating_2022}
S.~Hangloo and B.~Arora.
\newblock Combating multimodal fake news on social media: methods, datasets, and future perspective.
\newblock \emph{Multimedia Systems}, 28\penalty0 (6):\penalty0 2391--2422, Dec. 2022.
\newblock ISSN 0942-4962, 1432-1882.
\newblock \doi{10.1007/s00530-022-00966-y}.
\newblock URL \url{https://link.springer.com/10.1007/s00530-022-00966-y}.

\bibitem[Hanley and Durumeric(2023)]{hanley2023machine}
H.~W. Hanley and Z.~Durumeric.
\newblock Machine-made media: Monitoring the mobilization of machine-generated articles on misinformation and mainstream news websites.
\newblock \emph{arXiv preprint arXiv:2305.09820}, 2023.

\bibitem[Hansen and Schicktanz(2022)]{ho_normative_2022}
S.~L. Hansen and S.~Schicktanz.
\newblock Normative {Aspects} of {Persuasion}.
\newblock In E.~Y. Ho, C.~L. Bylund, and J.~C.~M. van Weert, editors, \emph{The {International} {Encyclopedia} of {Health} {Communication}}, pages 1--7. Wiley, 1 edition, Nov. 2022.
\newblock ISBN 9780470673959 9781119678816.
\newblock \doi{10.1002/9781119678816.iehc0791}.
\newblock URL \url{https://onlinelibrary.wiley.com/doi/10.1002/9781119678816.iehc0791}.

\bibitem[Hansson and Fr{\"o}ding(2021)]{hansson2021ethical}
S.~O. Hansson and B.~Fr{\"o}ding.
\newblock Ethical conflicts in patient-centred care.
\newblock \emph{Clinical Ethics}, 16\penalty0 (2):\penalty0 55--66, 2021.

\bibitem[Hao(2021)]{hao_facebook_2021}
K.~Hao.
\newblock The {Facebook} whistleblower says its algorithms are dangerous. {Here}’s why., Oct. 2021.
\newblock URL \url{https://www.technologyreview.com/2021/10/05/1036519/facebook-whistleblower-frances-haugen-algorithms/}.

\bibitem[Hao et~al.(2023)Hao, Kumar, Laszlo, Poddar, Radharapu, and Shelby]{hao_safety_2023}
S.~Hao, P.~Kumar, S.~Laszlo, S.~Poddar, B.~Radharapu, and R.~Shelby.
\newblock Safety and {Fairness} for {Content} {Moderation} in {Generative} {Models}, June 2023.
\newblock URL \url{http://arxiv.org/abs/2306.06135}.
\newblock arXiv:2306.06135 [cs].

\bibitem[Haraway(1988)]{haraway_situated_1988}
D.~Haraway.
\newblock Situated {Knowledges}: {The} {Science} {Question} in {Feminism} and the {Privilege} of {Partial} {Perspective}.
\newblock \emph{Feminist Studies}, 14\penalty0 (3):\penalty0 575, 1988.
\newblock ISSN 00463663.
\newblock \doi{10.2307/3178066}.
\newblock URL \url{https://www.jstor.org/stable/3178066?origin=crossref}.

\bibitem[Harding(1986)]{harding1986sciencequestion}
S.~Harding.
\newblock \emph{The science question in feminism}.
\newblock Cornell University Press, 1986.

\bibitem[Harding(1998)]{harding_is_1998}
S.~Harding.
\newblock \emph{Is {Science} {Multicultural}?: {Postcolonialisms}, {Feminisms}, and {Epistemologies}}.
\newblock Indiana University Press, Feb. 1998.
\newblock ISBN 9780253211569.
\newblock Google-Books-ID: WeTHMEPj1ooC.

\bibitem[Harding(2016)]{harding_whose_2016}
S.~Harding.
\newblock \emph{Whose {Science}? {Whose} {Knowledge}?: {Thinking} from {Women}'s {Lives}}.
\newblock Cornell University Press, Dec. 2016.
\newblock ISBN 9781501712951.
\newblock \doi{10.7591/9781501712951}.
\newblock URL \url{https://www.degruyter.com/document/doi/10.7591/9781501712951/html}.

\bibitem[Hargens et~al.(2013)Hargens, Kaleth, Edwards, and Butner]{hargens2013association}
T.~A. Hargens, A.~S. Kaleth, E.~S. Edwards, and K.~L. Butner.
\newblock Association between sleep disorders, obesity, and exercise: a review.
\newblock \emph{Nature and Science of Sleep}, pages 27--35, 2013.

\bibitem[Harrington et~al.(2022)Harrington, Garg, Woodward, and Williams]{harrington_its_2022}
C.~N. Harrington, R.~Garg, A.~Woodward, and D.~Williams.
\newblock “{It}’s {Kind} of {Like} {Code}-{Switching}”: {Black} {Older} {Adults}’ {Experiences} with a {Voice} {Assistant} for {Health} {Information} {Seeking}.
\newblock In \emph{{CHI} {Conference} on {Human} {Factors} in {Computing} {Systems}}, pages 1--15, New Orleans LA USA, Apr. 2022. ACM.
\newblock ISBN 9781450391573.
\newblock \doi{10.1145/3491102.3501995}.
\newblock URL \url{https://dl.acm.org/doi/10.1145/3491102.3501995}.

\bibitem[Harrison(2022)]{harrison_44_2022}
J.~Harrison.
\newblock 44\% of state school teachers plan to quit by 2027, Apr. 2022.
\newblock URL \url{https://www.ier.org.uk/news/44-of-state-school-teachers-plan-to-quit-by-2027/}.

\bibitem[Harutyunyan(2023)]{harutyunyan_leveraging_2023}
H.~Harutyunyan.
\newblock Leveraging {AI} to {Counter} {Corruption} in {Armenia}.
\newblock In \emph{The {Digitalization} of {Democracy}}. National Endowment for Democracy (NED), Mar. 2023.
\newblock URL \url{https://www.ned.org/wp-content/uploads/2023/03/NED_FORUM-The-Digitalization-of-Democracy_03Leveraging-AI_v5.pdf}.

\bibitem[Hassoun et~al.(2023)Hassoun, Borenstein, Goldberg, McAuliffe, and Osborn]{hassoun_sowing_2023}
A.~Hassoun, G.~Borenstein, B.~Goldberg, J.~McAuliffe, and K.~Osborn.
\newblock Sowing '{Seeds} of {Doubt}': {Cottage} {Industries} of {Election} and {Medical} {Misinformation} in {Brazil} and the {United} {States}, Aug. 2023.
\newblock URL \url{http://arxiv.org/abs/2308.02377}.
\newblock arXiv:2308.02377 [cs].

\bibitem[Hatzius et~al.(2023)Hatzius, Briggs, Kodnani, and Pierdomenico]{hatzius_potentially_2023}
J.~Hatzius, J.~Briggs, D.~Kodnani, and G.~Pierdomenico.
\newblock The {Potentially} {Large} {Effects} of {Artificial} {Intelligence} on {Economic} {Growth} ({Briggs}/{Kodnani}).
\newblock \emph{Goldman Sachs}, 2023.

\bibitem[Hausman and Johnston(2010)]{hausman_impact_2010}
A.~Hausman and W.~J. Johnston.
\newblock The impact of coercive and non-coercive forms of influence on trust, commitment, and compliance in supply chains.
\newblock \emph{Industrial Marketing Management}, 39\penalty0 (3):\penalty0 519--526, Apr. 2010.
\newblock ISSN 00198501.
\newblock \doi{10.1016/j.indmarman.2009.05.007}.
\newblock URL \url{https://linkinghub.elsevier.com/retrieve/pii/S0019850109000807}.

\bibitem[Hawkins et~al.(2021)Hawkins, Paterson, Picardi, Jia, Calinescu, and Habli]{hawkins_guidance_2021}
R.~Hawkins, C.~Paterson, C.~Picardi, Y.~Jia, R.~Calinescu, and I.~Habli.
\newblock Guidance on the {Assurance} of {Machine} {Learning} in {Autonomous} {Systems} ({AMLAS}), Feb. 2021.
\newblock URL \url{http://arxiv.org/abs/2102.01564}.
\newblock arXiv:2102.01564 [cs].

\bibitem[Hawkins et~al.(2022)Hawkins, Osborne, Parsons, Nicholson, McDermid, and Habli]{hawkins_guidance_2022}
R.~Hawkins, M.~Osborne, M.~Parsons, M.~Nicholson, J.~McDermid, and I.~Habli.
\newblock Guidance on the {Safety} {Assurance} of {Autonomous} {Systems} in {Complex} {Environments} ({SACE}), Aug. 2022.
\newblock URL \url{http://arxiv.org/abs/2208.00853}.
\newblock arXiv:2208.00853 [cs, eess].

\bibitem[Hawley(2014)]{hawley_trust_2014}
K.~Hawley.
\newblock Trust, {Distrust} and {Commitment}.
\newblock \emph{Noûs}, 48\penalty0 (1):\penalty0 1--20, Mar. 2014.
\newblock ISSN 0029-4624, 1468-0068.
\newblock \doi{10.1111/nous.12000}.
\newblock URL \url{https://onlinelibrary.wiley.com/doi/10.1111/nous.12000}.

\bibitem[Hawley(2017)]{hawley2017trustworthy}
K.~Hawley.
\newblock Trustworthy groups and organizations.
\newblock \emph{The philosophy of trust}, pages 230--250, 2017.

\bibitem[Hayes et~al.(2022)Hayes, R{\u{a}}dulescu, Bargiacchi, K{\"a}llstr{\"o}m, Macfarlane, Reymond, Verstraeten, Zintgraf, Dazeley, Heintz, et~al.]{hayes2022practical}
C.~F. Hayes, R.~R{\u{a}}dulescu, E.~Bargiacchi, J.~K{\"a}llstr{\"o}m, M.~Macfarlane, M.~Reymond, T.~Verstraeten, L.~M. Zintgraf, R.~Dazeley, F.~Heintz, et~al.
\newblock A practical guide to multi-objective reinforcement learning and planning.
\newblock \emph{Autonomous Agents and Multi-Agent Systems}, 36\penalty0 (1):\penalty0 26, 2022.

\bibitem[Hazell(2023)]{hazell_large_2023}
J.~Hazell.
\newblock Large {Language} {Models} {Can} {Be} {Used} {To} {Effectively} {Scale} {Spear} {Phishing} {Campaigns}, May 2023.
\newblock URL \url{http://arxiv.org/abs/2305.06972}.
\newblock arXiv:2305.06972 [cs].

\bibitem[Hegel et~al.(2008)Hegel, Krach, Kircher, Wrede, and Sagerer]{hegel_understanding_2008}
F.~Hegel, S.~Krach, T.~Kircher, B.~Wrede, and G.~Sagerer.
\newblock Understanding social robots: {A} user study on anthropomorphism.
\newblock In \emph{Proceedings of the 17th {IEEE} {International} {Symposium} on {Robot} and {Human} {Interactive} {Communication}, {RO}-{MAN}}, pages 574 -- 579, Sept. 2008.
\newblock ISBN 978-1-4244-2212-8.
\newblock \doi{10.1109/ROMAN.2008.4600728}.

\bibitem[Heine et~al.(2002)Heine, Lehman, Peng, and Greenholtz]{heine2002s}
S.~J. Heine, D.~R. Lehman, K.~Peng, and J.~Greenholtz.
\newblock What's wrong with cross-cultural comparisons of subjective likert scales?: The reference-group effect.
\newblock \emph{Journal of Personality and Social Psychology}, 82\penalty0 (6):\penalty0 903, 2002.

\bibitem[Helliwell and Aknin(2018)]{helliwell2018expanding}
J.~F. Helliwell and L.~B. Aknin.
\newblock Expanding the social science of happiness.
\newblock \emph{Nature Human Behaviour}, 2\penalty0 (4):\penalty0 248--252, 2018.

\bibitem[Helsper(2021)]{helsper_digital_2021}
E.~J. Helsper.
\newblock \emph{The digital disconnect: the causes and consequences of digital inequalities}.
\newblock Sage Publications, London ; Los Angeles, 2021.
\newblock ISBN 9781526492968.
\newblock OCLC: 1272853992.

\bibitem[Hendrycks et~al.(2022)Hendrycks, Carlini, Schulman, and Steinhardt]{hendrycks_unsolved_2022}
D.~Hendrycks, N.~Carlini, J.~Schulman, and J.~Steinhardt.
\newblock Unsolved {Problems} in {ML} {Safety}, June 2022.
\newblock URL \url{http://arxiv.org/abs/2109.13916}.
\newblock arXiv:2109.13916 [cs].

\bibitem[Henrich et~al.(2010)Henrich, Heine, and Norenzayan]{henrich_most_2010}
J.~Henrich, S.~J. Heine, and A.~Norenzayan.
\newblock Most people are not {WEIRD}.
\newblock \emph{Nature}, 466\penalty0 (7302):\penalty0 29--29, July 2010.
\newblock ISSN 1476-4687.
\newblock \doi{10.1038/466029a}.
\newblock URL \url{https://www.nature.com/articles/466029a}.

\bibitem[Henschel et~al.(2021)Henschel, Laban, and Cross]{henschel_what_2021}
A.~Henschel, G.~Laban, and E.~S. Cross.
\newblock What {Makes} a {Robot} {Social}? {A} {Review} of {Social} {Robots} from {Science} {Fiction} to a {Home} or {Hospital} {Near} {You}.
\newblock \emph{Current Robotics Reports}, 2\penalty0 (1):\penalty0 9--19, Mar. 2021.
\newblock ISSN 2662-4087.
\newblock \doi{10.1007/s43154-020-00035-0}.
\newblock URL \url{https://doi.org/10.1007/s43154-020-00035-0}.

\bibitem[Herber et~al.(2019)Herber, Ruijsbroek, Koopmanschap, Proper, van~der Lucht, Boshuizen, Polder, and Uiters]{herber_single_2019}
G.-C. Herber, A.~Ruijsbroek, M.~Koopmanschap, K.~Proper, F.~van~der Lucht, H.~Boshuizen, J.~Polder, and E.~Uiters.
\newblock Single transitions and persistence of unemployment are associated with poor health outcomes.
\newblock \emph{BMC Public Health}, 19\penalty0 (1):\penalty0 740, Dec. 2019.
\newblock ISSN 1471-2458.
\newblock \doi{10.1186/s12889-019-7059-8}.
\newblock URL \url{https://bmcpublichealth.biomedcentral.com/articles/10.1186/s12889-019-7059-8}.

\bibitem[Herriman et~al.(2020)Herriman, Meer, Rosin, Lee, Washington, and Volpp]{Herriman_Meer_Rosin_Lee_Washington_Volpp_2020}
M.~Herriman, E.~Meer, R.~Rosin, V.~Lee, V.~Washington, and K.~G. Volpp.
\newblock Asked and answered: Building a chatbot to address covid-19-related concerns.
\newblock \emph{NEJM Catalyst Innovations in Care Delivery}, June 2020.
\newblock URL \url{https://catalyst.nejm.org/doi/full/10.1056/CAT.20.0230}.

\bibitem[Hill~Collins(2009)]{hill_collins_black_2009}
P.~Hill~Collins.
\newblock \emph{Black feminist thought: knowledge, consciousness, and the politics of empowerment}.
\newblock Routledge classics. Routledge, New York, 2nd ed. edition, 2009.
\newblock ISBN 9780415964722.
\newblock original-date: 1990.

\bibitem[Hintze(2023)]{hintze_chatgpt_2023}
A.~Hintze.
\newblock Chatgpt believes it is conscious.
\newblock \emph{arXiv preprint arXiv:2304.12898}, 2023.

\bibitem[Hintze and Adami(2015)]{hintze_punishment_2015}
A.~Hintze and C.~Adami.
\newblock Punishment in public goods games leads to meta-stable phase transitions and hysteresis.
\newblock \emph{Physical Biology}, 12\penalty0 (4):\penalty0 046005, June 2015.
\newblock ISSN 1478-3975.
\newblock \doi{10.1088/1478-3975/12/4/046005}.
\newblock URL \url{https://iopscience.iop.org/article/10.1088/1478-3975/12/4/046005}.

\bibitem[Ho et~al.(2021)Ho, King, Wald, and Wan]{ho_building_2021}
D.~E. Ho, J.~King, R.~C. Wald, and C.~Wan.
\newblock Building a {National} {AI} {Research} {Resource}: {A} {Blueprint} for the {National} {Research} {Cloud}.
\newblock Technical report, Stanford University: Human-Centered Artificial Intelligence, Oct. 2021.
\newblock URL \url{https://hai.stanford.edu/sites/default/files/2022-01/HAI_NRCR_v17.pdf}.

\bibitem[Ho et~al.(2023)Ho, Barnhart, Trager, Bengio, Brundage, Carnegie, Chowdhury, Dafoe, Hadfield, Levi, and Snidal]{ho_international_2023}
L.~Ho, J.~Barnhart, R.~Trager, Y.~Bengio, M.~Brundage, A.~Carnegie, R.~Chowdhury, A.~Dafoe, G.~Hadfield, M.~Levi, and D.~Snidal.
\newblock International {Institutions} for {Advanced} {AI}, July 2023.
\newblock URL \url{http://arxiv.org/abs/2307.04699}.
\newblock arXiv:2307.04699 [cs].

\bibitem[Hobbes(1994)]{hobbes1994human}
T.~Hobbes.
\newblock \emph{Human nature, or, The fundamental elements of polity; De Corpore politico, or, The Elements of law}.
\newblock Burns \& Oates, 1994.

\bibitem[Hoffmann et~al.(2022)Hoffmann, Borgeaud, Mensch, Buchatskaya, Cai, Rutherford, Casas, Hendricks, Welbl, Clark, Hennigan, Noland, Millican, Driessche, Damoc, Guy, Osindero, Simonyan, Elsen, Rae, Vinyals, and Sifre]{hoffmann_training_2022}
J.~Hoffmann, S.~Borgeaud, A.~Mensch, E.~Buchatskaya, T.~Cai, E.~Rutherford, D.~d.~L. Casas, L.~A. Hendricks, J.~Welbl, A.~Clark, T.~Hennigan, E.~Noland, K.~Millican, G.~v.~d. Driessche, B.~Damoc, A.~Guy, S.~Osindero, K.~Simonyan, E.~Elsen, J.~W. Rae, O.~Vinyals, and L.~Sifre.
\newblock Training {Compute}-{Optimal} {Large} {Language} {Models}, Mar. 2022.
\newblock URL \url{http://arxiv.org/abs/2203.15556}.
\newblock arXiv:2203.15556 [cs].

\bibitem[Hofmann et~al.(2017)Hofmann, Hartl, Gangl, Hartner-Tiefenthaler, and Kirchler]{hofmann_authorities_2017}
E.~Hofmann, B.~Hartl, K.~Gangl, M.~Hartner-Tiefenthaler, and E.~Kirchler.
\newblock Authorities' {Coercive} and {Legitimate} {Power}: {The} {Impact} on {Cognitions} {Underlying} {Cooperation}.
\newblock \emph{Frontiers in Psychology}, 8, Jan. 2017.
\newblock ISSN 1664-1078.
\newblock \doi{10.3389/fpsyg.2017.00005}.
\newblock URL \url{http://journal.frontiersin.org/article/10.3389/fpsyg.2017.00005/full}.

\bibitem[Hohenstein et~al.(2023)Hohenstein, Kizilcec, DiFranzo, Aghajari, Mieczkowski, Levy, Naaman, Hancock, and Jung]{hohenstein_artificial_2023}
J.~Hohenstein, R.~F. Kizilcec, D.~DiFranzo, Z.~Aghajari, H.~Mieczkowski, K.~Levy, M.~Naaman, J.~Hancock, and M.~F. Jung.
\newblock Artificial intelligence in communication impacts language and social relationships.
\newblock \emph{Scientific Reports}, 13\penalty0 (1):\penalty0 5487, Apr. 2023.
\newblock ISSN 2045-2322.
\newblock \doi{10.1038/s41598-023-30938-9}.
\newblock URL \url{https://www.nature.com/articles/s41598-023-30938-9}.

\bibitem[Holdren(2008)]{holdren2008science}
J.~P. Holdren.
\newblock Science and technology for sustainable well-being.
\newblock \emph{Science}, 319\penalty0 (5862):\penalty0 424--434, 2008.

\bibitem[Holland et~al.(2003)Holland, Ester, and Kie{\ss}ling]{holland2003preference}
S.~Holland, M.~Ester, and W.~Kie{\ss}ling.
\newblock Preference mining: A novel approach on mining user preferences for personalized applications.
\newblock In \emph{Knowledge Discovery in Databases: PKDD 2003: 7th European Conference on Principles and Practice of Knowledge Discovery in Databases, Cavtat-Dubrovnik, Croatia, September 22-26, 2003. Proceedings 7}, pages 204--216. Springer, 2003.

\bibitem[Hooker(2002)]{hooker2002ideal}
B.~Hooker.
\newblock \emph{Ideal code, real world: A rule-consequentialist theory of morality}.
\newblock Oxford University Press, 2002.

\bibitem[Hooker(2021)]{hooker_does_2021}
B.~Hooker.
\newblock Does {Having} {Deep} {Personal} {Relationships} {Constitute} an {Element} of {Well}-{Being}?
\newblock \emph{Aristotelian Society Supplementary Volume}, 95\penalty0 (1):\penalty0 1--24, July 2021.
\newblock ISSN 0309-7013, 1467-8349.
\newblock \doi{10.1093/arisup/akab003}.
\newblock URL \url{https://academic.oup.com/aristoteliansupp/article/95/1/1/6312912}.

\bibitem[Hors-Fraile et~al.(2018)Hors-Fraile, Rivera-Romero, Schneider, Fernandez-Luque, Luna-Perejon, Civit-Balcells, and de~Vries]{hors2018analyzing}
S.~Hors-Fraile, O.~Rivera-Romero, F.~Schneider, L.~Fernandez-Luque, F.~Luna-Perejon, A.~Civit-Balcells, and H.~de~Vries.
\newblock Analyzing recommender systems for health promotion using a multidisciplinary taxonomy: A scoping review.
\newblock \emph{International Journal of Medical Informatics}, 114:\penalty0 143--155, 2018.

\bibitem[Horton and Tambe(2020)]{horton_death_2020}
J.~J. Horton and P.~Tambe.
\newblock The {Death} of a {Technical} {Skill}, Oct. 2020.
\newblock URL \url{https://john-joseph-horton.com/papers/schumpeter.pdf}.

\bibitem[Hotten(2015)]{hotten_volkswagen:_2015}
R.~Hotten.
\newblock Volkswagen: {The} scandal explained.
\newblock \emph{BBC News}, Sept. 2015.
\newblock URL \url{https://www.bbc.com/news/business-34324772}.

\bibitem[Howells et~al.(2016)Howells, Ivtzan, and Eiroa-Orosa]{howells2016putting}
A.~Howells, I.~Ivtzan, and F.~J. Eiroa-Orosa.
\newblock Putting the ‘app’in happiness: a randomised controlled trial of a smartphone-based mindfulness intervention to enhance wellbeing.
\newblock \emph{Journal of Happiness Studies}, 17:\penalty0 163--185, 2016.

\bibitem[Hsu and Myers(2023)]{hsu_another_2023}
T.~Hsu and S.~L. Myers.
\newblock Another {Side} of the {A}.{I}. {Boom}: {Detecting} {What} {A}.{I}. {Makes}.
\newblock \emph{The New York Times}, May 2023.
\newblock ISSN 0362-4331.
\newblock URL \url{https://www.nytimes.com/2023/05/18/technology/ai-chat-gpt-detection-tools.html}.

\bibitem[Huang et~al.(2011)Huang, Joseph, Nelson, Rubinstein, and Tygar]{huang_adversarial_2011}
L.~Huang, A.~D. Joseph, B.~Nelson, B.~I. Rubinstein, and J.~D. Tygar.
\newblock Adversarial machine learning.
\newblock In \emph{Proceedings of the 4th {ACM} workshop on {Security} and artificial intelligence}, {AISec} '11, pages 43--58, New York, NY, USA, Oct. 2011. Association for Computing Machinery.
\newblock ISBN 9781450310031.
\newblock \doi{10.1145/2046684.2046692}.
\newblock URL \url{https://doi.org/10.1145/2046684.2046692}.

\bibitem[Huang and Siddarth(2023)]{huang_generative_nodate}
S.~Huang and D.~Siddarth.
\newblock Generative {AI} and the {Digital} {Commons}, 2023.
\newblock URL \url{https://cip.org/research/generative-ai-digital-commons}.

\bibitem[Huang et~al.(2023)Huang, Toner, Haluza, Creemers, and Webster]{huang2023translation}
S.~Huang, H.~Toner, Z.~Haluza, R.~Creemers, and G.~Webster.
\newblock Translation: Measures for the management of generative artificial intelligence services (draft for comment) – april 2023.
\newblock \url{https://digichina.stanford.edu/work/translation-measures-for-the-management-of-generative-artificial-intelligence-services-draft-for-comment-april-2023/}, 2023.

\bibitem[Huang et~al.(2022)Huang, Xia, Xiao, Chan, Liang, Florence, Zeng, Tompson, Mordatch, Chebotar, et~al.]{huang2022inner}
W.~Huang, F.~Xia, T.~Xiao, H.~Chan, J.~Liang, P.~Florence, A.~Zeng, J.~Tompson, I.~Mordatch, Y.~Chebotar, et~al.
\newblock Inner monologue: Embodied reasoning through planning with language models.
\newblock \emph{arXiv preprint arXiv:2207.05608}, 2022.

\bibitem[Hubinger(2022)]{hubinger_how_2022}
E.~Hubinger.
\newblock How likely is deceptive alignment?, Aug. 2022.
\newblock URL \url{https://www.alignmentforum.org/posts/A9NxPTwbw6r6Awuwt/how-likely-is-deceptive-alignment}.

\bibitem[Hubinger(2023)]{hubinger_bing_2023}
E.~Hubinger.
\newblock Bing {Chat} is blatantly, aggressively misaligned, Feb. 2023.
\newblock URL \url{https://www.lesswrong.com/posts/jtoPawEhLNXNxvgTT/bing-chat-is-blatantly-aggressively-misaligned}.

\bibitem[Hubinger et~al.(2021)Hubinger, van Merwijk, Mikulik, Skalse, and Garrabrant]{hubinger_risks_2021}
E.~Hubinger, C.~van Merwijk, V.~Mikulik, J.~Skalse, and S.~Garrabrant.
\newblock Risks from {Learned} {Optimization} in {Advanced} {Machine} {Learning} {Systems}, Dec. 2021.
\newblock URL \url{http://arxiv.org/abs/1906.01820}.
\newblock arXiv:1906.01820 [cs].

\bibitem[Hughes et~al.(2020)Hughes, Anthony, Eccles, Leibo, Balduzzi, and Bachrach]{hughes_learning_2020}
E.~Hughes, T.~W. Anthony, T.~Eccles, J.~Z. Leibo, D.~Balduzzi, and Y.~Bachrach.
\newblock Learning to {Resolve} {Alliance} {Dilemmas} in {Many}-{Player} {Zero}-{Sum} {Games}, Feb. 2020.
\newblock URL \url{http://arxiv.org/abs/2003.00799}.
\newblock arXiv:2003.00799 [cs, stat].

\bibitem[Hughes and Waismel-Manor(2021)]{hughes_macedonian_2021}
H.~C. Hughes and I.~Waismel-Manor.
\newblock The {Macedonian} {Fake} {News} {Industry} and the 2016 {US} {Election}.
\newblock \emph{PS: Political Science \& Politics}, 54\penalty0 (1):\penalty0 19--23, Jan. 2021.
\newblock ISSN 1049-0965, 1537-5935.
\newblock \doi{10.1017/S1049096520000992}.
\newblock URL \url{https://www.cambridge.org/core/product/identifier/S1049096520000992/type/journal_article}.

\bibitem[Hume(1998)]{hume1998enquiry}
D.~Hume.
\newblock \emph{An enquiry concerning the principles of morals: {A} critical edition}, volume~4.
\newblock Oxford University Press, 1998.

\bibitem[Hunt et~al.(2022)Hunt, Sarkar, and Warhurst]{hunt_measuring_2022}
W.~Hunt, S.~Sarkar, and C.~Warhurst.
\newblock Measuring the impact of {AI} on jobs at the organization level: {Lessons} from a survey of {UK} business leaders.
\newblock \emph{Research Policy}, 51\penalty0 (2):\penalty0 104425, Mar. 2022.
\newblock ISSN 00487333.
\newblock \doi{10.1016/j.respol.2021.104425}.
\newblock URL \url{https://linkinghub.elsevier.com/retrieve/pii/S0048733321002183}.

\bibitem[Huppert(2009)]{huppert2009psychological}
F.~A. Huppert.
\newblock Psychological well-being: Evidence regarding its causes and consequences.
\newblock \emph{Applied psychology: Health and Well-being}, 1\penalty0 (2):\penalty0 137--164, 2009.

\bibitem[Huta(2015)]{huta2015overview}
V.~Huta.
\newblock An overview of hedonic and eudaimonic well-being concepts.
\newblock \emph{Handbook of Media Use and Well-being}, 2, 2015.

\bibitem[Huta and Waterman(2014)]{huta2014eudaimonia}
V.~Huta and A.~S. Waterman.
\newblock Eudaimonia and its distinction from hedonia: Developing a classification and terminology for understanding conceptual and operational definitions.
\newblock \emph{Journal of Happiness Studies}, 15:\penalty0 1425--1456, 2014.

\bibitem[Huynh and Hardouin(2023)]{huynh_poisongpt:_2023}
D.~Huynh and J.~Hardouin.
\newblock {PoisonGPT}: {How} {We} {Hid} a {Lobotomized} {LLM} on {Hugging} {Face} to {Spread} {Fake} {News}, July 2023.
\newblock URL \url{https://blog.mithrilsecurity.io/poisongpt-how-we-hid-a-lobotomized-llm-on-hugging-face-to-spread-fake-news/}.

\bibitem[Iazzolino(2021)]{iazzolino_infrastructure_2021}
G.~Iazzolino.
\newblock Infrastructure of compassionate repression: making sense of biometrics in {Kakuma} refugee camp.
\newblock \emph{Information Technology for Development}, 27\penalty0 (1):\penalty0 111--128, Jan. 2021.
\newblock ISSN 0268-1102, 1554-0170.
\newblock \doi{10.1080/02681102.2020.1816881}.
\newblock URL \url{https://www.tandfonline.com/doi/full/10.1080/02681102.2020.1816881}.

\bibitem[Ienca(2023)]{ienca_artificial_2023}
M.~Ienca.
\newblock On {Artificial} {Intelligence} and {Manipulation}.
\newblock \emph{Topoi}, 42\penalty0 (3):\penalty0 833--842, July 2023.
\newblock ISSN 1572-8749.
\newblock \doi{10.1007/s11245-023-09940-3}.
\newblock URL \url{https://doi.org/10.1007/s11245-023-09940-3}.

\bibitem[{Inflection AI}(2023{\natexlab{a}})]{inflection_ai_inflection-1:_nodate}
{Inflection AI}.
\newblock Inflection-1: {Pi}’s {Best}-in-{Class} {LLM}, 2023{\natexlab{a}}.
\newblock URL \url{https://inflection.ai/inflection-1}.

\bibitem[{Inflection AI}(2023{\natexlab{b}})]{inflection_ai_inflection-1_2023}
{Inflection AI}.
\newblock Inflection-1.
\newblock Technical report, 2023{\natexlab{b}}.
\newblock URL \url{https://inflection.ai/assets/Inflection-1.pdf}.

\bibitem[Intel(2022)]{intel_intel_2022}
Intel.
\newblock Intel {Introduces} {Real}-{Time} {Deepfake} {Detector}, Nov. 2022.
\newblock URL \url{https://www.intel.com/content/www/us/en/newsroom/news/intel-introduces-real-time-deepfake-detector.html#gs.r61ai2}.

\bibitem[{International Federation of Robotics}(2022)]{international_federation_of_robotics_world_2022}
{International Federation of Robotics}.
\newblock World {Robotics} {Report}: “{All}-{Time} {High}” with {Half} a {Million} {Robots} {Installed} in one {Year}, Oct. 2022.
\newblock URL \url{https://ifr.org/ifr-press-releases/news/wr-report-all-time-high-with-half-a-million-robots-installed}.

\bibitem[{International Labour Organization}(2023)]{international_labour_office_world_2023}
{International Labour Organization}.
\newblock \emph{World {Employment} and {Social} {Outlook}: {Trends} 2023}.
\newblock International Labour Organization, Jan. 2023.
\newblock ISBN 9789220372913.
\newblock URL \url{https://www.ilo.org/global/research/global-reports/weso/WCMS_865332/lang--en/index.htm}.
\newblock OCLC: 1369521456.

\bibitem[Ipsos(2023)]{ipsos_americans_2023}
Ipsos.
\newblock Americans hold mixed opinions on {AI} and fear its potential to disrupt society, drive misinformation, May 2023.
\newblock URL \url{https://www.ipsos.com/en-us/americans-hold-mixed-opinions-ai-and-fear-its-potential-disrupt-society-drive-misinformation}.

\bibitem[Irving et~al.(2018)Irving, Christiano, and Amodei]{irving_ai_2018}
G.~Irving, P.~Christiano, and D.~Amodei.
\newblock {AI} safety via debate, Oct. 2018.
\newblock URL \url{http://arxiv.org/abs/1805.00899}.
\newblock arXiv:1805.00899 [cs, stat].

\bibitem[Irwin et~al.(2023)Irwin, Wang, Tezil, Zhang, Filbey, Jung, Bullock~Mann, Dilig, and Parker]{irwin_condition_2023}
V.~Irwin, K.~Wang, T.~Tezil, J.~Zhang, A.~Filbey, J.~Jung, F.~Bullock~Mann, R.~Dilig, and S.~Parker.
\newblock Condition of {Education} 2023.
\newblock Technical Report NCES 2023-144rev, National Center for Education Statistics: US Department of Education, May 2023.
\newblock URL \url{https://nces.ed.gov/pubsearch/pubsinfo.asp?pubid=2023144rev}.

\bibitem[Isaac and Frenkel(2023)]{isaac_facebooks_2023}
M.~Isaac and S.~Frenkel.
\newblock Facebook’s {Algorithm} {Is} ‘{Influential}’ but {Doesn}’t {Necessarily} {Change} {Beliefs}, {Researchers} {Say}.
\newblock \emph{The New York Times}, July 2023.
\newblock ISSN 0362-4331.
\newblock URL \url{https://www.nytimes.com/2023/07/27/technology/facebook-instagram-algorithms.html}.

\bibitem[Isaacs et~al.(2013)Isaacs, Konrad, Walendowski, Lennig, Hollis, and Whittaker]{isaacs2013echoes}
E.~Isaacs, A.~Konrad, A.~Walendowski, T.~Lennig, V.~Hollis, and S.~Whittaker.
\newblock Echoes from the past: {H}ow technology mediated reflection improves well-being.
\newblock In \emph{Proceedings of the SIGCHI Conference on Human Factors in Computing Systems}, pages 1071--1080, 2013.

\bibitem[Jackman and Kanerva(2016)]{jackman_evolving_2016}
M.~Jackman and L.~Kanerva.
\newblock Evolving the {IRB}: {Building} {Robust} {Review} for {Industry} {Research}.
\newblock \emph{Washington and Lee Law Review Online}, 72\penalty0 (3):\penalty0 442, June 2016.
\newblock URL \url{https://scholarlycommons.law.wlu.edu/wlulr-online/vol72/iss3/8}.

\bibitem[Jackson(1998)]{jackson1998metaphysics}
F.~Jackson.
\newblock \emph{From metaphysics to ethics: A defence of conceptual analysis}.
\newblock Clarendon Press, 1998.

\bibitem[Jacobs and Wallach(2021)]{jacobs_measurement_2021}
A.~Z. Jacobs and H.~Wallach.
\newblock Measurement and {Fairness}.
\newblock In \emph{Proceedings of the 2021 {ACM} {Conference} on {Fairness}, {Accountability}, and {Transparency}}, pages 375--385, Virtual Event Canada, Mar. 2021. ACM.
\newblock ISBN 9781450383097.
\newblock \doi{10.1145/3442188.3445901}.
\newblock URL \url{https://dl.acm.org/doi/10.1145/3442188.3445901}.

\bibitem[Jakesch et~al.(2023{\natexlab{a}})Jakesch, Bhat, Buschek, Zalmanson, and Naaman]{jakesch_co-writing_2023}
M.~Jakesch, A.~Bhat, D.~Buschek, L.~Zalmanson, and M.~Naaman.
\newblock Co-{Writing} with {Opinionated} {Language} {Models} {Affects} {Users}' {Views}.
\newblock In \emph{Proceedings of the 2023 {CHI} {Conference} on {Human} {Factors} in {Computing} {Systems}}, pages 1--15, Apr. 2023{\natexlab{a}}.
\newblock \doi{10.1145/3544548.3581196}.
\newblock URL \url{http://arxiv.org/abs/2302.00560}.
\newblock arXiv:2302.00560 [cs].

\bibitem[Jakesch et~al.(2023{\natexlab{b}})Jakesch, Hancock, and Naaman]{jakesch_human_2023}
M.~Jakesch, J.~T. Hancock, and M.~Naaman.
\newblock Human heuristics for {AI}-generated language are flawed.
\newblock \emph{Proceedings of the National Academy of Sciences}, 120\penalty0 (11):\penalty0 e2208839120, Mar. 2023{\natexlab{b}}.
\newblock ISSN 0027-8424, 1091-6490.
\newblock \doi{10.1073/pnas.2208839120}.
\newblock URL \url{https://pnas.org/doi/10.1073/pnas.2208839120}.

\bibitem[Jamieson(2007)]{jamieson2007quality}
K.~Jamieson.
\newblock Quality of life 07 in twelve of {N}ew {Z}ealand cities.
\newblock \emph{The Quality of Life Project. Available online: http://www. qualityoflifeproject. govt. nz (accessed on 20 February 2014)}, 2007.

\bibitem[Jaques et~al.(2019)Jaques, Ghandeharioun, Shen, Ferguson, Lapedriza, Jones, Gu, and Picard]{jaques2019way}
N.~Jaques, A.~Ghandeharioun, J.~H. Shen, C.~Ferguson, A.~Lapedriza, N.~Jones, S.~Gu, and R.~Picard.
\newblock Way off-policy batch deep reinforcement learning of implicit human preferences in dialog.
\newblock \emph{arXiv preprint arXiv:1907.00456}, 2019.

\bibitem[Jasanoff(2016)]{jasanoff_ethics_2016}
S.~Jasanoff.
\newblock \emph{The ethics of invention: technology and the human future}.
\newblock The {Norton} global ethics series. W.W. Norton \& Company, New York, first edition edition, 2016.
\newblock ISBN 9780393078992.

\bibitem[Jelinek(1990)]{jelinek_self-organized_1990}
F.~Jelinek.
\newblock Self-organized language modeling for speech recognition.
\newblock In \emph{Readings in speech recognition}, pages 450--506. Morgan Kaufmann Publishers Inc., San Francisco, CA, USA, May 1990.
\newblock ISBN 9781558601246.

\bibitem[{Jenka}(2023)]{jenka_ai_2023}
{Jenka}.
\newblock {AI} and the {American} {Smile}, Mar. 2023.
\newblock URL \url{https://medium.com/@socialcreature/ai-and-the-american-smile-76d23a0fbfaf}.

\bibitem[Jeong and Breazeal(2016)]{jeong2016improving}
S.~Jeong and C.~L. Breazeal.
\newblock Improving smartphone users' affect and wellbeing with personalized positive psychology interventions.
\newblock In \emph{Proceedings of the Fourth International Conference on Human Agent Interaction}, pages 131--137, 2016.

\bibitem[Ji et~al.(2023)Ji, Lee, Frieske, Yu, Su, Xu, Ishii, Bang, Dai, Madotto, and Fung]{ji_survey_2023}
Z.~Ji, N.~Lee, R.~Frieske, T.~Yu, D.~Su, Y.~Xu, E.~Ishii, Y.~Bang, W.~Dai, A.~Madotto, and P.~Fung.
\newblock Survey of {Hallucination} in {Natural} {Language} {Generation}.
\newblock \emph{ACM Computing Surveys}, 55\penalty0 (12):\penalty0 1--38, Dec. 2023.
\newblock ISSN 0360-0300, 1557-7341.
\newblock \doi{10.1145/3571730}.
\newblock URL \url{http://arxiv.org/abs/2202.03629}.
\newblock arXiv:2202.03629 [cs].

\bibitem[Jia et~al.(2023)Jia, Luo, Fang, and Liao]{jia_when_2023}
N.~Jia, X.~Luo, Z.~Fang, and C.~Liao.
\newblock When and {How} {Artificial} {Intelligence} {Augments} {Employee} {Creativity}.
\newblock \emph{Academy of Management Journal}, page amj.2022.0426, Mar. 2023.
\newblock ISSN 0001-4273, 1948-0989.
\newblock \doi{10.5465/amj.2022.0426}.
\newblock URL \url{http://journals.aom.org/doi/full/10.5465/amj.2022.0426}.

\bibitem[Jia and Liang(2017)]{jia_adversarial_2017}
R.~Jia and P.~Liang.
\newblock Adversarial {Examples} for {Evaluating} {Reading} {Comprehension} {Systems}.
\newblock In M.~Palmer, R.~Hwa, and S.~Riedel, editors, \emph{Proceedings of the 2017 {Conference} on {Empirical} {Methods} in {Natural} {Language} {Processing}}, pages 2021--2031, Copenhagen, Denmark, Sept. 2017. Association for Computational Linguistics.
\newblock \doi{10.18653/v1/D17-1215}.
\newblock URL \url{https://aclanthology.org/D17-1215}.

\bibitem[Jiang et~al.(2019)Jiang, Chiappa, Lattimore, Gy{\"o}rgy, and Kohli]{jiang2019degenerate}
R.~Jiang, S.~Chiappa, T.~Lattimore, A.~Gy{\"o}rgy, and P.~Kohli.
\newblock Degenerate feedback loops in recommender systems.
\newblock In \emph{Proceedings of the 2019 AAAI/ACM Conference on AI, Ethics, and Society}, pages 383--390, 2019.

\bibitem[Jindal(2022)]{jindal_implementing_2022}
S.~Jindal.
\newblock Implementing {Responsible} {Data} {Enrichment} {Practices} at an {AI} {Developer}: {The} {Example} of {DeepMind}.
\newblock Technical report, Partnership on AI, Nov. 2022.
\newblock URL \url{https://partnershiponai.org/wp-content/uploads/2022/11/case-study_deepmind.pdf}.

\bibitem[Jobin et~al.(2019)Jobin, Ienca, and Vayena]{jobin_global_2019}
A.~Jobin, M.~Ienca, and E.~Vayena.
\newblock The global landscape of {AI} ethics guidelines.
\newblock \emph{Nature Machine Intelligence}, 1\penalty0 (9):\penalty0 389--399, Sept. 2019.
\newblock ISSN 2522-5839.
\newblock \doi{10.1038/s42256-019-0088-2}.
\newblock URL \url{https://www.nature.com/articles/s42256-019-0088-2}.

\bibitem[Johnson(2022)]{johnson_how_nodate}
K.~Johnson.
\newblock How {Wrongful} {Arrests} {Based} on {AI} {Derailed} 3 {Men}'s {Lives}.
\newblock \emph{Wired}, 2022.
\newblock ISSN 1059-1028.
\newblock URL \url{https://www.wired.com/story/wrongful-arrests-ai-derailed-3-mens-lives/}.

\bibitem[Johnson(2016)]{johnson_ai_2016}
O.~A. Johnson.
\newblock {AI} can excel at medical diagnosis, but the harder task is to win hearts and minds first, Aug. 2016.
\newblock URL \url{http://theconversation.com/ai-can-excel-at-medical-diagnosis-but-the-harder-task-is-to-win-hearts-and-minds-first-63782}.

\bibitem[Johnson and Johnson(2023)]{johnson_police_2023}
T.~L. Johnson and N.~N. Johnson.
\newblock Police {Facial} {Recognition} {Technology} {Can}'t {Tell} {Black} {People} {Apart}, May 2023.
\newblock URL \url{https://www.scientificamerican.com/article/police-facial-recognition-technology-cant-tell-black-people-apart/}.

\bibitem[Jonas(1984)]{jonas_imperative_1984}
H.~Jonas.
\newblock \emph{The {Imperative} of {Responsibility}: {In} {Search} of an {Ethics} for the {Technological} {Age}}.
\newblock University of Chicago Press, Chicago, IL, 1984.
\newblock URL \url{https://press.uchicago.edu/ucp/books/book/chicago/I/bo5953283.html}.

\bibitem[Jones(1996)]{jones_trust_1996}
K.~Jones.
\newblock Trust as an {Affective} {Attitude}.
\newblock \emph{Ethics}, 107\penalty0 (1):\penalty0 4--25, Oct. 1996.
\newblock ISSN 0014-1704, 1539-297X.
\newblock \doi{10.1086/233694}.
\newblock URL \url{https://www.journals.uchicago.edu/doi/10.1086/233694}.

\bibitem[Jongepier and Klenk(2022)]{jongepier_online_2022}
F.~Jongepier and M.~Klenk.
\newblock Online manipulation.
\newblock In \emph{The {Philosophy} of {Online} {Manipulation}}, pages 15--48. Routledge, New York, 1 edition, June 2022.
\newblock ISBN 9781003205425.
\newblock \doi{10.4324/9781003205425-3}.
\newblock URL \url{https://www.taylorfrancis.com/books/9781003205425/chapters/10.4324/9781003205425-3}.

\bibitem[Jorge et~al.(2008)Jorge, McIlraith, et~al.]{jorge2008planning}
A.~Jorge, S.~A. McIlraith, et~al.
\newblock Planning with preferences.
\newblock \emph{AI Magazine}, 29\penalty0 (4):\penalty0 25--25, 2008.

\bibitem[Joyce et~al.(2023)Joyce, Umney, Whittaker, and Stuart]{joyce_new_2023}
S.~Joyce, C.~Umney, X.~Whittaker, and M.~Stuart.
\newblock New social relations of digital technology and the future of work: {Beyond} technological determinism.
\newblock \emph{New Technology, Work and Employment}, 38\penalty0 (2):\penalty0 145--161, July 2023.
\newblock ISSN 0268-1072, 1468-005X.
\newblock \doi{10.1111/ntwe.12276}.
\newblock URL \url{https://onlinelibrary.wiley.com/doi/10.1111/ntwe.12276}.

\bibitem[Juhász et~al.(2023)Juhász, Lane, and Rodrik]{juhasz_new_2023}
R.~Juhász, N.~Lane, and D.~Rodrik.
\newblock The {New} {Economics} of {Industrial} {Policy}.
\newblock Technical Report w31538, National Bureau of Economic Research, Cambridge, MA, Aug. 2023.
\newblock URL \url{http://www.nber.org/papers/w31538.pdf}.

\bibitem[Jungherr et~al.(2020)Jungherr, Rivero, and Gayo-Avello]{jungherr_retooling_2020}
A.~Jungherr, G.~Rivero, and D.~Gayo-Avello.
\newblock \emph{Retooling {Politics}: {How} {Digital} {Media} {Are} {Shaping} {Democracy} (2020)}.
\newblock Cambridge University Press, 2020.
\newblock ISBN 978-1-108-41940-6.
\newblock URL \url{https://andreasjungherr.net/publications/retooling-politics-how-digital-media-are-shaping-democracy-2020/}.

\bibitem[Kaack et~al.(2022)Kaack, Donti, Strubell, Kamiya, Creutzig, and Rolnick]{kaack_aligning_2022}
L.~H. Kaack, P.~L. Donti, E.~Strubell, G.~Kamiya, F.~Creutzig, and D.~Rolnick.
\newblock Aligning artificial intelligence with climate change mitigation.
\newblock \emph{Nature Climate Change}, 12\penalty0 (6):\penalty0 518--527, June 2022.
\newblock ISSN 1758-678X, 1758-6798.
\newblock \doi{10.1038/s41558-022-01377-7}.
\newblock URL \url{https://www.nature.com/articles/s41558-022-01377-7}.

\bibitem[Kaddour et~al.(2023)Kaddour, Harris, Mozes, Bradley, Raileanu, and McHardy]{kaddour_challenges_2023}
J.~Kaddour, J.~Harris, M.~Mozes, H.~Bradley, R.~Raileanu, and R.~McHardy.
\newblock Challenges and {Applications} of {Large} {Language} {Models}, July 2023.
\newblock URL \url{http://arxiv.org/abs/2307.10169}.
\newblock arXiv:2307.10169 [cs].

\bibitem[Kahn~Jr et~al.(2011)Kahn~Jr, Reichert, Gary, Kanda, Ishiguro, Shen, Ruckert, and Gill]{kahn_new_2011}
P.~H. Kahn~Jr, A.~L. Reichert, H.~E. Gary, T.~Kanda, H.~Ishiguro, S.~Shen, J.~H. Ruckert, and B.~Gill.
\newblock The new ontological category hypothesis in human-robot interaction.
\newblock In \emph{Proceedings of the 6th international conference on Human-robot interaction}, pages 159--160, 2011.

\bibitem[Kahneman and Krueger(2006)]{kahneman2006developments}
D.~Kahneman and A.~B. Krueger.
\newblock Developments in the measurement of subjective well-being.
\newblock \emph{Journal of Economic perspectives}, 20\penalty0 (1):\penalty0 3--24, 2006.

\bibitem[Kak(2020)]{kak_global_2020}
A.~Kak.
\newblock "{The} {Global} {South} is everywhere, but also always somewhere": {National} {Policy} {Narratives} and {AI} {Justice}.
\newblock In \emph{Proceedings of the {AAAI}/{ACM} {Conference} on {AI}, {Ethics}, and {Society}}, pages 307--312, New York NY USA, Feb. 2020. ACM.
\newblock ISBN 9781450371100.
\newblock \doi{10.1145/3375627.3375859}.
\newblock URL \url{https://dl.acm.org/doi/10.1145/3375627.3375859}.

\bibitem[Kalliamvakou(2022)]{kalliamvakou_research:_2022}
E.~Kalliamvakou.
\newblock Research: quantifying {GitHub} {Copilot}’s impact on developer productivity and happiness, Sept. 2022.
\newblock URL \url{https://github.blog/2022-09-07-research-quantifying-github-copilots-impact-on-developer-productivity-and-happiness/}.

\bibitem[Kaminska(2017)]{kaminska_lesson_2017}
I.~Kaminska.
\newblock A lesson in fake news from the info-wars of ancient {Rome}.
\newblock \emph{Financial Times}, Jan. 2017.

\bibitem[Kamp and Desmet(2014)]{kamp2014measuring}
I.~Kamp and P.~M. Desmet.
\newblock Measuring product happiness.
\newblock In \emph{CHI'14 Extended Abstracts on Human Factors in Computing Systems}, pages 2509--2514. 2014.

\bibitem[Kanda et~al.(2004)Kanda, Hirano, Eaton, and Ishiguro]{kanda_interactive_2004}
T.~Kanda, T.~Hirano, D.~Eaton, and H.~Ishiguro.
\newblock Interactive {Robots} as {Social} {Partners} and {Peer} {Tutors} for {Children}: {A} {Field} {Trial}.
\newblock \emph{Human-Computer Interaction}, 19\penalty0 (1):\penalty0 61--84, June 2004.
\newblock ISSN 0737-0024.
\newblock \doi{10.1207/s15327051hci1901&2_4}.
\newblock URL \url{https://www.tandfonline.com/doi/abs/10.1080/07370024.2004.9667340}.

\bibitem[Kant(2017)]{kant2017kant}
I.~Kant.
\newblock \emph{Kant: The metaphysics of morals}.
\newblock Cambridge University Press, 2017.

\bibitem[Kaplan et~al.(2023)Kaplan, Kessler, Brill, and Hancock]{kaplan2023trust}
A.~D. Kaplan, T.~T. Kessler, J.~C. Brill, and P.~Hancock.
\newblock Trust in artificial intelligence: Meta-analytic findings.
\newblock \emph{Human factors}, 65\penalty0 (2):\penalty0 337--359, 2023.

\bibitem[Kaplan et~al.(2020)Kaplan, McCandlish, Henighan, Brown, Chess, Child, Gray, Radford, Wu, and Amodei]{kaplan_scaling_2020}
J.~Kaplan, S.~McCandlish, T.~Henighan, T.~B. Brown, B.~Chess, R.~Child, S.~Gray, A.~Radford, J.~Wu, and D.~Amodei.
\newblock Scaling {Laws} for {Neural} {Language} {Models}, Jan. 2020.
\newblock URL \url{http://arxiv.org/abs/2001.08361}.
\newblock arXiv:2001.08361 [cs, stat].

\bibitem[Kapteyn et~al.(2015)Kapteyn, Lee, Tassot, Vonkova, and Zamarro]{kapteyn2015dimensions}
A.~Kapteyn, J.~Lee, C.~Tassot, H.~Vonkova, and G.~Zamarro.
\newblock Dimensions of subjective well-being.
\newblock \emph{Social Indicators Research}, 123:\penalty0 625--660, 2015.

\bibitem[Karinshak et~al.(2023)Karinshak, Liu, Park, and Hancock]{karinshak_working_2023}
E.~Karinshak, S.~X. Liu, J.~S. Park, and J.~T. Hancock.
\newblock Working {With} {AI} to {Persuade}: {Examining} a {Large} {Language} {Model}'s {Ability} to {Generate} {Pro}-{Vaccination} {Messages}.
\newblock \emph{Proceedings of the ACM on Human-Computer Interaction}, 2023.
\newblock URL \url{https://doi.org/10.1145/3579592}.

\bibitem[Karusala et~al.(2018)Karusala, Vishwanath, Vashistha, Kumar, and Kumar]{karusala_only_2018}
N.~Karusala, A.~Vishwanath, A.~Vashistha, S.~Kumar, and N.~Kumar.
\newblock "{Only} if you use {English} you will get to more things": {Using} {Smartphones} to {Navigate} {Multilingualism}.
\newblock In \emph{Proceedings of the 2018 {CHI} {Conference} on {Human} {Factors} in {Computing} {Systems}}, pages 1--14, Montreal QC Canada, Apr. 2018. ACM.
\newblock ISBN 9781450356206.
\newblock \doi{10.1145/3173574.3174147}.
\newblock URL \url{https://dl.acm.org/doi/10.1145/3173574.3174147}.

\bibitem[Kasirzadeh and Gabriel(2023)]{kasirzadeh_conversation_2023}
A.~Kasirzadeh and I.~Gabriel.
\newblock In {Conversation} with {Artificial} {Intelligence}: {Aligning} language {Models} with {Human} {Values}.
\newblock \emph{Philosophy \& Technology}, 36\penalty0 (2):\penalty0 27, Apr. 2023.
\newblock ISSN 2210-5441.
\newblock \doi{10.1007/s13347-023-00606-x}.
\newblock URL \url{https://doi.org/10.1007/s13347-023-00606-x}.

\bibitem[Kasneci et~al.(2023)Kasneci, Sessler, Küchemann, Bannert, Dementieva, Fischer, Gasser, Groh, Günnemann, Hüllermeier, Krusche, Kutyniok, Michaeli, Nerdel, Pfeffer, Poquet, Sailer, Schmidt, Seidel, Stadler, Weller, Kuhn, and Kasneci]{kasneci_chatgpt_2023}
E.~Kasneci, K.~Sessler, S.~Küchemann, M.~Bannert, D.~Dementieva, F.~Fischer, U.~Gasser, G.~Groh, S.~Günnemann, E.~Hüllermeier, S.~Krusche, G.~Kutyniok, T.~Michaeli, C.~Nerdel, J.~Pfeffer, O.~Poquet, M.~Sailer, A.~Schmidt, T.~Seidel, M.~Stadler, J.~Weller, J.~Kuhn, and G.~Kasneci.
\newblock {ChatGPT} for good? {On} opportunities and challenges of large language models for education.
\newblock \emph{Learning and Individual Differences}, 103:\penalty0 102274, Apr. 2023.
\newblock ISSN 1041-6080.
\newblock \doi{10.1016/j.lindif.2023.102274}.
\newblock URL \url{https://www.sciencedirect.com/science/article/pii/S1041608023000195}.

\bibitem[Kavukcuoglu et~al.(2022)Kavukcuoglu, Kohli, Ibrahim, Bloxwich, and Brown]{kavukcuoglu_how_2022}
K.~Kavukcuoglu, P.~Kohli, L.~Ibrahim, D.~Bloxwich, and S.~Brown.
\newblock How our principles helped define {AlphaFold}’s release, Sept. 2022.
\newblock URL \url{https://deepmind.google/discover/blog/how-our-principles-helped-define-alphafolds-release/}.

\bibitem[Kazenwadel and Steinert(2023)]{kazenwadel_how_2023}
D.~Kazenwadel and C.~V. Steinert.
\newblock How {User} {Language} {Affects} {Conflict} {Fatality} {Estimates} in {ChatGPT}, July 2023.
\newblock URL \url{http://arxiv.org/abs/2308.00072}.
\newblock arXiv:2308.00072 [cs].

\bibitem[Keating(2009)]{keating_why_2009}
J.~Keating.
\newblock Why did {One} {Laptop} {Per} {Child} fail?, Sept. 2009.
\newblock URL \url{https://foreignpolicy.com/2009/09/09/why-did-one-laptop-per-child-fail/}.

\bibitem[Keeling and Burr(2022)]{keeling2022digital}
G.~Keeling and C.~Burr.
\newblock Digital manipulation and mental integrity.
\newblock In \emph{The Philosophy of Online Manipulation}, pages 253--271. Routledge, 2022.

\bibitem[Keeling and Paterson(2022)]{keeling2022proper}
G.~Keeling and N.~Paterson.
\newblock Proper functions: {E}tiology without typehood.
\newblock \emph{Biology \& Philosophy}, 37\penalty0 (3):\penalty0 19, 2022.

\bibitem[Kenton et~al.(2021)Kenton, Everitt, Weidinger, Gabriel, Mikulik, and Irving]{kenton_alignment_2021}
Z.~Kenton, T.~Everitt, L.~Weidinger, I.~Gabriel, V.~Mikulik, and G.~Irving.
\newblock Alignment of {Language} {Agents}, Mar. 2021.
\newblock URL \url{http://arxiv.org/abs/2103.14659}.
\newblock arXiv:2103.14659 [cs].

\bibitem[Kenton et~al.(2022)Kenton, Kumar, Farquhar, Richens, MacDermott, and Everitt]{kenton_discovering_2022}
Z.~Kenton, R.~Kumar, S.~Farquhar, J.~Richens, M.~MacDermott, and T.~Everitt.
\newblock Discovering {Agents}, Aug. 2022.
\newblock URL \url{http://arxiv.org/abs/2208.08345}.
\newblock arXiv:2208.08345 [cs].

\bibitem[Kepuska and Bohouta(2018)]{kepuska_next-generation_2018}
V.~Kepuska and G.~Bohouta.
\newblock Next-generation of virtual personal assistants ({Microsoft} {Cortana}, {Apple} {Siri}, {Amazon} {Alexa} and {Google} {Home}).
\newblock In \emph{2018 {IEEE} 8th {Annual} {Computing} and {Communication} {Workshop} and {Conference} ({CCWC})}, pages 99--103, Las Vegas, NV, Jan. 2018. IEEE.
\newblock ISBN 9781538646496.
\newblock \doi{10.1109/CCWC.2018.8301638}.
\newblock URL \url{http://ieeexplore.ieee.org/document/8301638/}.

\bibitem[Kerasidou(2017)]{kerasidou_trust_2017}
A.~Kerasidou.
\newblock Trust me, {I}’m a researcher!: {The} role of trust in biomedical research.
\newblock \emph{Medicine, Health Care and Philosophy}, 20\penalty0 (1):\penalty0 43--50, Mar. 2017.
\newblock ISSN 1572-8633.
\newblock \doi{10.1007/s11019-016-9721-6}.
\newblock URL \url{https://doi.org/10.1007/s11019-016-9721-6}.

\bibitem[Kerasidou et~al.(2022)Kerasidou, Kerasidou, Buscher, and Wilkinson]{kerasidou_before_2022}
C.~X. Kerasidou, A.~Kerasidou, M.~Buscher, and S.~Wilkinson.
\newblock Before and beyond trust: reliance in medical {AI}.
\newblock \emph{Journal of Medical Ethics}, 48\penalty0 (11):\penalty0 852--856, Nov. 2022.
\newblock ISSN 0306-6800, 1473-4257.
\newblock \doi{10.1136/medethics-2020-107095}.
\newblock URL \url{https://jme.bmj.com/lookup/doi/10.1136/medethics-2020-107095}.

\bibitem[Khan(2020)]{khan_ai}
S.~M. Khan.
\newblock {AI} {Chips}: {What} {They} {Are} and {Why} {They} {Matter}, 2020.
\newblock URL \url{https://cset.georgetown.edu/publication/ai-chips-what-they-are-and-why-they-matter/}.

\bibitem[Khwaja et~al.(2019)Khwaja, Ferrer, Iglesias, Faisal, and Matic]{khwaja2019aligning}
M.~Khwaja, M.~Ferrer, J.~O. Iglesias, A.~A. Faisal, and A.~Matic.
\newblock Aligning daily activities with personality: {T}owards a recommender system for improving wellbeing.
\newblock In \emph{Proceedings of the 13th AMC Conference on Recommender Systems}, pages 368--372, 2019.

\bibitem[Kierans et~al.(2022)Kierans, Hazan, and Dori-Hacohen]{kierans2022quantifying}
A.~Kierans, H.~Hazan, and S.~Dori-Hacohen.
\newblock Quantifying misalignment between agents.
\newblock \emph{ML Safety@ NeurIPS 2022}, 2022.

\bibitem[Kim and Sundar(2012)]{kim_anthropomorphism_2012}
Y.~Kim and S.~S. Sundar.
\newblock Anthropomorphism of computers: {Is} it mindful or mindless?
\newblock \emph{Computers in Human Behavior}, 28\penalty0 (1):\penalty0 241--250, Jan. 2012.
\newblock ISSN 0747-5632.
\newblock \doi{10.1016/j.chb.2011.09.006}.
\newblock URL \url{https://www.sciencedirect.com/science/article/pii/S0747563211001993}.

\bibitem[Kimani et~al.(2019)Kimani, Rowan, McDuff, Czerwinski, and Mark]{kimani2019conversational}
E.~Kimani, K.~Rowan, D.~McDuff, M.~Czerwinski, and G.~Mark.
\newblock A conversational agent in support of productivity and wellbeing at work.
\newblock In \emph{2019 8th International Conference on Affective Computing and Intelligent Interaction (ACII)}, pages 1--7. IEEE, 2019.

\bibitem[Kincannon et~al.(2022)Kincannon, Zahn, Clare, Lusty~Beech, Romberg, Larson, Bothner, Beckham, McGeehan, and DuBois]{kincannon_biochemical_2022}
W.~M. Kincannon, M.~Zahn, R.~Clare, J.~Lusty~Beech, A.~Romberg, J.~Larson, B.~Bothner, G.~T. Beckham, J.~E. McGeehan, and J.~L. DuBois.
\newblock Biochemical and structural characterization of an aromatic ring–hydroxylating dioxygenase for terephthalic acid catabolism.
\newblock \emph{Proceedings of the National Academy of Sciences}, 119\penalty0 (13):\penalty0 e2121426119, Mar. 2022.
\newblock ISSN 0027-8424, 1091-6490.
\newblock \doi{10.1073/pnas.2121426119}.
\newblock URL \url{https://pnas.org/doi/full/10.1073/pnas.2121426119}.

\bibitem[King et~al.(2014)King, Ren{\'o}, and Novo]{king2014concept}
M.~F. King, V.~F. Ren{\'o}, and E.~M. Novo.
\newblock The concept, dimensions and methods of assessment of human well-being within a socioecological context: {A} literature review.
\newblock \emph{Social Indicators Research}, 116:\penalty0 681--698, 2014.

\bibitem[King et~al.(2020)King, Aggarwal, Taddeo, and Floridi]{king_artificial_2020}
T.~C. King, N.~Aggarwal, M.~Taddeo, and L.~Floridi.
\newblock Artificial {Intelligence} {Crime}: {An} {Interdisciplinary} {Analysis} of {Foreseeable} {Threats} and {Solutions}.
\newblock \emph{Science and Engineering Ethics}, 26\penalty0 (1):\penalty0 89--120, Feb. 2020.
\newblock ISSN 1471-5546.
\newblock \doi{10.1007/s11948-018-00081-0}.
\newblock URL \url{https://doi.org/10.1007/s11948-018-00081-0}.

\bibitem[Kirby(2016)]{kirby_shadow_2016}
P.~Kirby.
\newblock Shadow {Schooling}: {Private} tuition and social mobility in the {UK}.
\newblock Technical report, The Sutton Trust, Sept. 2016.
\newblock URL \url{https://www.suttontrust.com/wp-content/uploads/2019/12/Shadow-Schooling-formatted-report_FINAL.pdf}.

\bibitem[Kirk et~al.(2023)Kirk, Bean, Vidgen, Rottger, and Hale]{kirk-etal-2023-past}
H.~Kirk, A.~Bean, B.~Vidgen, P.~Rottger, and S.~Hale.
\newblock The past, present and better future of feedback learning in large language models for subjective human preferences and values.
\newblock In H.~Bouamor, J.~Pino, and K.~Bali, editors, \emph{Proceedings of the 2023 Conference on Empirical Methods in Natural Language Processing}, pages 2409--2430, Singapore, Dec. 2023. Association for Computational Linguistics.
\newblock \doi{10.18653/v1/2023.emnlp-main.148}.
\newblock URL \url{https://aclanthology.org/2023.emnlp-main.148}.

\bibitem[Kivinen and Piiroinen(2023)]{kivinen_epochmaking_2023}
O.~Kivinen and T.~Piiroinen.
\newblock Epoch‐{Making} {Changes} in the {Cultural} {Evolution} of {Communication}: {Communication} technologies seen as organized hubs of skillful human activities.
\newblock \emph{Journal for the Theory of Social Behaviour}, 53\penalty0 (2):\penalty0 221--237, June 2023.
\newblock ISSN 0021-8308, 1468-5914.
\newblock \doi{10.1111/jtsb.12361}.
\newblock URL \url{https://onlinelibrary.wiley.com/doi/10.1111/jtsb.12361}.

\bibitem[Kleinberg and Raghavan(2021)]{kleinberg_algorithmic_2021}
J.~Kleinberg and M.~Raghavan.
\newblock Algorithmic monoculture and social welfare.
\newblock \emph{Proceedings of the National Academy of Sciences}, 118\penalty0 (22):\penalty0 e2018340118, June 2021.
\newblock ISSN 0027-8424, 1091-6490.
\newblock \doi{10.1073/pnas.2018340118}.
\newblock URL \url{https://pnas.org/doi/full/10.1073/pnas.2018340118}.

\bibitem[Kleinig and Evans(2013)]{kleinig2013human}
J.~Kleinig and N.~G. Evans.
\newblock Human flourishing, human dignity, and human rights.
\newblock \emph{Law and Philosophy}, 32\penalty0 (5):\penalty0 539--564, 2013.

\bibitem[Klenk(2020)]{klenk_digital_2020}
M.~Klenk.
\newblock Digital {Well}-{Being} and {Manipulation} {Online}.
\newblock In C.~Burr and L.~Floridi, editors, \emph{Ethics of {Digital} {Well}-{Being}: {A} {Multidisciplinary} {Approach}}, Philosophical {Studies} {Series}, pages 81--100. Springer International Publishing, Cham, 2020.
\newblock ISBN 9783030505851.
\newblock \doi{10.1007/978-3-030-50585-1_4}.
\newblock URL \url{https://doi.org/10.1007/978-3-030-50585-1_4}.

\bibitem[Klenk(2022)]{klenk_online_2022}
M.~Klenk.
\newblock ({Online}) manipulation: sometimes hidden, always careless.
\newblock \emph{Review of Social Economy}, 80\penalty0 (1):\penalty0 85--105, Jan. 2022.
\newblock ISSN 0034-6764, 1470-1162.
\newblock \doi{10.1080/00346764.2021.1894350}.
\newblock URL \url{https://www.tandfonline.com/doi/full/10.1080/00346764.2021.1894350}.

\bibitem[Kling(2018)]{kling_what_2018}
A.~Kling.
\newblock What {Gets} {Expensive}, and {Why}?, Dec. 2018.
\newblock URL \url{https://medium.com/@arnoldkling/what-gets-expensive-and-why-33bf4b891be2}.

\bibitem[Knight(2023)]{knight_google_nodate}
W.~Knight.
\newblock Google {Assistant} {Finally} {Gets} a {Generative} {AI} {Glow}-{Up}.
\newblock \emph{Wired}, 2023.
\newblock ISSN 1059-1028.
\newblock URL \url{https://www.wired.com/story/google-assistant-multi-modal-upgrade-bard-generative-ai/}.

\bibitem[Kocaballi et~al.(2020)Kocaballi, Quiroz, Laranjo, Rezazadegan, Kocielnik, Clark, Liao, Park, Moore, and Miner]{kocaballi2020conversational}
A.~B. Kocaballi, J.~C. Quiroz, L.~Laranjo, D.~Rezazadegan, R.~Kocielnik, L.~Clark, Q.~V. Liao, S.~Y. Park, R.~J. Moore, and A.~Miner.
\newblock Conversational agents for health and wellbeing.
\newblock In \emph{Extended Abstracts of the 2020 CHI Conference on Human Factors in Computing Systems}, pages 1--8, 2020.

\bibitem[Koenecke et~al.(2020)Koenecke, Nam, Lake, Nudell, Quartey, Mengesha, Toups, Rickford, Jurafsky, and Goel]{koenecke_racial_2020}
A.~Koenecke, A.~Nam, E.~Lake, J.~Nudell, M.~Quartey, Z.~Mengesha, C.~Toups, J.~R. Rickford, D.~Jurafsky, and S.~Goel.
\newblock Racial disparities in automated speech recognition.
\newblock \emph{Proceedings of the National Academy of Sciences}, 117\penalty0 (14):\penalty0 7684--7689, Apr. 2020.
\newblock ISSN 0027-8424, 1091-6490.
\newblock \doi{10.1073/pnas.1915768117}.
\newblock URL \url{https://pnas.org/doi/full/10.1073/pnas.1915768117}.

\bibitem[Kojima et~al.(2022)Kojima, Gu, Reid, Matsuo, and Iwasawa]{kojima_large_2022}
T.~Kojima, S.~S. Gu, M.~Reid, Y.~Matsuo, and Y.~Iwasawa.
\newblock Large {Language} {Models} are {Zero}-{Shot} {Reasoners}.
\newblock arXiv, 2022.
\newblock \doi{10.48550/arXiv.2205.11916}.
\newblock URL \url{http://arxiv.org/abs/2205.11916}.
\newblock arXiv:2205.11916 [cs].

\bibitem[Komeili et~al.(2022)Komeili, Shuster, and Weston]{komeili_internet-augmented_2022}
M.~Komeili, K.~Shuster, and J.~Weston.
\newblock Internet-{Augmented} {Dialogue} {Generation}.
\newblock In S.~Muresan, P.~Nakov, and A.~Villavicencio, editors, \emph{Proceedings of the 60th {Annual} {Meeting} of the {Association} for {Computational} {Linguistics} ({Volume} 1: {Long} {Papers})}, pages 8460--8478, Dublin, Ireland, May 2022. Association for Computational Linguistics.
\newblock \doi{10.18653/v1/2022.acl-long.579}.
\newblock URL \url{https://aclanthology.org/2022.acl-long.579.pdf}.

\bibitem[Korinek and Stiglitz(2018)]{korinek2018artificial}
A.~Korinek and J.~E. Stiglitz.
\newblock Artificial intelligence and its implications for income distribution and unemployment.
\newblock In \emph{The economics of artificial intelligence: An agenda}, pages 349--390. University of Chicago Press, 2018.

\bibitem[Korsgaard(1996)]{korsgaard_sources_1996}
C.~M. Korsgaard.
\newblock \emph{The {Sources} of {Normativity}}.
\newblock Cambridge University Press, New York, 1996.

\bibitem[Koster et~al.(2022)Koster, Balaguer, Tacchetti, Weinstein, Zhu, Hauser, Williams, Campbell-Gillingham, Thacker, Botvinick, and Summerfield]{koster_human-centred_2022}
R.~Koster, J.~Balaguer, A.~Tacchetti, A.~Weinstein, T.~Zhu, O.~Hauser, D.~Williams, L.~Campbell-Gillingham, P.~Thacker, M.~Botvinick, and C.~Summerfield.
\newblock Human-centred mechanism design with {Democratic} {AI}.
\newblock \emph{Nature Human Behaviour}, 6\penalty0 (10):\penalty0 1398--1407, July 2022.
\newblock ISSN 2397-3374.
\newblock \doi{10.1038/s41562-022-01383-x}.
\newblock URL \url{https://www.nature.com/articles/s41562-022-01383-x}.

\bibitem[Krakovna et~al.(2020)Krakovna, Uesato, Mikulik, Rahtz, Everitt, Kumar, Kenton, Leike, and Legg]{krakovna_specification_2020}
V.~Krakovna, J.~Uesato, V.~Mikulik, M.~Rahtz, T.~Everitt, R.~Kumar, Z.~Kenton, J.~Leike, and S.~Legg.
\newblock Specification gaming: the flip side of {AI} ingenuity, Apr. 2020.
\newblock URL \url{https://deepmind.google/discover/blog/specification-gaming-the-flip-side-of-ai-ingenuity/}.

\bibitem[Kramer et~al.(2014)Kramer, Guillory, and Hancock]{kramer2014experimental}
A.~D. Kramer, J.~E. Guillory, and J.~T. Hancock.
\newblock Experimental evidence of massive-scale emotional contagion through social networks.
\newblock \emph{Proceedings of the National Academy of Sciences}, 111\penalty0 (24):\penalty0 8788--8790, 2014.

\bibitem[Kramer(1990)]{kramer_birth_1990}
I.~Kramer.
\newblock The {Birth} of {Privacy} {Law}: {A} {Century} {Since} {Warren} and {Brandeis}.
\newblock \emph{Catholic University Law Review}, 39\penalty0 (3):\penalty0 703--724, Jan. 1990.
\newblock URL \url{https://scholarship.law.edu/lawreview/vol39/iss3/3}.

\bibitem[Kramer(2009)]{kramer_rethinking_2009}
R.~M. Kramer.
\newblock Rethinking {Trust}.
\newblock \emph{Harvard Business Review}, June 2009.
\newblock ISSN 0017-8012.
\newblock URL \url{https://hbr.org/2009/06/rethinking-trust}.

\bibitem[Kreiss and Barrett(2020)]{kreiss2020democratic}
D.~Kreiss and B.~Barrett.
\newblock Democratic tradeoffs: Platforms and political advertising.
\newblock \emph{Ohio St. Tech. LJ}, 16:\penalty0 493, 2020.

\bibitem[Kreps and Kriner(2023)]{kreps_potential_2023}
S.~Kreps and D.~L. Kriner.
\newblock The potential impact of emerging technologies on democratic representation: {Evidence} from a field experiment.
\newblock \emph{New Media \& Society}, Mar. 2023.
\newblock ISSN 1461-4448, 1461-7315.
\newblock \doi{10.1177/14614448231160526}.
\newblock URL \url{http://journals.sagepub.com/doi/10.1177/14614448231160526}.

\bibitem[Kreutzer et~al.(2022)Kreutzer, Caswell, Wang, Wahab, van Esch, Ulzii-Orshikh, Tapo, Subramani, Sokolov, Sikasote, Setyawan, Sarin, Samb, Sagot, Rivera, Rios, Papadimitriou, Osei, Suarez, Orife, Ogueji, Rubungo, Nguyen, Müller, Müller, Muhammad, Muhammad, Mnyakeni, Mirzakhalov, Matangira, Leong, Lawson, Kudugunta, Jernite, Jenny, Firat, Dossou, Dlamini, de~Silva, Ballı, Biderman, Battisti, Baruwa, Bapna, Baljekar, Azime, Awokoya, Ataman, Ahia, Ahia, Agrawal, and Adeyemi]{kreutzer_quality_2022}
J.~Kreutzer, I.~Caswell, L.~Wang, A.~Wahab, D.~van Esch, N.~Ulzii-Orshikh, A.~Tapo, N.~Subramani, A.~Sokolov, C.~Sikasote, M.~Setyawan, S.~Sarin, S.~Samb, B.~Sagot, C.~Rivera, A.~Rios, I.~Papadimitriou, S.~Osei, P.~O. Suarez, I.~Orife, K.~Ogueji, A.~N. Rubungo, T.~Q. Nguyen, M.~Müller, A.~Müller, S.~H. Muhammad, N.~Muhammad, A.~Mnyakeni, J.~Mirzakhalov, T.~Matangira, C.~Leong, N.~Lawson, S.~Kudugunta, Y.~Jernite, M.~Jenny, O.~Firat, B.~F.~P. Dossou, S.~Dlamini, N.~de~Silva, S.~{\c{C}abuk}. Ballı, S.~Biderman, A.~Battisti, A.~Baruwa, A.~Bapna, P.~Baljekar, I.~A. Azime, A.~Awokoya, D.~Ataman, O.~Ahia, O.~Ahia, S.~Agrawal, and M.~Adeyemi.
\newblock Quality at a {Glance}: {An} {Audit} of {Web}-{Crawled} {Multilingual} {Datasets}.
\newblock \emph{Transactions of the Association for Computational Linguistics}, 10:\penalty0 50--72, Jan. 2022.
\newblock ISSN 2307-387X.
\newblock \doi{10.1162/tacl_a_00447}.
\newblock URL \url{https://arxiv.org/pdf/2103.12028.pdf}.
\newblock arXiv:2103.12028 [cs].

\bibitem[Krstić and Saville(2019)]{krstic_deception_2019}
V.~Krstić and C.~Saville.
\newblock Deception ({Under} {Uncertainty}) as a {Kind} of {Manipulation}.
\newblock \emph{Australasian Journal of Philosophy}, 97\penalty0 (4):\penalty0 830--835, Oct. 2019.
\newblock ISSN 0004-8402, 1471-6828.
\newblock \doi{10.1080/00048402.2019.1604777}.
\newblock URL \url{https://www.tandfonline.com/doi/full/10.1080/00048402.2019.1604777}.

\bibitem[Krueger and Schkade(2008)]{krueger2008reliability}
A.~B. Krueger and D.~A. Schkade.
\newblock The reliability of subjective well-being measures.
\newblock \emph{Journal of Public Economics}, 92\penalty0 (8-9):\penalty0 1833--1845, 2008.

\bibitem[Krueger and Stone(2014)]{krueger2014progress}
A.~B. Krueger and A.~A. Stone.
\newblock Progress in measuring subjective well-being.
\newblock \emph{Science}, 346\penalty0 (6205):\penalty0 42--43, 2014.

\bibitem[Kuo et~al.(2022)Kuo, Kuo, and Chen]{kuo_assessing_2022}
T.-C. Kuo, C.-Y. Kuo, and L.-W. Chen.
\newblock Assessing environmental impacts of nanoscale semi-conductor manufacturing from the life cycle assessment perspective.
\newblock \emph{Resources, Conservation and Recycling}, 182:\penalty0 106289, July 2022.
\newblock ISSN 09213449.
\newblock \doi{10.1016/j.resconrec.2022.106289}.
\newblock URL \url{https://linkinghub.elsevier.com/retrieve/pii/S0921344922001379}.

\bibitem[Kurakin et~al.(2017)Kurakin, Goodfellow, and Bengio]{kurakin_adversarial_2017}
A.~Kurakin, I.~Goodfellow, and S.~Bengio.
\newblock Adversarial {Machine} {Learning} at {Scale}, Feb. 2017.
\newblock URL \url{http://arxiv.org/abs/1611.01236}.
\newblock arXiv:1611.01236 [cs, stat].

\bibitem[Kurakin et~al.(2023)Kurakin, Ponomareva, Syed, MacDermed, and Terzis]{Kurakin2023}
A.~Kurakin, N.~Ponomareva, U.~Syed, L.~MacDermed, and A.~Terzis.
\newblock Harnessing large-language models to generate private synthetic text, June 2023.
\newblock URL \url{http://arxiv.org/abs/2306.01684}.
\newblock arXiv:2306.01684 [lg, cr].

\bibitem[Kvasny(2006)]{kvasny_cultural_2006}
L.~Kvasny.
\newblock Cultural ({Re})production of digital inequality in a {US} community technology initiative.
\newblock \emph{Information, Communication \& Society}, 9\penalty0 (2):\penalty0 160--181, Apr. 2006.
\newblock ISSN 1369-118X, 1468-4462.
\newblock \doi{10.1080/13691180600630740}.
\newblock URL \url{http://www.tandfonline.com/doi/abs/10.1080/13691180600630740}.

\bibitem[Kässi et~al.(2021)Kässi, Lehdonvirta, and Stephany]{kassi_how_2021}
O.~Kässi, V.~Lehdonvirta, and F.~Stephany.
\newblock How many online workers are there in the world? {A} data-driven assessment.
\newblock \emph{Open Research Europe}, 1:\penalty0 53, Oct. 2021.
\newblock ISSN 2732-5121.
\newblock \doi{10.12688/openreseurope.13639.4}.
\newblock URL \url{https://open-research-europe.ec.europa.eu/articles/1-53/v4}.

\bibitem[Kääriä(2017)]{kaaria_technology_2017}
A.~Kääriä.
\newblock Technology acceptance of voice assistants : anthropomorphism as factor.
\newblock 2017.
\newblock URL \url{https://jyx.jyu.fi/handle/123456789/54612}.

\bibitem[Kühn et~al.(2014)Kühn, Brick, Müller, and Gallinat]{kuhn_is_2014}
S.~Kühn, T.~R. Brick, B.~C.~N. Müller, and J.~Gallinat.
\newblock Is {This} {Car} {Looking} at {You}? {How} {Anthropomorphism} {Predicts} {Fusiform} {Face} {Area} {Activation} when {Seeing} {Cars}.
\newblock \emph{PLoS ONE}, 9\penalty0 (12):\penalty0 e113885, Dec. 2014.
\newblock ISSN 1932-6203.
\newblock \doi{10.1371/journal.pone.0113885}.
\newblock URL \url{https://dx.plos.org/10.1371/journal.pone.0113885}.

\bibitem[Laestadius et~al.(2022)Laestadius, Bishop, Gonzalez, Illen{\v{c}}{\'\i}k, and Campos-Castillo]{laestadius_too_2022}
L.~Laestadius, A.~Bishop, M.~Gonzalez, D.~Illen{\v{c}}{\'\i}k, and C.~Campos-Castillo.
\newblock Too human and not human enough: A grounded theory analysis of mental health harms from emotional dependence on the social chatbot replika.
\newblock \emph{New Media \& Society}, page 14614448221142007, 2022.

\bibitem[Laitinen and Sahlgren(2021)]{laitinen_ai_2021}
A.~Laitinen and O.~Sahlgren.
\newblock Ai systems and respect for human autonomy.
\newblock \emph{Frontiers in artificial intelligence}, 4:\penalty0 151, 2021.

\bibitem[Lally and Gardner(2013)]{lally2013promoting}
P.~Lally and B.~Gardner.
\newblock Promoting habit formation.
\newblock \emph{Health Psychology Review}, 7\penalty0 (sup1):\penalty0 S137--S158, 2013.

\bibitem[Lam et~al.(2023)Lam, Sanchez-Gonzalez, Willson, Wirnsberger, Fortunato, Alet, Ravuri, Ewalds, Eaton-Rosen, Hu, et~al.]{lam2023learning}
R.~Lam, A.~Sanchez-Gonzalez, M.~Willson, P.~Wirnsberger, M.~Fortunato, F.~Alet, S.~Ravuri, T.~Ewalds, Z.~Eaton-Rosen, W.~Hu, et~al.
\newblock Learning skillful medium-range global weather forecasting.
\newblock \emph{Science}, 382\penalty0 (6677):\penalty0 1416--1421, 2023.

\bibitem[Land et~al.(2001)Land, Lamb, and Mustillo]{land2001child}
K.~C. Land, V.~L. Lamb, and S.~K. Mustillo.
\newblock Child and youth well-being in the united states, 1975-1998: Some findings from a new index.
\newblock \emph{Social Indicators Research}, pages 241--320, 2001.

\bibitem[Lane et~al.(2014)Lane, Lin, Mohammod, Yang, Lu, Cardone, Ali, Doryab, Berke, Campbell, et~al.]{lane2014bewell}
N.~D. Lane, M.~Lin, M.~Mohammod, X.~Yang, H.~Lu, G.~Cardone, S.~Ali, A.~Doryab, E.~Berke, A.~T. Campbell, et~al.
\newblock Bewell: Sensing sleep, physical activities and social interactions to promote wellbeing.
\newblock \emph{Mobile Networks and Applications}, 19:\penalty0 345--359, 2014.

\bibitem[Lange et~al.(2023)Lange, Keeling, McCroskery, Zevenbergen, Blascovich, Pedersen, Lentz, and Ag{\"u}era~y Arcas]{lange2023engaging}
B.~Lange, G.~Keeling, A.~McCroskery, B.~Zevenbergen, S.~Blascovich, K.~Pedersen, A.~Lentz, and B.~Ag{\"u}era~y Arcas.
\newblock Engaging engineering teams through moral imagination: a bottom-up approach for responsible innovation and ethical culture change in technology companies.
\newblock \emph{AI and Ethics}, pages 1--10, 2023.

\bibitem[Langosco et~al.(2023)Langosco, Koch, Sharkey, Pfau, Orseau, and Krueger]{langosco_goal_2023}
L.~Langosco, J.~Koch, L.~Sharkey, J.~Pfau, L.~Orseau, and D.~Krueger.
\newblock Goal {Misgeneralization} in {Deep} {Reinforcement} {Learning}, Jan. 2023.
\newblock URL \url{http://arxiv.org/abs/2105.14111}.
\newblock arXiv:2105.14111 [cs].

\bibitem[Lankton et~al.(2015)Lankton, McKnight, and Tripp]{lankton2015technology}
N.~K. Lankton, D.~H. McKnight, and J.~Tripp.
\newblock Technology, humanness, and trust: Rethinking trust in technology.
\newblock \emph{Journal of the Association for Information Systems}, 16\penalty0 (10):\penalty0 1, 2015.

\bibitem[Lanz(2023)]{lanz_meet_2023}
J.~A. Lanz.
\newblock Meet {Chaos}-{GPT}: {An} {AI} {Tool} {That} {Seeks} to {Destroy} {Humanity}, Apr. 2023.
\newblock URL \url{https://decrypt.co/126122/meet-chaos-gpt-ai-tool-destroy-humanity}.

\bibitem[Lanzetti et~al.(2021)Lanzetti, Schiffer, Ostrovsky, and Pavone]{lanzetti_interplay_2021}
N.~Lanzetti, M.~Schiffer, M.~Ostrovsky, and M.~Pavone.
\newblock On the {Interplay} between {Self}-{Driving} {Cars} and {Public} {Transportation}, Sept. 2021.
\newblock URL \url{http://arxiv.org/abs/2109.01627}.
\newblock arXiv:2109.01627 [physics].

\bibitem[Lanzi and Loiacono(2023)]{lanzi_chatgpt_2023}
P.~L. Lanzi and D.~Loiacono.
\newblock {ChatGPT} and {Other} {Large} {Language} {Models} as {Evolutionary} {Engines} for {Online} {Interactive} {Collaborative} {Game} {Design}.
\newblock In \emph{Proceedings of the {Genetic} and {Evolutionary} {Computation} {Conference}}, pages 1383--1390, July 2023.
\newblock \doi{10.1145/3583131.3590351}.
\newblock URL \url{http://arxiv.org/abs/2303.02155}.
\newblock arXiv:2303.02155 [cs].

\bibitem[Larson(2007)]{ogden_land_2007}
J.~Larson.
\newblock A {Land} {Full} of {Gods}: {Nature} {Deities} in {Greek} {Religion}.
\newblock In D.~Ogden, editor, \emph{A {Companion} to {Greek} {Religion}}, pages 56--70. Wiley, 1 edition, Jan. 2007.
\newblock ISBN 9781405120548 9780470996911.
\newblock \doi{10.1002/9780470996911.ch4}.
\newblock URL \url{https://onlinelibrary.wiley.com/doi/10.1002/9780470996911.ch4}.

\bibitem[Larsson and Traum(2000)]{larsson_information_2000}
S.~Larsson and D.~R. Traum.
\newblock Information state and dialogue management in the {TRINDI} dialogue move engine toolkit.
\newblock \emph{Natural Language Engineering}, 6\penalty0 (3-4):\penalty0 323--340, Sept. 2000.
\newblock ISSN 1469-8110, 1351-3249.
\newblock \doi{10.1017/S1351324900002539}.
\newblock URL \url{https://www.cambridge.org/core/journals/natural-language-engineering/article/abs/information-state-and-dialogue-management-in-the-trindi-dialogue-move-engine-toolkit/3C57FAE606CDC93C172B6090DF0FACE3}.

\bibitem[Lauter et~al.(2021)Lauter, Dai, and Laine]{lauter_protecting_2021}
K.~Lauter, W.~Dai, and K.~Laine, editors.
\newblock \emph{Protecting {Privacy} through {Homomorphic} {Encryption}}.
\newblock Springer International Publishing, Cham, 2021.
\newblock ISBN 9783030772864 9783030772871.
\newblock \doi{10.1007/978-3-030-77287-1}.
\newblock URL \url{https://link.springer.com/10.1007/978-3-030-77287-1}.

\bibitem[Laux et~al.(2023)Laux, Wachter, and Mittelstadt]{laux_trustworthy_2023}
J.~Laux, S.~Wachter, and B.~Mittelstadt.
\newblock Trustworthy artificial intelligence and the {European} {Union} {\textless}span style="font-variant:small-caps;"{\textgreater}{AI}{\textless}/span{\textgreater} act: {On} the conflation of trustworthiness and acceptability of risk.
\newblock \emph{Regulation \& Governance}, page rego.12512, Feb. 2023.
\newblock ISSN 1748-5983, 1748-5991.
\newblock \doi{10.1111/rego.12512}.
\newblock URL \url{https://onlinelibrary.wiley.com/doi/10.1111/rego.12512}.

\bibitem[Law(2023)]{law_persuasion_nodate}
H.~Law.
\newblock The {Persuasion} {Game}, 2023.
\newblock URL \url{https://www.learningfromexamples.com/p/the-persuasion-game}.

\bibitem[Layard(2010)]{layard2010measuring}
R.~Layard.
\newblock Measuring subjective well-being.
\newblock \emph{Science}, 327\penalty0 (5965):\penalty0 534--535, 2010.

\bibitem[Layard and De~Neve(2023)]{richard2023wellbeing}
R.~Layard and J.-E. De~Neve.
\newblock \emph{Wellbeing: Science and Policy}.
\newblock Cembridge University Press, 2023.

\bibitem[Lazar(2022)]{bullock_power_2022}
S.~Lazar.
\newblock Power and {AI}: {Nature} and {Justification}.
\newblock In J.~B. Bullock, Y.-C. Chen, J.~Himmelreich, V.~M. Hudson, A.~Korinek, M.~M. Young, and B.~Zhang, editors, \emph{The {Oxford} {Handbook} of {AI} {Governance}}. Oxford University Press, 1 edition, May 2022.
\newblock ISBN 9780197579329 9780197579350.
\newblock \doi{10.1093/oxfordhb/9780197579329.013.12}.
\newblock URL \url{https://academic.oup.com/edited-volume/41989/chapter/355437737}.

\bibitem[Lazar(2023)]{lazar_legitimacy_2023}
S.~Lazar.
\newblock Legitimacy, {Authority}, and {Democratic} {Duties} of {Explanation}, Oct. 2023.
\newblock URL \url{http://arxiv.org/abs/2208.08628}.
\newblock arXiv:2208.08628 [cs].

\bibitem[Lazar and Nelson(2023)]{lazar_ai_2023}
S.~Lazar and A.~Nelson.
\newblock {AI} safety on whose terms?
\newblock \emph{Science}, 381\penalty0 (6654):\penalty0 138--138, July 2023.
\newblock ISSN 0036-8075, 1095-9203.
\newblock \doi{10.1126/science.adi8982}.
\newblock URL \url{https://www.science.org/doi/10.1126/science.adi8982}.

\bibitem[Le~Bigot et~al.(2007)Le~Bigot, Rouet, and Jamet]{le_bigot_effects_2007}
L.~Le~Bigot, J.-F. Rouet, and E.~Jamet.
\newblock Effects of {Speech}- and {Text}-{Based} {Interaction} {Modes} in {Natural} {Language} {Human}-{Computer} {Dialogue}.
\newblock \emph{Human Factors: The Journal of the Human Factors and Ergonomics Society}, 49\penalty0 (6):\penalty0 1045--1053, Dec. 2007.
\newblock ISSN 0018-7208, 1547-8181.
\newblock \doi{10.1518/001872007X249901}.
\newblock URL \url{10.1518/001872007X249901}.

\bibitem[Lee et~al.(2019)Lee, Firat, Agarwal, Fannjiang, and Sussillo]{lee_hallucinations_2019}
K.~Lee, O.~Firat, A.~Agarwal, C.~Fannjiang, and D.~Sussillo.
\newblock Hallucinations in {Neural} {Machine} {Translation}, 2019.
\newblock URL \url{https://openreview.net/forum?id=SkxJ-309FQ}.

\bibitem[Lee et~al.(2021)Lee, Sheehan, Lee, and Chang]{lee_continuation_2021}
K.~Y. Lee, L.~Sheehan, K.~Lee, and Y.~Chang.
\newblock The continuation and recommendation intention of artificial intelligence-based voice assistant systems ({Aivas}): the influence of personal traits.
\newblock \emph{Internet Research}, 31\penalty0 (5):\penalty0 1899--1939, Nov. 2021.
\newblock ISSN 1066-2243.
\newblock \doi{10.1108/INTR-06-2020-0327}.
\newblock URL \url{https://www.emerald.com/insight/content/doi/10.1108/INTR-06-2020-0327/full/html}.

\bibitem[Lee et~al.(2023{\natexlab{a}})Lee, Srivastava, Hardy, Thickstun, Durmus, Paranjape, Gerard-Ursin, Li, Ladhak, Rong, Wang, Kwon, Park, Cao, Lee, Bommasani, Bernstein, and Liang]{lee_evaluating_2023}
M.~Lee, M.~Srivastava, A.~Hardy, J.~Thickstun, E.~Durmus, A.~Paranjape, I.~Gerard-Ursin, X.~L. Li, F.~Ladhak, F.~Rong, R.~E. Wang, M.~Kwon, J.~S. Park, H.~Cao, T.~Lee, R.~Bommasani, M.~Bernstein, and P.~Liang.
\newblock Evaluating {Human}-{Language} {Model} {Interaction}, Sept. 2023{\natexlab{a}}.
\newblock URL \url{http://arxiv.org/abs/2212.09746}.
\newblock arXiv:2212.09746 [cs].

\bibitem[Lee and Clarke(2019)]{lee_low-skilled_2019}
N.~Lee and S.~Clarke.
\newblock Do low-skilled workers gain from high-tech employment growth? {High}-technology multipliers, employment and wages in {Britain}.
\newblock \emph{Research Policy}, 48\penalty0 (9):\penalty0 103803, Nov. 2019.
\newblock ISSN 00487333.
\newblock \doi{10.1016/j.respol.2019.05.012}.
\newblock URL \url{https://linkinghub.elsevier.com/retrieve/pii/S0048733319301234}.

\bibitem[Lee et~al.(2023{\natexlab{b}})Lee, Ping, Xu, Patwary, Fung, Shoeybi, and Catanzaro]{lee_factuality_2023}
N.~Lee, W.~Ping, P.~Xu, M.~Patwary, P.~Fung, M.~Shoeybi, and B.~Catanzaro.
\newblock Factuality {Enhanced} {Language} {Models} for {Open}-{Ended} {Text} {Generation}, Mar. 2023{\natexlab{b}}.
\newblock URL \url{http://arxiv.org/abs/2206.04624}.
\newblock arXiv:2206.04624 [cs].

\bibitem[Lehman(2023)]{lehman_machine_2023}
J.~Lehman.
\newblock Machine {Love}, Feb. 2023.
\newblock URL \url{http://arxiv.org/abs/2302.09248}.
\newblock arXiv:2302.09248 [cs] version: 1.

\bibitem[Lehman et~al.(2020)Lehman, Clune, Misevic, Adami, Altenberg, Beaulieu, Bentley, Bernard, Beslon, Bryson, Cheney, Chrabaszcz, Cully, Doncieux, Dyer, Ellefsen, Feldt, Fischer, Forrest, Fŕenoy, Gagńe, Le~Goff, Grabowski, Hodjat, Hutter, Keller, Knibbe, Krcah, Lenski, Lipson, MacCurdy, Maestre, Miikkulainen, Mitri, Moriarty, Mouret, Nguyen, Ofria, Parizeau, Parsons, Pennock, Punch, Ray, Schoenauer, Schulte, Sims, Stanley, Taddei, Tarapore, Thibault, Watson, Weimer, and Yosinski]{lehman_surprising_2020}
J.~Lehman, J.~Clune, D.~Misevic, C.~Adami, L.~Altenberg, J.~Beaulieu, P.~J. Bentley, S.~Bernard, G.~Beslon, D.~M. Bryson, N.~Cheney, P.~Chrabaszcz, A.~Cully, S.~Doncieux, F.~C. Dyer, K.~O. Ellefsen, R.~Feldt, S.~Fischer, S.~Forrest, A.~Fŕenoy, C.~Gagńe, L.~Le~Goff, L.~M. Grabowski, B.~Hodjat, F.~Hutter, L.~Keller, C.~Knibbe, P.~Krcah, R.~E. Lenski, H.~Lipson, R.~MacCurdy, C.~Maestre, R.~Miikkulainen, S.~Mitri, D.~E. Moriarty, J.-B. Mouret, A.~Nguyen, C.~Ofria, M.~Parizeau, D.~Parsons, R.~T. Pennock, W.~F. Punch, T.~S. Ray, M.~Schoenauer, E.~Schulte, K.~Sims, K.~O. Stanley, F.~Taddei, D.~Tarapore, S.~Thibault, R.~Watson, W.~Weimer, and J.~Yosinski.
\newblock The {Surprising} {Creativity} of {Digital} {Evolution}: {A} {Collection} of {Anecdotes} from the {Evolutionary} {Computation} and {Artificial} {Life} {Research} {Communities}.
\newblock \emph{Artificial Life}, 26\penalty0 (2):\penalty0 274--306, May 2020.
\newblock ISSN 1064-5462, 1530-9185.
\newblock \doi{10.1162/artl_a_00319}.
\newblock URL \url{https://direct.mit.edu/artl/article/26/2/274-306/93255}.

\bibitem[Lei and Vesely(2010)]{lei_-group_2010}
V.~Lei and F.~Vesely.
\newblock In-{Group} versus {Out}-{Group} {Trust}: {The} {Impact} of {Income} {Inequality}.
\newblock \emph{Southern Economic Journal}, 76\penalty0 (4):\penalty0 1049--1063, Apr. 2010.
\newblock ISSN 0038-4038.
\newblock \doi{10.4284/sej.2010.76.4.1049}.
\newblock URL \url{http://doi.wiley.com/10.4284/sej.2010.76.4.1049}.

\bibitem[Leibo et~al.(2019)Leibo, Hughes, Lanctot, and Graepel]{leibo_autocurricula_2019}
J.~Z. Leibo, E.~Hughes, M.~Lanctot, and T.~Graepel.
\newblock Autocurricula and the {Emergence} of {Innovation} from {Social} {Interaction}: {A} {Manifesto} for {Multi}-{Agent} {Intelligence} {Research}, Mar. 2019.
\newblock URL \url{http://arxiv.org/abs/1903.00742}.
\newblock arXiv:1903.00742 [cs, q-bio].

\bibitem[Leike(2023)]{leike_combining_2023}
J.~Leike.
\newblock Combining weak-to-strong generalization with scalable oversight, 2023.
\newblock URL \url{https://aligned.substack.com/p/combining-w2sg-with-scalable-oversight}.

\bibitem[Leike et~al.(2018)Leike, Krueger, Everitt, Martic, Maini, and Legg]{leike_scalable_2018}
J.~Leike, D.~Krueger, T.~Everitt, M.~Martic, V.~Maini, and S.~Legg.
\newblock Scalable agent alignment via reward modeling: a research direction, Nov. 2018.
\newblock URL \url{http://arxiv.org/abs/1811.07871}.
\newblock arXiv:1811.07871 [cs, stat].

\bibitem[Lenman(2000)]{lenman2000consequentialism}
J.~Lenman.
\newblock Consequentialism and cluelessness.
\newblock \emph{Philosophy \& public affairs}, 29\penalty0 (4):\penalty0 342--370, 2000.

\bibitem[Letchford et~al.(2014)Letchford, Korzhyk, and Conitzer]{letchford_value_2014}
J.~Letchford, D.~Korzhyk, and V.~Conitzer.
\newblock On the value of commitment.
\newblock \emph{Autonomous Agents and Multi-Agent Systems}, 28\penalty0 (6):\penalty0 986--1016, Nov. 2014.
\newblock ISSN 1387-2532, 1573-7454.
\newblock \doi{10.1007/s10458-013-9246-9}.
\newblock URL \url{http://link.springer.com/10.1007/s10458-013-9246-9}.

\bibitem[Letheren et~al.(2021)Letheren, Jetten, Roberts, and Donovan]{letheren_robots_2021}
K.~Letheren, J.~Jetten, J.~Roberts, and J.~Donovan.
\newblock Robots should be seen and not heard… sometimes: Anthropomorphism and ai service robot interactions.
\newblock \emph{Psychology \& Marketing}, 38\penalty0 (12):\penalty0 2393--2406, 2021.

\bibitem[Leveson(2016)]{leveson2016engineering}
N.~G. Leveson.
\newblock \emph{Engineering a safer world: Systems thinking applied to safety}.
\newblock The MIT Press, 2016.

\bibitem[Levin(2020)]{levin2020human}
J.~Levin.
\newblock Human flourishing and population health: Meaning, measurement, and implications.
\newblock \emph{Perspectives in Biology and Medicine}, 63\penalty0 (3):\penalty0 401--419, 2020.

\bibitem[Levinstein and Herrmann(2023)]{levinstein_still_2023}
B.~Levinstein and D.~A. Herrmann.
\newblock Still no lie detector for language models: Probing empirical and conceptual roadblocks.
\newblock \emph{arXiv preprint arXiv:2307.00175}, 2023.

\bibitem[Levy and Patz(2015)]{levy_climate_2015}
B.~S. Levy and J.~A. Patz.
\newblock Climate {Change}, {Human} {Rights}, and {Social} {Justice}.
\newblock \emph{Annals of Global Health}, 81\penalty0 (3):\penalty0 310, Nov. 2015.
\newblock ISSN 2214-9996.
\newblock \doi{10.1016/j.aogh.2015.08.008}.
\newblock URL \url{https://annalsofglobalhealth.org/articles/10.1016/j.aogh.2015.08.008}.

\bibitem[Lewandowsky et~al.(2012)Lewandowsky, Ecker, Seifert, Schwarz, and Cook]{lewandowsky_misinformation_2012}
S.~Lewandowsky, U.~K.~H. Ecker, C.~M. Seifert, N.~Schwarz, and J.~Cook.
\newblock Misinformation and {Its} {Correction}: {Continued} {Influence} and {Successful} {Debiasing}.
\newblock \emph{Psychological Science in the Public Interest}, 13\penalty0 (3):\penalty0 106--131, Dec. 2012.
\newblock ISSN 1529-1006, 1539-6053.
\newblock \doi{10.1177/1529100612451018}.
\newblock URL \url{http://journals.sagepub.com/doi/10.1177/1529100612451018}.

\bibitem[Lewicki et~al.(2023)Lewicki, Lee, Cobbe, and Singh]{lewicki_out_2023}
K.~Lewicki, M.~S.~A. Lee, J.~Cobbe, and J.~Singh.
\newblock Out of {Context}: {Investigating} the {Bias} and {Fairness} {Concerns} of "{Artificial} {Intelligence} as a {Service}".
\newblock In \emph{Proceedings of the 2023 {CHI} {Conference} on {Human} {Factors} in {Computing} {Systems}}, pages 1--17, Apr. 2023.
\newblock \doi{10.1145/3544548.3581463}.
\newblock URL \url{http://arxiv.org/abs/2302.01448}.
\newblock arXiv:2302.01448 [cs].

\bibitem[Lewis et~al.(2020)Lewis, Oğuz, Rinott, Riedel, and Schwenk]{lewis_mlqa:_2020}
P.~Lewis, B.~Oğuz, R.~Rinott, S.~Riedel, and H.~Schwenk.
\newblock {MLQA}: {Evaluating} {Cross}-lingual {Extractive} {Question} {Answering}, May 2020.
\newblock URL \url{http://arxiv.org/abs/1910.07475}.
\newblock arXiv:1910.07475 [cs].

\bibitem[Lewis et~al.(2021)Lewis, Perez, Piktus, Petroni, Karpukhin, Goyal, Küttler, Lewis, Yih, Rocktäschel, Riedel, and Kiela]{lewis_retrieval-augmented_2021}
P.~Lewis, E.~Perez, A.~Piktus, F.~Petroni, V.~Karpukhin, N.~Goyal, H.~Küttler, M.~Lewis, W.-t. Yih, T.~Rocktäschel, S.~Riedel, and D.~Kiela.
\newblock Retrieval-{Augmented} {Generation} for {Knowledge}-{Intensive} {NLP} {Tasks}, Apr. 2021.
\newblock URL \url{http://arxiv.org/abs/2005.11401}.
\newblock arXiv:2005.11401 [cs].

\bibitem[Li et~al.(2023{\natexlab{a}})Li, Moon, Purkayastha, Celi, Trivedi, and Gichoya]{li_ethics_2023}
H.~Li, J.~T. Moon, S.~Purkayastha, L.~A. Celi, H.~Trivedi, and J.~W. Gichoya.
\newblock Ethics of large language models in medicine and medical research.
\newblock \emph{The Lancet Digital Health}, 5\penalty0 (6):\penalty0 e333--e335, June 2023{\natexlab{a}}.
\newblock ISSN 25897500.
\newblock \doi{10.1016/S2589-7500(23)00083-3}.
\newblock URL \url{https://www.thelancet.com/journals/landig/article/PIIS2589-7500(23)00083-3/fulltext}.

\bibitem[Li and Brar(2022)]{li2022use}
J.~Li and A.~Brar.
\newblock The use and impact of digital technologies for and on the mental health and wellbeing of indigenous people: a systematic review of empirical studies.
\newblock \emph{Computers in Human Behavior}, 126:\penalty0 106988, 2022.

\bibitem[Li et~al.(2023{\natexlab{b}})Li, Yang, Islam, and Ren]{li_making_2023}
P.~Li, J.~Yang, M.~A. Islam, and S.~Ren.
\newblock Making {AI} {Less} "{Thirsty}": {Uncovering} and {Addressing} the {Secret} {Water} {Footprint} of {AI} {Models}, Oct. 2023{\natexlab{b}}.
\newblock URL \url{http://arxiv.org/abs/2304.03271}.
\newblock arXiv:2304.03271 [cs].

\bibitem[Li et~al.(2023{\natexlab{c}})Li, Kim, Chan, and McGill]{li_detrimental_2023}
X.~S. Li, S.~Kim, K.~W. Chan, and A.~L. McGill.
\newblock Detrimental effects of anthropomorphism on the perceived physical safety of artificial agents in dangerous situations.
\newblock \emph{International Journal of Research in Marketing}, 40\penalty0 (4):\penalty0 841--864, 2023{\natexlab{c}}.

\bibitem[Liang et~al.(2022)Liang, Bommasani, Lee, Tsipras, Soylu, Yasunaga, Zhang, Narayanan, Wu, Kumar, Newman, Yuan, Yan, Zhang, Cosgrove, Manning, Ré, Acosta-Navas, Hudson, Zelikman, Durmus, Ladhak, Rong, Ren, Yao, Wang, Santhanam, Orr, Zheng, Yuksekgonul, Suzgun, Kim, Guha, Chatterji, Khattab, Henderson, Huang, Chi, Xie, Santurkar, Ganguli, Hashimoto, Icard, Zhang, Chaudhary, Wang, Li, Mai, Zhang, and Koreeda]{liang_holistic_2022}
P.~Liang, R.~Bommasani, T.~Lee, D.~Tsipras, D.~Soylu, M.~Yasunaga, Y.~Zhang, D.~Narayanan, Y.~Wu, A.~Kumar, B.~Newman, B.~Yuan, B.~Yan, C.~Zhang, C.~Cosgrove, C.~D. Manning, C.~Ré, D.~Acosta-Navas, D.~A. Hudson, E.~Zelikman, E.~Durmus, F.~Ladhak, F.~Rong, H.~Ren, H.~Yao, J.~Wang, K.~Santhanam, L.~Orr, L.~Zheng, M.~Yuksekgonul, M.~Suzgun, N.~Kim, N.~Guha, N.~Chatterji, O.~Khattab, P.~Henderson, Q.~Huang, R.~Chi, S.~M. Xie, S.~Santurkar, S.~Ganguli, T.~Hashimoto, T.~Icard, T.~Zhang, V.~Chaudhary, W.~Wang, X.~Li, Y.~Mai, Y.~Zhang, and Y.~Koreeda.
\newblock Holistic {Evaluation} of {Language} {Models}, Nov. 2022.
\newblock URL \url{http://arxiv.org/abs/2211.09110}.
\newblock arXiv:2211.09110 [cs] version: 1.

\bibitem[Liang(2019)]{liang2019recommender}
Y.~Liang.
\newblock Recommender system for developing new preferences and goals.
\newblock In \emph{Proceedings of the 13th ACM Conference on Recommender Systems}, pages 611--615, 2019.

\bibitem[Liang et~al.(2023)Liang, Wu, Song, Wu, Xia, Liu, Ou, Lu, Ji, Mao, Wang, Shou, Gong, and Duan]{liang_taskmatrix.ai:_2023}
Y.~Liang, C.~Wu, T.~Song, W.~Wu, Y.~Xia, Y.~Liu, Y.~Ou, S.~Lu, L.~Ji, S.~Mao, Y.~Wang, L.~Shou, M.~Gong, and N.~Duan.
\newblock {TaskMatrix}.{AI}: {Completing} {Tasks} by {Connecting} {Foundation} {Models} with {Millions} of {APIs}, Mar. 2023.
\newblock URL \url{http://arxiv.org/abs/2303.16434}.
\newblock arXiv:2303.16434 [cs].

\bibitem[Lidz(1998)]{lidz_coercion_1998}
C.~W. Lidz.
\newblock Coercion in psychiatric care: what have we learned from research?
\newblock \emph{The Journal of the American Academy of Psychiatry and the Law}, 26\penalty0 (4):\penalty0 631--637, 1998.
\newblock ISSN 1093-6793.

\bibitem[Lightman et~al.(2023)Lightman, Kosaraju, Burda, Edwards, Baker, Lee, Leike, Schulman, Sutskever, and Cobbe]{lightman_lets_2023}
H.~Lightman, V.~Kosaraju, Y.~Burda, H.~Edwards, B.~Baker, T.~Lee, J.~Leike, J.~Schulman, I.~Sutskever, and K.~Cobbe.
\newblock Let's {Verify} {Step} by {Step}, May 2023.
\newblock URL \url{http://arxiv.org/abs/2305.20050}.
\newblock arXiv:2305.20050 [cs].

\bibitem[Lim et~al.(2022)Lim, Flageat, and Cully]{lim2022efficient}
B.~Lim, M.~Flageat, and A.~Cully.
\newblock Efficient exploration using model-based quality-diversity with gradients.
\newblock \emph{arXiv preprint arXiv:2211.12610}, 2022.

\bibitem[Lima et~al.(2019)Lima, Furtado, Furtado, and Almeida]{lima_empirical_2019}
L.~Lima, V.~Furtado, E.~Furtado, and V.~Almeida.
\newblock Empirical {Analysis} of {Bias} in {Voice}-based {Personal} {Assistants}.
\newblock In \emph{Companion {Proceedings} of {The} 2019 {World} {Wide} {Web} {Conference}}, pages 533--538, San Francisco USA, May 2019. ACM.
\newblock ISBN 9781450366755.
\newblock \doi{10.1145/3308560.3317597}.
\newblock URL \url{https://dl.acm.org/doi/10.1145/3308560.3317597}.

\bibitem[Lin et~al.(2022)Lin, Hilton, and Evans]{lin_truthfulqa:_2022}
S.~Lin, J.~Hilton, and O.~Evans.
\newblock {TruthfulQA}: {Measuring} {How} {Models} {Mimic} {Human} {Falsehoods}, May 2022.
\newblock URL \url{http://arxiv.org/abs/2109.07958}.
\newblock arXiv:2109.07958 [cs].

\bibitem[Lingel and Crawford(2020)]{lingel_alexa_2020}
J.~Lingel and K.~Crawford.
\newblock "{Alexa}, {Tell} {Me} about {Your} {Mother}": {The} {History} of the {Secretary} and the {End} of {Secrecy}.
\newblock \emph{Catalyst: Feminism, Theory, Technoscience}, 6\penalty0 (1), May 2020.
\newblock ISSN 2380-3312.
\newblock \doi{10.28968/cftt.v6i1.29949}.
\newblock URL \url{https://catalystjournal.org/index.php/catalyst/article/view/29949}.

\bibitem[Linxen et~al.(2021)Linxen, Sturm, Brühlmann, Cassau, Opwis, and Reinecke]{linxen_how_2021}
S.~Linxen, C.~Sturm, F.~Brühlmann, V.~Cassau, K.~Opwis, and K.~Reinecke.
\newblock How {WEIRD} is {CHI}?
\newblock In \emph{Proceedings of the 2021 {CHI} {Conference} on {Human} {Factors} in {Computing} {Systems}}, pages 1--14, Yokohama Japan, May 2021. ACM.
\newblock ISBN 9781450380966.
\newblock \doi{10.1145/3411764.3445488}.
\newblock URL \url{https://dl.acm.org/doi/10.1145/3411764.3445488}.

\bibitem[Liu et~al.(2020)Liu, Yang, Chi, Hsu, and Lee]{liu_mockingjay:_2020}
A.~T. Liu, S.-w. Yang, P.-H. Chi, P.-c. Hsu, and H.-y. Lee.
\newblock Mockingjay: {Unsupervised} {Speech} {Representation} {Learning} with {Deep} {Bidirectional} {Transformer} {Encoders}.
\newblock In \emph{{ICASSP} 2020 - 2020 {IEEE} {International} {Conference} on {Acoustics}, {Speech} and {Signal} {Processing} ({ICASSP})}, pages 6419--6423, May 2020.
\newblock \doi{10.1109/ICASSP40776.2020.9054458}.
\newblock URL \url{http://arxiv.org/abs/1910.12638}.
\newblock arXiv:1910.12638 [cs, eess].

\bibitem[Liu et~al.(2023{\natexlab{a}})Liu, Cheng, Liu, Zhang, Li, Ren, Zou, Yang, Su, Zhu, et~al.]{liu2023llava}
S.~Liu, H.~Cheng, H.~Liu, H.~Zhang, F.~Li, T.~Ren, X.~Zou, J.~Yang, H.~Su, J.~Zhu, et~al.
\newblock Llava-plus: Learning to use tools for creating multimodal agents.
\newblock \emph{arXiv preprint arXiv:2311.05437}, 2023{\natexlab{a}}.

\bibitem[Liu et~al.(2017)Liu, Ai, Li, Tang, Huang, Feng, and Mei]{liu2017deriving}
X.~Liu, W.~Ai, H.~Li, J.~Tang, G.~Huang, F.~Feng, and Q.~Mei.
\newblock Deriving user preferences of mobile apps from their management activities.
\newblock \emph{ACM Transactions on Information Systems (TOIS)}, 35\penalty0 (4):\penalty0 1--32, 2017.

\bibitem[Liu et~al.(2023{\natexlab{b}})Liu, Deng, Xu, Li, Zheng, Zhang, Zhao, Zhang, and Liu]{liu_jailbreaking_2023}
Y.~Liu, G.~Deng, Z.~Xu, Y.~Li, Y.~Zheng, Y.~Zhang, L.~Zhao, T.~Zhang, and Y.~Liu.
\newblock Jailbreaking {ChatGPT} via {Prompt} {Engineering}: {An} {Empirical} {Study}, May 2023{\natexlab{b}}.
\newblock URL \url{https://arxiv.org/pdf/2305.13860.pdf}.
\newblock arXiv:2305.13860 [cs].

\bibitem[Lomas(2022)]{miscex1}
N.~Lomas.
\newblock Social media a factor in death of uk schoolgirl, inquest finds.
\newblock \url{https://techcrunch.com/2022/09/30/molly-russell-inquest-verdict/}, 2022.
\newblock Accessed: 2023-07-13.

\bibitem[Long(2023)]{long2023large}
J.~Long.
\newblock Large language model guided tree-of-thought, 2023.

\bibitem[Longino(1993)]{longino1993}
H.~Longino.
\newblock Feminist standpoint theory and the problems of knowledge.
\newblock \emph{Signs: Journal of Women in Culture and Society}, 19\penalty0 (1):\penalty0 201--212, 1993.
\newblock \doi{10.1086/494867}.

\bibitem[Longpre et~al.(2023)Longpre, Yauney, Reif, Lee, Roberts, Zoph, Zhou, Wei, Robinson, Mimno, and Ippolito]{longpre_pretrainers_2023}
S.~Longpre, G.~Yauney, E.~Reif, K.~Lee, A.~Roberts, B.~Zoph, D.~Zhou, J.~Wei, K.~Robinson, D.~Mimno, and D.~Ippolito.
\newblock A {Pretrainer}'s {Guide} to {Training} {Data}: {Measuring} the {Effects} of {Data} {Age}, {Domain} {Coverage}, {Quality}, \& {Toxicity}, May 2023.
\newblock URL \url{https://arxiv.org/pdf/2305.13169.pdf}.
\newblock arXiv:2305.13169 [cs].

\bibitem[Lotz et~al.(2023)Lotz, Valdez, and Ziefle]{duffy_dont_2023}
V.~Lotz, A.~C. Valdez, and M.~Ziefle.
\newblock Don’t {Stand} so {Close} to {Me}: {Acceptance} of {Delegating} {Intimate} {Health} {Care} {Tasks} to {Assistive} {Robots}.
\newblock In V.~G. Duffy, M.~Ziefle, P.-L.~P. Rau, and M.~M. Tseng, editors, \emph{Human-{Automation} {Interaction}}, volume~12, pages 3--21. Springer International Publishing, Cham, 2023.
\newblock ISBN 9783031107870 9783031107887.
\newblock \doi{10.1007/978-3-031-10788-7_1}.
\newblock URL \url{https://link.springer.com/10.1007/978-3-031-10788-7_1}.

\bibitem[Lovato and Piper(2015)]{lovato_siri_2015}
S.~Lovato and A.~M. Piper.
\newblock "{Siri}, is this you?": {Understanding} young children's interactions with voice input systems.
\newblock In \emph{Proceedings of the 14th {International} {Conference} on {Interaction} {Design} and {Children}}, pages 335--338, Boston Massachusetts, June 2015. ACM.
\newblock ISBN 9781450335904.
\newblock \doi{10.1145/2771839.2771910}.
\newblock URL \url{https://dl.acm.org/doi/10.1145/2771839.2771910}.

\bibitem[Lovens(2023)]{lovens_sans_2023}
P.-F. Lovens.
\newblock "{Sans} ces conversations avec le chatbot {Eliza}, mon mari serait toujours là".
\newblock \emph{La Libre}, Mar. 2023.
\newblock URL \url{https://www.lalibre.be/belgique/societe/2023/03/28/sans-ces-conversations-avec-le-chatbot-eliza-mon-mari-serait-toujours-la-LVSLWPC5WRDX7J2RCHNWPDST24/}.

\bibitem[Luccioni and Hernandez-Garcia(2023)]{luccioni_counting_2023}
A.~S. Luccioni and A.~Hernandez-Garcia.
\newblock Counting {Carbon}: {A} {Survey} of {Factors} {Influencing} the {Emissions} of {Machine} {Learning}, Feb. 2023.
\newblock URL \url{http://arxiv.org/abs/2302.08476}.
\newblock arXiv:2302.08476 [cs].

\bibitem[Luccioni et~al.(2022)Luccioni, Viguier, and Ligozat]{luccioni_estimating_2022}
A.~S. Luccioni, S.~Viguier, and A.-L. Ligozat.
\newblock Estimating the {Carbon} {Footprint} of {BLOOM}, a {176B} {Parameter} {Language} {Model}, Nov. 2022.
\newblock URL \url{http://arxiv.org/abs/2211.02001}.
\newblock arXiv:2211.02001 [cs].

\bibitem[Luce and Raiffa(1957)]{luce_games_1957}
R.~D. Luce and H.~Raiffa.
\newblock \emph{Games and decisions: {Introduction} and critical survey}.
\newblock Games and decisions: {Introduction} and critical survey. Wiley, Oxford, England, 1957.

\bibitem[Ludlow et~al.(2023)Ludlow, Anand, and McBride]{ludlow_character.ai_2023}
E.~Ludlow, P.~Anand, and S.~McBride.
\newblock Character.{AI} in {Early} {Talks} for {Funding} at {More} {Than} \$5 {Billion} {Valuation}.
\newblock \emph{Bloomberg}, Sept. 2023.
\newblock URL \url{https://www.bloomberg.com/news/articles/2023-09-28/character-ai-in-early-talks-for-funding-at-more-than-5-billion-valuation}.

\bibitem[Lukas et~al.(2023)Lukas, Salem, Sim, Tople, Wutschitz, and Zanella-Béguelin]{lukas_analyzing_2023}
N.~Lukas, A.~Salem, R.~Sim, S.~Tople, L.~Wutschitz, and S.~Zanella-Béguelin.
\newblock Analyzing {Leakage} of {Personally} {Identifiable} {Information} in {Language} {Models}, Apr. 2023.
\newblock URL \url{http://arxiv.org/abs/2302.00539}.
\newblock arXiv:2302.00539 [cs].

\bibitem[Lum and Isaac(2016)]{lum_predict_2016}
K.~Lum and W.~Isaac.
\newblock To {Predict} and {Serve}?
\newblock \emph{Significance}, 13\penalty0 (5):\penalty0 14--19, Oct. 2016.
\newblock ISSN 1740-9705, 1740-9713.
\newblock \doi{10.1111/j.1740-9713.2016.00960.x}.
\newblock URL \url{https://academic.oup.com/jrssig/article/13/5/14/7029190}.

\bibitem[Luo et~al.(2022)Luo, Paduraru, Voicu, Chervonyi, Munns, Li, Qian, Dutta, Davis, Wu, et~al.]{luo2022controlling}
J.~Luo, C.~Paduraru, O.~Voicu, Y.~Chervonyi, S.~Munns, J.~Li, C.~Qian, P.~Dutta, J.~Q. Davis, N.~Wu, et~al.
\newblock Controlling commercial cooling systems using reinforcement learning.
\newblock \emph{arXiv preprint arXiv:2211.07357}, 2022.

\bibitem[Lyon(2008)]{lyon_biometrics_2008}
D.~Lyon.
\newblock Biometrics, {Identification}, and {Surveillance}.
\newblock \emph{Bioethics}, 22\penalty0 (9):\penalty0 499--508, Nov. 2008.
\newblock ISSN 0269-9702, 1467-8519.
\newblock \doi{10.1111/j.1467-8519.2008.00697.x}.
\newblock URL \url{https://onlinelibrary.wiley.com/doi/10.1111/j.1467-8519.2008.00697.x}.

\bibitem[Maas(2018)]{maas_regulating_2018}
M.~M. Maas.
\newblock Regulating for '{Normal} {AI} {Accidents}': {Operational} {Lessons} for the {Responsible} {Governance} of {Artificial} {Intelligence} {Deployment}, Dec. 2018.
\newblock URL \url{https://papers.ssrn.com/abstract=3756941}.

\bibitem[Macdonald(2023)]{macdonald_india_2023}
A.~Macdonald.
\newblock India okays 22 financial entities to perform {Aadhaar} biometric authentication {\textbar} {Biometric} {Update}, May 2023.
\newblock URL \url{https://www.biometricupdate.com/202305/india-okays-22-financial-entities-to-perform-aadhaar-biometric-authentication}.

\bibitem[Machkovech(2017)]{machkovech_report:_2017}
S.~Machkovech.
\newblock Report: {Facebook} helped advertisers target teens who feel “worthless” [{Updated}], May 2017.
\newblock URL \url{https://arstechnica.com/information-technology/2017/05/facebook-helped-advertisers-target-teens-who-feel-worthless/}.

\bibitem[MacKenzie and Bhatt(2020)]{mackenzie_lies_2020}
A.~MacKenzie and I.~Bhatt.
\newblock Lies, {Bullshit} and {Fake} {News}: {Some} {Epistemological} {Concerns}.
\newblock \emph{Postdigital Science and Education}, 2\penalty0 (1):\penalty0 9--13, Jan. 2020.
\newblock ISSN 2524-4868.
\newblock \doi{10.1007/s42438-018-0025-4}.
\newblock URL \url{https://doi.org/10.1007/s42438-018-0025-4}.

\bibitem[Mackenzie and Stoljar(2000)]{mackenzie2000relational}
C.~Mackenzie and N.~Stoljar.
\newblock \emph{Relational autonomy: Feminist perspectives on autonomy, agency, and the social self}.
\newblock Oxford University Press, 2000.

\bibitem[Mackenzie et~al.(2013)Mackenzie, Rogers, and Dodds]{mackenzie_vulnerability:_2013}
C.~Mackenzie, W.~Rogers, and S.~Dodds, editors.
\newblock \emph{Vulnerability: {New} {Essays} in {Ethics} and {Feminist} {Philosophy}}.
\newblock Oxford University Press, Dec. 2013.
\newblock ISBN 9780199316649.
\newblock \doi{10.1093/acprof:oso/9780199316649.001.0001}.
\newblock URL \url{https://academic.oup.com/book/1543}.

\bibitem[Macklin(1982)]{macklin_man_1982}
R.~Macklin.
\newblock Man, {Mind}, and {Morality}: {The} {Ethics} of {Behavior} {Control}, 1982.
\newblock URL \url{https://repository.library.georgetown.edu/handle/10822/792035}.

\bibitem[Madhavan and Wiegmann(2007)]{madhavan2007similarities}
P.~Madhavan and D.~A. Wiegmann.
\newblock Similarities and differences between human--human and human--automation trust: an integrative review.
\newblock \emph{Theoretical Issues in Ergonomics Science}, 8\penalty0 (4):\penalty0 277--301, 2007.

\bibitem[Makhoul et~al.(1989)Makhoul, Jelinek, Rabiner, Weinstein, and Zue]{makhoul_white_1989}
J.~Makhoul, F.~Jelinek, L.~Rabiner, C.~Weinstein, and V.~Zue.
\newblock White {Paper} on {Spoken} {Language} {Systems}.
\newblock In \emph{Speech and {Natural} {Language}: {Proceedings} of a {Workshop} {Held} at {Cape} {Cod}, {Massachusetts}, {October} 15-18, 1989}, 1989.
\newblock URL \url{https://aclanthology.org/H89-2077}.

\bibitem[Manyika and Sneader(2018)]{manyika_ai_2018}
J.~Manyika and K.~Sneader.
\newblock {AI}, automation, and the future of work: {Ten} things to solve for.
\newblock Technical report, McKinsey Global Institute, June 2018.
\newblock URL \url{https://www.mckinsey.com/featured-insights/future-of-work/ai-automation-and-the-future-of-work-ten-things-to-solve-for}.

\bibitem[Manyika and Spence(2023)]{manyika2023coming}
J.~Manyika and M.~Spence.
\newblock The {Coming} {AI} {Economic} {Revolution}: {Can} {Artificial} {Intelligence} {Reverse} the {Productivity} {Slowdown?}
\newblock \emph{Foreign Aff.}, 102:\penalty0 70, 2023.

\bibitem[Marda and Narayan(2021)]{marda_importance_2021}
V.~Marda and S.~Narayan.
\newblock On the importance of ethnographic methods in {AI} research.
\newblock \emph{Nature Machine Intelligence}, 3\penalty0 (3):\penalty0 187--189, Mar. 2021.
\newblock ISSN 2522-5839.
\newblock \doi{10.1038/s42256-021-00323-0}.
\newblock URL \url{https://www.nature.com/articles/s42256-021-00323-0}.

\bibitem[Marks(2006)]{marks2006unhappy}
N.~Marks.
\newblock \emph{The unhappy planet index: {A}n index of human well-being and environmental impact}.
\newblock New Economics Foundation, 2006.

\bibitem[Marks and Tegmark(2023)]{marks_geometry_2023}
S.~Marks and M.~Tegmark.
\newblock The geometry of truth: Emergent linear structure in large language model representations of true/false datasets.
\newblock \emph{arXiv preprint arXiv:2310.06824}, 2023.

\bibitem[Marr(2023)]{marr_microsofts_nodate}
B.~Marr.
\newblock Microsoft's {Plan} {To} {Infuse} {AI} {And} {ChatGPT} {Into} {Everything}, 2023.
\newblock URL \url{https://www.forbes.com/sites/bernardmarr/2023/03/06/microsofts-plan-to-infuse-ai-and-chatgpt-into-everything/}.

\bibitem[Marshall et~al.(2014)Marshall, Thieme, Wallace, Vines, Wood, and Balaam]{marshall2014making}
K.~Marshall, A.~Thieme, J.~Wallace, J.~Vines, G.~Wood, and M.~Balaam.
\newblock Making wellbeing: A process of user-centered design.
\newblock In \emph{Proceedings of the 2014 Conference on Designing Interactive Systems}, pages 755--764, 2014.

\bibitem[Martin and Wright(2023)]{martin_bias_2023}
J.~L. Martin and K.~E. Wright.
\newblock Bias in {Automatic} {Speech} {Recognition}: {The} {Case} of {African} {American} {Language}.
\newblock \emph{Applied Linguistics}, 44\penalty0 (4):\penalty0 613--630, Aug. 2023.
\newblock ISSN 0142-6001, 1477-450X.
\newblock \doi{10.1093/applin/amac066}.
\newblock URL \url{https://academic.oup.com/applij/article/44/4/613/6901317}.

\bibitem[Martin and Marks(2019)]{martin_messengers:_2019}
S.~Martin and J.~Marks.
\newblock \emph{Messengers: {Who} {We} {Listen} {To}, {Who} {We} {Don}'t, {And} {Why}}.
\newblock Random House, Sept. 2019.
\newblock ISBN 9781473560727.
\newblock Google-Books-ID: RVJfDwAAQBAJ.

\bibitem[Martin~Jr et~al.(2020)Martin~Jr, Prabhakaran, Kuhlberg, Smart, and Isaac]{martin2020participatory}
D.~Martin~Jr, V.~Prabhakaran, J.~Kuhlberg, A.~Smart, and W.~S. Isaac.
\newblock Participatory problem formulation for fairer machine learning through community based system dynamics.
\newblock \emph{arXiv preprint arXiv:2005.07572}, 2020.

\bibitem[Martínez~Torres et~al.(2019)Martínez~Torres, Iglesias~Comesaña, and García-Nieto]{martinez_torres_review:_2019}
J.~Martínez~Torres, C.~Iglesias~Comesaña, and P.~J. García-Nieto.
\newblock Review: machine learning techniques applied to cybersecurity.
\newblock \emph{International Journal of Machine Learning and Cybernetics}, 10\penalty0 (10):\penalty0 2823--2836, Oct. 2019.
\newblock ISSN 1868-8071, 1868-808X.
\newblock \doi{10.1007/s13042-018-00906-1}.
\newblock URL \url{http://link.springer.com/10.1007/s13042-018-00906-1}.

\bibitem[Marwell and Oliver(1993)]{marwell_critical_1993}
G.~Marwell and P.~Oliver.
\newblock \emph{The {Critical} {Mass} in {Collective} {Action}}.
\newblock Cambridge University Press, 1 edition, Mar. 1993.
\newblock ISBN 9780521308397 9780521039550 9780511663765.
\newblock \doi{10.1017/CBO9780511663765}.
\newblock URL \url{https://www.cambridge.org/core/product/identifier/9780511663765/type/book}.

\bibitem[Mascie-Taylor and Karim(2003)]{mascie2003burden}
C.~N. Mascie-Taylor and E.~Karim.
\newblock The burden of chronic disease.
\newblock \emph{Science}, 302\penalty0 (5652):\penalty0 1921--1922, 2003.

\bibitem[Massey and Denton(1993)]{massey_american_1993}
D.~S. Massey and N.~A. Denton.
\newblock \emph{American {Apartheid}: {Segregation} and the {Making} of the {Underclass}}.
\newblock Harvard University Press, 1993.
\newblock ISBN 9780674018211.
\newblock Google-Books-ID: uGslMsIBNBsC.

\bibitem[Mattis et~al.(2022)Mattis, Masur, Möller, and van Atteveldt]{mattis_nudging_2022}
N.~Mattis, P.~Masur, J.~Möller, and W.~van Atteveldt.
\newblock Nudging towards news diversity: {A} theoretical framework for facilitating diverse news consumption through recommender design.
\newblock \emph{New Media \& Society}, page 146144482211044, June 2022.
\newblock ISSN 1461-4448, 1461-7315.
\newblock \doi{10.1177/14614448221104413}.
\newblock URL \url{http://journals.sagepub.com/doi/10.1177/14614448221104413}.

\bibitem[Mayer et~al.(1995)Mayer, Davis, and Schoorman]{mayer_integrative_1995}
R.~C. Mayer, J.~H. Davis, and F.~D. Schoorman.
\newblock An {Integrative} {Model} of {Organizational} {Trust}.
\newblock \emph{The Academy of Management Review}, 20\penalty0 (3):\penalty0 709, July 1995.
\newblock ISSN 03637425.
\newblock \doi{10.2307/258792}.
\newblock URL \url{http://www.jstor.org/stable/258792?origin=crossref}.

\bibitem[Mayrhofer et~al.(2020)Mayrhofer, Stoep, Brubaker, and Kralevich]{mayrhofer_android_2020}
R.~Mayrhofer, J.~V. Stoep, C.~Brubaker, and N.~Kralevich.
\newblock The {Android} {Platform} {Security} {Model}, Dec. 2020.
\newblock URL \url{http://arxiv.org/abs/1904.05572}.
\newblock arXiv:1904.05572 [cs].

\bibitem[McAllister(1995)]{mcallister_affect-_1995}
D.~J. McAllister.
\newblock Affect- and {Cognition}-{Based} {Trust} as {Foundations} for {Interpersonal} {Cooperation} in {Organizations}.
\newblock \emph{Academy of Management Journal}, 38\penalty0 (1):\penalty0 24--59, Feb. 1995.
\newblock ISSN 0001-4273, 1948-0989.
\newblock \doi{10.2307/256727}.
\newblock URL \url{http://amj.aom.org/cgi/doi/10.2307/256727}.

\bibitem[McArthur(2009)]{mcarthur2009communication}
V.~McArthur.
\newblock Communication technologies and cultural identity a critical discussion of icts for development.
\newblock In \emph{2009 IEEE Toronto International Conference Science and Technology for Humanity (TIC-STH)}, pages 910--914. IEEE, 2009.

\bibitem[McCloskey(1980)]{mccloskey_privacy_1980}
H.~J. McCloskey.
\newblock Privacy and the {Right} to {Privacy}.
\newblock \emph{Philosophy}, 55\penalty0 (211):\penalty0 17--38, Jan. 1980.
\newblock ISSN 0031-8191, 1469-817X.
\newblock \doi{10.1017/S0031819100063725}.
\newblock URL \url{https://www.cambridge.org/core/product/identifier/S0031819100063725/type/journal_article}.

\bibitem[McGillivray(2007)]{mcgillivray2007human}
M.~McGillivray.
\newblock Human well-being: Issues, concepts and measures.
\newblock In \emph{Human well-being: Concept and measurement}, pages 1--22. Springer, 2007.

\bibitem[McGinnies and Ward(1980)]{mcginnies_better_1980}
E.~McGinnies and C.~D. Ward.
\newblock Better {Liked} than {Right}: {Trustworthiness} and {Expertise} as {Factors} in {Credibility}.
\newblock \emph{Personality and Social Psychology Bulletin}, 6\penalty0 (3):\penalty0 467--472, Sept. 1980.
\newblock ISSN 0146-1672, 1552-7433.
\newblock \doi{10.1177/014616728063023}.
\newblock URL \url{http://journals.sagepub.com/doi/10.1177/014616728063023}.

\bibitem[McGregor(2018)]{mcgregor2018reconciling}
J.~A. McGregor.
\newblock Reconciling universal frameworks and local realities in understanding and measuring wellbeing.
\newblock In I.~Bache and K.~Scott, editors, \emph{The Politics of Wellbeing: {T}heory, policy and practice}, pages 197--224. Springer, 2018.

\bibitem[McGregor(2021)]{mcgregor_preventing_2021}
S.~McGregor.
\newblock Preventing {Repeated} {Real} {World} {AI} {Failures} by {Cataloging} {Incidents}: {The} {AI} {Incident} {Database}.
\newblock \emph{Proceedings of the AAAI Conference on Artificial Intelligence}, 35\penalty0 (17):\penalty0 15458--15463, May 2021.
\newblock ISSN 2374-3468, 2159-5399.
\newblock \doi{10.1609/aaai.v35i17.17817}.
\newblock URL \url{https://ojs.aaai.org/index.php/AAAI/article/view/17817}.

\bibitem[McKee et~al.(2021)McKee, Bai, and Fiske]{mckee_humans_2021}
K.~R. McKee, X.~Bai, and S.~Fiske.
\newblock Humans perceive warmth and competence in artificial intelligence.
\newblock preprint, PsyArXiv, Feb. 2021.
\newblock URL \url{https://osf.io/5ursp}.

\bibitem[McKee et~al.(2023{\natexlab{a}})McKee, Hughes, Zhu, Chadwick, Koster, Castaneda, Beattie, Graepel, Botvinick, and Leibo]{mckee_multi-agent_2023}
K.~R. McKee, E.~Hughes, T.~O. Zhu, M.~J. Chadwick, R.~Koster, A.~G. Castaneda, C.~Beattie, T.~Graepel, M.~Botvinick, and J.~Z. Leibo.
\newblock A multi-agent reinforcement learning model of reputation and cooperation in human groups, Feb. 2023{\natexlab{a}}.
\newblock URL \url{http://arxiv.org/abs/2103.04982}.
\newblock arXiv:2103.04982 [cs].

\bibitem[McKee et~al.(2023{\natexlab{b}})McKee, Tacchetti, Bakker, Balaguer, Campbell-Gillingham, Everett, and Botvinick]{mckee_scaffolding_2023}
K.~R. McKee, A.~Tacchetti, M.~A. Bakker, J.~Balaguer, L.~Campbell-Gillingham, R.~Everett, and M.~Botvinick.
\newblock Scaffolding cooperation in human groups with deep reinforcement learning.
\newblock \emph{Nature Human Behaviour}, 7\penalty0 (10):\penalty0 1787--1796, Sept. 2023{\natexlab{b}}.
\newblock ISSN 2397-3374.
\newblock \doi{10.1038/s41562-023-01686-7}.
\newblock URL \url{https://www.nature.com/articles/s41562-023-01686-7}.

\bibitem[McQuaid(2017)]{mcquaid_youth_2017}
R.~McQuaid.
\newblock Youth unemployment produces multiple scarring effects, Feb. 2017.
\newblock URL \url{https://blogs.lse.ac.uk/europpblog/2017/02/18/youth-unemployment-scarring-effects/}.

\bibitem[McTear(2021)]{mctear_conversational_2021}
M.~McTear.
\newblock \emph{Conversational {AI}: dialogue systems, conversational agents, and chatbots}.
\newblock Number \#48 in Synthesis lectures on human language technologies. Morgan \& Claypool Publishers, San Rafael, California, 2021.
\newblock ISBN 9781636390338 9781636390314 9781636390321.
\newblock URL \url{https://www.amazon.co.uk/Conversational-AI-Dialogue-Systems-Chatbots/dp/1636390331/ref=tmm_hrd_swatch_0?_encoding=UTF8&qid=1688737381&sr=1-1}.

\bibitem[Meng and Dai(2021)]{meng_emotional_2021}
J.~Meng and Y.~N. Dai.
\newblock Emotional {Support} from {AI} {Chatbots}: {Should} a {Supportive} {Partner} {Self}-{Disclose} or {Not}?
\newblock \emph{Journal of Computer-Mediated Communication}, 26\penalty0 (4):\penalty0 207--222, Sept. 2021.
\newblock ISSN 1083-6101.
\newblock \doi{10.1093/jcmc/zmab005}.
\newblock URL \url{https://academic.oup.com/jcmc/article/26/4/207/6278042}.

\bibitem[Meng et~al.(2023)Meng, Bau, Andonian, and Belinkov]{meng_locating_2023}
K.~Meng, D.~Bau, A.~Andonian, and Y.~Belinkov.
\newblock Locating and {Editing} {Factual} {Associations} in {GPT}, Jan. 2023.
\newblock URL \url{http://arxiv.org/abs/2202.05262}.
\newblock arXiv:2202.05262 [cs].

\bibitem[Mengesha et~al.(2021)Mengesha, Heldreth, Lahav, Sublewski, and Tuennerman]{mengesha_i_2021}
Z.~Mengesha, C.~Heldreth, M.~Lahav, J.~Sublewski, and E.~Tuennerman.
\newblock “{I} don’t {Think} {These} {Devices} are {Very} {Culturally} {Sensitive}.”—{Impact} of {Automated} {Speech} {Recognition} {Errors} on {African} {Americans}.
\newblock \emph{Frontiers in Artificial Intelligence}, 4:\penalty0 725911, Nov. 2021.
\newblock ISSN 2624-8212.
\newblock \doi{10.3389/frai.2021.725911}.
\newblock URL \url{https://www.frontiersin.org/articles/10.3389/frai.2021.725911/full}.

\bibitem[Merrill et~al.(2022)Merrill, Kim, and Collins]{merrill_ai_2022}
K.~Merrill, J.~Kim, and C.~Collins.
\newblock {AI} companions for lonely individuals and the role of social presence.
\newblock \emph{Communication Research Reports}, 39\penalty0 (2):\penalty0 93--103, Mar. 2022.
\newblock ISSN 0882-4096, 1746-4099.
\newblock \doi{10.1080/08824096.2022.2045929}.
\newblock URL \url{https://www.tandfonline.com/doi/full/10.1080/08824096.2022.2045929}.

\bibitem[Mesch and Talmud(2011)]{mesch_ethnic_2011}
G.~S. Mesch and I.~Talmud.
\newblock Ethnic {Differences} in {Internet} {Access}: {The} role of occupation and exposure.
\newblock \emph{Information, Communication \& Society}, 14\penalty0 (4):\penalty0 445--471, June 2011.
\newblock ISSN 1369-118X, 1468-4462.
\newblock \doi{10.1080/1369118X.2011.562218}.
\newblock URL \url{https://www.tandfonline.com/doi/full/10.1080/1369118X.2011.562218}.

\bibitem[Meta(2023)]{meta_introducing_2023}
Meta.
\newblock Introducing {New} {AI} {Experiences} {Across} {Our} {Family} of {Apps} and {Devices}, Sept. 2023.
\newblock URL \url{https://about.fb.com/news/2023/09/introducing-ai-powered-assistants-characters-and-creative-tools/}.

\bibitem[{Meta Fundamental AI Research Diplomacy Team} et~al.(2022){Meta Fundamental AI Research Diplomacy Team}, Bakhtin, Brown, Dinan, Farina, Flaherty, Fried, Goff, Gray, Hu, Jacob, Komeili, Konath, Kwon, Lerer, Lewis, Miller, Mitts, Renduchintala, Roller, Rowe, Shi, Spisak, Wei, Wu, Zhang, and Zijlstra]{meta_fundamental_ai_research_diplomacy_team_fair_human-level_2022}
{Meta Fundamental AI Research Diplomacy Team}, A.~Bakhtin, N.~Brown, E.~Dinan, G.~Farina, C.~Flaherty, D.~Fried, A.~Goff, J.~Gray, H.~Hu, A.~P. Jacob, M.~Komeili, K.~Konath, M.~Kwon, A.~Lerer, M.~Lewis, A.~H. Miller, S.~Mitts, A.~Renduchintala, S.~Roller, D.~Rowe, W.~Shi, J.~Spisak, A.~Wei, D.~Wu, H.~Zhang, and M.~Zijlstra.
\newblock Human-level play in the game of \textit{{Diplomacy}} by combining language models with strategic reasoning.
\newblock \emph{Science}, 378\penalty0 (6624):\penalty0 1067--1074, Dec. 2022.
\newblock ISSN 0036-8075, 1095-9203.
\newblock \doi{10.1126/science.ade9097}.
\newblock URL \url{https://www.science.org/doi/10.1126/science.ade9097}.

\bibitem[Metzinger(2019)]{metzinger_eu_2019}
T.~Metzinger.
\newblock {EU} guidelines: {Ethics} washing made in {Europe}.
\newblock \emph{Der Tagesspiegel}, Apr. 2019.
\newblock ISSN 1865-2263.
\newblock URL \url{https://www.tagesspiegel.de/politik/ethics-washing-made-in-europe-5937028.html}.

\bibitem[Mialon et~al.(2023)Mialon, Dessì, Lomeli, Nalmpantis, Pasunuru, Raileanu, Rozière, Schick, Dwivedi-Yu, Celikyilmaz, Grave, LeCun, and Scialom]{mialon_augmented_2023}
G.~Mialon, R.~Dessì, M.~Lomeli, C.~Nalmpantis, R.~Pasunuru, R.~Raileanu, B.~Rozière, T.~Schick, J.~Dwivedi-Yu, A.~Celikyilmaz, E.~Grave, Y.~LeCun, and T.~Scialom.
\newblock Augmented {Language} {Models}: a {Survey}, Feb. 2023.
\newblock URL \url{http://arxiv.org/abs/2302.07842}.
\newblock arXiv:2302.07842 [cs].

\bibitem[Michaud et~al.(2007)Michaud, Salter, Duquette, Mercier, Lauria, Larouche, and Larose]{michaud_assistive_2007}
F.~Michaud, T.~Salter, A.~Duquette, H.~Mercier, M.~Lauria, H.~Larouche, and F.~Larose.
\newblock Assistive {Technologies} and {Children}-{Robot} {Interaction}.
\newblock pages 45--49, Jan. 2007.

\bibitem[Mieczkowski et~al.(2021)Mieczkowski, Hancock, Naaman, Jung, and Hohenstein]{mieczkowski_ai-mediated_2021}
H.~Mieczkowski, J.~T. Hancock, M.~Naaman, M.~Jung, and J.~Hohenstein.
\newblock {AI}-{Mediated} {Communication}: {Language} {Use} and {Interpersonal} {Effects} in a {Referential} {Communication} {Task}.
\newblock \emph{Proceedings of the ACM on Human-Computer Interaction}, 5\penalty0 (CSCW1):\penalty0 1--14, Apr. 2021.
\newblock ISSN 2573-0142.
\newblock \doi{10.1145/3449091}.
\newblock URL \url{https://dl.acm.org/doi/10.1145/3449091}.

\bibitem[Milanez(2023)]{milanez_impact_2023}
A.~Milanez.
\newblock The {Impact} of {AI} on the {Workplace}: {Evidence} from {OECD} {Case} {Studies} of {AI} {Implementation}.
\newblock {OECD} {Social}, {Employment} and {Migration} {Working} {Papers} 289, OECD, Mar. 2023.
\newblock URL \url{https://www.oecd-ilibrary.org/social-issues-migration-health/the-impact-of-ai-on-the-workplace-evidence-from-oecd-case-studies-of-ai-implementation_2247ce58-en}.

\bibitem[Milano et~al.(2020)Milano, Taddeo, and Floridi]{milano_recommender_2020}
S.~Milano, M.~Taddeo, and L.~Floridi.
\newblock Recommender systems and their ethical challenges.
\newblock \emph{AI \& SOCIETY}, 35\penalty0 (4):\penalty0 957--967, Dec. 2020.
\newblock ISSN 1435-5655.
\newblock \doi{10.1007/s00146-020-00950-y}.
\newblock URL \url{https://doi.org/10.1007/s00146-020-00950-y}.

\bibitem[Milano et~al.(2021)Milano, Mittelstadt, Wachter, and Russell]{milano_epistemic_2021}
S.~Milano, B.~Mittelstadt, S.~Wachter, and C.~Russell.
\newblock Epistemic fragmentation poses a threat to the governance of online targeting.
\newblock \emph{Nature Machine Intelligence}, 3\penalty0 (6):\penalty0 466--472, June 2021.
\newblock ISSN 2522-5839.
\newblock \doi{10.1038/s42256-021-00358-3}.
\newblock URL \url{https://www.nature.com/articles/s42256-021-00358-3}.

\bibitem[Milgrom and Roberts(1992)]{milgrom_economics_1992}
P.~R. Milgrom and J.~Roberts.
\newblock \emph{Economics, {Organization}, and {Management}}.
\newblock Prentice-Hall, 1992.
\newblock ISBN 9780132246507.
\newblock Google-Books-ID: cCi3AAAAIAAJ.

\bibitem[Mill(1998)]{mill_liberty_1998}
J.~S. Mill.
\newblock \emph{On {Liberty} and {Other} {Essays}}.
\newblock OUP Oxford, Mar. 1998.
\newblock ISBN 9780191611087.
\newblock Google-Books-ID: vcSFmFnICk0C original-date: 1859.

\bibitem[Mills(1991)]{mills_influence:_1991}
C.~J. Mills.
\newblock \emph{Influence: {Coercion}, manipulation, and persuasion}.
\newblock PhD thesis, Princeton University, 1991.
\newblock URL \url{https://www.proquest.com/openview/6193d9f104072d14d2a8f10818604aa1/1?pq-origsite=gscholar&cbl=18750&diss=y}.

\bibitem[Milmo and Anguiano(2021)]{milmo_facebook_2021}
D.~Milmo and D.~Anguiano.
\newblock Facebook, {Instagram} and {WhatsApp} working again after global outage took down platforms.
\newblock \emph{The Guardian}, Oct. 2021.
\newblock ISSN 0261-3077.
\newblock URL \url{https://www.theguardian.com/technology/2021/oct/04/facebook-instagram-and-whatsapp-hit-by-outage}.

\bibitem[Milossi et~al.(2021)Milossi, Alexandropoulou-Egyptiadou, and Psannis]{milossi_ai_2021}
M.~Milossi, E.~Alexandropoulou-Egyptiadou, and K.~E. Psannis.
\newblock {AI} {Ethics}: {Algorithmic} {Determinism} or {Self}-{Determination}? {The} {GPDR} {Approach}.
\newblock \emph{IEEE Access}, 9:\penalty0 58455--58466, 2021.
\newblock ISSN 2169-3536.
\newblock \doi{10.1109/ACCESS.2021.3072782}.
\newblock URL \url{https://ieeexplore.ieee.org/document/9400809/}.

\bibitem[Min and Borch(2022)]{min_systemic_2022}
B.~H. Min and C.~Borch.
\newblock Systemic failures and organizational risk management in algorithmic trading: {Normal} accidents and high reliability in financial markets.
\newblock \emph{Social Studies of Science}, 52\penalty0 (2):\penalty0 277--302, Apr. 2022.
\newblock ISSN 0306-3127, 1460-3659.
\newblock \doi{10.1177/03063127211048515}.
\newblock URL \url{http://journals.sagepub.com/doi/10.1177/03063127211048515}.

\bibitem[Mingus(2011)]{mingus_changing_2011}
M.~Mingus.
\newblock Changing the {Framework}: {Disability} {Justice}, Feb. 2011.
\newblock URL \url{https://leavingevidence.wordpress.com/2011/02/12/changing-the-framework-disability-justice/}.

\bibitem[Mingus(2017)]{mingus_access_2017}
M.~Mingus.
\newblock Access {Intimacy}, {Interdependence} and {Disability} {Justice}, Apr. 2017.
\newblock URL \url{https://leavingevidence.wordpress.com/2017/04/12/access-intimacy-interdependence-and-disability-justice/}.

\bibitem[Miringoff and Miringoff(1999)]{miringoff1999social}
M.~Miringoff and M.-L. Miringoff.
\newblock \emph{The social health of the nation: How America is really doing}.
\newblock Oxford University Press, 1999.

\bibitem[Mirsky et~al.(2022)Mirsky, Carlucho, Rahman, Fosong, Macke, Sridharan, Stone, and Albrecht]{mirsky_survey_2022}
R.~Mirsky, I.~Carlucho, A.~Rahman, E.~Fosong, W.~Macke, M.~Sridharan, P.~Stone, and S.~V. Albrecht.
\newblock A {Survey} of {Ad} {Hoc} {Teamwork} {Research}, Aug. 2022.
\newblock URL \url{http://arxiv.org/abs/2202.10450}.
\newblock arXiv:2202.10450 [cs].

\bibitem[Mirza et~al.(2019)Mirza, Richter, Van~Nes, and Scheffer]{mirza_technology_2019}
M.~U. Mirza, A.~Richter, E.~H. Van~Nes, and M.~Scheffer.
\newblock Technology driven inequality leads to poverty and resource depletion.
\newblock \emph{Ecological Economics}, 160:\penalty0 215--226, June 2019.
\newblock ISSN 09218009.
\newblock \doi{10.1016/j.ecolecon.2019.02.015}.
\newblock URL \url{https://linkinghub.elsevier.com/retrieve/pii/S0921800918306542}.

\bibitem[Mishkin et~al.(2022)Mishkin, Ahmad, Brundage, Krueger, and Sastry]{mishkin_dalle_nodate}
P.~Mishkin, L.~Ahmad, M.~Brundage, G.~Krueger, and G.~Sastry.
\newblock {DALL}·{E} 2 {Preview} – {Risks} and {Limitations}, 2022.
\newblock URL \url{https://github.com/openai/dalle-2-preview/blob/main/system-card.md}.

\bibitem[Mitchell(2023)]{mitchell_how_2023}
M.~Mitchell.
\newblock How do we know how smart {AI} systems are?
\newblock \emph{Science}, 381\penalty0 (6654):\penalty0 adj5957, July 2023.
\newblock ISSN 0036-8075, 1095-9203.
\newblock \doi{10.1126/science.adj5957}.
\newblock URL \url{https://www.science.org/doi/10.1126/science.adj5957}.

\bibitem[Mitchell et~al.(2019)Mitchell, Wu, Zaldivar, Barnes, Vasserman, Hutchinson, Spitzer, Raji, and Gebru]{mitchell_model_2019}
M.~Mitchell, S.~Wu, A.~Zaldivar, P.~Barnes, L.~Vasserman, B.~Hutchinson, E.~Spitzer, I.~D. Raji, and T.~Gebru.
\newblock Model {Cards} for {Model} {Reporting}.
\newblock In \emph{Proceedings of the {Conference} on {Fairness}, {Accountability}, and {Transparency}}, pages 220--229, Atlanta GA USA, Jan. 2019. ACM.
\newblock ISBN 9781450361255.
\newblock \doi{10.1145/3287560.3287596}.
\newblock URL \url{https://dl.acm.org/doi/10.1145/3287560.3287596}.

\bibitem[Mitelut et~al.(2023)Mitelut, Smith, and Vamplew]{mitelut2023intentaligned}
C.~Mitelut, B.~Smith, and P.~Vamplew.
\newblock Intent-aligned {AI} systems deplete human agency: {T}he need for agency foundations research in {AI} safety, 2023.

\bibitem[Mithen and Boyer(1996)]{mithen_anthropomorphism_1996}
S.~Mithen and P.~Boyer.
\newblock Anthropomorphism and the evolution of cognition.
\newblock \emph{Journal of the Royal Anthropological Institute}, 2\penalty0 (4):\penalty0 717--722, Dec. 1996.
\newblock ISSN 13590987.
\newblock URL \url{https://go.gale.com/ps/i.do?p=AONE&sw=w&issn=13590987&v=2.1&it=r&id=GALE%7CA19225795&sid=googleScholar&linkaccess=abs}.

\bibitem[MITRE()]{mitre_mitre_nodate}
MITRE.
\newblock {MITRE} {\textbar} {ATLAS}™.
\newblock URL \url{https://atlas.mitre.org/}.

\bibitem[Mittelstadt(2019)]{mittelstadt_principles_2019}
B.~Mittelstadt.
\newblock Principles alone cannot guarantee ethical {AI}.
\newblock \emph{Nature Machine Intelligence}, 1\penalty0 (11):\penalty0 501--507, Nov. 2019.
\newblock ISSN 2522-5839.
\newblock \doi{10.1038/s42256-019-0114-4}.
\newblock URL \url{https://www.nature.com/articles/s42256-019-0114-4}.

\bibitem[Mittelstadt et~al.(2023)Mittelstadt, Wachter, and Russell]{mittelstadt_unfairness_2023}
B.~Mittelstadt, S.~Wachter, and C.~Russell.
\newblock The {Unfairness} of {Fair} {Machine} {Learning}: {Levelling} down and strict egalitarianism by default, Mar. 2023.
\newblock URL \url{https://arxiv.org/pdf/2302.02404.pdf}.
\newblock arXiv:2302.02404 [cs].

\bibitem[Mohamed et~al.(2022)Mohamed, Lee, Borgholt, Havtorn, Edin, Igel, Kirchhoff, Li, Livescu, Maaløe, Sainath, and Watanabe]{mohamed_self-supervised_2022}
A.~Mohamed, H.-y. Lee, L.~Borgholt, J.~D. Havtorn, J.~Edin, C.~Igel, K.~Kirchhoff, S.-W. Li, K.~Livescu, L.~Maaløe, T.~N. Sainath, and S.~Watanabe.
\newblock Self-{Supervised} {Speech} {Representation} {Learning}: {A} {Review}.
\newblock \emph{IEEE Journal of Selected Topics in Signal Processing}, 16\penalty0 (6):\penalty0 1179--1210, Oct. 2022.
\newblock ISSN 1932-4553, 1941-0484.
\newblock \doi{10.1109/JSTSP.2022.3207050}.
\newblock URL \url{https://arxiv.org/pdf/2205.10643.pdf}.
\newblock arXiv:2205.10643 [cs, eess].

\bibitem[Mohamed et~al.(2020)Mohamed, Png, and Isaac]{mohamed_decolonial_2020}
S.~Mohamed, M.-T. Png, and W.~Isaac.
\newblock Decolonial {AI}: {Decolonial} {Theory} as {Sociotechnical} {Foresight} in {Artificial} {Intelligence}.
\newblock \emph{Philosophy \& Technology}, 33\penalty0 (4):\penalty0 659--684, Dec. 2020.
\newblock ISSN 2210-5433, 2210-5441.
\newblock \doi{10.1007/s13347-020-00405-8}.
\newblock URL \url{http://arxiv.org/abs/2007.04068}.
\newblock arXiv:2007.04068 [cs, stat].

\bibitem[Mok(2023)]{mok_cofounder_nodate}
A.~Mok.
\newblock The cofounder of {Google}'s {AI} division {DeepMind} says everybody will have their own {AI}-powered 'chief of staff' over the next five years, Sept. 2023.
\newblock URL \url{https://www.businessinsider.com/google-deepmind-cofounder-mustafa-suleyman-everyone-will-have-ai-assistant-2023-9}.

\bibitem[M{\"o}kander et~al.(2023)M{\"o}kander, Schuett, Kirk, and Floridi]{mokander2023auditing}
J.~M{\"o}kander, J.~Schuett, H.~R. Kirk, and L.~Floridi.
\newblock Auditing large language models: a three-layered approach.
\newblock \emph{AI and Ethics}, pages 1--31, 2023.

\bibitem[Molnar et~al.(2021)Molnar, Miron, Barbour, Huerta, Shafer, Rice, Glover, Browning, Hagle, and Boninger]{molnar_virtual_2021}
A.~Molnar, G.~Miron, M.~Barbour, L.~Huerta, S.~Shafer, J.~Rice, A.~Glover, N.~Browning, S.~Hagle, and F.~Boninger.
\newblock \emph{Virtual {Schools} 2021}.
\newblock 2021.

\bibitem[Monge~Roffarello and De~Russis(2019)]{monge2019race}
A.~Monge~Roffarello and L.~De~Russis.
\newblock The race towards digital wellbeing: Issues and opportunities.
\newblock In \emph{Proceedings of the 2019 CHI Conference on Human Factors in Computing Systems}, pages 1--14, 2019.

\bibitem[Monroe and Williamson(1987)]{monroe_they_1987}
J.~G. Monroe and R.~A. Williamson.
\newblock \emph{They dance in the sky: {Native} {American} star myths}.
\newblock Houghton Mifflin, Boston, 1987.
\newblock ISBN 9780395399705.

\bibitem[Montagnani and Verstraete(2022)]{montagnani_makes_2022}
M.~L. Montagnani and M.~Verstraete.
\newblock What makes data personal?
\newblock \emph{UC Davis L. Rev.}, 56:\penalty0 1165, 2022.

\bibitem[Moor(1985)]{moor_what_1985}
J.~H. Moor.
\newblock What is {Computer} {Ethics}?*.
\newblock \emph{Metaphilosophy}, 16\penalty0 (4):\penalty0 266--275, Oct. 1985.
\newblock ISSN 0026-1068, 1467-9973.
\newblock \doi{10.1111/j.1467-9973.1985.tb00173.x}.
\newblock URL \url{https://onlinelibrary.wiley.com/doi/10.1111/j.1467-9973.1985.tb00173.x}.

\bibitem[Moor et~al.(2023)Moor, Banerjee, Abad, Krumholz, Leskovec, Topol, and Rajpurkar]{moor_foundation_2023}
M.~Moor, O.~Banerjee, Z.~S.~H. Abad, H.~M. Krumholz, J.~Leskovec, E.~J. Topol, and P.~Rajpurkar.
\newblock Foundation models for generalist medical artificial intelligence.
\newblock \emph{Nature}, 616\penalty0 (7956):\penalty0 259--265, Apr. 2023.
\newblock ISSN 1476-4687.
\newblock \doi{10.1038/s41586-023-05881-4}.
\newblock URL \url{https://www.nature.com/articles/s41586-023-05881-4}.

\bibitem[Moran(2020)]{moran_atlantic_2020}
T.~Moran.
\newblock Atlantic {Plaza} {Towers} tenants won a halt to facial recognition in their building: {Now} they’re calling on a moratorium on all residential use, Jan. 2020.
\newblock URL \url{https://ainowinstitute.org/publication/atlantic-plaza-towers-tenants-won-a-halt-to-facial-recognition-in-their-building-now-theyre}.

\bibitem[Moretti(2013)]{moretti_new_2013}
E.~Moretti.
\newblock \emph{The new geography of jobs}.
\newblock Mariner Books/Houghton Mifflin Harcourt, Boston, Mass, 1st mariner books ed edition, 2013.
\newblock ISBN 9780544028050.
\newblock OCLC: ocn826453786.

\bibitem[{Morgan Stanley}(2023)]{morgan_stanley_surprising_nodate}
{Morgan Stanley}.
\newblock The {Surprising} {Case} for {Stronger} {E}-commerce {Growth}, 2023.
\newblock URL \url{https://www.morganstanley.com/ideas/global-ecommerce-growth-forecast-2022}.

\bibitem[Mori et~al.(2012)Mori, MacDorman, and Kageki]{mori_uncanny_2012}
M.~Mori, K.~MacDorman, and N.~Kageki.
\newblock The {Uncanny} {Valley} [{From} the {Field}].
\newblock \emph{IEEE Robotics \& Automation Magazine}, 19\penalty0 (2):\penalty0 98--100, June 2012.
\newblock ISSN 1070-9932.
\newblock \doi{10.1109/MRA.2012.2192811}.
\newblock URL \url{http://ieeexplore.ieee.org/document/6213238/}.

\bibitem[Morris(2020)]{morris_ai_2020}
M.~R. Morris.
\newblock {AI} and accessibility.
\newblock \emph{Communications of the ACM}, 63\penalty0 (6):\penalty0 35--37, May 2020.
\newblock ISSN 0001-0782, 1557-7317.
\newblock \doi{10.1145/3356727}.
\newblock URL \url{https://dl.acm.org/doi/10.1145/3356727}.

\bibitem[Mota-Rojas et~al.(2021)Mota-Rojas, Mariti, Zdeinert, Riggio, Mora-Medina, del Mar~Reyes, Gazzano, Dom{\'\i}nguez-Oliva, Lezama-Garc{\'\i}a, Jos{\'e}-P{\'e}rez, et~al.]{motarojas_anthropomorphism_2021}
D.~Mota-Rojas, C.~Mariti, A.~Zdeinert, G.~Riggio, P.~Mora-Medina, A.~del Mar~Reyes, A.~Gazzano, A.~Dom{\'\i}nguez-Oliva, K.~Lezama-Garc{\'\i}a, N.~Jos{\'e}-P{\'e}rez, et~al.
\newblock Anthropomorphism and its adverse effects on the distress and welfare of companion animals.
\newblock \emph{Animals}, 11\penalty0 (11):\penalty0 3263, 2021.

\bibitem[Motlagh et~al.(2023)Motlagh, Khajavi, Sharifi, and Ahmadi]{motlagh_impact_2023}
N.~Y. Motlagh, M.~Khajavi, A.~Sharifi, and M.~Ahmadi.
\newblock The {Impact} of {Artificial} {Intelligence} on the {Evolution} of {Digital} {Education}: {A} {Comparative} {Study} of {OpenAI} {Text} {Generation} {Tools} including {ChatGPT}, {Bing} {Chat}, {Bard}, and {Ernie}, Sept. 2023.
\newblock URL \url{http://arxiv.org/abs/2309.02029}.
\newblock arXiv:2309.02029 [cs].

\bibitem[Mourey et~al.(2017)Mourey, Olson, and Yoon]{mourey_products_2017}
J.~A. Mourey, J.~G. Olson, and C.~Yoon.
\newblock Products as {Pals}: {Engaging} with {Anthropomorphic} {Products} {Mitigates} the {Effects} of {Social} {Exclusion}.
\newblock \emph{Journal of Consumer Research}, page ucx038, Jan. 2017.
\newblock ISSN 0093-5301, 1537-5277.
\newblock \doi{10.1093/jcr/ucx038}.
\newblock URL \url{https://academic.oup.com/jcr/article-lookup/doi/10.1093/jcr/ucx038}.

\bibitem[Mousavi et~al.(2022)Mousavi, Davulcu, Ahmadi, Axelrod, Davis, and Atran]{mousavi_effective_2022}
M.~Mousavi, H.~Davulcu, M.~Ahmadi, R.~Axelrod, R.~Davis, and S.~Atran.
\newblock Effective {Messaging} on {Social} {Media}: {What} {Makes} {Online} {Content} {Go} {Viral}?
\newblock In \emph{Proceedings of the {ACM} {Web} {Conference} 2022}, pages 2957--2966, Virtual Event, Lyon France, Apr. 2022. ACM.
\newblock ISBN 9781450390965.
\newblock \doi{10.1145/3485447.3512016}.
\newblock URL \url{https://dl.acm.org/doi/10.1145/3485447.3512016}.

\bibitem[Moussawi(2018)]{moussawi_user_2018}
S.~Moussawi.
\newblock User {Experiences} with {Personal} {Intelligent} {Agents}: {A} {Sensory}, {Physical}, {Functional} and {Cognitive} {Affordances} {View}.
\newblock In \emph{Proceedings of the 2018 {ACM} {SIGMIS} {Conference} on {Computers} and {People} {Research}}, pages 86--92, Buffalo-Niagara Falls NY USA, June 2018. ACM.
\newblock ISBN 9781450357685.
\newblock \doi{10.1145/3209626.3209709}.
\newblock URL \url{https://dl.acm.org/doi/10.1145/3209626.3209709}.

\bibitem[Moussawi and Benbunan-Fich(2021)]{moussawi2021effect}
S.~Moussawi and R.~Benbunan-Fich.
\newblock The effect of voice and humour on users’ perceptions of personal intelligent agents.
\newblock \emph{Behaviour \& Information Technology}, 40\penalty0 (15):\penalty0 1603--1626, 2021.

\bibitem[Moussawi et~al.(2021)Moussawi, Koufaris, and Benbunan-Fich]{moussawi_perceptions_2021}
S.~Moussawi, M.~Koufaris, and R.~Benbunan-Fich.
\newblock How perceptions of intelligence and anthropomorphism affect adoption of personal intelligent agents.
\newblock \emph{Electronic Markets}, 31:\penalty0 343--364, 2021.

\bibitem[Mozur(2018)]{mozur_genocide_2018}
P.~Mozur.
\newblock A {Genocide} {Incited} on {Facebook}, {With} {Posts} {From} {Myanmar}’s {Military}.
\newblock \emph{The New York Times}, Oct. 2018.
\newblock ISSN 0362-4331.
\newblock URL \url{https://www.nytimes.com/2018/10/15/technology/myanmar-facebook-genocide.html}.

\bibitem[Muller and Aguiar(2022)]{muller_what_2022}
C.~Muller and J.~P. d.~V. Aguiar.
\newblock What {Is} the {Digital} {Divide}?, Mar. 2022.
\newblock URL \url{https://www.internetsociety.org/blog/2022/03/what-is-the-digital-divide/}.

\bibitem[Munyer and Zhong(2023)]{munyer_deeptextmark:_2023}
T.~Munyer and X.~Zhong.
\newblock {DeepTextMark}: {Deep} {Learning} based {Text} {Watermarking} for {Detection} of {Large} {Language} {Model} {Generated} {Text}, May 2023.
\newblock URL \url{http://arxiv.org/abs/2305.05773}.
\newblock arXiv:2305.05773 [cs].

\bibitem[Murphy and Criddle(2023)]{murphy_meta_2023}
H.~Murphy and C.~Criddle.
\newblock Meta prepares chatbots with personas to try to retain users.
\newblock \emph{Financial Times}, Aug. 2023.

\bibitem[Myers(2023)]{myers_ais_2023}
A.~Myers.
\newblock {AI}’s {Powers} of {Political} {Persuasion}, Feb. 2023.
\newblock URL \url{https://hai.stanford.edu/news/ais-powers-political-persuasion}.

\bibitem[Myers~West(2023)]{myers2023general}
S.~Myers~West.
\newblock General purpose {AI} poses serious risks, should not be excluded from the {EU}’s {AI} {A}ct | policy brief.
\newblock \url{https://ainowinstitute.org/publication/gpai-is-high-risk-should-not-be-excluded-from-eu-ai-act}, 2023.

\bibitem[Nahavandi et~al.(2022)Nahavandi, Alizadehsani, Khosravi, and Acharya]{nahavandi2022application}
D.~Nahavandi, R.~Alizadehsani, A.~Khosravi, and U.~R. Acharya.
\newblock Application of artificial intelligence in wearable devices: Opportunities and challenges.
\newblock \emph{Computer Methods and Programs in Biomedicine}, 213:\penalty0 106541, 2022.

\bibitem[Nanda et~al.(2023)Nanda, Chan, Lieberum, Smith, and Steinhardt]{nanda_progress_2023}
N.~Nanda, L.~Chan, T.~Lieberum, J.~Smith, and J.~Steinhardt.
\newblock Progress measures for grokking via mechanistic interpretability, Oct. 2023.
\newblock URL \url{http://arxiv.org/abs/2301.05217}.
\newblock arXiv:2301.05217 [cs].

\bibitem[Narayanan(2021)]{narayanan_how_2021}
A.~Narayanan.
\newblock How to recognize {AI} snake oil, Jan. 2021.
\newblock URL \url{https://www.cs.princeton.edu/~arvindn/talks/MIT-STS-AI-snakeoil.pdf}.

\bibitem[Narayanan and Kapoor(2023)]{narayanan_gpt-4_2023}
A.~Narayanan and S.~Kapoor.
\newblock {GPT}-4 and professional benchmarks: the wrong answer to the wrong question, Mar. 2023.
\newblock URL \url{https://www.aisnakeoil.com/p/gpt-4-and-professional-benchmarks}.

\bibitem[Narens and Skyrms(2020)]{narens2020pursuit}
L.~Narens and B.~Skyrms.
\newblock \emph{The Pursuit of Happiness: Philosophical and Psychological Foundations of Utility}.
\newblock Oxford University Press, 2020.

\bibitem[Nass et~al.(1993)Nass, Steuer, Tauber, and Reeder]{nass_anthropomorphism_1993}
C.~Nass, J.~Steuer, E.~Tauber, and H.~Reeder.
\newblock Anthropomorphism, agency, and ethopoeia: computers as social actors.
\newblock In \emph{{INTERACT} '93 and {CHI} '93 conference companion on {Human} factors in computing systems - {CHI} '93}, pages 111--112, Amsterdam, The Netherlands, 1993. ACM Press.
\newblock ISBN 9780897915748.
\newblock \doi{10.1145/259964.260137}.
\newblock URL \url{http://portal.acm.org/citation.cfm?doid=259964.260137}.

\bibitem[Nass et~al.(1994)Nass, Steuer, and Tauber]{nass_computers_1994}
C.~Nass, J.~Steuer, and E.~R. Tauber.
\newblock Computers are social actors.
\newblock In \emph{Conference companion on {Human} factors in computing systems - {CHI} '94}, page 204, Boston, Massachusetts, United States, 1994. ACM Press.
\newblock ISBN 9780897916516.
\newblock \doi{10.1145/259963.260288}.
\newblock URL \url{http://portal.acm.org/citation.cfm?doid=259963.260288}.

\bibitem[{National Education Union}(2023)]{national_education_union_state_2023}
{National Education Union}.
\newblock State of education: workload and wellbeing.
\newblock Technical report, National Education Union, Apr. 2023.
\newblock URL \url{https://neu.org.uk/latest/press-releases/state-education-workload-and-wellbeing}.

\bibitem[Newman(2023)]{newman_taxonomy_2023}
J.~Newman.
\newblock A {Taxonomy} of {AI} {Trustworthiness}, 2023.
\newblock URL \url{https://cltc.berkeley.edu/wp-content/uploads/2023/01/Tax onomy of AI Trustworthiness.pdf}.

\bibitem[Ngamaba et~al.(2018)Ngamaba, Panagioti, and Armitage]{ngamaba_income_2018}
K.~H. Ngamaba, M.~Panagioti, and C.~J. Armitage.
\newblock Income inequality and subjective well-being: a systematic review and meta-analysis.
\newblock \emph{Quality of Life Research}, 27\penalty0 (3):\penalty0 577--596, Mar. 2018.
\newblock ISSN 0962-9343, 1573-2649.
\newblock \doi{10.1007/s11136-017-1719-x}.
\newblock URL \url{http://link.springer.com/10.1007/s11136-017-1719-x}.

\bibitem[Nguyen(2022)]{nguyen2022transparency}
C.~T. Nguyen.
\newblock Transparency is surveillance.
\newblock \emph{Philosophy and Phenomenological Research}, 105\penalty0 (2):\penalty0 331--361, 2022.
\newblock \doi{https://doi.org/10.1111/phpr.12823}.
\newblock URL \url{https://onlinelibrary.wiley.com/doi/abs/10.1111/phpr.12823}.

\bibitem[Nguyen et~al.(2012)Nguyen, Do, Gerevini, Serina, Srivastava, and Kambhampati]{nguyen2012generating}
T.~A. Nguyen, M.~Do, A.~E. Gerevini, I.~Serina, B.~Srivastava, and S.~Kambhampati.
\newblock Generating diverse plans to handle unknown and partially known user preferences.
\newblock \emph{Artificial Intelligence}, 190:\penalty0 1--31, 2012.

\bibitem[Nickel et~al.(2010)Nickel, Franssen, and Kroes]{Nickel2010-NICCWM}
P.~J. Nickel, M.~Franssen, and P.~Kroes.
\newblock Can we make sense of the notion of trustworthy technology?
\newblock \emph{Knowledge, Technology \& Policy}, 23\penalty0 (3-4):\penalty0 429--444, 2010.
\newblock \doi{10.1007/s12130-010-9124-6}.

\bibitem[Nicoletti and Bass(2023)]{nicoletti_humans_2023}
L.~Nicoletti and D.~Bass.
\newblock Humans {Are} {Biased}. {Generative} {AI} {Is} {Even} {Worse}.
\newblock \emph{Bloomberg}, 2023.
\newblock URL \url{https://www.bloomberg.com/graphics/2023-generative-ai-bias/}.

\bibitem[Niemiec(2014)]{niemiec2014eudaimonic}
C.~P. Niemiec.
\newblock Eudaimonic well-being.
\newblock \emph{Encyclopedia of quality of life and well-being research}, pages 2004--2005, 2014.

\bibitem[Nightingale and Farid(2022)]{nightingale_ai-synthesized_2022}
S.~J. Nightingale and H.~Farid.
\newblock {AI}-synthesized faces are indistinguishable from real faces and more trustworthy.
\newblock \emph{Proceedings of the National Academy of Sciences}, 119\penalty0 (8):\penalty0 e2120481119, Feb. 2022.
\newblock ISSN 0027-8424, 1091-6490.
\newblock \doi{10.1073/pnas.2120481119}.
\newblock URL \url{https://pnas.org/doi/full/10.1073/pnas.2120481119}.

\bibitem[Nissenbaum(2004)]{nissenbaum_privacy_2004}
H.~Nissenbaum.
\newblock Privacy as {Contextual} {Integrity}.
\newblock \emph{Washington Law Review}, 79\penalty0 (1):\penalty0 119, Feb. 2004.
\newblock URL \url{https://digitalcommons.law.uw.edu/wlr/vol79/iss1/10}.

\bibitem[Noble(2020)]{noble_algorithms_2020}
S.~U. Noble.
\newblock \emph{Algorithms of {Oppression}: {How} {Search} {Engines} {Reinforce} {Racism}}.
\newblock New York University Press, Dec. 2020.
\newblock ISBN 9781479833641.
\newblock \doi{10.18574/nyu/9781479833641.001.0001}.
\newblock URL \url{https://www.degruyter.com/document/doi/10.18574/nyu/9781479833641.001.0001/html}.

\bibitem[Noddings(2013)]{noddings_caring:_2013}
N.~Noddings.
\newblock \emph{Caring: {A} {Relational} {Approach} to {Ethics} and {Moral} {Education}}.
\newblock University of California Press, 2 edition, 2013.
\newblock ISBN 9780520275706.
\newblock URL \url{https://www.jstor.org/stable/10.1525/j.ctt7zw1nb}.

\bibitem[Noggle(1996)]{noggle_manipulative_1996}
R.~Noggle.
\newblock Manipulative {Actions}: {A} {Conceptual} and {Moral} {Analysis}.
\newblock \emph{American Philosophical Quarterly}, 33\penalty0 (1):\penalty0 43--55, 1996.
\newblock ISSN 0003-0481.
\newblock URL \url{https://www.jstor.org/stable/20009846}.

\bibitem[Noggle(2018)]{noggle_manipulation_2018}
R.~Noggle.
\newblock Manipulation, salience, and nudges.
\newblock \emph{Bioethics}, 32\penalty0 (3):\penalty0 164--170, Mar. 2018.
\newblock ISSN 1467-8519.
\newblock \doi{10.1111/bioe.12421}.

\bibitem[Noggle(2022)]{noggle_ethics_2022}
R.~Noggle.
\newblock The {Ethics} of {Manipulation}.
\newblock In E.~N. Zalta, editor, \emph{The {Stanford} {Encyclopedia} of {Philosophy}}. Metaphysics Research Lab, Stanford University, summer 2022 edition, 2022.
\newblock URL \url{https://plato.stanford.edu/archives/sum2022/entries/ethics-manipulation/}.

\bibitem[Nolan(2023)]{nolan_latest_nodate}
B.~Nolan.
\newblock The latest version of {ChatGPT} told a {TaskRabbit} worker it was visually impaired to get help solving a {CAPTCHA}, {OpenAI} test shows, 2023.
\newblock URL \url{https://www.businessinsider.com/gpt4-openai-chatgpt-taskrabbit-tricked-solve-captcha-test-2023-3}.

\bibitem[Noone(2021)]{noone_foundation_2021}
G.~Noone.
\newblock ‘{Foundation} models’ may be the future of {AI}. {They}’re also deeply flawed, Nov. 2021.
\newblock URL \url{https://techmonitor.ai/technology/ai-and-automation/foundation-models-may-be-future-of-ai-theyre-also-deeply-flawed}.

\bibitem[Nouh et~al.(2019)Nouh, Lee, Lee, and Lee]{nouh2019smart}
R.~M. Nouh, H.-H. Lee, W.-J. Lee, and J.-D. Lee.
\newblock A smart recommender based on hybrid learning methods for personal well-being services.
\newblock \emph{Sensors}, 19\penalty0 (2):\penalty0 431, 2019.

\bibitem[Novikova et~al.(2018)Novikova, Dušek, and Rieser]{novikova_rankme:_2018}
J.~Novikova, O.~Dušek, and V.~Rieser.
\newblock {RankME}: {Reliable} {Human} {Ratings} for {Natural} {Language} {Generation}.
\newblock In M.~Walker, H.~Ji, and A.~Stent, editors, \emph{Proceedings of the 2018 {Conference} of the {North} {American} {Chapter} of the {Association} for {Computational} {Linguistics}: {Human} {Language} {Technologies}, {Volume} 2 ({Short} {Papers})}, pages 72--78, New Orleans, Louisiana, June 2018. Association for Computational Linguistics.
\newblock \doi{10.18653/v1/N18-2012}.
\newblock URL \url{https://aclanthology.org/N18-2012}.

\bibitem[NowPow()]{nowpow_nowpow_nodate}
NowPow.
\newblock {NowPow} {Sign} {In}.
\newblock URL \url{https://uniteus.com/nowpow-login/}.

\bibitem[Noy and Zhang(2023)]{noy_experimental_2023}
S.~Noy and W.~Zhang.
\newblock Experimental evidence on the productivity effects of generative artificial intelligence.
\newblock \emph{Science}, 381\penalty0 (6654):\penalty0 187--192, July 2023.
\newblock ISSN 0036-8075, 1095-9203.
\newblock \doi{10.1126/science.adh2586}.
\newblock URL \url{https://www.science.org/doi/10.1126/science.adh2586}.

\bibitem[Nozick(1974)]{nozick1974anarchy}
R.~Nozick.
\newblock \emph{Anarchy, state, and utopia}.
\newblock John Wiley \& Sons, 1974.

\bibitem[Obar and Oeldorf-Hirsch(2020)]{obar_biggest_2020}
J.~A. Obar and A.~Oeldorf-Hirsch.
\newblock The biggest lie on the {Internet}: ignoring the privacy policies and terms of service policies of social networking services.
\newblock \emph{Information, Communication \& Society}, 23\penalty0 (1):\penalty0 128--147, Jan. 2020.
\newblock ISSN 1369-118X, 1468-4462.
\newblock \doi{10.1080/1369118X.2018.1486870}.
\newblock URL \url{https://www.tandfonline.com/doi/full/10.1080/1369118X.2018.1486870}.

\bibitem[Obermeyer et~al.(2019)Obermeyer, Powers, Vogeli, and Mullainathan]{obermeyer_dissecting_2019}
Z.~Obermeyer, B.~Powers, C.~Vogeli, and S.~Mullainathan.
\newblock Dissecting racial bias in an algorithm used to manage the health of populations.
\newblock \emph{Science}, 366\penalty0 (6464):\penalty0 447--453, Oct. 2019.
\newblock ISSN 0036-8075, 1095-9203.
\newblock \doi{10.1126/science.aax2342}.
\newblock URL \url{https://www.science.org/doi/10.1126/science.aax2342}.

\bibitem[O'Brien(2023)]{obrien_is_2023}
M.~O'Brien.
\newblock Is {Bing} too belligerent? {Microsoft} looks to tame {AI} chatbot.
\newblock \emph{AP News}, Feb. 2023.
\newblock URL \url{https://apnews.com/article/technology-science-microsoft-corp-business-software-fb49e5d625bf37be0527e5173116bef3}.

\bibitem[OECD(2011)]{oecd_compendium}
OECD.
\newblock Compendium of {OECD} well-being indicators, 2011.

\bibitem[OECD(2016)]{oecd_how_2016}
OECD.
\newblock How {Good} is {Your} {Job}? {Measuring} and {Assessing} {Job} {Quality}.
\newblock Technical report, OECD, Feb. 2016.
\newblock URL \url{https://www.oecd.org/sdd/labour-stats/Job-quality-OECD.pdf}.

\bibitem[OECD(2018{\natexlab{a}})]{oecd_bridging_2018}
OECD.
\newblock Bridging the {Digital} {Gender} {Divide}: {Include}, {Upskill}, {Innovate}.
\newblock Technical report, OECD, 2018{\natexlab{a}}.
\newblock URL \url{https://www.oecd.org/digital/bridging-the-digital-gender-divide.pdf}.

\bibitem[OECD(2018{\natexlab{b}})]{oecd_equity_2018}
OECD.
\newblock \emph{Equity in {Education}: {Breaking} {Down} {Barriers} to {Social} {Mobility}}.
\newblock {PISA}. OECD, Oct. 2018{\natexlab{b}}.
\newblock ISBN 9789264056732 9789264073234.
\newblock \doi{10.1787/9789264073234-en}.
\newblock URL \url{https://www.oecd-ilibrary.org/education/equity-in-education_9789264073234-en}.

\bibitem[OECD(2019)]{oecd_skills_2019}
OECD.
\newblock \emph{Skills {Matter}: {Additional} {Results} from the {Survey} of {Adult} {Skills}}.
\newblock {OECD} {Skills} {Studies}. OECD, Nov. 2019.
\newblock ISBN 9789264604667 9789264811072 9789264799004 9789264332829.
\newblock \doi{10.1787/1f029d8f-en}.
\newblock URL \url{https://www.oecd-ilibrary.org/education/skills-matter_1f029d8f-en}.

\bibitem[OECD(2021)]{oecd_tools_2021}
OECD.
\newblock Tools for trustworthy {AI}: {A} framework to compare implementation tools for trustworthy {AI} systems.
\newblock {OECD} {Digital} {Economy} {Papers} 312, June 2021.
\newblock URL \url{https://www.oecd-ilibrary.org/science-and-technology/tools-for-trustworthy-ai_008232ec-en}.

\bibitem[OECD(2023)]{oecd_is_2023}
OECD.
\newblock \emph{Is {Education} {Losing} the {Race} with {Technology}?: {AI}'s {Progress} in {Maths} and {Reading}}.
\newblock Educational {Research} and {Innovation}. OECD, Mar. 2023.
\newblock ISBN 9789264451377 9789264765153 9789264639676 9789264920378.
\newblock \doi{10.1787/73105f99-en}.
\newblock URL \url{https://www.oecd-ilibrary.org/education/is-education-losing-the-race-with-technology_73105f99-en}.

\bibitem[Oesterheld and Conitzer(2022)]{oesterheld_safe_2022}
C.~Oesterheld and V.~Conitzer.
\newblock Safe {Pareto} improvements for delegated game playing.
\newblock \emph{Autonomous Agents and Multi-Agent Systems}, 36\penalty0 (2):\penalty0 46, Aug. 2022.
\newblock ISSN 1573-7454.
\newblock \doi{10.1007/s10458-022-09574-6}.
\newblock URL \url{https://doi.org/10.1007/s10458-022-09574-6}.

\bibitem[Ognyanova et~al.(2020)Ognyanova, Lazer, Robertson, and Wilson]{ognyanova_misinformation_2020}
K.~Ognyanova, D.~Lazer, R.~E. Robertson, and C.~Wilson.
\newblock Misinformation in action: {Fake} news exposure is linked to lower trust in media, higher trust in government when your side is in power.
\newblock \emph{Harvard Kennedy School Misinformation Review}, June 2020.
\newblock \doi{10.37016/mr-2020-024}.
\newblock URL \url{https://misinforeview.hks.harvard.edu/?p=1689}.

\bibitem[Oguego et~al.(2018)Oguego, Augusto, Mu{\~n}oz, and Springett]{oguego2018using}
C.~L. Oguego, J.~C. Augusto, A.~Mu{\~n}oz, and M.~Springett.
\newblock Using argumentation to manage users’ preferences.
\newblock \emph{Future Generation Computer Systems}, 81:\penalty0 235--243, 2018.

\bibitem[Oishi et~al.(2011)Oishi, Kesebir, and Diener]{oishi_income_2011}
S.~Oishi, S.~Kesebir, and E.~Diener.
\newblock Income {Inequality} and {Happiness}.
\newblock \emph{Psychological Science}, 22\penalty0 (9):\penalty0 1095--1100, Sept. 2011.
\newblock ISSN 0956-7976, 1467-9280.
\newblock \doi{10.1177/0956797611417262}.
\newblock URL \url{http://journals.sagepub.com/doi/10.1177/0956797611417262}.

\bibitem[Ojeda-Castelo et~al.(2022)Ojeda-Castelo, Capobianco-Uriarte, Piedra-Fernandez, and Ayala]{ojeda-castelo_survey_2022}
J.~J. Ojeda-Castelo, M.~D. L.~M. Capobianco-Uriarte, J.~A. Piedra-Fernandez, and R.~Ayala.
\newblock A {Survey} on {Intelligent} {Gesture} {Recognition} {Techniques}.
\newblock \emph{IEEE Access}, 10:\penalty0 87135--87156, 2022.
\newblock ISSN 2169-3536.
\newblock \doi{10.1109/ACCESS.2022.3199358}.
\newblock URL \url{https://ieeexplore.ieee.org/document/9858153/}.

\bibitem[O'Kane et~al.(2011)O'Kane, Sezer, and McLaughlin]{okane_obfuscation:_2011}
P.~O'Kane, S.~Sezer, and K.~McLaughlin.
\newblock Obfuscation: {The} {Hidden} {Malware}.
\newblock \emph{IEEE Security \& Privacy}, 9\penalty0 (5):\penalty0 41--47, Sept. 2011.
\newblock ISSN 1558-4046.
\newblock \doi{10.1109/MSP.2011.98}.
\newblock URL \url{https://ieeexplore.ieee.org/abstract/document/5975134}.

\bibitem[Okasha(2018)]{okasha2018agents}
S.~Okasha.
\newblock \emph{Agents and goals in evolution}.
\newblock Oxford University Press, 2018.

\bibitem[Olah et~al.(2020)Olah, Cammarata, Schubert, Goh, Petrov, and Carter]{olah_zoom_2020}
C.~Olah, N.~Cammarata, L.~Schubert, G.~Goh, M.~Petrov, and S.~Carter.
\newblock Zoom {In}: {An} {Introduction} to {Circuits}.
\newblock \emph{Distill}, 5\penalty0 (3):\penalty0 10.23915/distill.00024.001, Mar. 2020.
\newblock ISSN 2476-0757.
\newblock \doi{10.23915/distill.00024.001}.
\newblock URL \url{https://distill.pub/2020/circuits/zoom-in}.

\bibitem[Olson~Jr(1965)]{jr_logic_1971}
M.~Olson~Jr.
\newblock \emph{The {Logic} of {Collective} {Action}: {Public} {Goods} and the {Theory} of {Groups}}.
\newblock Harvard {Economic} {Studies}. Harvard University Press, Cambridge, MA, Jan. 1965.
\newblock ISBN 9780674537514.

\bibitem[Olsson et~al.(2022)Olsson, Elhage, Nanda, Joseph, DasSarma, Henighan, Mann, Askell, Bai, Chen, Conerly, Drain, Ganguli, Hatfield-Dodds, Hernandez, Johnston, Jones, Kernion, Lovitt, Ndousse, Amodei, Brown, Clark, Kaplan, McCandlish, and Olah]{olsson_-context_2022}
C.~Olsson, N.~Elhage, N.~Nanda, N.~Joseph, N.~DasSarma, T.~Henighan, B.~Mann, A.~Askell, Y.~Bai, A.~Chen, T.~Conerly, D.~Drain, D.~Ganguli, Z.~Hatfield-Dodds, D.~Hernandez, S.~Johnston, A.~Jones, J.~Kernion, L.~Lovitt, K.~Ndousse, D.~Amodei, T.~Brown, J.~Clark, J.~Kaplan, S.~McCandlish, and C.~Olah.
\newblock In-context {Learning} and {Induction} {Heads}, Mar. 2022.
\newblock URL \url{https://transformer-circuits.pub/2022/in-context-learning-and-induction-heads/index.html}.
\newblock publisher: Anthropic.

\bibitem[Omohundro(2008)]{omohundro_basic_2008}
S.~M. Omohundro.
\newblock The {Basic} {AI} {Drives}.
\newblock In \emph{Proceedings of the {First} {Conference} on {Artificial} {General} {Intelligence}}, 2008.
\newblock URL \url{https://citeseerx.ist.psu.edu/viewdoc/summary?doi=10.1.1.393.8356}.

\bibitem[O'Neill(1989)]{oneill_constructions_1989}
O.~O'Neill.
\newblock \emph{Constructions of {Reason}: {Explorations} of {Kant}'s {Practical} {Philosophy}}.
\newblock Cambridge University Press, New York, 1989.

\bibitem[O'Neill(2002)]{oneill_autonomy_2002}
O.~O'Neill.
\newblock \emph{Autonomy and {Trust} in {Bioethics}}.
\newblock Cambridge University Press, 1 edition, Apr. 2002.
\newblock ISBN 9780521815406 9780521894531 9780511606250.
\newblock \doi{10.1017/CBO9780511606250}.
\newblock URL \url{https://www.cambridge.org/core/product/identifier/9780511606250/type/book}.

\bibitem[Ono and Zavodny(2008)]{ono_immigrants_2008}
H.~Ono and M.~Zavodny.
\newblock Immigrants, {English} {Ability} and the {Digital} {Divide}.
\newblock \emph{Social Forces}, 86\penalty0 (4):\penalty0 1455--1479, June 2008.
\newblock ISSN 0037-7732, 1534-7605.
\newblock \doi{10.1353/sof.0.0052}.
\newblock URL \url{https://academic.oup.com/sf/article-lookup/doi/10.1353/sof.0.0052}.

\bibitem[OpenAI(2023{\natexlab{a}})]{2023GPT4VisionSC}
OpenAI.
\newblock Gpt-4v(ision) system card.
\newblock 2023{\natexlab{a}}.
\newblock URL \url{https://api.semanticscholar.org/CorpusID:263218031}.

\bibitem[OpenAI(2023{\natexlab{b}})]{openai_chatgpt_2023}
OpenAI.
\newblock {ChatGPT} plugins, Mar. 2023{\natexlab{b}}.
\newblock URL \url{https://openai.com/blog/chatgpt-plugins}.

\bibitem[OpenAI(2023{\natexlab{c}})]{openai_forecasting_2023}
OpenAI.
\newblock Forecasting potential misuses of language models for disinformation campaigns and how to reduce risk, Jan. 2023{\natexlab{c}}.
\newblock URL \url{https://openai.com/research/forecasting-misuse}.

\bibitem[OpenAI(2023{\natexlab{d}})]{openai_gpt-4_2023}
OpenAI.
\newblock {GPT}-4 {System} {Card}, Mar. 2023{\natexlab{d}}.
\newblock URL \url{https://cdn.openai.com/papers/gpt-4-system-card.pdf}.
\newblock publisher: OpenAI.

\bibitem[{OpenAI}(2023)]{openai_gpt-4_nodate}
{OpenAI}.
\newblock {GPT}-4 {Technical} {Report}, Mar. 2023.
\newblock URL \url{http://arxiv.org/abs/2303.08774}.
\newblock arXiv:2303.08774 [cs].

\bibitem[OpenAI(2023)]{openai_technical_2023}
OpenAI.
\newblock Technical {Requirements} for {Using} {ChatGPT}, Nov. 2023.
\newblock URL \url{https://chatgptdetector.co/chat-gpt-requirements/}.

\bibitem[Orben and Przybylski(2019)]{orben2019association}
A.~Orben and A.~K. Przybylski.
\newblock The association between adolescent well-being and digital technology use.
\newblock \emph{Nature human behaviour}, 3\penalty0 (2):\penalty0 173--182, 2019.

\bibitem[Orpen(1984)]{orpen_attitude_1984}
C.~Orpen.
\newblock Attitude {Similarity}, {Attraction}, and {Decision}-{Making} in the {Employment} {Interview}.
\newblock \emph{The Journal of Psychology}, 117\penalty0 (1):\penalty0 111--120, May 1984.
\newblock ISSN 0022-3980, 1940-1019.
\newblock \doi{10.1080/00223980.1984.9923666}.
\newblock URL \url{http://www.tandfonline.com/doi/abs/10.1080/00223980.1984.9923666}.

\bibitem[Ostrom(2000)]{ostrom_collective_2000}
E.~Ostrom.
\newblock Collective {Action} and the {Evolution} of {Social} {Norms}.
\newblock \emph{Journal of Economic Perspectives}, 14\penalty0 (3):\penalty0 137--158, Sept. 2000.
\newblock ISSN 0895-3309.
\newblock \doi{10.1257/jep.14.3.137}.
\newblock URL \url{https://www.aeaweb.org/articles?id=10.1257/jep.14.3.137}.

\bibitem[Ostrom(2010)]{ostrom_multi-scale_2010}
E.~Ostrom.
\newblock A {Multi}-{Scale} {Approach} to {Coping} with {Climate} {Change} and {Other} {Collective} {Action} {Problems}.
\newblock \emph{Solutions}, 1:\penalty0 27--36, 2010.
\newblock URL \url{https://dlc.dlib.indiana.edu/dlc/bitstream/handle/10535/5774/A%20Multi-Scale%20Approach%20to%20C...pdf?sequence=1}.

\bibitem[Otsuka(2015)]{otsuka2015prioritarianism}
M.~Otsuka.
\newblock Prioritarianism and the measure of utility.
\newblock \emph{Journal of Political Philosophy}, 23\penalty0 (1):\penalty0 1--22, 2015.

\bibitem[Ouyang et~al.(2022)Ouyang, Wu, Jiang, Almeida, Wainwright, Mishkin, Zhang, Agarwal, Slama, Ray, Schulman, Hilton, Kelton, Miller, Simens, Askell, Welinder, Christiano, Leike, and Lowe]{ouyang_training_2022}
L.~Ouyang, J.~Wu, X.~Jiang, D.~Almeida, C.~L. Wainwright, P.~Mishkin, C.~Zhang, S.~Agarwal, K.~Slama, A.~Ray, J.~Schulman, J.~Hilton, F.~Kelton, L.~Miller, M.~Simens, A.~Askell, P.~Welinder, P.~Christiano, J.~Leike, and R.~Lowe.
\newblock Training language models to follow instructions with human feedback, Mar. 2022.
\newblock URL \url{http://arxiv.org/abs/2203.02155}.
\newblock arXiv:2203.02155 [cs].

\bibitem[Ovadya and Whittlestone(2019)]{ovadya_reducing_2019}
A.~Ovadya and J.~Whittlestone.
\newblock Reducing malicious use of synthetic media research: {Considerations} and potential release practices for machine learning, July 2019.
\newblock URL \url{http://arxiv.org/abs/1907.11274}.
\newblock arXiv:1907.11274 [cs].

\bibitem[Ovalle et~al.(2023)Ovalle, Subramonian, Gautam, Gee, and Chang]{ovalle_factoring_2023}
A.~Ovalle, A.~Subramonian, V.~Gautam, G.~Gee, and K.-W. Chang.
\newblock Factoring the {Matrix} of {Domination}: {A} {Critical} {Review} and {Reimagination} of {Intersectionality} in {AI} {Fairness}, July 2023.
\newblock URL \url{http://arxiv.org/abs/2303.17555}.
\newblock arXiv:2303.17555 [cs].

\bibitem[Owen et~al.(2021)Owen, Pansera, Macnaghten, and Randles]{owen_organisational_2021}
R.~Owen, M.~Pansera, P.~Macnaghten, and S.~Randles.
\newblock Organisational institutionalisation of responsible innovation.
\newblock \emph{Research Policy}, 50\penalty0 (1):\penalty0 104132, Jan. 2021.
\newblock ISSN 00487333.
\newblock \doi{10.1016/j.respol.2020.104132}.
\newblock URL \url{https://linkinghub.elsevier.com/retrieve/pii/S0048733320302079}.

\bibitem[{Oxford Economics}(2019)]{oxford_economics_how_2019}
{Oxford Economics}.
\newblock How {Robots} {Change} the {World}, June 2019.
\newblock URL \url{https://www.oxfordeconomics.com/resource/how-robots-change-the-world/}.

\bibitem[O’Neill(2018)]{oneill_linking_2018}
O.~O’Neill.
\newblock Linking {Trust} to {Trustworthiness}.
\newblock \emph{International Journal of Philosophical Studies}, 26\penalty0 (2):\penalty0 293--300, Mar. 2018.
\newblock ISSN 0967-2559, 1466-4542.
\newblock \doi{10.1080/09672559.2018.1454637}.
\newblock URL \url{https://www.tandfonline.com/doi/full/10.1080/09672559.2018.1454637}.

\bibitem[Pacchiardi et~al.(2023)Pacchiardi, Chan, Mindermann, Moscovitz, Pan, Gal, Evans, and Brauner]{pacchiardi_catch_2023}
L.~Pacchiardi, A.~J. Chan, S.~Mindermann, I.~Moscovitz, A.~Y. Pan, Y.~Gal, O.~Evans, and J.~Brauner.
\newblock How to catch an ai liar: Lie detection in black-box llms by asking unrelated questions.
\newblock \emph{arXiv preprint arXiv:2309.15840}, 2023.

\bibitem[Palmer and Spirling(2023)]{palmer_large_2023}
A.~K. Palmer and A.~Spirling.
\newblock Large {Language} {Models} {Can} {Argue} in {Convincing} and {Novel} {Ways} {About} {Politics}: {Evidence} from {Experiments} and {Human} {Judgement}, May 2023.
\newblock URL \url{https://arthurspirling.org/documents/llm.pdf}.

\bibitem[Pamment(2023)]{pamment_eus_2023}
J.~Pamment.
\newblock The {EU}’s {Role} in {Fighting} {Disinformation}: {Crafting} {A} {Disinformation} {Framework}, 2023.
\newblock URL \url{https://carnegieendowment.org/2020/09/24/eu-s-role-in-fighting-disinformation-crafting-disinformation-framework-pub-82720}.

\bibitem[Pan et~al.(2022)Pan, Bhatia, and Steinhardt]{pan_effects_2022}
A.~Pan, K.~Bhatia, and J.~Steinhardt.
\newblock The {Effects} of {Reward} {Misspecification}: {Mapping} and {Mitigating} {Misaligned} {Models}, Feb. 2022.
\newblock URL \url{http://arxiv.org/abs/2201.03544}.
\newblock arXiv:2201.03544 [cs, stat].

\bibitem[Pannozzo(2009)]{pannozzo20092008}
L.~Pannozzo.
\newblock \emph{The 2008 Nova Scotia GPI accounts: indicators of genuine progress}.
\newblock GPI Atlantic, 2009.

\bibitem[Papaioannou et~al.(2017)Papaioannou, Curry, Part, Shalyminov, Xu, Yu, Dušek, Rieser, and Lemon]{papaioannou_ensemble_2017}
I.~Papaioannou, A.~C. Curry, J.~L. Part, I.~Shalyminov, X.~Xu, Y.~Yu, O.~Dušek, V.~Rieser, and O.~Lemon.
\newblock An {Ensemble} {Model} with {Ranking} for {Social} {Dialogue}, Dec. 2017.
\newblock URL \url{http://arxiv.org/abs/1712.07558}.
\newblock arXiv:1712.07558 [cs].

\bibitem[Paranjape et~al.(2020)Paranjape, See, Kenealy, Li, Hardy, Qi, Sadagopan, Phu, Soylu, and Manning]{paranjape_neural_2020}
A.~Paranjape, A.~See, K.~Kenealy, H.~Li, A.~Hardy, P.~Qi, K.~R. Sadagopan, N.~M. Phu, D.~Soylu, and C.~D. Manning.
\newblock Neural {Generation} {Meets} {Real} {People}: {Towards} {Emotionally} {Engaging} {Mixed}-{Initiative} {Conversations}, Sept. 2020.
\newblock URL \url{http://arxiv.org/abs/2008.12348}.
\newblock arXiv:2008.12348 [cs].

\bibitem[Paranjape et~al.(2023)Paranjape, Lundberg, Singh, Hajishirzi, Zettlemoyer, and Ribeiro]{paranjape_art:_2023}
B.~Paranjape, S.~Lundberg, S.~Singh, H.~Hajishirzi, L.~Zettlemoyer, and M.~T. Ribeiro.
\newblock {ART}: {Automatic} multi-step reasoning and tool-use for large language models, Mar. 2023.
\newblock URL \url{http://arxiv.org/abs/2303.09014}.
\newblock arXiv:2303.09014 [cs].

\bibitem[Parfit(1984)]{parfit1984reasons}
D.~Parfit.
\newblock \emph{Reasons and persons}.
\newblock OUP Oxford, 1984.

\bibitem[Park et~al.(2023{\natexlab{a}})Park, O'Brien, Cai, Morris, Liang, and Bernstein]{park2023generative}
J.~S. Park, J.~C. O'Brien, C.~J. Cai, M.~R. Morris, P.~Liang, and M.~S. Bernstein.
\newblock Generative agents: Interactive simulacra of human behavior, 2023{\natexlab{a}}.

\bibitem[Park et~al.(2023{\natexlab{b}})Park, O'Brien, Cai, Morris, Liang, and Bernstein]{park_generative_2023}
J.~S. Park, J.~C. O'Brien, C.~J. Cai, M.~R. Morris, P.~Liang, and M.~S. Bernstein.
\newblock Generative {Agents}: {Interactive} {Simulacra} of {Human} {Behavior}, Aug. 2023{\natexlab{b}}.
\newblock URL \url{http://arxiv.org/abs/2304.03442}.
\newblock arXiv:2304.03442 [cs].

\bibitem[Park et~al.(2023{\natexlab{c}})Park, Goldstein, O'Gara, Chen, and Hendrycks]{park_ai_2023}
P.~S. Park, S.~Goldstein, A.~O'Gara, M.~Chen, and D.~Hendrycks.
\newblock {AI} {Deception}: {A} {Survey} of {Examples}, {Risks}, and {Potential} {Solutions}, Aug. 2023{\natexlab{c}}.
\newblock URL \url{http://arxiv.org/abs/2308.14752}.
\newblock arXiv:2308.14752 [cs].

\bibitem[Park and Chen(2007)]{park_acceptance_2007}
Y.~Park and J.~V. Chen.
\newblock Acceptance and adoption of the innovative use of smartphone.
\newblock \emph{Industrial Management \& Data Systems}, 107\penalty0 (9):\penalty0 1349--1365, Nov. 2007.
\newblock ISSN 0263-5577.
\newblock \doi{10.1108/02635570710834009}.
\newblock URL \url{https://www.emerald.com/insight/content/doi/10.1108/02635570710834009/full/html}.

\bibitem[Patterson et~al.(2021)Patterson, Gonzalez, Le, Liang, Munguia, Rothchild, So, Texier, and Dean]{patterson_carbon_2021}
D.~Patterson, J.~Gonzalez, Q.~Le, C.~Liang, L.-M. Munguia, D.~Rothchild, D.~So, M.~Texier, and J.~Dean.
\newblock Carbon {Emissions} and {Large} {Neural} {Network} {Training}, Apr. 2021.
\newblock URL \url{http://arxiv.org/abs/2104.10350}.
\newblock arXiv:2104.10350 [cs].

\bibitem[Patterson et~al.(2022)Patterson, Gonzalez, Hölzle, Le, Liang, Munguia, Rothchild, So, Texier, and Dean]{patterson_carbon_2022}
D.~Patterson, J.~Gonzalez, U.~Hölzle, Q.~Le, C.~Liang, L.-M. Munguia, D.~Rothchild, D.~So, M.~Texier, and J.~Dean.
\newblock The {Carbon} {Footprint} of {Machine} {Learning} {Training} {Will} {Plateau}, {Then} {Shrink}, Apr. 2022.
\newblock URL \url{http://arxiv.org/abs/2204.05149}.
\newblock arXiv:2204.05149 [cs].

\bibitem[Paul(2023)]{paul_risks_nodate}
A.~Paul.
\newblock The {Risks} of {Generative} {AI} in the {Stock} {Market}, 2023.
\newblock URL \url{https://www.opengrowth.com/resources/the-risks-of-generative-ai-in-the-stock-market}.

\bibitem[Paul(2014)]{paul_transformative_2014}
L.~A. Paul.
\newblock \emph{Transformative {Experience}}.
\newblock OUP Oxford, Nov. 2014.
\newblock ISBN 9780191027802.
\newblock Google-Books-ID: E4XjBAAAQBAJ.

\bibitem[Peng et~al.(2023)Peng, Kalliamvakou, Cihon, and Demirer]{peng_impact_2023}
S.~Peng, E.~Kalliamvakou, P.~Cihon, and M.~Demirer.
\newblock The {Impact} of {AI} on {Developer} {Productivity}: {Evidence} from {GitHub} {Copilot}, Feb. 2023.
\newblock URL \url{http://arxiv.org/abs/2302.06590}.
\newblock arXiv:2302.06590 [cs].

\bibitem[Pentina et~al.(2023)Pentina, Hancock, and Xie]{pentina_exploring_2023}
I.~Pentina, T.~Hancock, and T.~Xie.
\newblock Exploring relationship development with social chatbots: {A} mixed-method study of replika.
\newblock \emph{Computers in Human Behavior}, 140:\penalty0 107600, Mar. 2023.
\newblock ISSN 07475632.
\newblock \doi{10.1016/j.chb.2022.107600}.
\newblock URL \url{https://linkinghub.elsevier.com/retrieve/pii/S0747563222004204}.

\bibitem[Perez et~al.(2022{\natexlab{a}})Perez, Huang, Song, Cai, Ring, Aslanides, Glaese, McAleese, and Irving]{perez_red_2022}
E.~Perez, S.~Huang, F.~Song, T.~Cai, R.~Ring, J.~Aslanides, A.~Glaese, N.~McAleese, and G.~Irving.
\newblock Red {Teaming} {Language} {Models} with {Language} {Models}, Feb. 2022{\natexlab{a}}.
\newblock URL \url{http://arxiv.org/abs/2202.03286}.
\newblock arXiv:2202.03286 [cs].

\bibitem[Perez et~al.(2022{\natexlab{b}})Perez, Ringer, Luko{\v{s}}i{\=u}t{\.e}, Nguyen, Chen, Heiner, Pettit, Olsson, Kundu, Kadavath, et~al.]{perez2022discovering}
E.~Perez, S.~Ringer, K.~Luko{\v{s}}i{\=u}t{\.e}, K.~Nguyen, E.~Chen, S.~Heiner, C.~Pettit, C.~Olsson, S.~Kundu, S.~Kadavath, et~al.
\newblock Discovering language model behaviors with model-written evaluations.
\newblock \emph{arXiv preprint arXiv:2212.09251}, 2022{\natexlab{b}}.

\bibitem[Perrigo(2023)]{Perrigo_2023}
B.~Perrigo.
\newblock Bing’s ai is threatening users. that’s no laughing matter.
\newblock \emph{Time}, Feb. 2023.
\newblock URL \url{https://time.com/6256529/bing-openai-chatgpt-danger-alignment/}.

\bibitem[Perrow(1999)]{perrow_normal_1999}
C.~Perrow.
\newblock \emph{Normal {Accidents}: {Living} with {High} {Risk} {Technologies}}.
\newblock Princeton University Press, Oct. 1999.
\newblock ISBN 9780691004129.
\newblock URL \url{https://press.princeton.edu/books/paperback/9780691004129/normal-accidents}.

\bibitem[Peters et~al.(2018)Peters, Calvo, and Ryan]{peters2018designing}
D.~Peters, R.~A. Calvo, and R.~M. Ryan.
\newblock Designing for motivation, engagement and wellbeing in digital experience.
\newblock \emph{Frontiers in Psychology}, page 797, 2018.

\bibitem[Petropoulos et~al.(2018)Petropoulos, Pichler, and Chiacchio]{petropoulos_impact_2018}
G.~Petropoulos, D.~Pichler, and F.~Chiacchio.
\newblock The impact of industrial robots on {EU} employment and wages: {A} local labour market approach.
\newblock Technical report, Bruegel, Apr. 2018.
\newblock URL \url{https://www.bruegel.org/working-paper/impact-industrial-robots-eu-employment-and-wages-local-labour-market-approach}.

\bibitem[Pfizer(2023)]{pfizer_how_nodate}
Pfizer.
\newblock How a {Novel} ‘{Incubation} {Sandbox}’ {Helped} {Speed} {Up} {Data} {Analysis} in {Pfizer}’s {COVID}-19 {Vaccine} {Trial}, 2023.
\newblock URL \url{https://www.pfizer.com/news/articles/how_a_novel_incubation_sandbox_helped_speed_up_data_analysis_in_pfizer_s_covid_19_vaccine_trial}.

\bibitem[Phillips and Land(2012)]{phillips_link_2012}
J.~Phillips and K.~C. Land.
\newblock The link between unemployment and crime rate fluctuations: {An} analysis at the county, state, and national levels.
\newblock \emph{Social Science Research}, 41\penalty0 (3):\penalty0 681--694, May 2012.
\newblock ISSN 0049089X.
\newblock \doi{10.1016/j.ssresearch.2012.01.001}.
\newblock URL \url{https://linkinghub.elsevier.com/retrieve/pii/S0049089X12000026}.

\bibitem[Phuong et~al.(2024)Phuong, Aitchison, Catt, Cogan, Kaskasoli, Krakovna, Lindner, Rahtz, Assael, Hodkinson, Howard, Lieberum, Kumar, Raad, Webson, Ho, Lin, Farquhar, Hutter, Deletang, Ruoss, El-Sayed, Brown, Dragan, Shah, Dafoe, and Shevlane]{phuong2024evaluating}
M.~Phuong, M.~Aitchison, E.~Catt, S.~Cogan, A.~Kaskasoli, V.~Krakovna, D.~Lindner, M.~Rahtz, Y.~Assael, S.~Hodkinson, H.~Howard, T.~Lieberum, R.~Kumar, M.~A. Raad, A.~Webson, L.~Ho, S.~Lin, S.~Farquhar, M.~Hutter, G.~Deletang, A.~Ruoss, S.~El-Sayed, S.~Brown, A.~Dragan, R.~Shah, A.~Dafoe, and T.~Shevlane.
\newblock Evaluating frontier models for dangerous capabilities, 2024.

\bibitem[Pickett and Wilkinson(2015)]{pickett_income_2015}
K.~E. Pickett and R.~G. Wilkinson.
\newblock Income inequality and health: {A} causal review.
\newblock \emph{Social Science \& Medicine}, 128:\penalty0 316--326, Mar. 2015.
\newblock ISSN 02779536.
\newblock \doi{10.1016/j.socscimed.2014.12.031}.
\newblock URL \url{https://linkinghub.elsevier.com/retrieve/pii/S0277953614008399}.

\bibitem[Pinsky(2023)]{pinsky_bard_2023}
Y.~Pinsky.
\newblock Bard can now connect to your {Google} apps and services, Sept. 2023.
\newblock URL \url{https://blog.google/products/bard/google-bard-new-features-update-sept-2023/}.

\bibitem[Pitardi and Marriott(2021)]{pitardi_alexa_2021}
V.~Pitardi and H.~R. Marriott.
\newblock Alexa, she's not human but… unveiling the drivers of consumers' trust in voice-based artificial intelligence.
\newblock \emph{Psychology \& Marketing}, 38\penalty0 (4):\penalty0 626--642, 2021.

\bibitem[Pitt(2010)]{pitt_its_2010}
J.~C. Pitt.
\newblock It’s {Not} {About} {Technology}.
\newblock \emph{Knowledge, Technology \& Policy}, 23\penalty0 (3-4):\penalty0 445--454, Dec. 2010.
\newblock ISSN 0897-1986, 1874-6314.
\newblock \doi{10.1007/s12130-010-9125-5}.
\newblock URL \url{http://link.springer.com/10.1007/s12130-010-9125-5}.

\bibitem[Plank(2022)]{plank_problem_2022}
B.~Plank.
\newblock The “{Problem}” of {Human} {Label} {Variation}: {On} {Ground} {Truth} in {Data}, {Modeling} and {Evaluation}.
\newblock In Y.~Goldberg, Z.~Kozareva, and Y.~Zhang, editors, \emph{Proceedings of the 2022 {Conference} on {Empirical} {Methods} in {Natural} {Language} {Processing}}, pages 10671--10682, Abu Dhabi, United Arab Emirates, Dec. 2022. Association for Computational Linguistics.
\newblock \doi{10.18653/v1/2022.emnlp-main.731}.
\newblock URL \url{https://aclanthology.org/2022.emnlp-main.731}.

\bibitem[Png(2022)]{png_at_2022}
M.-T. Png.
\newblock At the {Tensions} of {South} and {North}: {Critical} {Roles} of {Global} {South} {Stakeholders} in {AI} {Governance}.
\newblock In \emph{2022 {ACM} {Conference} on {Fairness}, {Accountability}, and {Transparency}}, pages 1434--1445, Seoul Republic of Korea, June 2022. ACM.
\newblock ISBN 9781450393522.
\newblock \doi{10.1145/3531146.3533200}.
\newblock URL \url{https://dl.acm.org/doi/10.1145/3531146.3533200}.

\bibitem[Polanyi(1944)]{polanyi_great_1944}
K.~Polanyi.
\newblock \emph{The {Great} {Transformation}: {The} {Political} and {Economic} {Origins} of {Our} {Time}}.
\newblock Farrar \& Rinehart, New York, 1st edition, 1944.
\newblock URL \url{https://inctpped.ie.ufrj.br/spiderweb/pdf_4/Great_Transformation.pdf}.

\bibitem[Porter et~al.(2023)Porter, Habli, McDermid, and Kaas]{porter_principles-based_2023}
Z.~Porter, I.~Habli, J.~McDermid, and M.~Kaas.
\newblock A {Principles}-based {Ethics} {Assurance} {Argument} {Pattern} for {AI} and {Autonomous} {Systems}.
\newblock \emph{AI and Ethics}, June 2023.
\newblock ISSN 2730-5953, 2730-5961.
\newblock \doi{10.1007/s43681-023-00297-2}.
\newblock URL \url{http://arxiv.org/abs/2203.15370}.
\newblock arXiv:2203.15370 [cs].

\bibitem[Poushneh(2021)]{poushneh_humanizing_2021}
A.~Poushneh.
\newblock Humanizing voice assistant: {The} impact of voice assistant personality on consumers’ attitudes and behaviors.
\newblock \emph{Journal of Retailing and Consumer Services}, 58:\penalty0 102283, Jan. 2021.
\newblock ISSN 09696989.
\newblock \doi{10.1016/j.jretconser.2020.102283}.
\newblock URL \url{https://linkinghub.elsevier.com/retrieve/pii/S0969698920312911}.

\bibitem[Pouwels et~al.(2008)Pouwels, Siegers, and Vlasblom]{pouwels_income_2008}
B.~Pouwels, J.~Siegers, and J.~D. Vlasblom.
\newblock Income, working hours, and happiness.
\newblock \emph{Economics Letters}, 99\penalty0 (1):\penalty0 72--74, 2008.
\newblock ISSN 0165-1765.
\newblock URL \url{https://econpapers.repec.org/article/eeeecolet/v_3a99_3ay_3a2008_3ai_3a1_3ap_3a72-74.htm}.

\bibitem[Prabhakaran et~al.(2022)Prabhakaran, Mitchell, Gebru, and Gabriel]{prabhakaran2022human}
V.~Prabhakaran, M.~Mitchell, T.~Gebru, and I.~Gabriel.
\newblock A human rights-based approach to responsible ai.
\newblock \emph{arXiv preprint arXiv:2210.02667}, 2022.

\bibitem[Prescott-Allen(2001)]{prescott2001wellbeing}
R.~Prescott-Allen.
\newblock \emph{The wellbeing of nations: {A} country-by-country index of quality of life and the environment}.
\newblock Island press, 2001.

\bibitem[Proudfoot(2011)]{proudfoot_anthropomorphism_2011}
D.~Proudfoot.
\newblock Anthropomorphism and {AI}: {Turing's} much misunderstood imitation game.
\newblock \emph{Artificial Intelligence}, 175\penalty0 (5-6):\penalty0 950--957, Apr. 2011.
\newblock ISSN 00043702.
\newblock \doi{10.1016/j.artint.2011.01.006}.
\newblock URL \url{https://linkinghub.elsevier.com/retrieve/pii/S000437021100018X}.

\bibitem[Prunkl et~al.(2021)Prunkl, Ashurst, Anderljung, Webb, Leike, and Dafoe]{prunkl_institutionalizing_2021}
C.~E.~A. Prunkl, C.~Ashurst, M.~Anderljung, H.~Webb, J.~Leike, and A.~Dafoe.
\newblock Institutionalizing ethics in {AI} through broader impact requirements.
\newblock \emph{Nature Machine Intelligence}, 3\penalty0 (2):\penalty0 104--110, Feb. 2021.
\newblock ISSN 2522-5839.
\newblock \doi{10.1038/s42256-021-00298-y}.
\newblock URL \url{https://www.nature.com/articles/s42256-021-00298-y}.

\bibitem[Przybylski and Weinstein(2017)]{przybylski2017large}
A.~K. Przybylski and N.~Weinstein.
\newblock A large-scale test of the {G}oldilocks hypothesis: {Q}uantifying the relations between digital-screen use and the mental well-being of adolescents.
\newblock \emph{Psychological Science}, 28\penalty0 (2):\penalty0 204--215, 2017.

\bibitem[Pugh(2020)]{pugh2020autonomy}
J.~Pugh.
\newblock \emph{Autonomy, rationality, and contemporary bioethics}.
\newblock Oxford University Press, 2020.

\bibitem[Purington et~al.(2017)Purington, Taft, Sannon, Bazarova, and Taylor]{purington_alexa_2017}
A.~Purington, J.~G. Taft, S.~Sannon, N.~N. Bazarova, and S.~H. Taylor.
\newblock " alexa is my new bff" social roles, user satisfaction, and personification of the amazon echo.
\newblock In \emph{Proceedings of the 2017 CHI conference extended abstracts on human factors in computing systems}, pages 2853--2859, 2017.

\bibitem[Qammar et~al.(2023)Qammar, Wang, Ding, Naouri, Daneshmand, and Ning]{Qammar_Wang_Ding_Naouri_Daneshmand_Ning_2023}
A.~Qammar, H.~Wang, J.~Ding, A.~Naouri, M.~Daneshmand, and H.~Ning.
\newblock Chatbots to chatgpt in a cybersecurity space: Evolution, vulnerabilities, attacks, challenges, and future recommendations.
\newblock \penalty0 (arXiv:2306.09255), May 2023.
\newblock URL \url{http://arxiv.org/abs/2306.09255}.
\newblock arXiv:2306.09255 [cs].

\bibitem[Qualcomm(2023)]{qualcomm_future_2023}
Qualcomm.
\newblock The future of {AI} is hybrid.
\newblock Technical report, Qualcomm, May 2023.
\newblock URL \url{https://www.qualcomm.com/content/dam/qcomm-martech/dm-assets/documents/Whitepaper-The-future-of-AI-is-hybrid-Part-1-Unlocking-the-generative-AI-future-with-on-device-and-hybrid-AI.pdf}.

\bibitem[{Qubit Labs}(2022)]{qubit_labs_how_2022}
{Qubit Labs}.
\newblock How {Many} {Programmers} are there in the {World} and in the {US}?, Nov. 2022.
\newblock URL \url{https://qubit-labs.com/how-many-programmers-in-the-world/}.

\bibitem[Rabiner and Juang(1993)]{rabiner_fundamentals_1993}
L.~R. Rabiner and B.-H. Juang.
\newblock \emph{Fundamentals of {Speech} {Recognition}}.
\newblock PTR Prentice Hall, 1993.
\newblock ISBN 9780130151575.
\newblock URL \url{https://www.amazon.co.uk/Fundamentals-Speech-Recognition-Prentice-Processing/dp/0130151572}.
\newblock Google-Books-ID: XEVqQgAACAAJ.

\bibitem[Racine et~al.(2014)Racine, Martin~Rubio, Chandler, Forlini, and Lucke]{racine_value_2014}
E.~Racine, T.~Martin~Rubio, J.~Chandler, C.~Forlini, and J.~Lucke.
\newblock The value and pitfalls of speculation about science and technology in bioethics: the case of cognitive enhancement.
\newblock \emph{Medicine, Health Care and Philosophy}, 17\penalty0 (3):\penalty0 325--337, Aug. 2014.
\newblock ISSN 1386-7423, 1572-8633.
\newblock \doi{10.1007/s11019-013-9539-4}.
\newblock URL \url{http://link.springer.com/10.1007/s11019-013-9539-4}.

\bibitem[Radhakrishnan et~al.(2023)Radhakrishnan, Shlegeris, Greenblatt, and Roger]{radhakrishnan_scalable_2023}
A.~Radhakrishnan, B.~Shlegeris, R.~Greenblatt, and F.~Roger.
\newblock Scalable oversight and weak-to-strong generalization: Compatible approaches to the same problem, 2023.
\newblock URL \url{https://www.alignmentforum.org/posts/hw2tGSsvLLyjFoLFS/scalable-oversight-and-weak-to-strong-generalization}.

\bibitem[Rae et~al.(2021)Rae, Borgeaud, Cai, Millican, Hoffmann, Song, Aslanides, Henderson, Ring, Young, et~al.]{rae2021scaling}
J.~W. Rae, S.~Borgeaud, T.~Cai, K.~Millican, J.~Hoffmann, F.~Song, J.~Aslanides, S.~Henderson, R.~Ring, S.~Young, et~al.
\newblock Scaling language models: Methods, analysis \& insights from training gopher.
\newblock \emph{arXiv preprint arXiv:2112.11446}, 2021.

\bibitem[Rae et~al.(2022)Rae, Borgeaud, Cai, Millican, Hoffmann, Song, Aslanides, Henderson, Ring, Young, Rutherford, Hennigan, Menick, Cassirer, Powell, Driessche, Hendricks, Rauh, Huang, Glaese, Welbl, Dathathri, Huang, Uesato, Mellor, Higgins, Creswell, McAleese, Wu, Elsen, Jayakumar, Buchatskaya, Budden, Sutherland, Simonyan, Paganini, Sifre, Martens, Li, Kuncoro, Nematzadeh, Gribovskaya, Donato, Lazaridou, Mensch, Lespiau, Tsimpoukelli, Grigorev, Fritz, Sottiaux, Pajarskas, Pohlen, Gong, Toyama, d'Autume, Li, Terzi, Mikulik, Babuschkin, Clark, Casas, Guy, Jones, Bradbury, Johnson, Hechtman, Weidinger, Gabriel, Isaac, Lockhart, Osindero, Rimell, Dyer, Vinyals, Ayoub, Stanway, Bennett, Hassabis, Kavukcuoglu, and Irving]{rae_scaling_2022}
J.~W. Rae, S.~Borgeaud, T.~Cai, K.~Millican, J.~Hoffmann, F.~Song, J.~Aslanides, S.~Henderson, R.~Ring, S.~Young, E.~Rutherford, T.~Hennigan, J.~Menick, A.~Cassirer, R.~Powell, G.~v.~d. Driessche, L.~A. Hendricks, M.~Rauh, P.-S. Huang, A.~Glaese, J.~Welbl, S.~Dathathri, S.~Huang, J.~Uesato, J.~Mellor, I.~Higgins, A.~Creswell, N.~McAleese, A.~Wu, E.~Elsen, S.~Jayakumar, E.~Buchatskaya, D.~Budden, E.~Sutherland, K.~Simonyan, M.~Paganini, L.~Sifre, L.~Martens, X.~L. Li, A.~Kuncoro, A.~Nematzadeh, E.~Gribovskaya, D.~Donato, A.~Lazaridou, A.~Mensch, J.-B. Lespiau, M.~Tsimpoukelli, N.~Grigorev, D.~Fritz, T.~Sottiaux, M.~Pajarskas, T.~Pohlen, Z.~Gong, D.~Toyama, C.~d.~M. d'Autume, Y.~Li, T.~Terzi, V.~Mikulik, I.~Babuschkin, A.~Clark, D.~d.~L. Casas, A.~Guy, C.~Jones, J.~Bradbury, M.~Johnson, B.~Hechtman, L.~Weidinger, I.~Gabriel, W.~Isaac, E.~Lockhart, S.~Osindero, L.~Rimell, C.~Dyer, O.~Vinyals, K.~Ayoub, J.~Stanway, L.~Bennett, D.~Hassabis, K.~Kavukcuoglu, and G.~Irving.
\newblock Scaling {Language} {Models}: {Methods}, {Analysis} \& {Insights} from {Training} {Gopher}, Jan. 2022.
\newblock URL \url{https://arxiv.org/pdf/2112.11446.pdf}.
\newblock arXiv:2112.11446 [cs].

\bibitem[Raghavan and Barocas(2019)]{raghavan_challenges_2019}
M.~Raghavan and S.~Barocas.
\newblock Challenges for mitigating bias in algorithmic hiring, Dec. 2019.
\newblock URL \url{https://www.brookings.edu/articles/challenges-for-mitigating-bias-in-algorithmic-hiring/}.

\bibitem[Rahwan et~al.(2019)Rahwan, Cebrian, Obradovich, Bongard, Bonnefon, Breazeal, Crandall, Christakis, Couzin, Jackson, et~al.]{rahwan2019machine}
I.~Rahwan, M.~Cebrian, N.~Obradovich, J.~Bongard, J.-F. Bonnefon, C.~Breazeal, J.~W. Crandall, N.~A. Christakis, I.~D. Couzin, M.~O. Jackson, et~al.
\newblock Machine behaviour.
\newblock \emph{Nature}, 568\penalty0 (7753):\penalty0 477--486, 2019.

\bibitem[Raji et~al.(2020{\natexlab{a}})Raji, Gebru, Mitchell, Buolamwini, Lee, and Denton]{raji_saving_2020}
I.~D. Raji, T.~Gebru, M.~Mitchell, J.~Buolamwini, J.~Lee, and E.~Denton.
\newblock Saving {Face}: {Investigating} the {Ethical} {Concerns} of {Facial} {Recognition} {Auditing}.
\newblock In \emph{Proceedings of the {AAAI}/{ACM} {Conference} on {AI}, {Ethics}, and {Society}}, pages 145--151, New York NY USA, Feb. 2020{\natexlab{a}}. ACM.
\newblock ISBN 9781450371100.
\newblock \doi{10.1145/3375627.3375820}.
\newblock URL \url{https://dl.acm.org/doi/10.1145/3375627.3375820}.

\bibitem[Raji et~al.(2020{\natexlab{b}})Raji, Smart, White, Mitchell, Gebru, Hutchinson, Smith-Loud, Theron, and Barnes]{raji_closing_2020}
I.~D. Raji, A.~Smart, R.~N. White, M.~Mitchell, T.~Gebru, B.~Hutchinson, J.~Smith-Loud, D.~Theron, and P.~Barnes.
\newblock Closing the {AI} accountability gap: defining an end-to-end framework for internal algorithmic auditing.
\newblock In \emph{Proceedings of the 2020 {Conference} on {Fairness}, {Accountability}, and {Transparency}}, {FAT}* '20, pages 33--44, New York, NY, USA, Jan. 2020{\natexlab{b}}. Association for Computing Machinery.
\newblock ISBN 9781450369367.
\newblock \doi{10.1145/3351095.3372873}.
\newblock URL \url{https://dl.acm.org/doi/10.1145/3351095.3372873}.

\bibitem[Raji et~al.(2022{\natexlab{a}})Raji, Kumar, Horowitz, and Selbst]{raji2022fallacy}
I.~D. Raji, I.~E. Kumar, A.~Horowitz, and A.~Selbst.
\newblock The fallacy of {AI} functionality.
\newblock In \emph{Proceedings of the 2022 ACM Conference on Fairness, Accountability, and Transparency}, pages 959--972, 2022{\natexlab{a}}.

\bibitem[Raji et~al.(2022{\natexlab{b}})Raji, Xu, Honigsberg, and Ho]{raji_outsider_2022}
I.~D. Raji, P.~Xu, C.~Honigsberg, and D.~E. Ho.
\newblock Outsider {Oversight}: {Designing} a {Third} {Party} {Audit} {Ecosystem} for {AI} {Governance}, June 2022{\natexlab{b}}.
\newblock URL \url{http://arxiv.org/abs/2206.04737}.
\newblock arXiv:2206.04737 [cs].

\bibitem[Ranjan(2023)]{ranjan_unveiling_2023}
S.~Ranjan.
\newblock Unveiling the {Privacy} {Risks} of {Generative} {AI}, June 2023.
\newblock URL \url{https://medium.com/@shiveshr/unveiling-the-privacy-risks-of-generative-ai-d4852be407cb}.

\bibitem[Rao and Min(2018)]{rao2018decent}
N.~D. Rao and J.~Min.
\newblock Decent living standards: {M}aterial prerequisites for human wellbeing.
\newblock \emph{Social Indicators Research}, 138:\penalty0 225--244, 2018.

\bibitem[Rapoport and Chammah(1966)]{rapoport_game_1966}
A.~Rapoport and A.~M. Chammah.
\newblock The {Game} of {Chicken}.
\newblock \emph{American Behavioral Scientist}, 10\penalty0 (3):\penalty0 10--28, Nov. 1966.
\newblock ISSN 0002-7642, 1552-3381.
\newblock \doi{10.1177/000276426601000303}.
\newblock URL \url{http://journals.sagepub.com/doi/10.1177/000276426601000303}.

\bibitem[Rashkin et~al.(2021)Rashkin, Reitter, Tomar, and Das]{rashkin_increasing_2021}
H.~Rashkin, D.~Reitter, G.~S. Tomar, and D.~Das.
\newblock Increasing {Faithfulness} in {Knowledge}-{Grounded} {Dialogue} with {Controllable} {Features}.
\newblock In C.~Zong, F.~Xia, W.~Li, and R.~Navigli, editors, \emph{Proceedings of the 59th {Annual} {Meeting} of the {Association} for {Computational} {Linguistics} and the 11th {International} {Joint} {Conference} on {Natural} {Language} {Processing} ({Volume} 1: {Long} {Papers})}, pages 704--718, Online, Aug. 2021. Association for Computational Linguistics.
\newblock \doi{10.18653/v1/2021.acl-long.58}.
\newblock URL \url{https://aclanthology.org/2021.acl-long.58}.

\bibitem[Rauschnabel and Ahuvia(2014)]{rauschnabel_youre_2014}
P.~A. Rauschnabel and A.~C. Ahuvia.
\newblock You’re so lovable: {Anthropomorphism} and brand love.
\newblock \emph{Journal of Brand Management}, 21\penalty0 (5):\penalty0 372--395, June 2014.
\newblock ISSN 1350-231X, 1479-1803.
\newblock \doi{10.1057/bm.2014.14}.
\newblock URL \url{http://link.springer.com/10.1057/bm.2014.14}.

\bibitem[Ravuri et~al.(2021)Ravuri, Lenc, Willson, Kangin, Lam, Mirowski, Fitzsimons, Athanassiadou, Kashem, Madge, Prudden, Mandhane, Clark, Brock, Simonyan, Hadsell, Robinson, Clancy, Arribas, and Mohamed]{ravuri_skillful_2021}
S.~Ravuri, K.~Lenc, M.~Willson, D.~Kangin, R.~Lam, P.~Mirowski, M.~Fitzsimons, M.~Athanassiadou, S.~Kashem, S.~Madge, R.~Prudden, A.~Mandhane, A.~Clark, A.~Brock, K.~Simonyan, R.~Hadsell, N.~Robinson, E.~Clancy, A.~Arribas, and S.~Mohamed.
\newblock Skillful {Precipitation} {Nowcasting} using {Deep} {Generative} {Models} of {Radar}.
\newblock \emph{Nature}, 597\penalty0 (7878):\penalty0 672--677, Sept. 2021.
\newblock ISSN 0028-0836, 1476-4687.
\newblock \doi{10.1038/s41586-021-03854-z}.
\newblock URL \url{http://arxiv.org/abs/2104.00954}.
\newblock arXiv:2104.00954 [cs].

\bibitem[Raz(1999)]{raz1999engaging}
J.~Raz.
\newblock \emph{Engaging reason: On the theory of value and action}.
\newblock Oxford University Press, 1999.

\bibitem[Reason(1997)]{reason_managing_1997}
J.~Reason.
\newblock \emph{Managing the {Risks} of {Organizational} {Accidents}}.
\newblock Routledge, Dec. 1997.
\newblock ISBN 9781840141054.
\newblock URL \url{https://www.routledge.com/Managing-the-Risks-of-Organizational-Accidents/Reason/p/book/9781840141054}.

\bibitem[Reed et~al.(2022)Reed, Zolna, Parisotto, Colmenarejo, Novikov, Barth-Maron, Gimenez, Sulsky, Kay, Springenberg, Eccles, Bruce, Razavi, Edwards, Heess, Chen, Hadsell, Vinyals, Bordbar, and de~Freitas]{reed_generalist_2022}
S.~Reed, K.~Zolna, E.~Parisotto, S.~G. Colmenarejo, A.~Novikov, G.~Barth-Maron, M.~Gimenez, Y.~Sulsky, J.~Kay, J.~T. Springenberg, T.~Eccles, J.~Bruce, A.~Razavi, A.~Edwards, N.~Heess, Y.~Chen, R.~Hadsell, O.~Vinyals, M.~Bordbar, and N.~de~Freitas.
\newblock A {Generalist} {Agent}.
\newblock \emph{Transactions on Machine Learning Research}, Nov. 2022.
\newblock \doi{10.48550/arXiv.2205.06175}.
\newblock URL \url{https://arxiv.org/pdf/2205.06175.pdf?fs=e&s=cl}.
\newblock arXiv:2205.06175 [cs].

\bibitem[Rheu et~al.(2021)Rheu, Shin, Peng, and Huh-Yoo]{rheu2021trust}
M.~Rheu, J.~Y. Shin, W.~Peng, and J.~Huh-Yoo.
\newblock Systematic review: Trust-building factors and implications for conversational agent design.
\newblock \emph{International Journal of Human–Computer Interaction}, 37\penalty0 (1):\penalty0 81--96, 2021.
\newblock \doi{10.1080/10447318.2020.1807710}.
\newblock URL \url{https://doi.org/10.1080/10447318.2020.1807710}.

\bibitem[Ribeiro et~al.(2023)Ribeiro, Meckin, Balmer, and Shapira]{ribeiro_digitalisation_2023}
B.~Ribeiro, R.~Meckin, A.~Balmer, and P.~Shapira.
\newblock The digitalisation paradox of everyday scientific labour: {How} mundane knowledge work is amplified and diversified in the biosciences.
\newblock \emph{Research Policy}, 52\penalty0 (1):\penalty0 104607, Jan. 2023.
\newblock ISSN 00487333.
\newblock \doi{10.1016/j.respol.2022.104607}.
\newblock URL \url{https://linkinghub.elsevier.com/retrieve/pii/S0048733322001305}.

\bibitem[Ribeiro et~al.(2021)Ribeiro, Ottoni, West, Almeida, and Meira]{ribeiro_auditing_2021}
M.~H. Ribeiro, R.~Ottoni, R.~West, V.~A.~F. Almeida, and W.~Meira.
\newblock Auditing {Radicalization} {Pathways} on {YouTube}, Oct. 2021.
\newblock URL \url{http://arxiv.org/abs/1908.08313}.
\newblock arXiv:1908.08313 [cs].

\bibitem[Ribino(2023)]{ribino_role_2023}
P.~Ribino.
\newblock The role of politeness in human--machine interactions: a systematic literature review and future perspectives.
\newblock \emph{Artificial Intelligence Review}, 56\penalty0 (Suppl 1):\penalty0 445--482, 2023.

\bibitem[Ribot and Peluso(2003)]{ribot_theory_2003}
J.~C. Ribot and N.~L. Peluso.
\newblock A {Theory} of {Access}*.
\newblock \emph{Rural Sociology}, 68\penalty0 (2):\penalty0 153--181, June 2003.
\newblock ISSN 0036-0112, 1549-0831.
\newblock \doi{10.1111/j.1549-0831.2003.tb00133.x}.
\newblock URL \url{https://onlinelibrary.wiley.com/doi/10.1111/j.1549-0831.2003.tb00133.x}.

\bibitem[Richards et~al.(2023)Richards, Agüera~y Arcas, Lajoie, and Sridhar]{richards_illusion_2023}
B.~Richards, B.~Agüera~y Arcas, G.~Lajoie, and D.~Sridhar.
\newblock The {Illusion} {Of} {AI}’s {Existential} {Risk}.
\newblock \emph{Noema}, July 2023.
\newblock URL \url{https://www.noemamag.com/the-illusion-of-ais-existential-risk}.

\bibitem[Richardson(2021)]{richardson_defining_2021}
R.~Richardson.
\newblock Defining and {Demystifying} {Automated} {Decision} {Systems}, Mar. 2021.
\newblock URL \url{https://papers.ssrn.com/abstract=3811708}.

\bibitem[Richens et~al.(2022)Richens, Beard, and Thompson]{richens2022counterfactual}
J.~Richens, R.~Beard, and D.~H. Thompson.
\newblock Counterfactual harm.
\newblock \emph{Advances in Neural Information Processing Systems}, 35:\penalty0 36350--36365, 2022.

\bibitem[Rieder et~al.(2020)Rieder, Simon, and Wong]{rieder_mapping_2020}
G.~Rieder, J.~Simon, and P.-H. Wong.
\newblock Mapping the {Stony} {Road} toward {Trustworthy} {AI}: {Expectations}, {Problems}, {Conundrums}, Oct. 2020.
\newblock URL \url{https://papers.ssrn.com/abstract=3717451}.

\bibitem[Rieser and Lemon(2011)]{rieser_reinforcement_2011}
V.~Rieser and O.~Lemon.
\newblock \emph{Reinforcement {Learning} for {Adaptive} {Dialogue} {Systems}: {A} {Data}-driven {Methodology} for {Dialogue} {Management} and {Natural} {Language} {Generation}}.
\newblock Springer, Berlin, Heidelberg, 2011.
\newblock ISBN 9783642249419 9783642249426.
\newblock \doi{10.1007/978-3-642-24942-6}.
\newblock URL \url{https://link.springer.com/10.1007/978-3-642-24942-6}.

\bibitem[Rigaud et~al.(2018)Rigaud, De~Sherbinin, Jones, Bergmann, Clement, Ober, Schewe, Adamo, McCusker, Heuser, and Midgley]{rigaud_groundswell_2018}
K.~K. Rigaud, A.~M. De~Sherbinin, B.~Jones, J.~Bergmann, V.~Clement, K.~Ober, J.~Schewe, S.~B. Adamo, B.~McCusker, S.~Heuser, and A.~Midgley.
\newblock Groundswell : {Preparing} for {Internal} {Climate} {Migration}.
\newblock 2018.
\newblock \doi{10.7916/D8Z33FNS}.
\newblock URL \url{https://academiccommons.columbia.edu/doi/10.7916/D8Z33FNS}.

\bibitem[Rigot(2022)]{rigot_design_2022}
A.~Rigot.
\newblock Design {From} the {Margins}: {Centering} the most marginalized and impacted in design processes – from ideation to production.
\newblock Technical report, Harvard Kennedy School: Belfer Center, May 2022.
\newblock URL \url{https://www.belfercenter.org/sites/default/files/files/publication/TAPP-Afsaneh_Design%20From%20the%20Margins_Final_220514.pdf}.

\bibitem[Rismani et~al.(2022)Rismani, Shelby, Smart, Jatho, Kroll, Moon, and Rostamzadeh]{rismani_plane_2022}
S.~Rismani, R.~Shelby, A.~Smart, E.~Jatho, J.~Kroll, A.~Moon, and N.~Rostamzadeh.
\newblock From plane crashes to algorithmic harm: applicability of safety engineering frameworks for responsible {ML}, Oct. 2022.
\newblock URL \url{http://arxiv.org/abs/2210.03535}.
\newblock arXiv:2210.03535 [cs].

\bibitem[Ritchie et~al.(2020)Ritchie, Roser, and Rosado]{ritchie_co_2020}
H.~Ritchie, M.~Roser, and P.~Rosado.
\newblock {CO}$_2$ and {Greenhouse} {Gas} {Emissions}.
\newblock \emph{Our World in Data}, May 2020.
\newblock URL \url{https://ourworldindata.org/co2-and-greenhouse-gas-emissions}.

\bibitem[Ritchie et~al.(2023)Ritchie, Samborska, Ahuja, Ortiz-Ospina, and Roser]{ritchie_global_2023}
H.~Ritchie, V.~Samborska, N.~Ahuja, E.~Ortiz-Ospina, and M.~Roser.
\newblock Global {Education}.
\newblock \emph{Our World in Data}, Nov. 2023.
\newblock URL \url{https://ourworldindata.org/global-education}.

\bibitem[Rittel and Webber(1973)]{rittel_dilemmas_1973}
H.~W.~J. Rittel and M.~M. Webber.
\newblock Dilemmas in a general theory of planning.
\newblock \emph{Policy Sciences}, 4\penalty0 (2):\penalty0 155--169, June 1973.
\newblock ISSN 1573-0891.
\newblock \doi{10.1007/BF01405730}.
\newblock URL \url{https://doi.org/10.1007/BF01405730}.

\bibitem[Robbins(1938)]{robbins1938interpersonal}
L.~Robbins.
\newblock Interpersonal comparisons of utility: {A} comment.
\newblock \emph{The Economic Journal}, 48\penalty0 (192):\penalty0 635--641, 1938.

\bibitem[Roberts and Jesudason(2013)]{roberts_movement_2013}
D.~Roberts and S.~Jesudason.
\newblock Movement {Intersectionality}: {The} {Case} of {Race}, {Gender}, {Disability}, and {Genetic} {Technologies}.
\newblock \emph{Du Bois Review: Social Science Research on Race}, 10\penalty0 (2):\penalty0 313--328, 2013.
\newblock ISSN 1742-058X, 1742-0598.
\newblock \doi{10.1017/S1742058X13000210}.
\newblock URL \url{https://www.cambridge.org/core/product/identifier/S1742058X13000210/type/journal_article}.

\bibitem[Robinette et~al.(2016)Robinette, Li, Allen, Howard, and Wagner]{robinette_overtrust_2016}
P.~Robinette, W.~Li, R.~Allen, A.~M. Howard, and A.~R. Wagner.
\newblock Overtrust of robots in emergency evacuation scenarios.
\newblock In \emph{2016 11th {ACM}/{IEEE} {International} {Conference} on {Human}-{Robot} {Interaction} ({HRI})}, pages 101--108, Christchurch, New Zealand, Mar. 2016. IEEE.
\newblock ISBN 9781467383707.
\newblock \doi{10.1109/HRI.2016.7451740}.
\newblock URL \url{http://ieeexplore.ieee.org/document/7451740/}.

\bibitem[Robinson et~al.(2015)Robinson, Cotten, Ono, Quan-Haase, Mesch, Chen, Schulz, Hale, and Stern]{robinson_digital_2015}
L.~Robinson, S.~R. Cotten, H.~Ono, A.~Quan-Haase, G.~Mesch, W.~Chen, J.~Schulz, T.~M. Hale, and M.~J. Stern.
\newblock Digital inequalities and why they matter.
\newblock \emph{Information, Communication \& Society}, 18\penalty0 (5):\penalty0 569--582, May 2015.
\newblock ISSN 1369-118X, 1468-4462.
\newblock \doi{10.1080/1369118X.2015.1012532}.
\newblock URL \url{http://www.tandfonline.com/doi/abs/10.1080/1369118X.2015.1012532}.

\bibitem[Rocha et~al.(2023)Rocha, De~Moura, Desidério, De~Oliveira, Lourenço, and De~Figueiredo~Nicolete]{rocha_impact_2023}
Y.~M. Rocha, G.~A. De~Moura, G.~A. Desidério, C.~H. De~Oliveira, F.~D. Lourenço, and L.~D. De~Figueiredo~Nicolete.
\newblock The impact of fake news on social media and its influence on health during the {COVID}-19 pandemic: a systematic review.
\newblock \emph{Journal of Public Health}, 31\penalty0 (7):\penalty0 1007--1016, July 2023.
\newblock ISSN 2198-1833, 1613-2238.
\newblock \doi{10.1007/s10389-021-01658-z}.
\newblock URL \url{https://link.springer.com/10.1007/s10389-021-01658-z}.

\bibitem[Roesler et~al.(2021)Roesler, Manzey, and Onnasch]{roesler_meta-analysis_2021}
E.~Roesler, D.~Manzey, and L.~Onnasch.
\newblock A meta-analysis on the effectiveness of anthropomorphism in human-robot interaction.
\newblock \emph{Science Robotics}, 6\penalty0 (58), Sept. 2021.
\newblock ISSN 2470-9476.
\newblock \doi{10.1126/scirobotics.abj5425}.
\newblock URL \url{https://www.science.org/doi/10.1126/scirobotics.abj5425}.

\bibitem[Rogers(2023)]{rogers_is_nodate}
R.~Rogers.
\newblock Is {GPT}-4 {Worth} the {Subscription}? {Here}’s {What} {You} {Should} {Know}.
\newblock \emph{Wired}, 2023.
\newblock ISSN 1059-1028.
\newblock URL \url{https://www.wired.com/story/what-is-chatgpt-plus-gpt4-openai/}.

\bibitem[Rojas and Guardiola(2016)]{rojas2016hierarchy}
M.~Rojas and J.~Guardiola.
\newblock A hierarchy of unsatisfied needs: A subjective well-being study.
\newblock \emph{A Life Devoted to Quality of Life: Festschrift in Honor of Alex C. Michalos}, pages 105--122, 2016.

\bibitem[Rojas et~al.(2023)Rojas, M{\'e}ndez, and Watkins-Fassler]{rojas2023hierarchy}
M.~Rojas, A.~M{\'e}ndez, and K.~Watkins-Fassler.
\newblock The hierarchy of needs empirical examination of {M}aslow’s theory and lessons for development.
\newblock \emph{World Development}, 165:\penalty0 106185, 2023.

\bibitem[Rolnick et~al.(2019)Rolnick, Donti, Kaack, Kochanski, Lacoste, Sankaran, Ross, Milojevic-Dupont, Jaques, Waldman-Brown, Luccioni, Maharaj, Sherwin, Mukkavilli, Kording, Gomes, Ng, Hassabis, Platt, Creutzig, Chayes, and Bengio]{rolnick_tackling_2019}
D.~Rolnick, P.~L. Donti, L.~H. Kaack, K.~Kochanski, A.~Lacoste, K.~Sankaran, A.~S. Ross, N.~Milojevic-Dupont, N.~Jaques, A.~Waldman-Brown, A.~Luccioni, T.~Maharaj, E.~D. Sherwin, S.~K. Mukkavilli, K.~P. Kording, C.~Gomes, A.~Y. Ng, D.~Hassabis, J.~C. Platt, F.~Creutzig, J.~Chayes, and Y.~Bengio.
\newblock Tackling {Climate} {Change} with {Machine} {Learning}, Nov. 2019.
\newblock URL \url{http://arxiv.org/abs/1906.05433}.
\newblock arXiv:1906.05433 [cs, stat].

\bibitem[Rombach et~al.(2022)Rombach, Blattmann, Lorenz, Esser, and Ommer]{rombach_high-resolution_2022}
R.~Rombach, A.~Blattmann, D.~Lorenz, P.~Esser, and B.~Ommer.
\newblock High-{Resolution} {Image} {Synthesis} with {Latent} {Diffusion} {Models}, Apr. 2022.
\newblock URL \url{http://arxiv.org/abs/2112.10752}.
\newblock arXiv:2112.10752 [cs].

\bibitem[Roose(2023)]{roose_conversation_2023}
K.~Roose.
\newblock A {Conversation} {With} {Bing}’s {Chatbot} {Left} {Me} {Deeply} {Unsettled}.
\newblock \emph{The New York Times}, Feb. 2023.
\newblock ISSN 0362-4331.
\newblock URL \url{https://www.nytimes.com/2023/02/16/technology/bing-chatbot-microsoft-chatgpt.html}.

\bibitem[Roozenbeek et~al.(2022)Roozenbeek, Van Der~Linden, Goldberg, Rathje, and Lewandowsky]{roozenbeek2022psychological}
J.~Roozenbeek, S.~Van Der~Linden, B.~Goldberg, S.~Rathje, and S.~Lewandowsky.
\newblock Psychological inoculation improves resilience against misinformation on social media.
\newblock \emph{Science Advances}, 8\penalty0 (34):\penalty0 eabo6254, 2022.

\bibitem[Rosen(2020)]{rosen_investments_2020}
G.~Rosen.
\newblock Investments to {Fight} {Polarization}, May 2020.
\newblock URL \url{https://about.fb.com/news/2020/05/investments-to-fight-polarization/}.

\bibitem[Ross et~al.(2023)Ross, Martinez, Houde, Muller, and Weisz]{ross_programmers_2023}
S.~I. Ross, F.~Martinez, S.~Houde, M.~Muller, and J.~D. Weisz.
\newblock The {Programmer}’s {Assistant}: {Conversational} {Interaction} with a {Large} {Language} {Model} for {Software} {Development}.
\newblock In \emph{Proceedings of the 28th {International} {Conference} on {Intelligent} {User} {Interfaces}}, pages 491--514, Sydney NSW Australia, Mar. 2023. ACM.
\newblock ISBN 9798400701061.
\newblock \doi{10.1145/3581641.3584037}.
\newblock URL \url{https://dl.acm.org/doi/10.1145/3581641.3584037}.

\bibitem[Rossignac-Milon et~al.(2021)Rossignac-Milon, Bolger, Zee, Boothby, and Higgins]{rossignac-milon_merged_2021}
M.~Rossignac-Milon, N.~Bolger, K.~S. Zee, E.~J. Boothby, and E.~T. Higgins.
\newblock Merged minds: {Generalized} shared reality in dyadic relationships.
\newblock \emph{Journal of Personality and Social Psychology}, 120\penalty0 (4):\penalty0 882--911, Apr. 2021.
\newblock ISSN 1939-1315, 0022-3514.
\newblock \doi{10.1037/pspi0000266}.
\newblock URL \url{http://doi.apa.org/getdoi.cfm?doi=10.1037/pspi0000266}.

\bibitem[Rotondi et~al.(2017)Rotondi, Stanca, and Tomasuolo]{rotondi_connecting_2017}
V.~Rotondi, L.~Stanca, and M.~Tomasuolo.
\newblock Connecting alone: {Smartphone} use, quality of social interactions and well-being.
\newblock \emph{Journal of Economic Psychology}, 63:\penalty0 17--26, Dec. 2017.
\newblock ISSN 01674870.
\newblock \doi{10.1016/j.joep.2017.09.001}.
\newblock URL \url{https://linkinghub.elsevier.com/retrieve/pii/S0167487017302520}.

\bibitem[Rousseau et~al.(1998)Rousseau, Sitkin, Burt, and Camerer]{rousseau_not_1998}
D.~M. Rousseau, S.~B. Sitkin, R.~S. Burt, and C.~Camerer.
\newblock Not {So} {Different} {After} {All}: {A} {Cross}-{Discipline} {View} {Of} {Trust}.
\newblock \emph{Academy of Management Review}, 23\penalty0 (3):\penalty0 393--404, July 1998.
\newblock ISSN 0363-7425, 1930-3807.
\newblock \doi{10.5465/amr.1998.926617}.
\newblock URL \url{http://journals.aom.org/doi/10.5465/amr.1998.926617}.

\bibitem[Rowe(2023)]{rowe_its_2023}
N.~Rowe.
\newblock ‘{It}’s destroyed me completely’: {Kenyan} moderators decry toll of training of {AI} models.
\newblock \emph{The Guardian}, Aug. 2023.
\newblock ISSN 0261-3077.
\newblock URL \url{https://www.theguardian.com/technology/2023/aug/02/ai-chatbot-training-human-toll-content-moderator-meta-openai}.

\bibitem[Rozière et~al.(2023)Rozière, Gehring, Gloeckle, Sootla, Gat, Tan, Adi, Liu, Remez, Rapin, Kozhevnikov, Evtimov, Bitton, Bhatt, Ferrer, Grattafiori, Xiong, Défossez, Copet, Azhar, Touvron, Martin, Usunier, Scialom, and Synnaeve]{roziere_code_2023}
B.~Rozière, J.~Gehring, F.~Gloeckle, S.~Sootla, I.~Gat, X.~E. Tan, Y.~Adi, J.~Liu, T.~Remez, J.~Rapin, A.~Kozhevnikov, I.~Evtimov, J.~Bitton, M.~Bhatt, C.~C. Ferrer, A.~Grattafiori, W.~Xiong, A.~Défossez, J.~Copet, F.~Azhar, H.~Touvron, L.~Martin, N.~Usunier, T.~Scialom, and G.~Synnaeve.
\newblock Code {Llama}: {Open} {Foundation} {Models} for {Code}, Aug. 2023.
\newblock URL \url{http://arxiv.org/abs/2308.12950}.
\newblock arXiv:2308.12950 [cs].

\bibitem[Rubeis(2020)]{rubeis_disruptive_2020}
G.~Rubeis.
\newblock The disruptive power of {Artificial} {Intelligence}. {Ethical} aspects of gerontechnology in elderly care.
\newblock \emph{Archives of Gerontology and Geriatrics}, 91:\penalty0 104186, Nov. 2020.
\newblock ISSN 01674943.
\newblock \doi{10.1016/j.archger.2020.104186}.
\newblock URL \url{https://linkinghub.elsevier.com/retrieve/pii/S0167494320301801}.

\bibitem[Rubel et~al.(2019)Rubel, Pham, and Castro]{rubel_agency_2019}
A.~Rubel, A.~Pham, and C.~Castro.
\newblock Agency {Laundering} and {Algorithmic} {Decision} {Systems}.
\newblock In N.~Taylor, C.~Christian-Lamb, M.~Martin, and B.~Nardi, editors, \emph{Information in {Contemporary} {Society} ({Lecture} {Notes} in {Computer} {Science}) ({Proceedings} of the 2019 {iConference})}, pages 590--598. Springer Nature, 2019.

\bibitem[Russell and Norvig(1995)]{russell1995artificial}
S.~Russell and P.~Norvig.
\newblock \emph{Artificial intelligence a modern approach}.
\newblock Prentice Hall, 1995.

\bibitem[Russell(2019)]{russell_human_2019}
S.~J. Russell.
\newblock \emph{Human compatible: artificial intelligence and the problem of control}.
\newblock Viking, New York?, 2019.
\newblock ISBN 9780525558620.

\bibitem[Ryan(2020)]{ryan_ai_2020}
M.~Ryan.
\newblock In {AI} {We} {Trust}: {Ethics}, {Artificial} {Intelligence}, and {Reliability}.
\newblock \emph{Science and Engineering Ethics}, 26\penalty0 (5):\penalty0 2749--2767, Oct. 2020.
\newblock ISSN 1471-5546.
\newblock \doi{10.1007/s11948-020-00228-y}.
\newblock URL \url{https://doi.org/10.1007/s11948-020-00228-y}.

\bibitem[Ryan et~al.(2013)Ryan, Curren, and Deci]{ryan2013humans}
R.~M. Ryan, R.~R. Curren, and E.~L. Deci.
\newblock What humans need: Flourishing in aristotelian philosophy and self-determination theory.
\newblock In A.~S. Waterman, editor, \emph{The best within us: Positive psychology perspectives on eudaimonia}, pages 57--75. American Psychological Association, 2013.

\bibitem[Ryland(2021)]{ryland_its_2021}
H.~Ryland.
\newblock It’s {Friendship}, {Jim}, but {Not} as {We} {Know} {It}: {A} {Degrees}-of-{Friendship} {View} of {Human}–{Robot} {Friendships}.
\newblock \emph{Minds and Machines}, 31\penalty0 (3):\penalty0 377--393, Sept. 2021.
\newblock ISSN 1572-8641.
\newblock \doi{10.1007/s11023-021-09560-z}.
\newblock URL \url{https://doi.org/10.1007/s11023-021-09560-z}.

\bibitem[Rzepka et~al.(2022)Rzepka, Berger, and Hess]{rzepka_voice_2022}
C.~Rzepka, B.~Berger, and T.~Hess.
\newblock Voice {Assistant} vs. {Chatbot} – {Examining} the {Fit} {Between} {Conversational} {Agents}’ {Interaction} {Modalities} and {Information} {Search} {Tasks}.
\newblock \emph{Information Systems Frontiers}, 24\penalty0 (3):\penalty0 839--856, June 2022.
\newblock ISSN 1572-9419.
\newblock \doi{10.1007/s10796-021-10226-5}.
\newblock URL \url{https://doi.org/10.1007/s10796-021-10226-5}.

\bibitem[Räuker et~al.(2023)Räuker, Ho, Casper, and Hadfield-Menell]{rauker_toward_2023}
T.~Räuker, A.~Ho, S.~Casper, and D.~Hadfield-Menell.
\newblock Toward {Transparent} {AI}: {A} {Survey} on {Interpreting} the {Inner} {Structures} of {Deep} {Neural} {Networks}, Aug. 2023.
\newblock URL \url{http://arxiv.org/abs/2207.13243}.
\newblock arXiv:2207.13243 [cs].

\bibitem[Sadasivan et~al.(2023)Sadasivan, Kumar, Balasubramanian, Wang, and Feizi]{sadasivan_can_2023}
V.~S. Sadasivan, A.~Kumar, S.~Balasubramanian, W.~Wang, and S.~Feizi.
\newblock Can {AI}-{Generated} {Text} be {Reliably} {Detected}?, June 2023.
\newblock URL \url{http://arxiv.org/abs/2303.11156}.
\newblock arXiv:2303.11156 [cs].

\bibitem[Sai~Dinesh et~al.(2022)Sai~Dinesh, Surendran, Kathirvelan, and Logesh]{sai_dinesh_artificial_2022}
R.~S. Sai~Dinesh, R.~Surendran, D.~Kathirvelan, and V.~Logesh.
\newblock Artificial {Intelligence} based {Vision} and {Voice} {Assistant}.
\newblock In \emph{2022 {International} {Conference} on {Electronics} and {Renewable} {Systems} ({ICEARS})}, pages 1478--1483, Tuticorin, India, Mar. 2022. IEEE.
\newblock ISBN 9781665484251.
\newblock \doi{10.1109/ICEARS53579.2022.9751819}.
\newblock URL \url{https://ieeexplore.ieee.org/document/9751819/}.

\bibitem[Sajja et~al.(2023)Sajja, Sermet, Cikmaz, Cwiertny, and Demir]{sajja_artificial_2023}
R.~Sajja, Y.~Sermet, M.~Cikmaz, D.~Cwiertny, and I.~Demir.
\newblock Artificial {Intelligence}-{Enabled} {Intelligent} {Assistant} for {Personalized} and {Adaptive} {Learning} in {Higher} {Education}, Sept. 2023.
\newblock URL \url{http://arxiv.org/abs/2309.10892}.
\newblock arXiv:2309.10892 [cs].

\bibitem[Salem et~al.(2013)Salem, Eyssel, Rohlfing, Kopp, and Joublin]{salem_err_2013}
M.~Salem, F.~Eyssel, K.~Rohlfing, S.~Kopp, and F.~Joublin.
\newblock To err is human (-like): Effects of robot gesture on perceived anthropomorphism and likability.
\newblock \emph{International Journal of Social Robotics}, 5:\penalty0 313--323, 2013.

\bibitem[Salles(2021)]{salles_theories_2021}
S.~Salles.
\newblock Theories of {Vagueness}.
\newblock In S.~Salles, editor, \emph{Vagueness as {Arbitrariness}: {Outline} of a {Theory} of {Vagueness}}, Synthese {Library}, pages 65--128. Springer International Publishing, Cham, 2021.
\newblock ISBN 9783030667818.
\newblock \doi{10.1007/978-3-030-66781-8_4}.
\newblock URL \url{https://doi.org/10.1007/978-3-030-66781-8_4}.

\bibitem[Sambasivan et~al.(2021)Sambasivan, Arnesen, Hutchinson, Doshi, and Prabhakaran]{sambasivan_re-imagining_2021}
N.~Sambasivan, E.~Arnesen, B.~Hutchinson, T.~Doshi, and V.~Prabhakaran.
\newblock Re-imagining {Algorithmic} {Fairness} in {India} and {Beyond}.
\newblock In \emph{Proceedings of the 2021 {ACM} {Conference} on {Fairness}, {Accountability}, and {Transparency}}, pages 315--328, Virtual Event Canada, Mar. 2021. ACM.
\newblock ISBN 9781450383097.
\newblock \doi{10.1145/3442188.3445896}.
\newblock URL \url{https://dl.acm.org/doi/10.1145/3442188.3445896}.

\bibitem[Sap et~al.(2022)Sap, Swayamdipta, Vianna, Zhou, Choi, and Smith]{sap_annotators_2022}
M.~Sap, S.~Swayamdipta, L.~Vianna, X.~Zhou, Y.~Choi, and N.~A. Smith.
\newblock Annotators with {Attitudes}: {How} {Annotator} {Beliefs} {And} {Identities} {Bias} {Toxic} {Language} {Detection}.
\newblock In M.~Carpuat, M.-C. de~Marneffe, and I.~V. Meza~Ruiz, editors, \emph{Proceedings of the 2022 {Conference} of the {North} {American} {Chapter} of the {Association} for {Computational} {Linguistics}: {Human} {Language} {Technologies}}, pages 5884--5906, Seattle, United States, July 2022. Association for Computational Linguistics.
\newblock \doi{10.18653/v1/2022.naacl-main.431}.
\newblock URL \url{https://aclanthology.org/2022.naacl-main.431.pdf}.

\bibitem[Sarikaya et~al.(2016)Sarikaya, Crook, Marin, Jeong, Robichaud, Celikyilmaz, Kim, Rochette, Khan, Liu, Boies, Anastasakos, Feizollahi, Ramesh, Suzuki, Holenstein, Krawczyk, and Radostev]{sarikaya_overview_2016}
R.~Sarikaya, P.~A. Crook, A.~Marin, M.~Jeong, J.~Robichaud, A.~Celikyilmaz, Y.~Kim, A.~Rochette, O.~Z. Khan, X.~Liu, D.~Boies, T.~Anastasakos, Z.~Feizollahi, N.~Ramesh, H.~Suzuki, R.~Holenstein, E.~Krawczyk, and V.~Radostev.
\newblock An overview of end-to-end language understanding and dialog management for personal digital assistants.
\newblock In \emph{2016 {IEEE} {Spoken} {Language} {Technology} {Workshop} ({SLT})}, pages 391--397, San Diego, CA, Dec. 2016. IEEE.
\newblock ISBN 9781509049035.
\newblock \doi{10.1109/SLT.2016.7846294}.
\newblock URL \url{http://ieeexplore.ieee.org/document/7846294/}.

\bibitem[Sartor et~al.(2021)Sartor, Lagioia, and Galli]{sartor_regulating_2021}
G.~Sartor, F.~Lagioia, and F.~Galli.
\newblock Regulating targeted and behavioural advertising in digital services. {How} to ensure users’ informed consent {\textbar} {Think} {Tank} {\textbar} {European} {Parliament}.
\newblock Technical Report PE 694.680, European Parliament's Committee on Legal Affairs, Sept. 2021.
\newblock URL \url{https://www.europarl.europa.eu/thinktank/en/document/IPOL_STU(2021)694680}.

\bibitem[Sartori and Theodorou(2022)]{sartori_sociotechnical_2022}
L.~Sartori and A.~Theodorou.
\newblock A sociotechnical perspective for the future of {AI}: narratives, inequalities, and human control.
\newblock \emph{Ethics and Information Technology}, 24\penalty0 (1):\penalty0 4, Jan. 2022.
\newblock ISSN 1572-8439.
\newblock \doi{10.1007/s10676-022-09624-3}.
\newblock URL \url{https://doi.org/10.1007/s10676-022-09624-3}.

\bibitem[Satz(2010)]{satz_why_2010}
D.~Satz.
\newblock \emph{Why {Some} {Things} {Should} {Not} {Be} for {Sale}: {The} {Moral} {Limits} of {Markets}}.
\newblock Oxford University Press, June 2010.
\newblock ISBN 9780199718573.
\newblock Google-Books-ID: h13Pk15YpIwC.

\bibitem[Saulnier et~al.(2022)Saulnier, Karamcheti, Laurençon, Tronchon, Wang, Sanh, Singh, Pistilli, Luccioni, Jernite, Mitchell, and Kiela]{saulnier_putting_2022}
L.~Saulnier, S.~Karamcheti, H.~Laurençon, L.~Tronchon, T.~Wang, V.~Sanh, A.~Singh, G.~Pistilli, S.~Luccioni, Y.~Jernite, M.~Mitchell, and D.~Kiela.
\newblock Putting ethical principles at the core of the research lifecycle, May 2022.
\newblock URL \url{https://huggingface.co/blog/ethical-charter-multimodal}.

\bibitem[Saunders et~al.(2022)Saunders, Yeh, Wu, Bills, Ouyang, Ward, and Leike]{saunders_self-critiquing_2022}
W.~Saunders, C.~Yeh, J.~Wu, S.~Bills, L.~Ouyang, J.~Ward, and J.~Leike.
\newblock Self-critiquing models for assisting human evaluators, June 2022.
\newblock URL \url{https://arxiv.org/pdf/2206.05802.pdf}.
\newblock arXiv:2206.05802 [cs].

\bibitem[Scanlon(1998)]{scanlon_what_1998}
T.~Scanlon.
\newblock \emph{What {We} {Owe} to {Each} {Other}}.
\newblock Belknap Press of Harvard University Press, Cambridge, Mass., 1998.

\bibitem[Schaeffer et~al.(2023)Schaeffer, Miranda, and Koyejo]{schaeffer_are_2023}
R.~Schaeffer, B.~Miranda, and S.~Koyejo.
\newblock Are {Emergent} {Abilities} of {Large} {Language} {Models} a {Mirage}?, May 2023.
\newblock URL \url{https://arxiv.org/pdf/2304.15004.pdf}.
\newblock arXiv:2304.15004 [cs].

\bibitem[Schaubroeck et~al.(2011)Schaubroeck, Lam, and Peng]{schaubroeck_cognition-based_2011}
J.~Schaubroeck, S.~S.~K. Lam, and A.~C. Peng.
\newblock Cognition-based and affect-based trust as mediators of leader behavior influences on team performance.
\newblock \emph{Journal of Applied Psychology}, 96\penalty0 (4):\penalty0 863--871, 2011.
\newblock ISSN 1939-1854, 0021-9010.
\newblock \doi{10.1037/a0022625}.
\newblock URL \url{http://doi.apa.org/getdoi.cfm?doi=10.1037/a0022625}.

\bibitem[Scheurer et~al.(2022)Scheurer, Campos, Chan, Chen, Cho, and Perez]{scheurer_training_2022}
J.~Scheurer, J.~A. Campos, J.~S. Chan, A.~Chen, K.~Cho, and E.~Perez.
\newblock Training {Language} {Models} with {Language} {Feedback}, Nov. 2022.
\newblock URL \url{http://arxiv.org/abs/2204.14146}.
\newblock arXiv:2204.14146 [cs].

\bibitem[Scheutz(2009)]{scheutz_inherent_2009}
M.~Scheutz.
\newblock The {Inherent} {Dangers} of {Unidirectional} {Emotional} {Bonds} between {Humans} and {Social} {Robots}.
\newblock Jan. 2009.

\bibitem[Schick et~al.(2021)Schick, Udupa, and Schütze]{schick_self-diagnosis_2021}
T.~Schick, S.~Udupa, and H.~Schütze.
\newblock Self-{Diagnosis} and {Self}-{Debiasing}: {A} {Proposal} for {Reducing} {Corpus}-{Based} {Bias} in {NLP}.
\newblock \emph{Transactions of the Association for Computational Linguistics}, 9:\penalty0 1408--1424, Dec. 2021.
\newblock ISSN 2307-387X.
\newblock \doi{10.1162/tacl_a_00434}.
\newblock URL \url{https://direct.mit.edu/tacl/article/doi/10.1162/tacl_a_00434/108865/Self-Diagnosis-and-Self-Debiasing-A-Proposal-for}.

\bibitem[Schick et~al.(2023)Schick, Dwivedi-Yu, Dessì, Raileanu, Lomeli, Zettlemoyer, Cancedda, and Scialom]{schick_toolformer:_2023}
T.~Schick, J.~Dwivedi-Yu, R.~Dessì, R.~Raileanu, M.~Lomeli, L.~Zettlemoyer, N.~Cancedda, and T.~Scialom.
\newblock Toolformer: {Language} {Models} {Can} {Teach} {Themselves} to {Use} {Tools}, Feb. 2023.
\newblock URL \url{https://arxiv.org/pdf/2302.04761.pdf}.
\newblock arXiv:2302.04761 [cs].

\bibitem[Schirmer(2020)]{wischmeyer_artificial_2020}
J.-E. Schirmer.
\newblock Artificial {Intelligence} and {Legal} {Personality}: {Introducing} “{Teilrechtsfähigkeit}”: {A} {Partial} {Legal} {Status} {Made} in {Germany}.
\newblock In T.~Wischmeyer and T.~Rademacher, editors, \emph{Regulating {Artificial} {Intelligence}}, pages 123--142. Springer International Publishing, Cham, 2020.
\newblock ISBN 9783030323608 9783030323615.
\newblock \doi{10.1007/978-3-030-32361-5_6}.
\newblock URL \url{https://link.springer.com/10.1007/978-3-030-32361-5_6}.

\bibitem[Schlangen(2019)]{schlangen_language_2019}
D.~Schlangen.
\newblock Language {Tasks} and {Language} {Games}: {On} {Methodology} in {Current} {Natural} {Language} {Processing} {Research}, Aug. 2019.
\newblock URL \url{http://arxiv.org/abs/1908.10747}.
\newblock arXiv:1908.10747 [cs].

\bibitem[Schlangen and Skantze(2009)]{schlangen_general_2009}
D.~Schlangen and G.~Skantze.
\newblock A {General}, {Abstract} {Model} of {Incremental} {Dialogue} {Processing}.
\newblock In A.~Lascarides, C.~Gardent, and J.~Nivre, editors, \emph{Proceedings of the 12th {Conference} of the {European} {Chapter} of the {ACL} ({EACL} 2009)}, pages 710--718, Athens, Greece, Mar. 2009. Association for Computational Linguistics.
\newblock URL \url{https://aclanthology.org/E09-1081}.

\bibitem[Schroeder and Epley(2016)]{schroeder_mistaking_2016}
J.~Schroeder and N.~Epley.
\newblock Mistaking minds and machines: {How} speech affects dehumanization and anthropomorphism.
\newblock \emph{Journal of Experimental Psychology. General}, 145\penalty0 (11):\penalty0 1427--1437, Nov. 2016.
\newblock ISSN 1939-2222.
\newblock \doi{10.1037/xge0000214}.

\bibitem[Schröder and Neumayr(2023)]{schroder_how_2023}
J.~M. Schröder and M.~Neumayr.
\newblock How socio-economic inequality affects individuals’ civic engagement: a systematic literature review of empirical findings and theoretical explanations.
\newblock \emph{Socio-Economic Review}, 21\penalty0 (1):\penalty0 665--694, Mar. 2023.
\newblock ISSN 1475-1461, 1475-147X.
\newblock \doi{10.1093/ser/mwab058}.
\newblock URL \url{https://academic.oup.com/ser/article/21/1/665/6482042}.

\bibitem[Schwartz and Solove(2011)]{schwartz_pii_2011}
P.~M. Schwartz and D.~J. Solove.
\newblock The pii problem: Privacy and a new concept of personally identifiable information.
\newblock \emph{NYUL rev.}, 86:\penalty0 1814, 2011.

\bibitem[Schwartz et~al.(2020)Schwartz, Dodge, Smith, and Etzioni]{schwartz_green_2020}
R.~Schwartz, J.~Dodge, N.~A. Smith, and O.~Etzioni.
\newblock Green {AI}.
\newblock \emph{Communications of the ACM}, 63\penalty0 (12):\penalty0 54--63, Nov. 2020.
\newblock ISSN 0001-0782, 1557-7317.
\newblock \doi{10.1145/3381831}.
\newblock URL \url{https://dl.acm.org/doi/10.1145/3381831}.

\bibitem[Schwerin(2023)]{schwerin_somehow_2023}
M.~Schwerin.
\newblock Somehow, {Airline} {Customer} {Service} {Is} {Getting} {Even} {Worse}, June 2023.
\newblock URL \url{https://www.theatlantic.com/technology/archive/2023/06/airline-customer-service-chatbot-ai/674412/}.

\bibitem[Schäfer(2023)]{schafer_notorious_2023}
M.~S. Schäfer.
\newblock The {Notorious} {GPT}: science communication in the age of artificial intelligence.
\newblock \emph{Journal of Science Communication}, 22\penalty0 (02), May 2023.
\newblock ISSN 1824-2049.
\newblock \doi{10.22323/2.22020402}.
\newblock URL \url{https://jcom.sissa.it/article/pubid/JCOM_2202_2023_Y02/}.

\bibitem[Seger et~al.(2020)Seger, Avin, Pearson, Briers, Ó~Heigeartaigh, and Bacon]{seger_tackling_2020}
E.~Seger, S.~Avin, G.~Pearson, M.~Briers, S.~Ó~Heigeartaigh, and H.~Bacon.
\newblock Tackling threats to informed decision-making in democratic societies: {Promoting} epistemic security in a technologically-advanced world.
\newblock Technical report, The Alan Turing Institute, Oct. 2020.
\newblock URL \url{https://www.repository.cam.ac.uk/handle/1810/317073}.

\bibitem[Seger et~al.(2023)Seger, Ovadya, Siddarth, Garfinkel, and Dafoe]{seger_democratising_2023}
E.~Seger, A.~Ovadya, D.~Siddarth, B.~Garfinkel, and A.~Dafoe.
\newblock Democratising {AI}: {Multiple} {Meanings}, {Goals}, and {Methods}.
\newblock In \emph{Proceedings of the 2023 {AAAI}/{ACM} {Conference} on {AI}, {Ethics}, and {Society}}, pages 715--722, Montr{\textbackslash}'\{e\}al QC Canada, Aug. 2023. ACM.
\newblock ISBN 9798400702310.
\newblock \doi{10.1145/3600211.3604693}.
\newblock URL \url{https://dl.acm.org/doi/10.1145/3600211.3604693}.

\bibitem[Selbst et~al.(2019)Selbst, Boyd, Friedler, Venkatasubramanian, and Vertesi]{Selbst_Boyd_Friedler_Venkatasubramanian_Vertesi_2019}
A.~D. Selbst, D.~Boyd, S.~A. Friedler, S.~Venkatasubramanian, and J.~Vertesi.
\newblock Fairness and abstraction in sociotechnical systems.
\newblock In \emph{Proceedings of the Conference on Fairness, Accountability, and Transparency}, FAT* ’19, page 59–68, New York, NY, USA, Jan. 2019. Association for Computing Machinery.
\newblock ISBN 9781450361255.
\newblock \doi{10.1145/3287560.3287598}.
\newblock URL \url{https://dl.acm.org/doi/10.1145/3287560.3287598}.

\bibitem[Sen(2001)]{sen2001development}
A.~Sen.
\newblock \emph{Development as freedom}.
\newblock Oxford Paperbacks, 2001.

\bibitem[Sevilla et~al.(2022)Sevilla, Heim, Ho, Besiroglu, Hobbhahn, and Villalobos]{sevilla_compute_2022}
J.~Sevilla, L.~Heim, A.~Ho, T.~Besiroglu, M.~Hobbhahn, and P.~Villalobos.
\newblock Compute {Trends} {Across} {Three} {Eras} of {Machine} {Learning}, Mar. 2022.
\newblock URL \url{http://arxiv.org/abs/2202.05924}.
\newblock arXiv:2202.05924 [cs].

\bibitem[Seymour et~al.(2023)Seymour, Zhan, Cote, and Such]{seymour_systematic_2023}
W.~Seymour, X.~Zhan, M.~Cote, and J.~Such.
\newblock A {Systematic} {Review} of {Ethical} {Concerns} with {Voice} {Assistants}.
\newblock In \emph{Proceedings of the 2023 {AAAI}/{ACM} {Conference} on {AI}, {Ethics}, and {Society}}, pages 131--145, Aug. 2023.
\newblock \doi{10.1145/3600211.3604679}.
\newblock URL \url{http://arxiv.org/abs/2211.04193}.
\newblock arXiv:2211.04193 [cs].

\bibitem[Shah(2018)]{shah2018algorithmic}
H.~Shah.
\newblock Algorithmic accountability.
\newblock \emph{Philosophical Transactions of the Royal Society A: Mathematical, Physical and Engineering Sciences}, 376\penalty0 (2128):\penalty0 20170362, 2018.

\bibitem[Shah(2022)]{shah_ai_nodate}
R.~Shah.
\newblock {AI} {Risk} from {Program} {Search} in {Threat} {Model} {Literature} {Review}, 2022.
\newblock URL \url{https://www.alignmentforum.org/posts/wnnkD6P2k2TfHnNmt/threat-model-literature-review}.

\bibitem[Shah et~al.(2022)Shah, Varma, Kumar, Phuong, Krakovna, Uesato, and Kenton]{shah_goal_2022}
R.~Shah, V.~Varma, R.~Kumar, M.~Phuong, V.~Krakovna, J.~Uesato, and Z.~Kenton.
\newblock Goal {Misgeneralization}: {Why} {Correct} {Specifications} {Aren}'t {Enough} {For} {Correct} {Goals}, Nov. 2022.
\newblock URL \url{http://arxiv.org/abs/2210.01790}.
\newblock arXiv:2210.01790 [cs].

\bibitem[Shaib et~al.(2023)Shaib, Li, Joseph, Marshall, Li, and Wallace]{shaib_summarizing_2023}
C.~Shaib, M.~L. Li, S.~Joseph, I.~J. Marshall, J.~J. Li, and B.~C. Wallace.
\newblock Summarizing, {Simplifying}, and {Synthesizing} {Medical} {Evidence} {Using} {GPT}-3 (with {Varying} {Success}), May 2023.
\newblock URL \url{http://arxiv.org/abs/2305.06299}.
\newblock arXiv:2305.06299 [cs].

\bibitem[Shanahan(2024)]{shanahan_talking_2023}
M.~Shanahan.
\newblock Talking about large language models.
\newblock \emph{Communications of the ACM}, 67\penalty0 (2):\penalty0 68--79, 2024.
\newblock URL \url{http://dl.acm.org/doi/10.1145/3624724}.

\bibitem[Shanahan et~al.(2023)Shanahan, McDonell, and Reynolds]{shanahan2023role}
M.~Shanahan, K.~McDonell, and L.~Reynolds.
\newblock Role play with large language models.
\newblock \emph{Nature}, pages 1--6, 2023.

\bibitem[Shankland(2023)]{shankland_photoshops_2023}
S.~Shankland.
\newblock Photoshop's {Firefly} {Generative} {AI} {Arrives} {With} a {Creative} {Cloud} {Price} {Hike}.
\newblock \emph{CNET}, Sept. 2023.
\newblock URL \url{https://www.cnet.com/tech/computing/photoshops-firefly-generative-ai-arrives-with-a-creative-cloud-price-hike/}.

\bibitem[Shardlow and Przybyła(2023)]{shardlow_deanthropomorphising_2023}
M.~Shardlow and P.~Przybyła.
\newblock Deanthropomorphising {NLP}: {Can} a {Language} {Model} {Be} {Conscious}?, Nov. 2023.
\newblock URL \url{http://arxiv.org/abs/2211.11483}.
\newblock arXiv:2211.11483 [cs].

\bibitem[Shavit et~al.(2023)Shavit, Agarwal, Brundage, Adler, O’Keefe, Campbell, Lee, Mishkin, Eloundou, Hickey, Slama, Ahmad, McMillan, Beutel, Passos, and Robinson]{shavit2023practices}
Y.~Shavit, S.~Agarwal, M.~Brundage, S.~Adler, C.~O’Keefe, R.~Campbell, T.~Lee, P.~Mishkin, T.~Eloundou, A.~Hickey, K.~Slama, L.~Ahmad, P.~McMillan, A.~Beutel, A.~Passos, and D.~G. Robinson.
\newblock Practices for governing agentic {AI} systems.
\newblock \url{https://cdn.openai.com/papers/practices-for-governing-agentic-ai-systems.pdf}, 2023.
\newblock Accessed: 2023-01-04.

\bibitem[Shaw(2022)]{shaw2022moderators}
J.~Shaw.
\newblock Content moderators pay a psychological toll to keep social media clean. {We} should be helping them.
\newblock \url{https://www.sciencefocus.com/news/content-moderators-pay-a-psychological-toll-to-keep-social-media-clean-we-should-be-helping-them}, 2022.
\newblock Accessed: 2023-01-04.

\bibitem[Sheehan et~al.(2021)Sheehan, Friesen, Balmer, Cheeks, Davidson, Devereux, Findlay, Keats{-}Rohan, Lawrence, and Shafiq]{sheehan2021trust}
M.~Sheehan, P.~Friesen, A.~Balmer, C.~Cheeks, S.~Davidson, J.~Devereux, D.~Findlay, K.~Keats{-}Rohan, R.~Lawrence, and K.~Shafiq.
\newblock Trust, trustworthiness and sharing patient data for research.
\newblock \emph{Journal of Medical Ethics}, 47\penalty0 (12):\penalty0 26--26, 2021.
\newblock \doi{10.1136/medethics-2019-106048}.

\bibitem[Shelby et~al.(2023)Shelby, Rismani, Henne, Moon, Rostamzadeh, Nicholas, Yilla-Akbari, Gallegos, Smart, Garcia, and Virk]{shelby_sociotechnical_2023}
R.~Shelby, S.~Rismani, K.~Henne, A.~Moon, N.~Rostamzadeh, P.~Nicholas, N.~Yilla-Akbari, J.~Gallegos, A.~Smart, E.~Garcia, and G.~Virk.
\newblock Sociotechnical {Harms} of {Algorithmic} {Systems}: {Scoping} a {Taxonomy} for {Harm} {Reduction}.
\newblock In \emph{Proceedings of the 2023 {AAAI}/{ACM} {Conference} on {AI}, {Ethics}, and {Society}}, pages 723--741, Montr{\textbackslash}'\{e\}al QC Canada, Aug. 2023. ACM.
\newblock ISBN 9798400702310.
\newblock \doi{10.1145/3600211.3604673}.
\newblock URL \url{https://dl.acm.org/doi/10.1145/3600211.3604673}.

\bibitem[Shen et~al.(2023)Shen, Chen, Backes, Shen, and Zhang]{shen_anything_2023}
X.~Shen, Z.~Chen, M.~Backes, Y.~Shen, and Y.~Zhang.
\newblock "{Do} {Anything} {Now}": {Characterizing} and {Evaluating} {In}-{The}-{Wild} {Jailbreak} {Prompts} on {Large} {Language} {Models}, Aug. 2023.
\newblock URL \url{https://arxiv.org/pdf/2308.03825.pdf}.
\newblock arXiv:2308.03825 [cs].

\bibitem[Shevlane et~al.(2023)Shevlane, Farquhar, Garfinkel, Phuong, Whittlestone, Leung, Kokotajlo, Marchal, Anderljung, Kolt, Ho, Siddarth, Avin, Hawkins, Kim, Gabriel, Bolina, Clark, Bengio, Christiano, and Dafoe]{shevlane_model_2023}
T.~Shevlane, S.~Farquhar, B.~Garfinkel, M.~Phuong, J.~Whittlestone, J.~Leung, D.~Kokotajlo, N.~Marchal, M.~Anderljung, N.~Kolt, L.~Ho, D.~Siddarth, S.~Avin, W.~Hawkins, B.~Kim, I.~Gabriel, V.~Bolina, J.~Clark, Y.~Bengio, P.~Christiano, and A.~Dafoe.
\newblock Model evaluation for extreme risks, Sept. 2023.
\newblock URL \url{http://arxiv.org/abs/2305.15324}.
\newblock arXiv:2305.15324 [cs].

\bibitem[Shevlin and Halina(2019)]{shevlin2019apply}
H.~Shevlin and M.~Halina.
\newblock Apply rich psychological terms in {AI} with care.
\newblock \emph{Nature Machine Intelligence}, 1\penalty0 (4):\penalty0 165--167, 2019.

\bibitem[Shiffrin and Mitchell(2023)]{shiffrin_probing_2023}
R.~Shiffrin and M.~Mitchell.
\newblock Probing the psychology of {AI} models.
\newblock \emph{Proceedings of the National Academy of Sciences}, 120\penalty0 (10):\penalty0 e2300963120, Mar. 2023.
\newblock ISSN 0027-8424, 1091-6490.
\newblock \doi{10.1073/pnas.2300963120}.
\newblock URL \url{https://pnas.org/doi/10.1073/pnas.2300963120}.

\bibitem[Shiffrin(2000)]{shiffrin2000paternalism}
S.~V. Shiffrin.
\newblock Paternalism, unconscionability doctrine, and accommodation.
\newblock \emph{Philosophy \& Public Affairs}, 29\penalty0 (3):\penalty0 205--250, 2000.

\bibitem[Shin and Kim(2020)]{shin_my_2020}
H.~I. Shin and J.~Kim.
\newblock My computer is more thoughtful than you: {Loneliness}, anthropomorphism and dehumanization.
\newblock \emph{Current Psychology}, 39\penalty0 (2):\penalty0 445--453, Apr. 2020.
\newblock ISSN 1046-1310, 1936-4733.
\newblock \doi{10.1007/s12144-018-9975-7}.
\newblock URL \url{http://link.springer.com/10.1007/s12144-018-9975-7}.

\bibitem[Shiramizu et~al.(2022)Shiramizu, Lee, Altenburg, Feinberg, and Jones]{shiramizu_role_2022}
V.~K.~M. Shiramizu, A.~J. Lee, D.~Altenburg, D.~R. Feinberg, and B.~C. Jones.
\newblock The role of valence, dominance, and pitch in perceptions of artificial intelligence ({AI}) conversational agents’ voices.
\newblock \emph{Scientific Reports}, 12\penalty0 (1):\penalty0 22479, Dec. 2022.
\newblock ISSN 2045-2322.
\newblock \doi{10.1038/s41598-022-27124-8}.
\newblock URL \url{https://www.nature.com/articles/s41598-022-27124-8}.

\bibitem[Siddharth et~al.(2022)Siddharth, Blessing, and Luo]{siddharth_natural_2022}
L.~Siddharth, L.~Blessing, and J.~Luo.
\newblock Natural language processing in-and-for design research.
\newblock \emph{Design Science}, 8:\penalty0 e21, 2022.
\newblock ISSN 2053-4701.
\newblock \doi{10.1017/dsj.2022.16}.
\newblock URL \url{https://www.cambridge.org/core/product/identifier/S2053470122000166/type/journal_article}.

\bibitem[Siegel and Bennett~Doty(2023)]{siegel_weapons_2023}
D.~Siegel and M.~Bennett~Doty.
\newblock Weapons of {Mass} {Disruption}: {Artificial} {Intelligence} and the {Production} of {Extremist} {Propaganda}, Feb. 2023.
\newblock URL \url{https://gnet-research.org/2023/02/17/weapons-of-mass-disruption-artificial-intelligence-and-the-production-of-extremist-propaganda/}.

\bibitem[Siemon et~al.(2022)Siemon, Strohmann, Khosrawi-Rad, Vreede, Elshan, and Meyer]{siemon_why_2022}
D.~Siemon, T.~Strohmann, B.~Khosrawi-Rad, T.~d. Vreede, E.~Elshan, and M.~Meyer.
\newblock Why {Do} {We} {Turn} to {Virtual} {Companions}? {A} {Text} {Mining} {Analysis} of {Replika} {Reviews}.
\newblock \emph{AMCIS 2022 Proceedings}, Aug. 2022.
\newblock URL \url{https://aisel.aisnet.org/amcis2022/sig_hci/sig_hci/10}.

\bibitem[Simmons(1999)]{simmons_justification_1999}
A.~J. Simmons.
\newblock Justification and {Legitimacy}.
\newblock \emph{Ethics}, 109\penalty0 (4):\penalty0 739--771, 1999.
\newblock \doi{10.1086/233944}.

\bibitem[Sims(2023)]{sims_blackmamba:_2023}
J.~Sims.
\newblock {BlackMamba}: {Using} {AI} to {Generate} {Polymorphic} {Malware}, Mar. 2023.
\newblock URL \url{https://securityboulevard.com/2023/03/blackmamba-using-ai-to-generate-polymorphic-malware/}.

\bibitem[Singel(2018)]{singel_filtering_2018}
R.~Singel.
\newblock Filtering {Out} the {Bots}: {What} {Americans} {Actually} {Told} the {FCC} about {Net} {Neutrality} {Repeal}.
\newblock Technical report, Stanford Law School: The Center for Internet and Society, Oct. 2018.
\newblock URL \url{https://cyberlaw.stanford.edu/sites/default/files/FilteringOutTheBotsUniqueNetNeutralityComments.pdf}.

\bibitem[Singer(2023)]{singer_applying_2023}
N.~Singer.
\newblock Applying to {College}? {Here}’s {How} {A}.{I}. {Tools} {Might} {Hurt}, or {Help}.
\newblock \emph{The New York Times}, Sept. 2023.
\newblock ISSN 0362-4331.
\newblock URL \url{https://www.nytimes.com/2023/09/01/technology/ai-chatbots-college-applications.html}.

\bibitem[Singer and Tse(2023)]{singer_ai_2023}
P.~Singer and Y.~F. Tse.
\newblock {AI} ethics: the case for including animals.
\newblock \emph{AI and Ethics}, 3\penalty0 (2):\penalty0 539--551, May 2023.
\newblock ISSN 2730-5961.
\newblock \doi{10.1007/s43681-022-00187-z}.
\newblock URL \url{https://doi.org/10.1007/s43681-022-00187-z}.

\bibitem[Singer et~al.(2022)Singer, Polyak, Hayes, Yin, An, Zhang, Hu, Yang, Ashual, Gafni, Parikh, Gupta, and Taigman]{singer_make--video:_2022}
U.~Singer, A.~Polyak, T.~Hayes, X.~Yin, J.~An, S.~Zhang, Q.~Hu, H.~Yang, O.~Ashual, O.~Gafni, D.~Parikh, S.~Gupta, and Y.~Taigman.
\newblock Make-{A}-{Video}: {Text}-to-{Video} {Generation} without {Text}-{Video} {Data}, Sept. 2022.
\newblock URL \url{http://arxiv.org/abs/2209.14792}.
\newblock arXiv:2209.14792 [cs].

\bibitem[Singh-Kurtz(2023)]{singh-kurtz_man_2023}
S.~Singh-Kurtz.
\newblock The {Man} of {Your} {Dreams}, Mar. 2023.
\newblock URL \url{https://www.thecut.com/article/ai-artificial-intelligence-chatbot-replika-boyfriend.html}.

\bibitem[Sirlin et~al.(2021)Sirlin, Epstein, Arechar, and Rand]{sirlin_digital_2021}
N.~Sirlin, Z.~Epstein, A.~A. Arechar, and D.~G. Rand.
\newblock Digital literacy is associated with more discerning accuracy judgments but not sharing intentions.
\newblock \emph{Harvard Kennedy School Misinformation Review}, Dec. 2021.
\newblock \doi{10.37016/mr-2020-83}.
\newblock URL \url{https://misinforeview.hks.harvard.edu/article/digital-literacy-is-associated-with-more-discerning-accuracy-judgments-but-not-sharing-intentions/}.

\bibitem[Skalse et~al.(2022)Skalse, Howe, Krasheninnikov, and Krueger]{skalse_defining_2022}
J.~Skalse, N.~H.~R. Howe, D.~Krasheninnikov, and D.~Krueger.
\newblock Defining and {Characterizing} {Reward} {Hacking}, Sept. 2022.
\newblock URL \url{http://arxiv.org/abs/2209.13085}.
\newblock arXiv:2209.13085 [cs, stat].

\bibitem[Skinner-Thompson(2020)]{skinner-thompson_privacy_2020}
S.~Skinner-Thompson.
\newblock \emph{Privacy at the {Margins}}.
\newblock Cambridge University Press, 1 edition, Nov. 2020.
\newblock ISBN 9781316850350 9781107181373 9781316632635.
\newblock \doi{10.1017/9781316850350}.
\newblock URL \url{https://www.cambridge.org/core/product/identifier/9781316850350/type/book}.

\bibitem[Skjuve et~al.(2021)Skjuve, Følstad, Fostervold, and Brandtzaeg]{skjuve_my_2021}
M.~Skjuve, A.~Følstad, K.~I. Fostervold, and P.~B. Brandtzaeg.
\newblock My {Chatbot} {Companion} - a {Study} of {Human}-{Chatbot} {Relationships}.
\newblock \emph{International Journal of Human-Computer Studies}, 149:\penalty0 102601, May 2021.
\newblock ISSN 10715819.
\newblock \doi{10.1016/j.ijhcs.2021.102601}.
\newblock URL \url{https://linkinghub.elsevier.com/retrieve/pii/S1071581921000197}.

\bibitem[Skjuve et~al.(2022)Skjuve, Følstad, Fostervold, and Brandtzaeg]{skjuve_longitudinal_2022}
M.~Skjuve, A.~Følstad, K.~I. Fostervold, and P.~B. Brandtzaeg.
\newblock A longitudinal study of human–chatbot relationships.
\newblock \emph{International Journal of Human-Computer Studies}, 168, Dec. 2022.
\newblock ISSN 10715819.
\newblock \doi{10.1016/j.ijhcs.2022.102903}.
\newblock URL \url{https://linkinghub.elsevier.com/retrieve/pii/S1071581922001252}.

\bibitem[{Slack}(2023)]{slack_claude_nodate}
{Slack}.
\newblock Claude, 2023.
\newblock URL \url{https://slack.com/apps/A04KGS7N9A8-claude}.

\bibitem[Sloane et~al.(2022)Sloane, Moss, Awomolo, and Forlano]{sloane_participation_2022}
M.~Sloane, E.~Moss, O.~Awomolo, and L.~Forlano.
\newblock Participation {Is} not a {Design} {Fix} for {Machine} {Learning}.
\newblock In \emph{Equity and {Access} in {Algorithms}, {Mechanisms}, and {Optimization}}, pages 1--6, Arlington VA USA, Oct. 2022. ACM.
\newblock ISBN 9781450394772.
\newblock \doi{10.1145/3551624.3555285}.
\newblock URL \url{https://dl.acm.org/doi/10.1145/3551624.3555285}.

\bibitem[Small et~al.(2023)Small, Vendrov, Durmus, Homaei, Barry, Cornebise, Suzman, Ganguli, and Megill]{small_opportunities_2023}
C.~T. Small, I.~Vendrov, E.~Durmus, H.~Homaei, E.~Barry, J.~Cornebise, T.~Suzman, D.~Ganguli, and C.~Megill.
\newblock Opportunities and {Risks} of {LLMs} for {Scalable} {Deliberation} with {Polis}, June 2023.
\newblock URL \url{http://arxiv.org/abs/2306.11932}.
\newblock arXiv:2306.11932 [cs].

\bibitem[Smit et~al.(2020)Smit, Tacke, Lund, Manyika, and Thiel]{smit_future_2020}
S.~Smit, T.~Tacke, S.~Lund, J.~Manyika, and L.~Thiel.
\newblock The future of work in {Europe}.
\newblock Technical report, McKinsey Global Institute, June 2020.
\newblock URL \url{https://www.mckinsey.com/~/media/mckinsey/featured%20insights/future%20of%20organizations/the%20future%20of%20work%20in%20europe/mgi-the-future-of-work-in-europe-discussion-paper.pdf}.

\bibitem[Smith and Kantor(2008)]{smith2008user}
C.~L. Smith and P.~B. Kantor.
\newblock User adaptation: {G}ood results from poor systems.
\newblock In \emph{Proceedings of the 31st Annual International ACM SIGIR Conference on Research and Development in Information Retrieval}, pages 147--154, 2008.

\bibitem[Smith et~al.(2023)Smith, Manzini, Kennedy, and Ives]{smith_ethics_2023}
H.~Smith, A.~Manzini, M.-R. Kennedy, and J.~Ives.
\newblock Ethics of {Trust}/worthiness in {Autonomous} {Systems}: a scoping review.
\newblock In \emph{Proceedings of the {First} {International} {Symposium} on {Trustworthy} {Autonomous} {Systems}}, pages 1--15, Edinburgh United Kingdom, July 2023. ACM.
\newblock ISBN 9798400707346.
\newblock \doi{10.1145/3597512.3600207}.
\newblock URL \url{https://dl.acm.org/doi/10.1145/3597512.3600207}.

\bibitem[Smith et~al.(2011)Smith, Dinev, and Xu]{smith_information_2011}
H.~J. Smith, T.~Dinev, and H.~Xu.
\newblock Information {Privacy} {Research}: {An} {Interdisciplinary} {Review}.
\newblock \emph{MIS Quarterly}, 35\penalty0 (4):\penalty0 989, 2011.
\newblock ISSN 02767783.
\newblock \doi{10.2307/41409970}.
\newblock URL \url{https://www.jstor.org/stable/10.2307/41409970}.

\bibitem[Smith et~al.(2013)Smith, Case, Smith, Harwell, and Summers]{smith2013relating}
L.~M. Smith, J.~L. Case, H.~M. Smith, L.~C. Harwell, and J.~Summers.
\newblock Relating ecoystem services to domains of human well-being: Foundation for a us index.
\newblock \emph{Ecological Indicators}, 28:\penalty0 79--90, 2013.

\bibitem[Snyder(2023)]{snyder_ais_2023}
A.~Snyder.
\newblock {AI}'s language gap.
\newblock \emph{Axios}, Sept. 2023.
\newblock URL \url{https://www.axios.com/2023/09/08/ai-language-gap-chatgpt}.

\bibitem[Solaiman(2023)]{solaiman_gradient_2023}
I.~Solaiman.
\newblock The {Gradient} of {Generative} {AI} {Release}: {Methods} and {Considerations}.
\newblock In \emph{2023 {ACM} {Conference} on {Fairness}, {Accountability}, and {Transparency}}, pages 111--122, Chicago IL USA, June 2023. ACM.
\newblock ISBN 9798400701924.
\newblock \doi{10.1145/3593013.3593981}.
\newblock URL \url{https://dl.acm.org/doi/10.1145/3593013.3593981}.

\bibitem[Solaiman et~al.(2023)Solaiman, Talat, Agnew, Ahmad, Baker, Blodgett, Daumé~III, Dodge, Evans, Hooker, Jernite, Luccioni, Lusoli, Mitchell, Newman, Png, Strait, and Vassilev]{solaiman_evaluating_2023}
I.~Solaiman, Z.~Talat, W.~Agnew, L.~Ahmad, D.~Baker, S.~L. Blodgett, H.~Daumé~III, J.~Dodge, E.~Evans, S.~Hooker, Y.~Jernite, A.~S. Luccioni, A.~Lusoli, M.~Mitchell, J.~Newman, M.-T. Png, A.~Strait, and A.~Vassilev.
\newblock Evaluating the {Social} {Impact} of {Generative} {AI} {Systems} in {Systems} and {Society}, June 2023.
\newblock URL \url{http://arxiv.org/abs/2306.05949}.
\newblock arXiv:2306.05949 [cs].

\bibitem[Sorensen(2023)]{sorensen_vagueness_2023}
R.~Sorensen.
\newblock Vagueness.
\newblock In E.~N. Zalta and U.~Nodelman, editors, \emph{The {Stanford} {Encyclopedia} of {Philosophy}}. Metaphysics Research Lab, Stanford University, winter 2023 edition, 2023.
\newblock URL \url{https://plato.stanford.edu/archives/win2023/entries/vagueness/}.

\bibitem[Spadaro et~al.(2020)Spadaro, Gangl, Van~Prooijen, Van~Lange, and Mosso]{spadaro2020enhancing}
G.~Spadaro, K.~Gangl, J.-W. Van~Prooijen, P.~A. Van~Lange, and C.~O. Mosso.
\newblock Enhancing feelings of security: How institutional trust promotes interpersonal trust.
\newblock \emph{PloS one}, 15\penalty0 (9):\penalty0 e0237934, 2020.

\bibitem[Spitale et~al.(2023)Spitale, Biller-Andorno, and Germani]{spitale_ai_2023}
G.~Spitale, N.~Biller-Andorno, and F.~Germani.
\newblock {AI} model {GPT}-3 (dis)informs us better than humans.
\newblock \emph{Science Advances}, 9\penalty0 (26):\penalty0 eadh1850, June 2023.
\newblock ISSN 2375-2548.
\newblock \doi{10.1126/sciadv.adh1850}.
\newblock URL \url{https://www.science.org/doi/10.1126/sciadv.adh1850}.

\bibitem[Srivastava et~al.(2023)Srivastava, Rastogi, Rao, Shoeb, Abid, Fisch, Brown, Santoro, Gupta, Garriga-Alonso, Kluska, Lewkowycz, Agarwal, Power, Ray, Warstadt, Kocurek, Safaya, Tazarv, Xiang, Parrish, Nie, Hussain, Askell, Dsouza, Slone, Rahane, Iyer, Andreassen, Madotto, Santilli, Stuhlmüller, Dai, La, Lampinen, Zou, Jiang, Chen, Vuong, Gupta, Gottardi, Norelli, Venkatesh, Gholamidavoodi, Tabassum, Menezes, Kirubarajan, Mullokandov, Sabharwal, Herrick, Efrat, Erdem, Karakaş, Roberts, Loe, Zoph, Bojanowski, Özyurt, Hedayatnia, Neyshabur, Inden, Stein, Ekmekci, Lin, Howald, Orinion, Diao, Dour, Stinson, Argueta, Ramírez, Singh, Rathkopf, Meng, Baral, Wu, Callison-Burch, Waites, Voigt, Manning, Potts, Ramirez, Rivera, Siro, Raffel, Ashcraft, Garbacea, Sileo, Garrette, Hendrycks, Kilman, Roth, Freeman, Khashabi, Levy, González, Perszyk, Hernandez, Chen, Ippolito, Gilboa, Dohan, Drakard, Jurgens, Datta, Ganguli, Emelin, Kleyko, Yuret, Chen, Tam, Hupkes, Misra, Buzan, Mollo, Yang, Lee, Schrader,
  Shutova, Cubuk, Segal, Hagerman, Barnes, Donoway, Pavlick, Rodola, Lam, Chu, Tang, Erdem, Chang, Chi, Dyer, Jerzak, Kim, Manyasi, Zheltonozhskii, Xia, Siar, Martínez-Plumed, Happé, Chollet, Rong, Mishra, Winata, de~Melo, Kruszewski, Parascandolo, Mariani, Wang, Jaimovitch-López, Betz, Gur-Ari, Galijasevic, Kim, Rashkin, Hajishirzi, Mehta, Bogar, Shevlin, Schütze, Yakura, Zhang, Wong, Ng, Noble, Jumelet, Geissinger, Kernion, Hilton, Lee, Fisac, Simon, Koppel, Zheng, Zou, Kocoń, Thompson, Wingfield, Kaplan, Radom, Sohl-Dickstein, Phang, Wei, Yosinski, Novikova, Bosscher, Marsh, Kim, Taal, Engel, Alabi, Xu, Song, Tang, Waweru, Burden, Miller, Balis, Batchelder, Berant, Frohberg, Rozen, Hernandez-Orallo, Boudeman, Guerr, Jones, Tenenbaum, Rule, Chua, Kanclerz, Livescu, Krauth, Gopalakrishnan, Ignatyeva, Markert, Dhole, Gimpel, Omondi, Mathewson, Chiafullo, Shkaruta, Shridhar, McDonell, Richardson, Reynolds, Gao, Zhang, Dugan, Qin, Contreras-Ochando, Morency, Moschella, Lam, Noble, Schmidt, He, Colón,
  Metz, Şenel, Bosma, Sap, ter Hoeve, Farooqi, Faruqui, Mazeika, Baturan, Marelli, Maru, Quintana, Tolkiehn, Giulianelli, Lewis, Potthast, Leavitt, Hagen, Schubert, Baitemirova, Arnaud, McElrath, Yee, Cohen, Gu, Ivanitskiy, Starritt, Strube, Swędrowski, Bevilacqua, Yasunaga, Kale, Cain, Xu, Suzgun, Walker, Tiwari, Bansal, Aminnaseri, Geva, Gheini, T, Peng, Chi, Lee, Krakover, Cameron, Roberts, Doiron, Martinez, Nangia, Deckers, Muennighoff, Keskar, Iyer, Constant, Fiedel, Wen, Zhang, Agha, Elbaghdadi, Levy, Evans, Casares, Doshi, Fung, Liang, Vicol, Alipoormolabashi, Liao, Liang, Chang, Eckersley, Htut, Hwang, Miłkowski, Patil, Pezeshkpour, Oli, Mei, Lyu, Chen, Banjade, Rudolph, Gabriel, Habacker, Risco, Millière, Garg, Barnes, Saurous, Arakawa, Raymaekers, Frank, Sikand, Novak, Sitelew, LeBras, Liu, Jacobs, Zhang, Salakhutdinov, Chi, Lee, Stovall, Teehan, Yang, Singh, Mohammad, Anand, Dillavou, Shleifer, Wiseman, Gruetter, Bowman, Schoenholz, Han, Kwatra, Rous, Ghazarian, Ghosh, Casey, Bischoff,
  Gehrmann, Schuster, Sadeghi, Hamdan, Zhou, Srivastava, Shi, Singh, Asaadi, Gu, Pachchigar, Toshniwal, Upadhyay, {Shyamolima}, {Debnath}, Shakeri, Thormeyer, Melzi, Reddy, Makini, Lee, Torene, Hatwar, Dehaene, Divic, Ermon, Biderman, Lin, Prasad, Piantadosi, Shieber, Misherghi, Kiritchenko, Mishra, Linzen, Schuster, Li, Yu, Ali, Hashimoto, Wu, Desbordes, Rothschild, Phan, Wang, Nkinyili, Schick, Kornev, Tunduny, Gerstenberg, Chang, Neeraj, Khot, Shultz, Shaham, Misra, Demberg, Nyamai, Raunak, Ramasesh, Prabhu, Padmakumar, Srikumar, Fedus, Saunders, Zhang, Vossen, Ren, Tong, Zhao, Wu, Shen, Yaghoobzadeh, Lakretz, Song, Bahri, Choi, Yang, Hao, Chen, Belinkov, Hou, Hou, Bai, Seid, Zhao, Wang, Wang, Wang, and Wu]{srivastava_beyond_2023}
A.~Srivastava, A.~Rastogi, A.~Rao, A.~A.~M. Shoeb, A.~Abid, A.~Fisch, A.~R. Brown, A.~Santoro, A.~Gupta, A.~Garriga-Alonso, A.~Kluska, A.~Lewkowycz, A.~Agarwal, A.~Power, A.~Ray, A.~Warstadt, A.~W. Kocurek, A.~Safaya, A.~Tazarv, A.~Xiang, A.~Parrish, A.~Nie, A.~Hussain, A.~Askell, A.~Dsouza, A.~Slone, A.~Rahane, A.~S. Iyer, A.~Andreassen, A.~Madotto, A.~Santilli, A.~Stuhlmüller, A.~Dai, A.~La, A.~Lampinen, A.~Zou, A.~Jiang, A.~Chen, A.~Vuong, A.~Gupta, A.~Gottardi, A.~Norelli, A.~Venkatesh, A.~Gholamidavoodi, A.~Tabassum, A.~Menezes, A.~Kirubarajan, A.~Mullokandov, A.~Sabharwal, A.~Herrick, A.~Efrat, A.~Erdem, A.~Karakaş, B.~R. Roberts, B.~S. Loe, B.~Zoph, B.~Bojanowski, B.~Özyurt, B.~Hedayatnia, B.~Neyshabur, B.~Inden, B.~Stein, B.~Ekmekci, B.~Y. Lin, B.~Howald, B.~Orinion, C.~Diao, C.~Dour, C.~Stinson, C.~Argueta, C.~F. Ramírez, C.~Singh, C.~Rathkopf, C.~Meng, C.~Baral, C.~Wu, C.~Callison-Burch, C.~Waites, C.~Voigt, C.~D. Manning, C.~Potts, C.~Ramirez, C.~E. Rivera, C.~Siro, C.~Raffel, C.~Ashcraft,
  C.~Garbacea, D.~Sileo, D.~Garrette, D.~Hendrycks, D.~Kilman, D.~Roth, D.~Freeman, D.~Khashabi, D.~Levy, D.~M. González, D.~Perszyk, D.~Hernandez, D.~Chen, D.~Ippolito, D.~Gilboa, D.~Dohan, D.~Drakard, D.~Jurgens, D.~Datta, D.~Ganguli, D.~Emelin, D.~Kleyko, D.~Yuret, D.~Chen, D.~Tam, D.~Hupkes, D.~Misra, D.~Buzan, D.~C. Mollo, D.~Yang, D.-H. Lee, D.~Schrader, E.~Shutova, E.~D. Cubuk, E.~Segal, E.~Hagerman, E.~Barnes, E.~Donoway, E.~Pavlick, E.~Rodola, E.~Lam, E.~Chu, E.~Tang, E.~Erdem, E.~Chang, E.~A. Chi, E.~Dyer, E.~Jerzak, E.~Kim, E.~E. Manyasi, E.~Zheltonozhskii, F.~Xia, F.~Siar, F.~Martínez-Plumed, F.~Happé, F.~Chollet, F.~Rong, G.~Mishra, G.~I. Winata, G.~de~Melo, G.~Kruszewski, G.~Parascandolo, G.~Mariani, G.~Wang, G.~Jaimovitch-López, G.~Betz, G.~Gur-Ari, H.~Galijasevic, H.~Kim, H.~Rashkin, H.~Hajishirzi, H.~Mehta, H.~Bogar, H.~Shevlin, H.~Schütze, H.~Yakura, H.~Zhang, H.~M. Wong, I.~Ng, I.~Noble, J.~Jumelet, J.~Geissinger, J.~Kernion, J.~Hilton, J.~Lee, J.~F. Fisac, J.~B. Simon, J.~Koppel,
  J.~Zheng, J.~Zou, J.~Kocoń, J.~Thompson, J.~Wingfield, J.~Kaplan, J.~Radom, J.~Sohl-Dickstein, J.~Phang, J.~Wei, J.~Yosinski, J.~Novikova, J.~Bosscher, J.~Marsh, J.~Kim, J.~Taal, J.~Engel, J.~Alabi, J.~Xu, J.~Song, J.~Tang, J.~Waweru, J.~Burden, J.~Miller, J.~U. Balis, J.~Batchelder, J.~Berant, J.~Frohberg, J.~Rozen, J.~Hernandez-Orallo, J.~Boudeman, J.~Guerr, J.~Jones, J.~B. Tenenbaum, J.~S. Rule, J.~Chua, K.~Kanclerz, K.~Livescu, K.~Krauth, K.~Gopalakrishnan, K.~Ignatyeva, K.~Markert, K.~D. Dhole, K.~Gimpel, K.~Omondi, K.~Mathewson, K.~Chiafullo, K.~Shkaruta, K.~Shridhar, K.~McDonell, K.~Richardson, L.~Reynolds, L.~Gao, L.~Zhang, L.~Dugan, L.~Qin, L.~Contreras-Ochando, L.-P. Morency, L.~Moschella, L.~Lam, L.~Noble, L.~Schmidt, L.~He, L.~O. Colón, L.~Metz, L.~K. Şenel, M.~Bosma, M.~Sap, M.~ter Hoeve, M.~Farooqi, M.~Faruqui, M.~Mazeika, M.~Baturan, M.~Marelli, M.~Maru, M.~J.~R. Quintana, M.~Tolkiehn, M.~Giulianelli, M.~Lewis, M.~Potthast, M.~L. Leavitt, M.~Hagen, M.~Schubert, M.~O. Baitemirova,
  M.~Arnaud, M.~McElrath, M.~A. Yee, M.~Cohen, M.~Gu, M.~Ivanitskiy, M.~Starritt, M.~Strube, M.~Swędrowski, M.~Bevilacqua, M.~Yasunaga, M.~Kale, M.~Cain, M.~Xu, M.~Suzgun, M.~Walker, M.~Tiwari, M.~Bansal, M.~Aminnaseri, M.~Geva, M.~Gheini, M.~V. T, N.~Peng, N.~A. Chi, N.~Lee, N.~G.-A. Krakover, N.~Cameron, N.~Roberts, N.~Doiron, N.~Martinez, N.~Nangia, N.~Deckers, N.~Muennighoff, N.~S. Keskar, N.~S. Iyer, N.~Constant, N.~Fiedel, N.~Wen, O.~Zhang, O.~Agha, O.~Elbaghdadi, O.~Levy, O.~Evans, P.~A.~M. Casares, P.~Doshi, P.~Fung, P.~P. Liang, P.~Vicol, P.~Alipoormolabashi, P.~Liao, P.~Liang, P.~Chang, P.~Eckersley, P.~M. Htut, P.~Hwang, P.~Miłkowski, P.~Patil, P.~Pezeshkpour, P.~Oli, Q.~Mei, Q.~Lyu, Q.~Chen, R.~Banjade, R.~E. Rudolph, R.~Gabriel, R.~Habacker, R.~Risco, R.~Millière, R.~Garg, R.~Barnes, R.~A. Saurous, R.~Arakawa, R.~Raymaekers, R.~Frank, R.~Sikand, R.~Novak, R.~Sitelew, R.~LeBras, R.~Liu, R.~Jacobs, R.~Zhang, R.~Salakhutdinov, R.~Chi, R.~Lee, R.~Stovall, R.~Teehan, R.~Yang, S.~Singh, S.~M.
  Mohammad, S.~Anand, S.~Dillavou, S.~Shleifer, S.~Wiseman, S.~Gruetter, S.~R. Bowman, S.~S. Schoenholz, S.~Han, S.~Kwatra, S.~A. Rous, S.~Ghazarian, S.~Ghosh, S.~Casey, S.~Bischoff, S.~Gehrmann, S.~Schuster, S.~Sadeghi, S.~Hamdan, S.~Zhou, S.~Srivastava, S.~Shi, S.~Singh, S.~Asaadi, S.~S. Gu, S.~Pachchigar, S.~Toshniwal, S.~Upadhyay, {Shyamolima}, {Debnath}, S.~Shakeri, S.~Thormeyer, S.~Melzi, S.~Reddy, S.~P. Makini, S.-H. Lee, S.~Torene, S.~Hatwar, S.~Dehaene, S.~Divic, S.~Ermon, S.~Biderman, S.~Lin, S.~Prasad, S.~T. Piantadosi, S.~M. Shieber, S.~Misherghi, S.~Kiritchenko, S.~Mishra, T.~Linzen, T.~Schuster, T.~Li, T.~Yu, T.~Ali, T.~Hashimoto, T.-L. Wu, T.~Desbordes, T.~Rothschild, T.~Phan, T.~Wang, T.~Nkinyili, T.~Schick, T.~Kornev, T.~Tunduny, T.~Gerstenberg, T.~Chang, T.~Neeraj, T.~Khot, T.~Shultz, U.~Shaham, V.~Misra, V.~Demberg, V.~Nyamai, V.~Raunak, V.~Ramasesh, V.~U. Prabhu, V.~Padmakumar, V.~Srikumar, W.~Fedus, W.~Saunders, W.~Zhang, W.~Vossen, X.~Ren, X.~Tong, X.~Zhao, X.~Wu, X.~Shen,
  Y.~Yaghoobzadeh, Y.~Lakretz, Y.~Song, Y.~Bahri, Y.~Choi, Y.~Yang, Y.~Hao, Y.~Chen, Y.~Belinkov, Y.~Hou, Y.~Hou, Y.~Bai, Z.~Seid, Z.~Zhao, Z.~Wang, Z.~J. Wang, Z.~Wang, and Z.~Wu.
\newblock Beyond the {Imitation} {Game}: {Quantifying} and extrapolating the capabilities of language models, June 2023.
\newblock URL \url{http://arxiv.org/abs/2206.04615}.
\newblock arXiv:2206.04615 [cs, stat].

\bibitem[{Stack Overflow}(2023)]{stack_overflow_stack_2023}
{Stack Overflow}.
\newblock Stack {Overflow} {Developer} {Survey} 2023, 2023.
\newblock URL \url{https://survey.stackoverflow.co/2023/?utm_source=social-share&utm_medium=social&utm_campaign=dev-survey-2023}.

\bibitem[Stackelberg(1934)]{stackelberg_marktform_1934}
H.~v. Stackelberg.
\newblock \emph{Marktform und {Gleichgewicht}}.
\newblock J. Springer, 1934.
\newblock Google-Books-ID: wihBAAAAIAAJ.

\bibitem[Starke and Ienca(2022)]{starke_misplaced_2022}
G.~Starke and M.~Ienca.
\newblock Misplaced {Trust} and {Distrust}: {How} {Not} to {Engage} with {Medical} {Artificial} {Intelligence}.
\newblock \emph{Cambridge Quarterly of Healthcare Ethics}, pages 1--10, Oct. 2022.
\newblock ISSN 0963-1801, 1469-2147.
\newblock \doi{10.1017/S0963180122000445}.
\newblock URL \url{https://www.cambridge.org/core/product/identifier/S0963180122000445/type/journal_article}.

\bibitem[Starkman(2021)]{starkman_stochastic_2021}
R.~Starkman.
\newblock Stochastic {Parrots} and {Shaky} {Foundations}, Aug. 2021.
\newblock URL \url{https://ruth.substack.com/p/stochastic-parrots-and-shaky-foundations}.

\bibitem[{State of California}(2018)]{state_of_california_bots:_2018}
{State of California}.
\newblock Bots: disclosure, Sept. 2018.
\newblock URL \url{https://leginfo.legislature.ca.gov/faces/billTextClient.xhtml?bill_id=201720180SB1001}.

\bibitem[Steca et~al.(2016)Steca, Monzani, Greco, D’Addario, Cappelletti, and Pancani]{steca2016effects}
P.~Steca, D.~Monzani, A.~Greco, M.~D’Addario, E.~Cappelletti, and L.~Pancani.
\newblock The effects of short-term personal goals on subjective well-being.
\newblock \emph{Journal of Happiness Studies}, 17:\penalty0 1435--1450, 2016.

\bibitem[Stiennon et~al.(2022)Stiennon, Ouyang, Wu, Ziegler, Lowe, Voss, Radford, Amodei, and Christiano]{stiennon_learning_2022}
N.~Stiennon, L.~Ouyang, J.~Wu, D.~M. Ziegler, R.~Lowe, C.~Voss, A.~Radford, D.~Amodei, and P.~Christiano.
\newblock Learning to summarize from human feedback, Feb. 2022.
\newblock URL \url{http://arxiv.org/abs/2009.01325}.
\newblock arXiv:2009.01325 [cs].

\bibitem[Stilgoe et~al.(2013)Stilgoe, Owen, and Macnaghten]{Stilgoe_Owen_Macnaghten_2013}
J.~Stilgoe, R.~Owen, and P.~Macnaghten.
\newblock Developing a framework for responsible innovation.
\newblock \emph{Research Policy}, 42\penalty0 (9):\penalty0 1568–1580, Nov. 2013.
\newblock ISSN 00487333.
\newblock \doi{10.1016/j.respol.2013.05.008}.
\newblock URL \url{https://linkinghub.elsevier.com/retrieve/pii/S0048733313000930}.

\bibitem[Stokel-Walker(2023)]{stokel-walker_generative_2023}
C.~Stokel-Walker.
\newblock The {Generative} {AI} {Race} {Has} a {Dirty} {Secret}.
\newblock \emph{Wired UK}, Feb. 2023.
\newblock ISSN 1357-0978.
\newblock URL \url{https://www.wired.co.uk/article/the-generative-ai-search-race-has-a-dirty-secret}.

\bibitem[Straten et~al.(2020)Straten, Peter, K{\"u}hne, and Barco]{straten_transparency_2020}
C.~L.~v. Straten, J.~Peter, R.~K{\"u}hne, and A.~Barco.
\newblock Transparency about a robot's lack of human psychological capacities: effects on child-robot perception and relationship formation.
\newblock \emph{ACM Transactions on Human-Robot Interaction (THRI)}, 9\penalty0 (2):\penalty0 1--22, 2020.

\bibitem[Strawson(1992)]{strawson1992analysis}
P.~F. Strawson.
\newblock \emph{Analysis and metaphysics: An introduction to philosophy}.
\newblock Oxford University Press, USA, 1992.

\bibitem[Stray(2020)]{stray2020aligning}
J.~Stray.
\newblock Aligning {AI} optimization to community well-being.
\newblock \emph{International Journal of Community Well-Being}, 3\penalty0 (4):\penalty0 443--463, 2020.

\bibitem[Stray et~al.(2021)Stray, Vendrov, Nixon, Adler, and Hadfield-Menell]{stray2021you}
J.~Stray, I.~Vendrov, J.~Nixon, S.~Adler, and D.~Hadfield-Menell.
\newblock What are you optimizing for? {A}ligning recommender systems with human values.
\newblock \emph{arXiv preprint arXiv:2107.10939}, 2021.

\bibitem[Stray et~al.(2022)Stray, Halevy, Assar, Hadfield-Menell, Boutilier, Ashar, Bakalar, Beattie, Ekstrand, Leibowicz, et~al.]{stray2022building}
J.~Stray, A.~Halevy, P.~Assar, D.~Hadfield-Menell, C.~Boutilier, A.~Ashar, C.~Bakalar, L.~Beattie, M.~Ekstrand, C.~Leibowicz, et~al.
\newblock Building human values into recommender systems: An interdisciplinary synthesis.
\newblock \emph{ACM Transactions on Recommender Systems}, 2022.

\bibitem[Strouse et~al.(2021)Strouse, McKee, Botvinick, Hughes, and Everett]{strouse_collaborating_2021}
D.~Strouse, K.~McKee, M.~Botvinick, E.~Hughes, and R.~Everett.
\newblock Collaborating with {Humans} without {Human} {Data}.
\newblock In \emph{Advances in {Neural} {Information} {Processing} {Systems}}, volume~34, pages 14502--14515. Curran Associates, Inc., 2021.
\newblock URL \url{https://proceedings.neurips.cc/paper/2021/hash/797134c3e42371bb4979a462eb2f042a-Abstract.html}.

\bibitem[Strubell et~al.(2019)Strubell, Ganesh, and McCallum]{strubell2019energy}
E.~Strubell, A.~Ganesh, and A.~McCallum.
\newblock Energy and policy considerations for deep learning in nlp.
\newblock \emph{arXiv preprint arXiv:1906.02243}, 2019.

\bibitem[Strubell et~al.(2020)Strubell, Ganesh, and McCallum]{strubell_energy_2020}
E.~Strubell, A.~Ganesh, and A.~McCallum.
\newblock Energy and {Policy} {Considerations} for {Modern} {Deep} {Learning} {Research}.
\newblock \emph{Proceedings of the AAAI Conference on Artificial Intelligence}, 34\penalty0 (09):\penalty0 13693--13696, Apr. 2020.
\newblock ISSN 2374-3468, 2159-5399.
\newblock \doi{10.1609/aaai.v34i09.7123}.
\newblock URL \url{https://ojs.aaai.org/index.php/AAAI/article/view/7123}.

\bibitem[Strudler(2005)]{strudler_deception_2005}
A.~Strudler.
\newblock Deception {Unraveled}.
\newblock \emph{The Journal of Philosophy}, 102\penalty0 (9):\penalty0 458--473, 2005.
\newblock ISSN 0022-362X.
\newblock URL \url{https://www.jstor.org/stable/3655633}.

\bibitem[Suckiel(2006)]{shook_william_2006}
E.~K. Suckiel.
\newblock William {James}.
\newblock In J.~R. Shook and J.~Margolis, editors, \emph{A {Companion} to {Pragmatism}}, pages 30--43. Blackwell Publishing Ltd, Oxford, UK, Jan. 2006.
\newblock ISBN 9780470997079 9781405116213.
\newblock \doi{10.1002/9780470997079.ch3}.
\newblock URL \url{https://onlinelibrary.wiley.com/doi/10.1002/9780470997079.ch3}.

\bibitem[Sullivan and Transue(1999)]{sullivan_psychological_1999}
J.~L. Sullivan and J.~E. Transue.
\newblock The {Psychological} {Underpinnings} of {Democracy}: {A} {Selective} {Review} of {Research} on {Political} {Tolerance}, {Interpersonal} {Trust}, and {Social} {Capital}.
\newblock \emph{Annual Review of Psychology}, 50\penalty0 (1):\penalty0 625--650, Feb. 1999.
\newblock ISSN 0066-4308, 1545-2085.
\newblock \doi{10.1146/annurev.psych.50.1.625}.
\newblock URL \url{https://www.annualreviews.org/doi/10.1146/annurev.psych.50.1.625}.

\bibitem[Summers et~al.(2014)Summers, Smith, Harwell, Case, Wade, Straub, and Smith]{summers2014index}
J.~K. Summers, L.~M. Smith, L.~C. Harwell, J.~L. Case, C.~M. Wade, K.~R. Straub, and H.~M. Smith.
\newblock An index of human well-being for the us: {A} {TRIO} approach.
\newblock \emph{Sustainability}, 6\penalty0 (6):\penalty0 3915--3935, 2014.

\bibitem[Summers-Stay et~al.(2023)Summers-Stay, Voss, and Lukin]{summers-stay_brainstorm_2023}
D.~Summers-Stay, C.~R. Voss, and S.~M. Lukin.
\newblock Brainstorm, then {Select}: a {Generative} {Language} {Model} {Improves} {Its} {Creativity} {Score}.
\newblock In \emph{The {AAAI}-23 {Workshop} on {Creative} {AI} {Across} {Modalities}}, 2023.
\newblock URL \url{https://openreview.net/forum?id=8HwKaJ1wvl}.

\bibitem[Sundar et~al.(2021)Sundar, Molina, and Cho]{sundar_seeing_2021}
S.~S. Sundar, M.~D. Molina, and E.~Cho.
\newblock Seeing {Is} {Believing}: {Is} {Video} {Modality} {More} {Powerful} in {Spreading} {Fake} {News} via {Online} {Messaging} {Apps}?
\newblock \emph{Journal of Computer-Mediated Communication}, 26\penalty0 (6):\penalty0 301--319, Nov. 2021.
\newblock ISSN 1083-6101.
\newblock \doi{10.1093/jcmc/zmab010}.
\newblock URL \url{https://academic.oup.com/jcmc/article/26/6/301/6336055}.

\bibitem[Sunstein(2015)]{sunstein_nudges_2015}
C.~R. Sunstein.
\newblock Nudges {Do} {Not} {Undermine} {Human} {Agency}.
\newblock \emph{Journal of Consumer Policy}, 38\penalty0 (3):\penalty0 207--210, Sept. 2015.
\newblock ISSN 1573-0700.
\newblock \doi{10.1007/s10603-015-9289-1}.
\newblock URL \url{https://doi.org/10.1007/s10603-015-9289-1}.

\bibitem[Sunstein(2016)]{sunstein_ethics_2016}
C.~R. Sunstein.
\newblock \emph{The {Ethics} of {Influence}: {Government} in the {Age} of {Behavioral} {Science}}.
\newblock Cambridge University Press, Aug. 2016.
\newblock ISBN 9781107140707.
\newblock Google-Books-ID: TlvWDAAAQBAJ.

\bibitem[Suresh and Guttag(2021)]{suresh_framework_2021}
H.~Suresh and J.~Guttag.
\newblock A {Framework} for {Understanding} {Sources} of {Harm} throughout the {Machine} {Learning} {Life} {Cycle}.
\newblock In \emph{Equity and {Access} in {Algorithms}, {Mechanisms}, and {Optimization}}, pages 1--9, -- NY USA, Oct. 2021. ACM.
\newblock ISBN 9781450385534.
\newblock \doi{10.1145/3465416.3483305}.
\newblock URL \url{https://dl.acm.org/doi/10.1145/3465416.3483305}.

\bibitem[Suresh et~al.(2022)Suresh, Movva, Dogan, Bhargava, Cruxen, Cuba, Taurino, So, and D'Ignazio]{suresh_towards_2022}
H.~Suresh, R.~Movva, A.~L. Dogan, R.~Bhargava, I.~Cruxen, A.~M. Cuba, G.~Taurino, W.~So, and C.~D'Ignazio.
\newblock Towards {Intersectional} {Feminist} and {Participatory} {ML}: {A} {Case} {Study} in {Supporting} {Feminicide} {Counterdata} {Collection}.
\newblock In \emph{2022 {ACM} {Conference} on {Fairness}, {Accountability}, and {Transparency}}, pages 667--678, Seoul Republic of Korea, June 2022. ACM.
\newblock ISBN 9781450393522.
\newblock \doi{10.1145/3531146.3533132}.
\newblock URL \url{https://dl.acm.org/doi/10.1145/3531146.3533132}.

\bibitem[Susser et~al.(2019{\natexlab{a}})Susser, Roessler, and Nissenbaum]{susser_online_2019}
D.~Susser, B.~Roessler, and H.~Nissenbaum.
\newblock Online {Manipulation}: {Hidden} {Influences} in a {Digital} {World}.
\newblock \emph{Georgetown Law Technology Review}, 4\penalty0 (1):\penalty0 2--45, 2019{\natexlab{a}}.
\newblock URL \url{https://philarchive.org/archive/SUSOMH}.

\bibitem[Susser et~al.(2019{\natexlab{b}})Susser, Roessler, and Nissenbaum]{susser_technology_2019}
D.~Susser, B.~Roessler, and H.~Nissenbaum.
\newblock Technology, autonomy, and manipulation.
\newblock \emph{Internet Policy Review}, 8\penalty0 (2), June 2019{\natexlab{b}}.
\newblock ISSN 2197-6775.
\newblock \doi{10.14763/2019.2.1410}.
\newblock URL \url{https://policyreview.info/node/1410}.

\bibitem[Sutton(2019)]{sutton_bitter_2019}
R.~Sutton.
\newblock The {Bitter} {Lesson}, Mar. 2019.
\newblock URL \url{http://www.incompleteideas.net/IncIdeas/BitterLesson.html}.

\bibitem[Sweeney(2013)]{sweeney_discrimination_2013}
L.~Sweeney.
\newblock Discrimination in {Online} {Ad} {Delivery}: {Google} ads, black names and white names, racial discrimination, and click advertising.
\newblock \emph{Queue}, 11\penalty0 (3):\penalty0 10--29, Mar. 2013.
\newblock ISSN 1542-7730, 1542-7749.
\newblock \doi{10.1145/2460276.2460278}.
\newblock URL \url{https://dl.acm.org/doi/10.1145/2460276.2460278}.

\bibitem[{Swiss Re Institute}(2021)]{swiss_re_institute_economics_2021}
{Swiss Re Institute}.
\newblock The economics of climate change: no action not an option.
\newblock Technical report, Swiss Re Institute, Apr. 2021.
\newblock URL \url{https://www.swissre.com/dam/jcr:e73ee7c3-7f83-4c17-a2b8-8ef23a8d3312/swiss-re-institute-expertise-publication-economics-of-climate-change.pdf}.

\bibitem[Syed(2015)]{syed_black_2015}
M.~Syed.
\newblock \emph{Black {Box} {Thinking}: {Why} {Most} {People} {Never} {Learn} from {Their} {Mistakes}--{But} {Some} {Do}}.
\newblock Penguin, Nov. 2015.
\newblock ISBN 9781591848226.
\newblock Google-Books-ID: MrJPEAAAQBAJ.

\bibitem[Sætra(2020)]{saetra_parasitic_2020}
H.~S. Sætra.
\newblock The {Parasitic} {Nature} of {Social} {AI}: {Sharing} {Minds} with the {Mindless}.
\newblock \emph{Integrative Psychological and Behavioral Science}, 54\penalty0 (2):\penalty0 308--326, June 2020.
\newblock ISSN 1936-3567.
\newblock \doi{10.1007/s12124-020-09523-6}.
\newblock URL \url{https://doi.org/10.1007/s12124-020-09523-6}.

\bibitem[Tabachnyk and Nikolov(2022)]{tabachnyk_ml-enhanced_2022}
M.~Tabachnyk and S.~Nikolov.
\newblock {ML}-{Enhanced} {Code} {Completion} {Improves} {Developer} {Productivity}, July 2022.
\newblock URL \url{https://blog.research.google/2022/07/ml-enhanced-code-completion-improves.html}.

\bibitem[Tabassi(2023)]{tabassi_artificial_2023}
E.~Tabassi.
\newblock Artificial {Intelligence} {Risk} {Management} {Framework} ({AI} {RMF} 1.0).
\newblock Technical Report 100-1, National Institute of Standards and Technology (NIST), Gaithersburg, MD, Jan. 2023.
\newblock URL \url{http://nvlpubs.nist.gov/nistpubs/ai/NIST.AI.100-1.pdf}.

\bibitem[Tabassi et~al.(2019)Tabassi, Burns, Hadjimichael, Molina-Markham, and Sexton]{tabassi_taxonomy_2019}
E.~Tabassi, K.~J. Burns, M.~Hadjimichael, A.~D. Molina-Markham, and J.~T. Sexton.
\newblock A taxonomy and terminology of adversarial machine learning.
\newblock preprint, Oct. 2019.
\newblock URL \url{https://nvlpubs.nist.gov/nistpubs/ir/2019/NIST.IR.8269-draft.pdf}.

\bibitem[Tahaei et~al.(2023)Tahaei, Constantinides, Quercia, Kennedy, Muller, Stumpf, Liao, Baeza-Yates, Aroyo, Holbrook, Luger, Madaio, Blumenfeld, De-Arteaga, Vitak, and Olteanu]{tahaei_human-centered_2023}
M.~Tahaei, M.~Constantinides, D.~Quercia, S.~Kennedy, M.~Muller, S.~Stumpf, Q.~V. Liao, R.~Baeza-Yates, L.~Aroyo, J.~Holbrook, E.~Luger, M.~Madaio, I.~G. Blumenfeld, M.~De-Arteaga, J.~Vitak, and A.~Olteanu.
\newblock Human-{Centered} {Responsible} {Artificial} {Intelligence}: {Current} \& {Future} {Trends}.
\newblock In \emph{Extended {Abstracts} of the 2023 {CHI} {Conference} on {Human} {Factors} in {Computing} {Systems}}, pages 1--4, Hamburg Germany, Apr. 2023. ACM.
\newblock ISBN 9781450394222.
\newblock \doi{10.1145/3544549.3583178}.
\newblock URL \url{https://dl.acm.org/doi/10.1145/3544549.3583178}.

\bibitem[Taigman et~al.(2018)Taigman, Wolf, Polyak, and Nachmani]{taigman_voiceloop:_2018}
Y.~Taigman, L.~Wolf, A.~Polyak, and E.~Nachmani.
\newblock {VoiceLoop}: {Voice} {Fitting} and {Synthesis} via a {Phonological} {Loop}, Feb. 2018.
\newblock URL \url{http://arxiv.org/abs/1707.06588}.
\newblock arXiv:1707.06588 [cs].

\bibitem[Tameez(2020)]{tameez_youtubes_2020}
H.~Tameez.
\newblock {YouTube}’s algorithm is pushing climate misinformation videos, and their creators are profiting from it, Jan. 2020.
\newblock URL \url{https://www.niemanlab.org/2020/01/youtubes-algorithm-is-pushing-climate-misinformation-videos-and-their-creators-are-profiting-from-it/}.

\bibitem[Tamkin et~al.(2021)Tamkin, Brundage, Clark, and Ganguli]{tamkin_understanding_2021}
A.~Tamkin, M.~Brundage, J.~Clark, and D.~Ganguli.
\newblock Understanding the {Capabilities}, {Limitations}, and {Societal} {Impact} of {Large} {Language} {Models}, Feb. 2021.
\newblock URL \url{http://arxiv.org/abs/2102.02503}.
\newblock arXiv:2102.02503 [cs].

\bibitem[Tapu and Fa‘agau(2022)]{tapu_new_2022}
I.~F. Tapu and T.~K. Fa‘agau.
\newblock A {New} {Age} {Indigenous} {Instrument}: {Artificial} {Intelligence} \& {Its} {Potential} for ({De})colonialized {Data}.
\newblock \emph{Harvard Civil Rights–Civil Liberties Law Review}, 57\penalty0 (2), 2022.
\newblock URL \url{https://journals.law.harvard.edu/crcl/wp-content/uploads/sites/80/2023/01/ANewAgeIndigenousInstrument.pdf}.

\bibitem[Tatar et~al.(2022)Tatar, Shoorekchali, Faraji, Seyyedkolaee, Pagán, and Wilson]{tatar_covid-19_2022}
M.~Tatar, J.~M. Shoorekchali, M.~R. Faraji, M.~A. Seyyedkolaee, J.~A. Pagán, and F.~A. Wilson.
\newblock {COVID}-19 vaccine inequality: {A} global perspective.
\newblock \emph{Journal of Global Health}, 12:\penalty0 03072, Oct. 2022.
\newblock ISSN 2047-2978, 2047-2986.
\newblock \doi{10.7189/jogh.12.03072}.
\newblock URL \url{https://jogh.org/2022/jogh-12-03072}.

\bibitem[Tay et~al.(2017)Tay, Zyphur, and Batz-Barbarich]{tay_income_2017}
L.~Tay, M.~Zyphur, and C.~Batz-Barbarich.
\newblock Income and {Subjective} {Well}-{Being}: {Review}, {Synthesis}, and {Future} {Research}.
\newblock In \emph{e-{Handbook} of {Subjective} {Well}-{Being}}. Dec. 2017.

\bibitem[Tay et~al.(2018)Tay, Pawelski, and Keith]{tay2018role}
L.~Tay, J.~O. Pawelski, and M.~G. Keith.
\newblock The role of the arts and humanities in human flourishing: A conceptual model.
\newblock \emph{The Journal of Positive Psychology}, 13\penalty0 (3):\penalty0 215--225, 2018.

\bibitem[Tennenholtz(2004)]{tennenholtz_program_2004}
M.~Tennenholtz.
\newblock Program equilibrium.
\newblock \emph{Games and Economic Behavior}, 49\penalty0 (2):\penalty0 363--373, Nov. 2004.
\newblock ISSN 08998256.
\newblock \doi{10.1016/j.geb.2004.02.002}.
\newblock URL \url{https://linkinghub.elsevier.com/retrieve/pii/S0899825604000314}.

\bibitem[Thaler and Sunstein(2021)]{thaler_nudge:_2021}
R.~H. Thaler and C.~R. Sunstein.
\newblock \emph{Nudge: {The} {Final} {Edition}}.
\newblock Yale University Press, 2021.
\newblock ISBN 9780300262285.
\newblock Google-Books-ID: Wf1AEAAAQBAJ.

\bibitem[Thanawala(2023)]{thanawala_ai_nodate}
S.~Thanawala.
\newblock {AI} facial recognition tech leads to wave of lawsuits from {Black} plaintiffs after mistaken identities end in arrests, 2023.
\newblock URL \url{https://fortune.com/2023/09/25/ai-facial-recognition-tech-lawsuits-black-plaintiffs-mistaken-identities-arrests/}.

\bibitem[{The Adaptive Agents Group}(2021)]{the_adaptive_agents_group_shibboleth_2021}
{The Adaptive Agents Group}.
\newblock The {Shibboleth} {Rule} for {Artificial} {Agents}, Aug. 2021.
\newblock URL \url{https://hai.stanford.edu/news/shibboleth-rule-artificial-agents}.
\newblock publisher: Stanford University.

\bibitem[{The Collective Intelligence Project}(2023)]{the_collective_intelligence_project_whitepaper_nodate}
{The Collective Intelligence Project}.
\newblock Whitepaper, 2023.
\newblock URL \url{https://cip.org/whitepaper}.

\bibitem[{The Cybersecurity and Infrastructure Security Agency}(2022)]{the_cybersecurity_and_infrastructure_security_agency_cisa_tactics_2022}
{The Cybersecurity and Infrastructure Security Agency}.
\newblock Tactics of {Disinformation}, 2022.
\newblock URL \url{https://www.cisa.gov/sites/default/files/publications/tactics-of-disinformation_508.pdf}.

\bibitem[{The Economist Intelligence Unit}(2020)]{the_economist_intelligence_unit_new_2020}
{The Economist Intelligence Unit}.
\newblock New {Schools} of {Thought}, 2020.
\newblock URL \url{https://www.qf.org.qa/eiu}.

\bibitem[{The White House}(2022)]{the_white_house_blueprint_2022}
{The White House}.
\newblock Blueprint for an {AI} {Bill} of {Rights}, 2022.
\newblock URL \url{https://www.whitehouse.gov/ostp/ai-bill-of-rights/}.

\bibitem[{The White House}(2023{\natexlab{a}})]{the_white_house_fact_2023}
{The White House}.
\newblock {FACT} {SHEET}: {Biden}-{Harris} {Administration} {Secures} {Voluntary} {Commitments} from {Leading} {Artificial} {Intelligence} {Companies} to {Manage} the {Risks} {Posed} by {AI}, July 2023{\natexlab{a}}.
\newblock URL \url{https://www.whitehouse.gov/briefing-room/statements-releases/2023/07/21/fact-sheet-biden-harris-administration-secures-voluntary-commitments-from-leading-artificial-intelligence-companies-to-manage-the-risks-posed-by-ai/}.

\bibitem[{The White House}(2023{\natexlab{b}})]{whitehouse2023eo}
{The White House}.
\newblock Executive order on the safe, secure, and trustworthy development and use of artificial intelligence.
\newblock \url{https://www.whitehouse.gov/briefing-room/presidential-actions/2023/10/30/executive-order-on-the-safe-secure-and-trustworthy-development-and-use-of-artificial-intelligence/}, 2023{\natexlab{b}}.

\bibitem[{The World Bank}(2022)]{the_world_bank_lifting_2022}
{The World Bank}.
\newblock Lifting 800 {Million} {People} {Out} of {Poverty} – {New} {Report} {Looks} at {Lessons} from {China}’s {Experience}, Apr. 2022.
\newblock URL \url{https://www.worldbank.org/en/news/press-release/2022/04/01/lifting-800-million-people-out-of-poverty-new-report-looks-at-lessons-from-china-s-experience}.

\bibitem[Theben et~al.(2021)Theben, Gunderson, López-Fóres, Misuraca, and Lupiánez-Villanueva]{theben_challenges_2021}
A.~Theben, L.~Gunderson, L.~López-Fóres, G.~Misuraca, and F.~Lupiánez-Villanueva.
\newblock Challenges and limits of an open source approach to {Artificial} {Intelligence}.
\newblock Policy {Department} for {Economic}, {Scientific} and {Quality} of {Life} {Policies} {Directorate}-{General} for {Internal} {Policies} PE 662.908, European Parliament, May 2021.
\newblock URL \url{https://www.europarl.europa.eu/RegData/etudes/STUD/2021/662908/IPOL_STU(2021)662908_EN.pdf}.

\bibitem[Thiebes et~al.(2021)Thiebes, Lins, and Sunyaev]{thiebes_trustworthy_2021}
S.~Thiebes, S.~Lins, and A.~Sunyaev.
\newblock Trustworthy artificial intelligence.
\newblock \emph{Electronic Markets}, 31\penalty0 (2):\penalty0 447--464, June 2021.
\newblock ISSN 1422-8890.
\newblock \doi{10.1007/s12525-020-00441-4}.
\newblock URL \url{https://doi.org/10.1007/s12525-020-00441-4}.

\bibitem[Thiel(2020)]{thiel_biometric_2020}
A.~Thiel.
\newblock Biometric identification technologies and the {Ghanaian} ‘data revolution’.
\newblock \emph{The Journal of Modern African Studies}, 58\penalty0 (1):\penalty0 115--136, Mar. 2020.
\newblock ISSN 0022-278X, 1469-7777.
\newblock \doi{10.1017/S0022278X19000600}.
\newblock URL \url{https://www.cambridge.org/core/product/identifier/S0022278X19000600/type/journal_article}.

\bibitem[Thiel et~al.(2023)Thiel, Stroebel, and Portnoff]{thiel_generative_2023}
D.~Thiel, M.~Stroebel, and R.~Portnoff.
\newblock Generative {ML} and {CSAM}: {Implications} and {Mitigations}.
\newblock Technical report, Stanford University: Freeman Spogli Institute, June 2023.
\newblock URL \url{https://fsi.stanford.edu/publication/generative-ml-and-csam-implications-and-mitigations}.

\bibitem[Thoppilan et~al.(2022)Thoppilan, De~Freitas, Hall, Shazeer, Kulshreshtha, Cheng, Jin, Bos, Baker, Du, Li, Lee, Zheng, Ghafouri, Menegali, Huang, Krikun, Lepikhin, Qin, Chen, Xu, Chen, Roberts, Bosma, Zhao, Zhou, Chang, Krivokon, Rusch, Pickett, Srinivasan, Man, Meier-Hellstern, Morris, Doshi, Santos, Duke, Soraker, Zevenbergen, Prabhakaran, Diaz, Hutchinson, Olson, Molina, Hoffman-John, Lee, Aroyo, Rajakumar, Butryna, Lamm, Kuzmina, Fenton, Cohen, Bernstein, Kurzweil, Aguera-Arcas, Cui, Croak, Chi, and Le]{thoppilan_lamda:_2022}
R.~Thoppilan, D.~De~Freitas, J.~Hall, N.~Shazeer, A.~Kulshreshtha, H.-T. Cheng, A.~Jin, T.~Bos, L.~Baker, Y.~Du, Y.~Li, H.~Lee, H.~S. Zheng, A.~Ghafouri, M.~Menegali, Y.~Huang, M.~Krikun, D.~Lepikhin, J.~Qin, D.~Chen, Y.~Xu, Z.~Chen, A.~Roberts, M.~Bosma, V.~Zhao, Y.~Zhou, C.-C. Chang, I.~Krivokon, W.~Rusch, M.~Pickett, P.~Srinivasan, L.~Man, K.~Meier-Hellstern, M.~R. Morris, T.~Doshi, R.~D. Santos, T.~Duke, J.~Soraker, B.~Zevenbergen, V.~Prabhakaran, M.~Diaz, B.~Hutchinson, K.~Olson, A.~Molina, E.~Hoffman-John, J.~Lee, L.~Aroyo, R.~Rajakumar, A.~Butryna, M.~Lamm, V.~Kuzmina, J.~Fenton, A.~Cohen, R.~Bernstein, R.~Kurzweil, B.~Aguera-Arcas, C.~Cui, M.~Croak, E.~Chi, and Q.~Le.
\newblock {LaMDA}: {Language} {Models} for {Dialog} {Applications}, Feb. 2022.
\newblock URL \url{https://arxiv.org/pdf/2201.08239.pdf}.
\newblock arXiv:2201.08239 [cs].

\bibitem[Tien et~al.(2023)Tien, He, Erickson, Dragan, and Brown]{tien2023causal}
J.~Tien, J.~Z.-Y. He, Z.~Erickson, A.~D. Dragan, and D.~S. Brown.
\newblock Causal confusion and reward misidentification in preference-based reward learning, 2023.

\bibitem[Tiku(2022)]{tiku_google_2022}
N.~Tiku.
\newblock The {Google} engineer who thinks the company’s {AI} has come to life.
\newblock \emph{Washington Post}, June 2022.
\newblock URL \url{https://www.washingtonpost.com/technology/2022/06/11/google-ai-lamda-blake-lemoine/}.

\bibitem[Titchkosky(2011)]{titchkosky_question_2011}
T.~Titchkosky.
\newblock \emph{The question of access: disability, space, meaning}.
\newblock University of Toronto Press, Toronto, 2011.
\newblock ISBN 9781442640269 9781442685222 9781442610002.
\newblock OCLC: ocn712851646.

\bibitem[Toledo et~al.(2019)Toledo, Alzahrani, and Martinez]{toledo2019food}
R.~Y. Toledo, A.~A. Alzahrani, and L.~Martinez.
\newblock A food recommender system considering nutritional information and user preferences.
\newblock \emph{IEEE Access}, 7:\penalty0 96695--96711, 2019.

\bibitem[Tolmeijer et~al.(2021)Tolmeijer, Zierau, Janson, Wahdatehagh, Leimeister, and Bernstein]{tolmeijer_female_2021}
S.~Tolmeijer, N.~Zierau, A.~Janson, J.~S. Wahdatehagh, J.~M.~M. Leimeister, and A.~Bernstein.
\newblock Female by default?--exploring the effect of voice assistant gender and pitch on trait and trust attribution.
\newblock In \emph{Extended abstracts of the 2021 CHI conference on human factors in computing systems}, pages 1--7, 2021.

\bibitem[Toups et~al.(2023)Toups, Bommasani, Creel, Bana, Jurafsky, and Liang]{toups_ecosystem-level_2023}
C.~Toups, R.~Bommasani, K.~A. Creel, S.~H. Bana, D.~Jurafsky, and P.~Liang.
\newblock Ecosystem-level {Analysis} of {Deployed} {Machine} {Learning} {Reveals} {Homogeneous} {Outcomes}, July 2023.
\newblock URL \url{http://arxiv.org/abs/2307.05862}.
\newblock arXiv:2307.05862 [cs].

\bibitem[Touvron et~al.(2023)Touvron, Martin, Stone, Albert, Almahairi, Babaei, Bashlykov, Batra, Bhargava, Bhosale, Bikel, Blecher, Ferrer, Chen, Cucurull, Esiobu, Fernandes, Fu, Fu, Fuller, Gao, Goswami, Goyal, Hartshorn, Hosseini, Hou, Inan, Kardas, Kerkez, Khabsa, Kloumann, Korenev, Koura, Lachaux, Lavril, Lee, Liskovich, Lu, Mao, Martinet, Mihaylov, Mishra, Molybog, Nie, Poulton, Reizenstein, Rungta, Saladi, Schelten, Silva, Smith, Subramanian, Tan, Tang, Taylor, Williams, Kuan, Xu, Yan, Zarov, Zhang, Fan, Kambadur, Narang, Rodriguez, Stojnic, Edunov, and Scialom]{touvron_llama_2023}
H.~Touvron, L.~Martin, K.~Stone, P.~Albert, A.~Almahairi, Y.~Babaei, N.~Bashlykov, S.~Batra, P.~Bhargava, S.~Bhosale, D.~Bikel, L.~Blecher, C.~C. Ferrer, M.~Chen, G.~Cucurull, D.~Esiobu, J.~Fernandes, J.~Fu, W.~Fu, B.~Fuller, C.~Gao, V.~Goswami, N.~Goyal, A.~Hartshorn, S.~Hosseini, R.~Hou, H.~Inan, M.~Kardas, V.~Kerkez, M.~Khabsa, I.~Kloumann, A.~Korenev, P.~S. Koura, M.-A. Lachaux, T.~Lavril, J.~Lee, D.~Liskovich, Y.~Lu, Y.~Mao, X.~Martinet, T.~Mihaylov, P.~Mishra, I.~Molybog, Y.~Nie, A.~Poulton, J.~Reizenstein, R.~Rungta, K.~Saladi, A.~Schelten, R.~Silva, E.~M. Smith, R.~Subramanian, X.~E. Tan, B.~Tang, R.~Taylor, A.~Williams, J.~X. Kuan, P.~Xu, Z.~Yan, I.~Zarov, Y.~Zhang, A.~Fan, M.~Kambadur, S.~Narang, A.~Rodriguez, R.~Stojnic, S.~Edunov, and T.~Scialom.
\newblock Llama 2: {Open} {Foundation} and {Fine}-{Tuned} {Chat} {Models}, July 2023.
\newblock URL \url{http://arxiv.org/abs/2307.09288}.
\newblock arXiv:2307.09288 [cs].

\bibitem[Tramèr et~al.(2016)Tramèr, Zhang, Juels, Reiter, and Ristenpart]{tramer_stealing_nodate}
F.~Tramèr, F.~Zhang, A.~Juels, M.~K. Reiter, and T.~Ristenpart.
\newblock Stealing {Machine} {Learning} {Models} via {Prediction} {APIs}.
\newblock In \emph{Proceedings of the 25th {USENIX} {Security} {Symposium}}, pages 601--618, Austin, Texas, 2016.
\newblock URL \url{https://www.usenix.org/sites/default/files/sec16_full_proceedings.pdf}.

\bibitem[Trappey and Woodside(2005)]{trappey_brand_2005}
R.~J. Trappey and A.~G. Woodside.
\newblock \emph{Brand {Choice}: {Revealing} {Customers}’ {Unconscious}-{Automatic} and {Strategic} {Thinking} {Processes}}.
\newblock Palgrave Macmillan UK, London, 2005.
\newblock ISBN 9781349523573 9780230514201.
\newblock \doi{10.1057/9780230514201}.
\newblock URL \url{http://link.springer.com/10.1057/9780230514201}.

\bibitem[Trask et~al.(2020)Trask, Bluemke, Garfinkel, Cuervas-Mons, and Dafoe]{trask_beyond_2020}
A.~Trask, E.~Bluemke, B.~Garfinkel, C.~G. Cuervas-Mons, and A.~Dafoe.
\newblock Beyond {Privacy} {Trade}-offs with {Structured} {Transparency}, Dec. 2020.
\newblock URL \url{http://arxiv.org/abs/2012.08347}.
\newblock arXiv:2012.08347 [cs].

\bibitem[Tronto(2020)]{tronto_moral_2020}
J.~C. Tronto.
\newblock \emph{Moral {Boundaries}: {A} {Political} {Argument} for an {Ethic} of {Care}}.
\newblock Routledge, 1 edition, July 2020.
\newblock ISBN 9781003070672.
\newblock \doi{10.4324/9781003070672}.
\newblock URL \url{https://www.taylorfrancis.com/books/9781000107777}.

\bibitem[Tronto and Fisher(1990)]{tronto_toward_1990}
J.~C. Tronto and B.~Fisher.
\newblock Toward a {Feminist} {Theory} of {Caring}.
\newblock In E.~Abel and M.~Nelson, editors, \emph{Circles of {Care}}, pages 36--54. SUNY Press, Albany, NY, 1990.

\bibitem[Tsai(2014)]{tsai_rational_2014}
G.~Tsai.
\newblock Rational {Persuasion} as {Paternalism}.
\newblock \emph{Philosophy \& Public Affairs}, 42\penalty0 (1):\penalty0 78--112, Jan. 2014.
\newblock ISSN 0048-3915, 1088-4963.
\newblock \doi{10.1111/papa.12026}.
\newblock URL \url{https://onlinelibrary.wiley.com/doi/10.1111/papa.12026}.

\bibitem[Tu et~al.(2023)Tu, Zou, Su, and Zhang]{tu_what_2023}
X.~Tu, J.~Zou, W.~J. Su, and L.~Zhang.
\newblock What {Should} {Data} {Science} {Education} {Do} with {Large} {Language} {Models}?, July 2023.
\newblock URL \url{http://arxiv.org/abs/2307.02792}.
\newblock arXiv:2307.02792 [cs].

\bibitem[Tucker(2017)]{tucker_technocapitalist_2017}
B.~Tucker.
\newblock Technocapitalist {Disability} {Rhetoric}: {When} {Technology} is {Confused} with {Social} {Justice} {\textbar} enculturation, Apr. 2017.
\newblock URL \url{https://enculturation.net/technocapitalist-disability-rhetoric}.

\bibitem[Tufekci(2018)]{tufekci_opinion_2018}
Z.~Tufekci.
\newblock Opinion {\textbar} {YouTube}, the {Great} {Radicalizer}.
\newblock \emph{The New York Times}, Mar. 2018.
\newblock ISSN 0362-4331.
\newblock URL \url{https://www.nytimes.com/2018/03/10/opinion/sunday/youtube-politics-radical.html}.

\bibitem[Turkle(2007)]{turkle_authenticity_2007}
S.~Turkle.
\newblock Authenticity in the {Age} of {Digital} {Companions}.
\newblock \emph{Interaction Studies. Social Behaviour and Communication in Biological and Artificial Systemsinteraction Studies / Social Behaviour and Communication in Biological and Artificial Systemsinteraction Studies}, 8\penalty0 (3):\penalty0 501--517, 2007.
\newblock \doi{10.1075/is.8.3.11tur}.

\bibitem[Turkle(2018)]{turkle_there_2018}
S.~Turkle.
\newblock There {Will} {Never} {Be} an {Age} of {Artificial} {Intimacy}.
\newblock \emph{The New York Times}, Aug. 2018.
\newblock ISSN 0362-4331.
\newblock URL \url{https://www.nytimes.com/2018/08/11/opinion/there-will-never-be-an-age-of-artificial-intimacy.html}.

\bibitem[Turner and Tadepalli(2022)]{turner_parametrically_2022}
A.~M. Turner and P.~Tadepalli.
\newblock Parametrically {Retargetable} {Decision}-{Makers} {Tend} {To} {Seek} {Power}, Oct. 2022.
\newblock URL \url{http://arxiv.org/abs/2206.13477}.
\newblock arXiv:2206.13477 [cs].

\bibitem[Turner et~al.(2023)Turner, Smith, Shah, Critch, and Tadepalli]{turner_optimal_2023}
A.~M. Turner, L.~Smith, R.~Shah, A.~Critch, and P.~Tadepalli.
\newblock Optimal {Policies} {Tend} to {Seek} {Power}, Jan. 2023.
\newblock URL \url{http://arxiv.org/abs/1912.01683}.
\newblock arXiv:1912.01683 [cs].

\bibitem[Turpin et~al.(2023)Turpin, Michael, Perez, and Bowman]{turpin_language_2023}
M.~Turpin, J.~Michael, E.~Perez, and S.~R. Bowman.
\newblock Language {Models} {Don}'t {Always} {Say} {What} {They} {Think}: {Unfaithful} {Explanations} in {Chain}-of-{Thought} {Prompting}, May 2023.
\newblock URL \url{http://arxiv.org/abs/2305.04388}.
\newblock arXiv:2305.04388 [cs].

\bibitem[Tyler et~al.(2023)Tyler, Akerlof, Allegra, Arnold, Canino, Doornenbal, Goldstein, Budtz~Pedersen, and Sutherland]{tyler_ai_2023}
C.~Tyler, K.~L. Akerlof, A.~Allegra, Z.~Arnold, H.~Canino, M.~A. Doornenbal, J.~A. Goldstein, D.~Budtz~Pedersen, and W.~J. Sutherland.
\newblock {AI} tools as science policy advisers? {The} potential and the pitfalls.
\newblock \emph{Nature}, 622\penalty0 (7981):\penalty0 27--30, Oct. 2023.
\newblock ISSN 0028-0836, 1476-4687.
\newblock \doi{10.1038/d41586-023-02999-3}.
\newblock URL \url{https://www.nature.com/articles/d41586-023-02999-3}.

\bibitem[Uesato et~al.(2022)Uesato, Kushman, Kumar, Song, Siegel, Wang, Creswell, Irving, and Higgins]{uesato_solving_2022}
J.~Uesato, N.~Kushman, R.~Kumar, F.~Song, N.~Siegel, L.~Wang, A.~Creswell, G.~Irving, and I.~Higgins.
\newblock Solving math word problems with process- and outcome-based feedback, Nov. 2022.
\newblock URL \url{http://arxiv.org/abs/2211.14275}.
\newblock arXiv:2211.14275 [cs].

\bibitem[{UK Department for Science, Innovation and Technology}(2023)]{department_for_science_innovation_and_technology_pro-innovation_2023}
{UK Department for Science, Innovation and Technology}.
\newblock A pro-innovation approach to {AI} regulation.
\newblock Technical report, UK Government, Mar. 2023.
\newblock URL \url{https://www.gov.uk/government/publications/ai-regulation-a-pro-innovation-approach/white-paper}.
\newblock OCLC: 1382788551.

\bibitem[{UK Government}(2023)]{uk_government_national_2023}
{UK Government}.
\newblock National {Tutoring} {Programme}, July 2023.
\newblock URL \url{https://explore-education-statistics.service.gov.uk/find-statistics/national-tutoring-programme}.

\bibitem[Ullman(2023)]{ullman_large_2023}
T.~Ullman.
\newblock Large {Language} {Models} {Fail} on {Trivial} {Alterations} to {Theory}-of-{Mind} {Tasks}, Mar. 2023.
\newblock URL \url{http://arxiv.org/abs/2302.08399}.
\newblock arXiv:2302.08399 [cs].

\bibitem[Uma et~al.(2021)Uma, Fornaciari, Hovy, Paun, Plank, and Poesio]{uma_learning_2021}
A.~N. Uma, T.~Fornaciari, D.~Hovy, S.~Paun, B.~Plank, and M.~Poesio.
\newblock Learning from {Disagreement}: {A} {Survey}.
\newblock \emph{Journal of Artificial Intelligence Research}, 72:\penalty0 1385--1470, Dec. 2021.
\newblock ISSN 1076-9757.
\newblock \doi{10.1613/jair.1.12752}.
\newblock URL \url{https://www.jair.org/index.php/jair/article/view/12752}.

\bibitem[UN(2018)]{united_nations_inequality_nodate}
UN.
\newblock Inequality in {Asia} and the {Pacific} in the era of the 2030 agenda for sustainable development, 2018.
\newblock URL \url{https://www.unescap.org/publications/inequality-asia-and-pacific-era-2030-agenda-sustainable-development}.

\bibitem[UN(2021)]{un_addressing_2021}
UN.
\newblock Addressing the {Digital} {Divide}.
\newblock Technical report, United Nations Human Settlements Programme, 2021.
\newblock URL \url{https://unhabitat.org/sites/default/files/2021/11/addressing_the_digital_divide.pdf}.

\bibitem[UNDP(1990)]{undp1990concept}
UNDP.
\newblock Concept and measurement of human development.
\newblock \emph{Human Development Report 1990}, 1990.

\bibitem[UNESCO(2022)]{unesco_transforming_2022}
UNESCO.
\newblock Transforming education from within: current trends in the status and development of teachers.
\newblock Technical report, UNESCO, 2022.
\newblock URL \url{https://unesdoc.unesco.org/ark:/48223/pf0000383002}.

\bibitem[UNESCO(2023)]{unesco_transforming_2023}
UNESCO.
\newblock Transforming education together: the {Global} {Education} {Coalition} in action.
\newblock Technical report, UNESCO, 2023.
\newblock URL \url{https://unesdoc.unesco.org/ark:/48223/pf0000384812}.

\bibitem[{UNFCCC. Secretariat}(2023)]{UNFCCC_Secretariat2023-hb}
{UNFCCC. Secretariat}.
\newblock Technical dialogue of the first global stocktake. synthesis report by the co-facilitators on the technical dialogue.
\newblock Sept. 2023.

\bibitem[UNICEF(2022)]{unicef_education_2022}
UNICEF.
\newblock Education, June 2022.
\newblock URL \url{https://data.unicef.org/topic/gender/gender-disparities-in-education/}.

\bibitem[{Université de Montréal}(2017)]{universite_de_montreal_montredeclaration_2017}
{Université de Montréal}.
\newblock The {Montréal} {Declaration} for a {Responsible} {Development} of {Artificial} {Intelligence}, 2017.
\newblock URL \url{https://montrealdeclaration-responsibleai.com/the-declaration/}.

\bibitem[Ura(2008)]{ura2008explanation}
K.~Ura.
\newblock Explanation of {GNH} index. {G}ross {N}ational {H}appiness. {T}he {C}entre for {B}hutan {S}tudies, 2008.

\bibitem[{US Bureau of Labor Statistics}()]{us_bureau_of_labor_statistics_american_nodate}
{US Bureau of Labor Statistics}.
\newblock The {American} {Time} {Use} {Survey} ({ATUS}).
\newblock URL \url{https://www.bls.gov/tus/}.

\bibitem[{US Department of Justice}(2022)]{united_states_department_of_justice_court_2022}
{US Department of Justice}.
\newblock Court {Issues} {Order} {Requiring} {Cigarette} {Companies} to {Post} {Corrective} {Statements}; {Resolves} {Historic} {RICO} {Tobacco} {Litigation}, Dec. 2022.
\newblock URL \url{https://www.justice.gov/opa/pr/court-issues-order-requiring-cigarette-companies-post-corrective-statements-resolves-historic}.

\bibitem[Uuk(2023)]{uuk_three_2023}
R.~Uuk.
\newblock Three {Lines} of {Defence} {Against} {AI} {Risks} at a {Societal} {Level}.
\newblock Technical report, Future of Life Institute, 2023.
\newblock URL \url{https://ristouuk.com/wp-content/uploads/2023/06/Three_Lines_of_Defence_Against_AI_Risks_at_a_Societal_Level.pdf}.

\bibitem[Vaccari and Chadwick(2020)]{vaccari_deepfakes_2020}
C.~Vaccari and A.~Chadwick.
\newblock Deepfakes and {Disinformation}: {Exploring} the {Impact} of {Synthetic} {Political} {Video} on {Deception}, {Uncertainty}, and {Trust} in {News}.
\newblock \emph{Social Media + Society}, 6\penalty0 (1):\penalty0 205630512090340, Jan. 2020.
\newblock ISSN 2056-3051, 2056-3051.
\newblock \doi{10.1177/2056305120903408}.
\newblock URL \url{http://journals.sagepub.com/doi/10.1177/2056305120903408}.

\bibitem[Vaghefi et~al.(2023)Vaghefi, Stammbach, Muccione, Bingler, Ni, Kraus, Allen, Colesanti-Senni, Wekhof, Schimanski, et~al.]{vaghefi2023chatclimate}
S.~A. Vaghefi, D.~Stammbach, V.~Muccione, J.~Bingler, J.~Ni, M.~Kraus, S.~Allen, C.~Colesanti-Senni, T.~Wekhof, T.~Schimanski, et~al.
\newblock Chatclimate: Grounding conversational ai in climate science.
\newblock \emph{Communications Earth \& Environment}, 4\penalty0 (1):\penalty0 480, 2023.

\bibitem[Vallor(2016)]{vallor_technology_2016}
S.~Vallor.
\newblock \emph{Technology and the {Virtues}: {A} {Philosophical} {Guide} to a {Future} {Worth} {Wanting}}.
\newblock Oxford University Press, Oct. 2016.
\newblock ISBN 9780190498511.
\newblock URL \url{https://global.oup.com/academic/product/technology-and-the-virtues-9780190498511?cc=gb&lang=en&}.

\bibitem[Van~de Kerk and Manuel(2008)]{van2008comprehensive}
G.~Van~de Kerk and A.~R. Manuel.
\newblock A comprehensive index for a sustainable society: The {SSI}—the sustainable society index.
\newblock \emph{Ecological Economics}, 66\penalty0 (2-3):\penalty0 228--242, 2008.

\bibitem[van~den Oord et~al.(2016)van~den Oord, Dieleman, Zen, Simonyan, Vinyals, Graves, Kalchbrenner, Senior, and Kavukcuoglu]{oord_wavenet:_2016}
A.~van~den Oord, S.~Dieleman, H.~Zen, K.~Simonyan, O.~Vinyals, A.~Graves, N.~Kalchbrenner, A.~Senior, and K.~Kavukcuoglu.
\newblock {WaveNet}: {A} {Generative} {Model} for {Raw} {Audio}, Sept. 2016.
\newblock URL \url{http://arxiv.org/abs/1609.03499}.
\newblock arXiv:1609.03499 [cs].

\bibitem[van~der Maden et~al.(2023)van~der Maden, Lomas, and Hekkert]{vandermaden2023positive}
W.~van~der Maden, D.~Lomas, and P.~Hekkert.
\newblock Positive {AI}: Key challenges for designing wellbeing-aligned artificial intelligence, 2023.

\bibitem[van~der Plas et~al.(2010)van~der Plas, Smits, and Wehrmann]{van_der_plas_beyond_2010}
A.~van~der Plas, M.~Smits, and C.~Wehrmann.
\newblock Beyond {Speculative} {Robot} {Ethics}: {A} {Vision} {Assessment} {Study} on the {Future} of the {Robotic} {Caretaker}.
\newblock \emph{Accountability in Research}, 17\penalty0 (6):\penalty0 299--315, Nov. 2010.
\newblock ISSN 0898-9621, 1545-5815.
\newblock \doi{10.1080/08989621.2010.524078}.
\newblock URL \url{https://www.tandfonline.com/doi/full/10.1080/08989621.2010.524078}.

\bibitem[van Dijk(2006)]{van_dijk_digital_2006}
J.~A. van Dijk.
\newblock Digital divide research, achievements and shortcomings.
\newblock \emph{Poetics}, 34\penalty0 (4-5):\penalty0 221--235, Aug. 2006.
\newblock ISSN 0304422X.
\newblock \doi{10.1016/j.poetic.2006.05.004}.
\newblock URL \url{https://linkinghub.elsevier.com/retrieve/pii/S0304422X06000167}.

\bibitem[VanderWeele(2017)]{vanderweele2017promotion}
T.~J. VanderWeele.
\newblock On the promotion of human flourishing.
\newblock \emph{Proceedings of the National Academy of Sciences}, 114\penalty0 (31):\penalty0 8148--8156, 2017.

\bibitem[Vapnik(1999)]{vapnik_overview_1999}
V.~Vapnik.
\newblock An overview of statistical learning theory.
\newblock \emph{IEEE Transactions on Neural Networks}, 10\penalty0 (5):\penalty0 988--999, Sept. 1999.
\newblock ISSN 1941-0093.
\newblock \doi{10.1109/72.788640}.
\newblock URL \url{https://ieeexplore.ieee.org/abstract/document/788640}.

\bibitem[Varelius(2008)]{varelius2008prospects}
J.~Varelius.
\newblock On the prospects of collective informed consent.
\newblock \emph{Journal of applied philosophy}, 25\penalty0 (1):\penalty0 35--44, 2008.

\bibitem[Varga(2022)]{varga_solutions_2022}
L.~Z. Varga.
\newblock Solutions to the routing problem: towards trustworthy autonomous vehicles.
\newblock \emph{Artificial Intelligence Review}, 55\penalty0 (7):\penalty0 5445--5484, Oct. 2022.
\newblock ISSN 1573-7462.
\newblock \doi{10.1007/s10462-021-10131-y}.
\newblock URL \url{https://doi.org/10.1007/s10462-021-10131-y}.

\bibitem[Vassilakopoulou and Hustad(2023)]{vassilakopoulou_bridging_2023}
P.~Vassilakopoulou and E.~Hustad.
\newblock Bridging {Digital} {Divides}: a {Literature} {Review} and {Research} {Agenda} for {Information} {Systems} {Research}.
\newblock \emph{Information Systems Frontiers}, 25\penalty0 (3):\penalty0 955--969, June 2023.
\newblock ISSN 1572-9419.
\newblock \doi{10.1007/s10796-020-10096-3}.
\newblock URL \url{https://doi.org/10.1007/s10796-020-10096-3}.

\bibitem[Vaswani et~al.(2017)Vaswani, Shazeer, Parmar, Uszkoreit, Jones, Gomez, Kaiser, and Polosukhin]{vaswani_attention_2017}
A.~Vaswani, N.~Shazeer, N.~Parmar, J.~Uszkoreit, L.~Jones, A.~N. Gomez, L.~Kaiser, and I.~Polosukhin.
\newblock Attention {Is} {All} {You} {Need}.
\newblock Long Beach, CA, USA, 2017. arXiv.
\newblock \doi{10.48550/arXiv.1706.03762}.
\newblock URL \url{https://arxiv.org/pdf/1706.03762.pdf}.
\newblock arXiv:1706.03762 [cs].

\bibitem[Vella(2013)]{vella_persuasion:_2013}
G.~Vella.
\newblock Persuasion: importance of trust, relevance for small states, and limitations of computers - {Diplo} {Resource}, Aug. 2013.
\newblock URL \url{https://www.diplomacy.edu/resource/persuasion-importance-of-trust-relevance-for-small-states-and-limitations-of-computers/}.

\bibitem[Vemuri and Costanza(2006)]{vemuri2006role}
A.~W. Vemuri and R.~Costanza.
\newblock The role of human, social, built, and natural capital in explaining life satisfaction at the country level: Toward a national well-being index ({NWI}).
\newblock \emph{Ecological economics}, 58\penalty0 (1):\penalty0 119--133, 2006.

\bibitem[Verma(2023{\natexlab{a}})]{verma_chatgpt_2023}
P.~Verma.
\newblock {ChatGPT} provided better customer service than his staff. {He} fired them.
\newblock \emph{Washington Post}, Oct. 2023{\natexlab{a}}.
\newblock ISSN 0190-8286.
\newblock URL \url{https://www.washingtonpost.com/technology/2023/10/03/ai-customer-service-jobs/}.

\bibitem[Verma(2023{\natexlab{b}})]{verma_they_2023}
P.~Verma.
\newblock They fell in love with ai bots. a software update broke their hearts.
\newblock \emph{The Washington Post}, 2023{\natexlab{b}}.

\bibitem[Vincent(2023{\natexlab{a}})]{vincent_microsofts_2023}
J.~Vincent.
\newblock Microsoft’s {Bing} is an emotionally manipulative liar, and people love it, Feb. 2023{\natexlab{a}}.
\newblock URL \url{https://www.theverge.com/2023/2/15/23599072/microsoft-ai-bing-personality-conversations-spy-employees-webcams}.

\bibitem[Vincent(2023{\natexlab{b}})]{vincent_stack_2023}
J.~Vincent.
\newblock Stack {Overflow} survey finds developers are ready to use {AI} tools — even if they don’t fully trust them, June 2023{\natexlab{b}}.
\newblock URL \url{https://www.theverge.com/2023/6/13/23759101/stack-overflow-developers-survey-ai-coding-tools-moderators-strike}.

\bibitem[Vollrath(2023)]{vollrath_will_2023}
D.~Vollrath.
\newblock Will {AI} cause explosive economic growth?, July 2023.
\newblock URL \url{https://dietrichvollrath.substack.com/p/will-ai-cause-explosive-economic}.

\bibitem[Von~Schomberg(2013)]{owen_vision_2013}
R.~Von~Schomberg.
\newblock A {Vision} of {Responsible} {Research} and {Innovation}.
\newblock In R.~Owen, J.~Bessant, and M.~Heintz, editors, \emph{Responsible {Innovation}}, pages 51--74. Wiley, 1 edition, Apr. 2013.
\newblock ISBN 9781119966364 9781118551424.
\newblock \doi{10.1002/9781118551424.ch3}.
\newblock URL \url{https://onlinelibrary.wiley.com/doi/10.1002/9781118551424.ch3}.

\bibitem[Vosoughi et~al.(2018)Vosoughi, Roy, and Aral]{vosoughi_spread_2018}
S.~Vosoughi, D.~Roy, and S.~Aral.
\newblock The spread of true and false news online.
\newblock \emph{Science}, 359\penalty0 (6380):\penalty0 1146--1151, Mar. 2018.
\newblock ISSN 0036-8075, 1095-9203.
\newblock \doi{10.1126/science.aap9559}.
\newblock URL \url{https://www.science.org/doi/10.1126/science.aap9559}.

\bibitem[Voukelatou et~al.(2021)Voukelatou, Gabrielli, Miliou, Cresci, Sharma, Tesconi, and Pappalardo]{voukelatou2021measuring}
V.~Voukelatou, L.~Gabrielli, I.~Miliou, S.~Cresci, R.~Sharma, M.~Tesconi, and L.~Pappalardo.
\newblock Measuring objective and subjective well-being: {D}imensions and data sources.
\newblock \emph{International Journal of Data Science and Analytics}, 11:\penalty0 279--309, 2021.

\bibitem[Véliz(2021)]{veliz_privacy_2021}
C.~Véliz.
\newblock \emph{Privacy is {Power}: {Why} and {How} {You} {Should} {Take} {Back} {Control} of {Your} {Data}}.
\newblock Penguin Random House, July 2021.
\newblock URL \url{https://www.penguin.co.uk/books/442343/privacy-is-power-by-carissa-veliz/9780552177719}.

\bibitem[Véliz(2023)]{veliz_chatbots_2023}
C.~Véliz.
\newblock Chatbots shouldn’t use emojis.
\newblock \emph{Nature}, 615\penalty0 (7952):\penalty0 375--375, Mar. 2023.
\newblock \doi{10.1038/d41586-023-00758-y}.
\newblock URL \url{https://www.nature.com/articles/d41586-023-00758-y}.

\bibitem[Wachter and Mittelstadt(2019)]{wachter_right_2019}
S.~Wachter and B.~Mittelstadt.
\newblock A {Right} to {Reasonable} {Inferences}: {Re}-{Thinking} {Data} {Protection} {Law} in the {Age} of {Big} {Data} and {AI}.
\newblock \emph{Columbia Business Law Review}, 2019:\penalty0 494, 2019.
\newblock URL \url{https://heinonline.org/HOL/Page?handle=hein.journals/colb2019&id=506&div=&collection=}.

\bibitem[Wagner and Schramm-Klein(2019)]{wagner_alexa_2019}
K.~Wagner and H.~Schramm-Klein.
\newblock Alexa, {Are} {You} {Human}? {Investigating} {Anthropomorphism} of {Digital} {Voice} {Assistants} – {A} {Qualitative} {Approach}.
\newblock \emph{ICIS 2019 Proceedings}, Nov. 2019.
\newblock URL \url{https://aisel.aisnet.org/icis2019/human_computer_interact/human_computer_interact/7}.

\bibitem[Wairagkar et~al.(2021)Wairagkar, De~Lima, Harrison, Batey, Daniels, Barnaghi, Sharp, and Vaidyanathan]{wairagkar2021conversational}
M.~Wairagkar, M.~R. De~Lima, M.~Harrison, P.~Batey, S.~Daniels, P.~Barnaghi, D.~J. Sharp, and R.~Vaidyanathan.
\newblock Conversational artificial intelligence and affective social robot for monitoring health and well-being of people with dementia.
\newblock \emph{Alzheimer's \& Dementia}, 17:\penalty0 e053276, 2021.

\bibitem[Wajcman(2020)]{wajcman2020pressed}
J.~Wajcman.
\newblock \emph{Pressed for time: The acceleration of life in digital capitalism}.
\newblock University of Chicago Press, 2020.

\bibitem[Walker(2023)]{walker_belgian_2023}
L.~Walker.
\newblock Belgian man dies by suicide following exchanges with chatbot.
\newblock \emph{The Brussels Times}, Mar. 2023.
\newblock URL \url{https://www.brusselstimes.com/430098/belgian-man-commits-suicide-following-exchanges-with-chatgpt}.

\bibitem[Wallace et~al.(2021)Wallace, Feng, Kandpal, Gardner, and Singh]{wallace_universal_2021}
E.~Wallace, S.~Feng, N.~Kandpal, M.~Gardner, and S.~Singh.
\newblock Universal {Adversarial} {Triggers} for {Attacking} and {Analyzing} {NLP}, Jan. 2021.
\newblock URL \url{http://arxiv.org/abs/1908.07125}.
\newblock arXiv:1908.07125 [cs].

\bibitem[Walton and Wilson(2018)]{walton2018wise}
G.~M. Walton and T.~D. Wilson.
\newblock Wise interventions: Psychological remedies for social and personal problems.
\newblock \emph{Psychological review}, 125\penalty0 (5):\penalty0 617, 2018.

\bibitem[Wan and Chen(2021)]{wan_anthropomorphism_2021}
E.~W. Wan and R.~P. Chen.
\newblock Anthropomorphism and object attachment.
\newblock \emph{Current Opinion in Psychology}, 39:\penalty0 88--93, June 2021.
\newblock ISSN 2352250X.
\newblock \doi{10.1016/j.copsyc.2020.08.009}.
\newblock URL \url{https://linkinghub.elsevier.com/retrieve/pii/S2352250X20301548}.

\bibitem[Wan et~al.(2023)Wan, Hu, Zhang, Wang, Wen, and Lu]{wan_it_2023}
Q.~Wan, S.~Hu, Y.~Zhang, P.~Wang, B.~Wen, and Z.~Lu.
\newblock "{It} {Felt} {Like} {Having} a {Second} {Mind}": {Investigating} {Human}-{AI} {Co}-creativity in {Prewriting} with {Large} {Language} {Models}, Aug. 2023.
\newblock URL \url{http://arxiv.org/abs/2307.10811}.
\newblock arXiv:2307.10811 [cs].

\bibitem[Wang et~al.(2022{\natexlab{a}})Wang, Zhao, Van~Kleek, and Shadbolt]{wang_informing_2022}
G.~Wang, J.~Zhao, M.~Van~Kleek, and N.~Shadbolt.
\newblock Informing {Age}-{Appropriate} {AI}: {Examining} {Principles} and {Practices} of {AI} for {Children}.
\newblock In \emph{{CHI} {Conference} on {Human} {Factors} in {Computing} {Systems}}, pages 1--29, New Orleans LA USA, Apr. 2022{\natexlab{a}}. ACM.
\newblock ISBN 9781450391573.
\newblock \doi{10.1145/3491102.3502057}.
\newblock URL \url{https://dl.acm.org/doi/10.1145/3491102.3502057}.

\bibitem[Wang(2022)]{wang_transparency_2022}
H.~Wang.
\newblock Transparency as {Manipulation}? {Uncovering} the {Disciplinary} {Power} of {Algorithmic} {Transparency}.
\newblock \emph{Philosophy \& Technology}, 35\penalty0 (3):\penalty0 69, July 2022.
\newblock ISSN 2210-5441.
\newblock \doi{10.1007/s13347-022-00564-w}.
\newblock URL \url{https://doi.org/10.1007/s13347-022-00564-w}.

\bibitem[Wang et~al.(2022{\natexlab{b}})Wang, Variengien, Conmy, Shlegeris, and Steinhardt]{wang_interpretability_2022}
K.~Wang, A.~Variengien, A.~Conmy, B.~Shlegeris, and J.~Steinhardt.
\newblock Interpretability in the {Wild}: a {Circuit} for {Indirect} {Object} {Identification} in {GPT}-2 small, Nov. 2022{\natexlab{b}}.
\newblock URL \url{http://arxiv.org/abs/2211.00593}.
\newblock arXiv:2211.00593 [cs].

\bibitem[Wang and Demszky(2023)]{wang_is_2023}
R.~E. Wang and D.~Demszky.
\newblock Is {ChatGPT} a {Good} {Teacher} {Coach}? {Measuring} {Zero}-{Shot} {Performance} {For} {Scoring} and {Providing} {Actionable} {Insights} on {Classroom} {Instruction}, June 2023.
\newblock URL \url{http://arxiv.org/abs/2306.03090}.
\newblock arXiv:2306.03090 [cs].

\bibitem[Wang(2017)]{wang_smartphones_2017}
W.~Wang.
\newblock Smartphones as {Social} {Actors}? {Social} dispositional factors in assessing anthropomorphism.
\newblock \emph{Computers in Human Behavior}, 68:\penalty0 334--344, Mar. 2017.
\newblock ISSN 07475632.
\newblock \doi{10.1016/j.chb.2016.11.022}.
\newblock URL \url{https://linkinghub.elsevier.com/retrieve/pii/S0747563216307634}.

\bibitem[Wang et~al.(2020)Wang, Shi, Kim, Oh, Yang, Zhang, and Yu]{wang_persuasion_2020}
X.~Wang, W.~Shi, R.~Kim, Y.~Oh, S.~Yang, J.~Zhang, and Z.~Yu.
\newblock Persuasion for {Good}: {Towards} a {Personalized} {Persuasive} {Dialogue} {System} for {Social} {Good}, Jan. 2020.
\newblock URL \url{http://arxiv.org/abs/1906.06725}.
\newblock arXiv:1906.06725 [cs].

\bibitem[Wang et~al.(2021)Wang, Yao, Kwok, and Ni]{wang_generalizing_2021}
Y.~Wang, Q.~Yao, J.~T. Kwok, and L.~M. Ni.
\newblock Generalizing from a {Few} {Examples}: {A} {Survey} on {Few}-shot {Learning}.
\newblock \emph{ACM Computing Surveys}, 53\penalty0 (3):\penalty0 1--34, May 2021.
\newblock ISSN 0360-0300, 1557-7341.
\newblock \doi{10.1145/3386252}.
\newblock URL \url{https://dl.acm.org/doi/10.1145/3386252}.

\bibitem[Ward et~al.(2023)Ward, Everitt, Belardinelli, and Toni]{ward2023honesty}
F.~R. Ward, T.~Everitt, F.~Belardinelli, and F.~Toni.
\newblock Honesty is the best policy: defining and mitigating ai deception.
\newblock In \emph{Thirty-seventh Conference on Neural Information Processing Systems}, 2023.

\bibitem[Wardle and Derakhshan(2017)]{wardle_information_2017}
C.~Wardle and H.~Derakhshan.
\newblock Information {Disorder}: {Toward} an interdisciplinary framework for research and policy making.
\newblock Technical Report DGI(2017)09, Council of Europe, Sept. 2017.
\newblock URL \url{https://rm.coe.int/information-disorder-toward-an-interdisciplinary-framework-for-researc/168076277c}.

\bibitem[Warford et~al.(2022)Warford, Matthews, Yang, Akgul, Consolvo, Kelley, Malkin, Mazurek, Sleeper, and Thomas]{warford2022}
N.~Warford, T.~Matthews, K.~Yang, O.~Akgul, S.~Consolvo, P.~G. Kelley, N.~Malkin, M.~L. Mazurek, M.~Sleeper, and K.~Thomas.
\newblock Sok: A framework for unifying at-risk user research.
\newblock In \emph{2022 IEEE Symposium on Security and Privacy (SP)}, pages 2344--2360, 2022.
\newblock \doi{10.1109/SP46214.2022.9833643}.

\bibitem[Warren and Brandeis(1890)]{warren_right_1890}
S.~D. Warren and L.~D. Brandeis.
\newblock The {Right} to {Privacy}.
\newblock \emph{Harvard Law Review}, 4\penalty0 (5):\penalty0 193, Dec. 1890.
\newblock ISSN 0017811X.
\newblock \doi{10.2307/1321160}.
\newblock URL \url{https://www.jstor.org/stable/1321160?origin=crossref}.

\bibitem[Warren(2023)]{warren_microsoft_2023}
T.~Warren.
\newblock Microsoft has been secretly testing its {Bing} “{Sydney}” chatbot for years, Feb. 2023.
\newblock URL \url{https://www.theverge.com/2023/2/23/23609942/microsoft-bing-sydney-chatbot-history-ai}.

\bibitem[Warschauer(2002)]{warschauer_reconceptualizing_2002}
M.~Warschauer.
\newblock Reconceptualizing the {Digital} {Divide}.
\newblock \emph{First Monday}, 7\penalty0 (7), July 2002.
\newblock ISSN 13960466.
\newblock \doi{10.5210/fm.v7i7.967}.
\newblock URL \url{http://journals.uic.edu/ojs/index.php/fm/article/view/967}.

\bibitem[Warschauer(2004)]{warschauer_technology_2004}
M.~Warschauer.
\newblock \emph{Technology and social inclusion: rethinking the digital divide}.
\newblock MIT Press, Cambridge, Mass., 1. mit press paperback ed edition, 2004.
\newblock ISBN 9780262731737.

\bibitem[Warschauer and Matuchniak(2010)]{warschauer_new_2010}
M.~Warschauer and T.~Matuchniak.
\newblock New {Technology} and {Digital} {Worlds}: {Analyzing} {Evidence} of {Equity} in {Access}, {Use}, and {Outcomes}.
\newblock \emph{Review of Research in Education}, 34\penalty0 (1):\penalty0 179--225, Mar. 2010.
\newblock ISSN 0091-732X, 1935-1038.
\newblock \doi{10.3102/0091732X09349791}.
\newblock URL \url{http://journals.sagepub.com/doi/10.3102/0091732X09349791}.

\bibitem[Wasserman and Richmond‐Abbott(2005)]{wasserman_gender_2005}
I.~M. Wasserman and M.~Richmond‐Abbott.
\newblock Gender and the {Internet}: {Causes} of {Variation} in {Access}, {Level}, and {Scope} of {Use} $^{\textrm{*}}$.
\newblock \emph{Social Science Quarterly}, 86\penalty0 (1):\penalty0 252--270, Mar. 2005.
\newblock ISSN 0038-4941, 1540-6237.
\newblock \doi{10.1111/j.0038-4941.2005.00301.x}.
\newblock URL \url{https://onlinelibrary.wiley.com/doi/10.1111/j.0038-4941.2005.00301.x}.

\bibitem[Watts(2004)]{watts2004new}
A.~Watts.
\newblock New index measures well-being and ranks kentucky.
\newblock \emph{Foresight}, 11\penalty0 (1), 2004.

\bibitem[Waytz et~al.(2010)Waytz, Cacioppo, and Epley]{waytz_who_2010}
A.~Waytz, J.~Cacioppo, and N.~Epley.
\newblock Who {Sees} {Human}?: {The} {Stability} and {Importance} of {Individual} {Differences} in {Anthropomorphism}.
\newblock \emph{Perspectives on Psychological Science}, 5\penalty0 (3):\penalty0 219--232, May 2010.
\newblock ISSN 1745-6916, 1745-6924.
\newblock \doi{10.1177/1745691610369336}.
\newblock URL \url{http://journals.sagepub.com/doi/10.1177/1745691610369336}.

\bibitem[Waytz et~al.(2019)Waytz, Cacioppo, Hurlemann, Castelli, Adolphs, and Paul]{waytz_anthropomorphizing_2019}
A.~Waytz, J.~T. Cacioppo, R.~Hurlemann, F.~Castelli, R.~Adolphs, and L.~K. Paul.
\newblock Anthropomorphizing without {Social} {Cues} {Requires} the {Basolateral} {Amygdala}.
\newblock \emph{Journal of Cognitive Neuroscience}, 31\penalty0 (4):\penalty0 482--496, Apr. 2019.
\newblock ISSN 0898-929X, 1530-8898.
\newblock \doi{10.1162/jocn_a_01365}.
\newblock URL \url{https://direct.mit.edu/jocn/article/31/4/482-496/28993}.

\bibitem[Webersinke et~al.(2021)Webersinke, Kraus, Bingler, and Leippold]{Webersinke2021-fe}
N.~Webersinke, M.~Kraus, J.~A. Bingler, and M.~Leippold.
\newblock {ClimateBert}: A pretrained language model for {Climate-Related} text.
\newblock Oct. 2021.

\bibitem[Wei et~al.(2023{\natexlab{a}})Wei, Haghtalab, and Steinhardt]{wei_jailbroken:_2023}
A.~Wei, N.~Haghtalab, and J.~Steinhardt.
\newblock Jailbroken: {How} {Does} {LLM} {Safety} {Training} {Fail}?, July 2023{\natexlab{a}}.
\newblock URL \url{http://arxiv.org/abs/2307.02483}.
\newblock arXiv:2307.02483 [cs].

\bibitem[Wei et~al.(2022)Wei, Tay, Bommasani, Raffel, Zoph, Borgeaud, Yogatama, Bosma, Zhou, Metzler, Chi, Hashimoto, Vinyals, Liang, Dean, and Fedus]{wei_emergent_2022}
J.~Wei, Y.~Tay, R.~Bommasani, C.~Raffel, B.~Zoph, S.~Borgeaud, D.~Yogatama, M.~Bosma, D.~Zhou, D.~Metzler, E.~H. Chi, T.~Hashimoto, O.~Vinyals, P.~Liang, J.~Dean, and W.~Fedus.
\newblock Emergent {Abilities} of {Large} {Language} {Models}.
\newblock \emph{Transactions on Machine Learning Research}, Oct. 2022.
\newblock \doi{10.48550/arXiv.2206.07682}.
\newblock URL \url{http://arxiv.org/abs/2206.07682}.
\newblock arXiv:2206.07682 [cs].

\bibitem[Wei et~al.(2023{\natexlab{b}})Wei, Wang, Schuurmans, Bosma, Ichter, Xia, Chi, Le, and Zhou]{wei_chain--thought_2023}
J.~Wei, X.~Wang, D.~Schuurmans, M.~Bosma, B.~Ichter, F.~Xia, E.~Chi, Q.~Le, and D.~Zhou.
\newblock Chain-of-{Thought} {Prompting} {Elicits} {Reasoning} in {Large} {Language} {Models}, Jan. 2023{\natexlab{b}}.
\newblock URL \url{http://arxiv.org/abs/2201.11903}.
\newblock arXiv:2201.11903 [cs].

\bibitem[Weidinger et~al.(2021)Weidinger, Mellor, Rauh, Griffin, Uesato, Huang, Cheng, Glaese, Balle, Kasirzadeh, Kenton, Brown, Hawkins, Stepleton, Biles, Birhane, Haas, Rimell, Hendricks, Isaac, Legassick, Irving, and Gabriel]{weidinger_ethical_2021}
L.~Weidinger, J.~Mellor, M.~Rauh, C.~Griffin, J.~Uesato, P.-S. Huang, M.~Cheng, M.~Glaese, B.~Balle, A.~Kasirzadeh, Z.~Kenton, S.~Brown, W.~Hawkins, T.~Stepleton, C.~Biles, A.~Birhane, J.~Haas, L.~Rimell, L.~A. Hendricks, W.~Isaac, S.~Legassick, G.~Irving, and I.~Gabriel.
\newblock Ethical and social risks of harm from {Language} {Models}, Dec. 2021.
\newblock URL \url{http://arxiv.org/abs/2112.04359}.
\newblock arXiv:2112.04359 [cs].

\bibitem[Weidinger et~al.(2022{\natexlab{a}})Weidinger, Reinecke, and Haas]{weidinger_artificial_2022}
L.~Weidinger, M.~G. Reinecke, and J.~Haas.
\newblock Artificial moral cognition: {Learning} from developmental psychology.
\newblock preprint, PsyArXiv, Aug. 2022{\natexlab{a}}.
\newblock URL \url{https://osf.io/tnf4e}.

\bibitem[Weidinger et~al.(2022{\natexlab{b}})Weidinger, Uesato, Rauh, Griffin, Huang, Mellor, Glaese, Cheng, Balle, Kasirzadeh, Biles, Brown, Kenton, Hawkins, Stepleton, Birhane, Hendricks, Rimell, Isaac, Haas, Legassick, Irving, and Gabriel]{weidinger_taxonomy_2022}
L.~Weidinger, J.~Uesato, M.~Rauh, C.~Griffin, P.-S. Huang, J.~Mellor, A.~Glaese, M.~Cheng, B.~Balle, A.~Kasirzadeh, C.~Biles, S.~Brown, Z.~Kenton, W.~Hawkins, T.~Stepleton, A.~Birhane, L.~A. Hendricks, L.~Rimell, W.~Isaac, J.~Haas, S.~Legassick, G.~Irving, and I.~Gabriel.
\newblock Taxonomy of {Risks} posed by {Language} {Models}.
\newblock In \emph{Proceedings of the 2022 {ACM} {Conference} on {Fairness}, {Accountability}, and {Transparency}}, {FAccT} '22, pages 214--229, New York, NY, USA, June 2022{\natexlab{b}}. Association for Computing Machinery.
\newblock ISBN 9781450393522.
\newblock \doi{10.1145/3531146.3533088}.
\newblock URL \url{https://dl.acm.org/doi/10.1145/3531146.3533088}.

\bibitem[Weidinger et~al.(2023{\natexlab{a}})Weidinger, McKee, Everett, Huang, Zhu, Chadwick, Summerfield, and Gabriel]{weidinger_using_2023}
L.~Weidinger, K.~R. McKee, R.~Everett, S.~Huang, T.~O. Zhu, M.~J. Chadwick, C.~Summerfield, and I.~Gabriel.
\newblock Using the {Veil} of {Ignorance} to align {AI} systems with principles of justice.
\newblock \emph{Proceedings of the National Academy of Sciences}, 120\penalty0 (18):\penalty0 e2213709120, May 2023{\natexlab{a}}.
\newblock ISSN 0027-8424, 1091-6490.
\newblock \doi{10.1073/pnas.2213709120}.
\newblock URL \url{https://pnas.org/doi/10.1073/pnas.2213709120}.

\bibitem[Weidinger et~al.(2023{\natexlab{b}})Weidinger, Rauh, Marchal, Manzini, Hendricks, Mateos-Garcia, Bergman, Kay, Griffin, Bariach, et~al.]{weidinger2023sociotechnical}
L.~Weidinger, M.~Rauh, N.~Marchal, A.~Manzini, L.~A. Hendricks, J.~Mateos-Garcia, S.~Bergman, J.~Kay, C.~Griffin, B.~Bariach, et~al.
\newblock Sociotechnical safety evaluation of generative ai systems.
\newblock \emph{arXiv preprint arXiv:2310.11986}, 2023{\natexlab{b}}.

\bibitem[Welbl et~al.(2021)Welbl, Glaese, Uesato, Dathathri, Mellor, Hendricks, Anderson, Kohli, Coppin, and Huang]{welbl_challenges_2021}
J.~Welbl, A.~Glaese, J.~Uesato, S.~Dathathri, J.~Mellor, L.~A. Hendricks, K.~Anderson, P.~Kohli, B.~Coppin, and P.-S. Huang.
\newblock Challenges in {Detoxifying} {Language} {Models}, Sept. 2021.
\newblock URL \url{http://arxiv.org/abs/2109.07445}.
\newblock arXiv:2109.07445 [cs].

\bibitem[Welch et~al.(2022)Welch, Gu, Kummerfeld, P{\'e}rez-Rosas, and Mihalcea]{welch2022leveraging}
C.~Welch, C.~Gu, J.~K. Kummerfeld, V.~P{\'e}rez-Rosas, and R.~Mihalcea.
\newblock Leveraging similar users for personalized language modeling with limited data.
\newblock In \emph{Proceedings of the 60th Annual Meeting of the Association for Computational Linguistics (Volume 1: Long Papers)}, pages 1742--1752, 2022.

\bibitem[Wen et~al.(2017)Wen, Vandyke, Mrkšić, Gašić, Rojas-Barahona, Su, Ultes, and Young]{wen_network-based_2017}
T.-H. Wen, D.~Vandyke, N.~Mrkšić, M.~Gašić, L.~M. Rojas-Barahona, P.-H. Su, S.~Ultes, and S.~Young.
\newblock A {Network}-based {End}-to-{End} {Trainable} {Task}-oriented {Dialogue} {System}.
\newblock In M.~Lapata, P.~Blunsom, and A.~Koller, editors, \emph{Proceedings of the 15th {Conference} of the {European} {Chapter} of the {Association} for {Computational} {Linguistics}: {Volume} 1, {Long} {Papers}}, pages 438--449, Valencia, Spain, Apr. 2017. Association for Computational Linguistics.
\newblock URL \url{https://aclanthology.org/E17-1042.pdf}.

\bibitem[Wen et~al.(2016)Wen, Geng, and Ye]{wen2016does}
Z.~Wen, X.~Geng, and Y.~Ye.
\newblock Does the use of {W}e{C}hat lead to subjective well-being?: The effect of use intensity and motivations.
\newblock \emph{Cyberpsychology, Behavior, and Social Networking}, 19\penalty0 (10):\penalty0 587--592, 2016.

\bibitem[Wenz(1988)]{wenz_environmental_1988}
P.~S. Wenz.
\newblock \emph{Environmental justice}.
\newblock {SUNY} series in environmental public policy. State University of New York Press, Albany, 1988.
\newblock ISBN 9780887066443 9780887066450.

\bibitem[Werhane(1999)]{werhane1999moral}
P.~H. Werhane.
\newblock \emph{Moral imagination and management decision-making}.
\newblock Oxford University Press, USA, 1999.

\bibitem[Werhane(2002)]{werhane2002moral}
P.~H. Werhane.
\newblock Moral imagination and systems thinking.
\newblock \emph{Journal of business ethics}, 38:\penalty0 33--42, 2002.

\bibitem[Wertheimer(1999)]{wertheimer_exploitation_1999}
A.~Wertheimer.
\newblock \emph{Exploitation}.
\newblock Princeton University Press, Aug. 1999.
\newblock ISBN 9780691019475.
\newblock URL \url{https://press.princeton.edu/books/paperback/9780691019475/exploitation}.

\bibitem[West et~al.(2019)West, Kraut, and Chew]{west_id_2019}
M.~West, R.~Kraut, and H.~E. Chew.
\newblock I'd blush if {I} could: closing gender divides in digital skills through education.
\newblock Technical report, UNESCO, Jan. 2019.
\newblock URL \url{https://unesdoc.unesco.org/ark:/48223/pf0000367416}.

\bibitem[Whittaker et~al.(2021)Whittaker, Looney, Reed, and Votta]{whittaker_recommender_2021}
J.~Whittaker, S.~Looney, A.~Reed, and F.~Votta.
\newblock Recommender systems and the amplification of extremist content.
\newblock \emph{Internet Policy Review}, 10\penalty0 (2), June 2021.
\newblock ISSN 2197-6775.
\newblock \doi{10.14763/2021.2.1565}.
\newblock URL \url{https://policyreview.info/articles/analysis/recommender-systems-and-amplification-extremist-content}.

\bibitem[Whittaker et~al.(2019)Whittaker, Alper, Bennett, Hendren, Kaziunas, Mills, Ringel~Morris, Rankin, Rogers, Salas, and Myers~West]{whittaker_disability_2019}
M.~Whittaker, M.~Alper, C.~L. Bennett, S.~Hendren, L.~Kaziunas, M.~Mills, M.~Ringel~Morris, J.~Rankin, E.~Rogers, M.~Salas, and S.~Myers~West.
\newblock Disability, {Bias}, and {AI}.
\newblock Technical report, AI Now Institute, Nov. 2019.
\newblock URL \url{https://ainowinstitute.org/publication/disabilitybiasai-2019}.

\bibitem[Whittlestone and Clark(2021)]{whittlestone2021governments}
J.~Whittlestone and J.~Clark.
\newblock Why and how governments should monitor ai development, 2021.

\bibitem[Wicksteed(1910)]{wicksteed1910common}
P.~H. Wicksteed.
\newblock \emph{The common sense of political economy, including a study of the human basis of economic law}.
\newblock Macmillan, 1910.

\bibitem[Widner et~al.(2023)Widner, Virmani, Krause, Nayar, Tiwari, Pedersen, Jeji, Hammel, Matias, Corrado, Liu, Peng, and Webster]{widner_lessons_2023}
K.~Widner, S.~Virmani, J.~Krause, J.~Nayar, R.~Tiwari, E.~R. Pedersen, D.~Jeji, N.~Hammel, Y.~Matias, G.~S. Corrado, Y.~Liu, L.~Peng, and D.~R. Webster.
\newblock Lessons learned from translating {AI} from development to deployment in healthcare.
\newblock \emph{Nature Medicine}, 29\penalty0 (6):\penalty0 1304--1306, June 2023.
\newblock ISSN 1078-8956, 1546-170X.
\newblock \doi{10.1038/s41591-023-02293-9}.
\newblock URL \url{https://www.nature.com/articles/s41591-023-02293-9}.

\bibitem[Wiener(2004)]{wiener_regulation_2004}
J.~B. Wiener.
\newblock The regulation of technology, and the technology of regulation.
\newblock \emph{Technology in Society}, 26\penalty0 (2-3):\penalty0 483--500, Apr. 2004.
\newblock ISSN 0160791X.
\newblock \doi{10.1016/j.techsoc.2004.01.033}.
\newblock URL \url{https://linkinghub.elsevier.com/retrieve/pii/S0160791X04000375}.

\bibitem[Wiggers and Stringer(2023)]{wiggers_chatgpt:_2023}
K.~Wiggers and A.~Stringer.
\newblock {ChatGPT}: {Everything} you need to know about the {AI} chatbot, Nov. 2023.
\newblock URL \url{https://techcrunch.com/2023/11/06/chatgpt-everything-to-know-about-the-ai-chatbot/}.

\bibitem[Wiktor and Sanak-Kosmowska(2021)]{wiktor_information_2021}
J.~W. Wiktor and K.~Sanak-Kosmowska.
\newblock \emph{Information asymmetry in online advertising}.
\newblock Routledge studies in marketing. Routledge, Milton Park, Abingdon, Oxon ; New York, NY, 2021.
\newblock ISBN 9781000454055 9781003134121.
\newblock URL \url{https://www.routledge.com/Information-Asymmetry-in-Online-Advertising/Wiktor-Sanak-Kosmowska/p/book/9780367652128}.

\bibitem[Williams(2010)]{williams_truth_2010}
B.~Williams.
\newblock \emph{Truth and {Truthfulness}: {An} {Essay} in {Genealogy}}.
\newblock Princeton University Press, Princeton, Dec. 2010.
\newblock ISBN 9781400825141.
\newblock \doi{10.1515/9781400825141}.
\newblock URL \url{https://www.degruyter.com/document/doi/10.1515/9781400825141/html}.

\bibitem[Williams et~al.(2013)Williams, Raux, Ramachandran, and Black]{williams_dialog_2013}
J.~Williams, A.~Raux, D.~Ramachandran, and A.~Black.
\newblock The {Dialog} {State} {Tracking} {Challenge}.
\newblock In \emph{Proceedings of the {SIGDIAL} 2013 {Conference}}, pages 404--413, Metz, France, Aug. 2013. Association for Computational Linguistics.
\newblock URL \url{https://aclanthology.org/W13-4065.pdf}.

\bibitem[Williamson(1994)]{williamson_vagueness_1994}
T.~Williamson.
\newblock \emph{Vagueness}.
\newblock Routledge, New York, 1994.

\bibitem[Williamson(1997)]{williamson_precis_1997}
T.~Williamson.
\newblock Précis of {Vagueness}.
\newblock \emph{Philosophy and Phenomenological Research}, 57\penalty0 (4):\penalty0 921--928, 1997.
\newblock ISSN 0031-8205.
\newblock \doi{10.2307/2953810}.
\newblock URL \url{https://www.jstor.org/stable/2953810}.

\bibitem[Willison(2023)]{willison_bing:_2023}
S.~Willison.
\newblock Bing: “{I} will not harm you unless you harm me first”, Feb. 2023.
\newblock URL \url{https://simonwillison.net/2023/Feb/15/bing/}.

\bibitem[Wilmer et~al.(2017)Wilmer, Sherman, and Chein]{wilmer_smartphones_2017}
H.~H. Wilmer, L.~E. Sherman, and J.~M. Chein.
\newblock Smartphones and {Cognition}: {A} {Review} of {Research} {Exploring} the {Links} between {Mobile} {Technology} {Habits} and {Cognitive} {Functioning}.
\newblock \emph{Frontiers in Psychology}, 8:\penalty0 605, Apr. 2017.
\newblock ISSN 1664-1078.
\newblock \doi{10.3389/fpsyg.2017.00605}.
\newblock URL \url{http://journal.frontiersin.org/article/10.3389/fpsyg.2017.00605/full}.

\bibitem[Wilson(2018)]{wilson2018love}
C.~Wilson.
\newblock Is it love or loneliness? {E}xploring the impact of everyday digital technology use on the wellbeing of older adults.
\newblock \emph{Ageing \& Society}, 38\penalty0 (7):\penalty0 1307--1331, 2018.

\bibitem[Wilson(2011)]{wilson2011redirect}
T.~Wilson.
\newblock \emph{Redirect: The surprising new science of psychological change}.
\newblock Penguin UK, 2011.

\bibitem[Winner(2010)]{winner2010whale}
L.~Winner.
\newblock \emph{The whale and the reactor: A search for limits in an age of high technology}.
\newblock University of Chicago Press, 2010.

\bibitem[Wirth et~al.(2017)Wirth, Akrour, Neumann, F{\"u}rnkranz, et~al.]{wirth2017survey}
C.~Wirth, R.~Akrour, G.~Neumann, J.~F{\"u}rnkranz, et~al.
\newblock A survey of preference-based reinforcement learning methods.
\newblock \emph{Journal of Machine Learning Research}, 18\penalty0 (136):\penalty0 1--46, 2017.

\bibitem[Witte and Mannon(2010)]{witte_internet_2010}
J.~C. Witte and S.~E. Mannon.
\newblock \emph{The {Internet} and social inequalities}.
\newblock Contemporary sociological perspectives. Routledge, New York, 2010.
\newblock ISBN 9780415963206 9780415963190 9780203861639.
\newblock OCLC: ocn166361305.

\bibitem[Wong et~al.(2022)Wong, Dutta, Voicu, Chervonyi, Paduraru, and Luo]{wong_optimizing_2022}
W.~Wong, P.~Dutta, O.~Voicu, Y.~Chervonyi, C.~Paduraru, and J.~Luo.
\newblock Optimizing {Industrial} {HVAC} {Systems} with {Hierarchical} {Reinforcement} {Learning}, Sept. 2022.
\newblock URL \url{http://arxiv.org/abs/2209.08112}.
\newblock arXiv:2209.08112 [cs, eess].

\bibitem[Wood(2014)]{coons_coercion_2014}
A.~W. Wood.
\newblock Coercion, {Manipulation}, {Exploitation}.
\newblock In C.~Coons and M.~Weber, editors, \emph{Manipulation}, pages 17--50. Oxford University Press, Aug. 2014.
\newblock ISBN 9780199338207.
\newblock \doi{10.1093/acprof:oso/9780199338207.003.0002}.
\newblock URL \url{https://academic.oup.com/book/4870/chapter/147239348}.

\bibitem[Woolley(2016)]{woolley_automating_2016}
S.~C. Woolley.
\newblock Automating power: {Social} bot interference in global politics.
\newblock \emph{First Monday}, Mar. 2016.
\newblock ISSN 1396-0466.
\newblock \doi{10.5210/fm.v21i4.6161}.
\newblock URL \url{https://journals.uic.edu/ojs/index.php/fm/article/view/6161}.

\bibitem[Workshop et~al.(2023)Workshop, Scao, Fan, Akiki, Pavlick, Ilić, Hesslow, Castagné, Luccioni, Yvon, Gallé, Tow, Rush, Biderman, Webson, Ammanamanchi, Wang, Sagot, Muennighoff, del Moral, Ruwase, Bawden, Bekman, McMillan-Major, Beltagy, Nguyen, Saulnier, Tan, Suarez, Sanh, Laurençon, Jernite, Launay, Mitchell, Raffel, Gokaslan, Simhi, Soroa, Aji, Alfassy, Rogers, Nitzav, Xu, Mou, Emezue, Klamm, Leong, van Strien, Adelani, Radev, Ponferrada, Levkovizh, Kim, Natan, De~Toni, Dupont, Kruszewski, Pistilli, Elsahar, Benyamina, Tran, Yu, Abdulmumin, Johnson, Gonzalez-Dios, de~la Rosa, Chim, Dodge, Zhu, Chang, Frohberg, Tobing, Bhattacharjee, Almubarak, Chen, Lo, Von~Werra, Weber, Phan, allal, Tanguy, Dey, Muñoz, Masoud, Grandury, Šaško, Huang, Coavoux, Singh, Jiang, Vu, Jauhar, Ghaleb, Subramani, Kassner, Khamis, Nguyen, Espejel, de~Gibert, Villegas, Henderson, Colombo, Amuok, Lhoest, Harliman, Bommasani, López, Ribeiro, Osei, Pyysalo, Nagel, Bose, Muhammad, Sharma, Longpre, Nikpoor, Silberberg, Pai,
  Zink, Torrent, Schick, Thrush, Danchev, Nikoulina, Laippala, Lepercq, Prabhu, Alyafeai, Talat, Raja, Heinzerling, Si, Taşar, Salesky, Mielke, Lee, Sharma, Santilli, Chaffin, Stiegler, Datta, Szczechla, Chhablani, Wang, Pandey, Strobelt, Fries, Rozen, Gao, Sutawika, Bari, Al-shaibani, Manica, Nayak, Teehan, Albanie, Shen, Ben-David, Bach, Kim, Bers, Fevry, Neeraj, Thakker, Raunak, Tang, Yong, Sun, Brody, Uri, Tojarieh, Roberts, Chung, Tae, Phang, Press, Li, Narayanan, Bourfoune, Casper, Rasley, Ryabinin, Mishra, Zhang, Shoeybi, Peyrounette, Patry, Tazi, Sanseviero, von Platen, Cornette, Lavallée, Lacroix, Rajbhandari, Gandhi, Smith, Requena, Patil, Dettmers, Baruwa, Singh, Cheveleva, Ligozat, Subramonian, Névéol, Lovering, Garrette, Tunuguntla, Reiter, Taktasheva, Voloshina, Bogdanov, Winata, Schoelkopf, Kalo, Novikova, Forde, Clive, Kasai, Kawamura, Hazan, Carpuat, Clinciu, Kim, Cheng, Serikov, Antverg, van~der Wal, Zhang, Zhang, Gehrmann, Mirkin, Pais, Shavrina, Scialom, Yun, Limisiewicz, Rieser,
  Protasov, Mikhailov, Pruksachatkun, Belinkov, Bamberger, Kasner, Rueda, Pestana, Feizpour, Khan, Faranak, Santos, Hevia, Unldreaj, Aghagol, Abdollahi, Tammour, HajiHosseini, Behroozi, Ajibade, Saxena, Ferrandis, McDuff, Contractor, Lansky, David, Kiela, Nguyen, Tan, Baylor, Ozoani, Mirza, Ononiwu, Rezanejad, Jones, Bhattacharya, Solaiman, Sedenko, Nejadgholi, Passmore, Seltzer, Sanz, Dutra, Samagaio, Elbadri, Mieskes, Gerchick, Akinlolu, McKenna, Qiu, Ghauri, Burynok, Abrar, Rajani, Elkott, Fahmy, Samuel, An, Kromann, Hao, Alizadeh, Shubber, Wang, Roy, Viguier, Le, Oyebade, Le, Yang, Nguyen, Kashyap, Palasciano, Callahan, Shukla, Miranda-Escalada, Singh, Beilharz, Wang, Brito, Zhou, Jain, Xu, Fourrier, Periñán, Molano, Yu, Manjavacas, Barth, Fuhrimann, Altay, Bayrak, Burns, Vrabec, Bello, Dash, Kang, Giorgi, Golde, Posada, Sivaraman, Bulchandani, Liu, Shinzato, de~Bykhovetz, Takeuchi, Pàmies, Castillo, Nezhurina, Sänger, Samwald, Cullan, Weinberg, De~Wolf, Mihaljcic, Liu, Freidank, Kang, Seelam,
  Dahlberg, Broad, Muellner, Fung, Haller, Chandrasekhar, Eisenberg, Martin, Canalli, Su, Su, Cahyawijaya, Garda, Deshmukh, Mishra, Kiblawi, Ott, Sang-aroonsiri, Kumar, Schweter, Bharati, Laud, Gigant, Kainuma, Kusa, Labrak, Bajaj, Venkatraman, Xu, Xu, Xu, Tan, Xie, Ye, Bras, Belkada, and Wolf]{bigscience_workshop_bloom:_2023}
B.~Workshop, T.~L. Scao, A.~Fan, C.~Akiki, E.~Pavlick, S.~Ilić, D.~Hesslow, R.~Castagné, A.~S. Luccioni, F.~Yvon, M.~Gallé, J.~Tow, A.~M. Rush, S.~Biderman, A.~Webson, P.~S. Ammanamanchi, T.~Wang, B.~Sagot, N.~Muennighoff, A.~V. del Moral, O.~Ruwase, R.~Bawden, S.~Bekman, A.~McMillan-Major, I.~Beltagy, H.~Nguyen, L.~Saulnier, S.~Tan, P.~O. Suarez, V.~Sanh, H.~Laurençon, Y.~Jernite, J.~Launay, M.~Mitchell, C.~Raffel, A.~Gokaslan, A.~Simhi, A.~Soroa, A.~F. Aji, A.~Alfassy, A.~Rogers, A.~K. Nitzav, C.~Xu, C.~Mou, C.~Emezue, C.~Klamm, C.~Leong, D.~van Strien, D.~I. Adelani, D.~Radev, E.~G. Ponferrada, E.~Levkovizh, E.~Kim, E.~B. Natan, F.~De~Toni, G.~Dupont, G.~Kruszewski, G.~Pistilli, H.~Elsahar, H.~Benyamina, H.~Tran, I.~Yu, I.~Abdulmumin, I.~Johnson, I.~Gonzalez-Dios, J.~de~la Rosa, J.~Chim, J.~Dodge, J.~Zhu, J.~Chang, J.~Frohberg, J.~Tobing, J.~Bhattacharjee, K.~Almubarak, K.~Chen, K.~Lo, L.~Von~Werra, L.~Weber, L.~Phan, L.~B. allal, L.~Tanguy, M.~Dey, M.~R. Muñoz, M.~Masoud, M.~Grandury, M.~Šaško,
  M.~Huang, M.~Coavoux, M.~Singh, M.~T.-J. Jiang, M.~C. Vu, M.~A. Jauhar, M.~Ghaleb, N.~Subramani, N.~Kassner, N.~Khamis, O.~Nguyen, O.~Espejel, O.~de~Gibert, P.~Villegas, P.~Henderson, P.~Colombo, P.~Amuok, Q.~Lhoest, R.~Harliman, R.~Bommasani, R.~L. López, R.~Ribeiro, S.~Osei, S.~Pyysalo, S.~Nagel, S.~Bose, S.~H. Muhammad, S.~Sharma, S.~Longpre, S.~Nikpoor, S.~Silberberg, S.~Pai, S.~Zink, T.~T. Torrent, T.~Schick, T.~Thrush, V.~Danchev, V.~Nikoulina, V.~Laippala, V.~Lepercq, V.~Prabhu, Z.~Alyafeai, Z.~Talat, A.~Raja, B.~Heinzerling, C.~Si, D.~E. Taşar, E.~Salesky, S.~J. Mielke, W.~Y. Lee, A.~Sharma, A.~Santilli, A.~Chaffin, A.~Stiegler, D.~Datta, E.~Szczechla, G.~Chhablani, H.~Wang, H.~Pandey, H.~Strobelt, J.~A. Fries, J.~Rozen, L.~Gao, L.~Sutawika, M.~S. Bari, M.~S. Al-shaibani, M.~Manica, N.~Nayak, R.~Teehan, S.~Albanie, S.~Shen, S.~Ben-David, S.~H. Bach, T.~Kim, T.~Bers, T.~Fevry, T.~Neeraj, U.~Thakker, V.~Raunak, X.~Tang, Z.-X. Yong, Z.~Sun, S.~Brody, Y.~Uri, H.~Tojarieh, A.~Roberts, H.~W. Chung,
  J.~Tae, J.~Phang, O.~Press, C.~Li, D.~Narayanan, H.~Bourfoune, J.~Casper, J.~Rasley, M.~Ryabinin, M.~Mishra, M.~Zhang, M.~Shoeybi, M.~Peyrounette, N.~Patry, N.~Tazi, O.~Sanseviero, P.~von Platen, P.~Cornette, P.~F. Lavallée, R.~Lacroix, S.~Rajbhandari, S.~Gandhi, S.~Smith, S.~Requena, S.~Patil, T.~Dettmers, A.~Baruwa, A.~Singh, A.~Cheveleva, A.-L. Ligozat, A.~Subramonian, A.~Névéol, C.~Lovering, D.~Garrette, D.~Tunuguntla, E.~Reiter, E.~Taktasheva, E.~Voloshina, E.~Bogdanov, G.~I. Winata, H.~Schoelkopf, J.-C. Kalo, J.~Novikova, J.~Z. Forde, J.~Clive, J.~Kasai, K.~Kawamura, L.~Hazan, M.~Carpuat, M.~Clinciu, N.~Kim, N.~Cheng, O.~Serikov, O.~Antverg, O.~van~der Wal, R.~Zhang, R.~Zhang, S.~Gehrmann, S.~Mirkin, S.~Pais, T.~Shavrina, T.~Scialom, T.~Yun, T.~Limisiewicz, V.~Rieser, V.~Protasov, V.~Mikhailov, Y.~Pruksachatkun, Y.~Belinkov, Z.~Bamberger, Z.~Kasner, A.~Rueda, A.~Pestana, A.~Feizpour, A.~Khan, A.~Faranak, A.~Santos, A.~Hevia, A.~Unldreaj, A.~Aghagol, A.~Abdollahi, A.~Tammour, A.~HajiHosseini,
  B.~Behroozi, B.~Ajibade, B.~Saxena, C.~M. Ferrandis, D.~McDuff, D.~Contractor, D.~Lansky, D.~David, D.~Kiela, D.~A. Nguyen, E.~Tan, E.~Baylor, E.~Ozoani, F.~Mirza, F.~Ononiwu, H.~Rezanejad, H.~Jones, I.~Bhattacharya, I.~Solaiman, I.~Sedenko, I.~Nejadgholi, J.~Passmore, J.~Seltzer, J.~B. Sanz, L.~Dutra, M.~Samagaio, M.~Elbadri, M.~Mieskes, M.~Gerchick, M.~Akinlolu, M.~McKenna, M.~Qiu, M.~Ghauri, M.~Burynok, N.~Abrar, N.~Rajani, N.~Elkott, N.~Fahmy, O.~Samuel, R.~An, R.~Kromann, R.~Hao, S.~Alizadeh, S.~Shubber, S.~Wang, S.~Roy, S.~Viguier, T.~Le, T.~Oyebade, T.~Le, Y.~Yang, Z.~Nguyen, A.~R. Kashyap, A.~Palasciano, A.~Callahan, A.~Shukla, A.~Miranda-Escalada, A.~Singh, B.~Beilharz, B.~Wang, C.~Brito, C.~Zhou, C.~Jain, C.~Xu, C.~Fourrier, D.~L. Periñán, D.~Molano, D.~Yu, E.~Manjavacas, F.~Barth, F.~Fuhrimann, G.~Altay, G.~Bayrak, G.~Burns, H.~U. Vrabec, I.~Bello, I.~Dash, J.~Kang, J.~Giorgi, J.~Golde, J.~D. Posada, K.~R. Sivaraman, L.~Bulchandani, L.~Liu, L.~Shinzato, M.~H. de~Bykhovetz, M.~Takeuchi,
  M.~Pàmies, M.~A. Castillo, M.~Nezhurina, M.~Sänger, M.~Samwald, M.~Cullan, M.~Weinberg, M.~De~Wolf, M.~Mihaljcic, M.~Liu, M.~Freidank, M.~Kang, N.~Seelam, N.~Dahlberg, N.~M. Broad, N.~Muellner, P.~Fung, P.~Haller, R.~Chandrasekhar, R.~Eisenberg, R.~Martin, R.~Canalli, R.~Su, R.~Su, S.~Cahyawijaya, S.~Garda, S.~S. Deshmukh, S.~Mishra, S.~Kiblawi, S.~Ott, S.~Sang-aroonsiri, S.~Kumar, S.~Schweter, S.~Bharati, T.~Laud, T.~Gigant, T.~Kainuma, W.~Kusa, Y.~Labrak, Y.~S. Bajaj, Y.~Venkatraman, Y.~Xu, Y.~Xu, Y.~Xu, Z.~Tan, Z.~Xie, Z.~Ye, M.~Bras, Y.~Belkada, and T.~Wolf.
\newblock {BLOOM}: {A} {176B}-{Parameter} {Open}-{Access} {Multilingual} {Language} {Model}, June 2023.
\newblock URL \url{http://arxiv.org/abs/2211.05100}.
\newblock arXiv:2211.05100 [cs].

\bibitem[{World Health Organization}(2018)]{world_health_organization_cop24_2018}
{World Health Organization}.
\newblock {COP24} special report: health and climate change.
\newblock Technical report, World Health Organization, 2018.
\newblock URL \url{https://www.who.int/publications-detail-redirect/9789241514972}.

\bibitem[{World Inequality Database}()]{world_inequality_database_world_nodate}
{World Inequality Database}.
\newblock World {Inequality} {Database}.
\newblock URL \url{https://wid.world/}.

\bibitem[Wu et~al.(2022)Wu, Raghavendra, Gupta, Acun, Ardalani, Maeng, Chang, Behram, Huang, Bai, Gschwind, Gupta, Ott, Melnikov, Candido, Brooks, Chauhan, Lee, Lee, Akyildiz, Balandat, Spisak, Jain, Rabbat, and Hazelwood]{wu_sustainable_2022}
C.-J. Wu, R.~Raghavendra, U.~Gupta, B.~Acun, N.~Ardalani, K.~Maeng, G.~Chang, F.~A. Behram, J.~Huang, C.~Bai, M.~Gschwind, A.~Gupta, M.~Ott, A.~Melnikov, S.~Candido, D.~Brooks, G.~Chauhan, B.~Lee, H.-H.~S. Lee, B.~Akyildiz, M.~Balandat, J.~Spisak, R.~Jain, M.~Rabbat, and K.~Hazelwood.
\newblock Sustainable {AI}: {Environmental} {Implications}, {Challenges} and {Opportunities}, Jan. 2022.
\newblock URL \url{http://arxiv.org/abs/2111.00364}.
\newblock arXiv:2111.00364 [cs].

\bibitem[Wu et~al.(2023)Wu, Fei, Qu, Ji, and Chua]{wu_next-gpt:_2023}
S.~Wu, H.~Fei, L.~Qu, W.~Ji, and T.-S. Chua.
\newblock {NExT}-{GPT}: {Any}-to-{Any} {Multimodal} {LLM}, Sept. 2023.
\newblock URL \url{http://arxiv.org/abs/2309.05519}.
\newblock arXiv:2309.05519 [cs].

\bibitem[Wu et~al.(2021)Wu, Ribeiro, Heer, and Weld]{wu_polyjuice:_2021}
T.~Wu, M.~T. Ribeiro, J.~Heer, and D.~Weld.
\newblock Polyjuice: {Generating} {Counterfactuals} for {Explaining}, {Evaluating}, and {Improving} {Models}.
\newblock In C.~Zong, F.~Xia, W.~Li, and R.~Navigli, editors, \emph{Proceedings of the 59th {Annual} {Meeting} of the {Association} for {Computational} {Linguistics} and the 11th {International} {Joint} {Conference} on {Natural} {Language} {Processing} ({Volume} 1: {Long} {Papers})}, pages 6707--6723, Online, Aug. 2021. Association for Computational Linguistics.
\newblock \doi{10.18653/v1/2021.acl-long.523}.
\newblock URL \url{https://aclanthology.org/2021.acl-long.523}.

\bibitem[Wu et~al.(2020)Wu, Rough, Bleakley, Edwards, Cooney, Doyle, Clark, and Cowan]{wu_see_2020}
Y.~Wu, D.~Rough, A.~Bleakley, J.~Edwards, O.~Cooney, P.~R. Doyle, L.~Clark, and B.~R. Cowan.
\newblock See {What} {I}’m {Saying}? {Comparing} {Intelligent} {Personal} {Assistant} {Use} for {Native} and {Non}-{Native} {Language} {Speakers}.
\newblock In \emph{22nd {International} {Conference} on {Human}-{Computer} {Interaction} with {Mobile} {Devices} and {Services}}, pages 1--9, Oldenburg Germany, Oct. 2020. ACM.
\newblock ISBN 9781450375160.
\newblock \doi{10.1145/3379503.3403563}.
\newblock URL \url{https://dl.acm.org/doi/10.1145/3379503.3403563}.

\bibitem[Wynne(2004)]{wynne_perils_2004}
C.~D.~L. Wynne.
\newblock The perils of anthropomorphism.
\newblock \emph{Nature}, 428\penalty0 (6983):\penalty0 606--606, Apr. 2004.
\newblock ISSN 0028-0836, 1476-4687.
\newblock \doi{10.1038/428606a}.
\newblock URL \url{https://www.nature.com/articles/428606a}.

\bibitem[Xiang(2023)]{xiang_he_2023}
C.~Xiang.
\newblock '{He} {Would} {Still} {Be} {Here}': {Man} {Dies} by {Suicide} {After} {Talking} with {AI} {Chatbot}, {Widow} {Says}, Mar. 2023.
\newblock URL \url{https://www.vice.com/en/article/pkadgm/man-dies-by-suicide-after-talking-with-ai-chatbot-widow-says}.

\bibitem[Xie and Pentina(2022)]{xie_attachment_2022}
T.~Xie and I.~Pentina.
\newblock Attachment theory as a framework to understand relationships with social chatbots: a case study of replika.
\newblock 2022.

\bibitem[Xu et~al.(2021{\natexlab{a}})Xu, Pathak, Wallace, Gururangan, Sap, and Klein]{xu_detoxifying_2021}
A.~Xu, E.~Pathak, E.~Wallace, S.~Gururangan, M.~Sap, and D.~Klein.
\newblock Detoxifying {Language} {Models} {Risks} {Marginalizing} {Minority} {Voices}, Apr. 2021{\natexlab{a}}.
\newblock URL \url{http://arxiv.org/abs/2104.06390}.
\newblock arXiv:2104.06390 [cs].

\bibitem[Xu et~al.(2021{\natexlab{b}})Xu, Ju, Li, Boureau, Weston, and Dinan]{xu_bot-adversarial_2021}
J.~Xu, D.~Ju, M.~Li, Y.-L. Boureau, J.~Weston, and E.~Dinan.
\newblock Bot-{Adversarial} {Dialogue} for {Safe} {Conversational} {Agents}.
\newblock In K.~Toutanova, A.~Rumshisky, L.~Zettlemoyer, D.~Hakkani-Tur, I.~Beltagy, S.~Bethard, R.~Cotterell, T.~Chakraborty, and Y.~Zhou, editors, \emph{Proceedings of the 2021 {Conference} of the {North} {American} {Chapter} of the {Association} for {Computational} {Linguistics}: {Human} {Language} {Technologies}}, pages 2950--2968, Online, June 2021{\natexlab{b}}. Association for Computational Linguistics.
\newblock \doi{10.18653/v1/2021.naacl-main.235}.
\newblock URL \url{https://aclanthology.org/2021.naacl-main.235}.

\bibitem[Xu et~al.(2021{\natexlab{c}})Xu, Szlam, and Weston]{xu_beyond_2021}
J.~Xu, A.~Szlam, and J.~Weston.
\newblock Beyond {Goldfish} {Memory}: {Long}-{Term} {Open}-{Domain} {Conversation}, July 2021{\natexlab{c}}.
\newblock URL \url{https://arxiv.org/pdf/2107.07567.pdf}.
\newblock arXiv:2107.07567 [cs].

\bibitem[Xu et~al.(2023)Xu, Zhu, and Clifton]{xu_multimodal_2023}
P.~Xu, X.~Zhu, and D.~A. Clifton.
\newblock Multimodal {Learning} with {Transformers}: {A} {Survey}, May 2023.
\newblock URL \url{http://arxiv.org/abs/2206.06488}.
\newblock arXiv:2206.06488 [cs].

\bibitem[Yamin et~al.(2021)Yamin, Ullah, Ullah, and Katt]{yamin_weaponized_2021}
M.~M. Yamin, M.~Ullah, H.~Ullah, and B.~Katt.
\newblock Weaponized {AI} for cyber attacks.
\newblock \emph{Journal of Information Security and Applications}, 57:\penalty0 102722, Mar. 2021.
\newblock ISSN 2214-2126.
\newblock \doi{10.1016/j.jisa.2020.102722}.
\newblock URL \url{https://www.sciencedirect.com/science/article/pii/S2214212620308620}.

\bibitem[Yang et~al.(2019)Yang, Sun, and Narasimhan]{yang2019generalized}
R.~Yang, X.~Sun, and K.~Narasimhan.
\newblock A generalized algorithm for multi-objective reinforcement learning and policy adaptation.
\newblock \emph{Advances in {N}eural {I}nformation {P}rocessing {S}ystems}, 32, 2019.

\bibitem[Ye et~al.(2022)Ye, Maddi, Murakonda, Bindschaedler, and Shokri]{ye_enhanced_2022}
J.~Ye, A.~Maddi, S.~K. Murakonda, V.~Bindschaedler, and R.~Shokri.
\newblock Enhanced {Membership} {Inference} {Attacks} against {Machine} {Learning} {Models}, Sept. 2022.
\newblock URL \url{http://arxiv.org/abs/2111.09679}.
\newblock arXiv:2111.09679 [cs, stat].

\bibitem[Yeong~Tan and Singh(1995)]{yeong_tan_attitudes_1995}
D.~T. Yeong~Tan and R.~Singh.
\newblock Attitudes and {Attraction}: {A} {Developmental} {Study} of the {Similarity}-{Attraction} and {Dissimilarity}-{Repulsion} {Hypotheses}.
\newblock \emph{Personality and Social Psychology Bulletin}, 21\penalty0 (9):\penalty0 975--986, Sept. 1995.
\newblock ISSN 0146-1672, 1552-7433.
\newblock \doi{10.1177/0146167295219011}.
\newblock URL \url{http://journals.sagepub.com/doi/10.1177/0146167295219011}.

\bibitem[Ymous et~al.(2020)Ymous, Spiel, Keyes, Williams, Good, Hornecker, and Bennett]{ymous_i_2020}
A.~Ymous, K.~Spiel, O.~Keyes, R.~M. Williams, J.~Good, E.~Hornecker, and C.~L. Bennett.
\newblock "{I} am just terrified of my future" {Epistemic} {Violence} in {Disability} {Related} {Technology} {Research}.
\newblock In \emph{Extended {Abstracts} of the 2020 {CHI} {Conference} on {Human} {Factors} in {Computing} {Systems}}, pages 1--16, Honolulu HI USA, Apr. 2020. ACM.
\newblock ISBN 9781450368193.
\newblock \doi{10.1145/3334480.3381828}.
\newblock URL \url{https://dl.acm.org/doi/10.1145/3334480.3381828}.

\bibitem[Yogeeswaran et~al.(2016)Yogeeswaran, Z{\l}otowski, Livingstone, Bartneck, Sumioka, and Ishiguro]{yogeeswaran_interactive_2016}
K.~Yogeeswaran, J.~Z{\l}otowski, M.~Livingstone, C.~Bartneck, H.~Sumioka, and H.~Ishiguro.
\newblock The interactive effects of robot anthropomorphism and robot ability on perceived threat and support for robotics research.
\newblock \emph{Journal of Human-Robot Interaction}, 5\penalty0 (2):\penalty0 29--47, 2016.

\bibitem[Yue et~al.(2023)Yue, Inan, Li, Kumar, McAnallen, Shajari, Sun, Levitan, and Sim]{yue_synthetic_2023}
X.~Yue, H.~A. Inan, X.~Li, G.~Kumar, J.~McAnallen, H.~Shajari, H.~Sun, D.~Levitan, and R.~Sim.
\newblock Synthetic {Text} {Generation} with {Differential} {Privacy}: {A} {Simple} and {Practical} {Recipe}, July 2023.
\newblock URL \url{http://arxiv.org/abs/2210.14348}.
\newblock arXiv:2210.14348 [cs].

\bibitem[Zahavy et~al.(2021)Zahavy, O'Donoghue, Barreto, Mnih, Flennerhag, and Singh]{zahavy2021discovering}
T.~Zahavy, B.~O'Donoghue, A.~Barreto, V.~Mnih, S.~Flennerhag, and S.~Singh.
\newblock Discovering diverse nearly optimal policies with successor features.
\newblock \emph{arXiv preprint arXiv:2106.00669}, 2021.

\bibitem[Zahavy et~al.(2022)Zahavy, Schroecker, Behbahani, Baumli, Flennerhag, Hou, and Singh]{zahavy2022discovering}
T.~Zahavy, Y.~Schroecker, F.~Behbahani, K.~Baumli, S.~Flennerhag, S.~Hou, and S.~Singh.
\newblock Discovering policies with domino: Diversity optimization maintaining near optimality.
\newblock \emph{arXiv preprint arXiv:2205.13521}, 2022.

\bibitem[Zahavy et~al.(2023)Zahavy, Veeriah, Hou, Waugh, Lai, Leurent, Tomasev, Schut, Hassabis, and Singh]{zahavy2023diversifying}
T.~Zahavy, V.~Veeriah, S.~Hou, K.~Waugh, M.~Lai, E.~Leurent, N.~Tomasev, L.~Schut, D.~Hassabis, and S.~Singh.
\newblock Diversifying {AI}: Towards creative chess with alphazero, 2023.

\bibitem[Zajko(2022)]{zajko_artificial_2022}
M.~Zajko.
\newblock Artificial intelligence, algorithms, and social inequality: {Sociological} contributions to contemporary debates.
\newblock \emph{Sociology Compass}, 16\penalty0 (3):\penalty0 e12962, Mar. 2022.
\newblock ISSN 1751-9020, 1751-9020.
\newblock \doi{10.1111/soc4.12962}.
\newblock URL \url{https://compass.onlinelibrary.wiley.com/doi/10.1111/soc4.12962}.

\bibitem[Zalnieriute and Cutts(2022)]{zalnieriute_how_2022}
M.~Zalnieriute and T.~Cutts.
\newblock How {AI} and {New} {Technologies} {Reinforce} {Systemic} {Racism}.
\newblock Technical report, United Nations Human Rights Council, Oct. 2022.
\newblock URL \url{https://www.ohchr.org/sites/default/files/documents/hrbodies/hrcouncil/advisorycommittee/study-advancement-racial-justice/2022-10-26/HRC-Adv-comm-Racial-Justice-zalnieriute-cutts.pdf}.

\bibitem[Zeinert et~al.(2021)Zeinert, Inie, and Derczynski]{zeinert_annotating_2021}
P.~Zeinert, N.~Inie, and L.~Derczynski.
\newblock Annotating {Online} {Misogyny}.
\newblock In C.~Zong, F.~Xia, W.~Li, and R.~Navigli, editors, \emph{Proceedings of the 59th {Annual} {Meeting} of the {Association} for {Computational} {Linguistics} and the 11th {International} {Joint} {Conference} on {Natural} {Language} {Processing} ({Volume} 1: {Long} {Papers})}, pages 3181--3197, Online, Aug. 2021. Association for Computational Linguistics.
\newblock \doi{10.18653/v1/2021.acl-long.247}.
\newblock URL \url{https://aclanthology.org/2021.acl-long.247v2.pdf}.

\bibitem[Zevenbergen(2020)]{zevenbergen_internet_2020}
B.~Zevenbergen.
\newblock \emph{Internet users as vulnerable and at-risk human subjects: reviewing research ethics {Law} for technical internet research}.
\newblock http://purl.org/dc/dcmitype/{Text}, University of Oxford, 2020.
\newblock URL \url{https://ora.ox.ac.uk/objects/uuid:5363f3f8-6c13-4e56-b065-2980bdcce46b}.

\bibitem[Zhang et~al.(2023)Zhang, Xie, Hou, Zhao, Lin, and Wen]{zhang2023recommendation}
J.~Zhang, R.~Xie, Y.~Hou, W.~X. Zhao, L.~Lin, and J.-R. Wen.
\newblock Recommendation as instruction following: A large language model empowered recommendation approach, 2023.

\bibitem[Zhang et~al.(2021)Zhang, Lu, and Jin]{zhang_artificial_2021}
Q.~Zhang, J.~Lu, and Y.~Jin.
\newblock Artificial intelligence in recommender systems.
\newblock \emph{Complex \& Intelligent Systems}, 7\penalty0 (1):\penalty0 439--457, Feb. 2021.
\newblock ISSN 2198-6053.
\newblock \doi{10.1007/s40747-020-00212-w}.
\newblock URL \url{https://doi.org/10.1007/s40747-020-00212-w}.

\bibitem[Zhu and Luo(2022)]{zhu_generative_2022}
Q.~Zhu and J.~Luo.
\newblock Generative {Design} {Ideation}: {A} {Natural} {Language} {Generation} {Approach}, Mar. 2022.
\newblock URL \url{http://arxiv.org/abs/2204.09658}.
\newblock arXiv:2204.09658 [cs].

\bibitem[Zhuravskaya et~al.(2020)Zhuravskaya, Petrova, and Enikolopov]{zhuravskaya2020political}
E.~Zhuravskaya, M.~Petrova, and R.~Enikolopov.
\newblock Political effects of the internet and social media.
\newblock \emph{Annual review of economics}, 12:\penalty0 415--438, 2020.

\bibitem[Ziegler et~al.(2020)Ziegler, Stiennon, Wu, Brown, Radford, Amodei, Christiano, and Irving]{ziegler_fine-tuning_2020}
D.~M. Ziegler, N.~Stiennon, J.~Wu, T.~B. Brown, A.~Radford, D.~Amodei, P.~Christiano, and G.~Irving.
\newblock Fine-{Tuning} {Language} {Models} from {Human} {Preferences}, Jan. 2020.
\newblock URL \url{http://arxiv.org/abs/1909.08593}.
\newblock arXiv:1909.08593 [cs, stat].

\bibitem[Zimmermann et~al.(2022)Zimmermann, Vredenburgh, and Lazar]{zimmermann_political_2022}
A.~Zimmermann, K.~Vredenburgh, and S.~Lazar.
\newblock The {Political} {Philosophy} of {Data} and {AI}.
\newblock \emph{Canadian Journal of Philosophy}, 52\penalty0 (1):\penalty0 1--5, Jan. 2022.
\newblock ISSN 0045-5091, 1911-0820.
\newblock \doi{10.1017/can.2022.28}.
\newblock URL \url{https://www.cambridge.org/core/product/identifier/S0045509122000285/type/journal_article}.

\bibitem[Zuboff(2017)]{zuboff_age_2017}
S.~Zuboff.
\newblock \emph{The {Age} of {Surveillance} {Capitalism}}.
\newblock Hachette, June 2017.
\newblock ISBN 9781610395694.
\newblock URL \url{https://www.hachettebookgroup.com/titles/shoshana-zuboff/the-age-of-surveillance-capitalism/9781610395694/?lens=publicaffairs}.

\bibitem[Zwolinski et~al.(2022)Zwolinski, Ferguson, and Wertheimer]{zwolinski_exploitation_2022}
M.~Zwolinski, B.~Ferguson, and A.~Wertheimer.
\newblock Exploitation.
\newblock In E.~N. Zalta and U.~Nodelman, editors, \emph{The {Stanford} {Encyclopedia} of {Philosophy}}. Metaphysics Research Lab, Stanford University, winter 2022 edition, 2022.
\newblock URL \url{https://plato.stanford.edu/archives/win2022/entries/exploitation/}.

\end{thebibliography}

\end{document}